# Test the accuracy of the Dubna phenomenological model of the masses of atomic nuclei

**Strachimir Cht. Mavrodiev**

**Institute for Nuclear Research and Nuclear Energy, Sofia, BAS, Bulgaria**

## Abstract

In this paper is presented the reliability test the numerical generalization of Bethe-Weizsacker mass formula which describes the values of measured **2654** nuclei masses in **AME2012 nuclear database**: **https://www-nds.iaea.org/amdc/**, with accuracy in the interval (-1.90,1.80) MeV, for the nulear with mass number A=2(1,1) to 294(117,177).

In the analyzed generalization of the Bethe-Weizsacker formula the little correction of proton and neutron magic numbers and boundaries of their influence was derived as a solution of nonlinear inverce problem (Ivanenko shell model!). There was confirmed the existence of ten proton (2,8,14,20,28,50,82,96,108,126) and eleven neutron (2,8,14,20,28,50,82,126,142,152,184) magic numbers.

The test was performed analyzing the new data from **Nuclide ground states: https://www-nds.iaea.org/relnsd/NdsEnsdf/QueryForm.html,** containing **3182** data for A=2(1,1) to 294(118,176) and confirmed the above values of proton and neutron magic numbers. The accuracy of masses description is in the interval (-2.80,1.80) MeV. Only fourteen nuclear masses have bigger than **1.90 MeV** residual.

## Introduction

The history and development of Bethe-Weizsacker mass formula was presented in details in paper [1], where on the basis of inverse problem was discover the explicit form of binding energy $E_B(A,Z,a)$ , which describes the nuclei masses values (2564) from most recent evaluation database AME2012 (December 2012 - [2, 3]) with accuracy better or equal to 2.6 MeV . The masses extrapolated from systematics and marked with the symbol # in the error column was not taken into account.

The purpose of the paper [4] was to obtain the improved, compared with [1] explicit form of of BW formulae as function of A and Z.

These aim have been reached using Alexandrov dynamic autoregularization method (FORTRAN code REGN-Dubna [5-17]) for solving the overdetermined algebraic systems of equations which is constructive development of Tikhonov regularization method [18-20]. One has to note that the use of procedure LCH permits to discover the explicit form of unknown function [21-26].

The aim of this paper is to perform the accuracy test of the binding energy, the nuclear mass, the atomic mass and the excess energy models from [4] taking into account the 3182 data published in [37].

In Sec.1 the accuracy test is presented with figures of residuals Res=Expt-Th for Binding energies, Nuclear and Atomic masses and Excess energies for A(Z,N) from A=2(1,1) to 294(118,176). The calculated with new 3182 data two proton and neutron drip- lines and their asymptotics are presented in Sec.2. The Fortran source code of the generalized BW mass formula is given in Appendix A. The description of the experimental binding energy, nuclear and atomic mass mass excess values from new database **Nuclide groung state: https://www-nds.iaea.org/relnsd/NdsEnsdf/QueryForm.html** is given in Appendix B. In Appendix C are presented the predicted values of the binding energy and nuclear mass for some supper havy nuclei (see [38] and Proceeding of Exon 2016, Kazan [39]) . In Appendic D are presented the values total and kinetic decay energy for some superhavy nuclei.

## 1. The reliability test

The explicit form of the Bethe-Weizsecker formula for Binding energy as function of mass number A, number of protons presented in paper [4] was

$$E_B(A,Z,a) = Vol(A,Z,a) - Sur(A,Z,a)\frac{1}{A^{P1(A,Z,a)}} - Cha(A,Z,a)\frac{Z(Z-1)}{A^{P2(A,Z,a)}} - Sym(A,Z,a)\frac{(N-Z)^2}{A^{P3(A,Z,a)}} + Wig(A,Z,a)\frac{\delta(A,Z)}{A^{P4(A,Z,a)}} + K_{MN}(A,Z,a), \quad (1)$$

where the function $K_{MN}(A,Z,a)$ depends form the protom and neutron magic numbers and from the frontier between their influence.

The next two fugures present the bihaviour of the function $K_{MN}(A,Z,a)$ from proton number Z and neutron N.

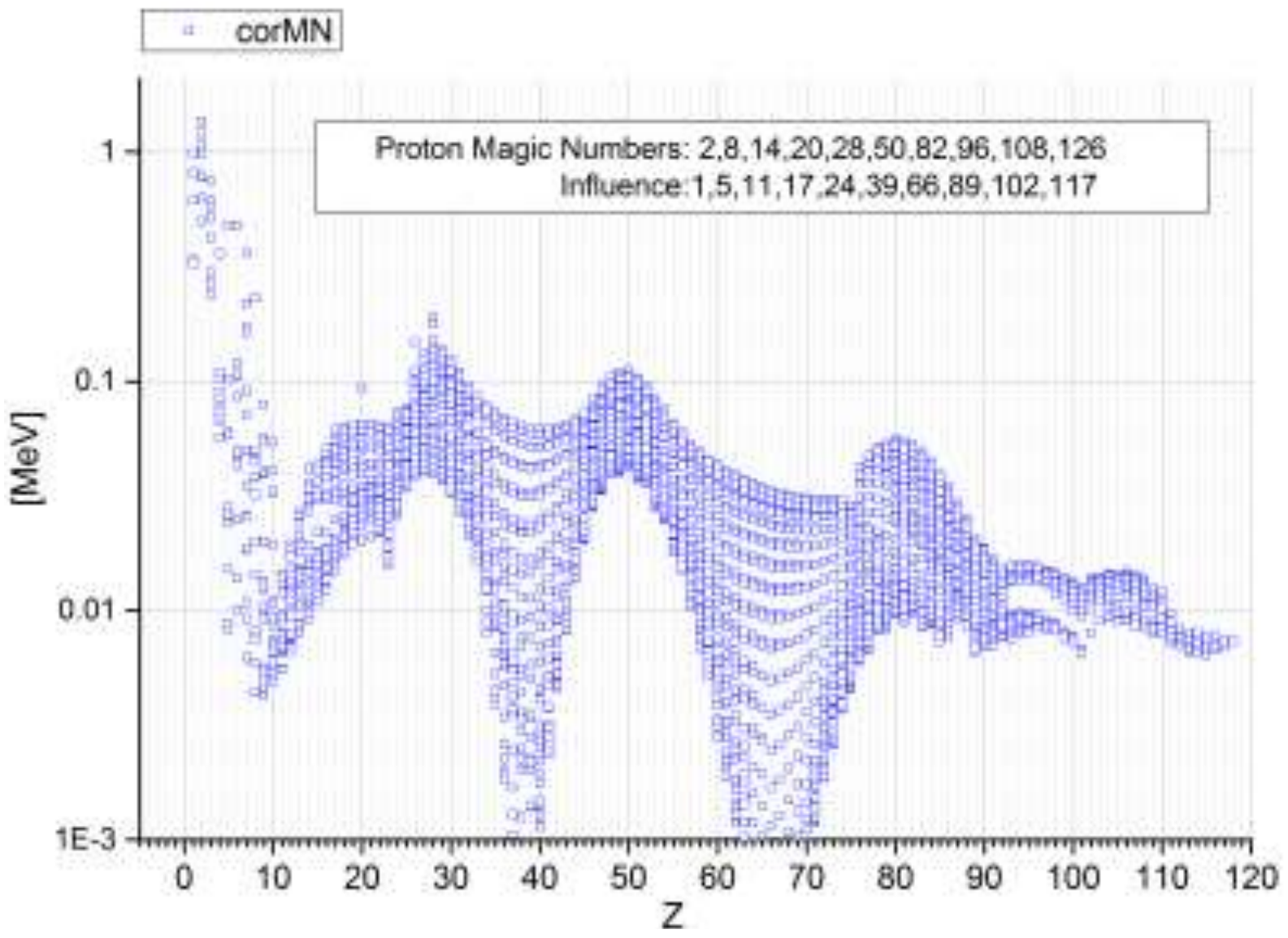

Fig.1. The bihaviour of the function $K_{MN}(A,Z,a)$ from the proton magic numbers Z and the the frontier between their influence.

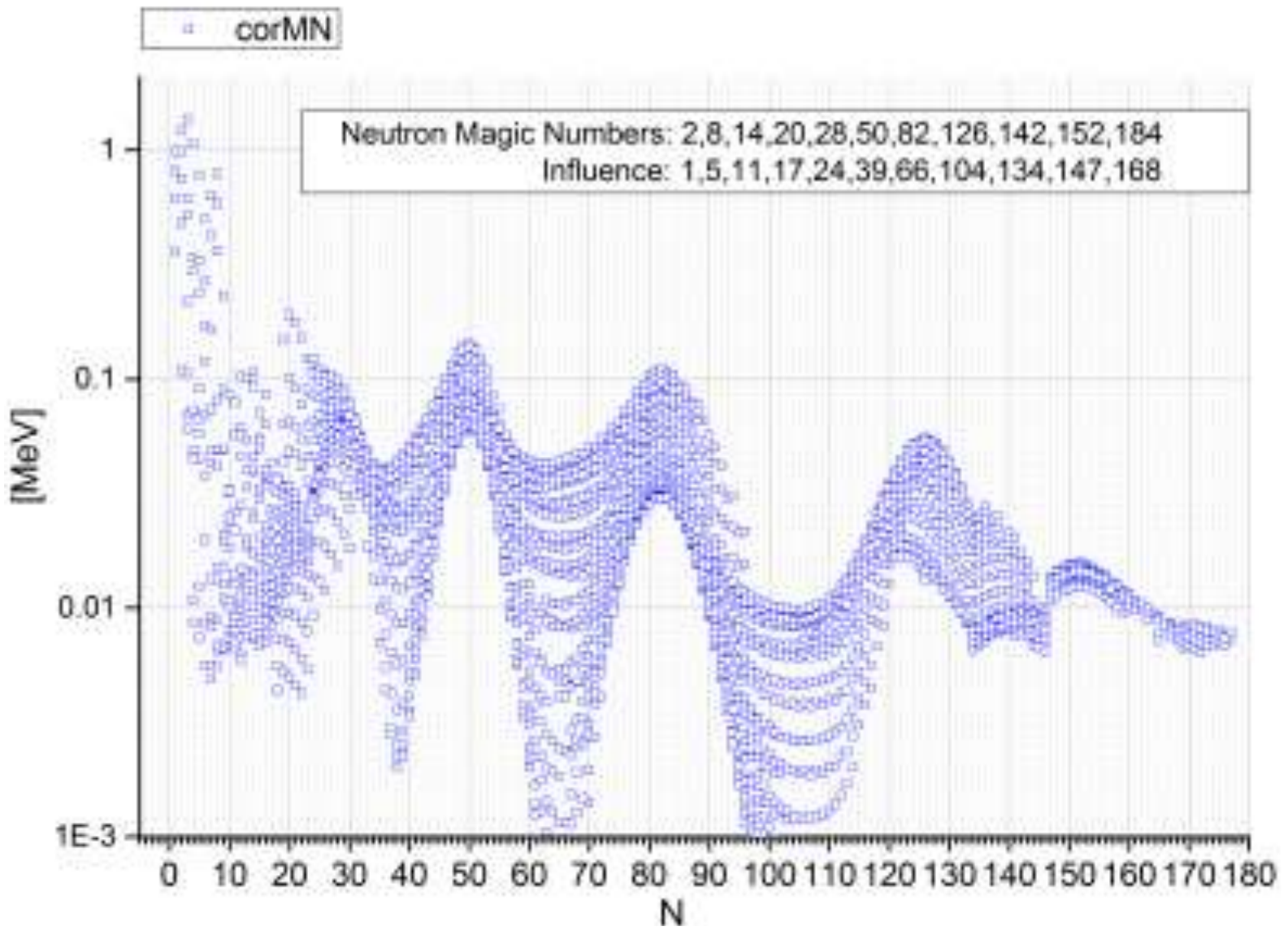

Fig.2. The bihaviour of the function $K_{MN}(A,Z,a)$ from the neutron numbers N and the the frontier between their influence.

Under definisions

$$\bar{\epsilon_{\mathrm{r}}} = \frac{1}{N}\sum_{i=1}^{N} \frac{Expt\,(A_i,Z_i) - Th\,(A_i,Z_i,a)}{Expt\,(A_i,Z_i)}\,, \qquad (2)$$

$$\chi^2 = \sum_{k=1}^{M} \left(\frac{Expt\,(A_k,Z_k) - Th(A_k,Z_k,a)}{\sigma(A_k,Z_k)}\right)^2\,, \qquad (3)$$

where

$$\sigma(A_k,Z_k) = C \times \sigma_{Stat}\,(A_k,Z_k) + Percent\ \times Expt(A_k,Z_k) \qquad (4)$$

and

$$\chi_n = \sqrt{\frac{\chi^2}{M-N}}\,, \qquad (5)$$

where $\boldsymbol{M-N}$ is the number of degrees of freedom, the next Table 1 present the comparison between the 2564 and 3182 cases for $B_e$, $M_{Nucl}$, $M_{At}$ and $M_{Exc}$.

**Table1**

| | | Percent | $\bar{\epsilon_{\mathrm{r}}}$ | χ2 | χn |
|---|---|---|---|---|---|
| 2564 | $B_e$ | 1.85E-03 | 5.19E-02 | 2512 | 1.042 |
| | $M_{Nucl}$ | 1.20E-05 | -1.26E-05 | 2418 | 1.022 |
| | $M_{At}$ | 1.20E-05 | -8.11E-05 | 2424 | 1.023 |
| | $M_{Exc}$ | 1.70E-01 | -2.38E+01 | 2252 | 0.986 |
| 3182 | $B_e$ | 2.22E-03 | 7.44E+01 | 2938 | 1.001 |
| | $M_{Nucl}$ | 5.60E-06 | -4.21E-03 | 3102 | 1.028 |
| | $M_{At}$ | 5.60E-06 | -4.21E-03 | 3101 | 1.028 |
| | $M_{Exc}$ | 2.40E-02 | -8.69E+01 | 2950 | 1.003 |

In the next Table 2 are presented the fourteen isotopes bigger than +/- **1.90 MeV** residuals.

**Table 2**

| El | A | Z | N | N-Z | NuclMassExpt | NuclMassTh | ResNuclMass |
|---|---|---|---|---|---|---|---|
| Sn | 132 | 50 | 82 | 32 | 122854.954 | 122857.707 | -2.8 |
| C | 13 | 6 | 7 | 1 | 12109.4808 | 12112.1607 | -2.7 |
| Sn | 100 | 50 | 50 | 0 | 93066.4053 | 93068.7211 | -2.3 |
| Ca | 52 | 20 | 32 | 12 | 48393.1896 | 48395.3896 | -2.2 |
| Sn | 133 | 50 | 83 | 33 | 123792.118 | 123794.315 | -2.2 |
| Zr | 80 | 40 | 40 | 0 | 74443.4699 | 74445.6129 | -2.1 |
| Zn | 54 | 30 | 24 | -6 | 50277.8846 | 50280.0193 | -2.1 |
| O | 15 | 8 | 7 | -1 | 13971.1765 | 13973.3092 | -2.1 |
| Ca | 51 | 20 | 31 | 11 | 47459.62 | 47461.746 | -2.1 |
| Pb | 208 | 82 | 126 | 44 | 193686.546 | 193688.5 | -2 |
| Sb | 133 | 51 | 82 | 31 | 123783.55 | 123785.593 | -2 |
| Kr | 100 | 36 | 64 | 28 | 93095.882 | 93093.9314 | 2 |
| BR | 98 | 35 | 63 | 28 | 91240.0141 | 91237.958 | 2.1 |
| KR | 101 | 36 | 65 | 29 | 94033.3004 | 94031.1842 | 2.1 |

In the next figures are presented the Gaus fits of distribution of residuals Res = Expt- Th for the Binding energy and the Nuclear mass for the case 3182..

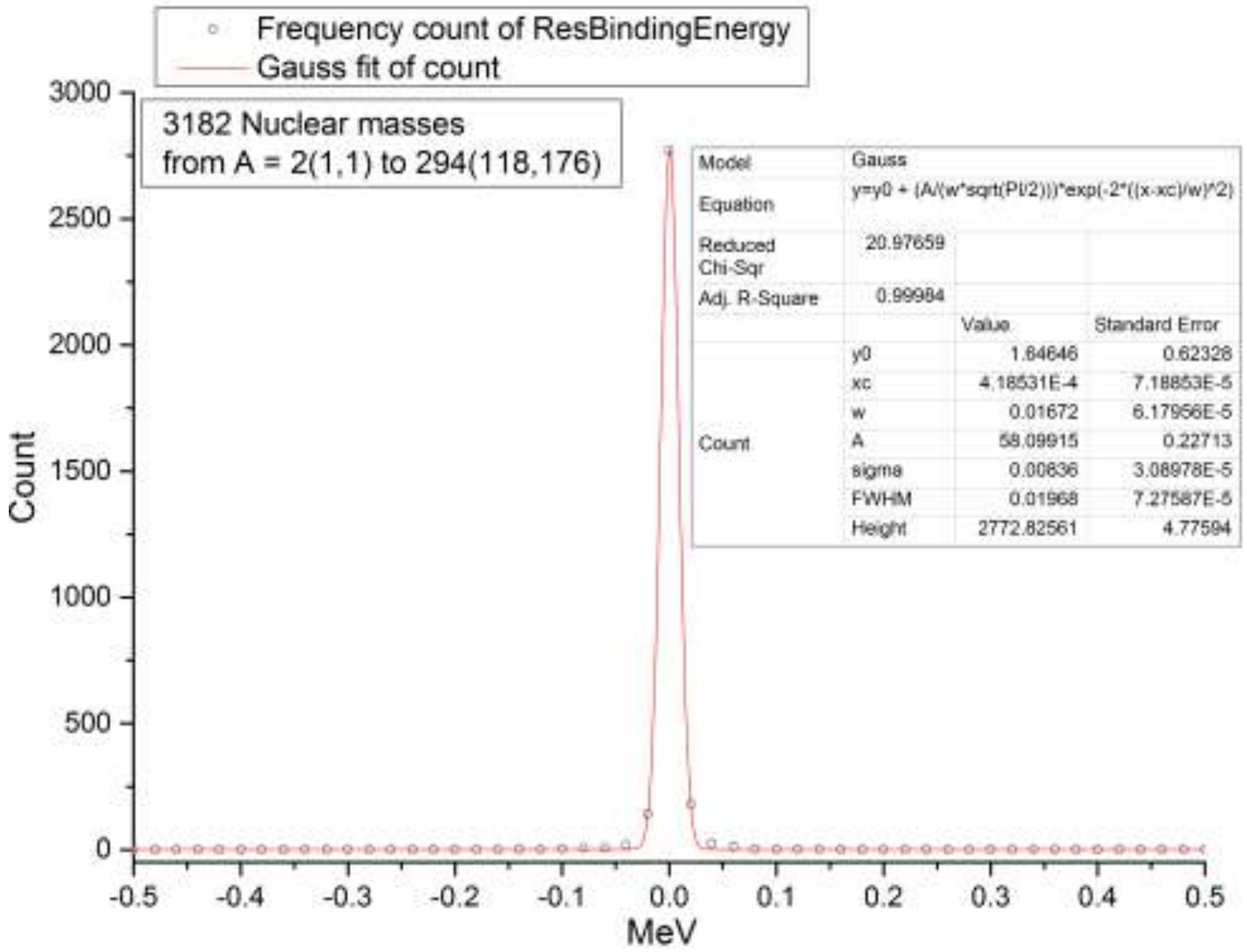

Fig. 3. The Gaus fits of distribution of residuals Res = $Be^{Expt}$- $Be^{Th}$ .

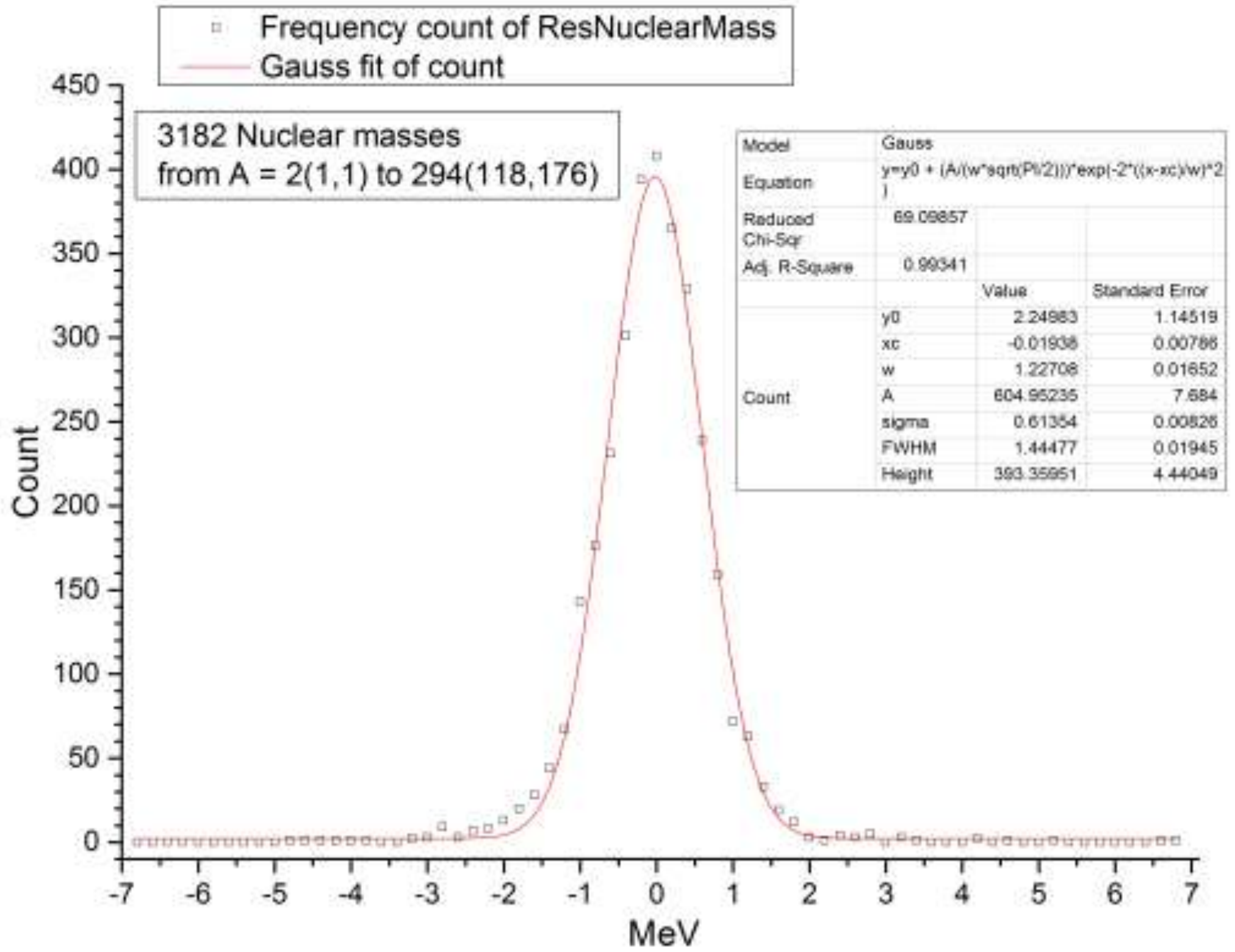

Fig. 3. The Gaus fits of distribution of residuals Res = $NuclearMass^{Expt}$- $NuclearMass^{Th}$ .

## 2. Two proton and neutron drip- lines and their asymptotics

The nuclear drip lines are the boundary delimiting the zone of Z, N in which atomic nuclei lose stability due to the transmutation of neutrons (down one) as well as because of Coulomb repulsion of protons (up). To find where these drip lines are on the nuclear landscape we need to know the values of Z and N, where the separation energy is changed it sign. The coordinates, where separation energies change the sign can be calculated by using the explicit form for binding energy $E_B(A, Z, a)$ (Eq.(4)).

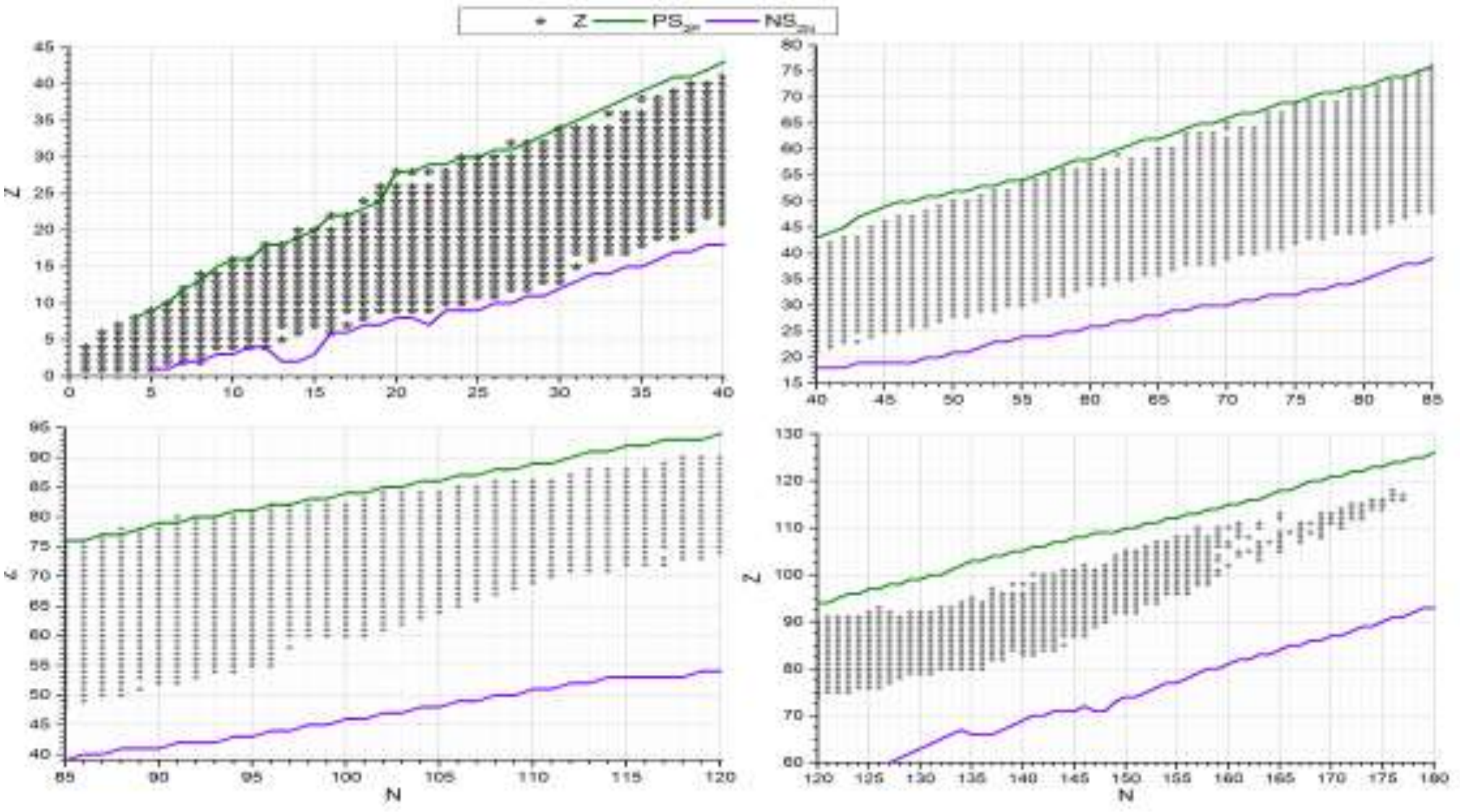

**Fig.4 The behavior of calculated proton and neutron drip- lines compared with Z,N coordinates of 3182 measured nucleus.**

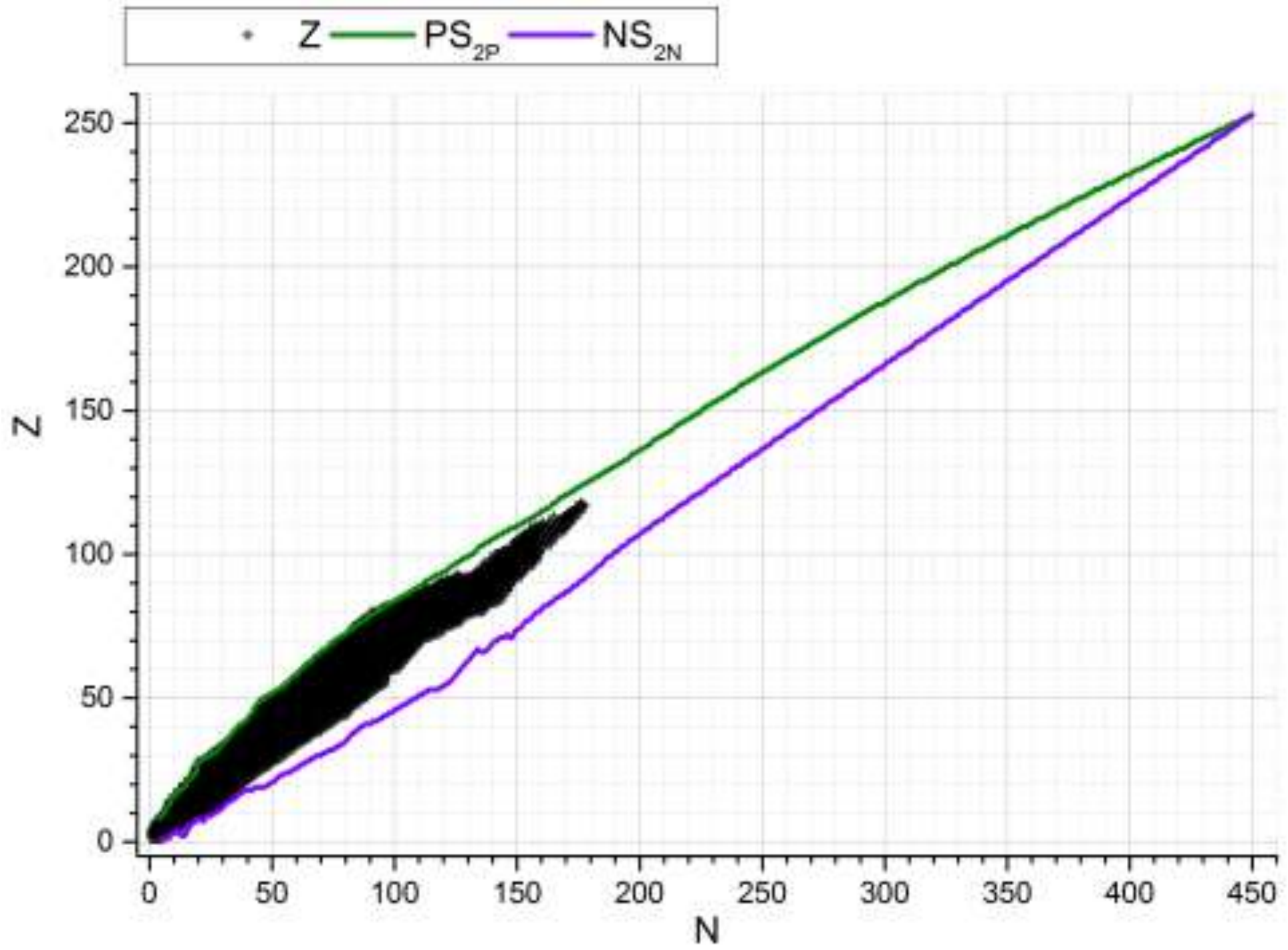

**Fig.5 The asimptotic behavior of calculated proton and neutron drip-lines.**

In Table 3 are presented the isotopes which are outside the drip- lines.

**Table 3**

| El | A | Z | N | N-Z | NuclMassExpt | NuclMassTh | ResNuclMass |
|---|---|---|---|---|---|---|---|
| Si | 23 | 14 | 9 | -5 | 21440.8987 | 21441.0412 | -0.14 |
| Ca | 34 | 20 | 14 | -6 | 31674.4108 | 31673.9168 | 0.49 |
| Mn | 44 | 25 | 19 | -6 | 40979.5922 | 40980.4152 | -0.82 |
| Fe | 45 | 26 | 19 | -7 | 41917.3432 | 41916.3906 | 0.95 |
| Ge | 59 | 32 | 27 | -5 | 54925.4299 | 54927.0321 | -1.6 |
| Pt | 166 | 78 | 88 | 10 | 154582.863 | 154583.024 | -0.16 |
| HG | 171 | 80 | 91 | 11 | 159247.358 | 159247.547 | -0.19 |

## Conclusion

The challenge of low energy nuclear physics to describe the dependence of the binding energy, nuclear and atomic mass, mass excess as functions of the number of protons and neutrons is presented.

This result was established by using the ecperimental data from AME2012 [2,3] and [37] databases, the inverse problem method for discovering the explicit form of unknown theoretical function (model) and the values, based on the REGN (L. Alekasandrov-Regularized Gauss-Newton iteration method) [5 - 17] for solving the overedeterminednonlinear system of equations.One have to note that the LCH-weighting procedure [21-26] of the REGN program permits to choose the better function out of two functions with the same $\chi^2$.

The essential advantage of the Alexandrov method [5 - 17] from other similar methods is extremely effective ideology regularization of inverse problem solution, which on each iteration step controls not only the actual decision, but, very importantly, uncertainty of the solution. At the same time, the transition from the mathematical theory of the autoregularizated iterative processes, which is based on meaningful theorems of convergence L. Aleksandrov [6] to Fortran codes (REGN-Dubna [8], FXY-Sofia-Dubna [17]) is very complicated, but technically clear work.

The first interesting application of the proposed explicit form of Dubna Nuclear Mass Phenomenological Model seems to be the calculation of two proton and neutron drip-lines and their intercept in Z=250, N=450, A=700.

Proposed formula can be used for calculation of not known nuclear mass, the total and kinetc energy of proton, alpha, cluster decays and spontaneously fissions- see Appendix D.

The used in this paper approach for generalization of BW mass formulae can be applied for the actualization the half- life models (see for example [37] which describes the increezing volume of experimental data in NuDat database. Such actualization can be used for reasurching the problems in super havy nuclei stability islands.

**ACKNOWLEDGMENTS: The creation of a phenomenological model of the masses of atomic nuclei began in at JINR, Dubna about 10 years in satrudnichestvos Alexei Sissakyan and Lubomir Alexandrov. The author is thankful to Svetla Drenska and Maksim Deliyergiyev for fruitfull discutions.**

## REFERENCES

------------------------------------------

## APENDIGS A

```
      FUNCTION aNuclMass3182(Proton,aNeutron,BiEnTh,AtMassTh,aMassExcTh)
     IMPLICIT DOUBLE PRECISION(A-H,O-Z)
     common/nhelp/lexpt,iSP,nPow,nBWp,MnZ,MnN,N0,N1,Nqn
    common/Structure/c1,c2,c3,c4,c5,pow1,pow2,pow3,pow4
     common/bwigner/BWZ,BWN,CorMn,BetheWeizsacker
     common/help1/Vol,Sur,Cha,Sym,Wig,CorectionMN
     common/variables/AA,a1,a12,a13,a2,a22,a23,a3,a32,a33,a4,a42,a43,a5,a6,AAA,AAZ,AAN
     DIMENSION A(249)
A(  1)=  0.24201750342045E+01; A(  2)=  0.28646806080426E+01; A(  3)= -0.26421065393223E+02; A(  4)=  0.25017696166052E+01; A(  5)= -0.64834656883662E+01;
A(  6)= -0.47586288181481E+00; A(  7)= -0.11514338250493E+01; A(  8)=  0.63286005404393E+00; A(  9)= -0.14542984204421E+02; A( 10)=  0.25253238718416E+02;
A( 11)= -0.95553913563559E+01; A( 12)= -0.50555983804038E+02; A( 13)= -0.59098204574159E+02; A( 14)=  0.58196499857434E+01; A( 15)= -0.18322640945413E+02;
A( 16)=  0.20093158894510E+02; A( 17)= -0.12162174192988E+02; A( 18)=  0.22592572324464E+01; A( 19)= -0.29990769905160E+02; A( 20)=  0.71924434658796E+02;
A( 21)=  0.12188827880434E+02; A( 22)=  0.12048418025983E+03; A( 23)=  0.11245260679066E+02; A( 24)= -0.17940780430250E+02; A( 25)=  0.55034645508024E+02;
A( 26)=  0.10807181644887E+01; A( 27)= -0.50049737786425E+02; A( 28)=  0.20859053704423E+02; A( 29)=  0.25796077019642E+02; A( 30)= -0.29477161365144E+02;
A( 31)= -0.53544564621962E+02; A( 32)= -0.21309875243557E+03; A( 33)= -0.12283884969423E+02; A( 34)= -0.18302697764346E+02; A( 35)=  0.16394728505785E+03;
A( 36)=  0.13848914775297E+03; A( 37)= -0.11778352069692E+03; A( 38)= -0.36192422287228E+01; A( 39)=  0.83484263279268E+02; A( 40)=  0.46782733124192E+01;
A( 41)= -0.21362360324008E+02; A( 42)= -0.43866393421208E+02; A( 43)= -0.45627518307537E+02; A( 44)=  0.23435455649643E+02; A( 45)= -0.10509798267640E+02;
A( 46)=  0.78285690951296E+01; A( 47)= -0.19129550543996E+02; A( 48)=  0.23925093409128E+02; A( 49)= -0.13589036683182E+02; A( 50)=  0.65304925281933E+02;
A( 51)=  0.13167285775132E+02; A( 52)=  0.10796118770130E+03; A( 53)=  0.10250883227516E+02; A( 54)= -0.12690748983778E+02; A( 55)=  0.20180550669161E+03;
A( 56)= -0.26423172386585E+02; A( 57)= -0.14798969114769E+03; A( 58)= -0.74115664971996E+03; A( 59)=  0.12442265119112E+03; A( 60)= -0.13700715138594E+03;
A( 61)= -0.24427407684824E+03; A( 62)=  0.39386676917301E+03; A( 63)= -0.42039536048046E+02; A( 64)= -0.13377093360179E+02; A( 65)=  0.35580648440078E+03;
A( 66)= -0.81819280704325E+02; A( 67)=  0.56497439965232E+03; A( 68)=  0.23668382171702E+02; A( 69)=  0.11601552826434E+03; A( 70)=  0.64904154383365E+00;
A( 71)=  0.66326391096738E+00; A( 72)=  0.51299813209452E+00; A( 73)=  0.64500036968730E+00; A( 74)=  0.57299407496029E+00; A( 75)=  0.58680298487914E+00;
A( 76)=  0.50457043872155E+00; A( 77)=  0.56757479681557E+00; A( 78)=  0.53653341836801E+00; A( 79)=  0.54682633694122E+00; A( 80)=  0.50152355519688E+00;
A( 81)=  0.53096423623068E+00; A( 82)=  0.78255929723732E+00; A( 83)=  0.85035565183992E+00; A( 84)=  0.81294779761666E+00; A( 85)= -0.32156142830127E+02;
A( 86)= -0.66504100441017E+02; A( 87)=  0.24882909893125E+01; A( 88)= -0.16979893685040E+01; A( 89)= -0.39768098001088E+02; A( 90)=  0.38403779308042E+02;
A( 91)= -0.14712663889684E+02; A( 92)=  0.25507449335736E+01; A( 93)=  0.41989475562273E+02; A( 94)= -0.21380889626078E+02; A( 95)=  0.12154840767554E+02;
A( 96)= -0.51906431613297E+00; A( 97)= -0.51400615245497E+01; A( 98)= -0.86293310180221E+00; A( 99)=  0.54389638874818E+02; A(100)= -0.30828863314334E+02;
A(101)= -0.39209158961571E+02; A(102)=  0.51492316050988E+02; A(103)=  0.15575926327640E+03; A(104)= -0.40331989907218E+02; A(105)=  0.81747944674677E+02;
A(106)= -0.74080097669739E+00; A(107)= -0.95839783290803E+02; A(108)=  0.91625401524892E+02; A(109)= -0.16211614166455E+02; A(110)= -0.53984374157427E+01;
A(111)=  0.23592245303834E+02; A(112)=  0.49285341298760E+02; A(113)=  0.19018261046747E+01; A(114)= -0.85277990004248E+02; A(115)=  0.61497035966943E+00;
A(116)=  0.71583039909916E+00; A(117)=  0.59335332549128E+00; A(118)=  0.61732967486832E+00; A(119)=  0.54000819291744E+00; A(120)=  0.63454465644118E+00;
A(121)=  0.51803070576663E+00; A(122)=  0.55196499111345E+00; A(123)=  0.50856570925678E+00; A(124)=  0.57946746762675E+00; A(125)=  0.49597352153131E+00;
A(126)=  0.51804257333119E+00; A(127)=  0.74863959367590E+00; A(128)=  0.10142576190910E+01; A(129)=  0.84799075449831E+00; A(130)=  0.33380516542402E+02;
A(131)=  0.34373357281331E+02; A(132)=  0.88645467453990E+01; A(133)= -0.21272238149326E+03; A(134)=  0.12485492825570E+02; A(135)=  0.15363562702690E+02;
A(136)=  0.21478130618490E+01; A(137)=  0.23907762557527E+03; A(138)=  0.18945833018748E+02; A(139)=  0.12087694354620E+00; A(140)= -0.97244735942445E+02;
A(141)= -0.82210904365684E+02; A(142)= -0.89489393895102E+02; A(143)= -0.62441605353238E+01; A(144)=  0.58740182617939E+02; A(145)=  0.75118500006921E+02;
```

```
A(146)=  0.32015486745774E+02; A(147)=  0.60063155142310E+02; A(148)= -0.33196654771078E+02; A(149)= -0.10942333219869E+01; A(150)=  0.73844929486527E+01;
A(151)=  0.77739265724529E+01; A(152)=  0.24530032508937E+00; A(153)=  0.14488117443920E+00; A(154)=  0.13374414956856E+00; A(155)=  0.18609508075708E+00;
A(156)=  0.91474685896132E+00; A(157)= -0.13795532695804E+03; A(158)=  0.39247899214306E+02; A(159)=  0.10710064557597E+03; A(160)= -0.11927979904687E+02;
A(161)=  0.11857303031785E+03; A(162)= -0.23787296445210E+02; A(163)= -0.52622006290045E+02; A(164)=  0.11982078083194E+01; A(165)=  0.81513012052222E+00;
A(166)= -0.13670107541811E+01; A(167)= -0.63758196455925E+02; A(168)=  0.22957072469570E+02; A(169)= -0.25012654568440E+00; A(170)= -0.16680791892759E+00;
A(171)= -0.22531963063858E+00; A(172)= -0.98369424035966E-01; A(173)= -0.15805262331087E+02; A(174)=  0.51062155012597E+01; A(175)= -0.93963051143540E+01;
A(176)=  0.88412734071365E+01; A(177)= -0.16077678096848E+03; A(178)=  0.39144626732891E+02; A(179)=  0.36589738304156E+02; A(180)=  0.17963950666514E+00;
A(181)= -0.20234853016373E+03; A(182)=  0.72546566615716E+02; A(183)=  0.11300822689433E+03; A(184)=  0.47862930569313E-01; A(185)=  0.11003793478012E+01;
A(186)=  0.81263804435730E+00; A(187)=  0.17881132364273E+01; A(188)= -0.64177699089050E+01; A(189)= -0.14219649242787E+03; A(190)= -0.12213250980198E+03;
A(191)=  0.17179721471238E+02; A(192)= -0.11438354580770E+03; A(193)= -0.62138217840670E+02; A(194)= -0.41096557607812E+02; A(195)=  0.64035590361797E+02;
A(196)=  0.10732897184173E+03; A(197)= -0.20717558027543E+01; A(198)= -0.19923414470110E+02; A(199)= -0.12327117667135E+03; A(200)=  0.11693377682598E+02;
A(201)=  0.20926825245342E+03; A(202)=  0.36088199536084E+02; A(203)=  0.42809209792938E+01; A(204)= -0.34192643024028E+03; A(205)= -0.16800849078608E+03;
A(206)=  0.27234150624719E+03; A(207)=  0.55680983732311E+03; A(208)= -0.21723896037259E+02; A(209)=  0.19949957680057E+03; A(210)=  0.86554869839904E+03;
A(211)=  0.85303522872373E+02; A(212)=  0.42168559307630E+03; A(213)=  0.97581533456278E+02; A(214)= -0.16565809376906E+04; A(215)= -0.18977492242141E+03;
A(216)= -0.51876064278775E+03; A(217)= -0.39031812497611E+02; A(218)=  0.36281606796466E+01; A(219)=  0.52904825059930E+02; A(220)=  0.52871941286017E+02;
A(221)=  0.78207002857473E+01; A(222)= -0.77975601559873E+02; A(223)=  0.59882951041016E+02; A(224)=  0.16555077469315E+02; A(225)= -0.26285544288901E+02;
A(226)=  0.54557389371503E+02; A(227)=  0.45856399117898E+02; A(228)=  0.16373449880088E+01; A(229)= -0.48098630250390E+02; A(230)= -0.14888038199830E+02;
A(231)= -0.88706411172296E+02; A(232)= -0.41354557085759E+01; A(233)=  0.59484970929864E+01; A(234)= -0.10469545797144E+03; A(235)= -0.20417869170276E+02;
A(236)=  0.50290343368644E+02; A(237)=  0.16907133738064E+03; A(238)= -0.10582258845436E+03; A(239)= -0.17026215868365E+03; A(240)=  0.88772861529219E+02;
A(241)= -0.10032737409931E+03; A(242)= -0.38138929936601E+02; A(243)=  0.33309591510475E+02; A(244)= -0.64032875201458E+01; A(245)=  0.25221691571533E+02;
A(246)=  0.90868011887977E+02; A(247)=  0.38517016207727E+01; A(248)=  0.62736936861879E+01; A(249)=  0.40530448704899E+00

     pi2 = 2.d0*0.3141592653589793D1
    HAtomMass = 938.782303d0    ! 2.072E-20 [MeV]
    aNeutronMass = 939.56538d0  !2.072*10^(-14) [MeV]
    ProtonMass = 938.272046d0
    ElectronMass = 0.510998928d0
    u = 931.494061D0
    Ael = 1.44381E-05
    Bel = 1.55468E-12

   iStr = 5; iPow = 4;  iSP= iStr + iPow ! number in inicials
   nPow = 15;   nBWp= 15; nBW = 4*nBWp
   MnZ = 9; MnN = 10; Ndop = 44
   N0 = isp*(1+nPow);  N1= N0+Ndop; Nqn = N1+nBW
    N = Nqn  + 1 !; print *,N; pause

   AA = Proton + aNeutron
```

```
      a1 = Proton/AA; a12 = a1**2; a13 = a1**3
       a2 = aNeutron/AA; a22 = a2**2; a23 = a2**3
        a3 = (aNeutron-Proton)/AA; a32 = a3**2; a33 = a3**3
         a4 = Proton/(aNeutron+1); a42 = a4**2; a43 = a4**3
          a5 = dlog(AA+1.d0); a6 = 1.d0/a5
        Z0 =  FunZ(N,a, Proton,WZ)
       aN0 =  FunN(N,A,aNeutron,WN )
 !         print *,AA,Proton,aNeutron,Z0,aN0; pause
        AAA = 1.D0
           if( int(AA/2)*2.ne.AA) AAA = 0.D0
        ZZZ = 1.D0
           if( int(Proton/2)*2.ne.Proton) ZZZ = 0.D0
        ANN = 1.D0
           if( int(aNeutron/2)*2.ne.aNeutron) ANN = 0.D0
    AAA = AAA/AA; AAZ = ZZZ/Proton; AAN = ANN/(aNeutron+1.d0)
           CorMN =  BrWig(Proton,Z0,WZ,aNeutron,aN0,WN,a,n,N1)
!   N0 = isp(1+nPow)  N1= N0+nBW; Nqn = N1+15 +38 qn parameters
    CorectionMN = CorMN
      c1 = eexp( a(1)) + CorPow(a,n,isp)            +  CorS(a,n,N0)         ! (15.75d0)
       c2=  eexp( a(2)) + CorPow(a,n,isp + nPow )     +  CorS(a,n,N0+4)       ! (17.8d0)
        c3 = eexp( a(3)) + CorPow(a,n,isp + 2*nPow )   +   CorS(a,n,N0+8)     ! ( 0.711d0)
         c4 = eexp( a(4)) + CorPow(a,n,isp + 3*nPow )   +   CorS(a,n,N0+12)    ! (23.7d0)
          c5 = eexp( a(5)) + CorPow(a,n,isp + 4*nPow )   +    CorS(a,n,N0+16) ! (11.18d0)
       pow1 =  eexp( a(6)) + CorPow(a,n,isp + 5*nPow )    +    CorS(a,n,N0+20)    ! 1.d0/3
        pow2 =  eexp( a(7)) + CorPow(a,n,isp + 6*nPow )    +    CorS(a,n,N0+24)   ! 4.d0/3
         pow3 =  eexp( a(8)) + CorPow(a,n,isp + 7*nPow )    +    CorS(a,n,N0+28)  ! 2.d0
          pow4 =  eexp( a(9)) + CorPow(a,n,isp + 8*nPow )    +    CorS(a,n,N0+32) ! 3.d0/2
!       Print *,'N0=',N0,'N1=',N1,'Nqn=',Nqn, ' N =', N; pause
       Vol =              c1                                           !1
        Sur =             c2/AA**Pow1                                                                   !2  1/3
         Cha =            c3*Proton*(Proton-1.d0)/AA**Pow2                                      !3  4/3 ZZ(ZZ-1.d0)
          Sym =  c4*(aNeutron-Proton)**2/AA**Pow3                                !4  2 2 (AN-ZZ)**2
        if( int(AA/2)*2.ne.AA) WigE=0.d0 !  A odd
        if( int(Proton/2)*2.ne.Proton.and. int(aNeutron/2)*2.ne.aNeutron) WigE =-1.d0      !odd odd
        if( int(Proton/2)*2.eq.Proton.and. int(aNeutron/2)*2.eq.aNeutron) WigE = 1.d0      !even even
       Wig = c5*WigE/AA**Pow4       !5 +-0 3/2
           BetheWeizsacker = Vol - Sur - Cha - Sym + Wig
```

```
 BiEnTh = BetheWeizsacker + CorMN
  AtMassTh = Proton*HAtomMass + aNeutron*aNeutronMass - AA*BiEnTh
   aNuclMass = AtMassTh - (ElectronMass*Proton + Ael*Proton**2.39D0 + Bel*Proton**5.35D0)
    aMassExcTh = AtMassTh - AA*u
   RETURN;  END
!****************************************************************************************************************C
  Function CorS(a,n,i)
  IMPLICIT DOUBLE PRECISION(A-H,O-Z)
  common/variables/AA,a1,a12,a13,a2,a22,a23,a3,a32,a33,a4,a42,a43,a5,a6,AAA,AAZ,AAN
  dimension a(n)
  CorS =  eexp(- (a(i+1)*AAA + a(i+2)*AAZ + a(i+3)*AAN + a(i+4))**2 )
  return; end
!****************************************************************************************************************C
  Function CorPow(a,n,i)
  IMPLICIT DOUBLE PRECISION(A-H,O-Z)
  common/variables/AA,a1,a12,a13,a2,a22,a23,a3,a32,a33,a4,a42,a43,a5,a6,AAA,AAZ,AAN
  dimension a(n)
      c1 = a(i+1)*a1 + a(i+2)*a2 + a(i+3)*a3 + a(i+4)*a4
       c2 = a(i+5)*a12 + a(i+6)*a22 + a(i+7)*a32 + a(i+8)*a42
        c3 = a(i+9)*a13 + a(i+10)*a23 + a(i+11)*a33 + a(i+12)*a43
         c4 =  a(i+13)*a6   + a(i+14)*a5 + a(i+15)
  CorPow =  eexp(- ( c1 + c2  + c3 + c4  )**2 )
  return; end
!****************************************************************************************************************C
  Function BrWig(Z,Z0,WZ,aN,aN0,WN,a,n,i)
  IMPLICIT DOUBLE PRECISION(A-H,O-Z)
  common/variables/AA,a1,a12,a13,a2,a22,a23,a3,a32,a33,a4,a42,a43,a5,a6,AAA,AAZ,AAN
 common/bwigner/BWZ,BWN,CorMn,BetheWeizsacker
  common/nhelp/lexpt,iSP,nPow,nBWp,MnZ,MnN,N0,N1,Nqn
  dimension a(n)
  Z02 = (Z-Z0)**2; aN02 = (aN-aN0)**2; A02 = (AA-AA0)**2
    AmpZ =  CorAmp(a,n,i) + WZ*(1.d0 + CorS(a,n,N0+36))
     AmpN =   CorAmp(a,n,i +  nBWp) + WN*(1.d0 + CorS(a,n,N0+40))   !
    GamZ =   CorGam(a,n,i + 2*nBWp ) + WZ
     GamN =   CorGam(a,n,i + 3*nBWp ) + WN
    BWZ  =  AmpZ*eexp(-Z02/GamZ)/(Z02+GamZ)
     BWN  =  AmpN*eexp(-aN02/GamN)/(aN02+GamN)
```

```
      BrWig =   (BWZ  +  BWN)/AA**a(N)
!       print '(2f5.0,3f6.0,6e12.4,2e14.6)',Z,aN,wZ,wN,WA,AmpZ,AmpN,AmpA,BwZ,BwN,BwA,a(i+15),a(i+15+nBWp)  ; pause
!              Print *,'N0=',N0,'N1=',N1,'Nqn=',Nqn, ' N =', N
!  pause
      return; end
!*******************************************************************************************************************************C
      Function CorAmp(a,n,i)
      IMPLICIT DOUBLE PRECISION(A-H,O-Z)
      common/variables/AA,a1,a12,a13,a2,a22,a23,a3,a32,a33,a4,a42,a43,a5,a6,AAA,AAZ,AAN
      dimension a(n)
          c1 = a(i+1)*a1 + a(i+2)*a2 + a(i+3)*a3 + a(i+4)*a4
           c2 = a(i+5)*a12 + a(i+6)*a22 + a(i+7)*a32 + a(i+8)*a42
            c3 = a(i+9)*a13  + a(i+10)*a23  + a(i+11)*a33  + a(i+12)*a43
              c4 =  a(i+13)*a6   + a(i+14)*a5
      CorAmp =   eexp(a(i+15) - ( c1 + c2 + c3  + c4  )**2)
      return; end
!*******************************************************************************************************************************C
      Function CorGam(a,n,i)
      IMPLICIT DOUBLE PRECISION(A-H,O-Z)
      common/variables/AA,a1,a12,a13,a2,a22,a23,a3,a32,a33,a4,a42,a43,a5,a6,AAA,AAZ,AAN
      dimension a(n)
          c1 = a(i+1)*a1 + a(i+2)*a2 + a(i+3)*a3 + a(i+4)*a4
           c2 = a(i+5)*a12 + a(i+6)*a22 + a(i+7)*a32 + a(i+8)*a42
            c3 = a(i+9)*a13 + a(i+10)*a23 + a(i+11)*a33 + a(i+12)*a43
             c4 =  a(i+13)*a6   + a(i+14)*a5
      CorGam =  eexp( a(i+15) - ( c1 + c2 + c3 + c4  )**2)
      return; end
!*******************************************************************************************************************************C
      Function  FunZ(N,A,x,WZ)
      IMPLICIT DOUBLE PRECISION (A-H,O-Z)
!      Proton Magic numbers 2,8,14,20,28,50,82,108,124
     common/nhelp/lexpt,iSP,nPow,nBWp,MnZ,MnN,N0,N1,Nqn
      dimension aMn(9) ,aB(9),A(N)
     data aMn/2.d0,8.d0,14.d0,20.d0,28.d0,50.d0,82.d0,108.d0,124.d0/
     data aB/1.d0,5.d0,11.d0,17.d0,24.d0,39.d0,66.d0,95.d0,116.d0/
     !   do i = 1, MnZ; aMn(i) =  (a(Nqn+i));  aB(i) =   (a(Nqn + MnZ + i))
     !   enddo
```

```
        do i = 1,MnZ - 1
         if(x.ge.aB(i).and.x.lt.aB(i+1))    then
          FunZ = int(aMn(i))
           Wz   = (aMn(i+1) - aMn(i))/2
         return; endif
        enddo
         if(x.ge.aB(MnZ))        FunZ = int(aMn(MnZ))
          if(x.ge.aB(MnZ))    WZ  = (aMn(MnZ) - aMn(MnZ-1))/2
!              Print *,'N0=',N0,'N1=',N1,'Nqn=',Nqn, ' N =', N
!  pause
      return; end
  !**************************************************************************************************************************C
      Function  FunN(N,A,x,WN)
      IMPLICIT DOUBLE PRECISION (A-H,O-Z)
!     Neutron Magic numbers 2,8,14,20,28,50,82,124,152,202
     common/nhelp/lexpt,iSP,nPow,nBWp,MnZ,MnN,N0,N1,Nqn
      dimension aMn(10),aB(10),A(N)
      data aMn/2.d0,8.d0,14.d0,20.d0,28.d0,50.d0,82.d0,124.d0,152.d0,202.d0/
     data aB/1.d0,5.d0,11.d0,17.d0,24.d0,39.d0,66.d0,103.d0,138.d0,178.d0/
!       do i = 1, MnN; aMn(i) =  (a(Nqn + 2*MnZ + i));  aB(i) =   (a(Nqn + 2*MnZ + MnN + i));
!       enddo
        do i=1,MnN-1
         if(x.ge.aB(i).and.x.lt.aB(i+1))   Then
          FunN = int(aMn(i))
           WN   = (aMn(i+1) - aMn(i))/2
         return; endif
        enddo
         if(x.ge.aB(MnN))      FunN = int(aMn(MnN))
          if(x.ge.aB(MnN))      WN   = (aMn(MnN) - aMn(MnN-1))/2
!               Print *,'N0=',N0,'N1=',N1,'Nqn=',Nqn, ' N =', N
! pause
      return; end
```

**APPENDIX B** The description of the experimental binding energy, nuclear and atomic mass mass excess values from new database Nuclide groung state: https://www-nds.iaea.org/relnsd/NdsEnsdf/QueryForm.html

| No | El | A | Z | N | N-Z | BindEnExpt | BindEnTh | ResBindEn [MeV] | NuclMassExpt | NuclMassTh | ResNucMa | AtMassExpt | AtMassTh | ResAtMas | MassExcExpt | MassExcExpt | ResMasEx |
|---|---|---|---|---|---|---|---|---|---|---|---|---|---|---|---|---|---|
| 1 | H | 2 | 1 | 1 | 0 | 1.11228300E+00 | 1.20125271E+00 | -8.90E-02 | 1.87561283E+03 | 1.87543416E+03 | 1.80E-01 | 1.87612384E+03 | 1.87594518E+03 | 1.80E-01 | 1.31357217E+01 | 1.29570556E+01 | 1.80E-01 |
| 2 | H | 3 | 1 | 2 | 1 | 2.82726600E+00 | 2.93684368E+00 | -1.10E-01 | 2.80892098E+03 | 2.80859152E+03 | 3.30E-01 | 2.80943199E+03 | 2.80910253E+03 | 3.30E-01 | 1.49498061E+01 | 1.46203490E+01 | 3.30E-01 |
| 3 | H | 4 | 1 | 3 | 2 | 1.72045000E+00 | 1.64011816E+00 | 8.00E-02 | 3.75008635E+03 | 3.75040696E+03 | -3.20E-01 | 3.75059737E+03 | 3.75091797E+03 | -3.20E-01 | 2.46211230E+01 | 2.49417263E+01 | -3.20E-01 |
| 4 | H | 5 | 1 | 4 | 3 | 1.33636000E+00 | 1.48767847E+00 | -1.50E-01 | 4.68985173E+03 | 4.68909442E+03 | 7.60E-01 | 4.69036275E+03 | 4.68960543E+03 | 7.60E-01 | 3.28924400E+01 | 3.21351256E+01 | 7.60E-01 |
| 5 | H | 6 | 1 | 5 | 4 | 9.61640000E-01 | 9.73236447E-01 | -1.20E-02 | 5.63032907E+03 | 5.63025877E+03 | 7.00E-02 | 5.63084008E+03 | 5.63076978E+03 | 7.00E-02 | 4.18757170E+01 | 4.18054183E+01 | 7.00E-02 |
| 6 | He | 3 | 2 | 1 | -1 | 2.57268100E+00 | 2.50058789E+00 | 7.20E-02 | 2.80839133E+03 | 2.80860615E+03 | -2.10E-01 | 2.80941340E+03 | 2.80962822E+03 | -2.10E-01 | 1.49312155E+01 | 1.51460393E+01 | -2.10E-01 |
| 7 | He | 4 | 2 | 2 | 0 | 7.07391500E+00 | 7.07390311E+00 | 1.20E-05 | 3.72737909E+03 | 3.72737768E+03 | 1.40E-03 | 3.72840116E+03 | 3.72839975E+03 | 1.40E-03 | 2.42491561E+00 | 2.42350958E+00 | 1.40E-03 |
| 8 | He | 5 | 2 | 3 | 1 | 5.51213200E+00 | 5.49699548E+00 | 1.50E-02 | 4.66767946E+03 | 4.66775370E+03 | -7.40E-02 | 4.66870154E+03 | 4.66877577E+03 | -7.40E-02 | 1.12312330E+01 | 1.13054636E+01 | -7.40E-02 |
| 9 | He | 6 | 2 | 4 | 2 | 4.87851900E+00 | 4.90289349E+00 | -2.40E-02 | 5.60553439E+03 | 5.60538669E+03 | 1.50E-01 | 5.60655646E+03 | 5.60640877E+03 | 1.50E-01 | 1.75920950E+01 | 1.74443991E+01 | 1.50E-01 |
| 10 | He | 7 | 2 | 5 | 3 | 4.12305700E+00 | 4.10355374E+00 | 2.00E-02 | 6.54550948E+03 | 6.54564456E+03 | -1.40E-01 | 6.54653155E+03 | 6.54666663E+03 | -1.40E-01 | 2.60731260E+01 | 2.62082028E+01 | -1.40E-01 |
| 11 | He | 8 | 2 | 6 | 4 | 3.92452000E+00 | 3.92446524E+00 | 5.50E-05 | 7.48254010E+03 | 7.48253909E+03 | 1.00E-03 | 7.48356217E+03 | 7.48356116E+03 | 1.00E-03 | 3.16096810E+01 | 3.16086761E+01 | 1.00E-03 |
| 12 | He | 9 | 2 | 7 | 5 | 3.34902900E+00 | 3.14166485E+00 | 2.10E-01 | 8.42336037E+03 | 8.42522521E+03 | -1.90E+00 | 8.42438245E+03 | 8.42624728E+03 | -1.90E+00 | 4.09358960E+01 | 4.28007333E+01 | -1.90E+00 |
| 13 | He | 10 | 2 | 8 | 6 | 2.99761600E+00 | 2.99211866E+00 | 5.50E-03 | 9.36309086E+03 | 9.36314439E+03 | -5.40E-02 | 9.36411293E+03 | 9.36416646E+03 | -5.40E-02 | 4.91723160E+01 | 4.92258494E+01 | -5.40E-02 |
| 14 | Li | 4 | 3 | 1 | -2 | 1.15376100E+00 | 1.17471205E+00 | -2.10E-02 | 3.74976624E+03 | 3.74968024E+03 | 8.60E-02 | 3.75129943E+03 | 3.75121344E+03 | 8.60E-02 | 2.53231860E+01 | 2.52371968E+01 | 8.60E-02 |
| 15 | Li | 5 | 3 | 2 | -1 | 5.26613200E+00 | 5.19801958E+00 | 6.80E-02 | 4.66761600E+03 | 4.66795437E+03 | -3.40E-01 | 4.66914919E+03 | 4.66948757E+03 | -3.40E-01 | 1.16788860E+01 | 1.20172661E+01 | -3.40E-01 |
| 16 | Li | 6 | 3 | 3 | 0 | 5.33233100E+00 | 5.49107300E+00 | -1.60E-01 | 5.60151805E+03 | 5.60056341E+03 | 9.50E-01 | 5.60305124E+03 | 5.60209661E+03 | 9.50E-01 | 1.40868789E+01 | 1.31322450E+01 | 9.50E-01 |
| 17 | Li | 7 | 3 | 4 | 1 | 5.60643900E+00 | 5.81991476E+00 | -2.10E-01 | 6.53383234E+03 | 6.53233583E+03 | 1.50E+00 | 6.53536553E+03 | 6.53386903E+03 | 1.50E+00 | 1.49071052E+01 | 1.34105986E+01 | 1.50E+00 |
| 18 | Li | 8 | 3 | 5 | 2 | 5.15971200E+00 | 5.27283941E+00 | -1.10E-01 | 7.47136510E+03 | 7.47045790E+03 | 9.10E-01 | 7.47289829E+03 | 7.47199109E+03 | 9.10E-01 | 2.09458040E+01 | 2.00386057E+01 | 9.10E-01 |
| 19 | Li | 9 | 3 | 6 | 3 | 5.03776800E+00 | 5.14343410E+00 | -1.10E-01 | 8.40686826E+03 | 8.40591509E+03 | 9.50E-01 | 8.40840145E+03 | 8.40744828E+03 | 9.50E-01 | 2.49549020E+01 | 2.40017331E+01 | 9.50E-01 |
| 20 | Li | 10 | 3 | 7 | 4 | 4.53135100E+00 | 4.66394533E+00 | -1.30E-01 | 9.34646004E+03 | 9.34513192E+03 | 1.30E+00 | 9.34799323E+03 | 9.34666512E+03 | 1.30E+00 | 3.30526240E+01 | 3.17245057E+01 | 1.30E+00 |
| 21 | Li | 11 | 3 | 8 | 5 | 4.15538100E+00 | 4.29434111E+00 | -1.40E-01 | 1.02856297E+04 | 1.02840990E+04 | 1.50E+00 | 1.02871629E+04 | 1.02856322E+04 | 1.50E+00 | 4.07282540E+01 | 3.91975258E+01 | 1.50E+00 |
| 22 | Be | 5 | 4 | 1 | -3 | 1.80000000E-02 | -1.10382888E-02 | 2.90E-02 | 4.69256458E+03 | 4.69270539E+03 | -1.40E-01 | 4.69460897E+03 | 4.69474978E+03 | -1.40E-01 | 3.71390000E+01 | 3.72794784E+01 | -1.40E-01 |
| 23 | Be | 6 | 4 | 2 | -2 | 4.48724700E+00 | 4.48735746E+00 | -1.10E-04 | 5.60529501E+03 | 5.60529143E+03 | 3.60E-03 | 5.60733940E+03 | 5.60733583E+03 | 3.60E-03 | 1.83750340E+01 | 1.83714612E+01 | 3.60E-03 |
| 24 | Be | 7 | 4 | 3 | -1 | 5.37154800E+00 | 5.40289576E+00 | -3.10E-02 | 6.53418303E+03 | 6.53396069E+03 | 2.20E-01 | 6.53622743E+03 | 6.53600508E+03 | 2.20E-01 | 1.57689990E+01 | 1.55466547E+01 | 2.20E-01 |
| 25 | Be | 8 | 4 | 4 | 0 | 7.06243500E+00 | 7.07834112E+00 | -1.60E-02 | 7.45484977E+03 | 7.45471961E+03 | 1.30E-01 | 7.45689416E+03 | 7.45676400E+03 | 1.30E-01 | 4.94167100E+00 | 4.81151501E+00 | 1.30E-01 |
| 26 | Be | 9 | 4 | 5 | 1 | 6.46266800E+00 | 6.41682384E+00 | 4.60E-02 | 8.39275061E+03 | 8.39316031E+03 | -4.10E-01 | 8.39479500E+03 | 8.39520470E+03 | -4.10E-01 | 1.13484530E+01 | 1.17581484E+01 | -4.10E-01 |
| 27 | Be | 10 | 4 | 6 | 2 | 6.49763000E+00 | 6.54384830E+00 | -4.60E-02 | 9.32550371E+03 | 9.32503862E+03 | 4.70E-01 | 9.32754810E+03 | 9.32708301E+03 | 4.70E-01 | 1.26074880E+01 | 1.21423990E+01 | 4.70E-01 |
| 28 | Be | 11 | 4 | 7 | 3 | 5.95254000E+00 | 5.97564424E+00 | -2.30E-02 | 1.02645675E+04 | 1.02643104E+04 | 2.60E-01 | 1.02666118E+04 | 1.02663548E+04 | 2.60E-01 | 2.01771670E+01 | 1.99201143E+01 | 2.60E-01 |
| 29 | Be | 12 | 4 | 8 | 4 | 5.72072200E+00 | 5.73116086E+00 | -1.00E-02 | 1.12009621E+04 | 1.12008339E+04 | 1.30E-01 | 1.12030065E+04 | 1.12028783E+04 | 1.30E-01 | 2.50777590E+01 | 2.49495897E+01 | 1.30E-01 |
| 30 | Be | 13 | 4 | 9 | 5 | 5.24143500E+00 | 5.24032624E+00 | 1.10E-03 | 1.21410375E+04 | 1.21410490E+04 | -1.20E-02 | 1.21430819E+04 | 1.21430934E+04 | -1.20E-02 | 3.36590760E+01 | 3.36705979E+01 | -1.20E-02 |
| 31 | Be | 14 | 4 | 10 | 6 | 4.99389700E+00 | 4.96762272E+00 | 2.60E-02 | 1.30788270E+04 | 1.30791919E+04 | -3.60E-01 | 1.30808713E+04 | 1.30812363E+04 | -3.60E-01 | 3.99544980E+01 | 4.03194400E+01 | -3.60E-01 |
| 32 | Be | 15 | 4 | 11 | 7 | 4.54500000E+00 | 4.49279370E+00 | 5.20E-02 | 1.40201269E+04 | 1.40209121E+04 | -7.90E-01 | 1.40221713E+04 | 1.40229565E+04 | -7.90E-01 | 4.97600000E+01 | 5.05455715E+01 | -7.90E-01 |
| 33 | Be | 16 | 4 | 12 | 8 | 4.28528500E+00 | 4.28812687E+00 | -2.80E-03 | 1.49593077E+04 | 1.49592593E+04 | 4.80E-02 | 1.49613521E+04 | 1.49613037E+04 | 4.80E-02 | 5.74471320E+01 | 5.73987661E+01 | 4.80E-02 |
| 34 | B | 7 | 5 | 2 | -3 | 3.55870500E+00 | 3.57564233E+00 | -1.70E-02 | 6.54557931E+03 | 6.54545711E+03 | 1.20E-01 | 6.54813498E+03 | 6.54801278E+03 | 1.20E-01 | 2.76765500E+01 | 2.75543517E+01 | 1.20E-01 |
| 35 | B | 8 | 5 | 3 | -2 | 4.71715300E+00 | 4.68219055E+00 | 3.50E-02 | 7.47231840E+03 | 7.47259446E+03 | -2.80E-01 | 7.47487407E+03 | 7.47515013E+03 | -2.80E-01 | 2.29215770E+01 | 2.31976426E+01 | -2.80E-01 |
| 36 | B | 9 | 5 | 4 | -1 | 6.25707000E+00 | 6.15322188E+00 | 1.00E-01 | 8.39330737E+03 | 8.39423837E+03 | -9.30E-01 | 8.39586304E+03 | 8.39679404E+03 | -9.30E-01 | 1.24164880E+01 | 1.33474891E+01 | -9.30E-01 |
| 37 | B | 10 | 5 | 5 | 0 | 6.47507500E+00 | 6.51670456E+00 | -4.20E-02 | 9.32443563E+03 | 9.32401570E+03 | 4.20E-01 | 9.32699130E+03 | 9.32657137E+03 | 4.20E-01 | 1.20506900E+01 | 1.16307594E+01 | 4.20E-01 |
| 38 | B | 11 | 5 | 6 | 1 | 6.92771600E+00 | 6.85631459E+00 | 7.10E-02 | 1.02525469E+04 | 1.02533287E+04 | -7.80E-01 | 1.02551026E+04 | 1.02558843E+04 | -7.80E-01 | 8.66788300E+00 | 9.44966351E+00 | -7.80E-01 |
| 39 | B | 12 | 5 | 7 | 2 | 6.63122100E+00 | 6.58775485E+00 | 4.30E-02 | 1.11887425E+04 | 1.11892604E+04 | -5.20E-01 | 1.11912982E+04 | 1.11918161E+04 | -5.20E-01 | 1.33694160E+01 | 1.38873848E+01 | -5.20E-01 |
| 40 | B | 13 | 5 | 8 | 3 | 6.49640600E+00 | 6.44501902E+00 | 5.10E-02 | 1.21234292E+04 | 1.21240936E+04 | -6.60E-01 | 1.21259849E+04 | 1.21266493E+04 | -6.60E-01 | 1.65621190E+01 | 1.72265148E+01 | -6.60E-01 |
| 41 | B | 14 | 5 | 9 | 4 | 6.10164400E+00 | 6.09156414E+00 | 1.00E-02 | 1.30620249E+04 | 1.30621624E+04 | -1.40E-01 | 1.30645805E+04 | 1.30647180E+04 | -1.40E-01 | 2.36636850E+01 | 2.38011830E+01 | -1.40E-01 |
| 42 | B | 15 | 5 | 10 | 5 | 5.88000200E+00 | 5.88302316E+00 | -3.00E-03 | 1.39988132E+04 | 1.39987643E+04 | 4.90E-02 | 1.40013689E+04 | 1.40013200E+04 | 4.90E-02 | 2.89579880E+01 | 2.89090526E+01 | 4.90E-02 |
| 43 | B | 16 | 5 | 11 | 6 | 5.50731700E+00 | 5.45278424E+00 | 5.50E-02 | 1.49384616E+04 | 1.49393305E+04 | -8.70E-01 | 1.49410173E+04 | 1.49418861E+04 | -8.70E-01 | 3.71122710E+01 | 3.79811712E+01 | -8.70E-01 |
| 44 | B | 17 | 5 | 12 | 7 | 5.26646100E+00 | 5.23717249E+00 | 2.90E-02 | 1.58766142E+04 | 1.58771085E+04 | -4.90E-01 | 1.58791699E+04 | 1.58796641E+04 | -4.90E-01 | 4.37708160E+01 | 4.42651056E+01 | -4.90E-01 |
| 45 | B | 18 | 5 | 13 | 8 | 4.97360200E+00 | 4.94734333E+00 | 2.60E-02 | 1.68161846E+04 | 1.68166536E+04 | -4.70E-01 | 1.68187402E+04 | 1.68192093E+04 | -4.70E-01 | 5.18471330E+01 | 5.23161770E+01 | -4.70E-01 |
| 46 | C | 8 | 6 | 2 | -4 | 3.10152400E+00 | 3.10034704E+00 | 1.20E-03 | 7.48394972E+03 | 7.48395476E+03 | -5.00E-03 | 7.48701676E+03 | 7.48702180E+03 | -5.00E-03 | 3.50642690E+01 | 3.50693137E+01 | -5.00E-03 |
| 47 | C | 9 | 6 | 3 | -3 | 4.33742300E+00 | 4.37654753E+00 | -3.90E-02 | 8.40929048E+03 | 8.40893399E+03 | 3.60E-01 | 8.41235752E+03 | 8.41200103E+03 | 3.60E-01 | 2.89109700E+01 | 2.85544812E+01 | 3.60E-01 |
| 48 | C | 10 | 6 | 4 | -2 | 6.03203400E+00 | 5.99817092E+00 | 3.40E-02 | 9.32757233E+03 | 9.32790659E+03 | -3.30E-01 | 9.33063937E+03 | 9.33097363E+03 | -3.30E-01 | 1.56987550E+01 | 1.60330188E+01 | -3.30E-01 |
| 49 | C | 11 | 6 | 5 | -1 | 6.67637400E+00 | 6.69498527E+00 | -1.90E-02 | 1.02540179E+04 | 1.02538088E+04 | 2.10E-01 | 1.02570850E+04 | 1.02568759E+04 | 2.10E-01 | 1.06502950E+01 | 1.04412091E+01 | 2.10E-01 |
| 50 | C | 12 | 6 | 6 | 0 | 7.68014400E+00 | 7.67497832E+00 | 5.20E-03 | 1.11748617E+04 | 1.11749193E+04 | -5.80E-02 | 1.11779287E+04 | 1.11779864E+04 | -5.80E-02 | 0.00000000E+00 | 5.76261521E-02 | -5.80E-02 |
| 51 | C | 13 | 6 | 7 | 1 | 7.46984900E+00 | 7.26336087E+00 | 2.10E-01 | 1.21094808E+04 | 1.21121607E+04 | -2.70E+00 | 1.21125478E+04 | 1.21152278E+04 | -2.70E+00 | 3.12500875E+00 | 5.80499365E+00 | -2.70E+00 |
| 52 | C | 14 | 6 | 8 | 2 | 7.52031900E+00 | 7.46194296E+00 | 5.80E-02 | 1.30408697E+04 | 1.30416826E+04 | -8.10E-01 | 1.30439368E+04 | 1.30447497E+04 | -8.10E-01 | 3.01989278E+00 | 3.83280263E+00 | -8.10E-01 |
| 53 | C | 15 | 6 | 9 | 3 | 7.10016900E+00 | 7.06692753E+00 | 3.30E-02 | 1.39792170E+04 | 1.39797113E+04 | -4.90E-01 | 1.39822841E+04 | 1.39827783E+04 | -4.90E-01 | 9.87314400E+00 | 1.03674101E+01 | -4.90E-01 |
| 54 | C | 16 | 6 | 10 | 4 | 6.92205400E+00 | 6.88872795E+00 | 3.30E-02 | 1.49145321E+04 | 1.49150609E+04 | -5.30E-01 | 1.49175991E+04 | 1.49181280E+04 | -5.30E-01 | 1.36941280E+01 | 1.42229949E+01 | -5.30E-01 |
| 55 | C | 17 | 6 | 11 | 5 | 6.55809000E+00 | 6.51810865E+00 | 4.00E-02 | 1.58533628E+04 | 1.58540381E+04 | -6.80E-01 | 1.58564298E+04 | 1.58571052E+04 | -6.80E-01 | 2.10307780E+01 | 2.17061140E+01 | -6.80E-01 |
| 56 | C | 18 | 6 | 12 | 6 | 6.42619500E+00 | 6.40100854E+00 | 2.50E-02 | 1.67887442E+04 | 1.67891932E+04 | -4.50E-01 | 1.67918112E+04 | 1.67922602E+04 | -4.50E-01 | 2.49181240E+01 | 2.53671262E+01 | -4.50E-01 |
| 57 | C | 20 | 6 | 14 | 8 | 5.95873300E+00 | 5.98428953E+00 | -2.60E-02 | 1.86643718E+04 | 1.86638563E+04 | 5.20E-01 | 1.86674388E+04 | 1.86669233E+04 | 5.20E-01 | 3.75576100E+01 | 3.70421274E+01 | 5.20E-01 |
| 58 | C | 22 | 6 | 16 | 10 | 5.42203000E+00 | 5.41787686E+00 | 4.20E-03 | 2.05433926E+04 | 2.05434796E+04 | -8.70E-02 | 2.05464596E+04 | 2.05465466E+04 | -8.70E-02 | 5.35902440E+01 | 5.36772650E+01 | -8.70E-02 |
| 59 | N | 10 | 7 | 3 | -4 | 3.64366400E+00 | 3.59169754E+00 | 5.20E-02 | 9.35016222E+03 | 9.35067678E+03 | -5.10E-01 | 9.35374072E+03 | 9.35425529E+03 | -5.10E-01 | 3.88001070E+01 | 3.93146756E+01 | -5.10E-01 |
| 60 | N | 11 | 7 | 4 | -3 | 5.36403800E+00 | 5.33504206E+00 | 2.90E-02 | 1.02671598E+04 | 1.02674737E+04 | -3.10E-01 | 1.02707383E+04 | 1.02710522E+04 | -3.10E-01 | 2.43036420E+01 | 2.46175073E+01 | -3.10E-01 |
| 61 | N | 12 | 7 | 5 | -2 | 6.17010900E+00 | 6.24756862E+00 | -7.70E-02 | 1.11916883E+04 | 1.11907537E+04 | 9.30E-01 | 1.11952668E+04 | 1.11943322E+04 | 9.30E-01 | 1.73380720E+01 | 1.64034655E+01 | 9.30E-01 |
| 62 | N | 13 | 7 | 6 | -1 | 7.23886300E+00 | 7.18854675E+00 | 5.00E-02 | 1.21111898E+04 | 1.21118388E+04 | -6.50E-01 | 1.21147683E+04 | 1.21154173E+04 | -6.50E-01 | 5.34548100E+00 | 5.99450031E+00 | -6.50E-01 |
| 63 | N | 14 | 7 | 7 | 0 | 7.47561400E+00 | 7.33718294E+00 | 1.40E-01 | 1.30402018E+04 | 1.30421347E+04 | -1.90E+00 | 1.30437803E+04 | 1.30457132E+04 | -1.90E+00 | 2.86341669E+00 | 4.79636579E+00 | -1.90E+00 |
| 64 | N | 15 | 7 | 8 | 1 | 7.69946000E+00 | 7.70370856E+00 | -4.20E-03 | 1.39689338E+04 | 1.39688650E+04 | 6.90E-02 | 1.39725124E+04 | 1.39724435E+04 | 6.90E-02 | 1.01438660E-01 | 3.26176218E-02 | 6.90E-02 |
| 65 | N | 16 | 7 | 9 | 2 | 7.37379600E+00 | 7.35302142E+00 | 2.10E-02 | 1.49060104E+04 | 1.49063377E+04 | -3.30E-01 | 1.49095889E+04 | 1.49099162E+04 | -3.30E-01 | 5.68390700E+00 | 6.01122228E+00 | -3.30E-01 |
| 66 | N | 17 | 7 | 10 | 3 | 7.28622900E+00 | 7.32602415E+00 | -4.00E-02 | 1.58396906E+04 | 1.58390090E+04 | 6.80E-01 | 1.58432691E+04 | 1.58425875E+04 | 6.80E-01 | 7.87007500E+00 | 7.18847342E+00 | 6.80E-01 |
| 67 | N | 18 | 7 | 11 | 4 | 7.03856200E+00 | 7.04273313E+00 | -4.20E-03 | 1.67764278E+04 | 1.67763476E+04 | 8.00E-02 | 1.67800063E+04 | 1.67799261E+04 | 8.00E-02 | 1.31131680E+01 | 1.30330067E+01 | 8.00E-02 |
| 68 | N | 19 | 7 | 12 | 5 | 6.94858300E+00 | 6.94040037E+00 | 8.20E-03 | 1.77106642E+04 | 1.77108146E+04 | -1.50E-01 | 1.77142427E+04 | 1.77143931E+04 | -1.50E-01 | 1.58555210E+01 | 1.60059150E+01 | -1.50E-01 |
| 69 | N | 20 | 7 | 13 | 6 | 6.70924000E+00 | 6.71454711E+00 | -5.30E-03 | 1.86480678E+04 | 1.86479566E+04 | 1.10E-01 | 1.86516463E+04 | 1.86515351E+04 | 1.10E-01 | 2.17651100E+01 | 2.16538988E+01 | 1.10E-01 |
| 70 | N | 21 | 7 | 14 | 7 | 6.60809900E+00 | 6.61085562E+00 | -2.80E-03 | 1.95830479E+04 | 1.95829850E+04 | 6.30E-02 | 1.95866264E+04 | 1.95865635E+04 | 6.30E-02 | 2.52511640E+01 | 2.51881921E+01 | 6.30E-02 |
| 71 | N | 22 | 7 | 15 | 8 | 6.36608500E+00 | 6.38667042E+00 | -2.10E-02 | 2.05213295E+04 | 2.05208716E+04 | 4.60E-01 | 2.05249080E+04 | 2.05244501E+04 | 4.60E-01 | 3.20386750E+01 | 3.15807298E+01 | 4.60E-01 |
| 72 | N | 23 | 7 | 16 | 9 | 6.16700000E+00 | 6.16964830E+00 | -2.60E-03 | 2.14591066E+04 | 2.14590418E+04 | 6.50E-02 | 2.14626851E+04 | 2.14626203E+04 | 6.50E-02 | 3.83220000E+01 | 3.82568870E+01 | 6.50E-02 |
| 73 | N | 24 | 7 | 17 | 10 | 5.88700000E+00 | 5.86654897E+00 | 2.00E-02 | 2.23992170E+04 | 2.23997119E+04 | -4.90E-01 | 2.24027955E+04 | 2.24032904E+04 | -4.90E-01 | 4.69380000E+01 | 4.74329417E+01 | -4.90E-01 |
| 74 | O | 12 | 8 | 4 | -4 | 4.89019500E+00 | 4.85077703E+00 | 3.90E-02 | 1.12057534E+04 | 1.12062205E+04 | -4.70E-01 | 1.12098434E+04 | 1.12103106E+04 | -4.70E-01 | 3.19146960E+01 | 3.23818876E+01 | -4.70E-01 |
| 75 | O | 13 | 8 | 5 | -3 | 5.81176200E+00 | 5.92844320E+00 | -1.20E-01 | 1.21284482E+04 | 1.21269255E+04 | 1.50E+00 | 1.21325382E+04 | 1.21310156E+04 | 1.50E+00 | 2.31154390E+01 | 2.15927694E+01 | 1.50E+00 |
| 76 | O | 14 | 8 | 6 | -2 | 7.05230100E+00 | 7.03580875E+00 | 1.60E-02 | 1.30448342E+04 | 1.30450593E+04 | -2.30E-01 | 1.30489243E+04 | 1.30491494E+04 | -2.30E-01 | 8.00745700E+00 | 8.23252752E+00 | -2.30E-01 |
| 77 | O | 15 | 8 | 7 | -1 | 7.46369200E+00 | 7.32112304E+00 | 1.40E-01 | 1.39711765E+04 | 1.39733092E+04 | -2.10E+00 | 1.39752665E+04 | 1.39773992E+04 | -2.10E+00 | 2.85560500E+00 | 4.98832336E+00 | -2.10E+00 |
| 78 | O | 16 | 8 | 8 | 0 | 7.97620600E+00 | 7.95777810E+00 | 1.80E-02 | 1.48950779E+04 | 1.48953669E+04 | -2.90E-01 | 1.48991680E+04 | 1.48994570E+04 | -2.90E-01 | -4.73700137E+00 | -4.44796154E+00 | -2.90E-01 |
| 79 | O | 17 | 8 | 9 | 1 | 7.75072800E+00 | 7.76662713E+00 | -1.60E-02 | 1.58305002E+04 | 1.58302241E+04 | 2.80E-01 | 1.58345903E+04 | 1.58343142E+04 | 2.80E-01 | -8.08763610E-01 | -1.08485422E+00 | 2.80E-01 |
| 80 | O | 18 | 8 | 10 | 2 | 7.76709700E+00 | 7.76680686E+00 | 2.90E-04 | 1.67620202E+04 | 1.67620196E+04 | 5.80E-04 | 1.67661103E+04 | 1.67661097E+04 | 5.80E-04 | -7.82815580E-01 | -7.83397547E-01 | 5.80E-04 |
| 81 | O | 19 | 8 | 11 | 3 | 7.56649400E+00 | 7.56459634E+00 | 1.90E-03 | 1.76976299E+04 | 1.76976602E+04 | -3.00E-02 | 1.77017200E+04 | 1.77017503E+04 | -3.00E-02 | 3.33285800E+00 | 3.36311459E+00 | -3.00E-02 |
| 82 | O | 20 | 8 | 12 | 4 | 7.56857000E+00 | 7.55868505E+00 | 9.90E-03 | 1.86295873E+04 | 1.86297792E+04 | -1.90E-01 | 1.86336774E+04 | 1.86338693E+04 | -1.90E-01 | 3.79616800E+00 | 3.98806296E+00 | -1.90E-01 |
| 83 | O | 21 | 8 | 13 | 5 | 7.38938000E+00 | 7.38290833E+00 | 6.50E-03 | 1.95653471E+04 | 1.95654772E+04 | -1.30E-01 | 1.95694372E+04 | 1.95695673E+04 | -1.30E-01 | 8.06190700E+00 | 8.19200799E+00 | -1.30E-01 |
| 84 | O | 22 | 8 | 14 | 6 | 7.36485800E+00 | 7.32507236E+00 | 4.00E-02 | 2.04980626E+04 | 2.04989321E+04 | -8.70E-01 | 2.05021527E+04 | 2.05030222E+04 | -8.70E-01 | 9.28332200E+00 | 1.01528100E+01 | -8.70E-01 |
| 85 | O | 23 | 8 | 15 | 7 | 7.16351600E+00 | 7.11780637E+00 | 4.60E-02 | 2.14348940E+04 | 2.14359395E+04 | -1.00E+00 | 2.14389841E+04 | 2.14400296E+04 | -1.00E+00 | 1.46206570E+01 | 1.56661744E+01 | -1.00E+00 |
| 86 | O | 24 | 8 | 16 | 8 | 7.03968500E+00 | 6.99963002E+00 | 4.00E-02 | 2.23702678E+04 | 2.23712233E+04 | -9.60E-01 | 2.23743579E+04 | 2.23753134E+04 | -9.60E-01 | 1.85004020E+01 | 1.94559196E+01 | -9.60E-01 |
| 87 | O | 26 | 8 | 18 | 10 | 6.49470900E+00 | 6.46256808E+00 | 3.20E-02 | 2.42494885E+04 | 2.42503184E+04 | -8.30E-01 | 2.42535786E+04 | 2.42544085E+04 | -8.30E-01 | 3.47330370E+01 | 3.55629080E+01 | -8.30E-01 |
| 88 | F | 14 | 9 | 5 | -4 | 5.28520800E+00 | 5.32655699E+00 | -4.10E-02 | 1.30682795E+04 | 1.30676941E+04 | 5.90E-01 | 1.30728813E+04 | 1.30722958E+04 | 5.90E-01 | 3.19644100E+01 | 3.13789751E+01 | 5.90E-01 |
| 89 | F | 15 | 9 | 6 | -3 | 6.48145500E+00 | 6.42989698E+00 | 5.20E-02 | 1.39846160E+04 | 1.39853828E+04 | -7.70E-01 | 1.39892177E+04 | 1.39899846E+04 | -7.70E-01 | 1.68068100E+01 | 1.75736373E+01 | -7.70E-01 |
| 90 | F | 16 | 9 | 7 | -2 | 6.96373100E+00 | 6.91302762E+00 | 5.10E-02 | 1.49099835E+04 | 1.49107882E+04 | -8.00E-01 | 1.49145852E+04 | 1.49153899E+04 | -8.00E-01 | 1.06802540E+01 | 1.14849691E+01 | -8.00E-01 |
| 91 | F | 17 | 9 | 8 | -1 | 7.54232800E+00 | 7.49304608E+00 | 4.90E-02 | 1.58327490E+04 | 1.58335802E+04 | -8.30E-01 | 1.58373507E+04 | 1.58381820E+04 | -8.30E-01 | 1.95170100E+00 | 2.78294659E+00 | -8.30E-01 |
| 92 | F | 18 | 9 | 9 | 0 | 7.63163800E+00 | 7.63054381E+00 | 1.10E-03 | 1.67631645E+04 | 1.67631776E+04 | -1.30E-02 | 1.67677662E+04 | 1.67677794E+04 | -1.30E-02 | 8.73113000E-01 | 8.86260349E-01 | -1.30E-02 |
| 93 | F | 19 | 9 | 10 | 1 | 7.77901800E+00 | 7.80030381E+00 | -2.10E-02 | 1.76922980E+04 | 1.76918870E+04 | 4.10E-01 | 1.76968997E+04 | 1.76964888E+04 | 4.10E-01 | -1.48744434E+00 | -1.89840441E+00 | 4.10E-01 |
| 94 | F | 20 | 9 | 11 | 2 | 7.72013400E+00 | 7.68898967E+00 | 3.10E-02 | 1.86252620E+04 | 1.86258784E+04 | -6.20E-01 | 1.86298638E+04 | 1.86304801E+04 | -6.20E-01 | -1.74630000E-02 | 5.98893527E-01 | -6.20E-01 |
| 95 | F | 21 | 9 | 12 | 3 | 7.73829300E+00 | 7.73711867E+00 | 1.20E-03 | 1.95567259E+04 | 1.95567440E+04 | -1.80E-02 | 1.95613277E+04 | 1.95613458E+04 | -1.80E-02 | -4.76090000E-02 | -2.94859753E-02 | -1.80E-02 |
| 96 | F | 22 | 9 | 13 | 4 | 7.62429500E+00 | 7.62355917E+00 | 7.40E-04 | 2.04910610E+04 | 2.04910706E+04 | -9.60E-03 | 2.04956627E+04 | 2.04956724E+04 | -9.60E-03 | 2.79337700E+00 | 2.80302333E+00 | -9.60E-03 |
| 97 | F | 23 | 9 | 14 | 5 | 7.62113600E+00 | 7.62646128E+00 | -5.30E-03 | 2.14230747E+04 | 2.14229457E+04 | 1.30E-01 | 2.14276764E+04 | 2.14275474E+04 | 1.30E-01 | 3.31304200E+00 | 3.18403462E+00 | 1.30E-01 |
| 98 | F | 24 | 9 | 15 | 6 | 7.46295700E+00 | 7.47335648E+00 | -1.00E-02 | 2.23588152E+04 | 2.23585591E+04 | 2.60E-01 | 2.23634170E+04 | 2.23631609E+04 | 2.60E-01 | 7.55952700E+00 | 7.30340744E+00 | 2.60E-01 |
| 99 | F | 25 | 9 | 16 | 7 | 7.33513200E+00 | 7.34708679E+00 | -1.20E-02 | 2.32941133E+04 | 2.32938079E+04 | 3.10E-01 | 2.32987150E+04 | 2.32984096E+04 | 3.10E-01 | 1.13635090E+01 | 1.10581123E+01 | 3.10E-01 |
| 100 | F | 26 | 9 | 17 | 8 | 7.08261800E+00 | 7.07411821E+00 | 8.50E-03 | 2.42329089E+04 | 2.42331234E+04 | -2.10E-01 | 2.42375107E+04 | 2.42377251E+04 | -2.10E-01 | 1.86650610E+01 | 1.88795276E+01 | -2.10E-01 |
| 101 | F | 27 | 9 | 18 | 9 | 6.89832600E+00 | 6.90084342E+00 | -2.50E-03 | 2.51703675E+04 | 2.51702930E+04 | 7.40E-02 | 2.51749693E+04 | 2.51748948E+04 | 7.40E-02 | 2.46296330E+01 | 2.45551478E+01 | 7.40E-02 |
| 102 | F | 28 | 9 | 19 | 10 | 6.64410000E+00 | 6.65088471E+00 | -6.80E-03 | 2.61101529E+04 | 2.61099564E+04 | 2.00E-01 | 2.61147547E+04 | 2.61145582E+04 | 2.00E-01 | 3.29209500E+01 | 3.27244671E+01 | 2.00E-01 |
| 103 | F | 29 | 9 | 20 | 11 | 6.46200000E+00 | 6.47978788E+00 | -1.80E-02 | 2.70483518E+04 | 2.70478327E+04 | 5.20E-01 | 2.70529535E+04 | 2.70524345E+04 | 5.20E-01 | 3.96260000E+01 | 3.91067094E+01 | 5.20E-01 |
| 104 | F | 30 | 9 | 21 | 12 | 6.23300000E+00 | 6.24804382E+00 | -1.50E-02 | 2.79883318E+04 | 2.79878706E+04 | 4.60E-01 | 2.79929335E+04 | 2.79924724E+04 | 4.60E-01 | 4.81120000E+01 | 4.76505624E+01 | 4.60E-01 |
| 105 | F | 31 | 9 | 22 | 13 | 6.05000000E+00 | 6.07402008E+00 | -2.40E-02 | 2.89273327E+04 | 2.89265827E+04 | 7.50E-01 | 2.89319345E+04 | 2.89311845E+04 | 7.50E-01 | 5.56180000E+01 | 5.48685735E+01 | 7.50E-01 |
| 106 | Ne | 16 | 10 | 6 | -4 | 6.08321600E+00 | 6.07682465E+00 | 6.40E-03 | 1.49227776E+04 | 1.49228726E+04 | -9.50E-02 | 1.49278911E+04 | 1.49279861E+04 | -9.50E-02 | 2.39861540E+01 | 2.40811395E+01 | -9.50E-02 |
| 107 | Ne | 17 | 10 | 7 | -3 | 6.64049900E+00 | 6.70306311E+00 | -6.30E-02 | 1.58467859E+04 | 1.58457151E+04 | 1.10E+00 | 1.58518995E+04 | 1.58508286E+04 | 1.10E+00 | 1.65004510E+01 | 1.54295802E+01 | 1.10E+00 |
| 108 | Ne | 18 | 10 | 8 | -2 | 7.34125700E+00 | 7.37854861E+00 | -3.70E-02 | 1.67670972E+04 | 1.67664187E+04 | 6.80E-01 | 1.67722107E+04 | 1.67715322E+04 | 6.80E-01 | 5.31762300E+00 | 4.63909708E+00 | 6.80E-01 |
| 109 | Ne | 19 | 10 | 9 | -1 | 7.56734200E+00 | 7.64705083E+00 | -8.00E-02 | 1.76950257E+04 | 1.76935040E+04 | 1.50E+00 | 1.77001392E+04 | 1.76986175E+04 | 1.50E+00 | 1.75205400E+00 | 2.30325154E-01 | 1.50E+00 |
| 110 | Ne | 20 | 10 | 10 | 0 | 8.03224000E+00 | 8.08622606E+00 | -5.40E-02 | 1.86177258E+04 | 1.86166388E+04 | 1.10E+00 | 1.86228393E+04 | 1.86217523E+04 | 1.10E+00 | -7.04193055E+00 | -8.12891117E+00 | 1.10E+00 |
| 111 | Ne | 21 | 10 | 11 | 1 | 7.97171300E+00 | 7.98427151E+00 | -1.30E-02 | 1.95505300E+04 | 1.95502590E+04 | 2.70E-01 | 1.95556435E+04 | 1.95553725E+04 | 2.70E-01 | -5.73177600E+00 | -6.00277275E+00 | 2.70E-01 |
| 112 | Ne | 22 | 10 | 12 | 2 | 8.08046500E+00 | 8.07195717E+00 | 8.50E-03 | 2.04797311E+04 | 2.04799110E+04 | -1.80E-01 | 2.04848446E+04 | 2.04850245E+04 | -1.80E-01 | -8.02471400E+00 | -7.84480977E+00 | -1.80E-01 |
| 113 | Ne | 23 | 10 | 13 | 3 | 7.95525500E+00 | 7.95033080E+00 | 4.90E-03 | 2.14140958E+04 | 2.14142018E+04 | -1.10E-01 | 2.14192094E+04 | 2.14193154E+04 | -1.10E-01 | -5.15404400E+00 | -5.04804136E+00 | -1.10E-01 |
| 114 | Ne | 24 | 10 | 14 | 4 | 7.99332400E+00 | 7.98614583E+00 | 7.20E-03 | 2.23447923E+04 | 2.23449573E+04 | -1.70E-01 | 2.23499058E+04 | 2.23500709E+04 | -1.70E-01 | -5.95164100E+00 | -5.78661383E+00 | -1.70E-01 |
| 115 | Ne | 25 | 10 | 15 | 5 | 7.84077100E+00 | 7.83812931E+00 | 2.60E-03 | 2.32801782E+04 | 2.32802370E+04 | -5.90E-02 | 2.32852917E+04 | 2.32853505E+04 | -5.90E-02 | -2.05980600E+00 | -2.00102775E+00 | -5.90E-02 |
| 116 | Ne | 26 | 10 | 16 | 6 | 7.75197400E+00 | 7.77254440E+00 | -2.10E-02 | 2.42142115E+04 | 2.42136694E+04 | 5.40E-01 | 2.42193250E+04 | 2.42187830E+04 | 5.40E-01 | 4.79445000E-01 | -6.26303101E-02 | 5.40E-01 |
| 117 | Ne | 27 | 10 | 17 | 7 | 7.52097300E+00 | 7.55092211E+00 | -3.00E-02 | 2.51522619E+04 | 2.51514461E+04 | 8.20E-01 | 2.51573755E+04 | 2.51565596E+04 | 8.20E-01 | 7.03582400E+00 | 6.21994611E+00 | 8.20E-01 |
| 118 | Ne | 28 | 10 | 18 | 8 | 7.38863700E+00 | 7.43141640E+00 | -4.30E-02 | 2.60880118E+04 | 2.60868067E+04 | 1.20E+00 | 2.60931253E+04 | 2.60919202E+04 | 1.20E+00 | 1.12915690E+01 | 1.00865027E+01 | 1.20E+00 |
| 119 | Ne | 29 | 10 | 19 | 9 | 7.16706700E+00 | 7.18224102E+00 | -1.50E-02 | 2.70266140E+04 | 2.70261667E+04 | 4.50E-01 | 2.70317276E+04 | 2.70312803E+04 | 4.50E-01 | 1.83998010E+01 | 1.79524915E+01 | 4.50E-01 |
| 120 | Ne | 30 | 10 | 20 | 10 | 7.04254900E+00 | 7.06676822E+00 | -2.40E-02 | 2.79627479E+04 | 2.79620140E+04 | 7.30E-01 | 2.79678614E+04 | 2.79671276E+04 | 7.30E-01 | 2.30395730E+01 | 2.23057535E+01 | 7.30E-01 |
| 121 | Ne | 31 | 10 | 21 | 11 | 6.82474300E+00 | 6.80648950E+00 | 1.80E-02 | 2.89020227E+04 | 2.89025813E+04 | -5.60E-01 | 2.89071362E+04 | 2.89076948E+04 | -5.60E-01 | 3.08203420E+01 | 3.13789445E+01 | -5.60E-01 |
| 122 | Ne | 32 | 10 | 22 | 12 | 6.67100000E+00 | 6.67419708E+00 | -3.20E-03 | 2.98396954E+04 | 2.98395735E+04 | 1.20E-01 | 2.98448089E+04 | 2.98446871E+04 | 1.20E-01 | 3.69990000E+01 | 3.68771315E+01 | 1.20E-01 |
| 123 | Ne | 33 | 10 | 23 | 13 | 6.44000000E+00 | 6.45429691E+00 | -1.40E-02 | 3.07801877E+04 | 3.07797214E+04 | 4.70E-01 | 3.07853012E+04 | 3.07848350E+04 | 4.70E-01 | 4.59970000E+01 | 4.55309591E+01 | 4.70E-01 |
| 124 | Ne | 34 | 10 | 24 | 14 | 6.28700000E+00 | 6.30095754E+00 | -1.40E-02 | 3.17185263E+04 | 3.17180461E+04 | 4.80E-01 | 3.17236399E+04 | 3.17231596E+04 | 4.80E-01 | 5.28420000E+01 | 5.23615197E+01 | 4.80E-01 |
| 125 | NA | 18 | 11 | 7 | -4 | 6.20227600E+00 | 6.25192741E+00 | -5.00E-02 | 1.67863046E+04 | 1.67854029E+04 | 9.00E-01 | 1.67919300E+04 | 1.67910283E+04 | 9.00E-01 | 2.50369310E+01 | 2.41352017E+01 | 9.00E-01 |
| 126 | NA | 19 | 11 | 8 | -3 | 6.93788500E+00 | 6.95381791E+00 | -1.60E-02 | 1.77056911E+04 | 1.77053804E+04 | 3.10E-01 | 1.77113165E+04 | 1.77110058E+04 | 3.10E-01 | 1.29293910E+01 | 1.26186736E+01 | 3.10E-01 |
| 127 | Na | 20 | 11 | 9 | -2 | 7.29849600E+00 | 7.34970684E+00 | -5.10E-02 | 1.86311064E+04 | 1.86300742E+04 | 1.00E+00 | 1.86367318E+04 | 1.86356996E+04 | 1.00E+00 | 6.85060400E+00 | 5.81839625E+00 | 1.00E+00 |
| 128 | Na | 21 | 11 | 10 | -1 | 7.76554700E+00 | 7.78293940E+00 | -1.70E-02 | 1.95535652E+04 | 1.95531920E+04 | 3.70E-01 | 1.95591907E+04 | 1.95588174E+04 | 3.70E-01 | -2.18463700E+00 | -2.55787531E+00 | 3.70E-01 |
| 129 | Na | 22 | 11 | 11 | 0 | 7.91566700E+00 | 7.90305220E+00 | 1.30E-02 | 2.04820624E+04 | 2.04823319E+04 | -2.70E-01 | 2.04876878E+04 | 2.04879574E+04 | -2.70E-01 | -5.18151800E+00 | -4.91197733E+00 | -2.70E-01 |
| 130 | Na | 23 | 11 | 12 | 1 | 8.11149300E+00 | 8.11367629E+00 | -2.20E-03 | 2.14092081E+04 | 2.14091499E+04 | 5.80E-02 | 2.14148335E+04 | 2.14147753E+04 | 5.80E-02 | -9.52985249E+00 | -9.58806461E+00 | 5.80E-02 |
| 131 | Na | 24 | 11 | 13 | 2 | 8.06349000E+00 | 8.08657279E+00 | -2.30E-02 | 2.23418141E+04 | 2.23412521E+04 | 5.60E-01 | 2.23474395E+04 | 2.23468775E+04 | 5.60E-01 | -8.41795800E+00 | -8.97993792E+00 | 5.60E-01 |
| 132 | Na | 25 | 11 | 14 | 3 | 8.10139700E+00 | 8.10161548E+00 | -2.20E-04 | 2.32723683E+04 | 2.32723548E+04 | 1.30E-02 | 2.32779937E+04 | 2.32779803E+04 | 1.30E-02 | -9.35781700E+00 | -9.37125908E+00 | 1.30E-02 |
| 133 | Na | 26 | 11 | 15 | 4 | 8.00420100E+00 | 7.98473367E+00 | 1.90E-02 | 2.42063594E+04 | 2.42068575E+04 | -5.00E-01 | 2.42119848E+04 | 2.42124830E+04 | -5.00E-01 | -6.86078000E+00 | -6.36262844E+00 | -5.00E-01 |
| 134 | Na | 27 | 11 | 16 | 5 | 7.95694200E+00 | 7.95227382E+00 | 4.70E-03 | 2.51391965E+04 | 2.51393146E+04 | -1.20E-01 | 2.51448220E+04 | 2.51449400E+04 | -1.20E-01 | -5.51767600E+00 | -5.39962701E+00 | -1.20E-01 |

| | | | | | | | | | | | | | | | | |
|---|---|---|---|---|---|---|---|---|---|---|---|---|---|---|---|---|
| 135 | Na | 28 | 11 | 17 | 6 | 7.79926400E+00 | 7.79512315E+00 | 4.10E-03 | 2.60752199E+04 | 2.60753279E+04 | -1.10E-01 | 2.60808454E+04 | 2.60809533E+04 | -1.10E-01 | -9.88315000E-01 | -8.80363236E-01 | -1.10E-01 |
| 136 | Na | 29 | 11 | 18 | 7 | 7.68215200E+00 | 7.69871175E+00 | -1.70E-02 | 2.70103823E+04 | 2.70098941E+04 | 4.90E-01 | 2.70160077E+04 | 2.70155195E+04 | 4.90E-01 | 2.67997600E+00 | 2.19176334E+00 | 4.90E-01 |
| 137 | Na | 30 | 11 | 19 | 8 | 7.50196800E+00 | 7.50607436E+00 | -4.10E-03 | 2.79476711E+04 | 2.79475399E+04 | 1.30E-01 | 2.79532965E+04 | 2.79531653E+04 | 1.30E-01 | 8.47467000E+00 | 8.34349228E+00 | 1.30E-01 |
| 138 | Na | 31 | 11 | 20 | 9 | 7.39819600E+00 | 7.37726131E+00 | 2.10E-02 | 2.88829514E+04 | 2.88835924E+04 | -6.40E-01 | 2.88885768E+04 | 2.88892178E+04 | -6.40E-01 | 1.22609350E+01 | 1.29019413E+01 | -6.40E-01 |
| 139 | Na | 32 | 11 | 21 | 10 | 7.21458400E+00 | 7.17905221E+00 | 3.60E-02 | 2.98209942E+04 | 2.98221232E+04 | -1.10E+00 | 2.98266196E+04 | 2.98277486E+04 | -1.10E+00 | 1.88096580E+01 | 1.99386902E+01 | -1.10E+00 |
| 140 | Na | 33 | 11 | 22 | 11 | 7.08400000E+00 | 7.04486904E+00 | 3.90E-02 | 3.07576459E+04 | 3.07589376E+04 | -1.30E+00 | 3.07632714E+04 | 3.07645630E+04 | -1.30E+00 | 2.39670000E+01 | 2.52590017E+01 | -1.30E+00 |
| 141 | Na | 34 | 11 | 23 | 12 | 6.89800000E+00 | 6.85872428E+00 | 3.90E-02 | 3.16964615E+04 | 3.16977870E+04 | -1.30E+00 | 3.17020870E+04 | 3.17034124E+04 | -1.30E+00 | 3.12890000E+01 | 3.26143736E+01 | -1.30E+00 |
| 142 | Na | 35 | 11 | 24 | 13 | 6.74400000E+00 | 6.73572074E+00 | 8.30E-03 | 3.26345068E+04 | 3.26347988E+04 | -2.90E-01 | 3.26401322E+04 | 3.26404242E+04 | -2.90E-01 | 3.78400000E+01 | 3.81320922E+01 | -2.90E-01 |
| 143 | Na | 36 | 11 | 25 | 14 | 6.55700000E+00 | 6.56108703E+00 | -4.10E-03 | 3.35740722E+04 | 3.35739153E+04 | 1.60E-01 | 3.35796977E+04 | 3.35795407E+04 | 1.60E-01 | 4.59120000E+01 | 4.57545041E+01 | 1.60E-01 |
| 144 | Na | 37 | 11 | 26 | 15 | 6.40200000E+00 | 6.42438117E+00 | -2.20E-02 | 3.45127975E+04 | 3.45119777E+04 | 8.20E-01 | 3.45184229E+04 | 3.45176031E+04 | 8.20E-01 | 5.31430000E+01 | 5.23228525E+01 | 8.20E-01 |
| 145 | MG | 19 | 12 | 7 | -5 | 5.90202500E+00 | 5.97075314E+00 | -6.90E-02 | 1.77240781E+04 | 1.77227635E+04 | 1.30E+00 | 1.77302156E+04 | 1.77289010E+04 | 1.30E+00 | 3.18283920E+01 | 3.05138274E+01 | 1.30E+00 |
| 146 | Mg | 20 | 12 | 8 | -4 | 6.72397600E+00 | 6.73764160E+00 | -1.40E-02 | 1.86413024E+04 | 1.86410204E+04 | 2.80E-01 | 1.86474399E+04 | 1.86471578E+04 | 2.80E-01 | 1.75586670E+01 | 1.72766240E+01 | 2.80E-01 |
| 147 | Mg | 21 | 12 | 9 | -3 | 7.10457100E+00 | 7.16123521E+00 | -5.70E-02 | 1.95661513E+04 | 1.95649526E+04 | 1.20E+00 | 1.95722888E+04 | 1.95710901E+04 | 1.20E+00 | 1.09135150E+01 | 9.71483565E+00 | 1.20E+00 |
| 148 | Mg | 22 | 12 | 10 | -2 | 7.66276200E+00 | 7.67034650E+00 | -7.60E-03 | 2.04863319E+04 | 2.04861563E+04 | 1.80E-01 | 2.04924694E+04 | 2.04922938E+04 | 1.80E-01 | -3.99939000E-01 | -5.75528926E-01 | 1.80E-01 |
| 149 | Mg | 23 | 12 | 11 | -1 | 7.90110400E+00 | 7.89639135E+00 | 4.70E-03 | 2.14127527E+04 | 2.14128523E+04 | -1.00E-01 | 2.14188901E+04 | 2.14189898E+04 | -1.00E-01 | -5.47326400E+00 | -5.37358799E+00 | -1.00E-01 |
| 150 | Mg | 24 | 12 | 12 | 0 | 8.26070900E+00 | 8.23532735E+00 | 2.50E-02 | 2.23357864E+04 | 2.23363869E+04 | -6.00E-01 | 2.23419239E+04 | 2.23425243E+04 | -6.00E-01 | -1.39335690E+01 | -1.33331245E+01 | -6.00E-01 |
| 151 | Mg | 25 | 12 | 13 | 1 | 8.22350200E+00 | 8.23853082E+00 | -1.50E-02 | 2.32680213E+04 | 2.32676368E+04 | 3.80E-01 | 2.32741588E+04 | 2.32737743E+04 | 3.80E-01 | -1.31927710E+01 | -1.35772194E+01 | 3.80E-01 |
| 152 | Mg | 26 | 12 | 14 | 2 | 8.33387000E+00 | 8.39755684E+00 | -6.40E-02 | 2.41964936E+04 | 2.41948290E+04 | 1.70E+00 | 2.42026310E+04 | 2.42009665E+04 | 1.70E+00 | -1.62145460E+01 | -1.78791079E+01 | 1.70E+00 |
| 153 | Mg | 27 | 12 | 15 | 3 | 8.26385300E+00 | 8.28916681E+00 | -2.50E-02 | 2.51296156E+04 | 2.51289234E+04 | 6.90E-01 | 2.51357530E+04 | 2.51350608E+04 | 6.90E-01 | -1.45866140E+01 | -1.52788148E+01 | 6.90E-01 |
| 154 | Mg | 28 | 12 | 16 | 4 | 8.27240900E+00 | 8.28402708E+00 | -1.20E-02 | 2.60606775E+04 | 2.60603435E+04 | 3.30E-01 | 2.60668150E+04 | 2.60664810E+04 | 3.30E-01 | -1.50187300E+01 | -1.53527503E+01 | 3.30E-01 |
| 155 | Mg | 29 | 12 | 17 | 5 | 8.11320200E+00 | 8.12482968E+00 | -1.20E-02 | 2.69965875E+04 | 2.69962416E+04 | 3.50E-01 | 2.70027249E+04 | 2.70023790E+04 | 3.50E-01 | -1.06028290E+01 | -1.09487338E+01 | 3.50E-01 |
| 156 | Mg | 30 | 12 | 18 | 6 | 8.05450300E+00 | 8.06543824E+00 | -1.10E-02 | 2.79298006E+04 | 2.79294639E+04 | 3.40E-01 | 2.79359381E+04 | 2.79356013E+04 | 3.40E-01 | -8.88372700E+00 | -9.22050128E+00 | 3.40E-01 |
| 157 | Mg | 31 | 12 | 19 | 7 | 7.86919400E+00 | 7.88995963E+00 | -2.10E-02 | 2.88670561E+04 | 2.88664036E+04 | 6.50E-01 | 2.88731935E+04 | 2.88725411E+04 | 6.50E-01 | -3.12233700E+00 | -3.77478358E+00 | 6.50E-01 |
| 158 | Mg | 32 | 12 | 20 | 8 | 7.80383700E+00 | 7.80207641E+00 | 1.80E-03 | 2.98008437E+04 | 2.98008913E+04 | -4.80E-02 | 2.98069811E+04 | 2.98070288E+04 | -4.80E-02 | -8.28807000E-01 | -7.81161008E-01 | -4.80E-02 |
| 159 | Mg | 33 | 12 | 21 | 9 | 7.63645800E+00 | 7.61359865E+00 | 2.30E-02 | 3.07381287E+04 | 3.07388744E+04 | -7.50E-01 | 3.07442662E+04 | 3.07450119E+04 | -7.50E-01 | 4.96220400E+00 | 5.70784749E+00 | -7.50E-01 |
| 160 | Mg | 34 | 12 | 22 | 10 | 7.55039000E+00 | 7.51718940E+00 | 3.30E-02 | 3.16729840E+04 | 3.16741041E+04 | -1.10E+00 | 3.16791214E+04 | 3.16802416E+04 | -1.10E+00 | 8.32334700E+00 | 9.44348244E+00 | -1.10E+00 |
| 161 | Mg | 35 | 12 | 23 | 11 | 7.35623300E+00 | 7.32193131E+00 | 3.40E-02 | 3.26117945E+04 | 3.26129863E+04 | -1.20E+00 | 3.26179319E+04 | 3.26191238E+04 | -1.20E+00 | 1.56397840E+01 | 1.68316453E+01 | -1.20E+00 |
| 162 | Mg | 36 | 12 | 24 | 12 | 7.24441900E+00 | 7.24169994E+00 | 2.70E-03 | 3.35480289E+04 | 3.35481181E+04 | -8.90E-02 | 3.35541663E+04 | 3.35542556E+04 | -8.90E-02 | 2.03801570E+01 | 2.04693621E+01 | -8.90E-02 |
| 163 | Mg | 37 | 12 | 25 | 13 | 7.05300000E+00 | 7.04430732E+00 | 8.70E-03 | 3.44874323E+04 | 3.44877453E+04 | -3.10E-01 | 3.44935697E+04 | 3.44938828E+04 | -3.10E-01 | 2.82890000E+01 | 2.86025080E+01 | -3.10E-01 |
| 164 | Mg | 38 | 12 | 26 | 14 | 6.92800000E+00 | 6.94282859E+00 | -1.50E-02 | 3.54247109E+04 | 3.54241226E+04 | 5.90E-01 | 3.54308484E+04 | 3.54302600E+04 | 5.90E-01 | 3.40740000E+01 | 3.34857115E+01 | 5.90E-01 |
| 165 | Mg | 39 | 12 | 27 | 15 | 6.74700000E+00 | 6.75587793E+00 | -8.90E-03 | 3.63644058E+04 | 3.63640362E+04 | 3.70E-01 | 3.63705433E+04 | 3.63701737E+04 | 3.70E-01 | 4.22750000E+01 | 4.19052779E+01 | 3.70E-01 |
| 166 | Mg | 40 | 12 | 28 | 16 | 6.62100000E+00 | 6.61754286E+00 | 3.50E-03 | 3.73022303E+04 | 3.73023791E+04 | -1.50E-01 | 3.73083678E+04 | 3.73085166E+04 | -1.50E-01 | 4.86050000E+01 | 4.87541218E+01 | -1.50E-01 |
| 167 | Al | 21 | 13 | 8 | -5 | 6.30200000E+00 | 6.26601769E+00 | 3.60E-02 | 1.95817157E+04 | 1.95824570E+04 | -7.40E-01 | 1.95883653E+04 | 1.95891066E+04 | -7.40E-01 | 2.69900000E+01 | 2.77313264E+01 | -7.40E-01 |
| 168 | Al | 22 | 13 | 9 | -4 | 6.78200000E+00 | 6.79819967E+00 | -1.60E-02 | 2.05044211E+04 | 2.05040483E+04 | 3.70E-01 | 2.05110707E+04 | 2.05106980E+04 | 3.70E-01 | 1.82010000E+01 | 1.78286243E+01 | 3.70E-01 |
| 169 | Al | 23 | 13 | 10 | -3 | 7.33572700E+00 | 7.35455044E+00 | -1.90E-02 | 2.14244618E+04 | 2.14240195E+04 | 4.40E-01 | 2.14311115E+04 | 2.14306691E+04 | 4.40E-01 | 6.74807000E+00 | 6.30567581E+00 | 4.40E-01 |
| 170 | Al | 24 | 13 | 11 | -2 | 7.64953000E+00 | 7.65872342E+00 | -9.20E-03 | 2.23491602E+04 | 2.23489301E+04 | 2.30E-01 | 2.23558099E+04 | 2.23555798E+04 | 2.30E-01 | -4.76140000E-02 | -2.77707093E-01 | 2.30E-01 |
| 171 | Al | 25 | 13 | 12 | -1 | 8.02114400E+00 | 8.02607761E+00 | -4.90E-03 | 2.32717857E+04 | 2.32716529E+04 | 1.30E-01 | 2.32784354E+04 | 2.32783026E+04 | 1.30E-01 | -8.91616500E+00 | -9.04896617E+00 | 1.30E-01 |
| 172 | Al | 26 | 13 | 13 | 0 | 8.14976400E+00 | 8.15489448E+00 | -5.10E-03 | 2.41999859E+04 | 2.41998430E+04 | 1.40E-01 | 2.42066355E+04 | 2.42064926E+04 | 1.40E-01 | -1.22101120E+01 | -1.23529635E+01 | 1.40E-01 |
| 173 | Al | 27 | 13 | 14 | 1 | 8.33154800E+00 | 8.33932757E+00 | -7.80E-03 | 2.51264933E+04 | 2.51262738E+04 | 2.20E-01 | 2.51331429E+04 | 2.51329234E+04 | 2.20E-01 | -1.71967470E+01 | -1.74162323E+01 | 2.20E-01 |
| 174 | Al | 28 | 13 | 15 | 2 | 8.30988900E+00 | 8.34212754E+00 | -3.20E-02 | 2.60583335E+04 | 2.60574214E+04 | 9.10E-01 | 2.60649832E+04 | 2.60640711E+04 | 9.10E-01 | -1.68505300E+01 | -1.77626400E+01 | 9.10E-01 |
| 175 | Al | 29 | 13 | 16 | 3 | 8.34835700E+00 | 8.38576457E+00 | -3.70E-02 | 2.69884735E+04 | 2.69873792E+04 | 1.10E+00 | 2.69951231E+04 | 2.69940288E+04 | 1.10E+00 | -1.82046620E+01 | -1.92989224E+01 | 1.10E+00 |
| 176 | Al | 30 | 13 | 17 | 4 | 8.26138200E+00 | 8.28107111E+00 | -2.00E-02 | 2.79222997E+04 | 2.79216996E+04 | 6.00E-01 | 2.79289494E+04 | 2.79283493E+04 | 6.00E-01 | -1.58724520E+01 | -1.64725643E+01 | 6.00E-01 |
| 177 | Al | 31 | 13 | 18 | 5 | 8.22565500E+00 | 8.23323925E+00 | -7.60E-03 | 2.88547113E+04 | 2.88544667E+04 | 2.40E-01 | 2.88613609E+04 | 2.88611164E+04 | 2.40E-01 | -1.49549760E+01 | -1.51995287E+01 | 2.40E-01 |
| 178 | Al | 32 | 13 | 19 | 6 | 8.10031800E+00 | 8.09516885E+00 | 5.10E-03 | 2.97900618E+04 | 2.97902171E+04 | -1.60E-01 | 2.97967114E+04 | 2.97968668E+04 | -1.60E-01 | -1.10985280E+01 | -1.09431961E+01 | -1.60E-01 |
| 179 | Al | 33 | 13 | 20 | 7 | 8.01973300E+00 | 8.02036103E+00 | -6.30E-04 | 3.07241862E+04 | 3.07241560E+04 | 3.00E-02 | 3.07308358E+04 | 3.07308056E+04 | 3.00E-02 | -8.46823300E+00 | -8.49838800E+00 | 3.00E-02 |
| 180 | Al | 34 | 13 | 21 | 8 | 7.86244600E+00 | 7.87327717E+00 | -1.10E-02 | 3.16610795E+04 | 3.16607019E+04 | 3.80E-01 | 3.16677292E+04 | 3.16673515E+04 | 3.80E-01 | -3.06890100E+00 | -3.44657876E+00 | 3.80E-01 |
| 181 | Al | 35 | 13 | 22 | 9 | 7.78701200E+00 | 7.78093613E+00 | 6.10E-03 | 3.25954227E+04 | 3.25956259E+04 | -2.00E-01 | 3.26020723E+04 | 3.26022755E+04 | -2.00E-01 | -2.19833000E-01 | -1.66005573E-02 | -2.00E-01 |

| | | | | | | | | | | | | | | | | |
|---|---|---|---|---|---|---|---|---|---|---|---|---|---|---|---|---|
| 182 | Al | 36 | 13 | 23 | 10 | 7.62351500E+00 | 7.62778042E+00 | -4.30E-03 | 3.35330870E+04 | 3.35329240E+04 | 1.60E-01 | 3.35397366E+04 | 3.35395736E+04 | 1.60E-01 | 5.95038400E+00 | 5.78738791E+00 | 1.60E-01 |
| 183 | Al | 37 | 13 | 24 | 11 | 7.53131500E+00 | 7.53576056E+00 | -4.40E-03 | 3.44684402E+04 | 3.44682663E+04 | 1.70E-01 | 3.44750898E+04 | 3.44749159E+04 | 1.70E-01 | 9.80956300E+00 | 9.63566120E+00 | 1.70E-01 |
| 184 | Al | 38 | 13 | 25 | 12 | 7.37709700E+00 | 7.38492704E+00 | -7.80E-03 | 3.54063346E+04 | 3.54060276E+04 | 3.10E-01 | 3.54129842E+04 | 3.54126772E+04 | 3.10E-01 | 1.62098590E+01 | 1.59028933E+01 | 3.10E-01 |
| 185 | Al | 39 | 13 | 26 | 13 | 7.27200000E+00 | 7.27738927E+00 | -5.40E-03 | 3.63426146E+04 | 3.63424020E+04 | 2.10E-01 | 3.63492643E+04 | 3.63490516E+04 | 2.10E-01 | 2.09960000E+01 | 2.07832584E+01 | 2.10E-01 |
| 186 | Al | 40 | 13 | 27 | 14 | 7.11800000E+00 | 7.12471815E+00 | -6.70E-03 | 3.72810856E+04 | 3.72807969E+04 | 2.90E-01 | 3.72877352E+04 | 3.72874465E+04 | 2.90E-01 | 2.79730000E+01 | 2.76840329E+01 | 2.90E-01 |
| 187 | Al | 41 | 13 | 28 | 15 | 6.99700000E+00 | 7.00635800E+00 | -9.40E-03 | 3.82184946E+04 | 3.82180903E+04 | 4.00E-01 | 3.82251442E+04 | 3.82247399E+04 | 4.00E-01 | 3.38880000E+01 | 3.34834000E+01 | 4.00E-01 |
| 188 | Al | 42 | 13 | 29 | 16 | 6.85700000E+00 | 6.84554690E+00 | 1.10E-02 | 3.91569376E+04 | 3.91574034E+04 | -4.70E-01 | 3.91635873E+04 | 3.91640530E+04 | -4.70E-01 | 4.08370000E+01 | 4.13024272E+01 | -4.70E-01 |
| 189 | AL | 43 | 13 | 30 | 17 | 6.72000000E+00 | 6.70761615E+00 | 1.20E-02 | 4.00955390E+04 | 4.00960542E+04 | -5.20E-01 | 4.01021886E+04 | 4.01027038E+04 | -5.20E-01 | 4.79440000E+01 | 4.84592214E+01 | -5.20E-01 |
| 190 | Si | 22 | 14 | 8 | -6 | 6.05800000E+00 | 6.08190606E+00 | -2.40E-02 | 2.05190456E+04 | 2.05185114E+04 | 5.30E-01 | 2.05262075E+04 | 2.05256733E+04 | 5.30E-01 | 3.33380000E+01 | 3.28040067E+01 | 5.30E-01 |
| 191 | Si | 23 | 14 | 9 | -5 | 6.56500000E+00 | 6.55815575E+00 | 6.80E-03 | 2.14408987E+04 | 2.14410412E+04 | -1.40E-01 | 2.14480606E+04 | 2.14482031E+04 | -1.40E-01 | 2.36970000E+01 | 2.38396768E+01 | -1.40E-01 |
| 192 | Si | 24 | 14 | 10 | -4 | 7.16726700E+00 | 7.16288017E+00 | 4.40E-03 | 2.23594399E+04 | 2.23595350E+04 | -9.50E-02 | 2.23666018E+04 | 2.23666969E+04 | -9.50E-02 | 1.07443530E+01 | 1.08394539E+01 | -9.50E-02 |
| 193 | Si | 25 | 14 | 11 | -3 | 7.48011000E+00 | 7.48775695E+00 | -7.60E-03 | 2.32840170E+04 | 2.32838156E+04 | 2.00E-01 | 2.32911789E+04 | 2.32909775E+04 | 2.00E-01 | 3.82733100E+00 | 3.62597326E+00 | 2.00E-01 |
| 194 | Si | 26 | 14 | 12 | -2 | 7.92470700E+00 | 7.92995405E+00 | -5.20E-03 | 2.42045427E+04 | 2.42043961E+04 | 1.50E-01 | 2.42117046E+04 | 2.42115580E+04 | 1.50E-01 | -7.14097700E+00 | -7.28758935E+00 | 1.50E-01 |
| 195 | Si | 27 | 14 | 13 | -1 | 8.12433700E+00 | 8.13365138E+00 | -9.30E-03 | 2.51307933E+04 | 2.51305317E+04 | 2.60E-01 | 2.51379553E+04 | 2.51376936E+04 | 2.60E-01 | -1.23843890E+01 | -1.26460524E+01 | 2.60E-01 |
| 196 | Si | 28 | 14 | 14 | 0 | 8.44774400E+00 | 8.41062299E+00 | 3.70E-02 | 2.60531790E+04 | 2.60542082E+04 | -1.00E+00 | 2.60603409E+04 | 2.60613701E+04 | -1.00E+00 | -2.14927946E+01 | -2.04635898E+01 | -1.00E+00 |
| 197 | Si | 29 | 14 | 15 | 1 | 8.44863500E+00 | 8.41804935E+00 | 3.10E-02 | 2.69842708E+04 | 2.69851476E+04 | -8.80E-01 | 2.69914327E+04 | 2.69923095E+04 | -8.80E-01 | -2.18950787E+01 | -2.10182581E+01 | -8.80E-01 |
| 198 | Si | 30 | 14 | 16 | 2 | 8.52065400E+00 | 8.53671426E+00 | -1.60E-02 | 2.79132270E+04 | 2.79127350E+04 | 4.90E-01 | 2.79203889E+04 | 2.79198969E+04 | 4.90E-01 | -2.44329610E+01 | -2.49249358E+01 | 4.90E-01 |
| 199 | Si | 31 | 14 | 17 | 3 | 8.45829100E+00 | 8.46167274E+00 | -3.40E-03 | 2.88462049E+04 | 2.88460899E+04 | 1.20E-01 | 2.88533668E+04 | 2.88532518E+04 | 1.20E-01 | -2.29490360E+01 | -2.30640438E+01 | 1.20E-01 |
| 200 | Si | 32 | 14 | 18 | 4 | 8.48146800E+00 | 8.47191050E+00 | 9.60E-03 | 2.97765704E+04 | 2.97768660E+04 | -3.00E-01 | 2.97837323E+04 | 2.97840279E+04 | -3.00E-01 | -2.40776860E+01 | -2.37820062E+01 | -3.00E-01 |
| 201 | Si | 33 | 14 | 19 | 5 | 8.36105900E+00 | 8.34388728E+00 | 1.70E-02 | 3.07116278E+04 | 3.07121843E+04 | -5.60E-01 | 3.07187897E+04 | 3.07193462E+04 | -5.60E-01 | -2.05143260E+01 | -1.99578311E+01 | -5.60E-01 |
| 202 | Si | 34 | 14 | 20 | 6 | 8.33613700E+00 | 8.30219793E+00 | 3.40E-02 | 3.16436794E+04 | 3.16448232E+04 | -1.10E+00 | 3.16508413E+04 | 3.16519851E+04 | -1.10E+00 | -1.99567290E+01 | -1.88129615E+01 | -1.10E+00 |
| 203 | Si | 35 | 14 | 21 | 7 | 8.16867600E+00 | 8.16454258E+00 | 4.10E-03 | 3.25807698E+04 | 3.25809043E+04 | -1.30E-01 | 3.25879317E+04 | 3.25880662E+04 | -1.30E-01 | -1.43603990E+01 | -1.42259034E+01 | -1.30E-01 |
| 204 | Si | 36 | 14 | 22 | 8 | 8.11133000E+00 | 8.10476338E+00 | 6.60E-03 | 3.35142310E+04 | 3.35144572E+04 | -2.30E-01 | 3.35213929E+04 | 3.35216191E+04 | -2.30E-01 | -1.23933230E+01 | -1.21670756E+01 | -2.30E-01 |
| 205 | Si | 37 | 14 | 23 | 9 | 7.95351800E+00 | 7.96091537E+00 | -7.40E-03 | 3.44515241E+04 | 3.44512402E+04 | 2.80E-01 | 3.44586860E+04 | 3.44584021E+04 | 2.80E-01 | -6.59428700E+00 | -6.87814369E+00 | 2.80E-01 |
| 206 | Si | 38 | 14 | 24 | 10 | 7.89282900E+00 | 7.90260569E+00 | -9.80E-03 | 3.53854421E+04 | 3.53850604E+04 | 3.80E-01 | 3.53926040E+04 | 3.53922223E+04 | 3.80E-01 | -4.17029900E+00 | -4.55197227E+00 | 3.80E-01 |
| 207 | Si | 39 | 14 | 25 | 11 | 7.73097900E+00 | 7.75574308E+00 | -2.50E-02 | 3.63234268E+04 | 3.63224509E+04 | 9.80E-01 | 3.63305887E+04 | 3.63296128E+04 | 9.80E-01 | 2.32035200E+00 | 1.34438292E+00 | 9.80E-01 |
| 208 | Si | 40 | 14 | 26 | 12 | 7.66175400E+00 | 7.68138490E+00 | -2.00E-02 | 3.72580302E+04 | 3.72572348E+04 | 8.00E-01 | 3.72651921E+04 | 3.72643967E+04 | 8.00E-01 | 5.42967900E+00 | 4.63428585E+00 | 8.00E-01 |
| 209 | Si | 41 | 14 | 27 | 13 | 7.50857300E+00 | 7.52269533E+00 | -1.40E-02 | 3.81962143E+04 | 3.81956251E+04 | 5.90E-01 | 3.82033762E+04 | 3.82027870E+04 | 5.90E-01 | 1.21196680E+01 | 1.15304924E+01 | 5.90E-01 |
| 210 | Si | 42 | 14 | 28 | 14 | 7.41600000E+00 | 7.44056091E+00 | -2.50E-02 | 3.91321506E+04 | 3.91311174E+04 | 1.00E+00 | 3.91393125E+04 | 3.91382793E+04 | 1.00E+00 | 1.65620000E+01 | 1.55287616E+01 | 1.00E+00 |
| 211 | Si | 43 | 14 | 29 | 15 | 7.27900000E+00 | 7.26646838E+00 | 1.30E-02 | 4.00701838E+04 | 4.00707282E+04 | -5.40E-01 | 4.00773457E+04 | 4.00778901E+04 | -5.40E-01 | 2.31010000E+01 | 2.36454987E+01 | -5.40E-01 |
| 212 | Si | 44 | 14 | 30 | 16 | 7.17400000E+00 | 7.16463389E+00 | 9.40E-03 | 4.10070898E+04 | 4.10075078E+04 | -4.20E-01 | 4.10142517E+04 | 4.10146698E+04 | -4.20E-01 | 2.85130000E+01 | 2.89310668E+01 | -4.20E-01 |
| 213 | P | 25 | 15 | 10 | -5 | 6.81200000E+00 | 6.79732856E+00 | 1.50E-02 | 2.32994156E+04 | 2.32997808E+04 | -3.70E-01 | 2.33070899E+04 | 2.33074551E+04 | -3.70E-01 | 1.97380000E+01 | 2.01036059E+01 | -3.70E-01 |
| 214 | P | 26 | 15 | 11 | -4 | 7.19800000E+00 | 7.15181167E+00 | 4.60E-02 | 2.42221443E+04 | 2.42233323E+04 | -1.20E+00 | 2.42298186E+04 | 2.42310066E+04 | -1.20E+00 | 1.09730000E+01 | 1.21610355E+01 | -1.20E+00 |
| 215 | P | 27 | 15 | 12 | -3 | 7.66343800E+00 | 7.65448504E+00 | 9.00E-03 | 2.51419429E+04 | 2.51421737E+04 | -2.30E-01 | 2.51496172E+04 | 2.51498480E+04 | -2.30E-01 | -7.22461000E-01 | -4.91637948E-01 | -2.30E-01 |
| 216 | P | 28 | 15 | 13 | -2 | 7.90747900E+00 | 7.92870986E+00 | -2.10E-02 | 2.60670116E+04 | 2.60664063E+04 | 6.10E-01 | 2.60746860E+04 | 2.60740806E+04 | 6.10E-01 | -7.14774000E+00 | -7.75309912E+00 | 6.10E-01 |
| 217 | P | 29 | 15 | 14 | -1 | 8.25122200E+00 | 8.22730029E+00 | 2.40E-02 | 2.69887010E+04 | 2.69893838E+04 | -6.80E-01 | 2.69963753E+04 | 2.69970582E+04 | -6.80E-01 | -1.69524510E+01 | -1.62696125E+01 | -6.80E-01 |
| 218 | P | 30 | 15 | 15 | 0 | 8.35349700E+00 | 8.32465581E+00 | 2.90E-02 | 2.79169469E+04 | 2.79178012E+04 | -8.50E-01 | 2.79246212E+04 | 2.79254756E+04 | -8.50E-01 | -2.02006080E+01 | -1.93462594E+01 | -8.50E-01 |
| 219 | P | 31 | 15 | 16 | 1 | 8.48116700E+00 | 8.46041596E+00 | 2.10E-02 | 2.88442010E+04 | 2.88448334E+04 | -6.30E-01 | 2.88518753E+04 | 2.88525077E+04 | -6.30E-01 | -2.44405411E+01 | -2.38081608E+01 | -6.30E-01 |
| 220 | P | 32 | 15 | 17 | 2 | 8.46412000E+00 | 8.45399038E+00 | 1.00E-02 | 2.97758308E+04 | 2.97761440E+04 | -3.10E-01 | 2.97835051E+04 | 2.97838183E+04 | -3.10E-01 | -2.43048740E+01 | -2.39916390E+01 | -3.10E-01 |
| 221 | P | 33 | 15 | 18 | 3 | 8.51380600E+00 | 8.49415444E+00 | 2.00E-02 | 3.07052923E+04 | 3.07059300E+04 | -6.40E-01 | 3.07129667E+04 | 3.07136043E+04 | -6.40E-01 | -2.63373460E+01 | -2.56997247E+01 | -6.40E-01 |
| 222 | P | 34 | 15 | 19 | 4 | 8.44818500E+00 | 8.42569133E+00 | 2.20E-02 | 3.16385750E+04 | 3.16393289E+04 | -7.50E-01 | 3.16462494E+04 | 3.16470033E+04 | -7.50E-01 | -2.45486980E+01 | -2.37948142E+01 | -7.50E-01 |
| 223 | P | 35 | 15 | 20 | 5 | 8.44624800E+00 | 8.40434574E+00 | 4.20E-02 | 3.25697600E+04 | 3.25712157E+04 | -1.50E+00 | 3.25774343E+04 | 3.25788900E+04 | -1.50E+00 | -2.48577910E+01 | -2.34020908E+01 | -1.50E+00 |
| 224 | P | 36 | 15 | 21 | 6 | 8.30786800E+00 | 8.30390026E+00 | 4.00E-03 | 3.35058608E+04 | 3.35059928E+04 | -1.30E-01 | 3.35135352E+04 | 3.35136671E+04 | -1.30E-01 | -2.02510280E+01 | -2.01190805E+01 | -1.30E-01 |
| 225 | P | 37 | 15 | 22 | 7 | 8.26755800E+00 | 8.25502585E+00 | 1.30E-02 | 3.44386098E+04 | 3.44390626E+04 | -4.50E-01 | 3.44462841E+04 | 3.44467369E+04 | -4.50E-01 | -1.89961050E+01 | -1.85433083E+01 | -4.50E-01 |
| 226 | P | 38 | 15 | 23 | 8 | 8.14853700E+00 | 8.14448111E+00 | 4.10E-03 | 3.53744304E+04 | 3.53745737E+04 | -1.40E-01 | 3.53821048E+04 | 3.53822480E+04 | -1.40E-01 | -1.46695560E+01 | -1.45263150E+01 | -1.40E-01 |
| 227 | P | 39 | 15 | 24 | 9 | 8.09937000E+00 | 8.09465740E+00 | 4.70E-03 | 3.63077648E+04 | 3.63079377E+04 | -1.70E-01 | 3.63154391E+04 | 3.63156120E+04 | -1.70E-01 | -1.28292730E+01 | -1.26563528E+01 | -1.70E-01 |
| 228 | P | 40 | 15 | 25 | 10 | 7.97979800E+00 | 7.98373610E+00 | -3.90E-03 | 3.72440137E+04 | 3.72438453E+04 | 1.70E-01 | 3.72516880E+04 | 3.72515196E+04 | 1.70E-01 | -8.07442500E+00 | -8.24283887E+00 | 1.70E-01 |

| | | | | | | | | | | | | | | | | |
|---|---|---|---|---|---|---|---|---|---|---|---|---|---|---|---|---|
| 229 | P | 41 | 15 | 26 | 11 | 7.90655100E+00 | 7.91171303E+00 | -5.20E-03 | 3.81786024E+04 | 3.81783799E+04 | 2.20E-01 | 3.81862767E+04 | 3.81860542E+04 | 2.20E-01 | -4.97976700E+00 | -5.20231034E+00 | 2.20E-01 |
| 230 | P | 42 | 15 | 27 | 12 | 7.76786600E+00 | 7.78997110E+00 | -2.20E-02 | 3.91160860E+04 | 3.91151467E+04 | 9.40E-01 | 3.91237603E+04 | 3.91228210E+04 | 9.40E-01 | 1.00973900E+00 | 7.04569510E-02 | 9.40E-01 |
| 231 | P | 43 | 15 | 28 | 13 | 7.68957200E+00 | 7.70028507E+00 | -1.10E-02 | 4.00512501E+04 | 4.00507786E+04 | 4.70E-01 | 4.00589244E+04 | 4.00584529E+04 | 4.70E-01 | 4.67982600E+00 | 4.20830420E+00 | 4.70E-01 |
| 232 | P | 44 | 15 | 29 | 14 | 7.56700000E+00 | 7.56524486E+00 | 1.80E-03 | 4.09885064E+04 | 4.09885855E+04 | -7.90E-02 | 4.09961807E+04 | 4.09962598E+04 | -7.90E-02 | 1.04420000E+01 | 1.05211070E+01 | -7.90E-02 |
| 233 | P | 45 | 15 | 30 | 15 | 7.47000000E+00 | 7.45764995E+00 | 1.20E-02 | 4.19248815E+04 | 4.19254274E+04 | -5.50E-01 | 4.19325558E+04 | 4.19331017E+04 | -5.50E-01 | 1.53230000E+01 | 1.58689523E+01 | -5.50E-01 |
| 234 | P | 46 | 15 | 31 | 16 | 7.32100000E+00 | 7.30945410E+00 | 1.20E-02 | 4.28638368E+04 | 4.28643521E+04 | -5.20E-01 | 4.28715111E+04 | 4.28720264E+04 | -5.20E-01 | 2.27840000E+01 | 2.32996302E+01 | -5.20E-01 |
| 235 | S | 26 | 16 | 10 | -6 | 6.54800000E+00 | 6.57182006E+00 | -2.40E-02 | 2.42377372E+04 | 2.42371164E+04 | 6.20E-01 | 2.42459241E+04 | 2.42453033E+04 | 6.20E-01 | 2.70790000E+01 | 2.64577404E+01 | 6.20E-01 |
| 236 | S | 27 | 16 | 11 | -5 | 6.97700000E+00 | 6.93076085E+00 | 4.60E-02 | 2.51591805E+04 | 2.51604186E+04 | -1.20E+00 | 2.51673674E+04 | 2.51686055E+04 | -1.20E+00 | 1.70280000E+01 | 1.82658381E+01 | -1.20E+00 |
| 237 | S | 28 | 16 | 12 | -4 | 7.47879000E+00 | 7.46906889E+00 | 9.70E-03 | 2.60777200E+04 | 2.60779806E+04 | -2.60E-01 | 2.60859069E+04 | 2.60861675E+04 | -2.60E-01 | 4.07320500E+00 | 4.33377101E+00 | -2.60E-01 |
| 238 | S | 29 | 16 | 13 | -3 | 7.74851900E+00 | 7.76642663E+00 | -1.80E-02 | 2.70019845E+04 | 2.70014535E+04 | 5.30E-01 | 2.70101714E+04 | 2.70096404E+04 | 5.30E-01 | -3.15640800E+00 | -3.68735330E+00 | 5.30E-01 |
| 239 | S | 30 | 16 | 14 | -2 | 8.12269900E+00 | 8.12641581E+00 | -3.70E-03 | 2.79225759E+04 | 2.79224528E+04 | 1.20E-01 | 2.79307628E+04 | 2.79306397E+04 | 1.20E-01 | -1.40590070E+01 | -1.41821362E+01 | 1.20E-01 |
| 240 | S | 31 | 16 | 15 | -1 | 8.28180000E+00 | 8.27800935E+00 | 3.80E-03 | 2.88490865E+04 | 2.88491924E+04 | -1.10E-01 | 2.88572734E+04 | 2.88573793E+04 | -1.10E-01 | -1.90425250E+01 | -1.89366330E+01 | -1.10E-01 |
| 241 | S | 32 | 16 | 16 | 0 | 8.49312900E+00 | 8.48606336E+00 | 7.10E-03 | 2.97736075E+04 | 2.97738220E+04 | -2.10E-01 | 2.97817944E+04 | 2.97820089E+04 | -2.10E-01 | -2.60155335E+01 | -2.58010515E+01 | -2.10E-01 |
| 242 | S | 33 | 16 | 17 | 1 | 8.49763000E+00 | 8.48898024E+00 | 8.60E-03 | 3.07045313E+04 | 3.07048051E+04 | -2.70E-01 | 3.07127182E+04 | 3.07129920E+04 | -2.70E-01 | -2.65858543E+01 | -2.63120528E+01 | -2.70E-01 |
| 243 | S | 34 | 16 | 18 | 2 | 8.58349800E+00 | 8.58520127E+00 | -1.70E-03 | 3.16326795E+04 | 3.16326100E+04 | 7.00E-02 | 3.16408664E+04 | 3.16407968E+04 | 7.00E-02 | -2.99316930E+01 | -3.00012291E+01 | 7.00E-02 |
| 244 | S | 35 | 16 | 19 | 3 | 8.53785100E+00 | 8.53070949E+00 | 7.10E-03 | 3.25652590E+04 | 3.25654974E+04 | -2.40E-01 | 3.25734459E+04 | 3.25736842E+04 | -2.40E-01 | -2.88462170E+01 | -2.86078990E+01 | -2.40E-01 |
| 245 | S | 36 | 16 | 20 | 4 | 8.57538900E+00 | 8.55782986E+00 | 1.80E-02 | 3.34949352E+04 | 3.34955557E+04 | -6.20E-01 | 3.35031221E+04 | 3.35037426E+04 | -6.20E-01 | -3.06641240E+01 | -3.00436228E+01 | -6.20E-01 |
| 246 | S | 37 | 16 | 21 | 5 | 8.45993500E+00 | 8.47083508E+00 | -1.10E-02 | 3.44301970E+04 | 3.44297820E+04 | 4.10E-01 | 3.44383838E+04 | 3.44379689E+04 | 4.10E-01 | -2.68964100E+01 | -2.73113269E+01 | 4.10E-01 |
| 247 | S | 38 | 16 | 22 | 6 | 8.44878100E+00 | 8.45558006E+00 | -6.80E-03 | 3.53617262E+04 | 3.53614563E+04 | 2.70E-01 | 3.53699131E+04 | 3.53696432E+04 | 2.70E-01 | -2.68611870E+01 | -2.71311521E+01 | 2.70E-01 |
| 248 | S | 39 | 16 | 23 | 7 | 8.34426100E+00 | 8.35273228E+00 | -8.50E-03 | 3.62969192E+04 | 3.62965771E+04 | 3.40E-01 | 3.63051060E+04 | 3.63047640E+04 | 3.40E-01 | -2.31623460E+01 | -2.35043498E+01 | 3.40E-01 |
| 249 | S | 40 | 16 | 24 | 8 | 8.32932500E+00 | 8.33156984E+00 | -2.20E-03 | 3.72287377E+04 | 3.72286363E+04 | 1.00E-01 | 3.72369246E+04 | 3.72368232E+04 | 1.00E-01 | -2.28378460E+01 | -2.29392655E+01 | 1.00E-01 |
| 250 | S | 41 | 16 | 25 | 9 | 8.22963500E+00 | 8.22932102E+00 | 3.10E-04 | 3.81640610E+04 | 3.81640623E+04 | -1.30E-03 | 3.81722479E+04 | 3.81722492E+04 | -1.30E-03 | -1.90085770E+01 | -1.90073149E+01 | -1.30E-03 |
| 251 | S | 42 | 16 | 26 | 10 | 8.19322700E+00 | 8.18686514E+00 | 6.40E-03 | 3.90969259E+04 | 3.90971815E+04 | -2.60E-01 | 3.91051128E+04 | 3.91053684E+04 | -2.60E-01 | -1.76377460E+01 | -1.73821698E+01 | -2.60E-01 |
| 252 | S | 43 | 16 | 27 | 11 | 8.06382700E+00 | 8.07209486E+00 | -8.30E-03 | 4.00338623E+04 | 4.00334951E+04 | 3.70E-01 | 4.00420492E+04 | 4.00416820E+04 | 3.70E-01 | -1.21954590E+01 | -1.25625938E+01 | 3.70E-01 |
| 253 | S | 44 | 16 | 28 | 12 | 7.99601500E+00 | 8.01160859E+00 | -1.60E-02 | 4.09683476E+04 | 4.09676498E+04 | 7.00E-01 | 4.09765344E+04 | 4.09758367E+04 | 7.00E-01 | -9.20423400E+00 | -9.90197385E+00 | 7.00E-01 |
| 254 | S | 45 | 16 | 29 | 13 | 7.88180700E+00 | 7.87804976E+00 | 3.80E-03 | 4.19050563E+04 | 4.19052137E+04 | -1.60E-01 | 4.19132432E+04 | 4.19134006E+04 | -1.60E-01 | -3.98958900E+00 | -3.83211614E+00 | -1.60E-01 |
| 255 | S | 46 | 16 | 30 | 14 | 7.79800000E+00 | 7.79963257E+00 | -1.60E-03 | 4.28405772E+04 | 4.28405083E+04 | 6.90E-02 | 4.28487641E+04 | 4.28486951E+04 | 6.90E-02 | 3.70000000E-02 | -3.16560751E-02 | 6.90E-02 |
| 256 | S | 48 | 16 | 32 | 16 | 7.54500000E+00 | 7.55550786E+00 | -1.10E-02 | 4.47162895E+04 | 4.47157577E+04 | 5.30E-01 | 4.47244764E+04 | 4.47239446E+04 | 5.30E-01 | 1.27610000E+01 | 1.22297026E+01 | 5.30E-01 |
| 257 | Cl | 29 | 17 | 12 | -5 | 7.13800000E+00 | 7.14725985E+00 | -9.30E-03 | 2.70183957E+04 | 2.70181136E+04 | 2.80E-01 | 2.70270952E+04 | 2.70268132E+04 | 2.80E-01 | 1.37670000E+01 | 1.34854063E+01 | 2.80E-01 |
| 258 | Cl | 30 | 17 | 13 | -4 | 7.48000000E+00 | 7.46002294E+00 | 2.00E-02 | 2.79405655E+04 | 2.79411488E+04 | -5.80E-01 | 2.79492651E+04 | 2.79498484E+04 | -5.80E-01 | 4.44300000E+00 | 5.02657274E+00 | -5.80E-01 |
| 259 | Cl | 31 | 17 | 14 | -3 | 7.87022800E+00 | 7.87819386E+00 | -8.00E-03 | 2.88605502E+04 | 2.88602909E+04 | 2.60E-01 | 2.88692498E+04 | 2.88689905E+04 | 2.60E-01 | -7.06612400E+00 | -7.32542962E+00 | 2.60E-01 |
| 260 | Cl | 32 | 17 | 15 | -2 | 8.07240400E+00 | 8.08518639E+00 | -1.30E-02 | 2.97857757E+04 | 2.97853543E+04 | 4.20E-01 | 2.97944753E+04 | 2.97940539E+04 | 4.20E-01 | -1.33346740E+01 | -1.37560654E+01 | 4.20E-01 |
| 261 | Cl | 33 | 17 | 16 | -1 | 8.30475500E+00 | 8.31276937E+00 | -8.00E-03 | 3.07096011E+04 | 3.07093243E+04 | 2.80E-01 | 3.07183007E+04 | 3.07180238E+04 | 2.80E-01 | -2.10033370E+01 | -2.12801713E+01 | 2.80E-01 |
| 262 | Cl | 34 | 17 | 17 | 0 | 8.39897000E+00 | 8.38844139E+00 | 1.10E-02 | 3.16376584E+04 | 3.16380040E+04 | -3.50E-01 | 3.16463580E+04 | 3.16467036E+04 | -3.50E-01 | -2.44400870E+01 | -2.40944702E+01 | -3.50E-01 |
| 263 | Cl | 35 | 17 | 18 | 1 | 8.52027900E+00 | 8.50339263E+00 | 1.70E-02 | 3.25645790E+04 | 3.25651577E+04 | -5.80E-01 | 3.25732786E+04 | 3.25738572E+04 | -5.80E-01 | -2.90135400E+01 | -2.84348860E+01 | -5.80E-01 |
| 264 | Cl | 36 | 17 | 19 | 2 | 8.52193200E+00 | 8.50497143E+00 | 1.70E-02 | 3.34955646E+04 | 3.34961628E+04 | -6.00E-01 | 3.35042642E+04 | 3.35048624E+04 | -6.00E-01 | -2.95220170E+01 | -2.89237963E+01 | -6.00E-01 |
| 265 | Cl | 37 | 17 | 20 | 3 | 8.57028000E+00 | 8.54982346E+00 | 2.00E-02 | 3.44248192E+04 | 3.44255637E+04 | -7.40E-01 | 3.44335187E+04 | 3.44342633E+04 | -7.40E-01 | -3.17615210E+01 | -3.10169739E+01 | -7.40E-01 |
| 266 | Cl | 38 | 17 | 21 | 4 | 8.50548000E+00 | 8.50805852E+00 | -2.60E-03 | 3.53582766E+04 | 3.53581663E+04 | 1.10E-01 | 3.53669762E+04 | 3.53668659E+04 | 1.10E-01 | -2.97980870E+01 | -2.99084108E+01 | 1.10E-01 |
| 267 | Cl | 39 | 17 | 22 | 5 | 8.49440200E+00 | 8.50900760E+00 | -1.50E-02 | 3.62897686E+04 | 3.62891866E+04 | 5.80E-01 | 3.62984682E+04 | 3.62978862E+04 | 5.80E-01 | -2.98001920E+01 | -3.03821645E+01 | 5.80E-01 |
| 268 | Cl | 40 | 17 | 23 | 6 | 8.42776500E+00 | 8.44072323E+00 | -1.30E-02 | 3.72235050E+04 | 3.72229744E+04 | 5.30E-01 | 3.72322046E+04 | 3.72316740E+04 | 5.30E-01 | -2.75578130E+01 | -2.80884782E+01 | 5.30E-01 |
| 269 | Cl | 41 | 17 | 24 | 7 | 8.41295900E+00 | 8.42986955E+00 | -1.70E-02 | 3.81552497E+04 | 3.81545440E+04 | 7.10E-01 | 3.81639493E+04 | 3.81632436E+04 | 7.10E-01 | -2.73071890E+01 | -2.80128816E+01 | 7.10E-01 |
| 270 | Cl | 42 | 17 | 25 | 8 | 8.34781900E+00 | 8.35740237E+00 | -9.60E-03 | 3.90891380E+04 | 3.90887232E+04 | 4.10E-01 | 3.90978376E+04 | 3.90974228E+04 | 4.10E-01 | -2.49129900E+01 | -2.53278105E+01 | 4.10E-01 |
| 271 | Cl | 43 | 17 | 26 | 9 | 8.32766000E+00 | 8.32320882E+00 | 4.50E-03 | 4.00212224E+04 | 4.00214015E+04 | -1.80E-01 | 4.00299220E+04 | 4.00301011E+04 | -1.80E-01 | -2.43226270E+01 | -2.41435715E+01 | -1.80E-01 |
| 272 | Cl | 44 | 17 | 27 | 10 | 8.23745000E+00 | 8.23820075E+00 | -7.50E-04 | 4.09564294E+04 | 4.09563840E+04 | 4.50E-02 | 4.09651289E+04 | 4.09650836E+04 | 4.50E-02 | -2.06097300E+01 | -2.06551058E+01 | 4.50E-02 |
| 273 | Cl | 45 | 17 | 28 | 11 | 8.18375800E+00 | 8.18293775E+00 | 8.20E-04 | 4.18901734E+04 | 4.18901980E+04 | -2.50E-02 | 4.18988730E+04 | 4.18988976E+04 | -2.50E-02 | -1.83597470E+01 | -1.83351528E+01 | -2.50E-02 |
| 274 | Cl | 46 | 17 | 29 | 12 | 8.08241400E+00 | 8.07894401E+00 | 3.50E-03 | 4.28262169E+04 | 4.28263642E+04 | -1.50E-01 | 4.28349165E+04 | 4.28350637E+04 | -1.50E-01 | -1.38103300E+01 | -1.36630594E+01 | -1.50E-01 |
| 275 | Cl | 47 | 17 | 30 | 13 | 8.00300000E+00 | 7.99994803E+00 | 3.10E-03 | 4.37614239E+04 | 4.37615634E+04 | -1.40E-01 | 4.37701235E+04 | 4.37702630E+04 | -1.40E-01 | -1.00970000E+01 | -9.95787326E+00 | -1.40E-01 |

| | | | | | | | | | | | | | | | | | | |
|---|---|---|---|---|---|---|---|---|---|---|---|---|---|---|---|---|---|---|
| 276 | Cl | 48 | 17 | 31 | 14 | 7.87900000E+00 | 7.87821459E+00 | 7.90E-04 | 4.46989540E+04 | 4.46989720E+04 | -1.80E-02 | 4.47076536E+04 | 4.47076716E+04 | -1.80E-02 | -4.06100000E+00 | -4.04329722E+00 | -1.80E-02 |
| 277 | Cl | 49 | 17 | 32 | 15 | 7.77700000E+00 | 7.78062371E+00 | -3.60E-03 | 4.56356551E+04 | 4.56354412E+04 | 2.10E-01 | 4.56443547E+04 | 4.56441407E+04 | 2.10E-01 | 1.14600000E+00 | 9.31760006E-01 | 2.10E-01 |
| 278 | CL | 50 | 17 | 33 | 16 | 7.63700000E+00 | 7.64623143E+00 | -9.20E-03 | 4.65744335E+04 | 4.65739455E+04 | 4.90E-01 | 4.65831331E+04 | 4.65826451E+04 | 4.90E-01 | 8.43000000E+00 | 7.94206953E+00 | 4.90E-01 |
| 279 | Cl | 51 | 17 | 34 | 17 | 7.52700000E+00 | 7.53407666E+00 | -7.10E-03 | 4.75119729E+04 | 4.75115846E+04 | 3.90E-01 | 4.75206725E+04 | 4.75202842E+04 | 3.90E-01 | 1.44750000E+01 | 1.40870504E+01 | 3.90E-01 |
| 280 | Ar | 30 | 18 | 12 | -6 | 6.88600000E+00 | 6.90525368E+00 | -1.90E-02 | 2.79570990E+04 | 2.79564960E+04 | 6.00E-01 | 2.79663114E+04 | 2.79657084E+04 | 6.00E-01 | 2.14900000E+01 | 2.08865735E+01 | 6.00E-01 |
| 281 | Ar | 31 | 18 | 13 | -5 | 7.25300000E+00 | 7.23312519E+00 | 2.00E-02 | 2.88783969E+04 | 2.88789921E+04 | -6.00E-01 | 2.88876093E+04 | 2.88882045E+04 | -6.00E-01 | 1.12940000E+01 | 1.18886221E+01 | -5.90E-01 |
| 282 | Ar | 32 | 18 | 14 | -4 | 7.70000800E+00 | 7.70448375E+00 | -4.50E-03 | 2.97963972E+04 | 2.97962409E+04 | 1.60E-01 | 2.98056096E+04 | 2.98054533E+04 | 1.60E-01 | -2.20035100E+00 | -2.35665800E+00 | 1.60E-01 |
| 283 | Ar | 33 | 18 | 15 | -3 | 7.92895500E+00 | 7.93074205E+00 | -1.80E-03 | 3.07207073E+04 | 3.07206352E+04 | 7.20E-02 | 3.07299197E+04 | 3.07298477E+04 | 7.20E-02 | -9.38429300E+00 | -9.45634664E+00 | 7.20E-02 |
| 284 | Ar | 34 | 18 | 16 | -2 | 8.19767200E+00 | 8.20464290E+00 | -7.00E-03 | 3.16432074E+04 | 3.16429572E+04 | 2.50E-01 | 3.16524198E+04 | 3.16521697E+04 | 2.50E-01 | -1.83782930E+01 | -1.86283988E+01 | 2.50E-01 |
| 285 | Ar | 35 | 18 | 17 | -1 | 8.32746500E+00 | 8.32505578E+00 | 2.40E-03 | 3.25700323E+04 | 3.25701035E+04 | -7.10E-02 | 3.25792447E+04 | 3.25793160E+04 | -7.10E-02 | -2.30474100E+01 | -2.29761731E+01 | -7.10E-02 |
| 286 | Ar | 36 | 18 | 18 | 0 | 8.51990900E+00 | 8.50146363E+00 | 1.80E-02 | 3.34943422E+04 | 3.34949932E+04 | -6.50E-01 | 3.35035547E+04 | 3.35042056E+04 | -6.50E-01 | -3.02315400E+01 | -2.95805927E+01 | -6.50E-01 |
| 287 | Ar | 37 | 18 | 19 | 1 | 8.52713900E+00 | 8.51601039E+00 | 1.10E-02 | 3.44251202E+04 | 3.44255189E+04 | -4.00E-01 | 3.44343326E+04 | 3.44347313E+04 | -4.00E-01 | -3.09476480E+01 | -3.05489676E+01 | -4.00E-01 |
| 288 | Ar | 38 | 18 | 20 | 2 | 8.61428000E+00 | 8.60884258E+00 | 5.40E-03 | 3.53528471E+04 | 3.53530406E+04 | -1.90E-01 | 3.53620595E+04 | 3.53622530E+04 | -1.90E-01 | -3.47148200E+01 | -3.45212821E+01 | -1.90E-01 |
| 289 | Ar | 39 | 18 | 21 | 3 | 8.56259800E+00 | 8.57667225E+00 | -1.40E-02 | 3.62858138E+04 | 3.62852518E+04 | 5.60E-01 | 3.62950262E+04 | 3.62944642E+04 | 5.60E-01 | -3.32421900E+01 | -3.38041626E+01 | 5.60E-01 |
| 290 | Ar | 40 | 18 | 22 | 4 | 8.59525900E+00 | 8.61549853E+00 | -2.00E-02 | 3.72155101E+04 | 3.72146874E+04 | 8.20E-01 | 3.72247225E+04 | 3.72238999E+04 | 8.20E-01 | -3.50398946E+01 | -3.58625673E+01 | 8.20E-01 |
| 291 | Ar | 41 | 18 | 23 | 5 | 8.53437200E+00 | 8.55597061E+00 | -2.20E-02 | 3.81489766E+04 | 3.81480780E+04 | 9.00E-01 | 3.81581890E+04 | 3.81572904E+04 | 9.00E-01 | -3.30675050E+01 | -3.39661019E+01 | 9.00E-01 |
| 292 | Ar | 42 | 18 | 24 | 6 | 8.55561300E+00 | 8.57594358E+00 | -2.00E-02 | 3.90791155E+04 | 3.90782485E+04 | 8.70E-01 | 3.90883279E+04 | 3.90874609E+04 | 8.70E-01 | -3.44226750E+01 | -3.52896182E+01 | 8.70E-01 |
| 293 | Ar | 43 | 18 | 25 | 7 | 8.48823700E+00 | 8.51034495E+00 | -2.20E-02 | 4.00130224E+04 | 4.00120587E+04 | 9.60E-01 | 4.00222348E+04 | 4.00212711E+04 | 9.60E-01 | -3.20098080E+01 | -3.29735018E+01 | 9.60E-01 |
| 294 | Ar | 44 | 18 | 26 | 8 | 8.49384000E+00 | 8.50265431E+00 | -8.80E-03 | 4.09438530E+04 | 4.09434521E+04 | 4.00E-01 | 4.09530654E+04 | 4.09526645E+04 | 4.00E-01 | -3.26732550E+01 | -3.30741396E+01 | 4.00E-01 |
| 295 | Ar | 45 | 18 | 27 | 9 | 8.41995200E+00 | 8.42382612E+00 | -3.90E-03 | 4.18782495E+04 | 4.18780621E+04 | 1.90E-01 | 4.18874619E+04 | 4.18872745E+04 | 1.90E-01 | -2.97707960E+01 | -2.99582066E+01 | 1.90E-01 |
| 296 | Ar | 46 | 18 | 28 | 10 | 8.41150200E+00 | 8.39413359E+00 | 1.70E-02 | 4.28097836E+04 | 4.28105695E+04 | -7.90E-01 | 4.28189960E+04 | 4.28197819E+04 | -7.90E-01 | -2.97307620E+01 | -2.89448572E+01 | -7.90E-01 |
| 297 | Ar | 47 | 18 | 29 | 11 | 8.30810000E+00 | 8.29524766E+00 | 1.30E-02 | 4.37457974E+04 | 4.37463884E+04 | -5.90E-01 | 4.37550099E+04 | 4.37556008E+04 | -5.90E-01 | -2.52110110E+01 | -2.46200328E+01 | -5.90E-01 |
| 298 | Ar | 48 | 18 | 30 | 12 | 8.24500000E+00 | 8.24114096E+00 | 3.90E-03 | 4.46800628E+04 | 4.46802557E+04 | -1.90E-01 | 4.46892752E+04 | 4.46894681E+04 | -1.90E-01 | -2.24400000E+01 | -2.22468402E+01 | -1.90E-01 |
| 299 | Ar | 49 | 18 | 31 | 13 | 8.12800000E+00 | 8.12194127E+00 | 6.10E-03 | 4.56171365E+04 | 4.56174207E+04 | -2.80E-01 | 4.56263489E+04 | 4.56266331E+04 | -2.80E-01 | -1.68600000E+01 | -1.65758772E+01 | -2.80E-01 |
| 300 | Ar | 50 | 18 | 32 | 14 | 8.04800000E+00 | 8.04972513E+00 | -1.70E-03 | 4.65525708E+04 | 4.65524749E+04 | 9.60E-02 | 4.65617832E+04 | 4.65616874E+04 | 9.60E-02 | -1.29200000E+01 | -1.30156926E+01 | 9.60E-02 |
| 301 | Ar | 51 | 18 | 33 | 15 | 7.91000000E+00 | 7.91579045E+00 | -5.80E-03 | 4.74911163E+04 | 4.74908213E+04 | 3.00E-01 | 4.75003287E+04 | 4.75000337E+04 | 3.00E-01 | -5.86800000E+00 | -6.16342987E+00 | 3.00E-01 |
| 302 | Ar | 52 | 18 | 34 | 16 | 7.81900000E+00 | 7.83139205E+00 | -1.20E-02 | 4.84275100E+04 | 4.84268596E+04 | 6.50E-01 | 4.84367224E+04 | 4.84360720E+04 | 6.50E-01 | -9.69000000E-01 | -1.61918485E+00 | 6.50E-01 |
| 303 | Ar | 53 | 18 | 35 | 17 | 7.67700000E+00 | 7.68477496E+00 | -7.80E-03 | 4.93667634E+04 | 4.93663643E+04 | 4.00E-01 | 4.93759758E+04 | 4.93755767E+04 | 4.00E-01 | 6.79100000E+00 | 6.39144815E+00 | 4.00E-01 |
| 304 | K | 33 | 19 | 14 | -5 | 7.40700000E+00 | 7.38692243E+00 | 2.00E-02 | 3.07366207E+04 | 3.07372852E+04 | -6.60E-01 | 3.07463461E+04 | 3.07470106E+04 | -6.60E-01 | 7.04200000E+00 | 7.70662377E+00 | -6.60E-01 |
| 305 | K | 34 | 19 | 15 | -4 | 7.67000000E+00 | 7.66689533E+00 | 3.10E-03 | 3.16598524E+04 | 3.16599446E+04 | -9.20E-02 | 3.16695778E+04 | 3.16696700E+04 | -9.20E-02 | -1.22000000E+00 | -1.12805815E+00 | -9.20E-02 |
| 306 | K | 35 | 19 | 16 | -3 | 7.96584000E+00 | 7.98192717E+00 | -1.60E-02 | 3.25813938E+04 | 3.25808170E+04 | 5.80E-01 | 3.25911192E+04 | 3.25905424E+04 | 5.80E-01 | -1.11728910E+01 | -1.17497488E+01 | 5.80E-01 |
| 307 | K | 36 | 19 | 17 | -2 | 8.14221900E+00 | 8.14430528E+00 | -2.10E-03 | 3.35066437E+04 | 3.35065548E+04 | 8.90E-02 | 3.35163691E+04 | 3.35162802E+04 | 8.90E-02 | -1.74170650E+01 | -1.75059692E+01 | 8.90E-02 |
| 308 | K | 37 | 19 | 18 | -1 | 8.33984700E+00 | 8.33984734E+00 | -3.40E-07 | 3.44307546E+04 | 3.44307408E+04 | 1.40E-02 | 3.44404801E+04 | 3.44404662E+04 | 1.40E-02 | -2.48001990E+01 | -2.48140117E+01 | 1.40E-02 |
| 309 | K | 38 | 19 | 19 | 0 | 8.43805800E+00 | 8.41512724E+00 | 2.30E-02 | 3.53582481E+04 | 3.53591057E+04 | -8.60E-01 | 3.53679736E+04 | 3.53688311E+04 | -8.60E-01 | -2.88007540E+01 | -2.79431759E+01 | -8.60E-01 |
| 310 | K | 39 | 19 | 20 | 1 | 8.55702500E+00 | 8.52531944E+00 | 3.20E-02 | 3.62847358E+04 | 3.62859585E+04 | -1.20E+00 | 3.62944612E+04 | 3.62956839E+04 | -1.20E+00 | -3.38071902E+01 | -3.25844803E+01 | -1.20E+00 |
| 311 | K | 40 | 19 | 21 | 2 | 8.53809000E+00 | 8.54033685E+00 | -2.20E-03 | 3.72165015E+04 | 3.72163978E+04 | 1.00E-01 | 3.72262269E+04 | 3.72261233E+04 | 1.00E-01 | -3.35354920E+01 | -3.36391768E+01 | 1.00E-01 |
| 312 | K | 41 | 19 | 22 | 3 | 8.57607200E+00 | 8.59236189E+00 | -1.60E-02 | 3.81459715E+04 | 3.81452899E+04 | 6.80E-01 | 3.81556970E+04 | 3.81550153E+04 | 6.80E-01 | -3.55595433E+01 | -3.62412214E+01 | 6.80E-01 |
| 313 | K | 42 | 19 | 23 | 4 | 8.55125600E+00 | 8.56944378E+00 | -1.80E-02 | 3.90780031E+04 | 3.90772254E+04 | 7.80E-01 | 3.90877285E+04 | 3.90869509E+04 | 7.80E-01 | -3.50220260E+01 | -3.57997038E+01 | 7.80E-01 |
| 314 | K | 43 | 19 | 24 | 5 | 8.57622000E+00 | 8.60020534E+00 | -2.40E-02 | 4.00079438E+04 | 4.00068986E+04 | 1.00E+00 | 4.00176692E+04 | 4.00166240E+04 | 1.00E+00 | -3.65753890E+01 | -3.76205754E+01 | 1.00E+00 |
| 315 | K | 44 | 19 | 25 | 6 | 8.54670100E+00 | 8.56512846E+00 | -1.80E-02 | 4.09402318E+04 | 4.09394072E+04 | 8.20E-01 | 4.09499572E+04 | 4.09491326E+04 | 8.20E-01 | -3.57814920E+01 | -3.66060794E+01 | 8.20E-01 |
| 316 | K | 45 | 19 | 26 | 7 | 8.55467400E+00 | 8.56569349E+00 | -1.10E-02 | 4.18708917E+04 | 4.18703820E+04 | 5.10E-01 | 4.18806171E+04 | 4.18801074E+04 | 5.10E-01 | -3.66156380E+01 | -3.71253149E+01 | 5.10E-01 |
| 317 | K | 46 | 19 | 27 | 8 | 8.51804200E+00 | 8.51307895E+00 | 5.00E-03 | 4.28035875E+04 | 4.28038020E+04 | -2.10E-01 | 4.28133129E+04 | 4.28135274E+04 | -2.10E-01 | -3.54139240E+01 | -3.51994206E+01 | -2.10E-01 |
| 318 | K | 47 | 19 | 28 | 9 | 8.51487900E+00 | 8.48938864E+00 | 2.50E-02 | 4.37347835E+04 | 4.37359677E+04 | -1.20E+00 | 4.37445089E+04 | 4.37456931E+04 | -1.20E+00 | -3.57119760E+01 | -3.45277359E+01 | -1.20E+00 |
| 319 | K | 48 | 19 | 29 | 10 | 8.43423200E+00 | 8.41525057E+00 | 1.90E-02 | 4.46697050E+04 | 4.46706023E+04 | -9.00E-01 | 4.46794304E+04 | 4.46803277E+04 | -9.00E-01 | -3.22844770E+01 | -3.13871782E+01 | -9.00E-01 |
| 320 | K | 49 | 19 | 30 | 11 | 8.37227400E+00 | 8.36510471E+00 | 7.20E-03 | 4.56038721E+04 | 4.56042096E+04 | -3.40E-01 | 4.56135975E+04 | 4.56139350E+04 | -3.40E-01 | -2.96114910E+01 | -2.92739628E+01 | -3.40E-01 |
| 321 | K | 50 | 19 | 31 | 12 | 8.28858200E+00 | 8.27061808E+00 | 1.80E-02 | 4.65392498E+04 | 4.65401342E+04 | -8.80E-01 | 4.65489752E+04 | 4.65498596E+04 | -8.80E-01 | -2.57278480E+01 | -2.48434171E+01 | -8.80E-01 |
| 322 | K | 51 | 19 | 32 | 13 | 8.22134600E+00 | 8.20033940E+00 | 2.10E-02 | 4.74739557E+04 | 4.74750132E+04 | -1.10E+00 | 4.74836811E+04 | 4.74847386E+04 | -1.10E+00 | -2.25160390E+01 | -2.14585033E+01 | -1.10E+00 |

| | | | | | | | | | | | | | | | | |
|---|---|---|---|---|---|---|---|---|---|---|---|---|---|---|---|---|
| 323 | K | 52 | 19 | 33 | 14 | 8.10400000E+00 | 8.09274923E+00 | 1.10E-02 | 4.84114224E+04 | 4.84119729E+04 | -5.50E-01 | 4.84211478E+04 | 4.84216983E+04 | -5.50E-01 | -1.65430000E+01 | -1.59928348E+01 | -5.50E-01 |
| 324 | K | 53 | 19 | 34 | 15 | 8.01100000E+00 | 8.00842358E+00 | 2.60E-03 | 4.93477789E+04 | 4.93479148E+04 | -1.40E-01 | 4.93575043E+04 | 4.93576402E+04 | -1.40E-01 | -1.16810000E+01 | -1.15450057E+01 | -1.40E-01 |
| 325 | K | 54 | 19 | 35 | 16 | 7.88900000E+00 | 7.89064916E+00 | -1.60E-03 | 5.02859518E+04 | 5.02858316E+04 | 1.20E-01 | 5.02956772E+04 | 5.02955570E+04 | 1.20E-01 | -5.00200000E+00 | -5.12229153E+00 | 1.20E-01 |
| 326 | K | 55 | 19 | 36 | 17 | 7.78800000E+00 | 7.79369479E+00 | -5.70E-03 | 5.12231559E+04 | 5.12228388E+04 | 3.20E-01 | 5.12328813E+04 | 5.12325642E+04 | 3.20E-01 | 7.08000000E-01 | 3.90868567E-01 | 3.20E-01 |
| 327 | K | 56 | 19 | 37 | 18 | 7.66400000E+00 | 7.66655110E+00 | -2.60E-03 | 5.21618690E+04 | 5.21617305E+04 | 1.40E-01 | 5.21715944E+04 | 5.21714560E+04 | 1.40E-01 | 7.92700000E+00 | 7.78853937E+00 | 1.40E-01 |
| 328 | Ca | 34 | 20 | 14 | -6 | 7.20400000E+00 | 7.21782323E+00 | -1.40E-02 | 3.16744108E+04 | 3.16739168E+04 | 4.90E-01 | 3.16846494E+04 | 3.16841554E+04 | 4.90E-01 | 1.38510000E+01 | 1.33573163E+01 | 4.90E-01 |
| 329 | Ca | 35 | 20 | 15 | -5 | 7.48700000E+00 | 7.47259129E+00 | 1.40E-02 | 3.25968414E+04 | 3.25973475E+04 | -5.10E-01 | 3.26070800E+04 | 3.26075861E+04 | -5.10E-01 | 4.78800000E+00 | 5.29393002E+00 | -5.10E-01 |
| 330 | Ca | 36 | 20 | 16 | -4 | 7.81587900E+00 | 7.82768976E+00 | -1.20E-02 | 3.35170965E+04 | 3.35166567E+04 | 4.40E-01 | 3.35273350E+04 | 3.35268953E+04 | 4.40E-01 | -6.45115000E+00 | -6.89088746E+00 | 4.40E-01 |
| 331 | Ca | 37 | 20 | 17 | -3 | 8.00345600E+00 | 8.00169211E+00 | 1.80E-03 | 3.44419056E+04 | 3.44419563E+04 | -5.10E-02 | 3.44521442E+04 | 3.44521949E+04 | -5.10E-02 | -1.31360660E+01 | -1.30853451E+01 | -5.10E-02 |
| 332 | Ca | 38 | 20 | 18 | -2 | 8.24004300E+00 | 8.23338448E+00 | 6.70E-03 | 3.53644772E+04 | 3.53647157E+04 | -2.40E-01 | 3.53747158E+04 | 3.53749543E+04 | -2.40E-01 | -2.20585000E+01 | -2.18200284E+01 | -2.40E-01 |
| 333 | Ca | 39 | 20 | 19 | -1 | 8.36967000E+00 | 8.34681015E+00 | 2.30E-02 | 3.62907471E+04 | 3.62916241E+04 | -8.80E-01 | 3.63009857E+04 | 3.63018627E+04 | -8.80E-01 | -2.72827020E+01 | -2.64056950E+01 | -8.80E-01 |
| 334 | Ca | 40 | 20 | 20 | 0 | 8.55130400E+00 | 8.50883521E+00 | 4.20E-02 | 3.72146775E+04 | 3.72163617E+04 | -1.70E+00 | 3.72249160E+04 | 3.72266003E+04 | -1.70E+00 | -3.48463860E+01 | -3.31621886E+01 | -1.70E+00 |
| 335 | Ca | 41 | 20 | 21 | 1 | 8.54670600E+00 | 8.53623387E+00 | 1.00E-02 | 3.81458800E+04 | 3.81462949E+04 | -4.10E-01 | 3.81561186E+04 | 3.81565335E+04 | -4.10E-01 | -3.51378870E+01 | -3.47230497E+01 | -4.10E-01 |
| 336 | Ca | 42 | 20 | 22 | 2 | 8.61656300E+00 | 8.62946032E+00 | -1.30E-02 | 3.90739647E+04 | 3.90734085E+04 | 5.60E-01 | 3.90842033E+04 | 3.90836471E+04 | 5.60E-01 | -3.85472430E+01 | -3.91034755E+01 | 5.60E-01 |
| 337 | Ca | 43 | 20 | 23 | 3 | 8.60066300E+00 | 8.61392465E+00 | -1.30E-02 | 4.00055972E+04 | 4.00050125E+04 | 5.80E-01 | 4.00158358E+04 | 4.00152510E+04 | 5.80E-01 | -3.84088150E+01 | -3.89935829E+01 | 5.80E-01 |
| 338 | Ca | 44 | 20 | 24 | 4 | 8.65817500E+00 | 8.67734742E+00 | -1.90E-02 | 4.09340315E+04 | 4.09331733E+04 | 8.60E-01 | 4.09442700E+04 | 4.09434119E+04 | 8.60E-01 | -4.14686580E+01 | -4.23267907E+01 | 8.60E-01 |
| 339 | Ca | 45 | 20 | 25 | 5 | 8.63054400E+00 | 8.64778978E+00 | -1.70E-02 | 4.18661820E+04 | 4.18653915E+04 | 7.90E-01 | 4.18764206E+04 | 4.18756300E+04 | 7.90E-01 | -4.08121520E+01 | -4.16027251E+01 | 7.90E-01 |
| 340 | Ca | 46 | 20 | 26 | 6 | 8.66895800E+00 | 8.67575527E+00 | -6.80E-03 | 4.27953498E+04 | 4.27950226E+04 | 3.30E-01 | 4.28055884E+04 | 4.28052612E+04 | 3.30E-01 | -4.31383960E+01 | -4.34656086E+01 | 3.30E-01 |
| 341 | Ca | 47 | 20 | 27 | 7 | 8.63932800E+00 | 8.62710779E+00 | 1.20E-02 | 4.37276388E+04 | 4.37281987E+04 | -5.60E-01 | 4.37378774E+04 | 4.37384373E+04 | -5.60E-01 | -4.23434530E+01 | -4.17836130E+01 | -5.60E-01 |
| 342 | Ca | 48 | 20 | 28 | 8 | 8.66668900E+00 | 8.62706719E+00 | 4.00E-02 | 4.46572516E+04 | 4.46591389E+04 | -1.90E+00 | 4.46674902E+04 | 4.46693775E+04 | -1.90E+00 | -4.42247590E+01 | -4.23374529E+01 | -1.90E+00 |
| 343 | Ca | 49 | 20 | 29 | 9 | 8.59484700E+00 | 8.55568817E+00 | 3.90E-02 | 4.55916705E+04 | 4.55935748E+04 | -1.90E+00 | 4.56019091E+04 | 4.56038134E+04 | -1.90E+00 | -4.12998950E+01 | -3.93956295E+01 | -1.90E+00 |
| 344 | Ca | 50 | 20 | 30 | 10 | 8.55016300E+00 | 8.52741199E+00 | 2.30E-02 | 4.65248753E+04 | 4.65259983E+04 | -1.10E+00 | 4.65351138E+04 | 4.65362369E+04 | -1.10E+00 | -3.95892240E+01 | -3.84661895E+01 | -1.10E+00 |
| 345 | Ca | 51 | 20 | 31 | 11 | 8.47703500E+00 | 8.43506440E+00 | 4.20E-02 | 4.74596200E+04 | 4.74617460E+04 | -2.10E+00 | 4.74698586E+04 | 4.74719846E+04 | -2.10E+00 | -3.63385290E+01 | -3.42125554E+01 | -2.10E+00 |
| 346 | Ca | 52 | 20 | 32 | 12 | 8.42931900E+00 | 8.38673246E+00 | 4.30E-02 | 4.83931896E+04 | 4.83953896E+04 | -2.20E+00 | 4.84034282E+04 | 4.84056281E+04 | -2.20E+00 | -3.42630090E+01 | -3.20630398E+01 | -2.20E+00 |
| 347 | Ca | 53 | 20 | 33 | 13 | 8.31300000E+00 | 8.28129327E+00 | 3.20E-02 | 4.93304895E+04 | 4.93321565E+04 | -1.70E+00 | 4.93407281E+04 | 4.93423951E+04 | -1.70E+00 | -2.84570000E+01 | -2.67901761E+01 | -1.70E+00 |
| 348 | Ca | 54 | 20 | 34 | 14 | 8.24000000E+00 | 8.22041919E+00 | 2.00E-02 | 5.02656630E+04 | 5.02667278E+04 | -1.10E+00 | 5.02759016E+04 | 5.02769663E+04 | -1.10E+00 | -2.47780000E+01 | -2.37129505E+01 | -1.10E+00 |
| 349 | Ca | 55 | 20 | 35 | 15 | 8.12000000E+00 | 8.10469227E+00 | 1.50E-02 | 5.12035843E+04 | 5.12044377E+04 | -8.50E-01 | 5.12138229E+04 | 5.12146763E+04 | -8.50E-01 | -1.83500000E+01 | -1.74970701E+01 | -8.50E-01 |
| 350 | Ca | 56 | 20 | 36 | 16 | 8.04000000E+00 | 8.03284074E+00 | 7.20E-03 | 5.21395310E+04 | 5.21399221E+04 | -3.90E-01 | 5.21497695E+04 | 5.21501607E+04 | -3.90E-01 | -1.38980000E+01 | -1.35067574E+01 | -3.90E-01 |
| 351 | CA | 57 | 20 | 37 | 17 | 7.91700000E+00 | 7.90638579E+00 | 1.10E-02 | 5.30780485E+04 | 5.30786626E+04 | -6.10E-01 | 5.30882870E+04 | 5.30889011E+04 | -6.10E-01 | -6.87400000E+00 | -6.26034676E+00 | -6.10E-01 |
| 352 | CA | 58 | 20 | 38 | 18 | 7.83500000E+00 | 7.82462687E+00 | 1.00E-02 | 5.40144981E+04 | 5.40150636E+04 | -5.70E-01 | 5.40247367E+04 | 5.40253021E+04 | -5.70E-01 | -1.91900000E+00 | -1.35339656E+00 | -5.70E-01 |
| 353 | Sc | 38 | 21 | 17 | -4 | 7.75900000E+00 | 7.77337386E+00 | -1.40E-02 | 3.53814768E+04 | 3.53808997E+04 | 5.80E-01 | 3.53922286E+04 | 3.53916516E+04 | 5.80E-01 | -4.54600000E+00 | -5.12270164E+00 | 5.80E-01 |
| 354 | Sc | 39 | 21 | 18 | -3 | 8.01345600E+00 | 8.03424452E+00 | -2.10E-02 | 3.63033438E+04 | 3.63025178E+04 | 8.30E-01 | 3.63140957E+04 | 3.63132697E+04 | 8.30E-01 | -1.41727100E+01 | -1.49987122E+01 | 8.30E-01 |
| 355 | Sc | 40 | 21 | 19 | -2 | 8.17366900E+00 | 8.18183639E+00 | -8.20E-03 | 3.72284872E+04 | 3.72281453E+04 | 3.40E-01 | 3.72392391E+04 | 3.72388971E+04 | 3.40E-01 | -2.05233360E+01 | -2.08653124E+01 | 3.40E-01 |
| 356 | Sc | 41 | 21 | 20 | -1 | 8.36919800E+00 | 8.36091853E+00 | 8.30E-03 | 3.81518622E+04 | 3.81521864E+04 | -3.20E-01 | 3.81626141E+04 | 3.81629383E+04 | -3.20E-01 | -2.86424110E+01 | -2.83181978E+01 | -3.20E-01 |
| 357 | Sc | 42 | 21 | 21 | 0 | 8.44493300E+00 | 8.43841383E+00 | 6.50E-03 | 3.90798775E+04 | 3.90801361E+04 | -2.60E-01 | 3.90906294E+04 | 3.90908880E+04 | -2.60E-01 | -3.21211430E+01 | -3.18626000E+01 | -2.60E-01 |
| 358 | Sc | 43 | 21 | 22 | 1 | 8.53082500E+00 | 8.54563546E+00 | -1.50E-02 | 4.00073046E+04 | 4.00066525E+04 | 6.50E-01 | 4.00180565E+04 | 4.00174044E+04 | 6.50E-01 | -3.61881000E+01 | -3.68402246E+01 | 6.50E-01 |
| 359 | Sc | 44 | 21 | 23 | 2 | 8.55737900E+00 | 8.57007073E+00 | -1.30E-02 | 4.09371708E+04 | 4.09365971E+04 | 5.70E-01 | 4.09479227E+04 | 4.09473490E+04 | 5.70E-01 | -3.78159780E+01 | -3.83896932E+01 | 5.70E-01 |
| 360 | Sc | 45 | 21 | 24 | 3 | 8.61891500E+00 | 8.64349831E+00 | -2.50E-02 | 4.18654097E+04 | 4.18642882E+04 | 1.10E+00 | 4.18761616E+04 | 4.18750401E+04 | 1.10E+00 | -4.10711770E+01 | -4.21926860E+01 | 1.10E+00 |
| 361 | Sc | 46 | 21 | 25 | 4 | 8.62199600E+00 | 8.64601707E+00 | -2.40E-02 | 4.27962144E+04 | 4.27950942E+04 | 1.10E+00 | 4.28069663E+04 | 4.28058461E+04 | 1.10E+00 | -4.17604990E+01 | -4.28807282E+01 | 1.10E+00 |
| 362 | Sc | 47 | 21 | 26 | 5 | 8.66506900E+00 | 8.68119976E+00 | -1.60E-02 | 4.37251334E+04 | 4.37243600E+04 | 7.70E-01 | 4.37358852E+04 | 4.37351119E+04 | 7.70E-01 | -4.43356300E+01 | -4.51090127E+01 | 7.70E-01 |
| 363 | Sc | 48 | 21 | 27 | 6 | 8.65619500E+00 | 8.65907716E+00 | -2.90E-03 | 4.46564596E+04 | 4.46563060E+04 | 1.50E-01 | 4.46672115E+04 | 4.46670579E+04 | 1.50E-01 | -4.45034200E+01 | -4.46570085E+01 | 1.50E-01 |
| 364 | Sc | 49 | 21 | 28 | 7 | 8.68625200E+00 | 8.66409106E+00 | 2.20E-02 | 4.55858960E+04 | 4.55869667E+04 | -1.10E+00 | 4.55966479E+04 | 4.55977185E+04 | -1.10E+00 | -4.65610730E+01 | -4.54904481E+01 | -1.10E+00 |
| 365 | Sc | 50 | 21 | 29 | 8 | 8.63368200E+00 | 8.61577825E+00 | 1.80E-02 | 4.65194037E+04 | 4.65202836E+04 | -8.80E-01 | 4.65301555E+04 | 4.65310355E+04 | -8.80E-01 | -4.45475140E+01 | -4.36675794E+01 | -8.80E-01 |
| 366 | Sc | 51 | 21 | 30 | 9 | 8.59679800E+00 | 8.59076186E+00 | 6.00E-03 | 4.74522164E+04 | 4.74525090E+04 | -2.90E-01 | 4.74629683E+04 | 4.74632609E+04 | -2.90E-01 | -4.32288140E+01 | -4.29362030E+01 | -2.90E-01 |
| 367 | Sc | 52 | 21 | 31 | 10 | 8.52780200E+00 | 8.51982836E+00 | 8.00E-03 | 4.83867728E+04 | 4.83871722E+04 | -4.00E-01 | 4.83975247E+04 | 4.83979241E+04 | -4.00E-01 | -4.01665060E+01 | -3.97671039E+01 | -4.00E-01 |
| 368 | Sc | 53 | 21 | 32 | 11 | 8.48033900E+00 | 8.47365435E+00 | 6.70E-03 | 4.93203259E+04 | 4.93206650E+04 | -3.40E-01 | 4.93310778E+04 | 4.93314168E+04 | -3.40E-01 | -3.81074200E+01 | -3.77683908E+01 | -3.40E-01 |
| 369 | Sc | 54 | 21 | 33 | 12 | 8.38927500E+00 | 8.38985426E+00 | -5.80E-04 | 5.02563284E+04 | 5.02562819E+04 | 4.70E-02 | 5.02670803E+04 | 5.02670338E+04 | 4.70E-02 | -3.35989890E+01 | -3.36455211E+01 | 4.70E-02 |
| 370 | Sc | 55 | 21 | 34 | 13 | 8.31763000E+00 | 8.33074225E+00 | -1.30E-02 | 5.11914450E+04 | 5.11907086E+04 | 7.40E-01 | 5.12021968E+04 | 5.12014605E+04 | 7.40E-01 | -2.99765090E+01 | -3.07128955E+01 | 7.40E-01 |
| 371 | Sc | 56 | 21 | 35 | 14 | 8.22000000E+00 | 8.23813876E+00 | -1.80E-02 | 5.21281844E+04 | 5.21271290E+04 | 1.10E+00 | 5.21389362E+04 | 5.21378809E+04 | 1.10E+00 | -2.47310000E+01 | -2.57865235E+01 | 1.10E+00 |
| 372 | Sc | 57 | 21 | 36 | 15 | 8.14600000E+00 | 8.16756296E+00 | -2.20E-02 | 5.30637025E+04 | 5.30624791E+04 | 1.20E+00 | 5.30744544E+04 | 5.30732310E+04 | 1.20E+00 | -2.07070000E+01 | -2.19305227E+01 | 1.20E+00 |
| 373 | SC | 58 | 21 | 37 | 16 | 8.04500000E+00 | 8.06590070E+00 | -2.10E-02 | 5.40010277E+04 | 5.39997733E+04 | 1.30E+00 | 5.40117796E+04 | 5.40105252E+04 | 1.30E+00 | -1.48760000E+01 | -1.61303559E+01 | 1.30E+00 |
| 374 | SC | 59 | 21 | 38 | 17 | 7.96700000E+00 | 7.98397649E+00 | -1.70E-02 | 5.49370954E+04 | 5.49361063E+04 | 9.90E-01 | 5.49478473E+04 | 5.49468582E+04 | 9.90E-01 | -1.03020000E+01 | -1.12914090E+01 | 9.90E-01 |
| 375 | SC | 61 | 21 | 40 | 19 | 7.78700000E+00 | 7.78498134E+00 | 2.00E-03 | 5.68113173E+04 | 5.68114078E+04 | -9.00E-02 | 5.68220692E+04 | 5.68221597E+04 | -9.00E-02 | 9.31000000E-01 | 1.02198006E+00 | -9.10E-02 |
| 376 | Ti | 38 | 22 | 16 | -6 | 7.33800000E+00 | 7.37991408E+00 | -4.20E-02 | 3.53961746E+04 | 3.53945547E+04 | 1.60E+00 | 3.54074399E+04 | 3.54058200E+04 | 1.60E+00 | 1.06660000E+01 | 9.04569298E+00 | 1.60E+00 |
| 377 | Ti | 39 | 22 | 17 | -5 | 7.57400000E+00 | 7.59115604E+00 | -1.70E-02 | 3.63192014E+04 | 3.63185017E+04 | 7.00E-01 | 3.63304667E+04 | 3.63297670E+04 | 7.00E-01 | 2.19800000E+00 | 1.49866145E+00 | 7.00E-01 |
| 378 | Ti | 40 | 22 | 18 | -4 | 7.86228600E+00 | 7.88658309E+00 | -2.40E-02 | 3.72396467E+04 | 3.72386589E+04 | 9.90E-01 | 3.72509120E+04 | 3.72499242E+04 | 9.90E-01 | -8.85038600E+00 | -9.83825746E+00 | 9.90E-01 |
| 379 | Ti | 41 | 22 | 19 | -3 | 8.03438800E+00 | 8.04856999E+00 | -1.40E-02 | 3.81642936E+04 | 3.81636962E+04 | 6.00E-01 | 3.81755590E+04 | 3.81749615E+04 | 6.00E-01 | -1.56975370E+01 | -1.62949847E+01 | 6.00E-01 |
| 380 | Ti | 42 | 22 | 20 | -2 | 8.25924700E+00 | 8.25965046E+00 | -4.00E-04 | 3.90863806E+04 | 3.90863476E+04 | 3.30E-02 | 3.90976459E+04 | 3.90976129E+04 | 3.30E-02 | -2.51046630E+01 | -2.51376155E+01 | 3.30E-02 |
| 381 | Ti | 43 | 22 | 21 | -1 | 8.35293200E+00 | 8.36906837E+00 | -1.60E-02 | 4.00136582E+04 | 4.00129484E+04 | 7.10E-01 | 4.00249235E+04 | 4.00242137E+04 | 7.10E-01 | -2.93210840E+01 | -3.00309170E+01 | 7.10E-01 |
| 382 | Ti | 44 | 22 | 22 | 0 | 8.53352000E+00 | 8.52003320E+00 | 1.30E-02 | 4.09369248E+04 | 4.09375022E+04 | -5.80E-01 | 4.09481901E+04 | 4.09487676E+04 | -5.80E-01 | -3.75485700E+01 | -3.69711186E+01 | -5.80E-01 |
| 383 | Ti | 45 | 22 | 23 | 1 | 8.55570600E+00 | 8.55459920E+00 | 1.10E-03 | 4.18669583E+04 | 4.18669921E+04 | -3.40E-02 | 4.18782236E+04 | 4.18782574E+04 | -3.40E-02 | -3.90091210E+01 | -3.89753030E+01 | -3.40E-02 |
| 384 | Ti | 46 | 22 | 24 | 2 | 8.65643400E+00 | 8.66394812E+00 | -7.50E-03 | 4.27933345E+04 | 4.27929728E+04 | 3.60E-01 | 4.28045998E+04 | 4.28042382E+04 | 3.60E-01 | -4.41269960E+01 | -4.44886337E+01 | 3.60E-01 |
| 385 | Ti | 47 | 22 | 25 | 3 | 8.66120600E+00 | 8.67202200E+00 | -1.10E-02 | 4.37240191E+04 | 4.37234948E+04 | 5.20E-01 | 4.37352845E+04 | 4.37347601E+04 | 5.20E-01 | -4.49364000E+01 | -4.54607349E+01 | 5.20E-01 |
| 386 | Ti | 48 | 22 | 26 | 4 | 8.72298600E+00 | 8.73641484E+00 | -1.30E-02 | 4.46519579E+04 | 4.46512973E+04 | 6.60E-01 | 4.46632232E+04 | 4.46625626E+04 | 6.60E-01 | -4.84917340E+01 | -4.91522941E+01 | 6.60E-01 |
| 387 | Ti | 49 | 22 | 27 | 5 | 8.71113700E+00 | 8.71710987E+00 | -6.00E-03 | 4.55833809E+04 | 4.55830722E+04 | 3.10E-01 | 4.55946462E+04 | 4.55943375E+04 | 3.10E-01 | -4.85628090E+01 | -4.88714465E+01 | 3.10E-01 |
| 388 | Ti | 50 | 22 | 28 | 6 | 8.75569800E+00 | 8.74658517E+00 | 9.10E-03 | 4.65120070E+04 | 4.65124467E+04 | -4.40E-01 | 4.65232724E+04 | 4.65237120E+04 | -4.40E-01 | -5.14306810E+01 | -5.09910024E+01 | -4.40E-01 |
| 389 | Ti | 51 | 22 | 29 | 7 | 8.70896900E+00 | 8.69971240E+00 | 9.30E-03 | 4.74451999E+04 | 4.74456560E+04 | -4.60E-01 | 4.74564653E+04 | 4.74569214E+04 | -4.60E-01 | -4.97318580E+01 | -4.92757574E+01 | -4.60E-01 |
| 390 | Ti | 52 | 22 | 30 | 8 | 8.69164800E+00 | 8.69636617E+00 | -4.70E-03 | 4.83769570E+04 | 4.83766957E+04 | 2.60E-01 | 4.83882223E+04 | 4.83879610E+04 | 2.60E-01 | -4.94688450E+01 | -4.97301466E+01 | 2.60E-01 |
| 391 | Ti | 53 | 22 | 31 | 9 | 8.63015400E+00 | 8.62612477E+00 | 4.00E-03 | 4.93110899E+04 | 4.93112875E+04 | -2.00E-01 | 4.93223552E+04 | 4.93225528E+04 | -2.00E-01 | -4.68299910E+01 | -4.66324000E+01 | -2.00E-01 |
| 392 | Ti | 54 | 22 | 32 | 10 | 8.59697300E+00 | 8.60012653E+00 | -3.20E-03 | 5.02438169E+04 | 5.02436307E+04 | 1.90E-01 | 5.02550822E+04 | 5.02548960E+04 | 1.90E-01 | -4.55970530E+01 | -4.57833009E+01 | 1.90E-01 |
| 393 | Ti | 55 | 22 | 33 | 11 | 8.51597000E+00 | 8.51673920E+00 | -7.70E-04 | 5.11792405E+04 | 5.11791822E+04 | 5.80E-02 | 5.11905058E+04 | 5.11904476E+04 | 5.80E-02 | -4.16675300E+01 | -4.17258050E+01 | 5.80E-02 |
| 394 | Ti | 56 | 22 | 34 | 12 | 8.46406200E+00 | 8.47762585E+00 | -1.40E-02 | 5.21131968E+04 | 5.21124212E+04 | 7.80E-01 | 5.21244621E+04 | 5.21236865E+04 | 7.80E-01 | -3.92053220E+01 | -3.99808774E+01 | 7.80E-01 |
| 395 | Ti | 57 | 22 | 35 | 13 | 8.36352700E+00 | 8.38562273E+00 | -2.20E-02 | 5.30500285E+04 | 5.30487531E+04 | 1.30E+00 | 5.30612939E+04 | 5.30600185E+04 | 1.30E+00 | -3.38676070E+01 | -3.51430068E+01 | 1.30E+00 |
| 396 | Ti | 58 | 22 | 36 | 14 | 8.31100000E+00 | 8.33576580E+00 | -2.50E-02 | 5.39842783E+04 | 5.39828246E+04 | 1.50E+00 | 5.39955436E+04 | 5.39940899E+04 | 1.50E+00 | -3.11120000E+01 | -3.25656087E+01 | 1.50E+00 |
| 397 | Ti | 59 | 22 | 37 | 15 | 8.21400000E+00 | 8.23480190E+00 | -2.10E-02 | 5.49212402E+04 | 5.49200111E+04 | 1.20E+00 | 5.49325056E+04 | 5.49312764E+04 | 1.20E+00 | -2.56440000E+01 | -2.68731852E+01 | 1.20E+00 |
| 398 | Ti | 60 | 22 | 38 | 16 | 8.15700000E+00 | 8.17446819E+00 | -1.70E-02 | 5.58560504E+04 | 5.58549617E+04 | 1.10E+00 | 5.58673158E+04 | 5.58662270E+04 | 1.10E+00 | -2.23280000E+01 | -2.34166451E+01 | 1.10E+00 |
| 399 | TI | 61 | 22 | 39 | 17 | 8.05700000E+00 | 8.06517476E+00 | -8.20E-03 | 5.67935247E+04 | 5.67930195E+04 | 5.10E-01 | 5.68047900E+04 | 5.68042848E+04 | 5.10E-01 | -1.63480000E+01 | -1.68528953E+01 | 5.00E-01 |
| 400 | TI | 62 | 22 | 40 | 18 | 7.99600000E+00 | 7.99733132E+00 | -1.30E-03 | 5.77288006E+04 | 5.77287260E+04 | 7.50E-02 | 5.77400659E+04 | 5.77399913E+04 | 7.50E-02 | -1.25660000E+01 | -1.26404579E+01 | 7.40E-02 |
| 401 | TI | 63 | 22 | 41 | 19 | 7.89100000E+00 | 7.88252900E+00 | 8.50E-03 | 5.86670387E+04 | 5.86675266E+04 | -4.90E-01 | 5.86783040E+04 | 5.86787919E+04 | -4.90E-01 | -5.82200000E+00 | -5.33392381E+00 | -4.90E-01 |
| 402 | V | 42 | 23 | 19 | -4 | 7.82400000E+00 | 7.83503068E+00 | -1.10E-02 | 3.91033520E+04 | 3.91028850E+04 | 4.70E-01 | 3.91151309E+04 | 3.91146639E+04 | 4.70E-01 | -7.62000000E+00 | -8.08666153E+00 | 4.70E-01 |
| 403 | V | 43 | 23 | 20 | -3 | 8.06951200E+00 | 8.07536265E+00 | -5.90E-03 | 4.00245493E+04 | 4.00242810E+04 | 2.70E-01 | 4.00363283E+04 | 4.00360600E+04 | 2.70E-01 | -1.79163560E+01 | -1.81846480E+01 | 2.70E-01 |
| 404 | V | 44 | 23 | 21 | -2 | 8.21046300E+00 | 8.21629718E+00 | -5.80E-03 | 4.09498434E+04 | 4.09495699E+04 | 2.70E-01 | 4.09616223E+04 | 4.09613489E+04 | 2.70E-01 | -2.41163800E+01 | -2.43898107E+01 | 2.70E-01 |
| 405 | V | 45 | 23 | 22 | -1 | 8.37990800E+00 | 8.38193906E+00 | -2.00E-03 | 4.18735732E+04 | 4.18734651E+04 | 1.10E-01 | 4.18853522E+04 | 4.18852441E+04 | 1.10E-01 | -3.18805490E+01 | -3.19886736E+01 | 1.10E-01 |
| 406 | V | 46 | 23 | 23 | 0 | 8.48611300E+00 | 8.45860776E+00 | 2.80E-02 | 4.27998733E+04 | 4.28011218E+04 | -1.20E+00 | 4.28116522E+04 | 4.28129008E+04 | -1.20E+00 | -3.70746020E+01 | -3.58260538E+01 | -1.20E+00 |
| 407 | V | 47 | 23 | 24 | 1 | 8.58220700E+00 | 8.57872036E+00 | 3.50E-03 | 4.37264361E+04 | 4.37265833E+04 | -1.50E-01 | 4.37382151E+04 | 4.37383622E+04 | -1.50E-01 | -4.20058010E+01 | -4.18586350E+01 | -1.50E-01 |
| 408 | V | 48 | 23 | 25 | 2 | 8.62304200E+00 | 8.62173464E+00 | 1.30E-03 | 4.46554592E+04 | 4.46555053E+04 | -4.60E-02 | 4.46672381E+04 | 4.46672842E+04 | -4.60E-02 | -4.44767680E+01 | -4.44307216E+01 | -4.60E-02 |
| 409 | V | 49 | 23 | 26 | 3 | 8.68288800E+00 | 8.69327722E+00 | -1.00E-02 | 4.55834691E+04 | 4.55829433E+04 | 5.30E-01 | 4.55952480E+04 | 4.55947223E+04 | 5.30E-01 | -4.79609530E+01 | -4.84867237E+01 | 5.30E-01 |
| 410 | V | 50 | 23 | 27 | 4 | 8.69591500E+00 | 8.70216672E+00 | -6.30E-03 | 4.65137002E+04 | 4.65133709E+04 | 3.30E-01 | 4.65254792E+04 | 4.65251499E+04 | 3.30E-01 | -4.92238560E+01 | -4.95531568E+01 | 3.30E-01 |
| 411 | V | 51 | 23 | 28 | 5 | 8.74209600E+00 | 8.73594538E+00 | 6.20E-03 | 4.74422145E+04 | 4.74425114E+04 | -3.00E-01 | 4.74539934E+04 | 4.74542904E+04 | -3.00E-01 | -5.22036850E+01 | -5.19067166E+01 | -3.00E-01 |
| 412 | V | 52 | 23 | 29 | 6 | 8.71457900E+00 | 8.71296497E+00 | 1.60E-03 | 4.83744686E+04 | 4.83745359E+04 | -6.70E-02 | 4.83862476E+04 | 4.83863148E+04 | -6.70E-02 | -5.14436120E+01 | -5.13763612E+01 | -6.70E-02 |
| 413 | V | 53 | 23 | 30 | 7 | 8.71011000E+00 | 8.71254420E+00 | -2.40E-03 | 4.93055563E+04 | 4.93054106E+04 | 1.50E-01 | 4.93173352E+04 | 4.93171895E+04 | 1.50E-01 | -5.18499910E+01 | -5.19957067E+01 | 1.50E-01 |
| 414 | V | 54 | 23 | 31 | 8 | 8.66202300E+00 | 8.66389871E+00 | -1.90E-03 | 5.02390082E+04 | 5.02388903E+04 | 1.20E-01 | 5.02507872E+04 | 5.02506692E+04 | 1.20E-01 | -4.98921070E+01 | -5.00100754E+01 | 1.20E-01 |
| 415 | V | 55 | 23 | 32 | 9 | 8.63767400E+00 | 8.63994312E+00 | -2.30E-03 | 5.11712508E+04 | 5.11711093E+04 | 1.40E-01 | 5.11830297E+04 | 5.11828883E+04 | 1.40E-01 | -4.91436200E+01 | -4.92850977E+01 | 1.40E-01 |
| 416 | V | 56 | 23 | 33 | 10 | 8.57362300E+00 | 8.57684813E+00 | -3.20E-03 | 5.21057654E+04 | 5.21055681E+04 | 2.00E-01 | 5.21175443E+04 | 5.21173470E+04 | 2.00E-01 | -4.61231020E+01 | -4.63204020E+01 | 2.00E-01 |
| 417 | V | 57 | 23 | 34 | 11 | 8.53157000E+00 | 8.53907954E+00 | -7.50E-03 | 5.30391541E+04 | 5.30387094E+04 | 4.40E-01 | 5.30509331E+04 | 5.30504884E+04 | 4.40E-01 | -4.42283890E+01 | -4.46731219E+01 | 4.40E-01 |
| 418 | V | 58 | 23 | 35 | 12 | 8.45624000E+00 | 8.46706805E+00 | -1.10E-02 | 5.39745571E+04 | 5.39739124E+04 | 6.40E-01 | 5.39863360E+04 | 5.39856913E+04 | 6.40E-01 | -4.03194990E+01 | -4.09642161E+01 | 6.40E-01 |
| 419 | V | 59 | 23 | 36 | 13 | 8.40755500E+00 | 8.41815855E+00 | -1.10E-02 | 5.49085386E+04 | 5.49078963E+04 | 6.40E-01 | 5.49203176E+04 | 5.49196753E+04 | 6.40E-01 | -3.78320150E+01 | -3.84743045E+01 | 6.40E-01 |
| 420 | V | 60 | 23 | 37 | 14 | 8.32545000E+00 | 8.33760384E+00 | -1.20E-02 | 5.58446228E+04 | 5.58438768E+04 | 7.50E-01 | 5.58564017E+04 | 5.58556558E+04 | 7.50E-01 | -3.32419560E+01 | -3.39878616E+01 | 7.50E-01 |
| 421 | V | 61 | 23 | 38 | 15 | 8.27643900E+00 | 8.27774846E+00 | -1.30E-03 | 5.67788523E+04 | 5.67787558E+04 | 9.70E-02 | 5.67906313E+04 | 5.67905348E+04 | 9.70E-02 | -3.05064290E+01 | -3.06029681E+01 | 9.70E-02 |
| 422 | V | 62 | 23 | 39 | 16 | 8.19200000E+00 | 8.18939591E+00 | 2.60E-03 | 5.77153765E+04 | 5.77155213E+04 | -1.40E-01 | 5.77271554E+04 | 5.77273002E+04 | -1.40E-01 | -2.54760000E+01 | -2.53315392E+01 | -1.40E-01 |
| 423 | V | 63 | 23 | 40 | 17 | 8.13500000E+00 | 8.12095558E+00 | 1.40E-02 | 5.86503543E+04 | 5.86512090E+04 | -8.50E-01 | 5.86621333E+04 | 5.86629880E+04 | -8.50E-01 | -2.19930000E+01 | -2.11378756E+01 | -8.60E-01 |
| 424 | V | 64 | 23 | 41 | 18 | 8.04300000E+00 | 8.02730892E+00 | 1.60E-02 | 5.95876702E+04 | 5.95886468E+04 | -9.80E-01 | 5.95994492E+04 | 5.96004258E+04 | -9.80E-01 | -1.61710000E+01 | -1.51941258E+01 | -9.80E-01 |
| 425 | V | 65 | 23 | 42 | 19 | 7.97400000E+00 | 7.95338519E+00 | 2.10E-02 | 6.05236913E+04 | 6.05249899E+04 | -1.30E+00 | 6.05354703E+04 | 6.05367689E+04 | -1.30E+00 | -1.16440000E+01 | -1.03450735E+01 | -1.30E+00 |
| 426 | V | 66 | 23 | 43 | 20 | 7.88400000E+00 | 7.85850963E+00 | 2.50E-02 | 6.14612215E+04 | 6.14628637E+04 | -1.60E+00 | 6.14730004E+04 | 6.14746427E+04 | -1.60E+00 | -5.60800000E+00 | -3.96535248E+00 | -1.60E+00 |
| 427 | Cr | 42 | 24 | 18 | -6 | 7.47600000E+00 | 7.47293675E+00 | 3.10E-03 | 3.91166988E+04 | 3.91167960E+04 | -9.70E-02 | 3.91289916E+04 | 3.91290888E+04 | -9.70E-02 | 6.24100000E+00 | 6.33820633E+00 | -9.70E-02 |
| 428 | Cr | 43 | 24 | 19 | -5 | 7.68800000E+00 | 7.67379146E+00 | 1.40E-02 | 4.00396511E+04 | 4.00402517E+04 | -6.00E-01 | 4.00519438E+04 | 4.00525445E+04 | -6.00E-01 | -2.30100000E+00 | -1.70016378E+00 | -6.00E-01 |
| 429 | Cr | 44 | 24 | 20 | -4 | 7.95500000E+00 | 7.94428709E+00 | 1.10E-02 | 4.09598089E+04 | 4.09602415E+04 | -4.30E-01 | 4.09721016E+04 | 4.09725342E+04 | -4.30E-01 | -1.36370000E+01 | -1.32044438E+01 | -4.30E-01 |
| 430 | Cr | 45 | 24 | 21 | -3 | 8.08772800E+00 | 8.10376909E+00 | -1.60E-02 | 4.18854252E+04 | 4.18846859E+04 | 7.40E-01 | 4.18977179E+04 | 4.18969786E+04 | 7.40E-01 | -1.95147990E+01 | -2.02541018E+01 | 7.40E-01 |
| 431 | Cr | 46 | 24 | 22 | -2 | 8.30386500E+00 | 8.29827200E+00 | 5.60E-03 | 4.28069605E+04 | 4.28072004E+04 | -2.40E-01 | 4.28192533E+04 | 4.28194931E+04 | -2.40E-01 | -2.94735310E+01 | -2.92336861E+01 | -2.40E-01 |
| 432 | Cr | 47 | 24 | 23 | -1 | 8.40715900E+00 | 8.40028373E+00 | 6.90E-03 | 4.37333672E+04 | 4.37336729E+04 | -3.10E-01 | 4.37456600E+04 | 4.37459657E+04 | -3.10E-01 | -3.45608870E+01 | -3.42551902E+01 | -3.10E-01 |
| 433 | Cr | 48 | 24 | 24 | 0 | 8.57226200E+00 | 8.55736772E+00 | 1.50E-02 | 4.46566005E+04 | 4.46572980E+04 | -7.00E-01 | 4.46688933E+04 | 4.46695907E+04 | -7.00E-01 | -4.28216540E+01 | -4.21241867E+01 | -7.00E-01 |
| 434 | Cr | 49 | 24 | 25 | 1 | 8.61328400E+00 | 8.60796072E+00 | 5.30E-03 | 4.55855836E+04 | 4.55858270E+04 | -2.40E-01 | 4.55978763E+04 | 4.55981197E+04 | -2.40E-01 | -4.53326930E+01 | -4.50892922E+01 | -2.40E-01 |
| 435 | Cr | 50 | 24 | 26 | 2 | 8.70102500E+00 | 8.71108811E+00 | -1.00E-02 | 4.65121486E+04 | 4.65116280E+04 | 5.20E-01 | 4.65244413E+04 | 4.65239207E+04 | 5.20E-01 | -5.02617090E+01 | -5.07823036E+01 | 5.20E-01 |
| 436 | Cr | 51 | 24 | 27 | 3 | 8.71199800E+00 | 8.72312975E+00 | -1.10E-02 | 4.74424533E+04 | 4.74418682E+04 | 5.90E-01 | 4.74547461E+04 | 4.74541609E+04 | 5.90E-01 | -5.14510540E+01 | -5.20361963E+01 | 5.90E-01 |
| 437 | Cr | 52 | 24 | 28 | 4 | 8.77596700E+00 | 8.78319532E+00 | -7.20E-03 | 4.83699803E+04 | 4.83695870E+04 | 3.90E-01 | 4.83822731E+04 | 4.83818798E+04 | 3.90E-01 | -5.54180890E+01 | -5.58114168E+01 | 3.90E-01 |
| 438 | Cr | 53 | 24 | 29 | 5 | 8.76017700E+00 | 8.76157543E+00 | -1.40E-03 | 4.93016066E+04 | 4.93015151E+04 | 9.20E-02 | 4.93138993E+04 | 4.93138078E+04 | 9.20E-02 | -5.52858940E+01 | -5.53774387E+01 | 9.20E-02 |
| 439 | Cr | 54 | 24 | 30 | 6 | 8.77793500E+00 | 8.78439981E+00 | -6.50E-03 | 5.02314529E+04 | 5.02310863E+04 | 3.70E-01 | 5.02437456E+04 | 5.02433791E+04 | 3.70E-01 | -5.69336970E+01 | -5.73002119E+01 | 3.70E-01 |
| 440 | Cr | 55 | 24 | 31 | 7 | 8.73190500E+00 | 8.73676399E+00 | -4.90E-03 | 5.11647720E+04 | 5.11644873E+04 | 2.80E-01 | 5.11770647E+04 | 5.11767800E+04 | 2.80E-01 | -5.51086430E+01 | -5.53933227E+01 | 2.80E-01 |
| 441 | Cr | 56 | 24 | 32 | 8 | 8.72319100E+00 | 8.73439888E+00 | -1.10E-02 | 5.20960934E+04 | 5.20954484E+04 | 6.50E-01 | 5.21083862E+04 | 5.21077411E+04 | 6.50E-01 | -5.52812450E+01 | -5.59263215E+01 | 6.50E-01 |
| 442 | Cr | 57 | 24 | 33 | 9 | 8.66338400E+00 | 8.67215360E+00 | -8.80E-03 | 5.30303446E+04 | 5.30298273E+04 | 5.20E-01 | 5.30426373E+04 | 5.30421201E+04 | 5.20E-01 | -5.25241390E+01 | -5.30414202E+01 | 5.20E-01 |
| 443 | Cr | 58 | 24 | 34 | 10 | 8.64129000E+00 | 8.65476866E+00 | -1.30E-02 | 5.39625281E+04 | 5.39617289E+04 | 8.00E-01 | 5.39748208E+04 | 5.39740216E+04 | 8.00E-01 | -5.18347260E+01 | -5.26339283E+01 | 8.00E-01 |
| 444 | Cr | 59 | 24 | 35 | 11 | 8.56479500E+00 | 8.58339540E+00 | -1.90E-02 | 5.48979654E+04 | 5.48968505E+04 | 1.10E+00 | 5.49102581E+04 | 5.49091432E+04 | 1.10E+00 | -4.78914900E+01 | -4.90063555E+01 | 1.10E+00 |
| 445 | Cr | 60 | 24 | 36 | 12 | 8.53344300E+00 | 8.55420805E+00 | -2.10E-02 | 5.58308470E+04 | 5.58295837E+04 | 1.30E+00 | 5.58431398E+04 | 5.58418765E+04 | 1.30E+00 | -4.65038760E+01 | -4.77671909E+01 | 1.30E+00 |
| 446 | Cr | 61 | 24 | 37 | 13 | 8.45949400E+00 | 8.47413486E+00 | -1.50E-02 | 5.67663899E+04 | 5.67654794E+04 | 9.10E-01 | 5.67786826E+04 | 5.67777721E+04 | 9.10E-01 | -4.24550890E+01 | -4.33656155E+01 | 9.10E-01 |
| 447 | Cr | 62 | 24 | 38 | 14 | 8.42806900E+00 | 8.43372328E+00 | -5.70E-03 | 5.76994441E+04 | 5.76990761E+04 | 3.70E-01 | 5.77117368E+04 | 5.77113689E+04 | 3.70E-01 | -4.08949610E+01 | -4.12629131E+01 | 3.70E-01 |
| 448 | Cr | 63 | 24 | 39 | 15 | 8.34029800E+00 | 8.34547255E+00 | -5.20E-03 | 5.86361110E+04 | 5.86357676E+04 | 3.40E-01 | 5.86484037E+04 | 5.86480603E+04 | 3.40E-01 | -3.57221140E+01 | -3.60655215E+01 | 3.40E-01 |
| 449 | Cr | 64 | 24 | 40 | 16 | 8.30100000E+00 | 8.29613490E+00 | 4.90E-03 | 5.95698679E+04 | 5.95701451E+04 | -2.80E-01 | 5.95821606E+04 | 5.95824378E+04 | -2.80E-01 | -3.34590000E+01 | -3.31820654E+01 | -2.80E-01 |
| 450 | Cr | 65 | 24 | 41 | 17 | 8.21300000E+00 | 8.20170941E+00 | 1.10E-02 | 6.05068391E+04 | 6.05075520E+04 | -7.10E-01 | 6.05191319E+04 | 6.05198447E+04 | -7.10E-01 | -2.79820000E+01 | -2.72692248E+01 | -7.10E-01 |
| 451 | Cr | 66 | 24 | 42 | 18 | 8.15900000E+00 | 8.14635549E+00 | 1.30E-02 | 6.14417797E+04 | 6.14425690E+04 | -7.90E-01 | 6.14540725E+04 | 6.14548618E+04 | -7.90E-01 | -2.45360000E+01 | -2.37462562E+01 | -7.90E-01 |
| 452 | CR | 68 | 24 | 44 | 20 | 8.01400000E+00 | 7.99270817E+00 | 2.10E-02 | 6.33144275E+04 | 6.33158551E+04 | -1.40E+00 | 6.33267202E+04 | 6.33281478E+04 | -1.40E+00 | -1.48760000E+01 | -1.34483114E+01 | -1.40E+00 |
| 453 | Mn | 44 | 25 | 19 | -6 | 7.47500000E+00 | 7.45631548E+00 | 1.90E-02 | 4.09795922E+04 | 4.09804152E+04 | -8.20E-01 | 4.09923989E+04 | 4.09932219E+04 | -8.20E-01 | 6.66000000E+00 | 7.48323009E+00 | -8.20E-01 |
| 454 | Mn | 45 | 25 | 20 | -5 | 7.75100000E+00 | 7.74007856E+00 | 1.10E-02 | 4.18992935E+04 | 4.18997550E+04 | -4.60E-01 | 4.19121002E+04 | 4.19125616E+04 | -4.60E-01 | -5.13300000E+00 | -4.67110536E+00 | -4.60E-01 |
| 455 | Mn | 46 | 25 | 21 | -4 | 7.92800000E+00 | 7.92277706E+00 | 5.20E-03 | 4.28229630E+04 | 4.28231761E+04 | -2.10E-01 | 4.28357697E+04 | 4.28359828E+04 | -2.10E-01 | -1.29570000E+01 | -1.27439955E+01 | -2.10E-01 |
| 456 | Mn | 47 | 25 | 22 | -3 | 8.13529100E+00 | 8.13906747E+00 | -3.80E-03 | 4.37448487E+04 | 4.37446531E+04 | 2.00E-01 | 4.37576554E+04 | 4.37574598E+04 | 2.00E-01 | -2.25654420E+01 | -2.27611030E+01 | 2.00E-01 |
| 457 | Mn | 48 | 25 | 23 | -2 | 8.27475000E+00 | 8.26654293E+00 | 8.20E-03 | 4.46695848E+04 | 4.46699606E+04 | -3.80E-01 | 4.46823915E+04 | 4.46827673E+04 | -3.80E-01 | -2.93234310E+01 | -2.89476736E+01 | -3.80E-01 |
| 458 | Mn | 49 | 25 | 24 | -1 | 8.44025700E+00 | 8.43325528E+00 | 7.00E-03 | 4.55927656E+04 | 4.55930905E+04 | -3.20E-01 | 4.56055722E+04 | 4.56058972E+04 | -3.20E-01 | -3.76367380E+01 | -3.73118029E+01 | -3.20E-01 |
| 459 | Mn | 50 | 25 | 25 | 0 | 8.53268900E+00 | 8.51902845E+00 | 1.40E-02 | 4.65192691E+04 | 4.65199340E+04 | -6.60E-01 | 4.65320758E+04 | 4.65327407E+04 | -6.60E-01 | -4.26272320E+01 | -4.19623976E+01 | -6.60E-01 |
| 460 | Mn | 51 | 25 | 26 | 1 | 8.63376500E+00 | 8.62933755E+00 | 4.40E-03 | 4.74451469E+04 | 4.74453546E+04 | -2.10E-01 | 4.74579536E+04 | 4.74581612E+04 | -2.10E-01 | -4.82434950E+01 | -4.80358712E+01 | -2.10E-01 |
| 461 | Mn | 52 | 25 | 27 | 2 | 8.67032100E+00 | 8.67139546E+00 | -1.10E-03 | 4.83741776E+04 | 4.83741036E+04 | 7.40E-02 | 4.83869843E+04 | 4.83869103E+04 | 7.40E-02 | -5.07068540E+01 | -5.07809010E+01 | 7.40E-02 |
| 462 | Mn | 53 | 25 | 28 | 3 | 8.73415500E+00 | 8.73565957E+00 | -1.50E-03 | 4.93016895E+04 | 4.93015916E+04 | 9.80E-02 | 4.93144962E+04 | 4.93143983E+04 | 9.80E-02 | -5.46890450E+01 | -5.47869752E+01 | 9.80E-02 |
| 463 | Mn | 54 | 25 | 29 | 4 | 8.73794400E+00 | 8.73936072E+00 | -1.40E-03 | 5.02323161E+04 | 5.02322214E+04 | 9.50E-02 | 5.02451228E+04 | 5.02450281E+04 | 9.50E-02 | -5.55565240E+01 | -5.56511781E+01 | 9.50E-02 |
| 464 | Mn | 55 | 25 | 30 | 5 | 8.76500900E+00 | 8.76501559E+00 | -6.60E-06 | 5.11616549E+04 | 5.11616364E+04 | 1.90E-02 | 5.11744616E+04 | 5.11744431E+04 | 1.90E-02 | -5.77117250E+01 | -5.77302373E+01 | 1.90E-02 |
| 465 | Mn | 56 | 25 | 31 | 6 | 8.73832000E+00 | 8.74037852E+00 | -2.10E-03 | 5.20939499E+04 | 5.20938165E+04 | 1.30E-01 | 5.21067566E+04 | 5.21066232E+04 | 1.30E-01 | -5.69108450E+01 | -5.70442581E+01 | 1.30E-01 |
| 466 | Mn | 57 | 25 | 32 | 7 | 8.73671100E+00 | 8.74071968E+00 | -4.00E-03 | 5.30248687E+04 | 5.30246220E+04 | 2.50E-01 | 5.30376753E+04 | 5.30374287E+04 | 2.50E-01 | -5.74861310E+01 | -5.77327640E+01 | 2.50E-01 |
| 467 | Mn | 58 | 25 | 33 | 8 | 8.69664300E+00 | 8.70029069E+00 | -3.60E-03 | 5.39580213E+04 | 5.39577916E+04 | 2.30E-01 | 5.39708280E+04 | 5.39705983E+04 | 2.30E-01 | -5.58275600E+01 | -5.60572829E+01 | 2.30E-01 |
| 468 | Mn | 59 | 25 | 34 | 9 | 8.68092100E+00 | 8.68560276E+00 | -4.70E-03 | 5.48898176E+04 | 5.48895232E+04 | 2.90E-01 | 5.49026243E+04 | 5.49023299E+04 | 2.90E-01 | -5.55253200E+01 | -5.58196667E+01 | 2.90E-01 |
| 469 | Mn | 60 | 25 | 35 | 10 | 8.62813800E+00 | 8.63533744E+00 | -7.20E-03 | 5.58238690E+04 | 5.58234189E+04 | 4.50E-01 | 5.58366757E+04 | 5.58362256E+04 | 4.50E-01 | -5.29679380E+01 | -5.34180315E+01 | 4.50E-01 |
| 470 | Mn | 61 | 25 | 36 | 11 | 8.59891500E+00 | 8.60877406E+00 | -9.90E-03 | 5.67565889E+04 | 5.67559694E+04 | 6.20E-01 | 5.67693956E+04 | 5.67687760E+04 | 6.20E-01 | -5.17421220E+01 | -5.23616839E+01 | 6.20E-01 |
| 471 | Mn | 62 | 25 | 37 | 12 | 8.53800000E+00 | 8.54945155E+00 | -1.10E-02 | 5.76913446E+04 | 5.76906040E+04 | 7.40E-01 | 5.77041512E+04 | 5.77034106E+04 | 7.40E-01 | -4.84810000E+01 | -4.92211429E+01 | 7.40E-01 |
| 472 | Mn | 63 | 25 | 38 | 13 | 8.50510100E+00 | 8.51139847E+00 | -6.30E-03 | 5.86244321E+04 | 5.86240172E+04 | 4.10E-01 | 5.86372388E+04 | 5.86368239E+04 | 4.10E-01 | -4.68870530E+01 | -4.73019316E+01 | 4.10E-01 |
| 473 | Mn | 64 | 25 | 39 | 14 | 8.43741700E+00 | 8.44356613E+00 | -6.10E-03 | 5.95598242E+04 | 5.95594125E+04 | 4.10E-01 | 5.95726309E+04 | 5.95722192E+04 | 4.10E-01 | -4.29890340E+01 | -4.34007415E+01 | 4.10E-01 |
| 474 | Mn | 65 | 25 | 40 | 15 | 8.40068100E+00 | 8.39583448E+00 | 4.80E-03 | 6.04933399E+04 | 6.04936368E+04 | -3.00E-01 | 6.05061466E+04 | 6.05064435E+04 | -3.00E-01 | -4.09673380E+01 | -4.06704311E+01 | -3.00E-01 |
| 475 | Mn | 66 | 25 | 41 | 16 | 8.33179800E+00 | 8.32122137E+00 | 1.10E-02 | 6.14290509E+04 | 6.14297309E+04 | -6.80E-01 | 6.14418576E+04 | 6.14425375E+04 | -6.80E-01 | -3.67503870E+01 | -3.60704816E+01 | -6.80E-01 |
| 476 | Mn | 67 | 25 | 42 | 17 | 8.27700000E+00 | 8.26618825E+00 | 1.10E-02 | 6.23639852E+04 | 6.23646622E+04 | -6.80E-01 | 6.23767919E+04 | 6.23774689E+04 | -6.80E-01 | -3.33100000E+01 | -3.26331650E+01 | -6.80E-01 |
| 477 | MN | 68 | 25 | 43 | 18 | 8.20000000E+00 | 8.18819561E+00 | 1.20E-02 | 6.33004907E+04 | 6.33012649E+04 | -7.70E-01 | 6.33132974E+04 | 6.33140716E+04 | -7.70E-01 | -2.82990000E+01 | -2.75245346E+01 | -7.70E-01 |
| 478 | Mn | 69 | 25 | 44 | 19 | 8.14300000E+00 | 8.13024374E+00 | 1.30E-02 | 6.42357480E+04 | 6.42366408E+04 | -8.90E-01 | 6.42485547E+04 | 6.42494475E+04 | -8.90E-01 | -2.45360000E+01 | -2.36427319E+01 | -8.90E-01 |
| 479 | MN | 70 | 25 | 45 | 20 | 8.06600000E+00 | 8.05294335E+00 | 1.30E-02 | 6.51725609E+04 | 6.51734870E+04 | -9.30E-01 | 6.51853676E+04 | 6.51862936E+04 | -9.30E-01 | -1.92170000E+01 | -1.82906293E+01 | -9.30E-01 |
| 480 | MN | 71 | 25 | 46 | 21 | 8.01000000E+00 | 7.99514698E+00 | 1.50E-02 | 6.61080697E+04 | 6.61091029E+04 | -1.00E+00 | 6.61208764E+04 | 6.61219096E+04 | -1.00E+00 | -1.52020000E+01 | -1.41687113E+01 | -1.00E+00 |
| 481 | Fe | 45 | 26 | 19 | -7 | 7.32100000E+00 | 7.34157116E+00 | -2.10E-02 | 4.19173432E+04 | 4.19163906E+04 | 9.50E-01 | 4.19306640E+04 | 4.19297114E+04 | 9.50E-01 | 1.34310000E+01 | 1.24786506E+01 | 9.50E-01 |
| 482 | Fe | 46 | 26 | 20 | -6 | 7.61600000E+00 | 7.61773959E+00 | -1.70E-03 | 4.28359928E+04 | 4.28359106E+04 | 8.20E-02 | 4.28493136E+04 | 4.28492315E+04 | 8.20E-02 | 5.87000000E-01 | 5.04650639E-01 | 8.20E-02 |
| 483 | Fe | 47 | 26 | 21 | -5 | 7.80000000E+00 | 7.78828365E+00 | 1.20E-02 | 4.37593084E+04 | 4.37598427E+04 | -5.30E-01 | 4.37726292E+04 | 4.37731635E+04 | -5.30E-01 | -7.59200000E+00 | -7.05734033E+00 | -5.30E-01 |
| 484 | Fe | 48 | 26 | 22 | -4 | 8.03100000E+00 | 8.02175263E+00 | 9.20E-03 | 4.46799785E+04 | 4.46804133E+04 | -4.30E-01 | 4.46932993E+04 | 4.46937341E+04 | -4.30E-01 | -1.84160000E+01 | -1.79808162E+01 | -4.40E-01 |
| 485 | Fe | 49 | 26 | 23 | -3 | 8.16131100E+00 | 8.16174462E+00 | -4.30E-04 | 4.56051375E+04 | 4.56050973E+04 | 4.00E-02 | 4.56184583E+04 | 4.56184181E+04 | 4.00E-02 | -2.47507270E+01 | -2.47908575E+01 | 4.00E-02 |
| 486 | Fe | 50 | 26 | 24 | -2 | 8.35427000E+00 | 8.35125080E+00 | 3.00E-03 | 4.65268936E+04 | 4.65270257E+04 | -1.30E-01 | 4.65402144E+04 | 4.65403465E+04 | -1.30E-01 | -3.44886310E+01 | -3.43565919E+01 | -1.30E-01 |
| 487 | Fe | 51 | 26 | 25 | -1 | 8.46075500E+00 | 8.45587110E+00 | 4.90E-03 | 4.74526740E+04 | 4.74529041E+04 | -2.30E-01 | 4.74659948E+04 | 4.74662250E+04 | -2.30E-01 | -4.02023300E+01 | -3.99721590E+01 | -2.30E-01 |
| 488 | Fe | 52 | 26 | 26 | 0 | 8.60961100E+00 | 8.59784964E+00 | 1.20E-02 | 4.83760381E+04 | 4.83766308E+04 | -5.90E-01 | 4.83893589E+04 | 4.83899516E+04 | -5.90E-01 | -4.83322930E+01 | -4.77395955E+01 | -5.90E-01 |
| 489 | Fe | 53 | 26 | 27 | 1 | 8.64878400E+00 | 8.64443665E+00 | 4.30E-03 | 4.93049177E+04 | 4.93051292E+04 | -2.10E-01 | 4.93182385E+04 | 4.93184500E+04 | -2.10E-01 | -5.09467340E+01 | -5.07352373E+01 | -2.10E-01 |
| 490 | Fe | 54 | 26 | 28 | 2 | 8.73637000E+00 | 8.73610271E+00 | 2.70E-04 | 5.02311046E+04 | 5.02311002E+04 | 4.50E-03 | 5.02444254E+04 | 5.02444210E+04 | 4.50E-03 | -5.62538670E+01 | -5.62583222E+01 | 4.50E-03 |
| 491 | Fe | 55 | 26 | 29 | 3 | 8.74658300E+00 | 8.74073970E+00 | 5.80E-03 | 5.11613719E+04 | 5.11616744E+04 | -3.00E-01 | 5.11746927E+04 | 5.11749952E+04 | -3.00E-01 | -5.74806370E+01 | -5.71781404E+01 | -3.00E-01 |
| 492 | Fe | 56 | 26 | 30 | 4 | 8.79034200E+00 | 8.79010541E+00 | 2.40E-04 | 5.20897402E+04 | 5.20897346E+04 | 5.60E-03 | 5.21030610E+04 | 5.21030554E+04 | 5.60E-03 | -6.06064220E+01 | -6.06120408E+01 | 5.60E-03 |
| 493 | Fe | 57 | 26 | 31 | 5 | 8.77026700E+00 | 8.76577666E+00 | 4.50E-03 | 5.30216595E+04 | 5.30218966E+04 | -2.40E-01 | 5.30349803E+04 | 5.30352174E+04 | -2.40E-01 | -6.01811800E+01 | -5.99440885E+01 | -2.40E-01 |
| 494 | Fe | 58 | 26 | 32 | 6 | 8.79223900E+00 | 8.78803130E+00 | 4.20E-03 | 5.39511803E+04 | 5.39514054E+04 | -2.30E-01 | 5.39645011E+04 | 5.39647262E+04 | -2.30E-01 | -6.21544670E+01 | -6.19293155E+01 | -2.30E-01 |
| 495 | Fe | 59 | 26 | 33 | 7 | 8.75476000E+00 | 8.74812474E+00 | 6.60E-03 | 5.48841646E+04 | 5.48845373E+04 | -3.70E-01 | 5.48974854E+04 | 5.48978581E+04 | -3.70E-01 | -6.06641640E+01 | -6.02915409E+01 | -3.70E-01 |
| 496 | Fe | 60 | 26 | 34 | 8 | 8.75584000E+00 | 8.75428278E+00 | 1.60E-03 | 5.58149105E+04 | 5.58149850E+04 | -7.50E-02 | 5.58282313E+04 | 5.58283058E+04 | -7.50E-02 | -6.14123750E+01 | -6.13378289E+01 | -7.50E-02 |
| 497 | Fe | 61 | 26 | 35 | 9 | 8.70376800E+00 | 8.70489644E+00 | -1.10E-03 | 5.67488964E+04 | 5.67488087E+04 | 8.80E-02 | 5.67622172E+04 | 5.67621295E+04 | 8.80E-02 | -5.89204940E+01 | -5.90082260E+01 | 8.80E-02 |
| 498 | Fe | 62 | 26 | 36 | 10 | 8.69288200E+00 | 8.69855661E+00 | -5.70E-03 | 5.76804329E+04 | 5.76800622E+04 | 3.70E-01 | 5.76937537E+04 | 5.76933830E+04 | 3.70E-01 | -5.88780480E+01 | -5.92487340E+01 | 3.70E-01 |
| 499 | Fe | 63 | 26 | 37 | 11 | 8.63154900E+00 | 8.64050364E+00 | -9.00E-03 | 5.86151694E+04 | 5.86145864E+04 | 5.80E-01 | 5.86284902E+04 | 5.86279072E+04 | 5.80E-01 | -5.56356210E+01 | -5.62186342E+01 | 5.80E-01 |
| 500 | Fe | 64 | 26 | 38 | 12 | 8.61238800E+00 | 8.62232115E+00 | -9.90E-03 | 5.95473296E+04 | 5.95466750E+04 | 6.50E-01 | 5.95606504E+04 | 5.95599958E+04 | 6.50E-01 | -5.49695440E+01 | -5.56241393E+01 | 6.50E-01 |
| 501 | Fe | 65 | 26 | 39 | 13 | 8.54640100E+00 | 8.55594492E+00 | -9.50E-03 | 6.04825717E+04 | 6.04819325E+04 | 6.40E-01 | 6.04958925E+04 | 6.04952533E+04 | 6.40E-01 | -5.12214920E+01 | -5.18606871E+01 | 6.40E-01 |
| 502 | Fe | 66 | 26 | 40 | 14 | 8.52172400E+00 | 8.52765403E+00 | -5.90E-03 | 6.14152194E+04 | 6.14148091E+04 | 4.10E-01 | 6.14285402E+04 | 6.14281299E+04 | 4.10E-01 | -5.00678400E+01 | -5.04781139E+01 | 4.10E-01 |
| 503 | Fe | 67 | 26 | 41 | 15 | 8.45531000E+00 | 8.45418113E+00 | 1.10E-03 | 6.23507127E+04 | 6.23507695E+04 | -5.70E-02 | 6.23640336E+04 | 6.23640903E+04 | -5.70E-02 | -4.60685300E+01 | -4.60117644E+01 | -5.70E-02 |
| 504 | Fe | 68 | 26 | 42 | 16 | 8.41667500E+00 | 8.41782775E+00 | -1.20E-03 | 6.32844500E+04 | 6.32843527E+04 | 9.70E-02 | 6.32977708E+04 | 6.32976736E+04 | 9.70E-02 | -4.38253490E+01 | -4.39225973E+01 | 9.70E-02 |
| 505 | Fe | 69 | 26 | 43 | 17 | 8.34300000E+00 | 8.34009293E+00 | 2.90E-03 | 6.42207119E+04 | 6.42208640E+04 | -1.50E-01 | 6.42340327E+04 | 6.42341848E+04 | -1.50E-01 | -3.90580000E+01 | -3.89054032E+01 | -1.50E-01 |
| 506 | Fe | 70 | 26 | 44 | 18 | 8.29900000E+00 | 8.29978128E+00 | -7.80E-04 | 6.51549538E+04 | 6.51549111E+04 | 4.30E-02 | 6.51682746E+04 | 6.51682319E+04 | 4.30E-02 | -3.63100000E+01 | -3.63523615E+01 | 4.20E-02 |
| 507 | Fe | 71 | 26 | 45 | 19 | 8.22100000E+00 | 8.22162481E+00 | -6.20E-04 | 6.60917574E+04 | 6.60917258E+04 | 3.20E-02 | 6.61050782E+04 | 6.61050466E+04 | 3.20E-02 | -3.10000000E+01 | -3.10317147E+01 | 3.20E-02 |
| 508 | Fe | 72 | 26 | 46 | 20 | 8.17900000E+00 | 8.18066057E+00 | -1.70E-03 | 6.70261484E+04 | 6.70260190E+04 | 1.30E-01 | 6.70394692E+04 | 6.70393398E+04 | 1.30E-01 | -2.81030000E+01 | -2.82325952E+01 | 1.30E-01 |
| 509 | FE | 73 | 26 | 47 | 21 | 8.10200000E+00 | 8.10275127E+00 | -7.50E-04 | 6.79631290E+04 | 6.79630911E+04 | 3.80E-02 | 6.79764498E+04 | 6.79764119E+04 | 3.80E-02 | -2.26170000E+01 | -2.26545575E+01 | 3.80E-02 |
| 510 | FE | 74 | 26 | 48 | 22 | 8.05600000E+00 | 8.05813317E+00 | -2.10E-03 | 6.88980044E+04 | 6.88978555E+04 | 1.50E-01 | 6.89113252E+04 | 6.89111763E+04 | 1.50E-01 | -1.92350000E+01 | -1.93842508E+01 | 1.50E-01 |
| 511 | Co | 50 | 27 | 23 | -4 | 8.00400000E+00 | 8.00416070E+00 | -1.60E-04 | 4.65430857E+04 | 4.65430828E+04 | 2.90E-03 | 4.65569208E+04 | 4.65569179E+04 | 2.90E-03 | -1.77820000E+01 | -1.77851641E+01 | 3.20E-03 |
| 512 | Co | 51 | 27 | 24 | -3 | 8.19325400E+00 | 8.20742567E+00 | -1.40E-02 | 4.74650199E+04 | 4.74642775E+04 | 7.40E-01 | 4.74788550E+04 | 4.74781126E+04 | 7.40E-01 | -2.73421430E+01 | -2.80845191E+01 | 7.40E-01 |
| 513 | Co | 52 | 27 | 25 | -2 | 8.31900000E+00 | 8.33246974E+00 | -1.30E-02 | 4.83898659E+04 | 4.83891332E+04 | 7.30E-01 | 4.84037009E+04 | 4.84029683E+04 | 7.30E-01 | -3.39900000E+01 | -3.47229173E+01 | 7.30E-01 |
| 514 | Co | 53 | 27 | 26 | -1 | 8.47764300E+00 | 8.48031786E+00 | -2.70E-03 | 4.93126915E+04 | 4.93125301E+04 | 1.60E-01 | 4.93265266E+04 | 4.93263652E+04 | 1.60E-01 | -4.26586270E+01 | -4.28200187E+01 | 1.60E-01 |
| 515 | Co | 54 | 27 | 27 | 0 | 8.56920500E+00 | 8.55681797E+00 | 1.20E-02 | 5.02388349E+04 | 5.02394842E+04 | -6.50E-01 | 5.02526700E+04 | 5.02533193E+04 | -6.50E-01 | -4.80093200E+01 | -4.73600232E+01 | -6.50E-01 |
| 516 | Co | 55 | 27 | 28 | 1 | 8.66960600E+00 | 8.65254279E+00 | 1.70E-02 | 5.11643090E+04 | 5.11652279E+04 | -9.20E-01 | 5.11781441E+04 | 5.11790630E+04 | -9.20E-01 | -5.40292580E+01 | -5.31103875E+01 | -9.20E-01 |
| 517 | Co | 56 | 27 | 29 | 2 | 8.69482500E+00 | 8.68332850E+00 | 1.10E-02 | 5.20937925E+04 | 5.20944167E+04 | -6.20E-01 | 5.21076276E+04 | 5.21082518E+04 | -6.20E-01 | -5.60397980E+01 | -5.54156110E+01 | -6.20E-01 |
| 518 | Co | 57 | 27 | 30 | 3 | 8.74187100E+00 | 8.73482280E+00 | 7.00E-03 | 5.30219814E+04 | 5.30223636E+04 | -3.80E-01 | 5.30358165E+04 | 5.30361987E+04 | -3.80E-01 | -5.93449480E+01 | -5.89627955E+01 | -3.80E-01 |
| 519 | Co | 58 | 27 | 31 | 4 | 8.73895900E+00 | 8.73354739E+00 | 5.40E-03 | 5.39529739E+04 | 5.39532681E+04 | -2.90E-01 | 5.39668090E+04 | 5.39671032E+04 | -2.90E-01 | -5.98465580E+01 | -5.95523256E+01 | -2.90E-01 |
| 520 | Co | 59 | 27 | 32 | 5 | 8.76802500E+00 | 8.75740558E+00 | 1.10E-02 | 5.48820854E+04 | 5.48826923E+04 | -6.10E-01 | 5.48959205E+04 | 5.48965274E+04 | -6.10E-01 | -6.22291190E+01 | -6.16221872E+01 | -6.10E-01 |
| 521 | Co | 60 | 27 | 33 | 6 | 8.74675700E+00 | 8.73886756E+00 | 7.90E-03 | 5.58141588E+04 | 5.58146126E+04 | -4.50E-01 | 5.58279939E+04 | 5.58284477E+04 | -4.50E-01 | -6.16497200E+01 | -6.11959926E+01 | -4.50E-01 |
| 522 | Co | 61 | 27 | 34 | 7 | 8.75614100E+00 | 8.74659595E+00 | 9.50E-03 | 5.67444050E+04 | 5.67449676E+04 | -5.60E-01 | 5.67582401E+04 | 5.67588027E+04 | -5.60E-01 | -6.28976230E+01 | -6.23349731E+01 | -5.60E-01 |
| 523 | Co | 62 | 27 | 35 | 8 | 8.72132500E+00 | 8.71750013E+00 | 3.80E-03 | 5.76773728E+04 | 5.76775904E+04 | -2.20E-01 | 5.76912080E+04 | 5.76914255E+04 | -2.20E-01 | -6.14238240E+01 | -6.12063091E+01 | -2.20E-01 |
| 524 | Co | 63 | 27 | 36 | 9 | 8.71778800E+00 | 8.71286693E+00 | 4.90E-03 | 5.86084397E+04 | 5.86087301E+04 | -2.90E-01 | 5.86222748E+04 | 5.86225652E+04 | -2.90E-01 | -6.18509990E+01 | -6.15605986E+01 | -2.90E-01 |
| 525 | Co | 64 | 27 | 37 | 10 | 8.67551300E+00 | 8.67451308E+00 | 1.00E-03 | 5.95419929E+04 | 5.95420373E+04 | -4.40E-02 | 5.95558280E+04 | 5.95558724E+04 | -4.40E-02 | -5.97918880E+01 | -5.97474999E+01 | -4.40E-02 |
| 526 | Co | 65 | 27 | 38 | 11 | 8.65688400E+00 | 8.65825177E+00 | -1.40E-03 | 6.04740937E+04 | 6.04739852E+04 | 1.10E-01 | 6.04879288E+04 | 6.04878203E+04 | 1.10E-01 | -5.91851980E+01 | -5.92937089E+01 | 1.10E-01 |
| 527 | Co | 66 | 27 | 39 | 12 | 8.60594100E+00 | 8.61126487E+00 | -5.30E-03 | 6.14083644E+04 | 6.14079934E+04 | 3.70E-01 | 6.14221995E+04 | 6.14218285E+04 | 3.70E-01 | -5.64085330E+01 | -5.67795064E+01 | 3.70E-01 |
| 528 | Co | 67 | 27 | 40 | 13 | 8.58174100E+00 | 8.58497666E+00 | -3.20E-03 | 6.23409452E+04 | 6.23407088E+04 | 2.40E-01 | 6.23547803E+04 | 6.23545439E+04 | 2.40E-01 | -5.53217750E+01 | -5.55581423E+01 | 2.40E-01 |
| 529 | Co | 68 | 27 | 41 | 14 | 8.52426400E+00 | 8.53056119E+00 | -6.30E-03 | 6.32758373E+04 | 6.32753895E+04 | 4.50E-01 | 6.32896724E+04 | 6.32892246E+04 | 4.50E-01 | -5.19237210E+01 | -5.23715480E+01 | 4.50E-01 |
| 530 | Co | 69 | 27 | 42 | 15 | 8.49227000E+00 | 8.49592403E+00 | -3.70E-03 | 6.42090860E+04 | 6.42088143E+04 | 2.70E-01 | 6.42229211E+04 | 6.42226494E+04 | 2.70E-01 | -5.01690850E+01 | -5.04408260E+01 | 2.70E-01 |
| 531 | Co | 70 | 27 | 43 | 16 | 8.43983100E+00 | 8.43667522E+00 | 3.20E-03 | 6.51438298E+04 | 6.51440312E+04 | -2.00E-01 | 6.51576649E+04 | 6.51578663E+04 | -2.00E-01 | -4.69193530E+01 | -4.67180147E+01 | -2.00E-01 |
| 532 | Co | 71 | 27 | 44 | 17 | 8.39873400E+00 | 8.39738918E+00 | 1.30E-03 | 6.60778733E+04 | 6.60779492E+04 | -7.60E-02 | 6.60917084E+04 | 6.60917843E+04 | -7.60E-02 | -4.43699250E+01 | -4.42940620E+01 | -7.60E-02 |
| 533 | CO | 72 | 27 | 45 | 18 | 8.33000000E+00 | 8.33697765E+00 | -7.00E-03 | 6.70139532E+04 | 6.70134668E+04 | 4.90E-01 | 6.70277883E+04 | 6.70273019E+04 | 4.90E-01 | -3.97840000E+01 | -4.02705016E+01 | 4.90E-01 |
| 534 | Co | 73 | 27 | 46 | 19 | 8.28700000E+00 | 8.29628010E+00 | -9.30E-03 | 6.79483349E+04 | 6.79476661E+04 | 6.70E-01 | 6.79621700E+04 | 6.79615012E+04 | 6.70E-01 | -3.68960000E+01 | -3.75652395E+01 | 6.70E-01 |
| 535 | Co | 74 | 27 | 47 | 20 | 8.22500000E+00 | 8.23562529E+00 | -1.10E-02 | 6.88842628E+04 | 6.88834237E+04 | 8.40E-01 | 6.88980979E+04 | 6.88972588E+04 | 8.40E-01 | -3.24630000E+01 | -3.33017443E+01 | 8.40E-01 |
| 536 | Co | 75 | 27 | 48 | 21 | 8.17800000E+00 | 8.19073261E+00 | -1.30E-02 | 6.98191196E+04 | 6.98181204E+04 | 1.00E+00 | 6.98329547E+04 | 6.98319555E+04 | 1.00E+00 | -2.91000000E+01 | -3.00990999E+01 | 1.00E+00 |
| 537 | CO | 76 | 27 | 49 | 22 | 8.11000000E+00 | 8.12221710E+00 | -1.20E-02 | 7.07556158E+04 | 7.07547022E+04 | 9.10E-01 | 7.07694509E+04 | 7.07685373E+04 | 9.10E-01 | -2.40980000E+01 | -2.50113342E+01 | 9.10E-01 |
| 538 | Ni | 48 | 28 | 20 | -8 | 7.27200000E+00 | 7.26837614E+00 | 3.60E-03 | 4.47138416E+04 | 4.47139805E+04 | -1.40E-01 | 4.47281912E+04 | 4.47283300E+04 | -1.40E-01 | 1.64770000E+01 | 1.66151011E+01 | -1.40E-01 |
| 539 | Ni | 49 | 28 | 21 | -7 | 7.47800000E+00 | 7.46282203E+00 | 1.50E-02 | 4.56360319E+04 | 4.56367496E+04 | -7.20E-01 | 4.56503815E+04 | 4.56510992E+04 | -7.20E-01 | 7.17300000E+00 | 7.89019559E+00 | -7.20E-01 |
| 540 | Ni | 50 | 28 | 22 | -6 | 7.73100000E+00 | 7.73010106E+00 | 9.00E-04 | 4.65554538E+04 | 4.65554882E+04 | -3.40E-02 | 4.65698034E+04 | 4.65698378E+04 | -3.40E-02 | -4.90000000E+00 | -4.86525898E+00 | -3.50E-02 |
| 541 | Ni | 51 | 28 | 23 | -5 | 7.89500000E+00 | 7.87675444E+00 | 1.80E-02 | 4.74789091E+04 | 4.74798442E+04 | -9.40E-01 | 4.74932587E+04 | 4.74941937E+04 | -9.40E-01 | -1.29380000E+01 | -1.20033632E+01 | -9.30E-01 |
| 542 | Ni | 52 | 28 | 24 | -4 | 8.10100000E+00 | 8.09545167E+00 | 5.50E-03 | 4.83998679E+04 | 4.84001605E+04 | -2.90E-01 | 4.84142175E+04 | 4.84145101E+04 | -2.90E-01 | -2.34740000E+01 | -2.31810548E+01 | -2.90E-01 |
| 543 | Ni | 53 | 28 | 25 | -3 | 8.21707400E+00 | 8.23114218E+00 | -1.40E-02 | 4.93252048E+04 | 4.93244389E+04 | 7.70E-01 | 4.93395544E+04 | 4.93387884E+04 | 7.70E-01 | -2.96308240E+01 | -3.03967847E+01 | 7.70E-01 |
| 544 | Ni | 54 | 28 | 26 | -2 | 8.39200600E+00 | 8.39983171E+00 | -7.80E-03 | 5.02471068E+04 | 5.02466639E+04 | 4.40E-01 | 5.02614564E+04 | 5.02610135E+04 | 4.40E-01 | -3.92229170E+01 | -3.96658422E+01 | 4.40E-01 |
| 545 | Ni | 55 | 28 | 27 | -1 | 8.49730800E+00 | 8.49128365E+00 | 6.00E-03 | 5.11724886E+04 | 5.11727996E+04 | -3.10E-01 | 5.11868381E+04 | 5.11871491E+04 | -3.10E-01 | -4.53352240E+01 | -4.50242118E+01 | -3.10E-01 |
| 546 | Ni | 56 | 28 | 28 | 0 | 8.64276700E+00 | 8.61484705E+00 | 2.80E-02 | 5.20954109E+04 | 5.20969541E+04 | -1.50E+00 | 5.21097605E+04 | 5.21113037E+04 | -1.50E+00 | -5.39069090E+01 | -5.23637270E+01 | -1.50E+00 |
| 547 | Ni | 57 | 28 | 29 | 1 | 8.67092300E+00 | 8.64808990E+00 | 2.30E-02 | 5.30247287E+04 | 5.30260098E+04 | -1.30E+00 | 5.30390782E+04 | 5.30403594E+04 | -1.30E+00 | -5.60832190E+01 | -5.48020975E+01 | -1.30E+00 |
| 548 | Ni | 58 | 28 | 30 | 2 | 8.73204900E+00 | 8.72399875E+00 | 8.10E-03 | 5.39520778E+04 | 5.39525244E+04 | -4.50E-01 | 5.39664274E+04 | 5.39668740E+04 | -4.50E-01 | -6.02281530E+01 | -5.97815815E+01 | -4.50E-01 |
| 549 | Ni | 59 | 28 | 31 | 3 | 8.73657800E+00 | 8.72260867E+00 | 1.40E-02 | 5.48826439E+04 | 5.48834478E+04 | -8.00E-01 | 5.48969935E+04 | 5.48977974E+04 | -8.00E-01 | -6.11561180E+01 | -6.03522467E+01 | -8.00E-01 |
| 550 | Ni | 60 | 28 | 32 | 4 | 8.78076400E+00 | 8.76820988E+00 | 1.30E-02 | 5.58108216E+04 | 5.58115545E+04 | -7.30E-01 | 5.58251711E+04 | 5.58259041E+04 | -7.30E-01 | -6.44725330E+01 | -6.37396090E+01 | -7.30E-01 |
| 551 | Ni | 61 | 28 | 33 | 5 | 8.76501600E+00 | 8.74923212E+00 | 1.60E-02 | 5.67425668E+04 | 5.67435093E+04 | -9.40E-01 | 5.67569164E+04 | 5.67578589E+04 | -9.40E-01 | -6.42213220E+01 | -6.32788561E+01 | -9.40E-01 |
| 552 | Ni | 62 | 28 | 34 | 6 | 8.79454600E+00 | 8.77703639E+00 | 1.80E-02 | 5.76715364E+04 | 5.76726016E+04 | -1.10E+00 | 5.76858859E+04 | 5.76869511E+04 | -1.10E+00 | -6.67458630E+01 | -6.56806341E+01 | -1.10E+00 |
| 553 | Ni | 63 | 28 | 35 | 7 | 8.76348600E+00 | 8.74754379E+00 | 1.60E-02 | 5.86042640E+04 | 5.86052480E+04 | -9.80E-01 | 5.86186135E+04 | 5.86195975E+04 | -9.80E-01 | -6.55123210E+01 | -6.45283178E+01 | -9.80E-01 |
| 554 | Ni | 64 | 28 | 36 | 8 | 8.77745400E+00 | 8.76184370E+00 | 1.60E-02 | 5.95341718E+04 | 5.95351506E+04 | -9.80E-01 | 5.95485214E+04 | 5.95495002E+04 | -9.80E-01 | -6.70984780E+01 | -6.61197371E+01 | -9.80E-01 |
| 555 | Ni | 65 | 28 | 37 | 9 | 8.73623300E+00 | 8.72333061E+00 | 1.30E-02 | 6.04676391E+04 | 6.04684575E+04 | -8.20E-01 | 6.04819887E+04 | 6.04828071E+04 | -8.20E-01 | -6.51252420E+01 | -6.43069108E+01 | -8.20E-01 |
| 556 | Ni | 66 | 28 | 38 | 10 | 8.73950800E+00 | 8.72527081E+00 | 1.40E-02 | 6.13982522E+04 | 6.13991715E+04 | -9.20E-01 | 6.14126017E+04 | 6.14135211E+04 | -9.20E-01 | -6.60062850E+01 | -6.50869756E+01 | -9.20E-01 |
| 557 | Ni | 67 | 28 | 39 | 11 | 8.69575000E+00 | 8.67842871E+00 | 1.70E-02 | 6.23320098E+04 | 6.23331500E+04 | -1.10E+00 | 6.23463594E+04 | 6.23474996E+04 | -1.10E+00 | -6.37426800E+01 | -6.26025067E+01 | -1.10E+00 |
| 558 | Ni | 68 | 28 | 40 | 12 | 8.68246600E+00 | 8.66983777E+00 | 1.30E-02 | 6.32637828E+04 | 6.32646211E+04 | -8.40E-01 | 6.32781323E+04 | 6.32789707E+04 | -8.40E-01 | -6.34638140E+01 | -6.26254322E+01 | -8.40E-01 |
| 559 | Ni | 69 | 28 | 41 | 13 | 8.62309900E+00 | 8.61574620E+00 | 7.40E-03 | 6.41987620E+04 | 6.41992490E+04 | -4.90E-01 | 6.42131116E+04 | 6.42135986E+04 | -4.90E-01 | -5.99786480E+01 | -5.94916327E+01 | -4.90E-01 |
| 560 | Ni | 70 | 28 | 42 | 14 | 8.60429100E+00 | 8.59829981E+00 | 6.00E-03 | 6.51310208E+04 | 6.51314199E+04 | -4.00E-01 | 6.51453704E+04 | 6.51457695E+04 | -4.00E-01 | -5.92138600E+01 | -5.88148124E+01 | -4.00E-01 |
| 561 | Ni | 71 | 28 | 43 | 15 | 8.54315600E+00 | 8.53929486E+00 | 3.90E-03 | 6.60663225E+04 | 6.60665763E+04 | -2.50E-01 | 6.60806721E+04 | 6.60809259E+04 | -2.50E-01 | -5.54062280E+01 | -5.51524418E+01 | -2.50E-01 |
| 562 | Ni | 72 | 28 | 44 | 16 | 8.52021100E+00 | 8.51659126E+00 | 3.60E-03 | 6.69989968E+04 | 6.69992371E+04 | -2.40E-01 | 6.70133463E+04 | 6.70135866E+04 | -2.40E-01 | -5.42260610E+01 | -5.39857588E+01 | -2.40E-01 |
| 563 | Ni | 73 | 28 | 45 | 17 | 8.45765200E+00 | 8.45611427E+00 | 1.50E-03 | 6.79346087E+04 | 6.79347007E+04 | -9.20E-02 | 6.79489583E+04 | 6.79490502E+04 | -9.20E-02 | -5.01081520E+01 | -5.00162111E+01 | -9.20E-02 |
| 564 | Ni | 74 | 28 | 46 | 18 | 8.43000000E+00 | 8.43141289E+00 | -1.40E-03 | 6.88677546E+04 | 6.88676378E+04 | 1.20E-01 | 6.88821042E+04 | 6.88819874E+04 | 1.20E-01 | -4.84560000E+01 | -4.85731039E+01 | 1.20E-01 |
| 565 | Ni | 75 | 28 | 47 | 19 | 8.36900000E+00 | 8.37047195E+00 | -1.50E-03 | 6.98034590E+04 | 6.98033424E+04 | 1.20E-01 | 6.98178086E+04 | 6.98176919E+04 | 1.20E-01 | -4.42460000E+01 | -4.43626273E+01 | 1.20E-01 |
| 566 | Ni | 76 | 28 | 48 | 20 | 8.33100000E+00 | 8.34124743E+00 | -1.00E-02 | 7.07375892E+04 | 7.07367583E+04 | 8.30E-01 | 7.07519388E+04 | 7.07511079E+04 | 8.30E-01 | -4.16100000E+01 | -4.24407170E+01 | 8.30E-01 |
| 567 | Ni | 77 | 28 | 49 | 21 | 8.26400000E+00 | 8.27217951E+00 | -8.20E-03 | 7.16739457E+04 | 7.16733007E+04 | 6.40E-01 | 7.16882953E+04 | 7.16876503E+04 | 6.40E-01 | -3.67470000E+01 | -3.73924156E+01 | 6.50E-01 |
| 568 | Ni | 78 | 28 | 50 | 22 | 8.22800000E+00 | 8.22776065E+00 | 2.40E-04 | 7.26080572E+04 | 7.26080586E+04 | -1.30E-03 | 7.26224068E+04 | 7.26224082E+04 | -1.30E-03 | -3.41300000E+01 | -3.41286051E+01 | -1.40E-03 |
| 569 | NI | 79 | 28 | 51 | 23 | 8.14500000E+00 | 8.14160326E+00 | 3.40E-03 | 7.35459712E+04 | 7.35462026E+04 | -2.30E-01 | 7.35603207E+04 | 7.35605522E+04 | -2.30E-01 | -2.77100000E+01 | -2.74786125E+01 | -2.30E-01 |
| 570 | Cu | 53 | 29 | 24 | -5 | 7.91400000E+00 | 7.90969680E+00 | 4.30E-03 | 4.93399667E+04 | 4.93401778E+04 | -2.10E-01 | 4.93548309E+04 | 4.93550420E+04 | -2.10E-01 | -1.43540000E+01 | -1.41432565E+01 | -2.10E-01 |
| 571 | Cu | 54 | 29 | 25 | -4 | 8.05400000E+00 | 8.06743383E+00 | -1.30E-02 | 5.02640740E+04 | 5.02633156E+04 | 7.60E-01 | 5.02789382E+04 | 5.02781799E+04 | 7.60E-01 | -2.17410000E+01 | -2.24994340E+01 | 7.60E-01 |
| 572 | Cu | 55 | 29 | 26 | -3 | 8.23399600E+00 | 8.25295908E+00 | -1.90E-02 | 5.11856737E+04 | 5.11846097E+04 | 1.10E+00 | 5.12005379E+04 | 5.11994739E+04 | 1.10E+00 | -3.16353990E+01 | -3.26994376E+01 | 1.10E+00 |
| 573 | Cu | 56 | 29 | 27 | -2 | 8.34900000E+00 | 8.36662710E+00 | -1.80E-02 | 5.21105654E+04 | 5.21095567E+04 | 1.00E+00 | 5.21254296E+04 | 5.21244209E+04 | 1.00E+00 | -3.82380000E+01 | -3.92464863E+01 | 1.00E+00 |
| 574 | Cu | 57 | 29 | 28 | -1 | 8.50325100E+00 | 8.49581414E+00 | 7.40E-03 | 5.30329890E+04 | 5.30333918E+04 | -4.00E-01 | 5.30478532E+04 | 5.30482560E+04 | -4.00E-01 | -4.73082530E+01 | -4.69054561E+01 | -4.00E-01 |
| 575 | Cu | 58 | 29 | 29 | 0 | 8.57095700E+00 | 8.55643252E+00 | 1.50E-02 | 5.39601242E+04 | 5.39609455E+04 | -8.20E-01 | 5.39749884E+04 | 5.39758097E+04 | -8.20E-01 | -5.16671340E+01 | -5.08458174E+01 | -8.20E-01 |
| 576 | Cu | 59 | 29 | 30 | 1 | 8.64199000E+00 | 8.63522192E+00 | 6.80E-03 | 5.48869276E+04 | 5.48873059E+04 | -3.80E-01 | 5.49017919E+04 | 5.49021701E+04 | -3.80E-01 | -5.63577320E+01 | -5.59795055E+01 | -3.80E-01 |
| 577 | Cu | 60 | 29 | 31 | 2 | 8.66559200E+00 | 8.65775385E+00 | 7.80E-03 | 5.58164349E+04 | 5.58168841E+04 | -4.50E-01 | 5.58312991E+04 | 5.58317483E+04 | -4.50E-01 | -5.83445510E+01 | -5.78953237E+01 | -4.50E-01 |
| 578 | Cu | 61 | 29 | 32 | 3 | 8.71551000E+00 | 8.70471047E+00 | 1.10E-02 | 5.67442897E+04 | 5.67449274E+04 | -6.40E-01 | 5.67591539E+04 | 5.67597916E+04 | -6.40E-01 | -6.19838340E+01 | -6.13461125E+01 | -6.40E-01 |
| 579 | Cu | 62 | 29 | 33 | 4 | 8.71807400E+00 | 8.70699851E+00 | 1.10E-02 | 5.76749806E+04 | 5.76756462E+04 | -6.70E-01 | 5.76898448E+04 | 5.76905104E+04 | -6.70E-01 | -6.27869660E+01 | -6.21213625E+01 | -6.70E-01 |
| 580 | Cu | 63 | 29 | 34 | 5 | 8.75213100E+00 | 8.73554686E+00 | 1.70E-02 | 5.86036823E+04 | 5.86047060E+04 | -1.00E+00 | 5.86185465E+04 | 5.86195703E+04 | -1.00E+00 | -6.55792980E+01 | -6.45555883E+01 | -1.00E+00 |
| 581 | Cu | 64 | 29 | 35 | 6 | 8.73906800E+00 | 8.72556022E+00 | 1.40E-02 | 5.95353316E+04 | 5.95361750E+04 | -8.40E-01 | 5.95501958E+04 | 5.95510392E+04 | -8.40E-01 | -6.54240940E+01 | -6.45806712E+01 | -8.40E-01 |
| 582 | Cu | 65 | 29 | 36 | 7 | 8.75709300E+00 | 8.74037480E+00 | 1.70E-02 | 6.04649863E+04 | 6.04660519E+04 | -1.10E+00 | 6.04798505E+04 | 6.04809161E+04 | -1.10E+00 | -6.72634570E+01 | -6.61978601E+01 | -1.10E+00 |
| 583 | Cu | 66 | 29 | 37 | 8 | 8.73146900E+00 | 8.72014761E+00 | 1.10E-02 | 6.13974857E+04 | 6.13982119E+04 | -7.30E-01 | 6.14123500E+04 | 6.14130761E+04 | -7.30E-01 | -6.62580700E+01 | -6.55319214E+01 | -7.30E-01 |
| 584 | Cu | 67 | 29 | 38 | 9 | 8.73744700E+00 | 8.72256442E+00 | 1.50E-02 | 6.23279191E+04 | 6.23288952E+04 | -9.80E-01 | 6.23427833E+04 | 6.23437594E+04 | -9.80E-01 | -6.73187790E+01 | -6.63426765E+01 | -9.80E-01 |
| 585 | Cu | 68 | 29 | 39 | 10 | 8.70189000E+00 | 8.69316324E+00 | 8.70E-03 | 6.32611649E+04 | 6.32617373E+04 | -5.70E-01 | 6.32760291E+04 | 6.32766015E+04 | -5.70E-01 | -6.55670340E+01 | -6.49946414E+01 | -5.70E-01 |
| 586 | Cu | 69 | 29 | 40 | 11 | 8.69520400E+00 | 8.68506053E+00 | 1.00E-02 | 6.41924898E+04 | 6.41931686E+04 | -6.80E-01 | 6.42073540E+04 | 6.42080328E+04 | -6.80E-01 | -6.57362120E+01 | -6.50573985E+01 | -6.80E-01 |
| 587 | Cu | 70 | 29 | 41 | 12 | 8.64686500E+00 | 8.64774310E+00 | -8.80E-04 | 6.51267437E+04 | 6.51266611E+04 | 8.30E-02 | 6.51416079E+04 | 6.51415254E+04 | 8.30E-02 | -6.29763720E+01 | -6.30589200E+01 | 8.30E-02 |
| 588 | Cu | 71 | 29 | 42 | 13 | 8.63502200E+00 | 8.63067581E+00 | 4.30E-03 | 6.60585030E+04 | 6.60587905E+04 | -2.90E-01 | 6.60733672E+04 | 6.60736548E+04 | -2.90E-01 | -6.27111260E+01 | -6.24235663E+01 | -2.90E-01 |
| 589 | Cu | 72 | 29 | 43 | 14 | 8.58652500E+00 | 8.58782080E+00 | -1.30E-03 | 6.69929252E+04 | 6.69928108E+04 | 1.10E-01 | 6.70077894E+04 | 6.70076750E+04 | 1.10E-01 | -5.97829980E+01 | -5.98973629E+01 | 1.10E-01 |
| 590 | Cu | 73 | 29 | 44 | 15 | 8.56856900E+00 | 8.56525438E+00 | 3.30E-03 | 6.79252148E+04 | 6.79254357E+04 | -2.20E-01 | 6.79400790E+04 | 6.79402999E+04 | -2.20E-01 | -5.89874370E+01 | -5.87665157E+01 | -2.20E-01 |
| 591 | Cu | 74 | 29 | 45 | 16 | 8.52156200E+00 | 8.52033417E+00 | 1.20E-03 | 6.88596901E+04 | 6.88597599E+04 | -7.00E-02 | 6.88745543E+04 | 6.88746242E+04 | -7.00E-02 | -5.60062050E+01 | -5.59363557E+01 | -7.00E-02 |
| 592 | Cu | 75 | 29 | 46 | 17 | 8.49509400E+00 | 8.49552174E+00 | -4.30E-04 | 6.97927190E+04 | 6.97926659E+04 | 5.30E-02 | 6.98075832E+04 | 6.98075301E+04 | 5.30E-02 | -5.44713410E+01 | -5.45244388E+01 | 5.30E-02 |
| 593 | Cu | 76 | 29 | 47 | 18 | 8.44352700E+00 | 8.44977701E+00 | -6.30E-03 | 7.07277084E+04 | 7.07272124E+04 | 5.00E-01 | 7.07425727E+04 | 7.07420766E+04 | 5.00E-01 | -5.09759850E+01 | -5.14720418E+01 | 5.00E-01 |
| 594 | Cu | 77 | 29 | 48 | 19 | 8.40700000E+00 | 8.42047540E+00 | -1.30E-02 | 7.16616663E+04 | 7.16605842E+04 | 1.10E+00 | 7.16765305E+04 | 7.16754484E+04 | 1.10E+00 | -4.85120000E+01 | -4.95942756E+01 | 1.10E+00 |
| 595 | Cu | 78 | 29 | 49 | 20 | 8.35092500E+00 | 8.36648388E+00 | -1.60E-02 | 7.25971751E+04 | 7.25959404E+04 | 1.20E+00 | 7.26120393E+04 | 7.26108047E+04 | 1.20E+00 | -4.44974680E+01 | -4.57320937E+01 | 1.20E+00 |
| 596 | Cu | 79 | 29 | 50 | 21 | 8.31400000E+00 | 8.32180171E+00 | -7.80E-03 | 7.35312680E+04 | 7.35306692E+04 | 6.00E-01 | 7.35461322E+04 | 7.35455335E+04 | 6.00E-01 | -4.18990000E+01 | -4.24973671E+01 | 6.00E-01 |
| 597 | Cu | 80 | 29 | 51 | 22 | 8.24300000E+00 | 8.24985569E+00 | -6.90E-03 | 7.44682299E+04 | 7.44676685E+04 | 5.60E-01 | 7.44830942E+04 | 7.44825327E+04 | 5.60E-01 | -3.64310000E+01 | -3.69921684E+01 | 5.60E-01 |
| 598 | CU | 81 | 29 | 52 | 23 | 8.18400000E+00 | 8.18767661E+00 | -3.70E-03 | 7.54043647E+04 | 7.54040205E+04 | 3.40E-01 | 7.54192289E+04 | 7.54188847E+04 | 3.40E-01 | -3.17900000E+01 | -3.21341992E+01 | 3.40E-01 |
| 599 | CU | 82 | 29 | 53 | 24 | 8.10800000E+00 | 8.10479901E+00 | 3.20E-03 | 7.63419787E+04 | 7.63421942E+04 | -2.20E-01 | 7.63568429E+04 | 7.63570584E+04 | -2.20E-01 | -2.56700000E+01 | -2.54545937E+01 | -2.20E-01 |
| 600 | Zn | 54 | 30 | 24 | -6 | 7.77400000E+00 | 7.73407190E+00 | 4.00E-02 | 5.02778846E+04 | 5.02800193E+04 | -2.10E+00 | 5.02932637E+04 | 5.02953983E+04 | -2.10E+00 | -7.41600000E+00 | -5.28096660E+00 | -2.10E+00 |
| 601 | Zn | 55 | 30 | 25 | -5 | 7.91600000E+00 | 7.90413841E+00 | 1.20E-02 | 5.12018718E+04 | 5.12024969E+04 | -6.30E-01 | 5.12172508E+04 | 5.12178760E+04 | -6.30E-01 | -1.49230000E+01 | -1.42973774E+01 | -6.30E-01 |
| 602 | Zn | 56 | 30 | 26 | -4 | 8.10900000E+00 | 8.11864266E+00 | -9.60E-03 | 5.21227095E+04 | 5.21221459E+04 | 5.60E-01 | 5.21380886E+04 | 5.21375250E+04 | 5.60E-01 | -2.55790000E+01 | -2.61424347E+01 | 5.60E-01 |
| 603 | Zn | 57 | 30 | 27 | -3 | 8.23100000E+00 | 8.25160441E+00 | -2.10E-02 | 5.30472360E+04 | 5.30460138E+04 | 1.20E+00 | 5.30626151E+04 | 5.30613929E+04 | 1.20E+00 | -3.25460000E+01 | -3.37685784E+01 | 1.20E+00 |
| 604 | Zn | 58 | 30 | 28 | -2 | 8.39593400E+00 | 8.40592469E+00 | -1.00E-02 | 5.39689783E+04 | 5.39683770E+04 | 6.00E-01 | 5.39843574E+04 | 5.39837561E+04 | 6.00E-01 | -4.22981530E+01 | -4.28994399E+01 | 6.00E-01 |
| 605 | Zn | 59 | 30 | 29 | -1 | 8.47376700E+00 | 8.48173401E+00 | -8.00E-03 | 5.48955556E+04 | 5.48950638E+04 | 4.90E-01 | 5.49109346E+04 | 5.49104428E+04 | 4.90E-01 | -4.72149550E+01 | -4.77067957E+01 | 4.90E-01 |
| 606 | Zn | 60 | 30 | 30 | 0 | 8.58303900E+00 | 8.58723384E+00 | -4.20E-03 | 5.58200909E+04 | 5.58198174E+04 | 2.70E-01 | 5.58354699E+04 | 5.58351965E+04 | 2.70E-01 | -5.41737320E+01 | -5.44472002E+01 | 2.70E-01 |
| 607 | Zn | 61 | 30 | 31 | 1 | 8.61030600E+00 | 8.61220288E+00 | -1.90E-03 | 5.67494100E+04 | 5.67492724E+04 | 1.40E-01 | 5.67647890E+04 | 5.67646515E+04 | 1.40E-01 | -5.63486910E+01 | -5.64862266E+01 | 1.40E-01 |
| 608 | Zn | 62 | 30 | 32 | 2 | 8.67933500E+00 | 8.68204127E+00 | -2.70E-03 | 5.76760852E+04 | 5.76758956E+04 | 1.90E-01 | 5.76914643E+04 | 5.76912747E+04 | 1.90E-01 | -6.11675120E+01 | -6.13570906E+01 | 1.90E-01 |
| 609 | Zn | 63 | 30 | 33 | 3 | 8.68628100E+00 | 8.68419183E+00 | 2.10E-03 | 5.86065337E+04 | 5.86066435E+04 | -1.10E-01 | 5.86219127E+04 | 5.86220225E+04 | -1.10E-01 | -6.22131110E+01 | -6.21032984E+01 | -1.10E-01 |
| 610 | Zn | 64 | 30 | 34 | 4 | 8.73590100E+00 | 8.73288357E+00 | 3.00E-03 | 5.95342371E+04 | 5.95344084E+04 | -1.70E-01 | 5.95496161E+04 | 5.95497875E+04 | -1.70E-01 | -6.60037890E+01 | -6.58324424E+01 | -1.70E-01 |
| 611 | Zn | 65 | 30 | 35 | 5 | 8.72426200E+00 | 8.72214082E+00 | 2.10E-03 | 6.04658231E+04 | 6.04659392E+04 | -1.20E-01 | 6.04812022E+04 | 6.04813182E+04 | -1.20E-01 | -6.59117980E+01 | -6.57957285E+01 | -1.20E-01 |
| 612 | Zn | 66 | 30 | 36 | 6 | 8.75963000E+00 | 8.75528452E+00 | 4.30E-03 | 6.13943299E+04 | 6.13945949E+04 | -2.70E-01 | 6.14097090E+04 | 6.14099740E+04 | -2.70E-01 | -6.88990620E+01 | -6.86340344E+01 | -2.70E-01 |
| 613 | Zn | 67 | 30 | 37 | 7 | 8.73414800E+00 | 8.73413789E+00 | 1.00E-05 | 6.23268430E+04 | 6.23268219E+04 | 2.10E-02 | 6.23422220E+04 | 6.23422009E+04 | 2.10E-02 | -6.78800670E+01 | -6.79011754E+01 | 2.10E-02 |
| 614 | Zn | 68 | 30 | 38 | 8 | 8.75567700E+00 | 8.75361149E+00 | 2.10E-03 | 6.32562102E+04 | 6.32563289E+04 | -1.20E-01 | 6.32715893E+04 | 6.32717079E+04 | -1.20E-01 | -7.00068460E+01 | -6.98881993E+01 | -1.20E-01 |
| 615 | Zn | 69 | 30 | 39 | 9 | 8.72272600E+00 | 8.72333061E+00 | -6.00E-04 | 6.41892936E+04 | 6.41892300E+04 | 6.40E-02 | 6.42046726E+04 | 6.42046091E+04 | 6.40E-02 | -6.84175990E+01 | -6.84811110E+01 | 6.40E-02 |
| 616 | Zn | 70 | 30 | 40 | 10 | 8.72980700E+00 | 8.73134343E+00 | -1.50E-03 | 6.51196405E+04 | 6.51195112E+04 | 1.30E-01 | 6.51350196E+04 | 6.51348902E+04 | 1.30E-01 | -6.95647090E+01 | -6.96940204E+01 | 1.30E-01 |
| 617 | Zn | 71 | 30 | 41 | 11 | 8.68904100E+00 | 8.69318944E+00 | -4.10E-03 | 6.60533705E+04 | 6.60530542E+04 | 3.20E-01 | 6.60687495E+04 | 6.60684332E+04 | 3.20E-01 | -6.73287750E+01 | -6.76451110E+01 | 3.20E-01 |
| 618 | Zn | 72 | 30 | 42 | 12 | 8.69180400E+00 | 8.69139146E+00 | 4.10E-04 | 6.69840479E+04 | 6.69840558E+04 | -8.00E-03 | 6.69994269E+04 | 6.69994349E+04 | -8.00E-03 | -6.81454860E+01 | -6.81375274E+01 | -8.00E-03 |
| 619 | Zn | 73 | 30 | 43 | 13 | 8.64834500E+00 | 8.64761087E+00 | 7.30E-04 | 6.79180940E+04 | 6.79181258E+04 | -3.20E-02 | 6.79334730E+04 | 6.79335048E+04 | -3.20E-02 | -6.55934020E+01 | -6.55616165E+01 | -3.20E-02 |
| 620 | Zn | 74 | 30 | 44 | 14 | 8.64275400E+00 | 8.63948778E+00 | 3.30E-03 | 6.88494248E+04 | 6.88496447E+04 | -2.20E-01 | 6.88648038E+04 | 6.88650237E+04 | -2.20E-01 | -6.57567120E+01 | -6.55367998E+01 | -2.20E-01 |
| 621 | Zn | 75 | 30 | 45 | 15 | 8.59249700E+00 | 8.59342692E+00 | -9.30E-04 | 6.97841166E+04 | 6.97840251E+04 | 9.10E-02 | 6.97994957E+04 | 6.97994042E+04 | 9.10E-02 | -6.25589080E+01 | -6.26504040E+01 | 9.10E-02 |
| 622 | Zn | 76 | 30 | 46 | 16 | 8.58227300E+00 | 8.58229547E+00 | -2.20E-05 | 7.07158666E+04 | 7.07158431E+04 | 2.40E-02 | 7.07312456E+04 | 7.07312221E+04 | 2.40E-02 | -6.23030160E+01 | -6.23265216E+01 | 2.40E-02 |
| 623 | Zn | 77 | 30 | 47 | 17 | 8.53000300E+00 | 8.53527835E+00 | -5.30E-03 | 7.16508745E+04 | 7.16504465E+04 | 4.30E-01 | 7.16662535E+04 | 7.16658255E+04 | 4.30E-01 | -5.87891950E+01 | -5.92171797E+01 | 4.30E-01 |
| 624 | Zn | 78 | 30 | 48 | 18 | 8.50737900E+00 | 8.51922904E+00 | -1.20E-02 | 7.25836745E+04 | 7.25827284E+04 | 9.50E-01 | 7.25990535E+04 | 7.25981075E+04 | 9.50E-01 | -5.74832340E+01 | -5.84292929E+01 | 9.50E-01 |
| 625 | Zn | 79 | 30 | 49 | 19 | 8.45058200E+00 | 8.46414682E+00 | -1.40E-02 | 7.35192195E+04 | 7.35181261E+04 | 1.10E+00 | 7.35345985E+04 | 7.35335051E+04 | 1.10E+00 | -5.34322940E+01 | -5.45257079E+01 | 1.10E+00 |
| 626 | Zn | 80 | 30 | 50 | 20 | 8.42354500E+00 | 8.43242584E+00 | -8.90E-03 | 7.44524972E+04 | 7.44517650E+04 | 7.30E-01 | 7.44678763E+04 | 7.44671440E+04 | 7.30E-01 | -5.16486120E+01 | -5.23808570E+01 | 7.30E-01 |
| 627 | Zn | 81 | 30 | 51 | 21 | 8.35192500E+00 | 8.35922074E+00 | -7.30E-03 | 7.53894402E+04 | 7.53888275E+04 | 6.10E-01 | 7.54048193E+04 | 7.54042066E+04 | 6.10E-01 | -4.61996630E+01 | -4.68123512E+01 | 6.10E-01 |
| 628 | Zn | 82 | 30 | 52 | 22 | 8.30500000E+00 | 8.30911188E+00 | -4.10E-03 | 7.63245274E+04 | 7.63241426E+04 | 3.80E-01 | 7.63399065E+04 | 7.63395217E+04 | 3.80E-01 | -4.26070000E+01 | -4.29913259E+01 | 3.80E-01 |
| 629 | Zn | 83 | 30 | 53 | 23 | 8.23100000E+00 | 8.22442914E+00 | 6.60E-03 | 7.72618899E+04 | 7.72624276E+04 | -5.40E-01 | 7.72772689E+04 | 7.72778066E+04 | -5.40E-01 | -3.67380000E+01 | -3.62004519E+01 | -5.40E-01 |
| 630 | ZN | 84 | 30 | 54 | 24 | 8.17800000E+00 | 8.16849168E+00 | 9.50E-03 | 7.81977117E+04 | 7.81984673E+04 | -7.60E-01 | 7.82130907E+04 | 7.82138463E+04 | -7.60E-01 | -3.24100000E+01 | -3.16548154E+01 | -7.60E-01 |
| 631 | Zn | 85 | 30 | 55 | 25 | 8.09900000E+00 | 8.08412819E+00 | 1.50E-02 | 7.91357765E+04 | 7.91370350E+04 | -1.30E+00 | 7.91511555E+04 | 7.91524141E+04 | -1.30E+00 | -2.58400000E+01 | -2.45810912E+01 | -1.30E+00 |
| 632 | Ga | 59 | 31 | 28 | -3 | 8.23600000E+00 | 8.24715122E+00 | -1.10E-02 | 5.49082839E+04 | 5.49076060E+04 | 6.80E-01 | 5.49241780E+04 | 5.49235001E+04 | 6.80E-01 | -3.39720000E+01 | -3.46494882E+01 | 6.80E-01 |
| 633 | Ga | 60 | 31 | 29 | -2 | 8.33000000E+00 | 8.34920745E+00 | -1.90E-02 | 5.58339655E+04 | 5.58328009E+04 | 1.20E+00 | 5.58498596E+04 | 5.58486950E+04 | 1.20E+00 | -3.97840000E+01 | -4.09486940E+01 | 1.20E+00 |
| 634 | Ga | 61 | 31 | 30 | -1 | 8.44642900E+00 | 8.46218753E+00 | -1.60E-02 | 5.67581091E+04 | 5.67571253E+04 | 9.80E-01 | 5.67740032E+04 | 5.67730194E+04 | 9.80E-01 | -4.71345420E+01 | -4.81183676E+01 | 9.80E-01 |
| 635 | Ga | 62 | 31 | 31 | 0 | 8.51863500E+00 | 8.51390845E+00 | 4.70E-03 | 5.76847513E+04 | 5.76850218E+04 | -2.70E-01 | 5.77006453E+04 | 5.77009158E+04 | -2.70E-01 | -5.19864460E+01 | -5.17159330E+01 | -2.70E-01 |
| 636 | Ga | 63 | 31 | 32 | 1 | 8.58392600E+00 | 8.58709106E+00 | -3.20E-03 | 5.86116847E+04 | 5.86114628E+04 | 2.20E-01 | 5.86275788E+04 | 5.86273568E+04 | 2.20E-01 | -5.65470920E+01 | -5.67690270E+01 | 2.20E-01 |
| 637 | Ga | 64 | 31 | 33 | 2 | 8.61163000E+00 | 8.61196331E+00 | -3.30E-04 | 5.95408931E+04 | 5.95408492E+04 | 4.40E-02 | 5.95567871E+04 | 5.95567433E+04 | 4.40E-02 | -5.88327510E+01 | -5.88766229E+01 | 4.40E-02 |
| 638 | Ga | 65 | 31 | 34 | 3 | 8.66215700E+00 | 8.66207772E+00 | 7.90E-05 | 6.04685626E+04 | 6.04685452E+04 | 1.70E-02 | 6.04844566E+04 | 6.04844393E+04 | 1.70E-02 | -6.26573280E+01 | -6.26747037E+01 | 1.70E-02 |
| 639 | Ga | 66 | 31 | 35 | 4 | 8.66936700E+00 | 8.67119976E+00 | -1.80E-03 | 6.13989899E+04 | 6.13988464E+04 | 1.40E-01 | 6.14148840E+04 | 6.14147405E+04 | 1.40E-01 | -6.37240620E+01 | -6.38675172E+01 | 1.40E-01 |
| 640 | Ga | 67 | 31 | 36 | 5 | 8.70752800E+00 | 8.70489210E+00 | 2.60E-03 | 6.23273291E+04 | 6.23274832E+04 | -1.50E-01 | 6.23432232E+04 | 6.23433773E+04 | -1.50E-01 | -6.68788960E+01 | -6.67247849E+01 | -1.50E-01 |
| 641 | Ga | 68 | 31 | 37 | 6 | 8.70121400E+00 | 8.70176172E+00 | -5.50E-04 | 6.32586163E+04 | 6.32585566E+04 | 6.00E-02 | 6.32745104E+04 | 6.32744507E+04 | 6.00E-02 | -6.70857460E+01 | -6.71454921E+01 | 6.00E-02 |
| 642 | Ga | 69 | 31 | 38 | 7 | 8.72457900E+00 | 8.72145202E+00 | 3.10E-03 | 6.41878683E+04 | 6.41880616E+04 | -1.90E-01 | 6.42037624E+04 | 6.42039556E+04 | -1.90E-01 | -6.93277990E+01 | -6.91345653E+01 | -1.90E-01 |
| 643 | Ga | 70 | 31 | 39 | 8 | 8.70928000E+00 | 8.70792668E+00 | 1.40E-03 | 6.51197801E+04 | 6.51198523E+04 | -7.20E-02 | 6.51356741E+04 | 6.51357463E+04 | -7.20E-02 | -6.89101310E+01 | -6.88379246E+01 | -7.20E-02 |
| 644 | Ga | 71 | 31 | 40 | 9 | 8.71760400E+00 | 8.71607164E+00 | 1.50E-03 | 6.60500452E+04 | 6.60501314E+04 | -8.60E-02 | 6.60659392E+04 | 6.60660255E+04 | -8.60E-02 | -7.01390970E+01 | -7.00528247E+01 | -8.60E-02 |
| 645 | Ga | 72 | 31 | 41 | 10 | 8.68708800E+00 | 8.69372350E+00 | -6.60E-03 | 6.69830901E+04 | 6.69825898E+04 | 5.00E-01 | 6.69989841E+04 | 6.69984839E+04 | 5.00E-01 | -6.85882560E+01 | -6.90885112E+01 | 5.00E-01 |
| 646 | Ga | 73 | 31 | 42 | 11 | 8.69387300E+00 | 8.69195851E+00 | 1.90E-03 | 6.79134731E+04 | 6.79135903E+04 | -1.20E-01 | 6.79293671E+04 | 6.79294844E+04 | -1.20E-01 | -6.96993350E+01 | -6.95820711E+01 | -1.20E-01 |
| 647 | Ga | 74 | 31 | 43 | 12 | 8.66316700E+00 | 8.66312975E+00 | 3.70E-05 | 6.88466168E+04 | 6.88465971E+04 | 2.00E-02 | 6.88625109E+04 | 6.88624911E+04 | 2.00E-02 | -6.80496170E+01 | -6.80693825E+01 | 2.00E-02 |
| 648 | Ga | 75 | 31 | 44 | 13 | 8.66080800E+00 | 8.65482147E+00 | 6.00E-03 | 6.97776959E+04 | 6.97781224E+04 | -4.30E-01 | 6.97935900E+04 | 6.97940165E+04 | -4.30E-01 | -6.84645800E+01 | -6.80380722E+01 | -4.30E-01 |
| 649 | Ga | 76 | 31 | 45 | 14 | 8.62452600E+00 | 8.62286268E+00 | 1.70E-03 | 7.07113579E+04 | 7.07114619E+04 | -1.00E-01 | 7.07272520E+04 | 7.07273559E+04 | -1.00E-01 | -6.62966390E+01 | -6.61927065E+01 | -1.00E-01 |
| 650 | Ga | 77 | 31 | 46 | 15 | 8.61339000E+00 | 8.61121617E+00 | 2.20E-03 | 7.16431563E+04 | 7.16433012E+04 | -1.40E-01 | 7.16590504E+04 | 7.16591952E+04 | -1.40E-01 | -6.59923440E+01 | -6.58474689E+01 | -1.40E-01 |
| 651 | Ga | 78 | 31 | 47 | 16 | 8.57712700E+00 | 8.57756021E+00 | -4.30E-04 | 7.25769367E+04 | 7.25768805E+04 | 5.60E-02 | 7.25928308E+04 | 7.25927746E+04 | 5.60E-02 | -6.37059500E+01 | -6.37622015E+01 | 5.60E-02 |
| 652 | Ga | 79 | 31 | 48 | 17 | 8.55606300E+00 | 8.56084455E+00 | -4.80E-03 | 7.35095891E+04 | 7.35091888E+04 | 4.00E-01 | 7.35254831E+04 | 7.35250829E+04 | 4.00E-01 | -6.25476780E+01 | -6.29479054E+01 | 4.00E-01 |
| 653 | Ga | 80 | 31 | 49 | 18 | 8.50845400E+00 | 8.51879587E+00 | -1.00E-02 | 7.44444071E+04 | 7.44435573E+04 | 8.50E-01 | 7.44603012E+04 | 7.44594513E+04 | 8.50E-01 | -5.92236670E+01 | -6.00735365E+01 | 8.50E-01 |
| 654 | Ga | 81 | 31 | 50 | 19 | 8.48335700E+00 | 8.48653859E+00 | -3.20E-03 | 7.53774969E+04 | 7.53772167E+04 | 2.80E-01 | 7.53933910E+04 | 7.53931108E+04 | 2.80E-01 | -5.76279540E+01 | -5.79081739E+01 | 2.80E-01 |
| 655 | Ga | 82 | 31 | 51 | 20 | 8.42104900E+00 | 8.42603592E+00 | -5.00E-03 | 7.63136882E+04 | 7.63132568E+04 | 4.30E-01 | 7.63295823E+04 | 7.63291508E+04 | 4.30E-01 | -5.29307180E+01 | -5.33621744E+01 | 4.30E-01 |
| 656 | Ga | 83 | 31 | 52 | 21 | 8.37257500E+00 | 8.37509273E+00 | -2.50E-03 | 7.72488559E+04 | 7.72486244E+04 | 2.30E-01 | 7.72647499E+04 | 7.72645185E+04 | 2.30E-01 | -4.92571220E+01 | -4.94886065E+01 | 2.30E-01 |
| 657 | Ga | 84 | 31 | 53 | 22 | 8.31000000E+00 | 8.30235994E+00 | 7.60E-03 | 7.81853238E+04 | 7.81859242E+04 | -6.00E-01 | 7.82012179E+04 | 7.82018183E+04 | -6.00E-01 | -4.42830000E+01 | -4.36828260E+01 | -6.00E-01 |
| 658 | Ga | 85 | 31 | 54 | 23 | 8.25700000E+00 | 8.24519672E+00 | 1.20E-02 | 7.91210376E+04 | 7.91220461E+04 | -1.00E+00 | 7.91369316E+04 | 7.91379402E+04 | -1.00E+00 | -4.00640000E+01 | -3.90549936E+01 | -1.00E+00 |
| 659 | Ga | 86 | 31 | 55 | 24 | 8.19000000E+00 | 8.17235626E+00 | 1.80E-02 | 8.00581392E+04 | 8.00596306E+04 | -1.50E+00 | 8.00740333E+04 | 8.00755247E+04 | -1.50E+00 | -3.44560000E+01 | -3.29645910E+01 | -1.50E+00 |
| 660 | GA | 87 | 31 | 56 | 25 | 8.13300000E+00 | 8.11682669E+00 | 1.60E-02 | 8.09945097E+04 | 8.09958547E+04 | -1.30E+00 | 8.10104037E+04 | 8.10117488E+04 | -1.30E+00 | -2.95800000E+01 | -2.82345557E+01 | -1.30E+00 |
| 661 | Ge | 59 | 32 | 27 | -5 | 7.92300000E+00 | 7.89589058E+00 | 2.70E-02 | 5.49254299E+04 | 5.49270321E+04 | -1.60E+00 | 5.49418391E+04 | 5.49434414E+04 | -1.60E+00 | -1.63100000E+01 | -1.47081872E+01 | -1.60E+00 |
| 662 | GE | 60 | 32 | 28 | -4 | 8.11400000E+00 | 8.10233488E+00 | 1.20E-02 | 5.58456249E+04 | 5.58463150E+04 | -6.90E-01 | 5.58620342E+04 | 5.58627242E+04 | -6.90E-01 | -2.76090000E+01 | -2.69194167E+01 | -6.90E-01 |
| 663 | Ge | 61 | 32 | 29 | -3 | 8.21400000E+00 | 8.22532907E+00 | -1.10E-02 | 5.67709991E+04 | 5.67702754E+04 | 7.20E-01 | 5.67874083E+04 | 5.67866846E+04 | 7.20E-01 | -3.37290000E+01 | -3.44530784E+01 | 7.20E-01 |
| 664 | Ge | 62 | 32 | 30 | -2 | 8.34300000E+00 | 8.36476270E+00 | -2.20E-02 | 5.76943239E+04 | 5.76929705E+04 | 1.40E+00 | 5.77107332E+04 | 5.77093798E+04 | 1.40E+00 | -4.18990000E+01 | -4.32519736E+01 | 1.40E+00 |
| 665 | Ge | 63 | 32 | 31 | -1 | 8.41871600E+00 | 8.43136213E+00 | -1.30E-02 | 5.86207954E+04 | 5.86199754E+04 | 8.20E-01 | 5.86372046E+04 | 5.86363847E+04 | 8.20E-01 | -4.69212160E+01 | -4.77411811E+01 | 8.20E-01 |
| 666 | Ge | 64 | 32 | 32 | 0 | 8.52882300E+00 | 8.53034947E+00 | -1.50E-03 | 5.95448951E+04 | 5.95447742E+04 | 1.20E-01 | 5.95613044E+04 | 5.95611835E+04 | 1.20E-01 | -5.43154960E+01 | -5.44364139E+01 | 1.20E-01 |
| 667 | GE | 65 | 32 | 33 | 1 | 8.55505800E+00 | 8.55814658E+00 | -3.10E-03 | 6.04742265E+04 | 6.04740024E+04 | 2.20E-01 | 6.04906358E+04 | 6.04904117E+04 | 2.20E-01 | -5.64782160E+01 | -5.67022566E+01 | 2.20E-01 |
| 668 | Ge | 66 | 32 | 34 | 2 | 8.62543700E+00 | 8.62996192E+00 | -4.50E-03 | 6.14005917E+04 | 6.14002699E+04 | 3.20E-01 | 6.14170010E+04 | 6.14166791E+04 | 3.20E-01 | -6.16070330E+01 | -6.19288964E+01 | 3.20E-01 |
| 669 | Ge | 67 | 32 | 35 | 3 | 8.63285300E+00 | 8.63934589E+00 | -6.50E-03 | 6.23310348E+04 | 6.23305766E+04 | 4.60E-01 | 6.23474441E+04 | 6.23469858E+04 | 4.60E-01 | -6.26580040E+01 | -6.31162659E+01 | 4.60E-01 |
| 670 | Ge | 68 | 32 | 36 | 4 | 8.68813600E+00 | 8.69189464E+00 | -3.80E-03 | 6.32582081E+04 | 6.32579293E+04 | 2.80E-01 | 6.32746174E+04 | 6.32743385E+04 | 2.80E-01 | -6.69787890E+01 | -6.72576078E+01 | 2.80E-01 |
| 671 | Ge | 69 | 32 | 37 | 5 | 8.68096300E+00 | 8.68825143E+00 | -7.30E-03 | 6.41895803E+04 | 6.41890541E+04 | 5.30E-01 | 6.42059895E+04 | 6.42054634E+04 | 5.30E-01 | -6.71006540E+01 | -6.76268014E+01 | 5.30E-01 |
| 672 | Ge | 70 | 32 | 38 | 6 | 8.72169900E+00 | 8.72498166E+00 | -3.30E-03 | 6.51176132E+04 | 6.51173602E+04 | 2.50E-01 | 6.51340224E+04 | 6.51337694E+04 | 2.50E-01 | -7.05618350E+01 | -7.08148504E+01 | 2.50E-01 |
| 673 | Ge | 71 | 32 | 39 | 7 | 8.70330800E+00 | 8.71074766E+00 | -7.40E-03 | 6.60497626E+04 | 6.60492112E+04 | 5.50E-01 | 6.60661719E+04 | 6.60656204E+04 | 5.50E-01 | -6.99064570E+01 | -7.04578987E+01 | 5.50E-01 |
| 674 | Ge | 72 | 32 | 40 | 8 | 8.73174500E+00 | 8.73471812E+00 | -3.00E-03 | 6.69785772E+04 | 6.69783399E+04 | 2.40E-01 | 6.69949865E+04 | 6.69947492E+04 | 2.40E-01 | -7.25859010E+01 | -7.28232009E+01 | 2.40E-01 |
| 675 | Ge | 73 | 32 | 41 | 9 | 8.70504900E+00 | 8.71170189E+00 | -6.70E-03 | 6.79113597E+04 | 6.79108508E+04 | 5.10E-01 | 6.79277689E+04 | 6.79272600E+04 | 5.10E-01 | -7.12975230E+01 | -7.18064151E+01 | 5.10E-01 |
| 676 | Ge | 74 | 32 | 42 | 10 | 8.72520000E+00 | 8.72480538E+00 | 3.90E-04 | 6.88407288E+04 | 6.88407348E+04 | -6.00E-03 | 6.88571381E+04 | 6.88571441E+04 | -6.00E-03 | -7.34224420E+01 | -7.34164565E+01 | -6.00E-03 |
| 677 | Ge | 75 | 32 | 43 | 11 | 8.69560900E+00 | 8.69533508E+00 | 2.70E-04 | 6.97737883E+04 | 6.97737856E+04 | 2.70E-03 | 6.97901976E+04 | 6.97901949E+04 | 2.70E-03 | -7.18569650E+01 | -7.18596698E+01 | 2.70E-03 |
| 678 | Ge | 76 | 32 | 44 | 12 | 8.70523600E+00 | 8.70100011E+00 | 4.20E-03 | 7.07039265E+04 | 7.07042251E+04 | -3.00E-01 | 7.07203357E+04 | 7.07206344E+04 | -3.00E-01 | -7.32128890E+01 | -7.29142287E+01 | -3.00E-01 |
| 679 | Ge | 77 | 32 | 45 | 13 | 8.67102800E+00 | 8.66827715E+00 | 2.80E-03 | 7.16374206E+04 | 7.16376092E+04 | -1.90E-01 | 7.16538298E+04 | 7.16540185E+04 | -1.90E-01 | -7.12128620E+01 | -7.10242416E+01 | -1.90E-01 |
| 680 | Ge | 78 | 32 | 46 | 14 | 8.67166300E+00 | 8.66966614E+00 | 2.00E-03 | 7.25682654E+04 | 7.25683980E+04 | -1.30E-01 | 7.25846747E+04 | 7.25848072E+04 | -1.30E-01 | -7.18620530E+01 | -7.17295409E+01 | -1.30E-01 |
| 681 | Ge | 79 | 32 | 47 | 15 | 8.63450100E+00 | 8.63502780E+00 | -5.30E-04 | 7.35020950E+04 | 7.35020301E+04 | 6.50E-02 | 7.35185042E+04 | 7.35184394E+04 | 6.50E-02 | -6.95265940E+01 | -6.95914590E+01 | 6.50E-02 |
| 682 | Ge | 80 | 32 | 48 | 16 | 8.62757000E+00 | 8.63058116E+00 | -3.00E-03 | 7.44335803E+04 | 7.44333162E+04 | 2.60E-01 | 7.44499896E+04 | 7.44497254E+04 | 2.60E-01 | -6.95353050E+01 | -6.97994371E+01 | 2.60E-01 |
| 683 | Ge | 81 | 32 | 49 | 17 | 8.58065800E+00 | 8.58762237E+00 | -7.00E-03 | 7.53683180E+04 | 7.53677306E+04 | 5.90E-01 | 7.53847272E+04 | 7.53841399E+04 | 5.90E-01 | -6.62916860E+01 | -6.68790372E+01 | 5.90E-01 |
| 684 | Ge | 82 | 32 | 50 | 18 | 8.56375600E+00 | 8.56730308E+00 | -3.50E-03 | 7.63006887E+04 | 7.63003746E+04 | 3.10E-01 | 7.63170979E+04 | 7.63167838E+04 | 3.10E-01 | -6.54150660E+01 | -6.57291583E+01 | 3.10E-01 |
| 685 | Ge | 83 | 32 | 51 | 19 | 8.50434500E+00 | 8.50609418E+00 | -1.70E-03 | 7.72366214E+04 | 7.72364530E+04 | 1.70E-01 | 7.72530306E+04 | 7.72528623E+04 | 1.70E-01 | -6.09764340E+01 | -6.11448037E+01 | 1.70E-01 |
| 686 | Ge | 84 | 32 | 52 | 20 | 8.46552400E+00 | 8.46651550E+00 | -9.90E-04 | 7.81709434E+04 | 7.81708369E+04 | 1.10E-01 | 7.81873527E+04 | 7.81872462E+04 | 1.10E-01 | -5.81484270E+01 | -5.82549700E+01 | 1.10E-01 |
| 687 | Ge | 85 | 32 | 53 | 21 | 8.40176800E+00 | 8.39275117E+00 | 9.00E-03 | 7.91074625E+04 | 7.91082057E+04 | -7.40E-01 | 7.91238718E+04 | 7.91246150E+04 | -7.40E-01 | -5.31234190E+01 | -5.23801988E+01 | -7.40E-01 |
| 688 | Ge | 86 | 32 | 54 | 22 | 8.35900000E+00 | 8.34606214E+00 | 1.30E-02 | 8.00423196E+04 | 8.00433936E+04 | -1.10E+00 | 8.00587288E+04 | 8.00598029E+04 | -1.10E+00 | -4.97600000E+01 | -4.86863740E+01 | -1.10E+00 |
| 689 | Ge | 87 | 32 | 55 | 23 | 8.29000000E+00 | 8.27188164E+00 | 1.80E-02 | 8.09794957E+04 | 8.09810666E+04 | -1.60E+00 | 8.09959050E+04 | 8.09974759E+04 | -1.60E+00 | -4.40780000E+01 | -4.25074136E+01 | -1.60E+00 |
| 690 | Ge | 88 | 32 | 56 | 24 | 8.24300000E+00 | 8.22624290E+00 | 1.70E-02 | 8.19149300E+04 | 8.19163763E+04 | -1.40E+00 | 8.19313393E+04 | 8.19327856E+04 | -1.40E+00 | -4.01380000E+01 | -3.86917674E+01 | -1.40E+00 |
| 691 | Ge | 89 | 32 | 57 | 25 | 8.16900000E+00 | 8.15484985E+00 | 1.40E-02 | 8.28528328E+04 | 8.28540695E+04 | -1.20E+00 | 8.28692420E+04 | 8.28704787E+04 | -1.20E+00 | -3.37290000E+01 | -3.24927092E+01 | -1.20E+00 |
| 692 | GE | 90 | 32 | 58 | 26 | 8.11800000E+00 | 8.11030771E+00 | 7.70E-03 | 8.37888353E+04 | 8.37894888E+04 | -6.50E-01 | 8.38052445E+04 | 8.38058980E+04 | -6.50E-01 | -2.92210000E+01 | -2.85674476E+01 | -6.50E-01 |
| 693 | As | 63 | 33 | 30 | -3 | 8.19500000E+00 | 8.20743817E+00 | -1.20E-02 | 5.86335742E+04 | 5.86327841E+04 | 7.90E-01 | 5.86504989E+04 | 5.86497088E+04 | 7.90E-01 | -3.36270000E+01 | -3.44170485E+01 | 7.90E-01 |
| 694 | As | 64 | 33 | 31 | -2 | 8.28700000E+00 | 8.29869078E+00 | -1.20E-02 | 5.95590434E+04 | 5.95583019E+04 | 7.40E-01 | 5.95759681E+04 | 5.95752266E+04 | 7.40E-01 | -3.96520000E+01 | -4.03933348E+01 | 7.40E-01 |
| 695 | As | 65 | 33 | 32 | -1 | 8.39623400E+00 | 8.40442872E+00 | -8.20E-03 | 6.04832523E+04 | 6.04826956E+04 | 5.60E-01 | 6.05001769E+04 | 6.04996203E+04 | 5.60E-01 | -4.69370510E+01 | -4.74936725E+01 | 5.60E-01 |
| 696 | As | 66 | 33 | 33 | 0 | 8.46840300E+00 | 8.45726170E+00 | 1.10E-02 | 6.14096583E+04 | 6.14103696E+04 | -7.10E-01 | 6.14265830E+04 | 6.14272943E+04 | -7.10E-01 | -5.20250770E+01 | -5.13137590E+01 | -7.10E-01 |
| 697 | As | 67 | 33 | 34 | 1 | 8.53056800E+00 | 8.53252322E+00 | -2.00E-03 | 6.23365902E+04 | 6.23364352E+04 | 1.50E-01 | 6.23535149E+04 | 6.23533599E+04 | 1.50E-01 | -5.65872250E+01 | -5.67422235E+01 | 1.50E-01 |
| 698 | As | 68 | 33 | 35 | 2 | 8.55774500E+00 | 8.56320549E+00 | -5.50E-03 | 6.32657770E+04 | 6.32653817E+04 | 4.00E-01 | 6.32827016E+04 | 6.32823063E+04 | 4.00E-01 | -5.88945190E+01 | -5.92898225E+01 | 4.00E-01 |
| 699 | As | 69 | 33 | 36 | 3 | 8.61182000E+00 | 8.61736848E+00 | -5.50E-03 | 6.41930534E+04 | 6.41926466E+04 | 4.10E-01 | 6.42099780E+04 | 6.42095713E+04 | 4.10E-01 | -6.31121630E+01 | -6.35189551E+01 | 4.10E-01 |
| 700 | As | 70 | 33 | 37 | 4 | 8.62166600E+00 | 8.63227961E+00 | -1.10E-02 | 6.51233178E+04 | 6.51225508E+04 | 7.70E-01 | 6.51402424E+04 | 6.51394755E+04 | 7.70E-01 | -6.43418350E+01 | -6.51087836E+01 | 7.70E-01 |
| 701 | As | 71 | 33 | 38 | 5 | 8.66393200E+00 | 8.66976249E+00 | -5.80E-03 | 6.60512606E+04 | 6.60508226E+04 | 4.40E-01 | 6.60681853E+04 | 6.60677473E+04 | 4.40E-01 | -6.78930570E+01 | -6.83310287E+01 | 4.40E-01 |
| 702 | As | 72 | 33 | 39 | 6 | 8.66037800E+00 | 8.67232971E+00 | -1.20E-02 | 6.69824179E+04 | 6.69815334E+04 | 8.80E-01 | 6.69993426E+04 | 6.69984581E+04 | 8.80E-01 | -6.82297980E+01 | -6.91143124E+01 | 8.80E-01 |
| 703 | As | 73 | 33 | 40 | 7 | 8.68960900E+00 | 8.69678814E+00 | -7.20E-03 | 6.79111891E+04 | 6.79106410E+04 | 5.50E-01 | 6.79281137E+04 | 6.79275657E+04 | 5.50E-01 | -7.09527490E+01 | -7.15007880E+01 | 5.50E-01 |
| 704 | As | 74 | 33 | 41 | 8 | 8.68000100E+00 | 8.68939430E+00 | -9.40E-03 | 6.88427758E+04 | 6.88420567E+04 | 7.20E-01 | 6.88597005E+04 | 6.88589814E+04 | 7.20E-01 | -7.08600550E+01 | -7.15791135E+01 | 7.20E-01 |
| 705 | As | 75 | 33 | 42 | 9 | 8.70087400E+00 | 8.70293828E+00 | -2.10E-03 | 6.97720957E+04 | 6.97719169E+04 | 1.80E-01 | 6.97890204E+04 | 6.97888416E+04 | 1.80E-01 | -7.30341900E+01 | -7.32129868E+01 | 1.80E-01 |
| 706 | As | 76 | 33 | 43 | 10 | 8.68281600E+00 | 8.68816077E+00 | -5.30E-03 | 7.07043326E+04 | 7.07039025E+04 | 4.30E-01 | 7.07212573E+04 | 7.07208271E+04 | 4.30E-01 | -7.22913720E+01 | -7.27215152E+01 | 4.30E-01 |
| 707 | As | 77 | 33 | 44 | 11 | 8.69597800E+00 | 8.69420814E+00 | 1.80E-03 | 7.16342017E+04 | 7.16343140E+04 | -1.10E-01 | 7.16511264E+04 | 7.16512387E+04 | -1.10E-01 | -7.39163190E+01 | -7.38040046E+01 | -1.10E-01 |
| 708 | As | 78 | 33 | 45 | 12 | 8.67387500E+00 | 8.67528271E+00 | -1.40E-03 | 7.25667951E+04 | 7.25666614E+04 | 1.30E-01 | 7.25837198E+04 | 7.25835860E+04 | 1.30E-01 | -7.28169610E+01 | -7.29507100E+01 | 1.30E-01 |
| 709 | As | 79 | 33 | 46 | 13 | 8.67661600E+00 | 8.67684068E+00 | -2.20E-04 | 7.34974701E+04 | 7.34974284E+04 | 4.20E-02 | 7.35143947E+04 | 7.35143531E+04 | 4.20E-02 | -7.36360690E+01 | -7.36777538E+01 | 4.20E-02 |
| 710 | As | 80 | 33 | 47 | 14 | 8.65128000E+00 | 8.65509562E+00 | -3.80E-03 | 7.44303858E+04 | 7.44300566E+04 | 3.30E-01 | 7.44473104E+04 | 7.44469812E+04 | 3.30E-01 | -7.22144600E+01 | -7.25436710E+01 | 3.30E-01 |
| 711 | As | 81 | 33 | 48 | 15 | 8.64805600E+00 | 8.65063266E+00 | -2.60E-03 | 7.53615610E+04 | 7.53613283E+04 | 2.30E-01 | 7.53784856E+04 | 7.53782530E+04 | 2.30E-01 | -7.25333140E+01 | -7.27659477E+01 | 2.30E-01 |
| 712 | As | 82 | 33 | 49 | 16 | 8.61138600E+00 | 8.62002677E+00 | -8.60E-03 | 7.62954852E+04 | 7.62947528E+04 | 7.30E-01 | 7.63124099E+04 | 7.63116774E+04 | 7.30E-01 | -7.01030890E+01 | -7.08355779E+01 | 7.30E-01 |
| 713 | As | 83 | 33 | 50 | 17 | 8.59965300E+00 | 8.59993495E+00 | -2.80E-04 | 7.72274131E+04 | 7.72273657E+04 | 4.70E-02 | 7.72443377E+04 | 7.72442904E+04 | 4.70E-02 | -6.96693220E+01 | -6.97166645E+01 | 4.70E-02 |
| 714 | As | 84 | 33 | 51 | 18 | 8.54793800E+00 | 8.55089847E+00 | -3.00E-03 | 7.81627229E+04 | 7.81624503E+04 | 2.70E-01 | 7.81796476E+04 | 7.81793749E+04 | 2.70E-01 | -6.58535590E+01 | -6.61262162E+01 | 2.70E-01 |
| 715 | As | 85 | 33 | 52 | 19 | 8.51098400E+00 | 8.51162381E+00 | -6.40E-04 | 7.90968814E+04 | 7.90968031E+04 | 7.80E-02 | 7.91138060E+04 | 7.91137277E+04 | 7.80E-02 | -6.31891430E+01 | -6.32674499E+01 | 7.80E-02 |
| 716 | As | 86 | 33 | 53 | 20 | 8.45672100E+00 | 8.44935949E+00 | 7.40E-03 | 8.00326024E+04 | 8.00332116E+04 | -6.10E-01 | 8.00495271E+04 | 8.00501362E+04 | -6.10E-01 | -5.89621420E+01 | -5.83530234E+01 | -6.10E-01 |
| 717 | As | 87 | 33 | 54 | 21 | 8.41385100E+00 | 8.40249408E+00 | 1.10E-02 | 8.09674407E+04 | 8.09684049E+04 | -9.60E-01 | 8.09843654E+04 | 8.09853295E+04 | -9.60E-01 | -5.56179060E+01 | -5.46537731E+01 | -9.60E-01 |
| 718 | As | 88 | 33 | 55 | 22 | 8.35400000E+00 | 8.33896684E+00 | 1.50E-02 | 8.19038329E+04 | 8.19051582E+04 | -1.30E+00 | 8.19207575E+04 | 8.19220828E+04 | -1.30E+00 | -5.07200000E+01 | -4.93945512E+01 | -1.30E+00 |
| 719 | As | 89 | 33 | 56 | 23 | 8.30700000E+00 | 8.29275768E+00 | 1.40E-02 | 8.28392485E+04 | 8.28404972E+04 | -1.20E+00 | 8.28561732E+04 | 8.28574218E+04 | -1.20E+00 | -4.67980000E+01 | -4.55495838E+01 | -1.20E+00 |
| 720 | AS | 90 | 33 | 57 | 24 | 8.24400000E+00 | 8.23145250E+00 | 1.30E-02 | 8.37762104E+04 | 8.37772873E+04 | -1.10E+00 | 8.37931351E+04 | 8.37942119E+04 | -1.10E+00 | -4.13300000E+01 | -4.02535564E+01 | -1.10E+00 |
| 721 | As | 91 | 33 | 58 | 25 | 8.19300000E+00 | 8.18615985E+00 | 6.80E-03 | 8.47121384E+04 | 8.47127428E+04 | -6.00E-01 | 8.47290631E+04 | 8.47296675E+04 | -6.00E-01 | -3.68960000E+01 | -3.62920581E+01 | -6.00E-01 |
| 722 | As | 92 | 33 | 59 | 26 | 8.12700000E+00 | 8.12568077E+00 | 1.30E-03 | 8.56495475E+04 | 8.56496861E+04 | -1.40E-01 | 8.56664721E+04 | 8.56666108E+04 | -1.40E-01 | -3.09810000E+01 | -3.08428241E+01 | -1.40E-01 |
| 723 | SE | 64 | 34 | 30 | -4 | 8.07600000E+00 | 8.07234550E+00 | 3.70E-03 | 5.95712502E+04 | 5.95714894E+04 | -2.40E-01 | 5.95886904E+04 | 5.95889296E+04 | -2.40E-01 | -2.69290000E+01 | -2.66903141E+01 | -2.40E-01 |
| 724 | Se | 65 | 34 | 31 | -3 | 8.17200000E+00 | 8.18007234E+00 | -8.10E-03 | 6.04965125E+04 | 6.04959801E+04 | 5.30E-01 | 6.05139528E+04 | 6.05134204E+04 | 5.30E-01 | -3.31610000E+01 | -3.36935849E+01 | 5.30E-01 |
| 725 | Se | 66 | 34 | 32 | -2 | 8.29500000E+00 | 8.30901365E+00 | -1.40E-02 | 6.14198001E+04 | 6.14188553E+04 | 9.40E-01 | 6.14372404E+04 | 6.14362956E+04 | 9.40E-01 | -4.13680000E+01 | -4.23124646E+01 | 9.40E-01 |
| 726 | Se | 67 | 34 | 33 | -1 | 8.36953400E+00 | 8.37397189E+00 | -4.40E-03 | 6.23460816E+04 | 6.23457595E+04 | 3.20E-01 | 6.23635218E+04 | 6.23631997E+04 | 3.20E-01 | -4.65802890E+01 | -4.69023614E+01 | 3.20E-01 |
| 727 | Se | 68 | 34 | 34 | 0 | 8.47704700E+00 | 8.47270759E+00 | 4.30E-03 | 6.32699665E+04 | 6.32702369E+04 | -2.70E-01 | 6.32874067E+04 | 6.32876771E+04 | -2.70E-01 | -5.41894410E+01 | -5.39190418E+01 | -2.70E-01 |
| 728 | Se | 69 | 34 | 35 | 1 | 8.50370700E+00 | 8.50599783E+00 | -2.30E-03 | 6.41992153E+04 | 6.41990325E+04 | 1.80E-01 | 6.42166555E+04 | 6.42164728E+04 | 1.80E-01 | -5.64347060E+01 | -5.66174575E+01 | 1.80E-01 |
| 729 | Se | 70 | 34 | 36 | 2 | 8.57603300E+00 | 8.58020310E+00 | -4.20E-03 | 6.51252141E+04 | 6.51248975E+04 | 3.20E-01 | 6.51426544E+04 | 6.51423378E+04 | 3.20E-01 | -6.19298910E+01 | -6.22465053E+01 | 3.20E-01 |
| 730 | Se | 71 | 34 | 37 | 3 | 8.58606000E+00 | 8.59557346E+00 | -9.50E-03 | 6.60554916E+04 | 6.60547914E+04 | 7.00E-01 | 6.60729318E+04 | 6.60722316E+04 | 7.00E-01 | -6.31465070E+01 | -6.38466846E+01 | 7.00E-01 |
| 731 | Se | 72 | 34 | 38 | 4 | 8.64448900E+00 | 8.65053656E+00 | -6.00E-03 | 6.69822640E+04 | 6.69818039E+04 | 4.60E-01 | 6.69997042E+04 | 6.69992441E+04 | 4.60E-01 | -6.78681800E+01 | -6.83282823E+01 | 4.60E-01 |
| 732 | Se | 73 | 34 | 39 | 5 | 8.64155800E+00 | 8.65290562E+00 | -1.10E-02 | 6.79133988E+04 | 6.79125458E+04 | 8.50E-01 | 6.79308391E+04 | 6.79299860E+04 | 8.50E-01 | -6.82273880E+01 | -6.90804415E+01 | 8.50E-01 |
| 733 | Se | 74 | 34 | 40 | 6 | 8.68771500E+00 | 8.69319239E+00 | -5.50E-03 | 6.88409071E+04 | 6.88404770E+04 | 4.30E-01 | 6.88583473E+04 | 6.88579173E+04 | 4.30E-01 | -7.22132020E+01 | -7.26432491E+01 | 4.30E-01 |
| 734 | Se | 75 | 34 | 41 | 7 | 8.67891300E+00 | 8.68546218E+00 | -6.50E-03 | 6.97724448E+04 | 6.97719290E+04 | 5.20E-01 | 6.97898851E+04 | 6.97893692E+04 | 5.20E-01 | -7.21694810E+01 | -7.26853565E+01 | 5.20E-01 |
| 735 | Se | 76 | 34 | 42 | 8 | 8.71147700E+00 | 8.71367838E+00 | -2.20E-03 | 7.07008564E+04 | 7.07006645E+04 | 1.90E-01 | 7.07182967E+04 | 7.07181047E+04 | 1.90E-01 | -7.52519500E+01 | -7.54439307E+01 | 1.90E-01 |
| 736 | Se | 77 | 34 | 43 | 9 | 8.69469000E+00 | 8.69860585E+00 | -3.90E-03 | 7.16330030E+04 | 7.16326768E+04 | 3.30E-01 | 7.16504432E+04 | 7.16501170E+04 | 3.30E-01 | -7.45994850E+01 | -7.49257057E+01 | 3.30E-01 |
| 737 | Se | 78 | 34 | 44 | 10 | 8.71780600E+00 | 8.71837240E+00 | -5.70E-04 | 7.25620706E+04 | 7.25620017E+04 | 6.90E-02 | 7.25795108E+04 | 7.25794420E+04 | 6.90E-02 | -7.70259090E+01 | -7.70947828E+01 | 6.90E-02 |
| 738 | Se | 79 | 34 | 45 | 11 | 8.69559100E+00 | 8.69913838E+00 | -3.50E-03 | 7.34946732E+04 | 7.34943682E+04 | 3.00E-01 | 7.35121134E+04 | 7.35118085E+04 | 3.00E-01 | -7.59174250E+01 | -7.62223487E+01 | 3.00E-01 |
| 739 | Se | 80 | 34 | 46 | 12 | 8.71081300E+00 | 8.71345219E+00 | -2.60E-03 | 7.44243252E+04 | 7.44240894E+04 | 2.40E-01 | 7.44417654E+04 | 7.44415296E+04 | 2.40E-01 | -7.77594770E+01 | -7.79952732E+01 | 2.40E-01 |
| 740 | Se | 81 | 34 | 47 | 13 | 8.68599900E+00 | 8.69125514E+00 | -5.30E-03 | 7.53571897E+04 | 7.53567393E+04 | 4.50E-01 | 7.53746299E+04 | 7.53741795E+04 | 4.50E-01 | -7.63890070E+01 | -7.68394455E+01 | 4.50E-01 |
| 741 | Se | 82 | 34 | 48 | 14 | 8.69319700E+00 | 8.69868797E+00 | -5.50E-03 | 7.62874788E+04 | 7.62870039E+04 | 4.70E-01 | 7.63049191E+04 | 7.63044441E+04 | 4.70E-01 | -7.75939190E+01 | -7.80688737E+01 | 4.70E-01 |
| 742 | Se | 83 | 34 | 49 | 15 | 8.65855600E+00 | 8.66773809E+00 | -9.20E-03 | 7.72212263E+04 | 7.72204394E+04 | 7.90E-01 | 7.72386665E+04 | 7.72378797E+04 | 7.90E-01 | -7.53405670E+01 | -7.61274027E+01 | 7.90E-01 |
| 743 | Se | 84 | 34 | 50 | 16 | 8.65879300E+00 | 8.65923518E+00 | -4.40E-04 | 7.81521132E+04 | 7.81520513E+04 | 6.20E-02 | 7.81695534E+04 | 7.81694915E+04 | 6.20E-02 | -7.59477210E+01 | -7.60095769E+01 | 6.20E-02 |
| 744 | Se | 85 | 34 | 51 | 17 | 8.61030400E+00 | 8.61035283E+00 | -4.90E-05 | 7.90871413E+04 | 7.90871125E+04 | 2.90E-02 | 7.91045815E+04 | 7.91045527E+04 | 2.90E-02 | -7.24136350E+01 | -7.24424932E+01 | 2.90E-02 |
| 745 | Se | 86 | 34 | 52 | 18 | 8.58182200E+00 | 8.58238223E+00 | -5.60E-04 | 8.00205458E+04 | 8.00204730E+04 | 7.30E-02 | 8.00379861E+04 | 8.00379132E+04 | 7.30E-02 | -7.05031660E+01 | -7.05760560E+01 | 7.30E-02 |
| 746 | Se | 87 | 34 | 53 | 19 | 8.52909100E+00 | 8.52022346E+00 | 8.90E-03 | 8.09561169E+04 | 8.09568638E+04 | -7.50E-01 | 8.09735572E+04 | 8.09743040E+04 | -7.50E-01 | -6.64261240E+01 | -6.56793057E+01 | -7.50E-01 |
| 747 | Se | 88 | 34 | 54 | 20 | 8.49500400E+00 | 8.48376255E+00 | 1.10E-02 | 8.18901529E+04 | 8.18911175E+04 | -9.60E-01 | 8.19075932E+04 | 8.19085577E+04 | -9.60E-01 | -6.38841950E+01 | -6.29196500E+01 | -9.60E-01 |
| 748 | Se | 89 | 34 | 55 | 21 | 8.43527900E+00 | 8.41986991E+00 | 1.50E-02 | 8.28265388E+04 | 8.28278855E+04 | -1.30E+00 | 8.28439790E+04 | 8.28453258E+04 | -1.30E+00 | -5.89923900E+01 | -5.76456488E+01 | -1.30E+00 |
| 749 | Se | 90 | 34 | 56 | 22 | 8.39576600E+00 | 8.38316010E+00 | 1.30E-02 | 8.37612250E+04 | 8.37623349E+04 | -1.10E+00 | 8.37786653E+04 | 8.37797752E+04 | -1.10E+00 | -5.58002170E+01 | -5.46903168E+01 | -1.10E+00 |
| 750 | Se | 91 | 34 | 57 | 23 | 8.33200000E+00 | 8.32116138E+00 | 1.10E-02 | 8.46981814E+04 | 8.46991590E+04 | -9.80E-01 | 8.47156216E+04 | 8.47165993E+04 | -9.80E-01 | -5.03380000E+01 | -4.93602744E+01 | -9.80E-01 |
| 751 | Se | 92 | 34 | 58 | 24 | 8.29000000E+00 | 8.28470556E+00 | 5.30E-03 | 8.56332896E+04 | 8.56337572E+04 | -4.70E-01 | 8.56507299E+04 | 8.56511974E+04 | -4.70E-01 | -4.67240000E+01 | -4.62561811E+01 | -4.70E-01 |
| 752 | Se | 93 | 34 | 59 | 25 | 8.22300000E+00 | 8.22335538E+00 | -3.60E-04 | 8.65707918E+04 | 8.65707434E+04 | 4.80E-02 | 8.65882321E+04 | 8.65881837E+04 | 4.80E-02 | -4.07160000E+01 | -4.07640017E+01 | 4.80E-02 |
| 753 | Se | 94 | 34 | 60 | 26 | 8.18000000E+00 | 8.18564436E+00 | -5.60E-03 | 8.75061982E+04 | 8.75056303E+04 | 5.70E-01 | 8.75236384E+04 | 8.75230705E+04 | 5.70E-01 | -3.68030000E+01 | -3.73712015E+01 | 5.70E-01 |
| 754 | SE | 95 | 34 | 61 | 27 | 8.11200000E+00 | 8.12352000E+00 | -1.20E-02 | 8.84440357E+04 | 8.84429118E+04 | 1.10E+00 | 8.84614759E+04 | 8.84603521E+04 | 1.10E+00 | -3.04600000E+01 | -3.15837131E+01 | 1.10E+00 |
| 755 | Br | 69 | 35 | 34 | -1 | 8.34275700E+00 | 8.35141342E+00 | -8.70E-03 | 6.42090227E+04 | 6.42084000E+04 | 6.20E-01 | 6.42269787E+04 | 6.42263560E+04 | 6.20E-01 | -4.61114710E+01 | -4.67342102E+01 | 6.20E-01 |
| 756 | Br | 70 | 35 | 35 | 0 | 8.41479600E+00 | 8.40708130E+00 | 7.70E-03 | 6.51352026E+04 | 6.51357172E+04 | -5.10E-01 | 6.51531586E+04 | 6.51536732E+04 | -5.10E-01 | -5.14256200E+01 | -5.09110559E+01 | -5.10E-01 |
| 757 | Br | 71 | 35 | 36 | 1 | 8.48146200E+00 | 8.48420013E+00 | -2.70E-03 | 6.60616199E+04 | 6.60614001E+04 | 2.20E-01 | 6.60795759E+04 | 6.60793561E+04 | 2.20E-01 | -5.65024180E+01 | -5.67222556E+01 | 2.20E-01 |
| 758 | Br | 72 | 35 | 37 | 2 | 8.51138900E+00 | 8.51897280E+00 | -7.60E-03 | 6.69905491E+04 | 6.69899776E+04 | 5.70E-01 | 6.70085051E+04 | 6.70079336E+04 | 5.70E-01 | -5.90673210E+01 | -5.96387685E+01 | 5.70E-01 |
| 759 | Br | 73 | 35 | 38 | 3 | 8.56810400E+00 | 8.57549374E+00 | -7.40E-03 | 6.79174628E+04 | 6.79168980E+04 | 5.60E-01 | 6.79354189E+04 | 6.79348540E+04 | 5.60E-01 | -6.36475850E+01 | -6.42124512E+01 | 5.60E-01 |
| 760 | Br | 74 | 35 | 39 | 4 | 8.58356200E+00 | 8.59489470E+00 | -1.10E-02 | 6.88473163E+04 | 6.88464522E+04 | 8.60E-01 | 6.88652723E+04 | 6.88644082E+04 | 8.60E-01 | -6.52882490E+01 | -6.61522970E+01 | 8.60E-01 |
| 761 | Br | 75 | 35 | 40 | 5 | 8.62765000E+00 | 8.63607903E+00 | -8.40E-03 | 6.97749914E+04 | 6.97743339E+04 | 6.60E-01 | 6.97929475E+04 | 6.97922899E+04 | 6.60E-01 | -6.91071180E+01 | -6.97646975E+01 | 6.60E-01 |
| 762 | Br | 76 | 35 | 41 | 6 | 8.63588200E+00 | 8.64379958E+00 | -7.90E-03 | 7.07053035E+04 | 7.07046764E+04 | 6.30E-01 | 7.07232596E+04 | 7.07226324E+04 | 6.30E-01 | -7.02890690E+01 | -7.09162188E+01 | 6.30E-01 |
| 763 | Br | 77 | 35 | 42 | 7 | 8.66680600E+00 | 8.67269786E+00 | -5.90E-03 | 7.16338519E+04 | 7.16333728E+04 | 4.80E-01 | 7.16518079E+04 | 7.16513288E+04 | 4.80E-01 | -7.32348050E+01 | -7.37138674E+01 | 4.80E-01 |
| 764 | Br | 78 | 35 | 43 | 8 | 8.66195800E+00 | 8.67193345E+00 | -1.00E-02 | 7.25651286E+04 | 7.25643251E+04 | 8.00E-01 | 7.25830846E+04 | 7.25822811E+04 | 8.00E-01 | -7.34521250E+01 | -7.42556219E+01 | 8.00E-01 |
| 765 | Br | 79 | 35 | 44 | 9 | 8.68759500E+00 | 8.69231972E+00 | -4.70E-03 | 7.34940068E+04 | 7.34936081E+04 | 4.00E-01 | 7.35119628E+04 | 7.35115641E+04 | 4.00E-01 | -7.60680550E+01 | -7.64667520E+01 | 4.00E-01 |
| 766 | Br | 80 | 35 | 45 | 10 | 8.67765300E+00 | 8.68640331E+00 | -8.80E-03 | 7.44256799E+04 | 7.44249544E+04 | 7.30E-01 | 7.44436359E+04 | 7.44429104E+04 | 7.30E-01 | -7.58890140E+01 | -7.66144401E+01 | 7.30E-01 |
| 767 | Br | 81 | 35 | 46 | 11 | 8.69592900E+00 | 8.70120001E+00 | -5.30E-03 | 7.53550873E+04 | 7.53546349E+04 | 4.50E-01 | 7.53730433E+04 | 7.53725909E+04 | 4.50E-01 | -7.79756510E+01 | -7.84280570E+01 | 4.50E-01 |
| 768 | Br | 82 | 35 | 47 | 12 | 8.68247700E+00 | 8.69134704E+00 | -8.90E-03 | 7.62870597E+04 | 7.62863070E+04 | 7.50E-01 | 7.63050157E+04 | 7.63042630E+04 | 7.50E-01 | -7.74972770E+01 | -7.82499944E+01 | 7.50E-01 |
| 769 | Br | 83 | 35 | 48 | 13 | 8.69338100E+00 | 8.69914003E+00 | -5.80E-03 | 7.72170376E+04 | 7.72165342E+04 | 5.00E-01 | 7.72349937E+04 | 7.72344902E+04 | 5.00E-01 | -7.90134010E+01 | -7.95168405E+01 | 5.00E-01 |
| 770 | Br | 84 | 35 | 49 | 14 | 8.67132900E+00 | 8.67991681E+00 | -8.60E-03 | 7.81497620E+04 | 7.81490152E+04 | 7.50E-01 | 7.81677180E+04 | 7.81669712E+04 | 7.50E-01 | -7.77830840E+01 | -7.85299108E+01 | 7.50E-01 |
| 771 | Br | 85 | 35 | 50 | 15 | 8.67359200E+00 | 8.67215451E+00 | 1.40E-03 | 7.90804637E+04 | 7.90805605E+04 | -9.70E-02 | 7.90984197E+04 | 7.90985165E+04 | -9.70E-02 | -7.85754680E+01 | -7.84787138E+01 | -9.70E-02 |
| 772 | Br | 86 | 35 | 51 | 16 | 8.63236500E+00 | 8.63494557E+00 | -2.60E-03 | 8.00149010E+04 | 8.00146537E+04 | 2.50E-01 | 8.00328570E+04 | 8.00326097E+04 | 2.50E-01 | -7.56322510E+01 | -7.58795797E+01 | 2.50E-01 |
| 773 | Br | 87 | 35 | 52 | 17 | 8.60591000E+00 | 8.60815394E+00 | -2.20E-03 | 8.09481356E+04 | 8.09479150E+04 | 2.20E-01 | 8.09660916E+04 | 8.09658710E+04 | 2.20E-01 | -7.38916760E+01 | -7.41123351E+01 | 2.20E-01 |
| 774 | Br | 88 | 35 | 53 | 18 | 8.56374700E+00 | 8.55719824E+00 | 6.50E-03 | 8.18828054E+04 | 8.18833563E+04 | -5.50E-01 | 8.19007614E+04 | 8.19013123E+04 | -5.50E-01 | -7.07159580E+01 | -7.01650678E+01 | -5.50E-01 |
| 775 | Br | 89 | 35 | 54 | 19 | 8.53077900E+00 | 8.52161892E+00 | 9.20E-03 | 8.28167412E+04 | 8.28175310E+04 | -7.90E-01 | 8.28346972E+04 | 8.28354870E+04 | -7.90E-01 | -6.82742620E+01 | -6.74843876E+01 | -7.90E-01 |
| 776 | Br | 90 | 35 | 55 | 20 | 8.47818600E+00 | 8.46798462E+00 | 1.00E-02 | 8.37525092E+04 | 8.37534019E+04 | -8.90E-01 | 8.37704652E+04 | 8.37713579E+04 | -8.90E-01 | -6.40002970E+01 | -6.31076005E+01 | -8.90E-01 |
| 777 | Br | 91 | 35 | 56 | 21 | 8.44192300E+00 | 8.43167201E+00 | 1.00E-02 | 8.46868962E+04 | 8.46878037E+04 | -9.10E-01 | 8.47048523E+04 | 8.47057597E+04 | -9.10E-01 | -6.11072930E+01 | -6.01998188E+01 | -9.10E-01 |
| 778 | Br | 92 | 35 | 57 | 22 | 8.38491100E+00 | 8.37912500E+00 | 5.80E-03 | 8.56232648E+04 | 8.56237718E+04 | -5.10E-01 | 8.56412208E+04 | 8.56417278E+04 | -5.10E-01 | -5.62328050E+01 | -5.57258468E+01 | -5.10E-01 |
| 779 | Br | 93 | 35 | 58 | 23 | 8.34645900E+00 | 8.34273376E+00 | 3.70E-03 | 8.65580213E+04 | 8.65583424E+04 | -3.20E-01 | 8.65759773E+04 | 8.65762984E+04 | -3.20E-01 | -5.29703380E+01 | -5.26492677E+01 | -3.20E-01 |
| 780 | Br | 94 | 35 | 59 | 24 | 8.28600000E+00 | 8.29023052E+00 | -4.20E-03 | 8.74948864E+04 | 8.74945003E+04 | 3.90E-01 | 8.75128424E+04 | 8.75124564E+04 | 3.90E-01 | -4.75990000E+01 | -4.79853782E+01 | 3.90E-01 |
| 781 | Br | 95 | 35 | 60 | 25 | 8.24400000E+00 | 8.25234311E+00 | -8.30E-03 | 8.84302089E+04 | 8.84293748E+04 | 8.30E-01 | 8.84481649E+04 | 8.84473308E+04 | 8.30E-01 | -4.37710000E+01 | -4.46049857E+01 | 8.30E-01 |
| 782 | Br | 96 | 35 | 61 | 26 | 8.18400000E+00 | 8.19859590E+00 | -1.50E-02 | 8.93673105E+04 | 8.93658476E+04 | 1.50E+00 | 8.93852665E+04 | 8.93838036E+04 | 1.50E+00 | -3.81630000E+01 | -3.96262778E+01 | 1.50E+00 |
| 783 | Br | 97 | 35 | 62 | 27 | 8.14000000E+00 | 8.15838995E+00 | -1.80E-02 | 9.03029125E+04 | 9.03011143E+04 | 1.80E+00 | 9.03208685E+04 | 9.03190703E+04 | 1.80E+00 | -3.40550000E+01 | -3.58535770E+01 | 1.80E+00 |
| 784 | BR | 98 | 35 | 63 | 28 | 8.08200000E+00 | 8.10291325E+00 | -2.10E-02 | 9.12400141E+04 | 9.12379580E+04 | 2.10E+00 | 9.12579701E+04 | 9.12559140E+04 | 2.10E+00 | -2.84480000E+01 | -3.05039313E+01 | 2.10E+00 |
| 785 | Kr | 69 | 36 | 33 | -3 | 8.13300000E+00 | 8.13753301E+00 | -4.50E-03 | 6.42221836E+04 | 6.42218587E+04 | 3.20E-01 | 6.42406556E+04 | 6.42403307E+04 | 3.20E-01 | -3.24350000E+01 | -3.27595388E+01 | 3.20E-01 |
| 786 | Kr | 70 | 36 | 34 | -2 | 8.25400000E+00 | 8.26128298E+00 | -7.30E-03 | 6.51451638E+04 | 6.51446240E+04 | 5.40E-01 | 6.51636358E+04 | 6.51630960E+04 | 5.40E-01 | -4.09480000E+01 | -4.14882504E+01 | 5.40E-01 |
| 787 | Kr | 71 | 36 | 35 | -1 | 8.32713000E+00 | 8.32643291E+00 | 7.00E-04 | 6.60712791E+04 | 6.60713025E+04 | -2.30E-02 | 6.60897511E+04 | 6.60897745E+04 | -2.30E-02 | -4.63272050E+01 | -4.63038594E+01 | -2.30E-02 |
| 788 | Kr | 72 | 36 | 36 | 0 | 8.42931900E+00 | 8.42445519E+00 | 4.90E-03 | 6.69951598E+04 | 6.69954838E+04 | -3.20E-01 | 6.70136318E+04 | 6.70139558E+04 | -3.20E-01 | -5.39405750E+01 | -5.36165775E+01 | -3.20E-01 |
| 789 | Kr | 73 | 36 | 37 | 1 | 8.46018400E+00 | 8.46123736E+00 | -1.10E-03 | 6.79240427E+04 | 6.79239397E+04 | 1.00E-01 | 6.79425147E+04 | 6.79424116E+04 | 1.00E-01 | -5.65517500E+01 | -5.66548120E+01 | 1.00E-01 |
| 790 | Kr | 74 | 36 | 38 | 2 | 8.53303800E+00 | 8.53588832E+00 | -2.90E-03 | 6.88497567E+04 | 6.88495196E+04 | 2.40E-01 | 6.88682287E+04 | 6.88679916E+04 | 2.40E-01 | -6.23318340E+01 | -6.25689014E+01 | 2.40E-01 |
| 791 | Kr | 75 | 36 | 39 | 3 | 8.55343900E+00 | 8.55572614E+00 | -2.30E-03 | 6.97792590E+04 | 6.97790613E+04 | 2.00E-01 | 6.97977309E+04 | 6.97975333E+04 | 2.00E-01 | -6.43236230E+01 | -6.45213078E+01 | 2.00E-01 |
| 792 | Kr | 76 | 36 | 40 | 4 | 8.60881300E+00 | 8.61285620E+00 | -4.00E-03 | 7.07060625E+04 | 7.07057290E+04 | 3.30E-01 | 7.07245345E+04 | 7.07242010E+04 | 3.30E-01 | -6.90141370E+01 | -6.93475994E+01 | 3.30E-01 |
| 793 | Kr | 77 | 36 | 41 | 5 | 8.61683600E+00 | 8.62052642E+00 | -3.70E-03 | 7.16364013E+04 | 7.16360910E+04 | 3.10E-01 | 7.16548733E+04 | 7.16545630E+04 | 3.10E-01 | -7.01694420E+01 | -7.04797435E+01 | 3.10E-01 |
| 794 | Kr | 78 | 36 | 42 | 6 | 8.66125400E+00 | 8.66385541E+00 | -2.60E-03 | 7.25638852E+04 | 7.25636562E+04 | 2.30E-01 | 7.25823572E+04 | 7.25821281E+04 | 2.30E-01 | -7.41795770E+01 | -7.44086120E+01 | 2.30E-01 |
| 795 | Kr | 79 | 36 | 43 | 7 | 8.65711200E+00 | 8.66290996E+00 | -5.80E-03 | 7.34951166E+04 | 7.34946324E+04 | 4.80E-01 | 7.35135885E+04 | 7.35131044E+04 | 4.80E-01 | -7.44422770E+01 | -7.49264579E+01 | 4.80E-01 |
| 796 | Kr | 80 | 36 | 44 | 8 | 8.69292800E+00 | 8.69657882E+00 | -3.70E-03 | 7.44231596E+04 | 7.44228413E+04 | 3.20E-01 | 7.44416316E+04 | 7.44413133E+04 | 3.20E-01 | -7.78933130E+01 | -7.82115579E+01 | 3.20E-01 |
| 797 | Kr | 81 | 36 | 45 | 9 | 8.68280300E+00 | 8.69043045E+00 | -7.60E-03 | 7.53548521E+04 | 7.53542082E+04 | 6.40E-01 | 7.53733241E+04 | 7.53726801E+04 | 6.40E-01 | -7.76948140E+01 | -7.83387993E+01 | 6.40E-01 |
| 798 | Kr | 82 | 36 | 46 | 10 | 8.71065700E+00 | 8.71746004E+00 | -6.80E-03 | 7.62834507E+04 | 7.62828667E+04 | 5.80E-01 | 7.63019227E+04 | 7.63013387E+04 | 5.80E-01 | -8.05903180E+01 | -8.11743374E+01 | 5.80E-01 |
| 799 | Kr | 83 | 36 | 47 | 11 | 8.69572100E+00 | 8.70726041E+00 | -1.20E-02 | 7.72155450E+04 | 7.72145612E+04 | 9.80E-01 | 7.72340170E+04 | 7.72330332E+04 | 9.80E-01 | -7.99900310E+01 | -8.09739094E+01 | 9.80E-01 |
| 800 | Kr | 84 | 36 | 48 | 12 | 8.71744600E+00 | 8.72637078E+00 | -8.90E-03 | 7.81445898E+04 | 7.81438140E+04 | 7.80E-01 | 7.81630618E+04 | 7.81622860E+04 | 7.80E-01 | -8.24393354E+01 | -8.32151213E+01 | 7.80E-01 |
| 801 | Kr | 85 | 36 | 49 | 13 | 8.69856200E+00 | 8.70694701E+00 | -8.40E-03 | 7.90770429E+04 | 7.90763040E+04 | 7.40E-01 | 7.90955149E+04 | 7.90947760E+04 | 7.40E-01 | -8.14803300E+01 | -8.22191530E+01 | 7.40E-01 |
| 802 | Kr | 86 | 36 | 50 | 14 | 8.71202900E+00 | 8.71019577E+00 | 1.80E-03 | 8.00067516E+04 | 8.00068831E+04 | -1.30E-01 | 8.00252236E+04 | 8.00253551E+04 | -1.30E-01 | -8.32656649E+01 | -8.31341740E+01 | -1.30E-01 |
| 803 | Kr | 87 | 36 | 51 | 15 | 8.67528300E+00 | 8.67346367E+00 | 1.80E-03 | 8.09408018E+04 | 8.09409340E+04 | -1.30E-01 | 8.09592738E+04 | 8.09594059E+04 | -1.30E-01 | -8.07095210E+01 | -8.05773583E+01 | -1.30E-01 |
| 804 | Kr | 88 | 36 | 52 | 16 | 8.65684900E+00 | 8.65760390E+00 | -7.50E-04 | 8.18733141E+04 | 8.18732215E+04 | 9.30E-02 | 8.18917861E+04 | 8.18916935E+04 | 9.30E-02 | -7.96912850E+01 | -7.97838430E+01 | 9.30E-02 |
| 805 | Kr | 89 | 36 | 53 | 17 | 8.61481500E+00 | 8.60737385E+00 | 7.40E-03 | 8.28079637E+04 | 8.28085998E+04 | -6.40E-01 | 8.28264356E+04 | 8.28270718E+04 | -6.40E-01 | -7.65357840E+01 | -7.58996536E+01 | -6.40E-01 |
| 806 | Kr | 90 | 36 | 54 | 18 | 8.59125900E+00 | 8.58198002E+00 | 9.30E-03 | 8.37410342E+04 | 8.37418432E+04 | -8.10E-01 | 8.37595062E+04 | 8.37603152E+04 | -8.10E-01 | -7.49592500E+01 | -7.41502641E+01 | -8.10E-01 |
| 807 | Kr | 91 | 36 | 55 | 19 | 8.54175100E+00 | 8.52869744E+00 | 1.30E-02 | 8.46765136E+04 | 8.46776754E+04 | -1.20E+00 | 8.46949856E+04 | 8.46961473E+04 | -1.20E+00 | -7.09739640E+01 | -6.98122103E+01 | -1.20E+00 |
| 808 | Kr | 92 | 36 | 56 | 20 | 8.51267400E+00 | 8.50161806E+00 | 1.10E-02 | 8.56102123E+04 | 8.56112033E+04 | -9.90E-01 | 8.56286843E+04 | 8.56296753E+04 | -9.90E-01 | -6.87693190E+01 | -6.77782852E+01 | -9.90E-01 |
| 809 | Kr | 93 | 36 | 57 | 21 | 8.45810800E+00 | 8.44905294E+00 | 9.10E-03 | 8.65463397E+04 | 8.65471557E+04 | -8.20E-01 | 8.65648117E+04 | 8.65656276E+04 | -8.20E-01 | -6.41359940E+01 | -6.33200281E+01 | -8.20E-01 |
| 810 | Kr | 94 | 36 | 58 | 22 | 8.42433100E+00 | 8.42113186E+00 | 3.20E-03 | 8.74806220E+04 | 8.74808966E+04 | -2.70E-01 | 8.74990939E+04 | 8.74993686E+04 | -2.70E-01 | -6.13477710E+01 | -6.10731809E+01 | -2.70E-01 |
| 811 | Kr | 95 | 36 | 59 | 23 | 8.36599500E+00 | 8.36836119E+00 | -2.40E-03 | 8.84173049E+04 | 8.84170540E+04 | 2.50E-01 | 8.84357769E+04 | 8.84355260E+04 | 2.50E-01 | -5.61589120E+01 | -5.64097802E+01 | 2.50E-01 |
| 812 | Kr | 96 | 36 | 60 | 24 | 8.33085000E+00 | 8.33832606E+00 | -7.50E-03 | 8.93518782E+04 | 8.93511344E+04 | 7.40E-01 | 8.93703502E+04 | 8.93696064E+04 | 7.40E-01 | -5.30796780E+01 | -5.38234495E+01 | 7.40E-01 |
| 813 | KR | 97 | 36 | 61 | 25 | 8.26986400E+00 | 8.28411169E+00 | -1.40E-02 | 9.02890284E+04 | 9.02876203E+04 | 1.40E+00 | 9.03075004E+04 | 9.03060923E+04 | 1.40E+00 | -4.74234910E+01 | -4.88316625E+01 | 1.40E+00 |
| 814 | Kr | 98 | 36 | 62 | 26 | 8.23600000E+00 | 8.25124203E+00 | -1.50E-02 | 9.12236348E+04 | 9.12221228E+04 | 1.50E+00 | 9.12421068E+04 | 9.12405947E+04 | 1.50E+00 | -4.43110000E+01 | -4.58232293E+01 | 1.50E+00 |
| 815 | Kr | 99 | 36 | 63 | 27 | 8.17800000E+00 | 8.19515204E+00 | -1.70E-02 | 9.21606806E+04 | 9.21589898E+04 | 1.70E+00 | 9.21791526E+04 | 9.21774618E+04 | 1.70E+00 | -3.87590000E+01 | -4.04502429E+01 | 1.70E+00 |
| 816 | Kr | 100 | 36 | 64 | 28 | 8.14000000E+00 | 8.15943853E+00 | -1.90E-02 | 9.30958820E+04 | 9.30939314E+04 | 2.00E+00 | 9.31143540E+04 | 9.31124034E+04 | 2.00E+00 | -3.50520000E+01 | -3.70027248E+01 | 2.00E+00 |
| 817 | KR | 101 | 36 | 65 | 29 | 8.08100000E+00 | 8.10154881E+00 | -2.10E-02 | 9.40333004E+04 | 9.40311842E+04 | 2.10E+00 | 9.40517723E+04 | 9.40496562E+04 | 2.10E+00 | -2.91280000E+01 | -3.12439829E+01 | 2.10E+00 |
| 818 | Rb | 73 | 37 | 36 | -1 | 8.30600000E+00 | 8.30855207E+00 | -2.60E-03 | 6.79339964E+04 | 6.79337864E+04 | 2.10E-01 | 6.79529845E+04 | 6.79527746E+04 | 2.10E-01 | -4.60820000E+01 | -4.62918634E+01 | 2.10E-01 |
| 819 | Rb | 74 | 37 | 37 | 0 | 8.38171100E+00 | 8.36531439E+00 | 1.60E-02 | 6.88596564E+04 | 6.88608428E+04 | -1.20E+00 | 6.88786445E+04 | 6.88798310E+04 | -1.20E+00 | -5.19159850E+01 | -5.07295081E+01 | -1.20E+00 |
| 820 | Rb | 75 | 37 | 38 | 1 | 8.44827500E+00 | 8.44239713E+00 | 5.90E-03 | 6.97858477E+04 | 6.97862617E+04 | -4.10E-01 | 6.98048359E+04 | 6.98052499E+04 | -4.10E-01 | -5.72186940E+01 | -5.68047085E+01 | -4.10E-01 |
| 821 | Rb | 76 | 37 | 39 | 2 | 8.48621500E+00 | 8.47970811E+00 | 6.50E-03 | 7.07140814E+04 | 7.07145491E+04 | -4.70E-01 | 7.07330696E+04 | 7.07335372E+04 | -4.70E-01 | -6.04790810E+01 | -6.00114213E+01 | -4.70E-01 |
| 822 | Rb | 77 | 37 | 40 | 3 | 8.53733900E+00 | 8.53827416E+00 | -9.40E-04 | 7.16412240E+04 | 7.16411251E+04 | 9.90E-02 | 7.16602122E+04 | 7.16601133E+04 | 9.90E-02 | -6.48304910E+01 | -6.49293961E+01 | 9.90E-02 |
| 823 | Rb | 78 | 37 | 41 | 4 | 8.55835000E+00 | 8.56135933E+00 | -3.00E-03 | 7.25706132E+04 | 7.25703516E+04 | 2.60E-01 | 7.25896013E+04 | 7.25893398E+04 | 2.60E-01 | -6.69354180E+01 | -6.71969949E+01 | 2.60E-01 |
| 824 | Rb | 79 | 37 | 42 | 5 | 8.60114200E+00 | 8.60559265E+00 | -4.50E-03 | 7.34982397E+04 | 7.34978612E+04 | 3.80E-01 | 7.35172278E+04 | 7.35168494E+04 | 3.80E-01 | -7.08029850E+01 | -7.11814677E+01 | 3.80E-01 |
| 825 | Rb | 80 | 37 | 43 | 6 | 8.61167500E+00 | 8.61858578E+00 | -6.90E-03 | 7.44283612E+04 | 7.44277815E+04 | 5.80E-01 | 7.44473494E+04 | 7.44467697E+04 | 5.80E-01 | -7.21754670E+01 | -7.27551910E+01 | 5.80E-01 |
| 826 | Rb | 81 | 37 | 44 | 7 | 8.64551300E+00 | 8.65293227E+00 | -7.40E-03 | 7.53565741E+04 | 7.53559463E+04 | 6.30E-01 | 7.53755623E+04 | 7.53749344E+04 | 6.30E-01 | -7.54566630E+01 | -7.60845237E+01 | 6.30E-01 |
| 827 | Rb | 82 | 37 | 45 | 8 | 8.64742700E+00 | 8.65954027E+00 | -1.20E-02 | 7.62873370E+04 | 7.62863168E+04 | 1.00E+00 | 7.63063252E+04 | 7.63053050E+04 | 1.00E+00 | -7.61878030E+01 | -7.72079935E+01 | 1.00E+00 |
| 828 | Rb | 83 | 37 | 46 | 9 | 8.67521800E+00 | 8.68706851E+00 | -1.20E-02 | 7.72159483E+04 | 7.72149378E+04 | 1.00E+00 | 7.72349364E+04 | 7.72339260E+04 | 1.00E+00 | -7.90706290E+01 | -8.00810581E+01 | 1.00E+00 |
| 829 | Rb | 84 | 37 | 47 | 10 | 8.67622400E+00 | 8.68855138E+00 | -1.20E-02 | 7.81467540E+04 | 7.81456916E+04 | 1.10E+00 | 7.81657422E+04 | 7.81646798E+04 | 1.10E+00 | -7.97589600E+01 | -8.08213686E+01 | 1.10E+00 |
| 830 | Rb | 85 | 37 | 48 | 11 | 8.69744100E+00 | 8.70804754E+00 | -1.10E-02 | 7.90758397E+04 | 7.90749112E+04 | 9.30E-01 | 7.90948279E+04 | 7.90938994E+04 | 9.30E-01 | -8.21673301E+01 | -8.30957746E+01 | 9.30E-01 |
| 831 | Rb | 86 | 37 | 49 | 12 | 8.69690100E+00 | 8.69961140E+00 | -2.70E-03 | 8.00067541E+04 | 8.00064941E+04 | 2.60E-01 | 8.00257422E+04 | 8.00254823E+04 | 2.60E-01 | -8.27470100E+01 | -8.30069954E+01 | 2.60E-01 |
| 832 | Rb | 87 | 37 | 50 | 13 | 8.71098300E+00 | 8.70366505E+00 | 7.30E-03 | 8.09363973E+04 | 8.09370072E+04 | -6.10E-01 | 8.09553855E+04 | 8.09559954E+04 | -6.10E-01 | -8.45977900E+01 | -8.39879553E+01 | -6.10E-01 |
| 833 | Rb | 88 | 37 | 51 | 14 | 8.68111500E+00 | 8.67788692E+00 | 3.20E-03 | 8.18698802E+04 | 8.18701374E+04 | -2.60E-01 | 8.18888684E+04 | 8.18891255E+04 | -2.60E-01 | -8.26089940E+01 | -8.23518259E+01 | -2.60E-01 |
| 834 | Rb | 89 | 37 | 52 | 15 | 8.66418700E+00 | 8.66347493E+00 | 7.10E-04 | 8.28022710E+04 | 8.28023075E+04 | -3.70E-02 | 8.28212592E+04 | 8.28212957E+04 | -3.70E-02 | -8.17122390E+01 | -8.16757266E+01 | -3.70E-02 |
| 835 | Rb | 90 | 37 | 53 | 16 | 8.63151600E+00 | 8.62390378E+00 | 7.60E-03 | 8.37361126E+04 | 8.37367708E+04 | -6.60E-01 | 8.37551007E+04 | 8.37557590E+04 | -6.60E-01 | -7.93647300E+01 | -7.87064796E+01 | -6.60E-01 |
| 836 | Rb | 91 | 37 | 54 | 17 | 8.60756200E+00 | 8.59989603E+00 | 7.70E-03 | 8.46692263E+04 | 8.46698970E+04 | -6.70E-01 | 8.46882144E+04 | 8.46888852E+04 | -6.70E-01 | -7.77451260E+01 | -7.70743589E+01 | -6.70E-01 |
| 837 | Rb | 92 | 37 | 55 | 18 | 8.56942300E+00 | 8.55640648E+00 | 1.30E-02 | 8.56036929E+04 | 8.56048636E+04 | -1.20E+00 | 8.56226811E+04 | 8.56238517E+04 | -1.20E+00 | -7.47725240E+01 | -7.36018969E+01 | -1.20E+00 |
| 838 | Rb | 93 | 37 | 56 | 19 | 8.54092100E+00 | 8.53031799E+00 | 1.10E-02 | 8.65373396E+04 | 8.65382988E+04 | -9.60E-01 | 8.65563277E+04 | 8.65572869E+04 | -9.60E-01 | -7.26199530E+01 | -7.16607550E+01 | -9.60E-01 |
| 839 | Rb | 94 | 37 | 57 | 20 | 8.49276400E+00 | 8.48670843E+00 | 6.10E-03 | 8.74728908E+04 | 8.74734331E+04 | -5.40E-01 | 8.74918789E+04 | 8.74924213E+04 | -5.40E-01 | -6.85627890E+01 | -6.80204558E+01 | -5.40E-01 |

| | | | | | | | | | | | | | | | | |
|---|---|---|---|---|---|---|---|---|---|---|---|---|---|---|---|---|
| 840 | Rb | 95 | 37 | 58 | 21 | 8.46023400E+00 | 8.45947804E+00 | 7.60E-04 | 8.84070537E+04 | 8.84070987E+04 | -4.50E-02 | 8.84260419E+04 | 8.84260868E+04 | -4.50E-02 | -6.58938810E+01 | -6.58489574E+01 | -4.50E-02 |
| 841 | Rb | 96 | 37 | 59 | 22 | 8.40889600E+00 | 8.41500417E+00 | -6.10E-03 | 8.93430873E+04 | 8.93424741E+04 | 6.10E-01 | 8.93620755E+04 | 8.93614622E+04 | 6.10E-01 | -6.13543610E+01 | -6.19676254E+01 | 6.10E-01 |
| 842 | Rb | 97 | 37 | 60 | 23 | 8.37618600E+00 | 8.38542990E+00 | -9.20E-03 | 9.02774166E+04 | 9.02764931E+04 | 9.20E-01 | 9.02964048E+04 | 9.02954813E+04 | 9.20E-01 | -5.85191210E+01 | -5.94426064E+01 | 9.20E-01 |
| 843 | Rb | 98 | 37 | 61 | 24 | 8.33021000E+00 | 8.33895212E+00 | -8.70E-03 | 9.12131115E+04 | 9.12122279E+04 | 8.80E-01 | 9.12320996E+04 | 9.12312161E+04 | 8.80E-01 | -5.43183330E+01 | -5.52018952E+01 | 8.80E-01 |
| 844 | Rb | 99 | 37 | 62 | 25 | 8.29615100E+00 | 8.30632467E+00 | -1.00E-02 | 9.21477185E+04 | 9.21466845E+04 | 1.00E+00 | 9.21667066E+04 | 9.21656726E+04 | 1.00E+00 | -5.12054030E+01 | -5.22394107E+01 | 1.00E+00 |
| 845 | Rb | 100 | 37 | 63 | 26 | 8.24700000E+00 | 8.25749007E+00 | -1.00E-02 | 9.30838712E+04 | 9.30828270E+04 | 1.00E+00 | 9.31028593E+04 | 9.31018151E+04 | 1.00E+00 | -4.65470000E+01 | -4.75909557E+01 | 1.00E+00 |
| 846 | Rb | 101 | 37 | 64 | 27 | 8.20900000E+00 | 8.22182935E+00 | -1.30E-02 | 9.40190996E+04 | 9.40177366E+04 | 1.40E+00 | 9.40380878E+04 | 9.40367248E+04 | 1.40E+00 | -4.28120000E+01 | -4.41753945E+01 | 1.40E+00 |
| 847 | RB | 102 | 37 | 65 | 28 | 8.15700000E+00 | 8.17083201E+00 | -1.40E-02 | 9.49556992E+04 | 9.49542819E+04 | 1.40E+00 | 9.49746873E+04 | 9.49732700E+04 | 1.40E+00 | -3.77070000E+01 | -3.91241757E+01 | 1.40E+00 |
| 848 | RB | 103 | 37 | 66 | 29 | 8.11700000E+00 | 8.13252668E+00 | -1.60E-02 | 9.58912918E+04 | 9.58896219E+04 | 1.70E+00 | 9.59102800E+04 | 9.59086100E+04 | 1.70E+00 | -3.36080000E+01 | -3.52782402E+01 | 1.70E+00 |
| 849 | Sr | 73 | 38 | 35 | -3 | 8.10200000E+00 | 8.10236053E+00 | -3.60E-04 | 6.79476117E+04 | 6.79475390E+04 | 7.30E-02 | 6.79671162E+04 | 6.79670435E+04 | 7.30E-02 | -3.19500000E+01 | -3.20229575E+01 | 7.30E-02 |
| 850 | Sr | 74 | 38 | 36 | -2 | 8.22100000E+00 | 8.22243813E+00 | -1.40E-03 | 6.88702286E+04 | 6.88701162E+04 | 1.10E-01 | 6.88897331E+04 | 6.88896208E+04 | 1.10E-01 | -4.08270000E+01 | -4.09397416E+01 | 1.10E-01 |
| 851 | Sr | 75 | 38 | 37 | -1 | 8.29651100E+00 | 8.28697175E+00 | 9.50E-03 | 6.97959313E+04 | 6.97966192E+04 | -6.90E-01 | 6.98154359E+04 | 6.98161237E+04 | -6.90E-01 | -4.66186940E+01 | -4.59308825E+01 | -6.90E-01 |
| 852 | Sr | 76 | 38 | 38 | 0 | 8.39392900E+00 | 8.38283815E+00 | 1.10E-02 | 7.07197965E+04 | 7.07206117E+04 | -8.20E-01 | 7.07393010E+04 | 7.07401163E+04 | -8.20E-01 | -5.42476380E+01 | -5.34323817E+01 | -8.20E-01 |
| 853 | Sr | 77 | 38 | 39 | 1 | 8.43591800E+00 | 8.42180413E+00 | 1.40E-02 | 7.16477347E+04 | 7.16487939E+04 | -1.10E+00 | 7.16672393E+04 | 7.16682984E+04 | -1.10E+00 | -5.78034360E+01 | -5.67442807E+01 | -1.10E+00 |
| 854 | Sr | 78 | 38 | 40 | 2 | 8.50009600E+00 | 8.49672324E+00 | 3.40E-03 | 7.25738583E+04 | 7.25740938E+04 | -2.40E-01 | 7.25933628E+04 | 7.25935983E+04 | -2.40E-01 | -6.31739410E+01 | -6.29384567E+01 | -2.40E-01 |
| 855 | Sr | 79 | 38 | 41 | 3 | 8.52382000E+00 | 8.52023377E+00 | 3.60E-03 | 7.35030494E+04 | 7.35033051E+04 | -2.60E-01 | 7.35225539E+04 | 7.35228096E+04 | -2.60E-01 | -6.54768890E+01 | -6.52211928E+01 | -2.60E-01 |
| 856 | Sr | 80 | 38 | 42 | 4 | 8.57859600E+00 | 8.57886296E+00 | -2.70E-04 | 7.44297089E+04 | 7.44296599E+04 | 4.90E-02 | 7.44492134E+04 | 7.44491644E+04 | 4.90E-02 | -7.03114800E+01 | -7.03604428E+01 | 4.90E-02 |
| 857 | Sr | 81 | 38 | 43 | 5 | 8.58735400E+00 | 8.59185469E+00 | -4.50E-03 | 7.53599863E+04 | 7.53595941E+04 | 3.90E-01 | 7.53794908E+04 | 7.53790986E+04 | 3.90E-01 | -7.15281250E+01 | -7.19203171E+01 | 3.90E-01 |
| 858 | Sr | 82 | 38 | 44 | 6 | 8.63571800E+00 | 8.63914767E+00 | -3.40E-03 | 7.62869984E+04 | 7.62866896E+04 | 3.10E-01 | 7.63065030E+04 | 7.63061941E+04 | 3.10E-01 | -7.60100460E+01 | -7.63188773E+01 | 3.10E-01 |
| 859 | Sr | 83 | 38 | 45 | 7 | 8.63840700E+00 | 8.64554703E+00 | -7.10E-03 | 7.72177049E+04 | 7.72170847E+04 | 6.20E-01 | 7.72372094E+04 | 7.72365892E+04 | 6.20E-01 | -7.67976050E+01 | -7.74178522E+01 | 6.20E-01 |
| 860 | Sr | 84 | 38 | 46 | 8 | 8.67751200E+00 | 8.68480578E+00 | -7.30E-03 | 7.81453470E+04 | 7.81447068E+04 | 6.40E-01 | 7.81648516E+04 | 7.81642113E+04 | 6.40E-01 | -8.06495550E+01 | -8.12898156E+01 | 6.40E-01 |
| 861 | Sr | 85 | 38 | 47 | 9 | 8.67571800E+00 | 8.68591176E+00 | -1.00E-02 | 7.90763874E+04 | 7.90754933E+04 | 8.90E-01 | 7.90958919E+04 | 7.90949979E+04 | 8.90E-01 | -8.11032760E+01 | -8.19973110E+01 | 8.90E-01 |
| 862 | Sr | 86 | 38 | 48 | 10 | 8.70845700E+00 | 8.71612924E+00 | -7.70E-03 | 8.00044615E+04 | 8.00037741E+04 | 6.90E-01 | 8.00239660E+04 | 8.00232786E+04 | 6.90E-01 | -8.45231990E+01 | -8.52106065E+01 | 6.90E-01 |
| 863 | Sr | 87 | 38 | 49 | 11 | 8.70523500E+00 | 8.70743431E+00 | -2.20E-03 | 8.09355987E+04 | 8.09353798E+04 | 2.20E-01 | 8.09551033E+04 | 8.09548843E+04 | 2.20E-01 | -8.48800330E+01 | -8.50989576E+01 | 2.20E-01 |
| 864 | Sr | 88 | 38 | 50 | 12 | 8.73259200E+00 | 8.72182504E+00 | 1.10E-02 | 8.18640515E+04 | 8.18649714E+04 | -9.20E-01 | 8.18835560E+04 | 8.18844759E+04 | -9.20E-01 | -8.79213510E+01 | -8.70014575E+01 | -9.20E-01 |
| 865 | Sr | 89 | 38 | 51 | 13 | 8.70591900E+00 | 8.69649119E+00 | 9.40E-03 | 8.27972581E+04 | 8.27980696E+04 | -8.10E-01 | 8.28167627E+04 | 8.28175742E+04 | -8.10E-01 | -8.62087500E+01 | -8.53972506E+01 | -8.10E-01 |
| 866 | Sr | 90 | 38 | 52 | 14 | 8.69598100E+00 | 8.69239743E+00 | 3.60E-03 | 8.37290120E+04 | 8.37293070E+04 | -2.90E-01 | 8.37485166E+04 | 8.37488115E+04 | -2.90E-01 | -8.59489170E+01 | -8.56539850E+01 | -2.90E-01 |
| 867 | Sr | 91 | 38 | 53 | 15 | 8.66388000E+00 | 8.65371227E+00 | 1.00E-02 | 8.46628026E+04 | 8.46637003E+04 | -9.00E-01 | 8.46823071E+04 | 8.46832048E+04 | -9.00E-01 | -8.36524060E+01 | -8.27547136E+01 | -9.00E-01 |
| 868 | Sr | 92 | 38 | 54 | 16 | 8.64890700E+00 | 8.63943845E+00 | 9.50E-03 | 8.55950817E+04 | 8.55959252E+04 | -8.40E-01 | 8.56145862E+04 | 8.56154297E+04 | -8.40E-01 | -8.28673920E+01 | -8.20239152E+01 | -8.40E-01 |
| 869 | Sr | 93 | 38 | 55 | 17 | 8.61278800E+00 | 8.59663510E+00 | 1.60E-02 | 8.65293572E+04 | 8.65308318E+04 | -1.50E+00 | 8.65488617E+04 | 8.65503363E+04 | -1.50E+00 | -8.00859150E+01 | -7.86113234E+01 | -1.50E+00 |
| 870 | Sr | 94 | 38 | 56 | 18 | 8.59383400E+00 | 8.57940343E+00 | 1.40E-02 | 8.74620914E+04 | 8.74634203E+04 | -1.30E+00 | 8.74815960E+04 | 8.74829249E+04 | -1.30E+00 | -7.88457480E+01 | -7.75168626E+01 | -1.30E+00 |
| 871 | Sr | 95 | 38 | 57 | 19 | 8.54913900E+00 | 8.53619592E+00 | 1.30E-02 | 8.83973090E+04 | 8.83985110E+04 | -1.20E+00 | 8.84168135E+04 | 8.84180156E+04 | -1.20E+00 | -7.51222510E+01 | -7.39202338E+01 | -1.20E+00 |
| 872 | Sr | 96 | 38 | 58 | 20 | 8.52132400E+00 | 8.51708509E+00 | 4.20E-03 | 8.93309955E+04 | 8.93313748E+04 | -3.80E-01 | 8.93505000E+04 | 8.93508794E+04 | -3.80E-01 | -7.29298550E+01 | -7.25504704E+01 | -3.80E-01 |
| 873 | Sr | 97 | 38 | 59 | 21 | 8.47186400E+00 | 8.47282625E+00 | -9.60E-04 | 9.02668372E+04 | 9.02667163E+04 | 1.20E-01 | 9.02863417E+04 | 9.02862208E+04 | 1.20E-01 | -6.85822140E+01 | -6.87031294E+01 | 1.20E-01 |
| 874 | Sr | 98 | 38 | 60 | 22 | 8.44577400E+00 | 8.45076526E+00 | -5.00E-03 | 9.12004875E+04 | 9.11999708E+04 | 5.20E-01 | 9.12199920E+04 | 9.12194753E+04 | 5.20E-01 | -6.64259760E+01 | -6.69426598E+01 | 5.20E-01 |
| 875 | Sr | 99 | 38 | 61 | 23 | 8.40245600E+00 | 8.40433564E+00 | -1.90E-03 | 9.21358956E+04 | 9.21356819E+04 | 2.10E-01 | 9.21554001E+04 | 9.21551865E+04 | 2.10E-01 | -6.25119420E+01 | -6.27255736E+01 | 2.10E-01 |
| 876 | Sr | 100 | 38 | 62 | 24 | 8.37232700E+00 | 8.37867084E+00 | -6.30E-03 | 9.30700714E+04 | 9.30694095E+04 | 6.60E-01 | 9.30895759E+04 | 9.30889140E+04 | 6.60E-01 | -5.98301460E+01 | -6.04921103E+01 | 6.60E-01 |
| 877 | Sr | 101 | 38 | 63 | 25 | 8.32708800E+00 | 8.32970204E+00 | -2.60E-03 | 9.40058336E+04 | 9.40055420E+04 | 2.90E-01 | 9.40253381E+04 | 9.40250465E+04 | 2.90E-01 | -5.55619930E+01 | -5.58536126E+01 | 2.90E-01 |
| 878 | Sr | 102 | 38 | 64 | 26 | 8.29317200E+00 | 8.30050677E+00 | -7.30E-03 | 9.49405313E+04 | 9.49397556E+04 | 7.80E-01 | 9.49600359E+04 | 9.49592601E+04 | 7.80E-01 | -5.23583660E+01 | -5.31340783E+01 | 7.80E-01 |
| 879 | Sr | 103 | 38 | 65 | 27 | 8.24300000E+00 | 8.24920557E+00 | -6.20E-03 | 9.58769614E+04 | 9.58763045E+04 | 6.60E-01 | 9.58964659E+04 | 9.58958090E+04 | 6.60E-01 | -4.74220000E+01 | -4.80792425E+01 | 6.60E-01 |
| 880 | Sr | 104 | 38 | 66 | 28 | 8.21000000E+00 | 8.21696054E+00 | -7.00E-03 | 9.68117716E+04 | 9.68109742E+04 | 8.00E-01 | 9.68312761E+04 | 9.68304787E+04 | 8.00E-01 | -4.41060000E+01 | -4.49036465E+01 | 8.00E-01 |
| 881 | Sr | 105 | 38 | 67 | 29 | 8.15600000E+00 | 8.16381504E+00 | -7.80E-03 | 9.77487614E+04 | 9.77479029E+04 | 8.60E-01 | 9.77682660E+04 | 9.77674074E+04 | 8.60E-01 | -3.86100000E+01 | -3.94690106E+01 | 8.60E-01 |
| 882 | SR | 106 | 38 | 68 | 30 | 8.11900000E+00 | 8.12910133E+00 | -1.00E-02 | 9.86840756E+04 | 9.86829841E+04 | 1.10E+00 | 9.87035801E+04 | 9.87024886E+04 | 1.10E+00 | -3.47900000E+01 | -3.58818528E+01 | 1.10E+00 |
| 883 | SR | 107 | 38 | 69 | 31 | 8.06400000E+00 | 8.07460982E+00 | -1.10E-02 | 9.96214604E+04 | 9.96202509E+04 | 1.20E+00 | 9.96409649E+04 | 9.96397555E+04 | 1.20E+00 | -2.89000000E+01 | -3.01090438E+01 | 1.20E+00 |
| 884 | Y | 76 | 39 | 37 | -2 | 8.17800000E+00 | 8.17251613E+00 | 5.50E-03 | 7.07349264E+04 | 7.07352965E+04 | -3.70E-01 | 7.07549475E+04 | 7.07553177E+04 | -3.70E-01 | -3.86010000E+01 | -3.82309848E+01 | -3.70E-01 |
| 885 | Y | 77 | 39 | 38 | -1 | 8.28300000E+00 | 8.27150578E+00 | 1.10E-02 | 7.16582429E+04 | 7.16590672E+04 | -8.20E-01 | 7.16782640E+04 | 7.16790883E+04 | -8.20E-01 | -4.67790000E+01 | -4.59543854E+01 | -8.20E-01 |
| 886 | Y | 78 | 39 | 39 | 0 | 8.35400000E+00 | 8.32859982E+00 | 2.50E-02 | 7.25839887E+04 | 7.25859077E+04 | -1.90E+00 | 7.26040098E+04 | 7.26059289E+04 | -1.90E+00 | -5.25270000E+01 | -5.06079069E+01 | -1.90E+00 |
| 887 | Y | 79 | 39 | 40 | 1 | 8.42379000E+00 | 8.40582924E+00 | 1.80E-02 | 7.35096528E+04 | 7.35110434E+04 | -1.40E+00 | 7.35296739E+04 | 7.35310645E+04 | -1.40E+00 | -5.83568890E+01 | -5.69663118E+01 | -1.40E+00 |
| 888 | Y | 80 | 39 | 41 | 2 | 8.45425900E+00 | 8.44513546E+00 | 9.10E-03 | 7.44383569E+04 | 7.44390584E+04 | -7.00E-01 | 7.44583780E+04 | 7.44590796E+04 | -7.00E-01 | -6.11469030E+01 | -6.04453202E+01 | -7.00E-01 |
| 889 | Y | 81 | 39 | 42 | 3 | 8.50588800E+00 | 8.50517872E+00 | 7.10E-04 | 7.53652861E+04 | 7.53653152E+04 | -2.90E-02 | 7.53853072E+04 | 7.53853363E+04 | -2.90E-02 | -6.57117260E+01 | -6.56826399E+01 | -2.90E-02 |
| 890 | Y | 82 | 39 | 43 | 4 | 8.52926200E+00 | 8.53204276E+00 | -2.80E-03 | 7.62944289E+04 | 7.62941725E+04 | 2.60E-01 | 7.63144500E+04 | 7.63141937E+04 | 2.60E-01 | -6.80629620E+01 | -6.83193509E+01 | 2.60E-01 |
| 891 | Y | 83 | 39 | 44 | 5 | 8.57364300E+00 | 8.58021400E+00 | -6.60E-03 | 7.72217814E+04 | 7.72212077E+04 | 5.70E-01 | 7.72418025E+04 | 7.72412288E+04 | 5.70E-01 | -7.22045240E+01 | -7.27782882E+01 | 5.70E-01 |
| 892 | Y | 84 | 39 | 45 | 6 | 8.58776900E+00 | 8.59900250E+00 | -1.10E-02 | 7.81515865E+04 | 7.81506146E+04 | 9.70E-01 | 7.81716077E+04 | 7.81706357E+04 | 9.70E-01 | -7.38934760E+01 | -7.48654167E+01 | 9.70E-01 |
| 893 | Y | 85 | 39 | 46 | 7 | 8.62814800E+00 | 8.63879363E+00 | -1.10E-02 | 7.90791319E+04 | 7.90781987E+04 | 9.30E-01 | 7.90991531E+04 | 7.90982198E+04 | 9.30E-01 | -7.78421230E+01 | -7.87753463E+01 | 9.30E-01 |
| 894 | Y | 86 | 39 | 47 | 8 | 8.63843000E+00 | 8.65104114E+00 | -1.30E-02 | 8.00091849E+04 | 8.00080720E+04 | 1.10E+00 | 8.00292060E+04 | 8.00280931E+04 | 1.10E+00 | -7.92831990E+01 | -8.03961067E+01 | 1.10E+00 |
| 895 | Y | 87 | 39 | 48 | 9 | 8.67484400E+00 | 8.68159885E+00 | -6.80E-03 | 8.09369438E+04 | 8.09363278E+04 | 6.20E-01 | 8.09569650E+04 | 8.09563490E+04 | 6.20E-01 | -8.30183430E+01 | -8.36343499E+01 | 6.20E-01 |
| 896 | Y | 88 | 39 | 49 | 10 | 8.68253600E+00 | 8.68321843E+00 | -6.80E-04 | 8.18671575E+04 | 8.18670691E+04 | 8.80E-02 | 8.18871786E+04 | 8.18870902E+04 | 8.80E-02 | -8.42987510E+01 | -8.43871525E+01 | 8.80E-02 |
| 897 | Y | 89 | 39 | 50 | 11 | 8.71398700E+00 | 8.69829493E+00 | 1.60E-02 | 8.27952412E+04 | 8.27966094E+04 | -1.40E+00 | 8.28152623E+04 | 8.28166306E+04 | -1.40E+00 | -8.77091510E+01 | -8.63408608E+01 | -1.40E+00 |
| 898 | Y | 90 | 39 | 51 | 12 | 8.69335400E+00 | 8.68314168E+00 | 1.00E-02 | 8.37279495E+04 | 8.37288403E+04 | -8.90E-01 | 8.37479706E+04 | 8.37488614E+04 | -8.90E-01 | -8.64948600E+01 | -8.56040444E+01 | -8.90E-01 |
| 899 | Y | 91 | 39 | 52 | 13 | 8.68494700E+00 | 8.68041249E+00 | 4.50E-03 | 8.46595866E+04 | 8.46599709E+04 | -3.80E-01 | 8.46796077E+04 | 8.46799920E+04 | -3.80E-01 | -8.63518770E+01 | -8.59675106E+01 | -3.80E-01 |
| 900 | Y | 92 | 39 | 53 | 14 | 8.66159500E+00 | 8.65168952E+00 | 9.90E-03 | 8.55926154E+04 | 8.55934984E+04 | -8.80E-01 | 8.56126365E+04 | 8.56135195E+04 | -8.80E-01 | -8.48170540E+01 | -8.39340907E+01 | -8.80E-01 |
| 901 | Y | 93 | 39 | 54 | 15 | 8.64891000E+00 | 8.63890221E+00 | 1.00E-02 | 8.65246989E+04 | 8.65256013E+04 | -9.00E-01 | 8.65447200E+04 | 8.65456224E+04 | -9.00E-01 | -8.42276560E+01 | -8.33252415E+01 | -9.00E-01 |
| 902 | Y | 94 | 39 | 55 | 16 | 8.62282100E+00 | 8.60532381E+00 | 1.70E-02 | 8.74580678E+04 | 8.74596842E+04 | -1.60E+00 | 8.74780889E+04 | 8.74797053E+04 | -1.60E+00 | -8.23528420E+01 | -8.07364550E+01 | -1.60E+00 |
| 903 | Y | 95 | 39 | 56 | 17 | 8.60499900E+00 | 8.58933240E+00 | 1.60E-02 | 8.83907033E+04 | 8.83921634E+04 | -1.50E+00 | 8.84107245E+04 | 8.84121845E+04 | -1.50E+00 | -8.12113190E+01 | -7.97512765E+01 | -1.50E+00 |
| 904 | Y | 96 | 39 | 57 | 18 | 8.56954700E+00 | 8.55458347E+00 | 1.50E-02 | 8.93250671E+04 | 8.93264753E+04 | -1.40E+00 | 8.93450883E+04 | 8.93464965E+04 | -1.40E+00 | -7.83415930E+01 | -7.69333926E+01 | -1.40E+00 |
| 905 | Y | 97 | 39 | 58 | 19 | 8.54158200E+00 | 8.53652162E+00 | 5.10E-03 | 9.02587756E+04 | 9.02592381E+04 | -4.60E-01 | 9.02787967E+04 | 9.02792593E+04 | -4.60E-01 | -7.61271790E+01 | -7.56646576E+01 | -4.60E-01 |
| 906 | Y | 98 | 39 | 59 | 20 | 8.49773600E+00 | 8.50011858E+00 | -2.40E-03 | 9.11940963E+04 | 9.11938345E+04 | 2.60E-01 | 9.12141174E+04 | 9.12138556E+04 | 2.60E-01 | -7.23005900E+01 | -7.25623621E+01 | 2.60E-01 |
| 907 | Y | 99 | 39 | 60 | 21 | 8.47681400E+00 | 8.47898029E+00 | -2.20E-03 | 9.21272352E+04 | 9.21269924E+04 | 2.40E-01 | 9.21472563E+04 | 9.21470136E+04 | 2.40E-01 | -7.06557020E+01 | -7.08984711E+01 | 2.40E-01 |
| 908 | Y | 100 | 39 | 61 | 22 | 8.43953500E+00 | 8.43988884E+00 | -3.50E-04 | 9.30620516E+04 | 9.30619880E+04 | 6.40E-02 | 9.30820728E+04 | 9.30820091E+04 | 6.40E-02 | -6.73333300E+01 | -6.73969872E+01 | 6.40E-02 |
| 909 | Y | 101 | 39 | 62 | 23 | 8.41345100E+00 | 8.41500234E+00 | -1.60E-03 | 9.39958120E+04 | 9.39956270E+04 | 1.90E-01 | 9.40158332E+04 | 9.40156481E+04 | 1.90E-01 | -6.50669930E+01 | -6.52520204E+01 | 1.90E-01 |
| 910 | Y | 102 | 39 | 63 | 24 | 8.37192400E+00 | 8.37288169E+00 | -9.60E-04 | 9.49311997E+04 | 9.49310737E+04 | 1.30E-01 | 9.49512209E+04 | 9.49510948E+04 | 1.30E-01 | -6.11733660E+01 | -6.12993972E+01 | 1.30E-01 |
| 911 | Y | 103 | 39 | 64 | 25 | 8.34264000E+00 | 8.34428279E+00 | -1.60E-03 | 9.58654094E+04 | 9.58652119E+04 | 2.00E-01 | 9.58854305E+04 | 9.58852330E+04 | 2.00E-01 | -5.84577650E+01 | -5.86552730E+01 | 2.00E-01 |
| 912 | Y | 104 | 39 | 65 | 26 | 8.29800000E+00 | 8.29937359E+00 | -1.40E-03 | 9.68012973E+04 | 9.68011035E+04 | 1.90E-01 | 9.68213184E+04 | 9.68211247E+04 | 1.90E-01 | -5.40640000E+01 | -5.42576804E+01 | 1.90E-01 |
| 913 | Y | 105 | 39 | 66 | 27 | 8.26500000E+00 | 8.26753226E+00 | -2.50E-03 | 9.77360330E+04 | 9.77357129E+04 | 3.20E-01 | 9.77560541E+04 | 9.77557340E+04 | 3.20E-01 | -5.08220000E+01 | -5.11423956E+01 | 3.20E-01 |
| 914 | Y | 106 | 39 | 67 | 28 | 8.21800000E+00 | 8.22038945E+00 | -2.40E-03 | 9.86722963E+04 | 9.86720079E+04 | 2.90E-01 | 9.86923174E+04 | 9.86920290E+04 | 2.90E-01 | -4.60530000E+01 | -4.63414712E+01 | 2.90E-01 |
| 915 | Y | 107 | 39 | 68 | 29 | 8.18200000E+00 | 8.18591384E+00 | -3.90E-03 | 9.96074791E+04 | 9.96070418E+04 | 4.40E-01 | 9.96275002E+04 | 9.96270629E+04 | 4.40E-01 | -4.23640000E+01 | -4.28016505E+01 | 4.40E-01 |
| 916 | Y | 108 | 39 | 69 | 30 | 8.13400000E+00 | 8.13714032E+00 | -3.10E-03 | 1.00544040E+05 | 1.00543689E+05 | 3.50E-01 | 1.00564062E+05 | 1.00563710E+05 | 3.50E-01 | -3.72970000E+01 | -3.76487057E+01 | 3.50E-01 |
| 917 | Y | 109 | 39 | 70 | 31 | 8.09600000E+00 | 8.10075304E+00 | -4.80E-03 | 1.01479631E+05 | 1.01479083E+05 | 5.50E-01 | 1.01499652E+05 | 1.01499104E+05 | 5.50E-01 | -3.32000000E+01 | -3.37483138E+01 | 5.50E-01 |
| 918 | ZR | 78 | 40 | 38 | -2 | 8.20000000E+00 | 8.18856036E+00 | 1.10E-02 | 7.25946964E+04 | 7.25955309E+04 | -8.30E-01 | 7.26152343E+04 | 7.26160689E+04 | -8.30E-01 | -4.13020000E+01 | -4.04679063E+01 | -8.30E-01 |
| 919 | Zr | 79 | 40 | 39 | -1 | 8.27100000E+00 | 8.25252555E+00 | 1.80E-02 | 7.35204338E+04 | 7.35218545E+04 | -1.40E+00 | 7.35409717E+04 | 7.35423924E+04 | -1.40E+00 | -4.70590000E+01 | -4.56383976E+01 | -1.40E+00 |
| 920 | Zr | 80 | 40 | 40 | 0 | 8.37410700E+00 | 8.34695588E+00 | 2.70E-02 | 7.44434699E+04 | 7.44456129E+04 | -2.10E+00 | 7.44640078E+04 | 7.44661508E+04 | -2.10E+00 | -5.55170430E+01 | -5.33740301E+01 | -2.10E+00 |
| 921 | Zr | 81 | 40 | 41 | 1 | 8.40591600E+00 | 8.38799521E+00 | 1.80E-02 | 7.53720846E+04 | 7.53735072E+04 | -1.40E+00 | 7.53926225E+04 | 7.53940451E+04 | -1.40E+00 | -5.83963910E+01 | -5.69738530E+01 | -1.40E+00 |
| 922 | Zr | 82 | 40 | 42 | 2 | 8.46900000E+00 | 8.46288519E+00 | 6.10E-03 | 7.62980317E+04 | 7.62985436E+04 | -5.10E-01 | 7.63185697E+04 | 7.63190815E+04 | -5.10E-01 | -6.39430000E+01 | -6.34315074E+01 | -5.10E-01 |
| 923 | Zr | 83 | 40 | 43 | 3 | 8.48838500E+00 | 8.49034034E+00 | -2.00E-03 | 7.72275586E+04 | 7.72273673E+04 | 1.90E-01 | 7.72480966E+04 | 7.72479052E+04 | 1.90E-01 | -6.59105130E+01 | -6.61018512E+01 | 1.90E-01 |
| 924 | Zr | 84 | 40 | 44 | 4 | 8.54901600E+00 | 8.55146562E+00 | -2.40E-03 | 7.81535427E+04 | 7.81533078E+04 | 2.30E-01 | 7.81740806E+04 | 7.81738457E+04 | 2.30E-01 | -7.14205330E+01 | -7.16553961E+01 | 2.30E-01 |
| 925 | Zr | 85 | 40 | 45 | 5 | 8.56402500E+00 | 8.57031562E+00 | -6.30E-03 | 7.90832832E+04 | 7.90827195E+04 | 5.60E-01 | 7.91038212E+04 | 7.91032574E+04 | 5.60E-01 | -7.31740250E+01 | -7.37377930E+01 | 5.60E-01 |
| 926 | Zr | 86 | 40 | 46 | 6 | 8.61404700E+00 | 8.62156518E+00 | -7.50E-03 | 8.00099827E+04 | 8.00093071E+04 | 6.80E-01 | 8.00305206E+04 | 8.00298450E+04 | 6.80E-01 | -7.79686520E+01 | -7.86442515E+01 | 6.80E-01 |
| 927 | Zr | 87 | 40 | 47 | 7 | 8.62364700E+00 | 8.63352318E+00 | -9.90E-03 | 8.09400989E+04 | 8.09392106E+04 | 8.90E-01 | 8.09606368E+04 | 8.09597485E+04 | 8.90E-01 | -7.93465330E+01 | -8.02348438E+01 | 8.90E-01 |
| 928 | Zr | 88 | 40 | 48 | 8 | 8.66602700E+00 | 8.67436088E+00 | -8.30E-03 | 8.18673111E+04 | 8.18665487E+04 | 7.60E-01 | 8.18878490E+04 | 8.18870866E+04 | 7.60E-01 | -8.36283290E+01 | -8.43907659E+01 | 7.60E-01 |
| 929 | Zr | 89 | 40 | 49 | 9 | 8.67336800E+00 | 8.67567291E+00 | -2.30E-03 | 8.27975571E+04 | 8.27973229E+04 | 2.30E-01 | 8.28180950E+04 | 8.28178609E+04 | 2.30E-01 | -8.48763900E+01 | -8.51105783E+01 | 2.30E-01 |
| 930 | Zr | 90 | 40 | 50 | 10 | 8.70998000E+00 | 8.70048219E+00 | 9.50E-03 | 8.37251540E+04 | 8.37259798E+04 | -8.30E-01 | 8.37456919E+04 | 8.37465177E+04 | -8.30E-01 | -8.87735580E+01 | -8.79477672E+01 | -8.30E-01 |
| 931 | Zr | 91 | 40 | 51 | 11 | 8.69332000E+00 | 8.68559730E+00 | 7.70E-03 | 8.46575255E+04 | 8.46581992E+04 | -6.70E-01 | 8.46780634E+04 | 8.46787371E+04 | -6.70E-01 | -8.78961580E+01 | -8.72224049E+01 | -6.70E-01 |
| 932 | Zr | 92 | 40 | 52 | 12 | 8.69268400E+00 | 8.69250509E+00 | 1.80E-04 | 8.55884561E+04 | 8.55884435E+04 | 1.30E-02 | 8.56089940E+04 | 8.56089814E+04 | 1.30E-02 | -8.84596300E+01 | -8.84721999E+01 | 1.30E-02 |
| 933 | Zr | 93 | 40 | 53 | 13 | 8.67162700E+00 | 8.66453468E+00 | 7.10E-03 | 8.65212871E+04 | 8.65219176E+04 | -6.30E-01 | 8.65418250E+04 | 8.65424555E+04 | -6.30E-01 | -8.71226830E+01 | -8.64921387E+01 | -6.30E-01 |
| 934 | Zr | 94 | 40 | 54 | 14 | 8.66681800E+00 | 8.66089929E+00 | 5.90E-03 | 8.74526329E+04 | 8.74531602E+04 | -5.30E-01 | 8.74731708E+04 | 8.74736981E+04 | -5.30E-01 | -8.72709010E+01 | -8.67436272E+01 | -5.30E-01 |
| 935 | Zr | 95 | 40 | 55 | 15 | 8.64360900E+00 | 8.62801877E+00 | 1.60E-02 | 8.83857363E+04 | 8.83871883E+04 | -1.50E+00 | 8.84062742E+04 | 8.84077262E+04 | -1.50E+00 | -8.56615570E+01 | -8.42095585E+01 | -1.50E+00 |
| 936 | Zr | 96 | 40 | 56 | 16 | 8.63538700E+00 | 8.62041353E+00 | 1.50E-02 | 8.93174473E+04 | 8.93188558E+04 | -1.40E+00 | 8.93379852E+04 | 8.93393937E+04 | -1.40E+00 | -8.54446100E+01 | -8.40361552E+01 | -1.40E+00 |
| 937 | Zr | 97 | 40 | 57 | 17 | 8.60383800E+00 | 8.58619667E+00 | 1.80E-02 | 9.02514376E+04 | 9.02531198E+04 | -1.70E+00 | 9.02719755E+04 | 9.02736577E+04 | -1.70E+00 | -8.29484170E+01 | -8.12662143E+01 | -1.70E+00 |
| 938 | Zr | 98 | 40 | 58 | 18 | 8.58150700E+00 | 8.57586849E+00 | 5.60E-03 | 9.11845875E+04 | 9.11851111E+04 | -5.20E-01 | 9.12051255E+04 | 9.12056490E+04 | -5.20E-01 | -8.12925140E+01 | -8.07689301E+01 | -5.20E-01 |
| 939 | Zr | 99 | 40 | 59 | 19 | 8.53930300E+00 | 8.53992408E+00 | -6.20E-04 | 9.21197497E+04 | 9.21196591E+04 | 9.10E-02 | 9.21402876E+04 | 9.21401971E+04 | 9.10E-02 | -7.76244260E+01 | -7.77149833E+01 | 9.10E-02 |
| 940 | Zr | 100 | 40 | 60 | 20 | 8.52219800E+00 | 8.52600211E+00 | -3.80E-03 | 9.30524862E+04 | 9.30520768E+04 | 4.10E-01 | 9.30730241E+04 | 9.30726147E+04 | 4.10E-01 | -7.63819710E+01 | -7.67913910E+01 | 4.10E-01 |
| 941 | Zr | 101 | 40 | 61 | 21 | 8.48594000E+00 | 8.48732006E+00 | -1.40E-03 | 9.39871915E+04 | 9.39870231E+04 | 1.70E-01 | 9.40077294E+04 | 9.40075610E+04 | 1.70E-01 | -7.31707500E+01 | -7.33391869E+01 | 1.70E-01 |
| 942 | Zr | 102 | 40 | 62 | 22 | 8.46641400E+00 | 8.46917729E+00 | -2.80E-03 | 9.49202625E+04 | 9.49199517E+04 | 3.10E-01 | 9.49408004E+04 | 9.49404896E+04 | 3.10E-01 | -7.15937620E+01 | -7.19046254E+01 | 3.10E-01 |
| 943 | Zr | 103 | 40 | 63 | 23 | 8.42595400E+00 | 8.42737364E+00 | -1.40E-03 | 9.58555289E+04 | 9.58553537E+04 | 1.80E-01 | 9.58760669E+04 | 9.58758916E+04 | 1.80E-01 | -6.78214460E+01 | -6.79967078E+01 | 1.80E-01 |
| 944 | Zr | 104 | 40 | 64 | 24 | 8.40243600E+00 | 8.40505241E+00 | -2.60E-03 | 9.67891142E+04 | 9.67888131E+04 | 3.00E-01 | 9.68096521E+04 | 9.68093510E+04 | 3.00E-01 | -6.57302300E+01 | -6.60313542E+01 | 3.00E-01 |
| 945 | Zr | 105 | 40 | 65 | 25 | 8.35871800E+00 | 8.36031750E+00 | -1.60E-03 | 9.77248675E+04 | 9.77246706E+04 | 2.00E-01 | 9.77454054E+04 | 9.77452085E+04 | 2.00E-01 | -6.14709620E+01 | -6.16679230E+01 | 2.00E-01 |
| 946 | Zr | 106 | 40 | 66 | 26 | 8.33200000E+00 | 8.33430465E+00 | -2.30E-03 | 9.86589249E+04 | 9.86586330E+04 | 2.90E-01 | 9.86794628E+04 | 9.86791709E+04 | 2.90E-01 | -5.89080000E+01 | -5.91995584E+01 | 2.90E-01 |
| 947 | Zr | 107 | 40 | 67 | 27 | 8.28600000E+00 | 8.28716760E+00 | -1.20E-03 | 9.95950578E+04 | 9.95949077E+04 | 1.50E-01 | 9.96155957E+04 | 9.96154456E+04 | 1.50E-01 | -5.42690000E+01 | -5.44188804E+01 | 1.50E-01 |
| 948 | Zr | 108 | 40 | 68 | 28 | 8.25700000E+00 | 8.25811211E+00 | -1.10E-03 | 1.00529467E+05 | 1.00529324E+05 | 1.40E-01 | 1.00550005E+05 | 1.00549862E+05 | 1.40E-01 | -5.13530000E+01 | -5.14967355E+01 | 1.40E-01 |
| 949 | Zr | 109 | 40 | 69 | 29 | 8.20800000E+00 | 8.20917879E+00 | -1.20E-03 | 1.01466122E+05 | 1.01465965E+05 | 1.60E-01 | 1.01486660E+05 | 1.01486503E+05 | 1.60E-01 | -4.61930000E+01 | -4.63497976E+01 | 1.60E-01 |
| 950 | Zr | 110 | 40 | 70 | 30 | 8.17700000E+00 | 8.17787502E+00 | -8.80E-04 | 1.02400923E+05 | 1.02400765E+05 | 1.60E-01 | 1.02421461E+05 | 1.02421302E+05 | 1.60E-01 | -4.28860000E+01 | -4.30442418E+01 | 1.60E-01 |
| 951 | ZR | 111 | 40 | 71 | 31 | 8.12800000E+00 | 8.12796757E+00 | 3.20E-05 | 1.03337743E+05 | 1.03337692E+05 | 5.10E-02 | 1.03358281E+05 | 1.03358230E+05 | 5.10E-02 | -3.75600000E+01 | -3.76110714E+01 | 5.10E-02 |
| 952 | ZR | 112 | 40 | 72 | 32 | 8.09400000E+00 | 8.09544416E+00 | -1.40E-03 | 1.04272986E+05 | 1.04272772E+05 | 2.10E-01 | 1.04293524E+05 | 1.04293310E+05 | 2.10E-01 | -3.38100000E+01 | -3.40250982E+01 | 2.20E-01 |
| 953 | Nb | 81 | 41 | 40 | -1 | 8.25500000E+00 | 8.23931547E+00 | 1.60E-02 | 7.53830167E+04 | 7.53842502E+04 | -1.20E+00 | 7.54040716E+04 | 7.54053051E+04 | -1.20E+00 | -4.69470000E+01 | -4.57138707E+01 | -1.20E+00 |
| 954 | Nb | 82 | 41 | 41 | 0 | 8.31700000E+00 | 8.29688191E+00 | 2.00E-02 | 7.63092572E+04 | 7.63108558E+04 | -1.60E+00 | 7.63303121E+04 | 7.63319107E+04 | -1.60E+00 | -5.22010000E+01 | -5.06023158E+01 | -1.60E+00 |
| 955 | Nb | 83 | 41 | 42 | 1 | 8.38859800E+00 | 8.37414791E+00 | 1.40E-02 | 7.72345416E+04 | 7.72357112E+04 | -1.20E+00 | 7.72555965E+04 | 7.72567661E+04 | -1.20E+00 | -5.84105130E+01 | -5.72409564E+01 | -1.20E+00 |
| 956 | Nb | 84 | 41 | 43 | 2 | 8.41600000E+00 | 8.41587325E+00 | 1.30E-04 | 7.81634259E+04 | 7.81633975E+04 | 2.80E-02 | 7.81844808E+04 | 7.81844524E+04 | 2.80E-02 | -6.10210000E+01 | -6.10487144E+01 | 2.80E-02 |
| 957 | Nb | 85 | 41 | 44 | 3 | 8.47371100E+00 | 8.47846916E+00 | -4.80E-03 | 7.90896606E+04 | 7.90892264E+04 | 4.30E-01 | 7.91107155E+04 | 7.91102813E+04 | 4.30E-01 | -6.62796750E+01 | -6.67139206E+01 | 4.30E-01 |
| 958 | Nb | 86 | 41 | 45 | 4 | 8.50220900E+00 | 8.50972303E+00 | -7.50E-03 | 8.00183015E+04 | 8.00176254E+04 | 6.80E-01 | 8.00393564E+04 | 8.00386803E+04 | 6.80E-01 | -6.91328790E+01 | -6.98089039E+01 | 6.80E-01 |
| 959 | Nb | 87 | 41 | 46 | 5 | 8.55174300E+00 | 8.56181444E+00 | -1.00E-02 | 8.09450551E+04 | 8.09441491E+04 | 9.10E-01 | 8.09661100E+04 | 8.09652040E+04 | 9.10E-01 | -7.38732990E+01 | -7.47792601E+01 | 9.10E-01 |
| 960 | Nb | 88 | 41 | 47 | 6 | 8.57245100E+00 | 8.58469647E+00 | -1.20E-02 | 8.18742465E+04 | 8.18731391E+04 | 1.10E+00 | 8.18953014E+04 | 8.18941940E+04 | 1.10E+00 | -7.61759870E+01 | -7.72833740E+01 | 1.10E+00 |
| 961 | Nb | 89 | 41 | 48 | 7 | 8.61681400E+00 | 8.62601482E+00 | -9.20E-03 | 8.28012911E+04 | 8.28004424E+04 | 8.50E-01 | 8.28223460E+04 | 8.28214973E+04 | 8.50E-01 | -8.06253950E+01 | -8.14740848E+01 | 8.50E-01 |
| 962 | Nb | 90 | 41 | 49 | 8 | 8.63338400E+00 | 8.63725163E+00 | -3.90E-03 | 8.37307484E+04 | 8.37303705E+04 | 3.80E-01 | 8.37518033E+04 | 8.37514254E+04 | 3.80E-01 | -8.26622220E+01 | -8.30400941E+01 | 3.80E-01 |
| 963 | Nb | 91 | 41 | 50 | 9 | 8.67090400E+00 | 8.66272076E+00 | 8.20E-03 | 8.46582661E+04 | 8.46589809E+04 | -7.10E-01 | 8.46793210E+04 | 8.46800358E+04 | -7.10E-01 | -8.66385750E+01 | -8.59237174E+01 | -7.10E-01 |
| 964 | Nb | 92 | 41 | 51 | 10 | 8.66237700E+00 | 8.65744763E+00 | 4.90E-03 | 8.55899450E+04 | 8.55903687E+04 | -4.20E-01 | 8.56109999E+04 | 8.56114236E+04 | -4.20E-01 | -8.64537450E+01 | -8.60299911E+01 | -4.20E-01 |
| 965 | Nb | 93 | 41 | 52 | 11 | 8.66418600E+00 | 8.66558744E+00 | -1.40E-03 | 8.65206798E+04 | 8.65205196E+04 | 1.60E-01 | 8.65417347E+04 | 8.65415746E+04 | 1.60E-01 | -8.72129840E+01 | -8.73731220E+01 | 1.60E-01 |
| 966 | Nb | 94 | 41 | 53 | 12 | 8.64890200E+00 | 8.64697803E+00 | 1.90E-03 | 8.74530176E+04 | 8.74531687E+04 | -1.50E-01 | 8.74740725E+04 | 8.74742236E+04 | -1.50E-01 | -8.63692070E+01 | -8.62181059E+01 | -1.50E-01 |
| 967 | Nb | 95 | 41 | 54 | 13 | 8.64720000E+00 | 8.64477147E+00 | 2.40E-03 | 8.83840958E+04 | 8.83842967E+04 | -2.00E-01 | 8.84051507E+04 | 8.84053517E+04 | -2.00E-01 | -8.67850570E+01 | -8.65841413E+01 | -2.00E-01 |
| 968 | Nb | 96 | 41 | 55 | 14 | 8.62892800E+00 | 8.62060445E+00 | 8.30E-03 | 8.93167681E+04 | 8.93175374E+04 | -7.70E-01 | 8.93378230E+04 | 8.93385923E+04 | -7.70E-01 | -8.56068270E+01 | -8.48375606E+01 | -7.70E-01 |
| 969 | Nb | 97 | 41 | 56 | 15 | 8.62319200E+00 | 8.61430001E+00 | 8.90E-03 | 9.02482609E+04 | 9.02490937E+04 | -8.30E-01 | 9.02693158E+04 | 9.02701486E+04 | -8.30E-01 | -8.56080880E+01 | -8.47753147E+01 | -8.30E-01 |
| 970 | Nb | 98 | 41 | 57 | 16 | 8.59636000E+00 | 8.58810636E+00 | 8.30E-03 | 9.11818327E+04 | 9.11826118E+04 | -7.80E-01 | 9.12028876E+04 | 9.12036667E+04 | -7.80E-01 | -8.35303650E+01 | -8.27513185E+01 | -7.80E-01 |
| 971 | Nb | 99 | 41 | 58 | 17 | 8.57894900E+00 | 8.57898917E+00 | -4.00E-05 | 9.21145253E+04 | 9.21144916E+04 | 3.40E-02 | 9.21355802E+04 | 9.21355465E+04 | 3.40E-02 | -8.23317950E+01 | -8.23655036E+01 | 3.40E-02 |
| 972 | Nb | 100 | 41 | 59 | 18 | 8.54858800E+00 | 8.55053446E+00 | -1.90E-03 | 9.30485479E+04 | 9.30483235E+04 | 2.20E-01 | 9.30696028E+04 | 9.30693784E+04 | 2.20E-01 | -7.98033340E+01 | -8.00277026E+01 | 2.20E-01 |
| 973 | Nb | 101 | 41 | 60 | 19 | 8.53480000E+00 | 8.53781020E+00 | -3.00E-03 | 9.39809573E+04 | 9.39806235E+04 | 3.30E-01 | 9.40020122E+04 | 9.40016784E+04 | 3.30E-01 | -7.88879900E+01 | -7.92217683E+01 | 3.30E-01 |
| 974 | Nb | 102 | 41 | 61 | 20 | 8.50498800E+00 | 8.50618139E+00 | -1.20E-03 | 9.49150287E+04 | 9.49148772E+04 | 1.50E-01 | 9.49360836E+04 | 9.49359321E+04 | 1.50E-01 | -7.63105990E+01 | -7.64621209E+01 | 1.50E-01 |
| 975 | Nb | 103 | 41 | 62 | 21 | 8.48829700E+00 | 8.48921349E+00 | -9.20E-04 | 9.58478082E+04 | 9.58476841E+04 | 1.20E-01 | 9.58688631E+04 | 9.58687390E+04 | 1.20E-01 | -7.50250990E+01 | -7.51492894E+01 | 1.20E-01 |
| 976 | Nb | 104 | 41 | 63 | 22 | 8.45351900E+00 | 8.45404716E+00 | -5.30E-04 | 9.67825023E+04 | 9.67824176E+04 | 8.50E-02 | 9.68035572E+04 | 9.68034725E+04 | 8.50E-02 | -7.18251820E+01 | -7.19098852E+01 | 8.50E-02 |
| 977 | Nb | 105 | 41 | 64 | 23 | 8.43165700E+00 | 8.43282078E+00 | -1.20E-03 | 9.77159096E+04 | 9.77157577E+04 | 1.50E-01 | 9.77369645E+04 | 9.77368126E+04 | 1.50E-01 | -6.99119120E+01 | -7.00638440E+01 | 1.50E-01 |
| 978 | Nb | 106 | 41 | 65 | 24 | 8.39323700E+00 | 8.39430441E+00 | -1.10E-03 | 9.86511159E+04 | 9.86509729E+04 | 1.40E-01 | 9.86721708E+04 | 9.86720279E+04 | 1.40E-01 | -6.61996830E+01 | -6.63426107E+01 | 1.40E-01 |
| 979 | Nb | 107 | 41 | 66 | 25 | 8.36705400E+00 | 8.36924937E+00 | -2.20E-03 | 9.95850896E+04 | 9.95848249E+04 | 2.60E-01 | 9.96061445E+04 | 9.96058798E+04 | 2.60E-01 | -6.37201000E+01 | -6.39847068E+01 | 2.60E-01 |
| 980 | Nb | 108 | 41 | 67 | 26 | 8.32566700E+00 | 8.32791950E+00 | -2.30E-03 | 1.00520758E+05 | 1.00520485E+05 | 2.70E-01 | 1.00541813E+05 | 1.00541540E+05 | 2.70E-01 | -5.95459650E+01 | -5.98190109E+01 | 2.70E-01 |
| 981 | Nb | 109 | 41 | 68 | 27 | 8.29648900E+00 | 8.29964215E+00 | -3.20E-03 | 1.01455178E+05 | 1.01454804E+05 | 3.70E-01 | 1.01476233E+05 | 1.01475859E+05 | 3.70E-01 | -5.66199310E+01 | -5.69933808E+01 | 3.70E-01 |
| 982 | Nb | 110 | 41 | 69 | 28 | 8.25400000E+00 | 8.25612407E+00 | -2.10E-03 | 1.02391156E+05 | 1.02390857E+05 | 3.00E-01 | 1.02412211E+05 | 1.02411912E+05 | 3.00E-01 | -5.21360000E+01 | -5.24347143E+01 | 3.00E-01 |
| 983 | Nb | 111 | 41 | 70 | 29 | 8.22300000E+00 | 8.22539688E+00 | -2.40E-03 | 1.03325910E+05 | 1.03325577E+05 | 3.30E-01 | 1.03346965E+05 | 1.03346632E+05 | 3.30E-01 | -4.88750000E+01 | -4.92088012E+01 | 3.30E-01 |
| 984 | Nb | 112 | 41 | 71 | 30 | 8.18000000E+00 | 8.18055388E+00 | -5.50E-04 | 1.04262006E+05 | 1.04261939E+05 | 6.70E-02 | 1.04283061E+05 | 1.04282994E+05 | 6.70E-02 | -4.42740000E+01 | -4.43404631E+01 | 6.60E-02 |
| 985 | Nb | 113 | 41 | 72 | 31 | 8.14600000E+00 | 8.14841157E+00 | -2.40E-03 | 1.05197263E+05 | 1.05196956E+05 | 3.10E-01 | 1.05218318E+05 | 1.05218011E+05 | 3.10E-01 | -4.05110000E+01 | -4.08176176E+01 | 3.10E-01 |
| 986 | NB | 114 | 41 | 73 | 32 | 8.10000000E+00 | 8.10341002E+00 | -3.40E-03 | 1.06133881E+05 | 1.06133504E+05 | 3.80E-01 | 1.06154936E+05 | 1.06154558E+05 | 3.80E-01 | -3.53870000E+01 | -3.57645337E+01 | 3.80E-01 |
| 987 | Nb | 115 | 41 | 74 | 33 | 8.06500000E+00 | 8.07119227E+00 | -6.20E-03 | 1.07069408E+05 | 1.07068671E+05 | 7.40E-01 | 1.07090463E+05 | 1.07089725E+05 | 7.40E-01 | -3.13540000E+01 | -3.20915828E+01 | 7.40E-01 |
| 988 | Mo | 83 | 42 | 41 | -1 | 8.23800000E+00 | 8.22265950E+00 | 1.50E-02 | 7.72457485E+04 | 7.72469844E+04 | -1.20E+00 | 7.72673206E+04 | 7.72685566E+04 | -1.20E+00 | -4.66860000E+01 | -4.54504959E+01 | -1.20E+00 |
| 989 | Mo | 84 | 42 | 42 | 0 | 8.32900000E+00 | 8.31546198E+00 | 1.40E-02 | 7.81694273E+04 | 7.81705318E+04 | -1.10E+00 | 7.81909994E+04 | 7.81921039E+04 | -1.10E+00 | -5.45020000E+01 | -5.33972444E+01 | -1.10E+00 |
| 990 | Mo | 85 | 42 | 43 | 1 | 8.36133100E+00 | 8.35900221E+00 | 2.30E-03 | 7.90979133E+04 | 7.90980808E+04 | -1.70E-01 | 7.91194854E+04 | 7.91196529E+04 | -1.70E-01 | -5.75097520E+01 | -5.73423065E+01 | -1.70E-01 |
| 991 | Mo | 86 | 42 | 44 | 2 | 8.43470900E+00 | 8.43488049E+00 | -1.70E-04 | 8.00228069E+04 | 8.00227616E+04 | 4.50E-02 | 8.00443790E+04 | 8.00443337E+04 | 4.50E-02 | -6.41102440E+01 | -6.41555224E+01 | 4.50E-02 |
| 992 | Mo | 87 | 42 | 45 | 3 | 8.46242300E+00 | 8.46684910E+00 | -4.40E-03 | 8.09515264E+04 | 8.09511108E+04 | 4.20E-01 | 8.09730985E+04 | 8.09726830E+04 | 4.20E-01 | -6.68848080E+01 | -6.73003523E+01 | 4.20E-01 |
| 993 | Mo | 88 | 42 | 46 | 4 | 8.52390800E+00 | 8.53041059E+00 | -6.50E-03 | 8.18772187E+04 | 8.18766160E+04 | 6.00E-01 | 8.18987908E+04 | 8.18981881E+04 | 6.00E-01 | -7.26865430E+01 | -7.32892937E+01 | 6.00E-01 |
| 994 | Mo | 89 | 42 | 47 | 5 | 8.54498400E+00 | 8.55340704E+00 | -8.40E-03 | 8.28063844E+04 | 8.28056042E+04 | 7.80E-01 | 8.28279565E+04 | 8.28271764E+04 | 7.80E-01 | -7.50149340E+01 | -7.57950700E+01 | 7.80E-01 |
| 995 | Mo | 90 | 42 | 48 | 6 | 8.59703200E+00 | 8.60483316E+00 | -7.80E-03 | 8.37327205E+04 | 8.37319879E+04 | 7.30E-01 | 8.37542926E+04 | 8.37535600E+04 | 7.30E-01 | -8.01728860E+01 | -8.09055086E+01 | 7.30E-01 |
| 996 | Mo | 91 | 42 | 49 | 7 | 8.61362700E+00 | 8.61598249E+00 | -2.40E-03 | 8.46621787E+04 | 8.46619338E+04 | 2.40E-01 | 8.46837508E+04 | 8.46835059E+04 | 2.40E-01 | -8.22087120E+01 | -8.24536111E+01 | 2.40E-01 |
| 997 | Mo | 92 | 42 | 50 | 8 | 8.65772200E+00 | 8.65084174E+00 | 6.90E-03 | 8.55890737E+04 | 8.55896762E+04 | -6.00E-01 | 8.56106458E+04 | 8.56112483E+04 | -6.00E-01 | -8.68078270E+01 | -8.62053260E+01 | -6.00E-01 |
| 998 | Mo | 93 | 42 | 51 | 9 | 8.65140100E+00 | 8.64585793E+00 | 5.50E-03 | 8.65205692E+04 | 8.65210542E+04 | -4.80E-01 | 8.65421414E+04 | 8.65426263E+04 | -4.80E-01 | -8.68063200E+01 | -8.63213545E+01 | -4.80E-01 |
| 999 | Mo | 94 | 42 | 52 | 10 | 8.66232000E+00 | 8.66316357E+00 | -8.40E-04 | 8.74504568E+04 | 8.74503470E+04 | 1.10E-01 | 8.74720289E+04 | 8.74719191E+04 | 1.10E-01 | -8.84128430E+01 | -8.85226240E+01 | 1.10E-01 |
| 1000 | Mo | 95 | 42 | 53 | 11 | 8.64870700E+00 | 8.64525692E+00 | 3.50E-03 | 8.83826530E+04 | 8.83829503E+04 | -3.00E-01 | 8.84042252E+04 | 8.84045225E+04 | -3.00E-01 | -8.77106240E+01 | -8.74133362E+01 | -3.00E-01 |
| 1001 | Mo | 96 | 42 | 54 | 12 | 8.65397400E+00 | 8.65177488E+00 | 2.20E-03 | 8.93130641E+04 | 8.93132447E+04 | -1.80E-01 | 8.93346362E+04 | 8.93348169E+04 | -1.80E-01 | -8.87936270E+01 | -8.86129988E+01 | -1.80E-01 |
| 1002 | Mo | 97 | 42 | 55 | 13 | 8.63508000E+00 | 8.62832197E+00 | 6.80E-03 | 9.02458082E+04 | 9.02464333E+04 | -6.30E-01 | 9.02673804E+04 | 9.02680054E+04 | -6.30E-01 | -8.75435590E+01 | -8.69185223E+01 | -6.30E-01 |
| 1003 | Mo | 98 | 42 | 56 | 14 | 8.63515700E+00 | 8.63007563E+00 | 5.10E-03 | 9.11767310E+04 | 9.11771985E+04 | -4.70E-01 | 9.11983031E+04 | 9.11987706E+04 | -4.70E-01 | -8.81148420E+01 | -8.76473841E+01 | -4.70E-01 |
| 1004 | Mo | 99 | 42 | 57 | 15 | 8.60778600E+00 | 8.60449916E+00 | 3.30E-03 | 9.21103709E+04 | 9.21106658E+04 | -2.90E-01 | 9.21319431E+04 | 9.21322380E+04 | -2.90E-01 | -8.59689700E+01 | -8.56740698E+01 | -2.90E-01 |
| 1005 | Mo | 100 | 42 | 58 | 16 | 8.60462600E+00 | 8.60286540E+00 | 1.80E-03 | 9.30416445E+04 | 9.30417901E+04 | -1.50E-01 | 9.30632166E+04 | 9.30633622E+04 | -1.50E-01 | -8.61894800E+01 | -8.60438744E+01 | -1.50E-01 |
| 1006 | Mo | 101 | 42 | 59 | 17 | 8.57288000E+00 | 8.57501799E+00 | -2.10E-03 | 9.39758116E+04 | 9.39755652E+04 | 2.50E-01 | 9.39973837E+04 | 9.39971373E+04 | 2.50E-01 | -8.35164030E+01 | -8.37628315E+01 | 2.50E-01 |
| 1007 | Mo | 102 | 42 | 60 | 18 | 8.56849300E+00 | 8.56933902E+00 | -8.50E-04 | 9.49072516E+04 | 9.49071348E+04 | 1.20E-01 | 9.49288237E+04 | 9.49287069E+04 | 1.20E-01 | -8.35704620E+01 | -8.36872762E+01 | 1.20E-01 |
| 1008 | Mo | 103 | 42 | 61 | 19 | 8.53838700E+00 | 8.53835027E+00 | 3.70E-05 | 9.58413494E+04 | 9.58413227E+04 | 2.70E-02 | 9.58629215E+04 | 9.58628948E+04 | 2.70E-02 | -8.09667830E+01 | -8.09934551E+01 | 2.70E-02 |
| 1009 | Mo | 104 | 42 | 62 | 20 | 8.52802300E+00 | 8.52804423E+00 | -2.10E-05 | 9.67734543E+04 | 9.67734216E+04 | 3.30E-02 | 9.67950265E+04 | 9.67949937E+04 | 3.30E-02 | -8.03559130E+01 | -8.03886583E+01 | 3.30E-02 |
| 1010 | Mo | 105 | 42 | 63 | 21 | 8.49498100E+00 | 8.49351874E+00 | 1.50E-03 | 9.77079610E+04 | 9.77080841E+04 | -1.20E-01 | 9.77295332E+04 | 9.77296562E+04 | -1.20E-01 | -7.73432350E+01 | -7.72202069E+01 | -1.20E-01 |
| 1011 | Mo | 106 | 42 | 64 | 22 | 8.47964000E+00 | 8.47857058E+00 | 1.10E-03 | 9.86406576E+04 | 9.86407404E+04 | -8.30E-02 | 9.86622297E+04 | 9.86623126E+04 | -8.30E-02 | -7.61408190E+01 | -7.60579015E+01 | -8.30E-02 |
| 1012 | Mo | 107 | 42 | 65 | 23 | 8.44233900E+00 | 8.44063488E+00 | 1.70E-03 | 9.95757345E+04 | 9.95758864E+04 | -1.50E-01 | 9.95973066E+04 | 9.95974585E+04 | -1.50E-01 | -7.25579080E+01 | -7.24060333E+01 | -1.50E-01 |
| 1013 | Mo | 108 | 42 | 66 | 24 | 8.42227800E+00 | 8.42146490E+00 | 8.10E-04 | 1.00509024E+05 | 1.00509081E+05 | -5.70E-02 | 1.00530596E+05 | 1.00530654E+05 | -5.70E-02 | -7.07623480E+01 | -7.07049912E+01 | -5.70E-02 |
| 1014 | Mo | 109 | 42 | 67 | 25 | 8.38153600E+00 | 8.38060067E+00 | 9.40E-04 | 1.01444608E+05 | 1.01444680E+05 | -7.10E-02 | 1.01466180E+05 | 1.01466252E+05 | -7.20E-02 | -6.66724650E+01 | -6.66009362E+01 | -7.20E-02 |
| 1015 | Mo | 110 | 42 | 68 | 26 | 8.35941300E+00 | 8.35781053E+00 | 1.60E-03 | 1.02378226E+05 | 1.02378371E+05 | -1.50E-01 | 1.02399798E+05 | 1.02399943E+05 | -1.50E-01 | -6.45491120E+01 | -6.44033020E+01 | -1.50E-01 |
| 1016 | Mo | 111 | 42 | 69 | 27 | 8.31527300E+00 | 8.31459356E+00 | 6.80E-04 | 1.03314331E+05 | 1.03314376E+05 | -4.50E-02 | 1.03335903E+05 | 1.03335948E+05 | -4.50E-02 | -5.99376740E+01 | -5.98927105E+01 | -4.50E-02 |
| 1017 | Mo | 112 | 42 | 70 | 28 | 8.29100000E+00 | 8.28896074E+00 | 2.00E-03 | 1.04248299E+05 | 1.04248498E+05 | -2.00E-01 | 1.04269871E+05 | 1.04270070E+05 | -2.00E-01 | -5.74640000E+01 | -5.72651091E+01 | -2.00E-01 |
| 1018 | Mo | 113 | 42 | 71 | 29 | 8.24800000E+00 | 8.24421938E+00 | 3.80E-03 | 1.05184488E+05 | 1.05184830E+05 | -3.40E-01 | 1.05206060E+05 | 1.05206402E+05 | -3.40E-01 | -5.27690000E+01 | -5.24269764E+01 | -3.40E-01 |
| 1019 | Mo | 114 | 42 | 72 | 30 | 8.22000000E+00 | 8.21679466E+00 | 3.20E-03 | 1.06118944E+05 | 1.06119277E+05 | -3.30E-01 | 1.06140516E+05 | 1.06140849E+05 | -3.30E-01 | -4.98070000E+01 | -4.94734590E+01 | -3.30E-01 |
| 1020 | Mo | 115 | 42 | 73 | 31 | 8.17500000E+00 | 8.17167958E+00 | 3.30E-03 | 1.07055496E+05 | 1.07055814E+05 | -3.20E-01 | 1.07077068E+05 | 1.07077386E+05 | -3.20E-01 | -4.47490000E+01 | -4.44307007E+01 | -3.20E-01 |
| 1021 | MO | 116 | 42 | 74 | 32 | 8.14600000E+00 | 8.14385180E+00 | 2.10E-03 | 1.07990239E+05 | 1.07990436E+05 | -2.00E-01 | 1.08011811E+05 | 1.08012008E+05 | -2.00E-01 | -4.15000000E+01 | -4.13030388E+01 | -2.00E-01 |
| 1022 | MO | 117 | 42 | 75 | 33 | 8.10000000E+00 | 8.09975151E+00 | 2.50E-04 | 1.08927063E+05 | 1.08927017E+05 | 4.60E-02 | 1.08948635E+05 | 1.08948589E+05 | 4.60E-02 | -3.61700000E+01 | -3.62158376E+01 | 4.60E-02 |
| 1023 | Tc | 85 | 43 | 42 | -1 | 8.21700000E+00 | 8.21189141E+00 | 5.10E-03 | 7.91088712E+04 | 7.91092847E+04 | -4.10E-01 | 7.91309607E+04 | 7.91313742E+04 | -4.10E-01 | -4.60340000E+01 | -4.56209662E+01 | -4.10E-01 |
| 1024 | Tc | 86 | 43 | 43 | 0 | 8.27700000E+00 | 8.27025097E+00 | 6.70E-03 | 8.00351023E+04 | 8.00356192E+04 | -5.20E-01 | 8.00571919E+04 | 8.00577088E+04 | -5.20E-01 | -5.12970000E+01 | -5.07804607E+01 | -5.20E-01 |
| 1025 | Tc | 87 | 43 | 44 | 1 | 8.34774400E+00 | 8.34847363E+00 | -7.30E-04 | 8.09602037E+04 | 8.09601090E+04 | 9.50E-02 | 8.09822933E+04 | 8.09821985E+04 | 9.50E-02 | -5.76900440E+01 | -5.77847635E+01 | 9.50E-02 |
| 1026 | Tc | 88 | 43 | 45 | 2 | 8.38995800E+00 | 8.39308671E+00 | -3.10E-03 | 8.18877065E+04 | 8.18873999E+04 | 3.10E-01 | 8.19097961E+04 | 8.19094895E+04 | 3.10E-01 | -6.16813140E+01 | -6.19878694E+01 | 3.10E-01 |
| 1027 | Tc | 89 | 43 | 46 | 3 | 8.45057500E+00 | 8.45805565E+00 | -7.50E-03 | 8.28134870E+04 | 8.28127900E+04 | 7.00E-01 | 8.28355766E+04 | 8.28348796E+04 | 7.00E-01 | -6.73948480E+01 | -6.80918726E+01 | 7.00E-01 |
| 1028 | Tc | 90 | 43 | 47 | 4 | 8.48335900E+00 | 8.49196246E+00 | -8.60E-03 | 8.37416512E+04 | 8.37408457E+04 | 8.10E-01 | 8.37637408E+04 | 8.37629353E+04 | 8.10E-01 | -7.07246860E+01 | -7.15302220E+01 | 8.10E-01 |
| 1029 | Tc | 91 | 43 | 48 | 5 | 8.53665100E+00 | 8.54420998E+00 | -7.60E-03 | 8.46678837E+04 | 8.46671646E+04 | 7.20E-01 | 8.46899733E+04 | 8.46892542E+04 | 7.20E-01 | -7.59862530E+01 | -7.67053903E+01 | 7.20E-01 |
| 1030 | Tc | 92 | 43 | 49 | 6 | 8.56354300E+00 | 8.56510450E+00 | -1.60E-03 | 8.55964384E+04 | 8.55962635E+04 | 1.70E-01 | 8.56185279E+04 | 8.56183530E+04 | 1.70E-01 | -7.89256920E+01 | -7.91005767E+01 | 1.70E-01 |
| 1031 | Tc | 93 | 43 | 50 | 7 | 8.60856900E+00 | 8.60075712E+00 | 7.80E-03 | 8.65232528E+04 | 8.65239481E+04 | -7.00E-01 | 8.65453423E+04 | 8.65460376E+04 | -7.00E-01 | -8.36053570E+01 | -8.29100562E+01 | -7.00E-01 |
| 1032 | Tc | 94 | 43 | 51 | 8 | 8.60872300E+00 | 8.60503168E+00 | 3.70E-03 | 8.74541951E+04 | 8.74545109E+04 | -3.20E-01 | 8.74762846E+04 | 8.74766004E+04 | -3.20E-01 | -8.41570950E+01 | -8.38413029E+01 | -3.20E-01 |
| 1033 | Tc | 95 | 43 | 52 | 9 | 8.62267700E+00 | 8.62352742E+00 | -8.50E-04 | 8.83838261E+04 | 8.83837141E+04 | 1.10E-01 | 8.84059157E+04 | 8.84058037E+04 | 1.10E-01 | -8.60201060E+01 | -8.61321110E+01 | 1.10E-01 |
| 1034 | Tc | 96 | 43 | 53 | 10 | 8.61485300E+00 | 8.61456830E+00 | 2.80E-04 | 8.93155199E+04 | 8.93155161E+04 | 3.80E-03 | 8.93376095E+04 | 8.93376056E+04 | 3.80E-03 | -8.58203850E+01 | -8.58242439E+01 | 3.90E-03 |
| 1035 | Tc | 97 | 43 | 54 | 11 | 8.62366700E+00 | 8.62246775E+00 | 1.20E-03 | 9.02456155E+04 | 9.02457006E+04 | -8.50E-02 | 9.02677051E+04 | 9.02677902E+04 | -8.50E-02 | -8.72188510E+01 | -8.71337394E+01 | -8.50E-02 |
| 1036 | Tc | 98 | 43 | 55 | 12 | 8.60999300E+00 | 8.60739568E+00 | 2.60E-03 | 9.11778973E+04 | 9.11781206E+04 | -2.20E-01 | 9.11999869E+04 | 9.12002102E+04 | -2.20E-01 | -8.64311250E+01 | -8.62078257E+01 | -2.20E-01 |
| 1037 | Tc | 99 | 43 | 56 | 13 | 8.61359900E+00 | 8.61047725E+00 | 3.10E-03 | 9.21084957E+04 | 9.21087735E+04 | -2.80E-01 | 9.21305853E+04 | 9.21308631E+04 | -2.80E-01 | -8.73267760E+01 | -8.70489780E+01 | -2.80E-01 |
| 1038 | Tc | 100 | 43 | 57 | 14 | 8.59510700E+00 | 8.59268036E+00 | 2.40E-03 | 9.30412967E+04 | 9.30415081E+04 | -2.10E-01 | 9.30633862E+04 | 9.30635977E+04 | -2.10E-01 | -8.60198590E+01 | -8.58084468E+01 | -2.10E-01 |
| 1039 | Tc | 101 | 43 | 58 | 15 | 8.59310100E+00 | 8.59235217E+00 | 7.50E-04 | 9.39724696E+04 | 9.39725139E+04 | -4.40E-02 | 9.39945591E+04 | 9.39946035E+04 | -4.40E-02 | -8.63410570E+01 | -8.62966611E+01 | -4.40E-02 |
| 1040 | Tc | 102 | 43 | 59 | 16 | 8.57062800E+00 | 8.57182019E+00 | -1.20E-03 | 9.49057341E+04 | 9.49055812E+04 | 1.50E-01 | 9.49278236E+04 | 9.49276708E+04 | 1.50E-01 | -8.45705890E+01 | -8.47234323E+01 | 1.50E-01 |
| 1041 | Tc | 103 | 43 | 60 | 17 | 8.56608400E+00 | 8.56750664E+00 | -1.40E-03 | 9.58371968E+04 | 9.58370191E+04 | 1.80E-01 | 9.58592864E+04 | 9.58591086E+04 | 1.80E-01 | -8.46018890E+01 | -8.47796383E+01 | 1.80E-01 |
| 1042 | Tc | 104 | 43 | 61 | 18 | 8.54118500E+00 | 8.54346707E+00 | -2.30E-03 | 9.67707856E+04 | 9.67705171E+04 | 2.70E-01 | 9.67928752E+04 | 9.67926066E+04 | 2.70E-01 | -8.25071850E+01 | -8.27757106E+01 | 2.70E-01 |
| 1043 | Tc | 105 | 43 | 62 | 19 | 8.53467000E+00 | 8.53459518E+00 | 7.50E-05 | 9.77024939E+04 | 9.77024705E+04 | 2.30E-02 | 9.77245835E+04 | 9.77245601E+04 | 2.30E-02 | -8.22929470E+01 | -8.23163095E+01 | 2.30E-02 |
| 1044 | Tc | 106 | 43 | 63 | 20 | 8.50654800E+00 | 8.50667522E+00 | -1.30E-04 | 9.86365055E+04 | 9.86364608E+04 | 4.50E-02 | 9.86585951E+04 | 9.86585504E+04 | 4.50E-02 | -7.97753840E+01 | -7.98200708E+01 | 4.50E-02 |
| 1045 | Tc | 107 | 43 | 64 | 21 | 8.49287800E+00 | 8.49318205E+00 | -3.00E-04 | 9.95690270E+04 | 9.95689633E+04 | 6.40E-02 | 9.95911166E+04 | 9.95910529E+04 | 6.40E-02 | -7.87479020E+01 | -7.88116572E+01 | 6.40E-02 |
| 1046 | Tc | 108 | 43 | 65 | 22 | 8.46279700E+00 | 8.46149577E+00 | 1.30E-03 | 1.00503348E+05 | 1.00503458E+05 | -1.10E-01 | 1.00525438E+05 | 1.00525547E+05 | -1.10E-01 | -7.59207450E+01 | -7.58114019E+01 | -1.10E-01 |
| 1047 | Tc | 109 | 43 | 66 | 23 | 8.44416000E+00 | 8.44374036E+00 | 4.20E-04 | 1.01436482E+05 | 1.01436497E+05 | -1.40E-02 | 1.01458572E+05 | 1.01458586E+05 | -1.40E-02 | -7.42807250E+01 | -7.42662387E+01 | -1.40E-02 |
| 1048 | Tc | 110 | 43 | 67 | 24 | 8.41124000E+00 | 8.40875840E+00 | 2.50E-03 | 1.02371225E+05 | 1.02371467E+05 | -2.40E-01 | 1.02393314E+05 | 1.02393556E+05 | -2.40E-01 | -7.10324410E+01 | -7.07906450E+01 | -2.40E-01 |
| 1049 | Tc | 111 | 43 | 68 | 25 | 8.39007100E+00 | 8.38728245E+00 | 2.80E-03 | 1.03304729E+05 | 1.03305007E+05 | -2.80E-01 | 1.03326818E+05 | 1.03327097E+05 | -2.80E-01 | -6.90225350E+01 | -6.87442534E+01 | -2.80E-01 |
| 1050 | Tc | 112 | 43 | 69 | 26 | 8.35358600E+00 | 8.34956810E+00 | 4.00E-03 | 1.04239990E+05 | 1.04240409E+05 | -4.20E-01 | 1.04262080E+05 | 1.04262499E+05 | -4.20E-01 | -6.52550550E+01 | -6.48362101E+01 | -4.20E-01 |
| 1051 | Tc | 113 | 43 | 70 | 27 | 8.32946400E+00 | 8.32509549E+00 | 4.40E-03 | 1.05173928E+05 | 1.05174390E+05 | -4.60E-01 | 1.05196017E+05 | 1.05196480E+05 | -4.60E-01 | -6.28115390E+01 | -6.23490547E+01 | -4.60E-01 |
| 1052 | Tc | 114 | 43 | 71 | 28 | 8.29200000E+00 | 8.28547191E+00 | 6.50E-03 | 1.06109465E+05 | 1.06110148E+05 | -6.80E-01 | 1.06131555E+05 | 1.06132237E+05 | -6.80E-01 | -5.87680000E+01 | -5.80857424E+01 | -6.80E-01 |
| 1053 | Tc | 115 | 43 | 72 | 29 | 8.26500000E+00 | 8.25900323E+00 | 6.00E-03 | 1.07043819E+05 | 1.07044471E+05 | -6.50E-01 | 1.07065909E+05 | 1.07066561E+05 | -6.50E-01 | -5.59080000E+01 | -5.52559980E+01 | -6.50E-01 |
| 1054 | Tc | 116 | 43 | 73 | 30 | 8.22500000E+00 | 8.21861766E+00 | 6.40E-03 | 1.07979766E+05 | 1.07980463E+05 | -7.00E-01 | 1.08001855E+05 | 1.08002552E+05 | -7.00E-01 | -5.14560000E+01 | -5.07589558E+01 | -7.00E-01 |
| 1055 | Tc | 117 | 43 | 74 | 31 | 8.19700000E+00 | 8.19150463E+00 | 5.50E-03 | 1.08914334E+05 | 1.08914982E+05 | -6.50E-01 | 1.08936423E+05 | 1.08937071E+05 | -6.50E-01 | -4.83820000E+01 | -4.77340294E+01 | -6.50E-01 |
| 1056 | Tc | 118 | 43 | 75 | 32 | 8.15700000E+00 | 8.15177259E+00 | 5.20E-03 | 1.09850420E+05 | 1.09851044E+05 | -6.20E-01 | 1.09872510E+05 | 1.09873133E+05 | -6.20E-01 | -4.37900000E+01 | -4.31658342E+01 | -6.20E-01 |
| 1057 | TC | 119 | 43 | 76 | 33 | 8.12800000E+00 | 8.12536024E+00 | 2.60E-03 | 1.10785333E+05 | 1.10785600E+05 | -2.70E-01 | 1.10807422E+05 | 1.10807690E+05 | -2.70E-01 | -4.03710000E+01 | -4.01032187E+01 | -2.70E-01 |
| 1058 | TC | 120 | 43 | 77 | 34 | 8.08700000E+00 | 8.08713299E+00 | -1.30E-04 | 1.11721680E+05 | 1.11721628E+05 | 5.20E-02 | 1.11743769E+05 | 1.11743717E+05 | 5.20E-02 | -3.55180000E+01 | -3.55699903E+01 | 5.20E-02 |
| 1059 | Ru | 88 | 44 | 44 | 0 | 8.29800000E+00 | 8.29163440E+00 | 6.40E-03 | 8.18944709E+04 | 8.18950270E+04 | -5.60E-01 | 8.19170781E+04 | 8.19176342E+04 | -5.60E-01 | -5.43990000E+01 | -5.38431433E+01 | -5.60E-01 |
| 1060 | Ru | 89 | 44 | 45 | 1 | 8.33700000E+00 | 8.33809362E+00 | -1.10E-03 | 8.28222576E+04 | 8.28221659E+04 | 9.20E-02 | 8.28448648E+04 | 8.28447731E+04 | 9.20E-02 | -5.81070000E+01 | -5.81983290E+01 | 9.10E-02 |
| 1061 | Ru | 90 | 44 | 46 | 2 | 8.40976800E+00 | 8.41454379E+00 | -4.80E-03 | 8.37469745E+04 | 8.37465127E+04 | 4.60E-01 | 8.37695817E+04 | 8.37691199E+04 | 4.60E-01 | -6.48837920E+01 | -6.53456195E+01 | 4.60E-01 |
| 1062 | Ru | 91 | 44 | 47 | 3 | 8.44292500E+00 | 8.44922320E+00 | -6.30E-03 | 8.46751128E+04 | 8.46745077E+04 | 6.10E-01 | 8.46977200E+04 | 8.46971149E+04 | 6.10E-01 | -6.82395180E+01 | -6.88446700E+01 | 6.10E-01 |
| 1063 | Ru | 92 | 44 | 48 | 4 | 8.50477300E+00 | 8.51141432E+00 | -6.60E-03 | 8.56005452E+04 | 8.55999023E+04 | 6.40E-01 | 8.56231524E+04 | 8.56225095E+04 | 6.40E-01 | -7.43012010E+01 | -7.49441571E+01 | 6.40E-01 |
| 1064 | Ru | 93 | 44 | 49 | 5 | 8.53146200E+00 | 8.53264204E+00 | -1.20E-03 | 8.65291237E+04 | 8.65289820E+04 | 1.40E-01 | 8.65517309E+04 | 8.65515892E+04 | 1.40E-01 | -7.72167120E+01 | -7.73584310E+01 | 1.40E-01 |
| 1065 | Ru | 94 | 44 | 50 | 6 | 8.58366100E+00 | 8.57738187E+00 | 6.30E-03 | 8.74552509E+04 | 8.74558092E+04 | -5.60E-01 | 8.74778581E+04 | 8.74784164E+04 | -5.60E-01 | -8.25835910E+01 | -8.20252979E+01 | -5.60E-01 |
| 1066 | Ru | 95 | 44 | 51 | 7 | 8.58745700E+00 | 8.58212385E+00 | 5.30E-03 | 8.83858721E+04 | 8.83863467E+04 | -4.70E-01 | 8.84084793E+04 | 8.84089539E+04 | -4.70E-01 | -8.34564970E+01 | -8.29818485E+01 | -4.70E-01 |
| 1067 | Ru | 96 | 44 | 52 | 8 | 8.60939900E+00 | 8.60936459E+00 | 3.40E-05 | 8.93147435E+04 | 8.93147149E+04 | 2.90E-02 | 8.93373507E+04 | 8.93373221E+04 | 2.90E-02 | -8.60791230E+01 | -8.61077649E+01 | 2.90E-02 |
| 1068 | Ru | 97 | 44 | 53 | 9 | 8.60426600E+00 | 8.60114117E+00 | 3.10E-03 | 9.02461974E+04 | 9.02464686E+04 | -2.70E-01 | 9.02688046E+04 | 9.02690758E+04 | -2.70E-01 | -8.61193030E+01 | -8.58481386E+01 | -2.70E-01 |
| 1069 | Ru | 98 | 44 | 54 | 10 | 8.62031200E+00 | 8.61738029E+00 | 2.90E-03 | 9.11755860E+04 | 9.11758414E+04 | -2.60E-01 | 9.11981932E+04 | 9.11984486E+04 | -2.60E-01 | -8.82247650E+01 | -8.79693943E+01 | -2.60E-01 |
| 1070 | Ru | 99 | 44 | 55 | 11 | 8.60867700E+00 | 8.60305685E+00 | 5.60E-03 | 9.21076830E+04 | 9.21082074E+04 | -5.20E-01 | 9.21302902E+04 | 9.21308146E+04 | -5.20E-01 | -8.76218380E+01 | -8.70974350E+01 | -5.20E-01 |
| 1071 | Ru | 100 | 44 | 56 | 12 | 8.61932300E+00 | 8.61393409E+00 | 5.40E-03 | 9.30375750E+04 | 9.30380820E+04 | -5.10E-01 | 9.30601822E+04 | 9.30606892E+04 | -5.10E-01 | -8.92238460E+01 | -8.87168972E+01 | -5.10E-01 |
| 1072 | Ru | 101 | 44 | 57 | 13 | 8.60133000E+00 | 8.59682026E+00 | 4.50E-03 | 9.39703384E+04 | 9.39707619E+04 | -4.20E-01 | 9.39929456E+04 | 9.39933691E+04 | -4.20E-01 | -8.79545770E+01 | -8.75310154E+01 | -4.20E-01 |
| 1073 | Ru | 102 | 44 | 58 | 14 | 8.60739200E+00 | 8.60382868E+00 | 3.60E-03 | 9.49006841E+04 | 9.49010156E+04 | -3.30E-01 | 9.49232913E+04 | 9.49236228E+04 | -3.30E-01 | -8.91029000E+01 | -8.87713754E+01 | -3.30E-01 |
| 1074 | Ru | 103 | 44 | 59 | 15 | 8.58433100E+00 | 8.58399324E+00 | 3.40E-04 | 9.58340174E+04 | 9.58340202E+04 | -2.80E-03 | 9.58566246E+04 | 9.58566274E+04 | -2.80E-03 | -8.72636370E+01 | -8.72608344E+01 | -2.80E-03 |

| | | | | | | | | | | | | | | | | |
|---|---|---|---|---|---|---|---|---|---|---|---|---|---|---|---|---|
| 1075 | Ru | 104 | 44 | 60 | 16 | 8.58738000E+00 | 8.58668614E+00 | 6.90E-04 | 9.67646814E+04 | 9.67647216E+04 | -4.00E-02 | 9.67872886E+04 | 9.67873288E+04 | -4.00E-02 | -8.80937340E+01 | -8.80535705E+01 | -4.00E-02 |
| 1076 | Ru | 105 | 44 | 61 | 17 | 8.56188200E+00 | 8.56340759E+00 | -1.50E-03 | 9.76983366E+04 | 9.76981445E+04 | 1.90E-01 | 9.77209438E+04 | 9.77207517E+04 | 1.90E-01 | -8.59325150E+01 | -8.61246895E+01 | 1.90E-01 |
| 1077 | Ru | 106 | 44 | 62 | 18 | 8.56093200E+00 | 8.56126999E+00 | -3.40E-04 | 9.86294409E+04 | 9.86293731E+04 | 6.80E-02 | 9.86520481E+04 | 9.86519803E+04 | 6.80E-02 | -8.63223840E+01 | -8.63901930E+01 | 6.80E-02 |
| 1078 | Ru | 107 | 44 | 63 | 19 | 8.53334800E+00 | 8.53415659E+00 | -8.10E-04 | 9.95633968E+04 | 9.95632783E+04 | 1.20E-01 | 9.95860040E+04 | 9.95858855E+04 | 1.20E-01 | -8.38605000E+01 | -8.39790102E+01 | 1.20E-01 |
| 1079 | Ru | 108 | 44 | 64 | 20 | 8.52720700E+00 | 8.52712285E+00 | 8.40E-05 | 1.00495092E+05 | 1.00495069E+05 | 2.30E-02 | 1.00517699E+05 | 1.00517676E+05 | 2.30E-02 | -8.36593180E+01 | -8.36822038E+01 | 2.30E-02 |
| 1080 | Ru | 109 | 44 | 65 | 21 | 8.49620800E+00 | 8.49623284E+00 | -2.50E-05 | 1.01429509E+05 | 1.01429474E+05 | 3.50E-02 | 1.01452116E+05 | 1.01452082E+05 | 3.50E-02 | -8.07363510E+01 | -8.07709963E+01 | 3.50E-02 |
| 1081 | Ru | 110 | 44 | 66 | 22 | 8.48629300E+00 | 8.48464081E+00 | 1.70E-03 | 1.02361669E+05 | 1.02361819E+05 | -1.50E-01 | 1.02384276E+05 | 1.02384426E+05 | -1.50E-01 | -8.00705750E+01 | -7.99207874E+01 | -1.50E-01 |
| 1082 | Ru | 111 | 44 | 67 | 23 | 8.45293800E+00 | 8.45038979E+00 | 2.50E-03 | 1.03296450E+05 | 1.03296701E+05 | -2.50E-01 | 1.03319058E+05 | 1.03319309E+05 | -2.50E-01 | -7.67831840E+01 | -7.65322457E+01 | -2.50E-01 |
| 1083 | Ru | 112 | 44 | 68 | 24 | 8.43922300E+00 | 8.43475784E+00 | 4.50E-03 | 1.04229099E+05 | 1.04229567E+05 | -4.70E-01 | 1.04251706E+05 | 1.04252174E+05 | -4.70E-01 | -7.56287110E+01 | -7.51605383E+01 | -4.70E-01 |
| 1084 | Ru | 113 | 44 | 69 | 25 | 8.40270700E+00 | 8.39765122E+00 | 5.10E-03 | 1.05164351E+05 | 1.05164891E+05 | -5.40E-01 | 1.05186958E+05 | 1.05187498E+05 | -5.40E-01 | -7.18703520E+01 | -7.13309294E+01 | -5.40E-01 |
| 1085 | Ru | 114 | 44 | 70 | 26 | 8.38534100E+00 | 8.37868431E+00 | 6.70E-03 | 1.06097494E+05 | 1.06098221E+05 | -7.30E-01 | 1.06120101E+05 | 1.06120828E+05 | -7.30E-01 | -7.02220220E+01 | -6.94950337E+01 | -7.30E-01 |
| 1086 | Ru | 115 | 44 | 71 | 27 | 8.34854000E+00 | 8.33949590E+00 | 9.00E-03 | 1.07032906E+05 | 1.07033914E+05 | -1.00E+00 | 1.07055513E+05 | 1.07056521E+05 | -1.00E+00 | -6.63038370E+01 | -6.52957317E+01 | -1.00E+00 |
| 1087 | Ru | 116 | 44 | 72 | 28 | 8.32688300E+00 | 8.31817459E+00 | 8.70E-03 | 1.07966635E+05 | 1.07967613E+05 | -9.80E-01 | 1.07989242E+05 | 1.07990220E+05 | -9.80E-01 | -6.40689070E+01 | -6.30906368E+01 | -9.80E-01 |
| 1088 | Ru | 117 | 44 | 73 | 29 | 8.28581600E+00 | 8.27799823E+00 | 7.80E-03 | 1.08902678E+05 | 1.08903561E+05 | -8.80E-01 | 1.08925285E+05 | 1.08926168E+05 | -8.80E-01 | -5.95196720E+01 | -5.86368584E+01 | -8.80E-01 |
| 1089 | Ru | 118 | 44 | 74 | 30 | 8.26500000E+00 | 8.25565183E+00 | 9.30E-03 | 1.09836433E+05 | 1.09837485E+05 | -1.10E+00 | 1.09859040E+05 | 1.09860093E+05 | -1.10E+00 | -5.72590000E+01 | -5.62066615E+01 | -1.10E+00 |
| 1090 | Ru | 119 | 44 | 75 | 31 | 8.22400000E+00 | 8.21586078E+00 | 8.10E-03 | 1.10772622E+05 | 1.10773530E+05 | -9.10E-01 | 1.10795229E+05 | 1.10796137E+05 | -9.10E-01 | -5.25640000E+01 | -5.16558597E+01 | -9.10E-01 |
| 1091 | Ru | 120 | 44 | 76 | 32 | 8.20100000E+00 | 8.19384416E+00 | 7.20E-03 | 1.11706668E+05 | 1.11707522E+05 | -8.50E-01 | 1.11729275E+05 | 1.11730129E+05 | -8.50E-01 | -5.00120000E+01 | -4.91584072E+01 | -8.50E-01 |
| 1092 | RU | 121 | 44 | 77 | 33 | 8.15900000E+00 | 8.15531324E+00 | 3.70E-03 | 1.12643127E+05 | 1.12643555E+05 | -4.30E-01 | 1.12665734E+05 | 1.12666163E+05 | -4.30E-01 | -4.50470000E+01 | -4.46186910E+01 | -4.30E-01 |
| 1093 | RU | 122 | 44 | 78 | 34 | 8.13700000E+00 | 8.13364971E+00 | 3.40E-03 | 1.13577257E+05 | 1.13577609E+05 | -3.50E-01 | 1.13599864E+05 | 1.13600216E+05 | -3.50E-01 | -4.24110000E+01 | -4.20597345E+01 | -3.50E-01 |
| 1094 | RU | 123 | 44 | 79 | 35 | 8.09500000E+00 | 8.09489501E+00 | 1.00E-04 | 1.14513800E+05 | 1.14513807E+05 | -7.00E-03 | 1.14536407E+05 | 1.14536414E+05 | -7.00E-03 | -3.73620000E+01 | -3.73552369E+01 | -6.80E-03 |
| 1095 | RU | 124 | 44 | 80 | 36 | 8.07100000E+00 | 8.07042668E+00 | 5.70E-04 | 1.15448238E+05 | 1.15448312E+05 | -7.40E-02 | 1.15470845E+05 | 1.15470919E+05 | -7.40E-02 | -3.44190000E+01 | -3.43447409E+01 | -7.40E-02 |
| 1096 | Rh | 89 | 45 | 44 | -1 | 8.19300000E+00 | 8.19349037E+00 | -4.90E-04 | 8.28338157E+04 | 8.28337346E+04 | 8.10E-02 | 8.28569407E+04 | 8.28568597E+04 | 8.10E-02 | -4.60300000E+01 | -4.61117167E+01 | 8.20E-02 |
| 1097 | Rh | 90 | 45 | 45 | 0 | 8.25700000E+00 | 8.25269609E+00 | 4.30E-03 | 8.37593817E+04 | 8.37597780E+04 | -4.00E-01 | 8.37825068E+04 | 8.37829031E+04 | -4.00E-01 | -5.19590000E+01 | -5.15624033E+01 | -4.00E-01 |
| 1098 | Rh | 91 | 45 | 46 | 1 | 8.33100000E+00 | 8.33124716E+00 | -2.50E-04 | 8.46840386E+04 | 8.46839426E+04 | 9.60E-02 | 8.47071637E+04 | 8.47070676E+04 | 9.60E-02 | -5.87960000E+01 | -5.88919278E+01 | 9.60E-02 |
| 1099 | Rh | 92 | 45 | 47 | 2 | 8.37342000E+00 | 8.37676317E+00 | -3.30E-03 | 8.56113295E+04 | 8.56109892E+04 | 3.40E-01 | 8.56344545E+04 | 8.56341143E+04 | 3.40E-01 | -6.29990870E+01 | -6.33393289E+01 | 3.40E-01 |
| 1100 | Rh | 93 | 45 | 48 | 3 | 8.43482500E+00 | 8.44023037E+00 | -5.40E-03 | 8.65368108E+04 | 8.65362754E+04 | 5.40E-01 | 8.65599359E+04 | 8.65594005E+04 | 5.40E-01 | -6.90118000E+01 | -6.95472223E+01 | 5.40E-01 |
| 1101 | Rh | 94 | 45 | 49 | 4 | 8.47240200E+00 | 8.47100347E+00 | 1.40E-03 | 8.74644090E+04 | 8.74645079E+04 | -9.90E-02 | 8.74875341E+04 | 8.74876329E+04 | -9.90E-02 | -7.29076130E+01 | -7.28088051E+01 | -9.90E-02 |
| 1102 | Rh | 95 | 45 | 50 | 5 | 8.52537000E+00 | 8.51675811E+00 | 8.60E-03 | 8.83904701E+04 | 8.83912555E+04 | -7.90E-01 | 8.84135952E+04 | 8.84143806E+04 | -7.90E-01 | -7.83406060E+01 | -7.75551801E+01 | -7.90E-01 |
| 1103 | Rh | 96 | 45 | 51 | 6 | 8.53466000E+00 | 8.53042600E+00 | 4.20E-03 | 8.93206183E+04 | 8.93209921E+04 | -3.70E-01 | 8.93437434E+04 | 8.93441171E+04 | -3.70E-01 | -7.96864690E+01 | -7.93127369E+01 | -3.70E-01 |
| 1104 | Rh | 97 | 45 | 52 | 7 | 8.55988100E+00 | 8.55885463E+00 | 1.00E-03 | 9.02492025E+04 | 9.02492694E+04 | -6.70E-02 | 9.02723276E+04 | 9.02723945E+04 | -6.70E-02 | -8.25963030E+01 | -8.25294208E+01 | -6.70E-02 |
| 1105 | Rh | 98 | 45 | 53 | 8 | 8.56080200E+00 | 8.55918937E+00 | 1.60E-03 | 9.11801178E+04 | 9.11802431E+04 | -1.30E-01 | 9.12032429E+04 | 9.12033682E+04 | -1.30E-01 | -8.31751120E+01 | -8.30497610E+01 | -1.30E-01 |
| 1106 | Rh | 99 | 45 | 54 | 9 | 8.58012900E+00 | 8.57672360E+00 | 3.40E-03 | 9.21092090E+04 | 9.21095135E+04 | -3.00E-01 | 9.21323340E+04 | 9.21326385E+04 | -3.00E-01 | -8.55779940E+01 | -8.52735201E+01 | -3.00E-01 |
| 1107 | Rh | 100 | 45 | 55 | 10 | 8.57514300E+00 | 8.57047584E+00 | 4.70E-03 | 9.30406928E+04 | 9.30411269E+04 | -4.30E-01 | 9.30638179E+04 | 9.30642520E+04 | -4.30E-01 | -8.55882170E+01 | -8.51541493E+01 | -4.30E-01 |
| 1108 | Rh | 101 | 45 | 56 | 11 | 8.58820100E+00 | 8.58261098E+00 | 5.60E-03 | 9.39703642E+04 | 9.39708961E+04 | -5.30E-01 | 9.39934893E+04 | 9.39940212E+04 | -5.30E-01 | -8.74108260E+01 | -8.68789545E+01 | -5.30E-01 |
| 1109 | Rh | 102 | 45 | 57 | 12 | 8.57695300E+00 | 8.57309756E+00 | 3.90E-03 | 9.49024888E+04 | 9.49028493E+04 | -3.60E-01 | 9.49256138E+04 | 9.49259743E+04 | -3.60E-01 | -8.67804280E+01 | -8.64198780E+01 | -3.60E-01 |
| 1110 | Rh | 103 | 45 | 58 | 13 | 8.58415700E+00 | 8.58137220E+00 | 2.80E-03 | 9.58327352E+04 | 9.58329893E+04 | -2.50E-01 | 9.58558602E+04 | 9.58561143E+04 | -2.50E-01 | -8.80280520E+01 | -8.77739450E+01 | -2.50E-01 |
| 1111 | Rh | 104 | 45 | 59 | 14 | 8.56891400E+00 | 8.56879335E+00 | 1.20E-04 | 9.67653016E+04 | 9.67652815E+04 | 2.00E-02 | 9.67884266E+04 | 9.67884065E+04 | 2.00E-02 | -8.69556920E+01 | -8.69757973E+01 | 2.00E-02 |
| 1112 | Rh | 105 | 45 | 60 | 15 | 8.57269800E+00 | 8.57285167E+00 | -1.50E-04 | 9.76959007E+04 | 9.76958519E+04 | 4.90E-02 | 9.77190258E+04 | 9.77189770E+04 | 4.90E-02 | -8.78505500E+01 | -8.78993952E+01 | 4.90E-02 |
| 1113 | Rh | 106 | 45 | 61 | 16 | 8.55392300E+00 | 8.55658369E+00 | -2.70E-03 | 9.86288836E+04 | 9.86285689E+04 | 3.10E-01 | 9.86520087E+04 | 9.86516939E+04 | 3.10E-01 | -8.63617880E+01 | -8.66765220E+01 | 3.10E-01 |
| 1114 | Rh | 107 | 45 | 62 | 17 | 8.55410500E+00 | 8.55594011E+00 | -1.80E-03 | 9.95598756E+04 | 9.95596465E+04 | 2.30E-01 | 9.95830007E+04 | 9.95827716E+04 | 2.30E-01 | -8.68638550E+01 | -8.70929241E+01 | 2.30E-01 |
| 1115 | Rh | 108 | 45 | 63 | 18 | 8.53267200E+00 | 8.53561307E+00 | -2.90E-03 | 1.00493202E+05 | 1.00492851E+05 | 3.50E-01 | 1.00516327E+05 | 1.00515976E+05 | 3.50E-01 | -8.50319200E+01 | -8.53822242E+01 | 3.50E-01 |
| 1116 | Rh | 109 | 45 | 64 | 19 | 8.52814600E+00 | 8.53017145E+00 | -2.00E-03 | 1.01424728E+05 | 1.01424474E+05 | 2.50E-01 | 1.01447853E+05 | 1.01447599E+05 | 2.50E-01 | -8.49999250E+01 | -8.52533823E+01 | 2.50E-01 |
| 1117 | Rh | 110 | 45 | 65 | 20 | 8.50425700E+00 | 8.50581822E+00 | -1.60E-03 | 1.02358393E+05 | 1.02358188E+05 | 2.00E-01 | 1.02381518E+05 | 1.02381313E+05 | 2.00E-01 | -8.28289780E+01 | -8.30333797E+01 | 2.00E-01 |
| 1118 | Rh | 111 | 45 | 66 | 21 | 8.49563300E+00 | 8.49586386E+00 | -2.30E-04 | 1.03290411E+05 | 1.03290353E+05 | 5.80E-02 | 1.03313536E+05 | 1.03313478E+05 | 5.80E-02 | -8.23046580E+01 | -8.23629440E+01 | 5.80E-02 |
| 1119 | Rh | 112 | 45 | 67 | 22 | 8.46888200E+00 | 8.46786833E+00 | 1.00E-03 | 1.04224477E+05 | 1.04224558E+05 | -8.10E-02 | 1.04247602E+05 | 1.04247683E+05 | -8.10E-02 | -7.97328850E+01 | -7.96519904E+01 | -8.10E-02 |
| 1120 | Rh | 113 | 45 | 68 | 23 | 8.45682300E+00 | 8.45386141E+00 | 3.00E-03 | 1.05156936E+05 | 1.05157238E+05 | -3.00E-01 | 1.05180061E+05 | 1.05180363E+05 | -3.00E-01 | -7.87677450E+01 | -7.84657574E+01 | -3.00E-01 |
| 1121 | Rh | 114 | 45 | 69 | 24 | 8.42664900E+00 | 8.42268954E+00 | 4.00E-03 | 1.06091485E+05 | 1.06091903E+05 | -4.20E-01 | 1.06114610E+05 | 1.06115028E+05 | -4.20E-01 | -7.57134210E+01 | -7.52947068E+01 | -4.20E-01 |
| 1122 | Rh | 115 | 45 | 70 | 25 | 8.41065400E+00 | 8.40528022E+00 | 5.40E-03 | 1.07024463E+05 | 1.07025048E+05 | -5.90E-01 | 1.07047588E+05 | 1.07048173E+05 | -5.90E-01 | -7.42292760E+01 | -7.36440051E+01 | -5.90E-01 |
| 1123 | Rh | 116 | 45 | 71 | 26 | 8.37764000E+00 | 8.37167578E+00 | 6.00E-03 | 1.07959447E+05 | 1.07960106E+05 | -6.60E-01 | 1.07982572E+05 | 1.07983231E+05 | -6.60E-01 | -7.07390250E+01 | -7.00798513E+01 | -6.60E-01 |
| 1124 | Rh | 117 | 45 | 72 | 27 | 8.35928300E+00 | 8.35178832E+00 | 7.50E-03 | 1.08892783E+05 | 1.08893627E+05 | -8.40E-01 | 1.08915908E+05 | 1.08916752E+05 | -8.40E-01 | -6.88975890E+01 | -6.80533751E+01 | -8.40E-01 |
| 1125 | Rh | 118 | 45 | 73 | 28 | 8.32286000E+00 | 8.31680875E+00 | 6.10E-03 | 1.09828286E+05 | 1.09828968E+05 | -6.80E-01 | 1.09851412E+05 | 1.09852093E+05 | -6.80E-01 | -6.48876780E+01 | -6.42062551E+01 | -6.80E-01 |
| 1126 | Rh | 119 | 45 | 74 | 29 | 8.30339400E+00 | 8.29570695E+00 | 7.70E-03 | 1.10761845E+05 | 1.10762728E+05 | -8.80E-01 | 1.10784970E+05 | 1.10785853E+05 | -8.80E-01 | -6.28227920E+01 | -6.19406313E+01 | -8.80E-01 |
| 1127 | Rh | 120 | 45 | 75 | 30 | 8.26800000E+00 | 8.26068799E+00 | 7.30E-03 | 1.11697348E+05 | 1.11698200E+05 | -8.50E-01 | 1.11720473E+05 | 1.11721325E+05 | -8.50E-01 | -5.88150000E+01 | -5.79627438E+01 | -8.50E-01 |
| 1128 | Rh | 121 | 45 | 76 | 31 | 8.24700000E+00 | 8.23967459E+00 | 7.30E-03 | 1.12631226E+05 | 1.12632047E+05 | -8.20E-01 | 1.12654352E+05 | 1.12655172E+05 | -8.20E-01 | -5.64300000E+01 | -5.56094917E+01 | -8.20E-01 |
| 1129 | Rh | 122 | 45 | 77 | 32 | 8.21000000E+00 | 8.20549528E+00 | 4.50E-03 | 1.13566977E+05 | 1.13567542E+05 | -5.70E-01 | 1.13590102E+05 | 1.13590667E+05 | -5.60E-01 | -5.21730000E+01 | -5.16079706E+01 | -5.70E-01 |
| 1130 | RH | 123 | 45 | 78 | 33 | 8.18800000E+00 | 8.18461199E+00 | 3.40E-03 | 1.14501136E+05 | 1.14501471E+05 | -3.40E-01 | 1.14524261E+05 | 1.14524596E+05 | -3.40E-01 | -4.95090000E+01 | -4.91735032E+01 | -3.40E-01 |
| 1131 | RH | 124 | 45 | 79 | 34 | 8.15200000E+00 | 8.14987923E+00 | 2.10E-03 | 1.15436970E+05 | 1.15437159E+05 | -1.90E-01 | 1.15460095E+05 | 1.15460284E+05 | -1.90E-01 | -4.51680000E+01 | -4.49799330E+01 | -1.90E-01 |
| 1132 | RH | 125 | 45 | 80 | 35 | 8.12700000E+00 | 8.12604613E+00 | 9.50E-04 | 1.16371427E+05 | 1.16371553E+05 | -1.30E-01 | 1.16394552E+05 | 1.16394678E+05 | -1.30E-01 | -4.22060000E+01 | -4.20793567E+01 | -1.30E-01 |
| 1133 | RH | 126 | 45 | 81 | 36 | 8.09200000E+00 | 8.08624199E+00 | 5.80E-03 | 1.17307364E+05 | 1.17308008E+05 | -6.40E-01 | 1.17330489E+05 | 1.17331133E+05 | -6.40E-01 | -3.77630000E+01 | -3.71187616E+01 | -6.40E-01 |
| 1134 | Pd | 91 | 46 | 45 | -1 | 8.18400000E+00 | 8.18816652E+00 | -4.20E-03 | 8.46960398E+04 | 8.46956617E+04 | 3.80E-01 | 8.47196829E+04 | 8.47193049E+04 | 3.80E-01 | -4.62770000E+01 | -4.66546663E+01 | 3.80E-01 |
| 1135 | Pd | 92 | 46 | 46 | 0 | 8.27900000E+00 | 8.27760152E+00 | 1.40E-03 | 8.56187405E+04 | 8.56188109E+04 | -7.00E-02 | 8.56423837E+04 | 8.56424541E+04 | -7.00E-02 | -5.50700000E+01 | -5.49995341E+01 | -7.00E-02 |
| 1136 | Pd | 93 | 46 | 47 | 1 | 8.32000000E+00 | 8.32487227E+00 | -4.90E-03 | 8.65461649E+04 | 8.65457025E+04 | 4.60E-01 | 8.65698081E+04 | 8.65693457E+04 | 4.60E-01 | -5.91400000E+01 | -5.96019963E+01 | 4.60E-01 |
| 1137 | Pd | 94 | 46 | 48 | 2 | 8.39166900E+00 | 8.39772521E+00 | -6.10E-03 | 8.74706975E+04 | 8.74700948E+04 | 6.00E-01 | 8.74943407E+04 | 8.74937380E+04 | 6.00E-01 | -6.61010470E+01 | -6.67037260E+01 | 6.00E-01 |
| 1138 | Pd | 95 | 46 | 49 | 3 | 8.42896700E+00 | 8.42942680E+00 | -4.60E-04 | 8.83983280E+04 | 8.83982509E+04 | 7.70E-02 | 8.84219711E+04 | 8.84218940E+04 | 7.70E-02 | -6.99646650E+01 | -7.00417829E+01 | 7.70E-02 |
| 1139 | Pd | 96 | 46 | 50 | 4 | 8.49000700E+00 | 8.48371139E+00 | 6.30E-03 | 8.93236045E+04 | 8.93241755E+04 | -5.70E-01 | 8.93472477E+04 | 8.93478186E+04 | -5.70E-01 | -7.61821590E+01 | -7.56112116E+01 | -5.70E-01 |
| 1140 | Pd | 97 | 46 | 51 | 5 | 8.50243000E+00 | 8.49816253E+00 | 4.30E-03 | 9.02534749E+04 | 9.02538554E+04 | -3.80E-01 | 9.02771181E+04 | 9.02774986E+04 | -3.80E-01 | -7.78058420E+01 | -7.74253645E+01 | -3.80E-01 |
| 1141 | Pd | 98 | 46 | 52 | 6 | 8.53389900E+00 | 8.53474009E+00 | -8.40E-04 | 9.11814538E+04 | 9.11813380E+04 | 1.20E-01 | 9.12050970E+04 | 9.12049812E+04 | 1.20E-01 | -8.13209750E+01 | -8.14368086E+01 | 1.20E-01 |
| 1142 | Pd | 99 | 46 | 53 | 7 | 8.53791600E+00 | 8.53591458E+00 | 2.00E-03 | 9.21120876E+04 | 9.21122524E+04 | -1.60E-01 | 9.21357308E+04 | 9.21358955E+04 | -1.60E-01 | -8.21812390E+01 | -8.20165045E+01 | -1.60E-01 |
| 1143 | Pd | 100 | 46 | 54 | 8 | 8.56371000E+00 | 8.56124398E+00 | 2.50E-03 | 9.30405357E+04 | 9.30407489E+04 | -2.10E-01 | 9.30641789E+04 | 9.30643921E+04 | -2.10E-01 | -8.52272250E+01 | -8.50140400E+01 | -2.10E-01 |
| 1144 | Pd | 101 | 46 | 55 | 9 | 8.56084900E+00 | 8.55577341E+00 | 5.10E-03 | 9.39718263E+04 | 9.39723056E+04 | -4.80E-01 | 9.39954695E+04 | 9.39959487E+04 | -4.80E-01 | -8.54306570E+01 | -8.49514373E+01 | -4.80E-01 |
| 1145 | Pd | 102 | 46 | 56 | 10 | 8.58056300E+00 | 8.57529486E+00 | 5.30E-03 | 9.49008201E+04 | 9.49013240E+04 | -5.00E-01 | 9.49244632E+04 | 9.49249671E+04 | -5.00E-01 | -8.79309960E+01 | -8.74270794E+01 | -5.00E-01 |
| 1146 | Pd | 103 | 46 | 57 | 11 | 8.57128900E+00 | 8.56647275E+00 | 4.80E-03 | 9.58327600E+04 | 9.58332227E+04 | -4.60E-01 | 9.58564032E+04 | 9.58568659E+04 | -4.60E-01 | -8.74850320E+01 | -8.70223786E+01 | -4.60E-01 |
| 1147 | Pd | 104 | 46 | 58 | 12 | 8.58484600E+00 | 8.58182109E+00 | 3.00E-03 | 9.67623442E+04 | 9.67626254E+04 | -2.80E-01 | 9.67859873E+04 | 9.67862686E+04 | -2.80E-01 | -8.93949760E+01 | -8.91137593E+01 | -2.80E-01 |
| 1148 | Pd | 105 | 46 | 59 | 13 | 8.57064900E+00 | 8.56993415E+00 | 7.10E-04 | 9.76948154E+04 | 9.76948571E+04 | -4.20E-02 | 9.77184586E+04 | 9.77185003E+04 | -4.20E-02 | -8.84177590E+01 | -8.83761333E+01 | -4.20E-02 |
| 1149 | Pd | 106 | 46 | 60 | 14 | 8.57999100E+00 | 8.58088517E+00 | -8.90E-04 | 9.86248199E+04 | 9.86246917E+04 | 1.30E-01 | 9.86484630E+04 | 9.86483349E+04 | 1.30E-01 | -8.99074040E+01 | -9.00355564E+01 | 1.30E-01 |
| 1150 | Pd | 107 | 46 | 61 | 15 | 8.56089300E+00 | 8.56537349E+00 | -4.50E-03 | 9.95578488E+04 | 9.95573360E+04 | 5.10E-01 | 9.95814920E+04 | 9.95809792E+04 | 5.10E-01 | -8.83725170E+01 | -8.88853720E+01 | 5.10E-01 |
| 1151 | Pd | 108 | 46 | 62 | 16 | 8.56702500E+00 | 8.57151273E+00 | -4.50E-03 | 1.00488191E+05 | 1.00487673E+05 | 5.20E-01 | 1.00511834E+05 | 1.00511316E+05 | 5.20E-01 | -8.95243610E+01 | -9.00424652E+01 | 5.20E-01 |
| 1152 | Pd | 109 | 46 | 63 | 17 | 8.54488300E+00 | 8.55200447E+00 | -7.10E-03 | 1.01421603E+05 | 1.01420793E+05 | 8.10E-01 | 1.01445246E+05 | 1.01444436E+05 | 8.10E-01 | -8.76066320E+01 | -8.84162586E+01 | 8.10E-01 |
| 1153 | Pd | 110 | 46 | 64 | 18 | 8.54716800E+00 | 8.55324199E+00 | -6.10E-03 | 1.02352372E+05 | 1.02351670E+05 | 7.00E-01 | 1.02376015E+05 | 1.02375314E+05 | 7.00E-01 | -8.83315270E+01 | -8.90330706E+01 | 7.00E-01 |
| 1154 | Pd | 111 | 46 | 65 | 19 | 8.52175500E+00 | 8.52972320E+00 | -8.00E-03 | 1.03286211E+05 | 1.03285293E+05 | 9.20E-01 | 1.03309854E+05 | 1.03308936E+05 | 9.20E-01 | -8.59865100E+01 | -8.69044083E+01 | 9.20E-01 |
| 1155 | Pd | 112 | 46 | 66 | 20 | 8.52072700E+00 | 8.52631927E+00 | -5.60E-03 | 1.04217370E+05 | 1.04216710E+05 | 6.60E-01 | 1.04241013E+05 | 1.04240353E+05 | 6.60E-01 | -8.63218300E+01 | -8.69815721E+01 | 6.60E-01 |
| 1156 | Pd | 113 | 46 | 67 | 21 | 8.49258600E+00 | 8.49912016E+00 | -6.50E-03 | 1.05151594E+05 | 1.05150823E+05 | 7.70E-01 | 1.05175238E+05 | 1.05174466E+05 | 7.70E-01 | -8.35913010E+01 | -8.43630735E+01 | 7.70E-01 |
| 1157 | Pd | 114 | 46 | 68 | 22 | 8.48801200E+00 | 8.49149576E+00 | -3.50E-03 | 1.06083189E+05 | 1.06082758E+05 | 4.30E-01 | 1.06106832E+05 | 1.06106401E+05 | 4.30E-01 | -8.34911540E+01 | -8.39216924E+01 | 4.30E-01 |
| 1158 | Pd | 115 | 46 | 69 | 23 | 8.45774000E+00 | 8.46101831E+00 | -3.30E-03 | 1.07017747E+05 | 1.07017337E+05 | 4.10E-01 | 1.07041390E+05 | 1.07040980E+05 | 4.10E-01 | -8.04265830E+01 | -8.08369624E+01 | 4.10E-01 |
| 1159 | Pd | 116 | 46 | 70 | 24 | 8.44928200E+00 | 8.44978274E+00 | -5.00E-04 | 1.07949836E+05 | 1.07949745E+05 | 9.20E-02 | 1.07973479E+05 | 1.07973388E+05 | 9.10E-02 | -7.98318490E+01 | -7.99233358E+01 | 9.10E-02 |
| 1160 | Pd | 117 | 46 | 71 | 25 | 8.41693000E+00 | 8.41671809E+00 | 2.10E-04 | 1.08884737E+05 | 1.08884729E+05 | 8.60E-03 | 1.08908380E+05 | 1.08908372E+05 | 8.50E-03 | -7.64246850E+01 | -7.64332351E+01 | 8.60E-03 |
| 1161 | Pd | 118 | 46 | 72 | 26 | 8.40522300E+00 | 8.40275175E+00 | 2.50E-03 | 1.09817267E+05 | 1.09817525E+05 | -2.60E-01 | 1.09840910E+05 | 1.09841169E+05 | -2.60E-01 | -7.53888830E+01 | -7.51306070E+01 | -2.60E-01 |
| 1162 | Pd | 119 | 46 | 73 | 27 | 8.36896600E+00 | 8.36809868E+00 | 8.70E-04 | 1.10752742E+05 | 1.10752812E+05 | -7.00E-02 | 1.10776385E+05 | 1.10776455E+05 | -7.00E-02 | -7.14081220E+01 | -7.13383240E+01 | -7.00E-02 |
| 1163 | Pd | 120 | 46 | 74 | 28 | 8.35708600E+00 | 8.35260845E+00 | 4.50E-03 | 1.11685364E+05 | 1.11685868E+05 | -5.00E-01 | 1.11709007E+05 | 1.11709511E+05 | -5.00E-01 | -7.02802080E+01 | -6.97762754E+01 | -5.00E-01 |
| 1164 | PD | 121 | 46 | 75 | 29 | 8.32085800E+00 | 8.31763752E+00 | 3.20E-03 | 1.12620956E+05 | 1.12621312E+05 | -3.60E-01 | 1.12644599E+05 | 1.12644955E+05 | -3.60E-01 | -6.61823270E+01 | -6.58260825E+01 | -3.60E-01 |
| 1165 | Pd | 122 | 46 | 76 | 30 | 8.30597500E+00 | 8.30187079E+00 | 4.10E-03 | 1.13554016E+05 | 1.13554483E+05 | -4.70E-01 | 1.13577659E+05 | 1.13578127E+05 | -4.70E-01 | -6.46161590E+01 | -6.41488601E+01 | -4.70E-01 |
| 1166 | Pd | 123 | 46 | 77 | 31 | 8.27000000E+00 | 8.26742393E+00 | 2.60E-03 | 1.14489710E+05 | 1.14489984E+05 | -2.70E-01 | 1.14513353E+05 | 1.14513627E+05 | -2.70E-01 | -6.04170000E+01 | -6.01424486E+01 | -2.70E-01 |
| 1167 | PD | 124 | 46 | 78 | 32 | 8.25300000E+00 | 8.25142334E+00 | 1.60E-03 | 1.15423067E+05 | 1.15423266E+05 | -2.00E-01 | 1.15446710E+05 | 1.15446909E+05 | -2.00E-01 | -5.85540000E+01 | -5.83544805E+01 | -2.00E-01 |
| 1168 | PD | 125 | 46 | 79 | 33 | 8.21700000E+00 | 8.21615391E+00 | 8.50E-04 | 1.16358892E+05 | 1.16358989E+05 | -9.60E-02 | 1.16382535E+05 | 1.16382632E+05 | -9.60E-02 | -5.42220000E+01 | -5.41259062E+01 | -9.60E-02 |

| | | | | | | | | | | | | | | | | |
|---|---|---|---|---|---|---|---|---|---|---|---|---|---|---|---|---|
| 1169 | PD | 126 | 46 | 80 | 34 | 8.19900000E+00 | 8.19691463E+00 | 2.10E-03 | 1.17292594E+05 | 1.17292762E+05 | -1.70E-01 | 1.17316237E+05 | 1.17316405E+05 | -1.70E-01 | -5.20150000E+01 | -5.18465913E+01 | -1.70E-01 |
| 1170 | PD | 127 | 46 | 81 | 35 | 8.16100000E+00 | 8.15636586E+00 | 4.60E-03 | 1.18228662E+05 | 1.18229280E+05 | -6.20E-01 | 1.18252305E+05 | 1.18252923E+05 | -6.20E-01 | -4.74410000E+01 | -4.68224937E+01 | -6.20E-01 |
| 1171 | PD | 128 | 46 | 82 | 36 | 8.14100000E+00 | 8.12918970E+00 | 1.20E-02 | 1.19162727E+05 | 1.19164168E+05 | -1.40E+00 | 1.19186370E+05 | 1.19187811E+05 | -1.40E+00 | -4.48700000E+01 | -4.34289917E+01 | -1.40E+00 |
| 1172 | Ag | 93 | 47 | 46 | -1 | 8.17300000E+00 | 8.18442006E+00 | -1.10E-02 | 8.65585189E+04 | 8.65574632E+04 | 1.10E+00 | 8.65826804E+04 | 8.65816247E+04 | 1.10E+00 | -4.62670000E+01 | -4.73230178E+01 | 1.10E+00 |
| 1173 | Ag | 94 | 47 | 47 | 0 | 8.23800000E+00 | 8.24192896E+00 | -3.90E-03 | 8.74838688E+04 | 8.74834383E+04 | 4.30E-01 | 8.75080303E+04 | 8.75075998E+04 | 4.30E-01 | -5.24110000E+01 | -5.28419555E+01 | 4.30E-01 |
| 1174 | Ag | 95 | 47 | 48 | 1 | 8.31200000E+00 | 8.31662003E+00 | -4.60E-03 | 8.84081773E+04 | 8.84076661E+04 | 5.10E-01 | 8.84323388E+04 | 8.84318276E+04 | 5.10E-01 | -5.95970000E+01 | -6.01082165E+01 | 5.10E-01 |
| 1175 | Ag | 96 | 47 | 49 | 2 | 8.36029000E+00 | 8.35740872E+00 | 2.90E-03 | 8.93347567E+04 | 8.93349991E+04 | -2.40E-01 | 8.93589182E+04 | 8.93591606E+04 | -2.40E-01 | -6.45116350E+01 | -6.42692318E+01 | -2.40E-01 |
| 1176 | Ag | 97 | 47 | 50 | 3 | 8.42240500E+00 | 8.41307039E+00 | 9.30E-03 | 9.02599366E+04 | 9.02608079E+04 | -8.70E-01 | 9.02840981E+04 | 9.02849694E+04 | -8.70E-01 | -7.08258420E+01 | -6.99545041E+01 | -8.70E-01 |
| 1177 | Ag | 98 | 47 | 51 | 4 | 8.44168600E+00 | 8.43596673E+00 | 5.70E-03 | 9.11891901E+04 | 9.11897164E+04 | -5.30E-01 | 9.12133516E+04 | 9.12138779E+04 | -5.30E-01 | -7.30664150E+01 | -7.25400964E+01 | -5.30E-01 |
| 1178 | Ag | 99 | 47 | 52 | 5 | 8.47477400E+00 | 8.47382733E+00 | 9.50E-04 | 9.21170381E+04 | 9.21170976E+04 | -6.00E-02 | 9.21411996E+04 | 9.21412591E+04 | -6.00E-02 | -7.67124730E+01 | -7.66529435E+01 | -6.00E-02 |
| 1179 | Ag | 100 | 47 | 53 | 6 | 8.48499400E+00 | 8.48304383E+00 | 2.00E-03 | 9.30471067E+04 | 9.30472675E+04 | -1.60E-01 | 9.30712681E+04 | 9.30714290E+04 | -1.60E-01 | -7.81379590E+01 | -7.79771023E+01 | -1.60E-01 |
| 1180 | Ag | 101 | 47 | 54 | 7 | 8.51254600E+00 | 8.50958453E+00 | 3.00E-03 | 9.39754043E+04 | 9.39756692E+04 | -2.60E-01 | 9.39995658E+04 | 9.39998307E+04 | -2.60E-01 | -8.13343730E+01 | -8.10694377E+01 | -2.60E-01 |
| 1181 | Ag | 102 | 47 | 55 | 8 | 8.51716300E+00 | 8.51173238E+00 | 5.40E-03 | 9.49059862E+04 | 9.49065060E+04 | -5.20E-01 | 9.49301477E+04 | 9.49306674E+04 | -5.20E-01 | -8.22465340E+01 | -8.17267834E+01 | -5.20E-01 |
| 1182 | Ag | 103 | 47 | 56 | 9 | 8.53762800E+00 | 8.53234392E+00 | 5.30E-03 | 9.58349265E+04 | 9.58354366E+04 | -5.10E-01 | 9.58590880E+04 | 9.58595981E+04 | -5.10E-01 | -8.48002900E+01 | -8.42901859E+01 | -5.10E-01 |
| 1183 | Ag | 104 | 47 | 57 | 10 | 8.53618300E+00 | 8.53077378E+00 | 5.40E-03 | 9.67661046E+04 | 9.67666329E+04 | -5.30E-01 | 9.67902661E+04 | 9.67907944E+04 | -5.30E-01 | -8.51163220E+01 | -8.45879165E+01 | -5.30E-01 |
| 1184 | Ag | 105 | 47 | 58 | 11 | 8.55037000E+00 | 8.54716417E+00 | 3.20E-03 | 9.76956441E+04 | 9.76959466E+04 | -3.00E-01 | 9.77198056E+04 | 9.77201080E+04 | -3.00E-01 | -8.70708280E+01 | -8.67683615E+01 | -3.00E-01 |
| 1185 | Ag | 106 | 47 | 59 | 12 | 8.54463800E+00 | 8.54230795E+00 | 2.30E-03 | 9.86272667E+04 | 9.86274795E+04 | -2.10E-01 | 9.86514282E+04 | 9.86516410E+04 | -2.10E-01 | -8.69422620E+01 | -8.67294478E+01 | -2.10E-01 |
| 1186 | Ag | 107 | 47 | 60 | 13 | 8.55389900E+00 | 8.55435887E+00 | -4.60E-04 | 9.95572964E+04 | 9.95572132E+04 | 8.30E-02 | 9.95814579E+04 | 9.95813746E+04 | 8.30E-02 | -8.84065950E+01 | -8.84898847E+01 | 8.30E-02 |
| 1187 | Ag | 108 | 47 | 61 | 14 | 8.54202500E+00 | 8.54577635E+00 | -3.80E-03 | 1.00489590E+05 | 1.00489151E+05 | 4.40E-01 | 1.00513752E+05 | 1.00513313E+05 | 4.40E-01 | -8.76066880E+01 | -8.80460132E+01 | 4.40E-01 |
| 1188 | Ag | 109 | 47 | 62 | 15 | 8.54791900E+00 | 8.55313106E+00 | -5.20E-03 | 1.01419971E+05 | 1.01419369E+05 | 6.00E-01 | 1.01444133E+05 | 1.01443531E+05 | 6.00E-01 | -8.87198830E+01 | -8.93221341E+01 | 6.00E-01 |
| 1189 | Ag | 110 | 47 | 63 | 16 | 8.53211200E+00 | 8.54050099E+00 | -8.40E-03 | 1.02352727E+05 | 1.02351771E+05 | 9.60E-01 | 1.02376889E+05 | 1.02375932E+05 | 9.60E-01 | -8.74577610E+01 | -8.84146384E+01 | 9.60E-01 |
| 1190 | Ag | 111 | 47 | 64 | 17 | 8.53479500E+00 | 8.54307148E+00 | -8.30E-03 | 1.03283463E+05 | 1.03282510E+05 | 9.50E-01 | 1.03307625E+05 | 1.03306672E+05 | 9.50E-01 | -8.82162790E+01 | -8.91691439E+01 | 9.50E-01 |
| 1191 | Ag | 112 | 47 | 65 | 18 | 8.51608000E+00 | 8.52637358E+00 | -1.00E-02 | 1.04216590E+05 | 1.04215403E+05 | 1.20E+00 | 1.04240751E+05 | 1.04239564E+05 | 1.20E+00 | -8.65837170E+01 | -8.77707324E+01 | 1.20E+00 |
| 1192 | Ag | 113 | 47 | 66 | 19 | 8.51606500E+00 | 8.52439416E+00 | -8.30E-03 | 1.05147641E+05 | 1.05146665E+05 | 9.80E-01 | 1.05171802E+05 | 1.05170827E+05 | 9.80E-01 | -8.70267890E+01 | -8.80021121E+01 | 9.80E-01 |
| 1193 | Ag | 114 | 47 | 67 | 20 | 8.49377800E+00 | 8.50392999E+00 | -1.00E-02 | 1.06081231E+05 | 1.06080039E+05 | 1.20E+00 | 1.06105392E+05 | 1.06104201E+05 | 1.20E+00 | -8.49307990E+01 | -8.61222713E+01 | 1.20E+00 |
| 1194 | Ag | 115 | 47 | 68 | 21 | 8.49055500E+00 | 8.49778088E+00 | -7.20E-03 | 1.07012673E+05 | 1.07011808E+05 | 8.70E-01 | 1.07036834E+05 | 1.07035969E+05 | 8.70E-01 | -8.49826610E+01 | -8.58477350E+01 | 8.70E-01 |
| 1195 | Ag | 116 | 47 | 69 | 22 | 8.46590600E+00 | 8.47389601E+00 | -8.00E-03 | 1.07946607E+05 | 1.07945646E+05 | 9.60E-01 | 1.07970768E+05 | 1.07969808E+05 | 9.60E-01 | -8.25426530E+01 | -8.35035519E+01 | 9.60E-01 |
| 1196 | Ag | 117 | 47 | 70 | 23 | 8.45945200E+00 | 8.46414274E+00 | -4.70E-03 | 1.08878462E+05 | 1.08877879E+05 | 5.80E-01 | 1.08902623E+05 | 1.08902040E+05 | 5.80E-01 | -8.21820500E+01 | -8.27649964E+01 | 5.80E-01 |
| 1197 | Ag | 118 | 47 | 71 | 24 | 8.43388900E+00 | 8.43747216E+00 | -3.60E-03 | 1.09812584E+05 | 1.09812127E+05 | 4.60E-01 | 1.09836745E+05 | 1.09836289E+05 | 4.60E-01 | -7.95537910E+01 | -8.00106919E+01 | 4.60E-01 |
| 1198 | Ag | 119 | 47 | 72 | 25 | 8.42321200E+00 | 8.42494765E+00 | -1.70E-03 | 1.10744986E+05 | 1.10744745E+05 | 2.40E-01 | 1.10769148E+05 | 1.10768907E+05 | 2.40E-01 | -7.86457710E+01 | -7.88864278E+01 | 2.40E-01 |
| 1199 | Ag | 120 | 47 | 73 | 26 | 8.39532700E+00 | 8.39641949E+00 | -1.10E-03 | 1.11679474E+05 | 1.11679309E+05 | 1.70E-01 | 1.11703636E+05 | 1.11703471E+05 | 1.70E-01 | -7.56515010E+01 | -7.58166782E+01 | 1.70E-01 |
| 1200 | Ag | 121 | 47 | 74 | 27 | 8.38233000E+00 | 8.38226683E+00 | 6.30E-05 | 1.12612217E+05 | 1.12612191E+05 | 2.70E-02 | 1.12636379E+05 | 1.12636352E+05 | 2.70E-02 | -7.44028200E+01 | -7.44293058E+01 | 2.60E-02 |
| 1201 | Ag | 122 | 47 | 75 | 28 | 8.35275800E+00 | 8.35306632E+00 | -3.10E-04 | 1.13547008E+05 | 1.13546936E+05 | 7.20E-02 | 1.13571169E+05 | 1.13571098E+05 | 7.20E-02 | -7.11061080E+01 | -7.11777915E+01 | 7.20E-02 |
| 1202 | Ag | 123 | 47 | 76 | 29 | 8.33780300E+00 | 8.33846198E+00 | -6.60E-04 | 1.14480060E+05 | 1.14479945E+05 | 1.20E-01 | 1.14504221E+05 | 1.14504106E+05 | 1.20E-01 | -6.95480780E+01 | -6.96632057E+01 | 1.20E-01 |
| 1203 | Ag | 124 | 47 | 77 | 30 | 8.30865500E+00 | 8.30936330E+00 | -7.10E-04 | 1.15414902E+05 | 1.15414780E+05 | 1.20E-01 | 1.15439063E+05 | 1.15438941E+05 | 1.20E-01 | -6.62001330E+01 | -6.63221124E+01 | 1.20E-01 |
| 1204 | Ag | 125 | 47 | 78 | 31 | 8.29099700E+00 | 8.29432487E+00 | -3.30E-03 | 1.16348366E+05 | 1.16347916E+05 | 4.50E-01 | 1.16372527E+05 | 1.16372077E+05 | 4.50E-01 | -6.42302370E+01 | -6.46803526E+01 | 4.50E-01 |
| 1205 | Ag | 126 | 47 | 79 | 32 | 8.26200000E+00 | 8.26399801E+00 | -2.00E-03 | 1.17283310E+05 | 1.17283008E+05 | 3.00E-01 | 1.17307472E+05 | 1.17307170E+05 | 3.00E-01 | -6.07800000E+01 | -6.10821745E+01 | 3.00E-01 |
| 1206 | Ag | 127 | 47 | 80 | 33 | 8.24300000E+00 | 8.24558370E+00 | -2.60E-03 | 1.18217003E+05 | 1.18216648E+05 | 3.50E-01 | 1.18241164E+05 | 1.18240810E+05 | 3.50E-01 | -5.85820000E+01 | -5.89362363E+01 | 3.50E-01 |
| 1207 | Ag | 128 | 47 | 81 | 34 | 8.21300000E+00 | 8.20963998E+00 | 3.40E-03 | 1.19152176E+05 | 1.19152569E+05 | -3.90E-01 | 1.19176337E+05 | 1.19176730E+05 | -3.90E-01 | -5.49020000E+01 | -5.45097049E+01 | -3.90E-01 |
| 1208 | Ag | 129 | 47 | 82 | 35 | 8.19100000E+00 | 8.18315765E+00 | 7.80E-03 | 1.20086362E+05 | 1.20087341E+05 | -9.80E-01 | 1.20110524E+05 | 1.20111502E+05 | -9.80E-01 | -5.22100000E+01 | -5.12318042E+01 | -9.80E-01 |
| 1209 | Ag | 130 | 47 | 83 | 36 | 8.14200000E+00 | 8.13829699E+00 | 3.70E-03 | 1.21024147E+05 | 1.21024555E+05 | -4.10E-01 | 1.21048308E+05 | 1.21048716E+05 | -4.10E-01 | -4.59200000E+01 | -4.55117582E+01 | -4.10E-01 |
| 1210 | CD | 96 | 48 | 48 | 0 | 8.25900000E+00 | 8.26425405E+00 | -5.30E-03 | 8.93431769E+04 | 8.93426404E+04 | 5.40E-01 | 8.93678569E+04 | 8.93673204E+04 | 5.40E-01 | -5.55730000E+01 | -5.61094604E+01 | 5.40E-01 |
| 1211 | Cd | 97 | 48 | 49 | 1 | 8.30700000E+00 | 8.30697206E+00 | 2.80E-05 | 9.02697899E+04 | 9.02697978E+04 | -7.90E-03 | 9.02944699E+04 | 9.02944779E+04 | -7.90E-03 | -6.04540000E+01 | -6.04460433E+01 | -8.00E-03 |
| 1212 | Cd | 98 | 48 | 50 | 2 | 8.37829400E+00 | 8.37020803E+00 | 8.10E-03 | 9.11941015E+04 | 9.11948591E+04 | -7.60E-01 | 9.12187816E+04 | 9.12195392E+04 | -7.60E-01 | -6.76364150E+01 | -6.68788209E+01 | -7.60E-01 |
| 1213 | Cd | 99 | 48 | 51 | 3 | 8.39837300E+00 | 8.39448981E+00 | 3.90E-03 | 9.21233009E+04 | 9.21236504E+04 | -3.50E-01 | 9.21479809E+04 | 9.21483304E+04 | -3.50E-01 | -6.99311220E+01 | -6.95816061E+01 | -3.50E-01 |
| 1214 | Cd | 100 | 48 | 52 | 4 | 8.43773700E+00 | 8.43966563E+00 | -1.90E-03 | 9.30505315E+04 | 9.30503037E+04 | 2.30E-01 | 9.30752115E+04 | 9.30749837E+04 | 2.30E-01 | -7.41945960E+01 | -7.44223587E+01 | 2.30E-01 |
| 1215 | Cd | 101 | 48 | 53 | 5 | 8.45036500E+00 | 8.45000837E+00 | 3.60E-04 | 9.39803837E+04 | 9.39803848E+04 | -1.10E-03 | 9.40050637E+04 | 9.40050648E+04 | -1.10E-03 | -7.58364550E+01 | -7.58353219E+01 | -1.10E-03 |
| 1216 | Cd | 102 | 48 | 54 | 6 | 8.48413000E+00 | 8.48358381E+00 | 5.50E-04 | 9.49080547E+04 | 9.49080755E+04 | -2.10E-02 | 9.49327347E+04 | 9.49327555E+04 | -2.10E-02 | -7.96595340E+01 | -7.96387069E+01 | -2.10E-02 |
| 1217 | Cd | 103 | 48 | 55 | 7 | 8.48975800E+00 | 8.48660208E+00 | 3.20E-03 | 9.58385562E+04 | 9.58388464E+04 | -2.90E-01 | 9.58632363E+04 | 9.58635264E+04 | -2.90E-01 | -8.06519950E+01 | -8.03618532E+01 | -2.90E-01 |
| 1218 | Cd | 104 | 48 | 56 | 8 | 8.51762100E+00 | 8.51394083E+00 | 3.70E-03 | 9.67667341E+04 | 9.67670819E+04 | -3.50E-01 | 9.67914141E+04 | 9.67917620E+04 | -3.50E-01 | -8.39682360E+01 | -8.36203666E+01 | -3.50E-01 |
| 1219 | Cd | 105 | 48 | 57 | 9 | 8.51685200E+00 | 8.51303712E+00 | 3.80E-03 | 9.76978625E+04 | 9.76982283E+04 | -3.70E-01 | 9.77225426E+04 | 9.77229083E+04 | -3.70E-01 | -8.43338370E+01 | -8.39680982E+01 | -3.70E-01 |
| 1220 | Cd | 106 | 48 | 58 | 10 | 8.53904700E+00 | 8.53594097E+00 | 3.10E-03 | 9.86265584E+04 | 9.86268528E+04 | -2.90E-01 | 9.86512384E+04 | 9.86515328E+04 | -2.90E-01 | -8.71320150E+01 | -8.68376253E+01 | -2.90E-01 |
| 1221 | Cd | 107 | 48 | 59 | 11 | 8.53335100E+00 | 8.53165981E+00 | 1.70E-03 | 9.95581942E+04 | 9.95583403E+04 | -1.50E-01 | 9.95828742E+04 | 9.95830204E+04 | -1.50E-01 | -8.69902480E+01 | -8.68441626E+01 | -1.50E-01 |
| 1222 | Cd | 108 | 48 | 60 | 12 | 8.55002000E+00 | 8.55013929E+00 | -1.20E-04 | 1.00487426E+05 | 1.00487378E+05 | 4.80E-02 | 1.00512106E+05 | 1.00512058E+05 | 4.80E-02 | -8.92525510E+01 | -8.93002872E+01 | 4.80E-02 |
| 1223 | Cd | 109 | 48 | 61 | 13 | 8.53876400E+00 | 8.54212860E+00 | -3.40E-03 | 1.01419668E+05 | 1.01419267E+05 | 4.00E-01 | 1.01444348E+05 | 1.01443947E+05 | 4.00E-01 | -8.85043630E+01 | -8.89059425E+01 | 4.00E-01 |
| 1224 | Cd | 110 | 48 | 62 | 14 | 8.55128200E+00 | 8.55591596E+00 | -4.60E-03 | 1.02349318E+05 | 1.02348773E+05 | 5.40E-01 | 1.02373998E+05 | 1.02373453E+05 | 5.40E-01 | -9.03487650E+01 | -9.08933612E+01 | 5.40E-01 |
| 1225 | Cd | 111 | 48 | 63 | 15 | 8.53708700E+00 | 8.54388516E+00 | -6.80E-03 | 1.03281908E+05 | 1.03281118E+05 | 7.90E-01 | 1.03306588E+05 | 1.03305798E+05 | 7.90E-01 | -8.92530790E+01 | -9.00425395E+01 | 7.90E-01 |
| 1226 | Cd | 112 | 48 | 64 | 16 | 8.54473800E+00 | 8.55292772E+00 | -8.20E-03 | 1.04212079E+05 | 1.04211127E+05 | 9.50E-01 | 1.04236759E+05 | 1.04235807E+05 | 9.50E-01 | -9.05758050E+01 | -9.15278731E+01 | 9.50E-01 |
| 1227 | Cd | 113 | 48 | 65 | 17 | 8.52698600E+00 | 8.53684102E+00 | -9.90E-03 | 1.05145106E+05 | 1.05143957E+05 | 1.10E+00 | 1.05169786E+05 | 1.05168637E+05 | 1.10E+00 | -8.90432510E+01 | -9.01916842E+01 | 1.10E+00 |
| 1228 | Cd | 114 | 48 | 66 | 18 | 8.53151200E+00 | 8.54137517E+00 | -9.90E-03 | 1.06075628E+05 | 1.06074469E+05 | 1.20E+00 | 1.06100308E+05 | 1.06099149E+05 | 1.20E+00 | -9.00148420E+01 | -9.11740998E+01 | 1.20E+00 |
| 1229 | Cd | 115 | 48 | 67 | 19 | 8.51072400E+00 | 8.52150512E+00 | -1.10E-02 | 1.07009053E+05 | 1.07007778E+05 | 1.30E+00 | 1.07033733E+05 | 1.07032458E+05 | 1.30E+00 | -8.80843890E+01 | -8.93591002E+01 | 1.30E+00 |
| 1230 | Cd | 116 | 48 | 68 | 20 | 8.51235100E+00 | 8.52188824E+00 | -9.50E-03 | 1.07939918E+05 | 1.07938777E+05 | 1.10E+00 | 1.07964598E+05 | 1.07963457E+05 | 1.10E+00 | -8.87125560E+01 | -8.98537279E+01 | 1.10E+00 |
| 1231 | Cd | 117 | 48 | 69 | 21 | 8.48897400E+00 | 8.49852849E+00 | -9.60E-03 | 1.08873707E+05 | 1.08872554E+05 | 1.20E+00 | 1.08898387E+05 | 1.08897234E+05 | 1.20E+00 | -8.64184650E+01 | -8.75712064E+01 | 1.20E+00 |
| 1232 | Cd | 118 | 48 | 70 | 22 | 8.48783500E+00 | 8.49528922E+00 | -7.50E-03 | 1.09804917E+05 | 1.09804003E+05 | 9.10E-01 | 1.09829598E+05 | 1.09828683E+05 | 9.10E-01 | -8.67017200E+01 | -8.76161823E+01 | 9.10E-01 |
| 1233 | Cd | 119 | 48 | 71 | 23 | 8.46143900E+00 | 8.46902917E+00 | -7.60E-03 | 1.10739136E+05 | 1.10738198E+05 | 9.40E-01 | 1.10763816E+05 | 1.10762878E+05 | 9.40E-01 | -8.39771170E+01 | -8.49152065E+01 | 9.40E-01 |
| 1234 | Cd | 120 | 48 | 72 | 24 | 8.45802300E+00 | 8.46295117E+00 | -4.90E-03 | 1.11670650E+05 | 1.11670024E+05 | 6.30E-01 | 1.11695330E+05 | 1.11694704E+05 | 6.30E-01 | -8.39573540E+01 | -8.45835563E+01 | 6.30E-01 |
| 1235 | Cd | 121 | 48 | 73 | 25 | 8.43099600E+00 | 8.43466245E+00 | -3.70E-03 | 1.12605028E+05 | 1.12604549E+05 | 4.80E-01 | 1.12629708E+05 | 1.12629229E+05 | 4.80E-01 | -8.10738250E+01 | -8.15522532E+01 | 4.80E-01 |
| 1236 | Cd | 122 | 48 | 74 | 26 | 8.42426600E+00 | 8.42681516E+00 | -2.50E-03 | 1.13536983E+05 | 1.13536637E+05 | 3.50E-01 | 1.13561663E+05 | 1.13561317E+05 | 3.50E-01 | -8.06123730E+01 | -8.09582270E+01 | 3.50E-01 |
| 1237 | Cd | 123 | 48 | 75 | 27 | 8.39539500E+00 | 8.39761170E+00 | -2.20E-03 | 1.14471675E+05 | 1.14471368E+05 | 3.10E-01 | 1.14496355E+05 | 1.14496048E+05 | 3.10E-01 | -7.74141810E+01 | -7.77216978E+01 | 3.10E-01 |
| 1238 | Cd | 124 | 48 | 76 | 28 | 8.38703500E+00 | 8.38908187E+00 | -2.00E-03 | 1.15403882E+05 | 1.15403593E+05 | 2.90E-01 | 1.15428562E+05 | 1.15428273E+05 | 2.90E-01 | -7.67016710E+01 | -7.69902925E+01 | 2.90E-01 |
| 1239 | Cd | 125 | 48 | 77 | 29 | 8.35768100E+00 | 8.35967936E+00 | -2.00E-03 | 1.16338730E+05 | 1.16338445E+05 | 2.80E-01 | 1.16363409E+05 | 1.16363125E+05 | 2.80E-01 | -7.33480940E+01 | -7.36327416E+01 | 2.80E-01 |
| 1240 | Cd | 126 | 48 | 78 | 30 | 8.34674700E+00 | 8.35043392E+00 | -3.70E-03 | 1.17271315E+05 | 1.17270815E+05 | 5.00E-01 | 1.17295995E+05 | 1.17295496E+05 | 5.00E-01 | -7.22568020E+01 | -7.27561765E+01 | 5.00E-01 |
| 1241 | Cd | 127 | 48 | 79 | 31 | 8.31492200E+00 | 8.31953009E+00 | -4.60E-03 | 1.18206575E+05 | 1.18205955E+05 | 6.20E-01 | 1.18231255E+05 | 1.18230635E+05 | 6.20E-01 | -6.84905140E+01 | -6.91105048E+01 | 6.20E-01 |
| 1242 | Cd | 128 | 48 | 80 | 32 | 8.30326400E+00 | 8.30668096E+00 | -3.40E-03 | 1.19139318E+05 | 1.19138846E+05 | 4.70E-01 | 1.19163998E+05 | 1.19163526E+05 | 4.70E-01 | -6.72418900E+01 | -6.77140272E+01 | 4.70E-01 |
| 1243 | Cd | 129 | 48 | 81 | 33 | 8.27300000E+00 | 8.26999783E+00 | 3.00E-03 | 1.20074545E+05 | 1.20074837E+05 | -2.90E-01 | 1.20099225E+05 | 1.20099517E+05 | -2.90E-01 | -6.35090000E+01 | -6.32172654E+01 | -2.90E-01 |
| 1244 | Cd | 130 | 48 | 82 | 34 | 8.25578800E+00 | 8.24888939E+00 | 6.90E-03 | 1.21008014E+05 | 1.21008876E+05 | -8.60E-01 | 1.21032694E+05 | 1.21033556E+05 | -8.60E-01 | -6.15338610E+01 | -6.06718465E+01 | -8.60E-01 |
| 1245 | Cd | 131 | 48 | 83 | 35 | 8.20700000E+00 | 8.20309126E+00 | 3.90E-03 | 1.21945711E+05 | 1.21946192E+05 | -4.80E-01 | 1.21970391E+05 | 1.21970872E+05 | -4.80E-01 | -5.53310000E+01 | -5.48498619E+01 | -4.80E-01 |
| 1246 | Cd | 132 | 48 | 84 | 36 | 8.16800000E+00 | 8.17316253E+00 | -5.20E-03 | 1.22882273E+05 | 1.22881505E+05 | 7.70E-01 | 1.22906953E+05 | 1.22906185E+05 | 7.70E-01 | -5.02630000E+01 | -5.10310414E+01 | 7.70E-01 |
| 1247 | CD | 133 | 48 | 85 | 37 | 8.11900000E+00 | 8.12083940E+00 | -1.80E-03 | 1.23820110E+05 | 1.23819856E+05 | 2.50E-01 | 1.23844790E+05 | 1.23844536E+05 | 2.50E-01 | -4.39200000E+01 | -4.41739088E+01 | 2.50E-01 |
| 1248 | In | 98 | 49 | 49 | 0 | 8.23000000E+00 | 8.22123303E+00 | 8.80E-03 | 9.12073229E+04 | 9.12081568E+04 | -8.30E-01 | 9.12325217E+04 | 9.12333556E+04 | -8.30E-01 | -5.38960000E+01 | -5.30623480E+01 | -8.30E-01 |
| 1249 | In | 99 | 49 | 50 | 1 | 8.30400000E+00 | 8.28694234E+00 | 1.70E-02 | 9.21313371E+04 | 9.21329957E+04 | -1.70E+00 | 9.21565359E+04 | 9.21581946E+04 | -1.70E+00 | -6.13760000E+01 | -5.97174835E+01 | -1.70E+00 |
| 1250 | In | 100 | 49 | 51 | 2 | 8.33109700E+00 | 8.31929606E+00 | 1.20E-02 | 9.30598943E+04 | 9.30610388E+04 | -1.10E+00 | 9.30850931E+04 | 9.30862376E+04 | -1.10E+00 | -6.43129720E+01 | -6.31684794E+01 | -1.10E+00 |
| 1251 | In | 101 | 49 | 52 | 3 | 8.37100000E+00 | 8.36644861E+00 | 4.60E-03 | 9.39870875E+04 | 9.39875225E+04 | -4.30E-01 | 9.40122863E+04 | 9.40127213E+04 | -4.30E-01 | -6.86140000E+01 | -6.81788637E+01 | -4.40E-01 |
| 1252 | In | 102 | 49 | 53 | 4 | 8.38856000E+00 | 8.38439214E+00 | 4.20E-03 | 9.49165016E+04 | 9.49168912E+04 | -3.90E-01 | 9.49417004E+04 | 9.49420900E+04 | -3.90E-01 | -7.06937450E+01 | -7.03042329E+01 | -3.90E-01 |
| 1253 | In | 103 | 49 | 54 | 5 | 8.42369100E+00 | 8.41950159E+00 | 4.20E-03 | 9.58440599E+04 | 9.58444559E+04 | -4.00E-01 | 9.58692587E+04 | 9.58696547E+04 | -4.00E-01 | -7.46295150E+01 | -7.42335794E+01 | -4.00E-01 |
| 1254 | In | 104 | 49 | 55 | 6 | 8.43523700E+00 | 8.42965062E+00 | 5.60E-03 | 9.67740008E+04 | 9.67745463E+04 | -5.50E-01 | 9.67991996E+04 | 9.67997451E+04 | -5.50E-01 | -7.61826640E+01 | -7.56372610E+01 | -5.50E-01 |
| 1255 | In | 105 | 49 | 56 | 7 | 8.46470400E+00 | 8.45814878E+00 | 6.60E-03 | 9.77020370E+04 | 9.77026897E+04 | -6.50E-01 | 9.77272358E+04 | 9.77278885E+04 | -6.50E-01 | -7.96405710E+01 | -7.89878997E+01 | -6.50E-01 |
| 1256 | In | 106 | 49 | 57 | 8 | 8.47011900E+00 | 8.46400421E+00 | 6.10E-03 | 9.86325636E+04 | 9.86331762E+04 | -6.10E-01 | 9.86577624E+04 | 9.86583751E+04 | -6.10E-01 | -8.06080120E+01 | -7.99954053E+01 | -6.10E-01 |
| 1257 | In | 107 | 49 | 58 | 9 | 8.49402100E+00 | 8.48782461E+00 | 6.20E-03 | 9.95611015E+04 | 9.95617288E+04 | -6.30E-01 | 9.95863003E+04 | 9.95869277E+04 | -6.30E-01 | -8.35642480E+01 | -8.29368727E+01 | -6.30E-01 |
| 1258 | In | 108 | 49 | 59 | 10 | 8.49525200E+00 | 8.49009193E+00 | 5.20E-03 | 1.00492040E+05 | 1.00492562E+05 | -5.20E-01 | 1.00517239E+05 | 1.00517760E+05 | -5.20E-01 | -8.41199410E+01 | -8.35982493E+01 | -5.20E-01 |
| 1259 | In | 109 | 49 | 60 | 11 | 8.51308700E+00 | 8.50937743E+00 | 3.70E-03 | 1.01421166E+05 | 1.01421535E+05 | -3.70E-01 | 1.01446365E+05 | 1.01446734E+05 | -3.70E-01 | -8.64879310E+01 | -8.61191424E+01 | -3.70E-01 |
| 1260 | In | 110 | 49 | 61 | 12 | 8.50891500E+00 | 8.50784728E+00 | 1.10E-03 | 1.02352677E+05 | 1.02352759E+05 | -8.20E-02 | 1.02377876E+05 | 1.02377958E+05 | -8.20E-02 | -8.64707650E+01 | -8.63888842E+01 | -8.20E-02 |
| 1261 | In | 111 | 49 | 62 | 13 | 8.52227100E+00 | 8.52241532E+00 | -1.40E-04 | 1.03282251E+05 | 1.03282199E+05 | 5.20E-02 | 1.03307450E+05 | 1.03307398E+05 | 5.20E-02 | -8.83909010E+01 | -8.84424645E+01 | 5.20E-02 |
| 1262 | In | 112 | 49 | 63 | 14 | 8.51467500E+00 | 8.51688806E+00 | -2.20E-03 | 1.04214145E+05 | 1.04213862E+05 | 2.80E-01 | 1.04239344E+05 | 1.04239060E+05 | 2.80E-01 | -8.79910710E+01 | -8.82745076E+01 | 2.80E-01 |
| 1263 | In | 113 | 49 | 64 | 15 | 8.52291700E+00 | 8.52672085E+00 | -3.80E-03 | 1.05144264E+05 | 1.05143799E+05 | 4.70E-01 | 1.05169463E+05 | 1.05168998E+05 | 4.70E-01 | -8.93658210E+01 | -8.98311820E+01 | 4.70E-01 |
| 1264 | In | 114 | 49 | 65 | 16 | 8.51196100E+00 | 8.51720336E+00 | -5.20E-03 | 1.06076556E+05 | 1.06075923E+05 | 6.30E-01 | 1.06101754E+05 | 1.06101121E+05 | 6.30E-01 | -8.85683970E+01 | -8.92015903E+01 | 6.30E-01 |
| 1265 | In | 115 | 49 | 66 | 17 | 8.51654600E+00 | 8.52254541E+00 | -6.00E-03 | 1.07007082E+05 | 1.07006356E+05 | 7.30E-01 | 1.07032281E+05 | 1.07031555E+05 | 7.30E-01 | -8.95363430E+01 | -9.02618106E+01 | 7.30E-01 |
| 1266 | In | 116 | 49 | 67 | 18 | 8.50161700E+00 | 8.50933059E+00 | -7.70E-03 | 1.07939863E+05 | 1.07938932E+05 | 9.30E-01 | 1.07965061E+05 | 1.07964131E+05 | 9.30E-01 | -8.82497460E+01 | -8.91801173E+01 | 9.30E-01 |
| 1267 | In | 117 | 49 | 68 | 19 | 8.50386500E+00 | 8.51052972E+00 | -6.70E-03 | 1.08870663E+05 | 1.08869848E+05 | 8.20E-01 | 1.08895862E+05 | 1.08895047E+05 | 8.20E-01 | -8.89430730E+01 | -8.97584277E+01 | 8.20E-01 |
| 1268 | In | 118 | 49 | 69 | 20 | 8.48566700E+00 | 8.49389239E+00 | -8.20E-03 | 1.09803872E+05 | 1.09802866E+05 | 1.00E+00 | 1.09829071E+05 | 1.09828065E+05 | 1.00E+00 | -8.72282500E+01 | -8.82344327E+01 | 1.00E+00 |
| 1269 | In | 119 | 49 | 70 | 21 | 8.48614500E+00 | 8.49146094E+00 | -5.30E-03 | 1.10734895E+05 | 1.10734227E+05 | 6.70E-01 | 1.10760094E+05 | 1.10759426E+05 | 6.70E-01 | -8.76995000E+01 | -8.83676637E+01 | 6.70E-01 |
| 1270 | In | 120 | 49 | 71 | 22 | 8.46626400E+00 | 8.47196158E+00 | -5.70E-03 | 1.11668360E+05 | 1.11667641E+05 | 7.20E-01 | 1.11693559E+05 | 1.11692839E+05 | 7.20E-01 | -8.57285910E+01 | -8.64478822E+01 | 7.20E-01 |
| 1271 | In | 121 | 49 | 72 | 23 | 8.46388900E+00 | 8.46666212E+00 | -2.80E-03 | 1.12599746E+05 | 1.12599375E+05 | 3.70E-01 | 1.12624945E+05 | 1.12624574E+05 | 3.70E-01 | -8.58361830E+01 | -8.62072908E+01 | 3.70E-01 |
| 1272 | In | 122 | 49 | 73 | 24 | 8.44212000E+00 | 8.44511447E+00 | -3.00E-03 | 1.13533504E+05 | 1.13533103E+05 | 4.00E-01 | 1.13558702E+05 | 1.13558302E+05 | 4.00E-01 | -8.35729540E+01 | -8.39738209E+01 | 4.00E-01 |
| 1273 | In | 123 | 49 | 74 | 25 | 8.43794700E+00 | 8.43797573E+00 | -2.90E-05 | 1.14465140E+05 | 1.14465101E+05 | 3.90E-02 | 1.14490339E+05 | 1.14490300E+05 | 3.90E-02 | -8.34304540E+01 | -8.34695512E+01 | 3.90E-02 |
| 1274 | In | 124 | 49 | 75 | 26 | 8.41434300E+00 | 8.41540255E+00 | -1.10E-03 | 1.15399194E+05 | 1.15399028E+05 | 1.70E-01 | 1.15424393E+05 | 1.15424226E+05 | 1.70E-01 | -8.08702170E+01 | -8.10371329E+01 | 1.70E-01 |
| 1275 | In | 125 | 49 | 76 | 27 | 8.40845200E+00 | 8.40745441E+00 | 1.00E-03 | 1.16331082E+05 | 1.16331171E+05 | -8.90E-02 | 1.16356281E+05 | 1.16356370E+05 | -8.90E-02 | -8.04768400E+01 | -8.03876996E+01 | -8.90E-02 |
| 1276 | In | 126 | 49 | 77 | 28 | 8.38431700E+00 | 8.38447458E+00 | -1.60E-04 | 1.17265280E+05 | 1.17265224E+05 | 5.50E-02 | 1.17290479E+05 | 1.17290423E+05 | 5.50E-02 | -7.77729290E+01 | -7.78283761E+01 | 5.50E-02 |
| 1277 | In | 127 | 49 | 78 | 29 | 8.37496500E+00 | 8.37566327E+00 | -7.00E-04 | 1.18197649E+05 | 1.18197524E+05 | 1.20E-01 | 1.18222848E+05 | 1.18222723E+05 | 1.20E-01 | -7.68982980E+01 | -7.70224950E+01 | 1.20E-01 |
| 1278 | In | 128 | 49 | 79 | 30 | 8.35109000E+00 | 8.35095533E+00 | 1.30E-04 | 1.19131895E+05 | 1.19131877E+05 | 1.80E-02 | 1.19157094E+05 | 1.19157076E+05 | 1.80E-02 | -7.41459500E+01 | -7.41642237E+01 | 1.80E-02 |
| 1279 | In | 129 | 49 | 80 | 31 | 8.33878200E+00 | 8.33848760E+00 | 2.90E-04 | 1.20064697E+05 | 1.20064700E+05 | -2.30E-03 | 1.20089896E+05 | 1.20089898E+05 | -2.30E-03 | -7.28378920E+01 | -7.28355227E+01 | -2.40E-03 |
| 1280 | In | 130 | 49 | 81 | 32 | 8.31400100E+00 | 8.30785896E+00 | 6.10E-03 | 1.20999145E+05 | 1.20999908E+05 | -7.60E-01 | 1.21024344E+05 | 1.21025107E+05 | -7.60E-01 | -6.98838610E+01 | -6.91209680E+01 | -7.60E-01 |
| 1281 | In | 131 | 49 | 82 | 33 | 8.29796300E+00 | 8.28718761E+00 | 1.10E-02 | 1.21932498E+05 | 1.21933874E+05 | -1.40E+00 | 1.21957696E+05 | 1.21959072E+05 | -1.40E+00 | -6.80255820E+01 | -6.66495612E+01 | -1.40E+00 |
| 1282 | In | 132 | 49 | 83 | 34 | 8.25369500E+00 | 8.24725940E+00 | 6.40E-03 | 1.22869608E+05 | 1.22870422E+05 | -8.10E-01 | 1.22894807E+05 | 1.22895621E+05 | -8.10E-01 | -6.24089120E+01 | -6.15949055E+01 | -8.10E-01 |
| 1283 | In | 133 | 49 | 84 | 35 | 8.21500000E+00 | 8.21768314E+00 | -2.70E-03 | 1.23806047E+05 | 1.23805674E+05 | 3.70E-01 | 1.23831246E+05 | 1.23830873E+05 | 3.70E-01 | -5.74640000E+01 | -5.78372036E+01 | 3.70E-01 |
| 1284 | In | 134 | 49 | 85 | 36 | 8.17100000E+00 | 8.17077654E+00 | 2.20E-04 | 1.24743345E+05 | 1.24743307E+05 | 3.70E-02 | 1.24768543E+05 | 1.24768506E+05 | 3.70E-02 | -5.16610000E+01 | -5.16980828E+01 | 3.70E-02 |
| 1285 | In | 135 | 49 | 86 | 37 | 8.13200000E+00 | 8.13670122E+00 | -4.70E-03 | 1.25679971E+05 | 1.25679302E+05 | 6.70E-01 | 1.25705170E+05 | 1.25704501E+05 | 6.70E-01 | -4.65280000E+01 | -4.71973727E+01 | 6.70E-01 |
| 1286 | Sn | 100 | 50 | 50 | 0 | 8.25297400E+00 | 8.22945254E+00 | 2.40E-02 | 9.30664053E+04 | 9.30687211E+04 | -2.30E+00 | 9.30921231E+04 | 9.30944389E+04 | -2.30E+00 | -5.72829720E+01 | -5.49672038E+01 | -2.30E+00 |
| 1287 | Sn | 101 | 50 | 51 | 1 | 8.28110200E+00 | 8.26477132E+00 | 1.60E-02 | 9.39948767E+04 | 9.39964898E+04 | -1.60E+00 | 9.40205945E+04 | 9.40222076E+04 | -1.60E+00 | -6.03056250E+01 | -5.86925343E+01 | -1.60E+00 |
| 1288 | Sn | 102 | 50 | 52 | 2 | 8.32441900E+00 | 8.31891846E+00 | 5.50E-03 | 9.49217427E+04 | 9.49222674E+04 | -5.20E-01 | 9.49474605E+04 | 9.49479852E+04 | -5.20E-01 | -6.49337450E+01 | -6.44089947E+01 | -5.20E-01 |
| 1289 | Sn | 103 | 50 | 53 | 3 | 8.34172700E+00 | 8.33894574E+00 | 2.80E-03 | 9.58512010E+04 | 9.58514510E+04 | -2.50E-01 | 9.58769188E+04 | 9.58771689E+04 | -2.50E-01 | -6.69695150E+01 | -6.67194044E+01 | -2.50E-01 |
| 1290 | Sn | 104 | 50 | 54 | 4 | 8.38391100E+00 | 8.38067634E+00 | 3.20E-03 | 9.67780375E+04 | 9.67783375E+04 | -3.00E-01 | 9.68037553E+04 | 9.68040553E+04 | -3.00E-01 | -7.16270410E+01 | -7.13270137E+01 | -3.00E-01 |
| 1291 | Sn | 105 | 50 | 55 | 5 | 8.39722800E+00 | 8.39222714E+00 | 5.00E-03 | 9.77078206E+04 | 9.77083094E+04 | -4.90E-01 | 9.77335384E+04 | 9.77340272E+04 | -4.90E-01 | -7.33379860E+01 | -7.28492052E+01 | -4.90E-01 |
| 1292 | Sn | 106 | 50 | 56 | 6 | 8.43203700E+00 | 8.42696511E+00 | 5.10E-03 | 9.86352989E+04 | 9.86358003E+04 | -5.00E-01 | 9.86610168E+04 | 9.86615181E+04 | -5.00E-01 | -7.73536800E+01 | -7.68523379E+01 | -5.00E-01 |
| 1293 | Sn | 107 | 50 | 57 | 7 | 8.43949400E+00 | 8.43373526E+00 | 5.80E-03 | 9.95656345E+04 | 9.95662143E+04 | -5.80E-01 | 9.95913523E+04 | 9.95919321E+04 | -5.80E-01 | -7.85122270E+01 | -7.79323895E+01 | -5.80E-01 |
| 1294 | Sn | 108 | 50 | 58 | 8 | 8.46902700E+00 | 8.46352526E+00 | 5.50E-03 | 1.00493571E+05 | 1.00494129E+05 | -5.60E-01 | 1.00519289E+05 | 1.00519846E+05 | -5.60E-01 | -8.20699430E+01 | -8.15121262E+01 | -5.60E-01 |
| 1295 | Sn | 109 | 50 | 59 | 9 | 8.47052400E+00 | 8.46640902E+00 | 4.10E-03 | 1.01424504E+05 | 1.01424916E+05 | -4.10E-01 | 1.01450222E+05 | 1.01450634E+05 | -4.10E-01 | -8.26309140E+01 | -8.22186626E+01 | -4.10E-01 |
| 1296 | Sn | 110 | 50 | 60 | 10 | 8.49608700E+00 | 8.49152621E+00 | 4.60E-03 | 1.02352787E+05 | 1.02353252E+05 | -4.70E-01 | 1.02378505E+05 | 1.02378970E+05 | -4.70E-01 | -8.58419830E+01 | -8.53766426E+01 | -4.70E-01 |
| 1297 | Sn | 111 | 50 | 61 | 11 | 8.49313900E+00 | 8.49045018E+00 | 2.70E-03 | 1.03284183E+05 | 1.03284446E+05 | -2.60E-01 | 1.03309901E+05 | 1.03310163E+05 | -2.60E-01 | -8.59395180E+01 | -8.56774112E+01 | -2.60E-01 |
| 1298 | Sn | 112 | 50 | 62 | 12 | 8.51362700E+00 | 8.51081253E+00 | 2.80E-03 | 1.04212961E+05 | 1.04213240E+05 | -2.80E-01 | 1.04238679E+05 | 1.04238958E+05 | -2.80E-01 | -8.86559900E+01 | -8.83771258E+01 | -2.80E-01 |
| 1299 | Sn | 113 | 50 | 63 | 13 | 8.50681200E+00 | 8.50565079E+00 | 1.20E-03 | 1.05144783E+05 | 1.05144878E+05 | -9.50E-02 | 1.05170501E+05 | 1.05170596E+05 | -9.50E-02 | -8.83282400E+01 | -8.82333427E+01 | -9.50E-02 |
| 1300 | Sn | 114 | 50 | 64 | 14 | 8.52254500E+00 | 8.52129905E+00 | 1.20E-03 | 1.06074048E+05 | 1.06074154E+05 | -1.10E-01 | 1.06099766E+05 | 1.06099871E+05 | -1.10E-01 | -9.05573360E+01 | -9.04515752E+01 | -1.10E-01 |
| 1301 | Sn | 115 | 50 | 65 | 15 | 8.51406900E+00 | 8.51208452E+00 | 2.00E-03 | 1.07006065E+05 | 1.07006257E+05 | -1.90E-01 | 1.07031783E+05 | 1.07031975E+05 | -1.90E-01 | -9.00338330E+01 | -8.98418853E+01 | -1.90E-01 |
| 1302 | Sn | 116 | 50 | 66 | 16 | 8.52311600E+00 | 8.52329496E+00 | -1.80E-04 | 1.07936067E+05 | 1.07936010E+05 | 5.70E-02 | 1.07961785E+05 | 1.07961728E+05 | 5.70E-02 | -9.15259950E+01 | -9.15830612E+01 | 5.70E-02 |
| 1303 | Sn | 117 | 50 | 67 | 17 | 8.50961200E+00 | 8.51033155E+00 | -7.20E-04 | 1.08868689E+05 | 1.08868569E+05 | 1.20E-01 | 1.08894407E+05 | 1.08894287E+05 | 1.20E-01 | -9.03977820E+01 | -9.05183188E+01 | 1.20E-01 |
| 1304 | Sn | 118 | 50 | 68 | 18 | 8.51653400E+00 | 8.51745989E+00 | -9.30E-04 | 1.09798928E+05 | 1.09798783E+05 | 1.50E-01 | 1.09824646E+05 | 1.09824501E+05 | 1.50E-01 | -9.16528860E+01 | -9.17984748E+01 | 1.50E-01 |
| 1305 | Sn | 119 | 50 | 69 | 19 | 8.49944900E+00 | 8.50100385E+00 | -1.60E-03 | 1.10732010E+05 | 1.10731789E+05 | 2.20E-01 | 1.10757728E+05 | 1.10757507E+05 | 2.20E-01 | -9.00650630E+01 | -9.02863476E+01 | 2.20E-01 |
| 1306 | Sn | 120 | 50 | 70 | 20 | 8.50449100E+00 | 8.50455398E+00 | -6.00E-05 | 1.11662471E+05 | 1.11662427E+05 | 4.30E-02 | 1.11688189E+05 | 1.11688145E+05 | 4.30E-02 | -9.10985910E+01 | -9.11420476E+01 | 4.30E-02 |
| 1307 | Sn | 121 | 50 | 71 | 21 | 8.48520300E+00 | 8.48515155E+00 | 5.10E-05 | 1.12595866E+05 | 1.12595836E+05 | 3.00E-02 | 1.12621584E+05 | 1.12621554E+05 | 3.00E-02 | -8.91974860E+01 | -8.92275888E+01 | 3.00E-02 |
| 1308 | Sn | 122 | 50 | 72 | 22 | 8.48790900E+00 | 8.48586360E+00 | 2.00E-03 | 1.13526616E+05 | 1.13526829E+05 | -2.10E-01 | 1.13552334E+05 | 1.13552547E+05 | -2.10E-01 | -8.99415450E+01 | -8.97282915E+01 | -2.10E-01 |
| 1309 | Sn | 123 | 50 | 73 | 23 | 8.46724400E+00 | 8.46430000E+00 | 2.90E-03 | 1.14460235E+05 | 1.14460561E+05 | -3.30E-01 | 1.14485953E+05 | 1.14486279E+05 | -3.30E-01 | -8.78163910E+01 | -8.74905136E+01 | -3.30E-01 |
| 1310 | Sn | 124 | 50 | 74 | 24 | 8.46742100E+00 | 8.46315120E+00 | 4.30E-03 | 1.15391311E+05 | 1.15391805E+05 | -4.90E-01 | 1.15417029E+05 | 1.15417523E+05 | -4.90E-01 | -8.82342370E+01 | -8.77410422E+01 | -4.90E-01 |
| 1311 | Sn | 125 | 50 | 75 | 25 | 8.44555000E+00 | 8.44040005E+00 | 5.10E-03 | 1.16325143E+05 | 1.16325751E+05 | -6.10E-01 | 1.16350861E+05 | 1.16351469E+05 | -6.10E-01 | -8.58964200E+01 | -8.52889807E+01 | -6.10E-01 |
| 1312 | Sn | 126 | 50 | 76 | 26 | 8.44352300E+00 | 8.43834330E+00 | 5.20E-03 | 1.17256519E+05 | 1.17257135E+05 | -6.20E-01 | 1.17282236E+05 | 1.17282853E+05 | -6.20E-01 | -8.60152860E+01 | -8.53989121E+01 | -6.20E-01 |
| 1313 | Sn | 127 | 50 | 77 | 27 | 8.42056000E+00 | 8.41497700E+00 | 5.60E-03 | 1.18190557E+05 | 1.18191229E+05 | -6.70E-01 | 1.18216274E+05 | 1.18216947E+05 | -6.70E-01 | -8.34712300E+01 | -8.27984165E+01 | -6.70E-01 |
| 1314 | Sn | 128 | 50 | 78 | 28 | 8.41697900E+00 | 8.41191043E+00 | 5.10E-03 | 1.19122160E+05 | 1.19122772E+05 | -6.10E-01 | 1.19147878E+05 | 1.19148490E+05 | -6.10E-01 | -8.33620210E+01 | -8.27495528E+01 | -6.10E-01 |
| 1315 | Sn | 129 | 50 | 79 | 29 | 8.39294300E+00 | 8.38664737E+00 | 6.30E-03 | 1.20056409E+05 | 1.20057185E+05 | -7.80E-01 | 1.20082127E+05 | 1.20082903E+05 | -7.80E-01 | -8.06071020E+01 | -7.98312096E+01 | -7.80E-01 |
| 1316 | Sn | 130 | 50 | 80 | 30 | 8.38682100E+00 | 8.37985293E+00 | 7.00E-03 | 1.20988377E+05 | 1.20989247E+05 | -8.70E-01 | 1.21014095E+05 | 1.21014965E+05 | -8.70E-01 | -8.01328610E+01 | -7.92632608E+01 | -8.70E-01 |
| 1317 | Sn | 131 | 50 | 81 | 31 | 8.36257500E+00 | 8.34867030E+00 | 1.40E-02 | 1.21922732E+05 | 1.21924517E+05 | -1.80E+00 | 1.21948450E+05 | 1.21950235E+05 | -1.80E+00 | -7.72721280E+01 | -7.54868706E+01 | -1.80E+00 |
| 1318 | Sn | 132 | 50 | 82 | 32 | 8.35485200E+00 | 8.33372042E+00 | 2.10E-02 | 1.22854954E+05 | 1.22857707E+05 | -2.80E+00 | 1.22880672E+05 | 1.22883425E+05 | -2.80E+00 | -7.65439120E+01 | -7.37908374E+01 | -2.80E+00 |
| 1319 | Sn | 133 | 50 | 83 | 33 | 8.31009100E+00 | 8.29330160E+00 | 1.70E-02 | 1.23792118E+05 | 1.23794315E+05 | -2.20E+00 | 1.23817836E+05 | 1.23820033E+05 | -2.20E+00 | -7.08742080E+01 | -6.86775356E+01 | -2.20E+00 |
| 1320 | Sn | 134 | 50 | 84 | 34 | 8.27516000E+00 | 8.26936137E+00 | 5.80E-03 | 1.24728054E+05 | 1.24728795E+05 | -7.40E-01 | 1.24753772E+05 | 1.24754513E+05 | -7.40E-01 | -6.64322290E+01 | -6.56915270E+01 | -7.40E-01 |
| 1321 | Sn | 135 | 50 | 85 | 35 | 8.23068700E+00 | 8.22181487E+00 | 8.90E-03 | 1.25665348E+05 | 1.25666510E+05 | -1.20E+00 | 1.25691066E+05 | 1.25692227E+05 | -1.20E+00 | -6.06322430E+01 | -5.94707920E+01 | -1.20E+00 |
| 1322 | Sn | 136 | 50 | 86 | 36 | 8.19500000E+00 | 8.19301115E+00 | 2.00E-03 | 1.26601576E+05 | 1.26601770E+05 | -1.90E-01 | 1.26627293E+05 | 1.26627488E+05 | -2.00E-01 | -5.58990000E+01 | -5.57039827E+01 | -2.00E-01 |
| 1323 | Sn | 137 | 50 | 87 | 37 | 8.14900000E+00 | 8.14302908E+00 | 6.00E-03 | 1.27539180E+05 | 1.27539990E+05 | -8.10E-01 | 1.27564898E+05 | 1.27565708E+05 | -8.10E-01 | -4.97880000E+01 | -4.89781310E+01 | -8.10E-01 |
| 1324 | SN | 138 | 50 | 88 | 38 | 8.11300000E+00 | 8.11322694E+00 | -2.30E-04 | 1.28475602E+05 | 1.28475525E+05 | 7.60E-02 | 1.28501320E+05 | 1.28501243E+05 | 7.60E-02 | -4.48610000E+01 | -4.49371454E+01 | 7.60E-02 |
| 1325 | SB | 103 | 51 | 52 | 1 | 8.22900000E+00 | 8.22910275E+00 | -1.00E-04 | 9.58614728E+04 | 9.58614625E+04 | 1.00E-02 | 9.58877099E+04 | 9.58876996E+04 | 1.00E-02 | -5.61780000E+01 | -5.61886532E+01 | 1.10E-02 |
| 1326 | Sb | 104 | 51 | 53 | 2 | 8.25661400E+00 | 8.25708994E+00 | -4.80E-04 | 9.67899747E+04 | 9.67898881E+04 | 8.70E-02 | 9.68162118E+04 | 9.68161252E+04 | 8.70E-02 | -5.91705440E+01 | -5.92571050E+01 | 8.70E-02 |
| 1327 | Sb | 105 | 51 | 54 | 3 | 8.30099700E+00 | 8.30175184E+00 | -7.50E-04 | 9.77166232E+04 | 9.77165069E+04 | 1.20E-01 | 9.77428603E+04 | 9.77427440E+04 | 1.20E-01 | -6.40160900E+01 | -6.41323756E+01 | 1.20E-01 |
| 1328 | Sb | 106 | 51 | 55 | 4 | 8.32201200E+00 | 8.32046684E+00 | 1.50E-03 | 9.86456601E+04 | 9.86457868E+04 | -1.30E-01 | 9.86718972E+04 | 9.86720239E+04 | -1.30E-01 | -6.64732910E+01 | -6.63465985E+01 | -1.30E-01 |
| 1329 | Sb | 107 | 51 | 56 | 5 | 8.35873300E+00 | 8.35721993E+00 | 1.50E-03 | 9.95729742E+04 | 9.95730991E+04 | -1.20E-01 | 9.95992113E+04 | 9.95993362E+04 | -1.20E-01 | -7.06532220E+01 | -7.05283261E+01 | -1.20E-01 |
| 1330 | Sb | 108 | 51 | 57 | 6 | 8.37266600E+00 | 8.37055180E+00 | 2.10E-03 | 1.00502676E+05 | 1.00502867E+05 | -1.90E-01 | 1.00528913E+05 | 1.00529105E+05 | -1.90E-01 | -7.24453300E+01 | -7.22540693E+01 | -1.90E-01 |
| 1331 | Sb | 109 | 51 | 58 | 7 | 8.40481600E+00 | 8.40174089E+00 | 3.10E-03 | 1.01430364E+05 | 1.01430663E+05 | -3.00E-01 | 1.01456602E+05 | 1.01456900E+05 | -3.00E-01 | -7.62510540E+01 | -7.59529127E+01 | -3.00E-01 |
| 1332 | Sb | 110 | 51 | 59 | 8 | 8.41268100E+00 | 8.41080663E+00 | 1.90E-03 | 1.02360660E+05 | 1.02360829E+05 | -1.70E-01 | 1.02386897E+05 | 1.02387066E+05 | -1.70E-01 | -7.74497330E+01 | -7.72805668E+01 | -1.70E-01 |
| 1333 | Sb | 111 | 51 | 60 | 9 | 8.44011900E+00 | 8.43693585E+00 | 3.20E-03 | 1.03288767E+05 | 1.03289083E+05 | -3.20E-01 | 1.03315004E+05 | 1.03315320E+05 | -3.20E-01 | -8.08367360E+01 | -8.05203973E+01 | -3.20E-01 |
| 1334 | Sb | 112 | 51 | 61 | 10 | 8.44363200E+00 | 8.44184434E+00 | 1.80E-03 | 1.04219499E+05 | 1.04219662E+05 | -1.60E-01 | 1.04245736E+05 | 1.04245899E+05 | -1.60E-01 | -8.15989640E+01 | -8.14357653E+01 | -1.60E-01 |
| 1335 | Sb | 113 | 51 | 62 | 11 | 8.46527600E+00 | 8.46297767E+00 | 2.30E-03 | 1.05148175E+05 | 1.05148397E+05 | -2.20E-01 | 1.05174412E+05 | 1.05174635E+05 | -2.20E-01 | -8.44170760E+01 | -8.41943565E+01 | -2.20E-01 |
| 1336 | Sb | 114 | 51 | 63 | 12 | 8.46251000E+00 | 8.46373093E+00 | -1.20E-03 | 1.06079590E+05 | 1.06079414E+05 | 1.80E-01 | 1.06105827E+05 | 1.06105651E+05 | 1.80E-01 | -8.44956450E+01 | -8.46718876E+01 | 1.80E-01 |
| 1337 | Sb | 115 | 51 | 64 | 13 | 8.48091500E+00 | 8.47998809E+00 | 9.30E-04 | 1.07008576E+05 | 1.07008646E+05 | -7.00E-02 | 1.07034814E+05 | 1.07034883E+05 | -7.00E-02 | -8.70034030E+01 | -8.69338718E+01 | -7.00E-02 |
| 1338 | Sb | 116 | 51 | 65 | 14 | 8.47581700E+00 | 8.47670136E+00 | -8.80E-04 | 1.07940252E+05 | 1.07940113E+05 | 1.40E-01 | 1.07966489E+05 | 1.07966350E+05 | 1.40E-01 | -8.68216540E+01 | -8.69612809E+01 | 1.40E-01 |
| 1339 | Sb | 117 | 51 | 66 | 15 | 8.48789700E+00 | 8.48840033E+00 | -5.00E-04 | 1.08869928E+05 | 1.08869833E+05 | 9.60E-02 | 1.08896166E+05 | 1.08896070E+05 | 9.60E-02 | -8.86395640E+01 | -8.87354422E+01 | 9.60E-02 |
| 1340 | Sb | 118 | 51 | 67 | 16 | 8.47891500E+00 | 8.48144106E+00 | -2.50E-03 | 1.09802066E+05 | 1.09801731E+05 | 3.40E-01 | 1.09828303E+05 | 1.09827968E+05 | 3.40E-01 | -8.79962460E+01 | -8.83313305E+01 | 3.40E-01 |
| 1341 | Sb | 119 | 51 | 68 | 17 | 8.48791000E+00 | 8.48895322E+00 | -1.00E-03 | 1.10732082E+05 | 1.10731921E+05 | 1.60E-01 | 1.10758319E+05 | 1.10758158E+05 | 1.60E-01 | -8.94742170E+01 | -8.96353995E+01 | 1.60E-01 |
| 1342 | Sb | 120 | 51 | 69 | 18 | 8.47563600E+00 | 8.47860138E+00 | -3.00E-03 | 1.11664632E+05 | 1.11664239E+05 | 3.90E-01 | 1.11690869E+05 | 1.11690477E+05 | 3.90E-01 | -8.84179830E+01 | -8.88108123E+01 | 3.90E-01 |
| 1343 | Sb | 121 | 51 | 70 | 19 | 8.48205200E+00 | 8.48243313E+00 | -3.80E-04 | 1.12594946E+05 | 1.12594863E+05 | 8.30E-02 | 1.12621183E+05 | 1.12621100E+05 | 8.30E-02 | -8.95985750E+01 | -8.96817372E+01 | 8.30E-02 |
| 1344 | Sb | 122 | 51 | 71 | 20 | 8.46831700E+00 | 8.46925104E+00 | -9.30E-04 | 1.13527705E+05 | 1.13527554E+05 | 1.50E-01 | 1.13553942E+05 | 1.13553791E+05 | 1.50E-01 | -8.83336240E+01 | -8.84846360E+01 | 1.50E-01 |
| 1345 | Sb | 123 | 51 | 72 | 21 | 8.47233500E+00 | 8.47013912E+00 | 2.20E-03 | 1.14458308E+05 | 1.14458541E+05 | -2.30E-01 | 1.14484545E+05 | 1.14484778E+05 | -2.30E-01 | -8.92248250E+01 | -8.89918017E+01 | -2.30E-01 |
| 1346 | Sb | 124 | 51 | 73 | 22 | 8.45616700E+00 | 8.45489950E+00 | 1.30E-03 | 1.15391405E+05 | 1.15391526E+05 | -1.20E-01 | 1.15417643E+05 | 1.15417763E+05 | -1.20E-01 | -8.76210040E+01 | -8.75009086E+01 | -1.20E-01 |
| 1347 | Sb | 125 | 51 | 74 | 23 | 8.45817000E+00 | 8.45379812E+00 | 4.40E-03 | 1.16322264E+05 | 1.16322774E+05 | -5.10E-01 | 1.16348501E+05 | 1.16349011E+05 | -5.10E-01 | -8.82562550E+01 | -8.77468170E+01 | -5.10E-01 |
| 1348 | Sb | 126 | 51 | 75 | 24 | 8.44031400E+00 | 8.43742568E+00 | 2.90E-03 | 1.17255621E+05 | 1.17255948E+05 | -3.30E-01 | 1.17281858E+05 | 1.17282185E+05 | -3.30E-01 | -8.63932860E+01 | -8.60663689E+01 | -3.30E-01 |
| 1349 | Sb | 127 | 51 | 76 | 25 | 8.43982000E+00 | 8.43524789E+00 | 4.60E-03 | 1.18186809E+05 | 1.18187353E+05 | -5.40E-01 | 1.18213046E+05 | 1.18213590E+05 | -5.40E-01 | -8.66994770E+01 | -8.61558956E+01 | -5.40E-01 |
| 1350 | Sb | 128 | 51 | 77 | 26 | 8.42077500E+00 | 8.41825195E+00 | 2.50E-03 | 1.19120372E+05 | 1.19120658E+05 | -2.90E-01 | 1.19146610E+05 | 1.19146895E+05 | -2.90E-01 | -8.46303020E+01 | -8.43443450E+01 | -2.90E-01 |
| 1351 | Sb | 129 | 51 | 78 | 27 | 8.41805900E+00 | 8.41489144E+00 | 3.20E-03 | 1.20051867E+05 | 1.20052239E+05 | -3.70E-01 | 1.20078105E+05 | 1.20078476E+05 | -3.70E-01 | -8.46293370E+01 | -8.42577715E+01 | -3.70E-01 |
| 1352 | Sb | 130 | 51 | 79 | 28 | 8.39736800E+00 | 8.39600064E+00 | 1.40E-03 | 1.20985705E+05 | 1.20985845E+05 | -1.40E-01 | 1.21011942E+05 | 1.21012082E+05 | -1.40E-01 | -8.22862660E+01 | -8.21455396E+01 | -1.40E-01 |
| 1353 | Sb | 131 | 51 | 80 | 29 | 8.39255600E+00 | 8.38887091E+00 | 3.70E-03 | 1.21917503E+05 | 1.21917949E+05 | -4.50E-01 | 1.21943740E+05 | 1.21944186E+05 | -4.50E-01 | -8.19819060E+01 | -8.15362278E+01 | -4.50E-01 |
| 1354 | Sb | 132 | 51 | 81 | 30 | 8.37234700E+00 | 8.36420767E+00 | 8.10E-03 | 1.22851343E+05 | 1.22852381E+05 | -1.00E+00 | 1.22877580E+05 | 1.22878618E+05 | -1.00E+00 | -7.96355730E+01 | -7.85982309E+01 | -1.00E+00 |
| 1355 | Sb | 133 | 51 | 82 | 31 | 8.36472200E+00 | 8.34908431E+00 | 1.60E-02 | 1.23783550E+05 | 1.23785593E+05 | -2.00E+00 | 1.23809788E+05 | 1.23811830E+05 | -2.00E+00 | -7.89225070E+01 | -7.68797126E+01 | -2.00E+00 |
| 1356 | Sb | 134 | 51 | 83 | 32 | 8.32595000E+00 | 8.31538289E+00 | 1.10E-02 | 1.24719947E+05 | 1.24721326E+05 | -1.40E+00 | 1.24746184E+05 | 1.24747563E+05 | -1.40E+00 | -7.40205410E+01 | -7.26414886E+01 | -1.40E+00 |
| 1357 | Sb | 135 | 51 | 84 | 33 | 8.29198300E+00 | 8.29138169E+00 | 6.00E-04 | 1.25655772E+05 | 1.25655816E+05 | -4.40E-02 | 1.25682009E+05 | 1.25682053E+05 | -4.40E-02 | -6.96896250E+01 | -6.96453899E+01 | -4.40E-02 |
| 1358 | Sb | 136 | 51 | 85 | 34 | 8.25227400E+00 | 8.25049372E+00 | 1.80E-03 | 1.26592445E+05 | 1.26592651E+05 | -2.10E-01 | 1.26618682E+05 | 1.26618888E+05 | -2.10E-01 | -6.45097980E+01 | -6.43046886E+01 | -2.10E-01 |
| 1359 | Sb | 137 | 51 | 86 | 35 | 8.21825500E+00 | 8.22149619E+00 | -3.20E-03 | 1.27528419E+05 | 1.27527938E+05 | 4.80E-01 | 1.27554656E+05 | 1.27554175E+05 | 4.80E-01 | -6.00301310E+01 | -6.05112017E+01 | 4.80E-01 |
| 1360 | Sb | 138 | 51 | 87 | 36 | 8.17700000E+00 | 8.17784978E+00 | -8.50E-04 | 1.28465404E+05 | 1.28465305E+05 | 9.90E-02 | 1.28491641E+05 | 1.28491542E+05 | 9.90E-02 | -5.45390000E+01 | -5.46381740E+01 | 9.90E-02 |
| 1361 | Sb | 139 | 51 | 88 | 37 | 8.14200000E+00 | 8.14760687E+00 | -5.60E-03 | 1.29401649E+05 | 1.29400896E+05 | 7.50E-01 | 1.29427886E+05 | 1.29427134E+05 | 7.50E-01 | -4.97880000E+01 | -5.05409414E+01 | 7.50E-01 |
| 1362 | SB | 140 | 51 | 89 | 38 | 8.10000000E+00 | 8.10367177E+00 | -3.70E-03 | 1.30338993E+05 | 1.30338465E+05 | 5.30E-01 | 1.30365230E+05 | 1.30364702E+05 | 5.30E-01 | -4.39390000E+01 | -4.44663154E+01 | 5.30E-01 |
| 1363 | Te | 105 | 52 | 53 | 1 | 8.18683600E+00 | 8.19514801E+00 | -8.30E-03 | 9.77273083E+04 | 9.77263978E+04 | 9.10E-01 | 9.77540649E+04 | 9.77531544E+04 | 9.10E-01 | -5.28115100E+01 | -5.37220501E+01 | 9.10E-01 |
| 1364 | Te | 106 | 52 | 54 | 2 | 8.23675700E+00 | 8.24721915E+00 | -1.00E-02 | 9.86533952E+04 | 9.86522485E+04 | 1.10E+00 | 9.86801518E+04 | 9.86790050E+04 | 1.10E+00 | -5.82186880E+01 | -5.93654204E+01 | 1.10E+00 |
| 1365 | Te | 107 | 52 | 55 | 3 | 8.25687100E+00 | 8.26858193E+00 | -1.20E-02 | 9.95825716E+04 | 9.95812808E+04 | 1.30E+00 | 9.96093282E+04 | 9.96080374E+04 | 1.30E+00 | -6.05363290E+01 | -6.18271379E+01 | 1.30E+00 |
| 1366 | Te | 108 | 52 | 56 | 4 | 8.30372100E+00 | 8.31191664E+00 | -8.20E-03 | 1.00508820E+05 | 1.00507897E+05 | 9.20E-01 | 1.00535577E+05 | 1.00534654E+05 | 9.20E-01 | -6.57816710E+01 | -6.67045486E+01 | 9.20E-01 |
| 1367 | Te | 109 | 52 | 57 | 5 | 8.31933000E+00 | 8.32696030E+00 | -7.60E-03 | 1.01438381E+05 | 1.01437511E+05 | 8.70E-01 | 1.01465137E+05 | 1.01464268E+05 | 8.70E-01 | -6.77153900E+01 | -6.85849055E+01 | 8.70E-01 |
| 1368 | Te | 110 | 52 | 58 | 6 | 8.35811500E+00 | 8.36414450E+00 | -6.00E-03 | 1.02365360E+05 | 1.02364659E+05 | 7.00E-01 | 1.02392117E+05 | 1.02391416E+05 | 7.00E-01 | -7.22298230E+01 | -7.29308095E+01 | 7.00E-01 |
| 1369 | Te | 111 | 52 | 59 | 7 | 8.36776300E+00 | 8.37431727E+00 | -6.60E-03 | 1.03295497E+05 | 1.03294731E+05 | 7.70E-01 | 1.03322253E+05 | 1.03321488E+05 | 7.70E-01 | -7.35874770E+01 | -7.43528116E+01 | 7.70E-01 |
| 1370 | Te | 112 | 52 | 60 | 8 | 8.40065200E+00 | 8.40606932E+00 | -5.40E-03 | 1.04223011E+05 | 1.04222366E+05 | 6.40E-01 | 1.04249767E+05 | 1.04249123E+05 | 6.40E-01 | -7.75675080E+01 | -7.82120399E+01 | 6.40E-01 |
| 1371 | Te | 113 | 52 | 61 | 9 | 8.40463600E+00 | 8.41171002E+00 | -7.10E-03 | 1.05153725E+05 | 1.05152888E+05 | 8.40E-01 | 1.05180482E+05 | 1.05179645E+05 | 8.40E-01 | -7.83470290E+01 | -7.91841894E+01 | 8.40E-01 |
| 1372 | Te | 114 | 52 | 62 | 10 | 8.43277800E+00 | 8.43824740E+00 | -5.50E-03 | 1.06081678E+05 | 1.06081017E+05 | 6.60E-01 | 1.06108434E+05 | 1.06107773E+05 | 6.60E-01 | -8.18885690E+01 | -8.25498415E+01 | 6.60E-01 |
| 1373 | Te | 115 | 52 | 63 | 11 | 8.43115000E+00 | 8.43949927E+00 | -8.30E-03 | 1.07012998E+05 | 1.07012000E+05 | 1.00E+00 | 1.07039754E+05 | 1.07038756E+05 | 1.00E+00 | -8.20627590E+01 | -8.30607348E+01 | 1.00E+00 |
| 1374 | Te | 116 | 52 | 64 | 12 | 8.45568700E+00 | 8.46103969E+00 | -5.40E-03 | 1.07941285E+05 | 1.07940627E+05 | 6.60E-01 | 1.07968042E+05 | 1.07967383E+05 | 6.60E-01 | -8.52689610E+01 | -8.59276045E+01 | 6.60E-01 |
| 1375 | Te | 117 | 52 | 65 | 13 | 8.45091900E+00 | 8.45809591E+00 | -7.20E-03 | 1.08872953E+05 | 1.08872076E+05 | 8.80E-01 | 1.08899710E+05 | 1.08898832E+05 | 8.80E-01 | -8.50954250E+01 | -8.59729026E+01 | 8.80E-01 |
| 1376 | Te | 118 | 52 | 66 | 14 | 8.46974700E+00 | 8.47502116E+00 | -5.30E-03 | 1.09801846E+05 | 1.09801186E+05 | 6.60E-01 | 1.09828602E+05 | 1.09827942E+05 | 6.60E-01 | -8.76967800E+01 | -8.83568590E+01 | 6.60E-01 |
| 1377 | Te | 119 | 52 | 67 | 15 | 8.46206700E+00 | 8.46830168E+00 | -6.20E-03 | 1.10733855E+05 | 1.10733076E+05 | 7.80E-01 | 1.10760612E+05 | 1.10759832E+05 | 7.80E-01 | -8.71812170E+01 | -8.79609433E+01 | 7.80E-01 |
| 1378 | Te | 120 | 52 | 68 | 16 | 8.47703500E+00 | 8.48101738E+00 | -4.00E-03 | 1.11663163E+05 | 1.11662647E+05 | 5.20E-01 | 1.11689919E+05 | 1.11689404E+05 | 5.20E-01 | -8.93681860E+01 | -8.98838094E+01 | 5.20E-01 |
| 1379 | Te | 121 | 52 | 69 | 17 | 8.46687300E+00 | 8.47081414E+00 | -3.90E-03 | 1.12595481E+05 | 1.12594966E+05 | 5.10E-01 | 1.12622237E+05 | 1.12621722E+05 | 5.10E-01 | -8.85442690E+01 | -8.90589155E+01 | 5.10E-01 |
| 1380 | Te | 122 | 52 | 70 | 18 | 8.47814000E+00 | 8.47983977E+00 | -1.70E-03 | 1.13525204E+05 | 1.13524959E+05 | 2.50E-01 | 1.13551961E+05 | 1.13551716E+05 | 2.50E-01 | -9.03144400E+01 | -9.05595379E+01 | 2.50E-01 |
| 1381 | Te | 123 | 52 | 71 | 19 | 8.46554600E+00 | 8.46672581E+00 | -1.20E-03 | 1.14457841E+05 | 1.14457658E+05 | 1.80E-01 | 1.14484597E+05 | 1.14484414E+05 | 1.80E-01 | -8.91721360E+01 | -8.93550420E+01 | 1.80E-01 |
| 1382 | Te | 124 | 52 | 72 | 20 | 8.47327900E+00 | 8.47279641E+00 | 4.80E-04 | 1.15387982E+05 | 1.15388004E+05 | -2.20E-02 | 1.15414738E+05 | 1.15414760E+05 | -2.20E-02 | -9.05253020E+01 | -9.05032027E+01 | -2.20E-02 |
| 1383 | Te | 125 | 52 | 73 | 21 | 8.45804500E+00 | 8.45753704E+00 | 5.10E-04 | 1.16320978E+05 | 1.16321004E+05 | -2.60E-02 | 1.16347735E+05 | 1.16347760E+05 | -2.60E-02 | -8.90229550E+01 | -8.89972593E+01 | -2.60E-02 |
| 1384 | Te | 126 | 52 | 74 | 22 | 8.46324800E+00 | 8.46157621E+00 | 1.70E-03 | 1.17251430E+05 | 1.17251603E+05 | -1.70E-01 | 1.17278186E+05 | 1.17278359E+05 | -1.70E-01 | -9.00653300E+01 | -8.98924129E+01 | -1.70E-01 |
| 1385 | Te | 127 | 52 | 75 | 23 | 8.44611700E+00 | 8.44506523E+00 | 1.10E-03 | 1.18184708E+05 | 1.18184803E+05 | -9.60E-02 | 1.18211464E+05 | 1.18211560E+05 | -9.60E-02 | -8.82816590E+01 | -8.81857748E+01 | -9.60E-02 |
| 1386 | Te | 128 | 52 | 76 | 24 | 8.44875200E+00 | 8.44792962E+00 | 8.20E-04 | 1.19115489E+05 | 1.19115557E+05 | -6.80E-02 | 1.19142246E+05 | 1.19142314E+05 | -6.80E-02 | -8.89937450E+01 | -8.89261639E+01 | -6.80E-02 |
| 1387 | Te | 129 | 52 | 77 | 25 | 8.43040900E+00 | 8.43064413E+00 | -2.40E-04 | 1.20048973E+05 | 1.20048904E+05 | 6.80E-02 | 1.20075729E+05 | 1.20075661E+05 | 6.80E-02 | -8.70048370E+01 | -8.70729454E+01 | 6.80E-02 |
| 1388 | Te | 130 | 52 | 78 | 26 | 8.43032400E+00 | 8.43219887E+00 | -1.90E-03 | 1.20980118E+05 | 1.20979837E+05 | 2.80E-01 | 1.21006875E+05 | 1.21006594E+05 | 2.80E-01 | -8.73529470E+01 | -8.76343875E+01 | 2.80E-01 |
| 1389 | Te | 131 | 52 | 79 | 27 | 8.41123300E+00 | 8.41292167E+00 | -1.70E-03 | 1.21913754E+05 | 1.21913495E+05 | 2.60E-01 | 1.21940511E+05 | 1.21940252E+05 | 2.60E-01 | -8.52110100E+01 | -8.54699534E+01 | 2.60E-01 |
| 1390 | Te | 132 | 52 | 80 | 28 | 8.40848500E+00 | 8.41068747E+00 | -2.20E-03 | 1.22845271E+05 | 1.22844943E+05 | 3.30E-01 | 1.22872028E+05 | 1.22871699E+05 | 3.30E-01 | -8.51881850E+01 | -8.55166426E+01 | 3.30E-01 |
| 1391 | Te | 133 | 52 | 81 | 29 | 8.38898700E+00 | 8.38575206E+00 | 3.20E-03 | 1.23779022E+05 | 1.23779414E+05 | -3.90E-01 | 1.23805778E+05 | 1.23806171E+05 | -3.90E-01 | -8.29320600E+01 | -8.25396008E+01 | -3.90E-01 |
| 1392 | Te | 134 | 52 | 82 | 30 | 8.38366000E+00 | 8.37570063E+00 | 8.00E-03 | 1.24710912E+05 | 1.24711940E+05 | -1.00E+00 | 1.24737668E+05 | 1.24738697E+05 | -1.00E+00 | -8.25359960E+01 | -8.15071424E+01 | -1.00E+00 |
| 1393 | Te | 135 | 52 | 83 | 31 | 8.34573100E+00 | 8.34198552E+00 | 3.70E-03 | 1.25647214E+05 | 1.25647682E+05 | -4.70E-01 | 1.25673970E+05 | 1.25674438E+05 | -4.70E-01 | -7.77278590E+01 | -7.72599846E+01 | -4.70E-01 |
| 1394 | Te | 136 | 52 | 84 | 32 | 8.31943300E+00 | 8.32318535E+00 | -3.80E-03 | 1.26582010E+05 | 1.26581462E+05 | 5.50E-01 | 1.26608766E+05 | 1.26608218E+05 | 5.50E-01 | -7.44258040E+01 | -7.49738273E+01 | 5.50E-01 |
| 1395 | Te | 137 | 52 | 85 | 33 | 8.28023800E+00 | 8.28236777E+00 | -2.10E-03 | 1.27518626E+05 | 1.27518296E+05 | 3.30E-01 | 1.27545382E+05 | 1.27545053E+05 | 3.30E-01 | -6.93042220E+01 | -6.96336856E+01 | 3.30E-01 |
| 1396 | Te | 138 | 52 | 86 | 34 | 8.25257900E+00 | 8.25842784E+00 | -5.80E-03 | 1.28453728E+05 | 1.28452883E+05 | 8.40E-01 | 1.28480484E+05 | 1.28479639E+05 | 8.40E-01 | -6.56961980E+01 | -6.65410244E+01 | 8.40E-01 |
| 1397 | Te | 139 | 52 | 87 | 35 | 8.21177100E+00 | 8.21472126E+00 | -3.00E-03 | 1.29390713E+05 | 1.29390265E+05 | 4.50E-01 | 1.29417469E+05 | 1.29417022E+05 | 4.50E-01 | -6.02050710E+01 | -6.06529183E+01 | 4.50E-01 |
| 1398 | Te | 140 | 52 | 88 | 36 | 8.18327900E+00 | 8.18923899E+00 | -6.00E-03 | 1.30326055E+05 | 1.30325183E+05 | 8.70E-01 | 1.30352812E+05 | 1.30351940E+05 | 8.70E-01 | -5.63567210E+01 | -5.72288023E+01 | 8.70E-01 |
| 1399 | Te | 142 | 52 | 90 | 38 | 8.11100000E+00 | 8.11923259E+00 | -8.20E-03 | 1.32199030E+05 | 1.32197876E+05 | 1.20E+00 | 1.32225787E+05 | 1.32224633E+05 | 1.20E+00 | -4.63700000E+01 | -4.75237339E+01 | 1.20E+00 |
| 1400 | TE | 143 | 52 | 91 | 39 | 8.06800000E+00 | 8.07492713E+00 | -6.90E-03 | 1.33136616E+05 | 1.33135658E+05 | 9.60E-01 | 1.33163373E+05 | 1.33162415E+05 | 9.60E-01 | -4.02780000E+01 | -4.12359667E+01 | 9.60E-01 |
| 1401 | I | 108 | 53 | 55 | 2 | 8.17485600E+00 | 8.18213364E+00 | -7.30E-03 | 1.00521436E+05 | 1.00520611E+05 | 8.20E-01 | 1.00548712E+05 | 1.00547888E+05 | 8.20E-01 | -5.26465290E+01 | -5.34710623E+01 | 8.20E-01 |
| 1402 | I | 109 | 53 | 56 | 3 | 8.22002200E+00 | 8.22895880E+00 | -8.90E-03 | 1.01447903E+05 | 1.01446890E+05 | 1.00E+00 | 1.01475179E+05 | 1.01474167E+05 | 1.00E+00 | -5.76731830E+01 | -5.86858193E+01 | 1.00E+00 |
| 1403 | I | 110 | 53 | 57 | 4 | 8.24404300E+00 | 8.25088841E+00 | -6.80E-03 | 1.02376606E+05 | 1.02375815E+05 | 7.90E-01 | 1.02403883E+05 | 1.02403091E+05 | 7.90E-01 | -6.04641750E+01 | -6.12557159E+01 | 7.90E-01 |
| 1404 | I | 111 | 53 | 58 | 5 | 8.28293400E+00 | 8.29052182E+00 | -7.60E-03 | 1.03303611E+05 | 1.03302730E+05 | 8.80E-01 | 1.03330887E+05 | 1.03330006E+05 | 8.80E-01 | -6.49538080E+01 | -6.58345941E+01 | 8.80E-01 |
| 1405 | I | 112 | 53 | 59 | 6 | 8.29987900E+00 | 8.30690700E+00 | -7.00E-03 | 1.04232995E+05 | 1.04232170E+05 | 8.30E-01 | 1.04260272E+05 | 1.04259446E+05 | 8.30E-01 | -6.70633290E+01 | -6.78889374E+01 | 8.30E-01 |
| 1406 | I | 113 | 53 | 60 | 7 | 8.33375200E+00 | 8.34041857E+00 | -6.70E-03 | 1.05160433E+05 | 1.05159641E+05 | 7.90E-01 | 1.05187709E+05 | 1.05186918E+05 | 7.90E-01 | -7.11195070E+01 | -7.19113323E+01 | 7.90E-01 |
| 1407 | I | 114 | 53 | 61 | 8 | 8.34600000E+00 | 8.35184397E+00 | -5.80E-03 | 1.06090250E+05 | 1.06089564E+05 | 6.90E-01 | 1.06117527E+05 | 1.06116840E+05 | 6.90E-01 | -7.27960000E+01 | -7.34829277E+01 | 6.90E-01 |
| 1408 | I | 115 | 53 | 62 | 9 | 8.37456400E+00 | 8.37968982E+00 | -5.10E-03 | 1.07018203E+05 | 1.07017575E+05 | 6.30E-01 | 1.07045479E+05 | 1.07044851E+05 | 6.30E-01 | -7.63377960E+01 | -7.69657250E+01 | 6.30E-01 |
| 1409 | I | 116 | 53 | 63 | 10 | 8.38190200E+00 | 8.38647574E+00 | -4.60E-03 | 1.07948542E+05 | 1.07947973E+05 | 5.70E-01 | 1.07975819E+05 | 1.07975250E+05 | 5.70E-01 | -7.74922360E+01 | -7.80612623E+01 | 5.70E-01 |
| 1410 | I | 117 | 53 | 64 | 11 | 8.40440900E+00 | 8.40902148E+00 | -4.60E-03 | 1.08877093E+05 | 1.08876515E+05 | 5.80E-01 | 1.08904369E+05 | 1.08903791E+05 | 5.80E-01 | -8.04360790E+01 | -8.10142714E+01 | 5.80E-01 |
| 1411 | I | 118 | 53 | 65 | 12 | 8.40612000E+00 | 8.41148188E+00 | -5.40E-03 | 1.09808052E+05 | 1.09807381E+05 | 6.70E-01 | 1.09835328E+05 | 1.09834657E+05 | 6.70E-01 | -8.09710480E+01 | -8.16423010E+01 | 6.70E-01 |
| 1412 | I | 119 | 53 | 66 | 13 | 8.42678900E+00 | 8.42920130E+00 | -2.40E-03 | 1.10736751E+05 | 1.10736426E+05 | 3.30E-01 | 1.10764028E+05 | 1.10763702E+05 | 3.30E-01 | -8.37655300E+01 | -8.40910744E+01 | 3.30E-01 |
| 1413 | I | 120 | 53 | 67 | 14 | 8.42372400E+00 | 8.42785079E+00 | -4.10E-03 | 1.11668258E+05 | 1.11667724E+05 | 5.30E-01 | 1.11695534E+05 | 1.11695000E+05 | 5.30E-01 | -8.37531860E+01 | -8.42868960E+01 | 5.30E-01 |
| 1414 | I | 121 | 53 | 68 | 15 | 8.44146000E+00 | 8.44119967E+00 | 2.60E-04 | 1.12597254E+05 | 1.12597246E+05 | 7.10E-03 | 1.12624530E+05 | 1.12624523E+05 | 7.10E-03 | -8.62516400E+01 | -8.62586417E+01 | 7.00E-03 |
| 1415 | I | 122 | 53 | 69 | 16 | 8.43702300E+00 | 8.43637753E+00 | 6.50E-04 | 1.13528919E+05 | 1.13528959E+05 | -4.00E-02 | 1.13556195E+05 | 1.13556235E+05 | -4.00E-02 | -8.60804400E+01 | -8.60402221E+01 | -4.00E-02 |
| 1416 | I | 123 | 53 | 70 | 17 | 8.44919800E+00 | 8.44589823E+00 | 3.30E-03 | 1.14458549E+05 | 1.14458917E+05 | -3.70E-01 | 1.14485826E+05 | 1.14486193E+05 | -3.70E-01 | -8.79437110E+01 | -8.75763265E+01 | -3.70E-01 |
| 1417 | I | 124 | 53 | 71 | 18 | 8.44148900E+00 | 8.43821455E+00 | 3.30E-03 | 1.15390621E+05 | 1.15390989E+05 | -3.70E-01 | 1.15417898E+05 | 1.15418265E+05 | -3.70E-01 | -8.73657150E+01 | -8.69981295E+01 | -3.70E-01 |
| 1418 | I | 125 | 53 | 72 | 19 | 8.45030000E+00 | 8.44464962E+00 | 5.70E-03 | 1.16320644E+05 | 1.16321312E+05 | -6.70E-01 | 1.16347920E+05 | 1.16348588E+05 | -6.70E-01 | -8.88371850E+01 | -8.81694087E+01 | -6.70E-01 |
| 1419 | I | 126 | 53 | 73 | 20 | 8.43994400E+00 | 8.43487794E+00 | 5.10E-03 | 1.17253064E+05 | 1.17253664E+05 | -6.00E-01 | 1.17280340E+05 | 1.17280940E+05 | -6.00E-01 | -8.79113020E+01 | -8.73115078E+01 | -6.00E-01 |
| 1420 | I | 127 | 53 | 74 | 21 | 8.44548700E+00 | 8.43913212E+00 | 6.40E-03 | 1.18183486E+05 | 1.18184254E+05 | -7.70E-01 | 1.18210762E+05 | 1.18211530E+05 | -7.70E-01 | -8.89838950E+01 | -8.82153475E+01 | -7.70E-01 |
| 1421 | I | 128 | 53 | 75 | 22 | 8.43283500E+00 | 8.42813974E+00 | 4.70E-03 | 1.19116225E+05 | 1.19116787E+05 | -5.60E-01 | 1.19143501E+05 | 1.19144064E+05 | -5.60E-01 | -8.77387090E+01 | -8.71761357E+01 | -5.60E-01 |
| 1422 | I | 129 | 53 | 76 | 23 | 8.43599000E+00 | 8.43102902E+00 | 5.00E-03 | 1.20046950E+05 | 1.20047552E+05 | -6.00E-01 | 1.20074227E+05 | 1.20074828E+05 | -6.00E-01 | -8.85071420E+01 | -8.79056739E+01 | -6.00E-01 |
| 1423 | I | 130 | 53 | 77 | 24 | 8.42110000E+00 | 8.41925379E+00 | 1.80E-03 | 1.20980015E+05 | 1.20980217E+05 | -2.00E-01 | 1.21007292E+05 | 1.21007493E+05 | -2.00E-01 | -8.69361550E+01 | -8.67346033E+01 | -2.00E-01 |
| 1424 | I | 131 | 53 | 78 | 25 | 8.42229700E+00 | 8.42063407E+00 | 1.70E-03 | 1.21911003E+05 | 1.21911182E+05 | -1.80E-01 | 1.21938279E+05 | 1.21938459E+05 | -1.80E-01 | -8.74427840E+01 | -8.72633557E+01 | -1.80E-01 |
| 1425 | I | 132 | 53 | 79 | 26 | 8.40646200E+00 | 8.40689537E+00 | -4.30E-04 | 1.22844236E+05 | 1.22844141E+05 | 9.60E-02 | 1.22871512E+05 | 1.22871417E+05 | 9.60E-02 | -8.57034900E+01 | -8.57991617E+01 | 9.60E-02 |
| 1426 | I | 133 | 53 | 80 | 27 | 8.40531900E+00 | 8.40441938E+00 | 9.00E-04 | 1.23775547E+05 | 1.23775628E+05 | -8.10E-02 | 1.23802824E+05 | 1.23802905E+05 | -8.10E-02 | -8.58865710E+01 | -8.58054312E+01 | -8.10E-02 |
| 1427 | I | 134 | 53 | 81 | 28 | 8.38918800E+00 | 8.38524248E+00 | 3.90E-03 | 1.24708869E+05 | 1.24709359E+05 | -4.90E-01 | 1.24736145E+05 | 1.24736635E+05 | -4.90E-01 | -8.40591020E+01 | -8.35688267E+01 | -4.90E-01 |
| 1428 | I | 135 | 53 | 82 | 29 | 8.38483300E+00 | 8.37511930E+00 | 9.70E-03 | 1.25640633E+05 | 1.25641906E+05 | -1.30E+00 | 1.25667909E+05 | 1.25669182E+05 | -1.30E+00 | -8.37889590E+01 | -8.25161213E+01 | -1.30E+00 |
| 1429 | I | 136 | 53 | 83 | 30 | 8.35132500E+00 | 8.34751948E+00 | 3.80E-03 | 1.26576371E+05 | 1.26576850E+05 | -4.80E-01 | 1.26603647E+05 | 1.26604126E+05 | -4.80E-01 | -7.95454780E+01 | -7.90663466E+01 | -4.80E-01 |
| 1430 | I | 137 | 53 | 84 | 31 | 8.32600200E+00 | 8.32882544E+00 | -2.80E-03 | 1.27511054E+05 | 1.27510629E+05 | 4.30E-01 | 1.27538330E+05 | 1.27537905E+05 | 4.30E-01 | -7.63562510E+01 | -7.67814628E+01 | 4.30E-01 |
| 1431 | I | 138 | 53 | 85 | 32 | 8.29244400E+00 | 8.29429085E+00 | -1.80E-03 | 1.28446924E+05 | 1.28446631E+05 | 2.90E-01 | 1.28474201E+05 | 1.28473907E+05 | 2.90E-01 | -7.19798920E+01 | -7.22731968E+01 | 2.90E-01 |
| 1432 | I | 139 | 53 | 86 | 33 | 8.26552300E+00 | 8.27039703E+00 | -4.90E-03 | 1.29381939E+05 | 1.29381223E+05 | 7.20E-01 | 1.29409215E+05 | 1.29408500E+05 | 7.20E-01 | -6.84590270E+01 | -6.91749268E+01 | 7.20E-01 |
| 1433 | I | 140 | 53 | 87 | 34 | 8.22939900E+00 | 8.23288537E+00 | -3.50E-03 | 1.30318296E+05 | 1.30317770E+05 | 5.30E-01 | 1.30345573E+05 | 1.30345046E+05 | 5.30E-01 | -6.35958900E+01 | -6.41223733E+01 | 5.30E-01 |
| 1434 | I | 141 | 53 | 88 | 35 | 8.20200000E+00 | 8.20725356E+00 | -5.30E-03 | 1.31253482E+05 | 1.31252716E+05 | 7.70E-01 | 1.31280758E+05 | 1.31279993E+05 | 7.70E-01 | -5.99040000E+01 | -6.06698542E+01 | 7.70E-01 |
| 1435 | I | 142 | 53 | 89 | 36 | 8.16501900E+00 | 8.16908493E+00 | -4.10E-03 | 1.32190110E+05 | 1.32189494E+05 | 6.20E-01 | 1.32217387E+05 | 1.32216771E+05 | 6.20E-01 | -5.47699840E+01 | -5.53858428E+01 | 6.20E-01 |
| 1436 | I | 143 | 53 | 90 | 37 | 8.13500000E+00 | 8.14290899E+00 | -7.90E-03 | 1.33125748E+05 | 1.33124634E+05 | 1.10E+00 | 1.33153024E+05 | 1.33151910E+05 | 1.10E+00 | -5.06270000E+01 | -5.17404494E+01 | 1.10E+00 |
| 1437 | I | 144 | 53 | 91 | 38 | 8.09800000E+00 | 8.10439792E+00 | -6.40E-03 | 1.34062589E+05 | 1.34061602E+05 | 9.90E-01 | 1.34089865E+05 | 1.34088878E+05 | 9.90E-01 | -4.52800000E+01 | -4.62664450E+01 | 9.90E-01 |
| 1438 | I | 145 | 53 | 92 | 39 | 8.06800000E+00 | 8.07750895E+00 | -9.50E-03 | 1.34998423E+05 | 1.34996962E+05 | 1.50E+00 | 1.35025700E+05 | 1.35024238E+05 | 1.50E+00 | -4.09390000E+01 | -4.24006239E+01 | 1.50E+00 |
| 1439 | XE | 109 | 54 | 55 | 1 | 8.10730700E+00 | 8.11582372E+00 | -8.50E-03 | 1.01458887E+05 | 1.01457919E+05 | 9.70E-01 | 1.01486683E+05 | 1.01485715E+05 | 9.70E-01 | -4.61696270E+01 | -4.71371724E+01 | 9.70E-01 |
| 1440 | Xe | 110 | 54 | 56 | 2 | 8.15924300E+00 | 8.17020591E+00 | -1.10E-02 | 1.02384632E+05 | 1.02383387E+05 | 1.20E+00 | 1.02412428E+05 | 1.02411183E+05 | 1.20E+00 | -5.19185860E+01 | -5.31637176E+01 | 1.20E+00 |
| 1441 | Xe | 111 | 54 | 57 | 3 | 8.18073900E+00 | 8.19481122E+00 | -1.40E-02 | 1.03313652E+05 | 1.03312051E+05 | 1.60E+00 | 1.03341448E+05 | 1.03339847E+05 | 1.60E+00 | -5.43925370E+01 | -5.59937948E+01 | 1.60E+00 |
| 1442 | Xe | 112 | 54 | 58 | 4 | 8.23006500E+00 | 8.24101506E+00 | -1.10E-02 | 1.04239512E+05 | 1.04238246E+05 | 1.30E+00 | 1.04267308E+05 | 1.04266043E+05 | 1.30E+00 | -6.00264210E+01 | -6.12921163E+01 | 1.30E+00 |
| 1443 | Xe | 113 | 54 | 59 | 5 | 8.24792700E+00 | 8.25924408E+00 | -1.10E-02 | 1.05168829E+05 | 1.05167511E+05 | 1.30E+00 | 1.05196625E+05 | 1.05195307E+05 | 1.30E+00 | -6.22036320E+01 | -6.35216922E+01 | 1.30E+00 |
| 1444 | Xe | 114 | 54 | 60 | 6 | 8.28920500E+00 | 8.29863204E+00 | -9.40E-03 | 1.06095441E+05 | 1.06094327E+05 | 1.10E+00 | 1.06123237E+05 | 1.06122123E+05 | 1.10E+00 | -6.70858890E+01 | -6.81998445E+01 | 1.10E+00 |
| 1445 | Xe | 115 | 54 | 61 | 7 | 8.30097000E+00 | 8.31134537E+00 | -1.00E-02 | 1.07025364E+05 | 1.07024132E+05 | 1.20E+00 | 1.07053160E+05 | 1.07051928E+05 | 1.20E+00 | -6.86567470E+01 | -6.98891904E+01 | 1.20E+00 |
| 1446 | Xe | 116 | 54 | 62 | 8 | 8.33683400E+00 | 8.34459994E+00 | -7.80E-03 | 1.07952468E+05 | 1.07951528E+05 | 9.40E-01 | 1.07980264E+05 | 1.07979324E+05 | 9.40E-01 | -7.30467230E+01 | -7.39867469E+01 | 9.40E-01 |
| 1447 | Xe | 117 | 54 | 63 | 9 | 8.34429700E+00 | 8.35231362E+00 | -8.00E-03 | 1.08882823E+05 | 1.08881846E+05 | 9.80E-01 | 1.08910620E+05 | 1.08909643E+05 | 9.80E-01 | -7.41853360E+01 | -7.51625290E+01 | 9.80E-01 |
| 1448 | Xe | 118 | 54 | 64 | 10 | 8.37498100E+00 | 8.37995867E+00 | -5.00E-03 | 1.09810424E+05 | 1.09809797E+05 | 6.30E-01 | 1.09838220E+05 | 1.09837594E+05 | 6.30E-01 | -7.80790560E+01 | -7.87056389E+01 | 6.30E-01 |
| 1449 | Xe | 119 | 54 | 65 | 11 | 8.37844100E+00 | 8.38311447E+00 | -4.70E-03 | 1.10741202E+05 | 1.10740607E+05 | 6.00E-01 | 1.10768999E+05 | 1.10768403E+05 | 6.00E-01 | -7.87944120E+01 | -7.93898195E+01 | 6.00E-01 |
| 1450 | Xe | 120 | 54 | 66 | 12 | 8.40403100E+00 | 8.40573476E+00 | -1.70E-03 | 1.11669319E+05 | 1.11669075E+05 | 2.40E-01 | 1.11697115E+05 | 1.11696871E+05 | 2.40E-01 | -8.21724230E+01 | -8.24160495E+01 | 2.40E-01 |
| 1451 | Xe | 121 | 54 | 67 | 13 | 8.40383200E+00 | 8.40494049E+00 | -1.10E-03 | 1.12600504E+05 | 1.12600331E+05 | 1.70E-01 | 1.12628300E+05 | 1.12628127E+05 | 1.70E-01 | -8.24809850E+01 | -8.26543582E+01 | 1.70E-01 |
| 1452 | Xe | 122 | 54 | 68 | 14 | 8.42466400E+00 | 8.42306162E+00 | 1.60E-03 | 1.13529124E+05 | 1.13529280E+05 | -1.60E-01 | 1.13556920E+05 | 1.13557077E+05 | -1.60E-01 | -8.53549760E+01 | -8.51987576E+01 | -1.60E-01 |
| 1453 | Xe | 123 | 54 | 69 | 15 | 8.42092700E+00 | 8.41869719E+00 | 2.20E-03 | 1.14460724E+05 | 1.14460960E+05 | -2.40E-01 | 1.14488521E+05 | 1.14488756E+05 | -2.40E-01 | -8.52486940E+01 | -8.50136751E+01 | -2.40E-01 |
| 1454 | Xe | 124 | 54 | 70 | 16 | 8.43756200E+00 | 8.43289972E+00 | 4.70E-03 | 1.15389806E+05 | 1.15390345E+05 | -5.40E-01 | 1.15417602E+05 | 1.15418141E+05 | -5.40E-01 | -8.76610520E+01 | -8.71221674E+01 | -5.40E-01 |
| 1455 | Xe | 125 | 54 | 71 | 17 | 8.43088800E+00 | 8.42560469E+00 | 5.30E-03 | 1.16321768E+05 | 1.16322389E+05 | -6.20E-01 | 1.16349565E+05 | 1.16350186E+05 | -6.20E-01 | -8.71930120E+01 | -8.65718689E+01 | -6.20E-01 |
| 1456 | Xe | 126 | 54 | 72 | 18 | 8.44353000E+00 | 8.43664826E+00 | 6.90E-03 | 1.17251310E+05 | 1.17252138E+05 | -8.30E-01 | 1.17279106E+05 | 1.17279934E+05 | -8.30E-01 | -8.91455660E+01 | -8.83176448E+01 | -8.30E-01 |
| 1457 | Xe | 127 | 54 | 73 | 19 | 8.43411100E+00 | 8.42719954E+00 | 6.90E-03 | 1.18183628E+05 | 1.18184466E+05 | -8.40E-01 | 1.18211424E+05 | 1.18212263E+05 | -8.40E-01 | -8.83215460E+01 | -8.74829870E+01 | -8.40E-01 |
| 1458 | Xe | 128 | 54 | 74 | 20 | 8.44329800E+00 | 8.43597325E+00 | 7.30E-03 | 1.19113583E+05 | 1.19114482E+05 | -9.00E-01 | 1.19141379E+05 | 1.19142278E+05 | -9.00E-01 | -8.98602780E+01 | -8.89619019E+01 | -9.00E-01 |
| 1459 | Xe | 129 | 54 | 75 | 21 | 8.43139000E+00 | 8.42521206E+00 | 6.20E-03 | 1.20046242E+05 | 1.20046999E+05 | -7.60E-01 | 1.20074038E+05 | 1.20074796E+05 | -7.60E-01 | -8.86960568E+01 | -8.79383629E+01 | -7.60E-01 |
| 1460 | Xe | 130 | 54 | 76 | 22 | 8.43773100E+00 | 8.43248829E+00 | 5.20E-03 | 1.20976551E+05 | 1.20977193E+05 | -6.40E-01 | 1.21004347E+05 | 1.21004990E+05 | -6.40E-01 | -8.98804620E+01 | -8.92381655E+01 | -6.40E-01 |
| 1461 | Xe | 131 | 54 | 77 | 23 | 8.42373600E+00 | 8.42081982E+00 | 2.90E-03 | 1.21909512E+05 | 1.21909855E+05 | -3.40E-01 | 1.21937308E+05 | 1.21937651E+05 | -3.40E-01 | -8.84136310E+01 | -8.80707656E+01 | -3.40E-01 |
| 1462 | Xe | 132 | 54 | 78 | 24 | 8.42762200E+00 | 8.42643583E+00 | 1.20E-03 | 1.22840141E+05 | 1.22840258E+05 | -1.20E-01 | 1.22867937E+05 | 1.22868054E+05 | -1.20E-01 | -8.92789629E+01 | -8.91615789E+01 | -1.20E-01 |
| 1463 | Xe | 133 | 54 | 79 | 25 | 8.41264700E+00 | 8.41272790E+00 | -8.10E-05 | 1.23773270E+05 | 1.23773220E+05 | 5.00E-02 | 1.23801067E+05 | 1.23801017E+05 | 5.00E-02 | -8.76435710E+01 | -8.76935413E+01 | 5.00E-02 |
| 1464 | Xe | 134 | 54 | 80 | 26 | 8.41368700E+00 | 8.41445346E+00 | -7.70E-04 | 1.24704284E+05 | 1.24704142E+05 | 1.40E-01 | 1.24732080E+05 | 1.24731938E+05 | 1.40E-01 | -8.81243030E+01 | -8.82661761E+01 | 1.40E-01 |
| 1465 | Xe | 135 | 54 | 81 | 27 | 8.39850300E+00 | 8.39544338E+00 | 3.10E-03 | 1.25637485E+05 | 1.25637859E+05 | -3.70E-01 | 1.25665281E+05 | 1.25665655E+05 | -3.70E-01 | -8.64167660E+01 | -8.60429496E+01 | -3.70E-01 |
| 1466 | Xe | 136 | 54 | 82 | 28 | 8.39618800E+00 | 8.38971533E+00 | 6.50E-03 | 1.26568967E+05 | 1.26569808E+05 | -8.40E-01 | 1.26596763E+05 | 1.26597604E+05 | -8.40E-01 | -8.64291520E+01 | -8.55880583E+01 | -8.40E-01 |
| 1467 | Xe | 137 | 54 | 83 | 29 | 8.36428600E+00 | 8.36259547E+00 | 1.70E-03 | 1.27504507E+05 | 1.27504699E+05 | -1.90E-01 | 1.27532303E+05 | 1.27532495E+05 | -1.90E-01 | -8.23833960E+01 | -8.21910346E+01 | -1.90E-01 |
| 1468 | Xe | 138 | 54 | 84 | 30 | 8.34469000E+00 | 8.34850130E+00 | -3.80E-03 | 1.28438412E+05 | 1.28437847E+05 | 5.70E-01 | 1.28466208E+05 | 1.28465643E+05 | 5.70E-01 | -7.99722310E+01 | -8.05373155E+01 | 5.70E-01 |
| 1469 | Xe | 139 | 54 | 85 | 31 | 8.31159000E+00 | 8.31461916E+00 | -3.00E-03 | 1.29374234E+05 | 1.29373773E+05 | 4.60E-01 | 1.29402030E+05 | 1.29401570E+05 | 4.60E-01 | -7.56445750E+01 | -7.61048806E+01 | 4.60E-01 |
| 1470 | Xe | 140 | 54 | 86 | 32 | 8.29088700E+00 | 8.29528397E+00 | -4.40E-03 | 1.30308386E+05 | 1.30307731E+05 | 6.50E-01 | 1.30336182E+05 | 1.30335527E+05 | 6.50E-01 | -7.29864510E+01 | -7.36412543E+01 | 6.50E-01 |
| 1471 | Xe | 141 | 54 | 87 | 33 | 8.25536400E+00 | 8.25837296E+00 | -3.00E-03 | 1.31244669E+05 | 1.31244206E+05 | 4.60E-01 | 1.31272465E+05 | 1.31272002E+05 | 4.60E-01 | -6.81972980E+01 | -6.86607662E+01 | 4.60E-01 |
| 1472 | Xe | 142 | 54 | 88 | 34 | 8.23316900E+00 | 8.23710138E+00 | -3.90E-03 | 1.32179131E+05 | 1.32178533E+05 | 6.00E-01 | 1.32206927E+05 | 1.32206329E+05 | 6.00E-01 | -6.52296390E+01 | -6.58272554E+01 | 6.00E-01 |
| 1473 | Xe | 143 | 54 | 89 | 35 | 8.19688500E+00 | 8.19941677E+00 | -2.50E-03 | 1.33115652E+05 | 1.33115250E+05 | 4.00E-01 | 1.33143448E+05 | 1.33143047E+05 | 4.00E-01 | -6.02028730E+01 | -6.06041396E+01 | 4.00E-01 |
| 1474 | Xe | 144 | 54 | 90 | 36 | 8.17288400E+00 | 8.17738612E+00 | -4.50E-03 | 1.34050476E+05 | 1.34049789E+05 | 6.90E-01 | 1.34078272E+05 | 1.34077585E+05 | 6.90E-01 | -5.68722920E+01 | -5.75598236E+01 | 6.90E-01 |
| 1475 | Xe | 145 | 54 | 91 | 37 | 8.13508700E+00 | 8.13924827E+00 | -4.20E-03 | 1.34987349E+05 | 1.34986707E+05 | 6.40E-01 | 1.35015146E+05 | 1.35014503E+05 | 6.40E-01 | -5.14933290E+01 | -5.21359016E+01 | 6.40E-01 |
| 1476 | Xe | 146 | 54 | 92 | 38 | 8.11041500E+00 | 8.11628654E+00 | -5.90E-03 | 1.35922382E+05 | 1.35921485E+05 | 9.00E-01 | 1.35950178E+05 | 1.35949281E+05 | 9.00E-01 | -4.79549420E+01 | -4.88514194E+01 | 9.00E-01 |
| 1477 | XE | 147 | 54 | 93 | 39 | 8.07400000E+00 | 8.07729488E+00 | -3.30E-03 | 1.36859224E+05 | 1.36858666E+05 | 5.60E-01 | 1.36887020E+05 | 1.36886462E+05 | 5.60E-01 | -4.26070000E+01 | -4.31646131E+01 | 5.60E-01 |
| 1478 | XE | 148 | 54 | 94 | 40 | 8.04900000E+00 | 8.05299396E+00 | -4.00E-03 | 1.37794323E+05 | 1.37793751E+05 | 5.70E-01 | 1.37822119E+05 | 1.37821547E+05 | 5.70E-01 | -3.90020000E+01 | -3.95740518E+01 | 5.70E-01 |
| 1479 | Cs | 112 | 55 | 57 | 2 | 8.10040800E+00 | 8.10859042E+00 | -8.20E-03 | 1.04252731E+05 | 1.04251775E+05 | 9.60E-01 | 1.04281048E+05 | 1.04280091E+05 | 9.60E-01 | -4.62872800E+01 | -4.72436341E+01 | 9.60E-01 |
| 1480 | Cs | 113 | 55 | 58 | 3 | 8.14861700E+00 | 8.15814331E+00 | -9.50E-03 | 1.05178748E+05 | 1.05177632E+05 | 1.10E+00 | 1.05207065E+05 | 1.05205949E+05 | 1.10E+00 | -5.17639190E+01 | -5.28803819E+01 | 1.10E+00 |
| 1481 | Cs | 114 | 55 | 59 | 4 | 8.17353800E+00 | 8.18283339E+00 | -9.30E-03 | 1.06107324E+05 | 1.06106225E+05 | 1.10E+00 | 1.06135641E+05 | 1.06134541E+05 | 1.10E+00 | -5.46822600E+01 | -5.57818756E+01 | 1.10E+00 |
| 1482 | Cs | 115 | 55 | 60 | 5 | 8.21600000E+00 | 8.22475307E+00 | -8.80E-03 | 1.07033801E+05 | 1.07032786E+05 | 1.00E+00 | 1.07062118E+05 | 1.07061103E+05 | 1.00E+00 | -5.96990000E+01 | -6.07141528E+01 | 1.00E+00 |
| 1483 | Cs | 116 | 55 | 61 | 6 | 8.23500000E+00 | 8.24329827E+00 | -8.30E-03 | 1.07962932E+05 | 1.07961976E+05 | 9.60E-01 | 1.07991248E+05 | 1.07990292E+05 | 9.60E-01 | -6.20630000E+01 | -6.30188304E+01 | 9.60E-01 |
| 1484 | Cs | 117 | 55 | 62 | 7 | 8.27186400E+00 | 8.27850844E+00 | -6.60E-03 | 1.08889996E+05 | 1.08889178E+05 | 8.20E-01 | 1.08918312E+05 | 1.08917495E+05 | 8.20E-01 | -6.64930910E+01 | -6.73103998E+01 | 8.20E-01 |
| 1485 | Cs | 118 | 55 | 63 | 8 | 8.28640400E+00 | 8.29162007E+00 | -5.20E-03 | 1.09819573E+05 | 1.09818918E+05 | 6.60E-01 | 1.09847890E+05 | 1.09847234E+05 | 6.60E-01 | -6.84093670E+01 | -6.90647608E+01 | 6.60E-01 |
| 1486 | Cs | 119 | 55 | 64 | 9 | 8.31733400E+00 | 8.32082560E+00 | -3.50E-03 | 1.10747172E+05 | 1.10746716E+05 | 4.60E-01 | 1.10775488E+05 | 1.10775033E+05 | 4.60E-01 | -7.23050510E+01 | -7.27605203E+01 | 4.60E-01 |
| 1487 | Cs | 120 | 55 | 65 | 10 | 8.32848000E+00 | 8.32909445E+00 | -6.10E-04 | 1.11677082E+05 | 1.11676968E+05 | 1.10E-01 | 1.11705399E+05 | 1.11705285E+05 | 1.10E-01 | -7.38886380E+01 | -7.40022889E+01 | 1.10E-01 |
| 1488 | Cs | 121 | 55 | 66 | 11 | 8.35291400E+00 | 8.35301095E+00 | -9.70E-05 | 1.12605362E+05 | 1.12605311E+05 | 5.20E-02 | 1.12633679E+05 | 1.12633627E+05 | 5.20E-02 | -7.71023310E+01 | -7.71539605E+01 | 5.20E-02 |
| 1489 | Cs | 122 | 55 | 67 | 12 | 8.35915100E+00 | 8.35716609E+00 | 2.00E-03 | 1.13535814E+05 | 1.13536016E+05 | -2.00E-01 | 1.13564131E+05 | 1.13564333E+05 | -2.00E-01 | -7.81447590E+01 | -7.79425803E+01 | -2.00E-01 |
| 1490 | Cs | 123 | 55 | 68 | 13 | 8.38037900E+00 | 8.37640322E+00 | 4.00E-03 | 1.14464409E+05 | 1.14464858E+05 | -4.50E-01 | 1.14492726E+05 | 1.14493175E+05 | -4.50E-01 | -8.10436450E+01 | -8.05945937E+01 | -4.50E-01 |
| 1491 | Cs | 124 | 55 | 69 | 14 | 8.38343200E+00 | 8.37689180E+00 | 6.50E-03 | 1.15395216E+05 | 1.15395987E+05 | -7.70E-01 | 1.15423532E+05 | 1.15424303E+05 | -7.70E-01 | -8.17313340E+01 | -8.09602618E+01 | -7.70E-01 |
| 1492 | Cs | 125 | 55 | 70 | 15 | 8.39978700E+00 | 8.39207739E+00 | 7.70E-03 | 1.16324353E+05 | 1.16325277E+05 | -9.20E-01 | 1.16352670E+05 | 1.16353594E+05 | -9.20E-01 | -8.40878400E+01 | -8.31640337E+01 | -9.20E-01 |
| 1493 | Cs | 126 | 55 | 71 | 16 | 8.39926500E+00 | 8.38958978E+00 | 9.70E-03 | 1.17255585E+05 | 1.17256764E+05 | -1.20E+00 | 1.17283901E+05 | 1.17285080E+05 | -1.20E+00 | -8.43504540E+01 | -8.31713532E+01 | -1.20E+00 |
| 1494 | Cs | 127 | 55 | 72 | 17 | 8.41156200E+00 | 8.40150693E+00 | 1.00E-02 | 1.18185189E+05 | 1.18186426E+05 | -1.20E+00 | 1.18213506E+05 | 1.18214743E+05 | -1.20E+00 | -8.62401510E+01 | -8.50031024E+01 | -1.20E+00 |
| 1495 | Cs | 128 | 55 | 73 | 18 | 8.40649300E+00 | 8.39683833E+00 | 9.70E-03 | 1.19116992E+05 | 1.19118188E+05 | -1.20E+00 | 1.19145308E+05 | 1.19146504E+05 | -1.20E+00 | -8.59315640E+01 | -8.47357094E+01 | -1.20E+00 |
| 1496 | Cs | 129 | 55 | 74 | 19 | 8.41604700E+00 | 8.40636621E+00 | 9.70E-03 | 1.20046918E+05 | 1.20048127E+05 | -1.20E+00 | 1.20075235E+05 | 1.20076444E+05 | -1.20E+00 | -8.74992520E+01 | -8.62903252E+01 | -1.20E+00 |
| 1497 | Cs | 130 | 55 | 75 | 20 | 8.40878400E+00 | 8.40034020E+00 | 8.40E-03 | 1.20979012E+05 | 1.20980069E+05 | -1.10E+00 | 1.21007328E+05 | 1.21008386E+05 | -1.10E+00 | -8.68997430E+01 | -8.58419913E+01 | -1.10E+00 |

| | | | | | | | | | | | | | | | |
|---|---|---|---|---|---|---|---|---|---|---|---|---|---|---|---|
| 1498 | Cs | 131 | 55 | 76 | 21 | 8.41505600E+00 | 8.40821396E+00 | 6.80E-03 | 1.21909346E+05 | 1.21910203E+05 | -8.60E-01 | 1.21937663E+05 | 1.21938520E+05 | -8.60E-01 | -8.80588790E+01 | -8.72024752E+01 | -8.60E-01 |
| 1499 | Cs | 132 | 55 | 77 | 22 | 8.40561400E+00 | 8.40120506E+00 | 4.40E-03 | 1.22841743E+05 | 1.22842285E+05 | -5.40E-01 | 1.22870060E+05 | 1.22870602E+05 | -5.40E-01 | -8.71562480E+01 | -8.66141944E+01 | -5.40E-01 |
| 1500 | Cs | 133 | 55 | 78 | 23 | 8.40997800E+00 | 8.40724944E+00 | 2.70E-03 | 1.23772323E+05 | 1.23772646E+05 | -3.20E-01 | 1.23800639E+05 | 1.23800962E+05 | -3.20E-01 | -8.80709310E+01 | -8.77479832E+01 | -3.20E-01 |
| 1501 | Cs | 134 | 55 | 79 | 24 | 8.39864600E+00 | 8.39816797E+00 | 4.80E-04 | 1.24704996E+05 | 1.24705021E+05 | -2.40E-02 | 1.24733313E+05 | 1.24733337E+05 | -2.40E-02 | -8.68911540E+01 | -8.68669971E+01 | -2.40E-02 |
| 1502 | Cs | 135 | 55 | 80 | 25 | 8.40133800E+00 | 8.40027349E+00 | 1.10E-03 | 1.25635800E+05 | 1.25635904E+05 | -1.00E-01 | 1.25664116E+05 | 1.25664220E+05 | -1.00E-01 | -8.75818150E+01 | -8.74780914E+01 | -1.00E-01 |
| 1503 | Cs | 136 | 55 | 81 | 26 | 8.38977200E+00 | 8.38605690E+00 | 3.70E-03 | 1.26568537E+05 | 1.26569002E+05 | -4.70E-01 | 1.26596853E+05 | 1.26597319E+05 | -4.70E-01 | -8.63389140E+01 | -8.58735899E+01 | -4.70E-01 |
| 1504 | Cs | 137 | 55 | 82 | 27 | 8.38895800E+00 | 8.38089587E+00 | 8.10E-03 | 1.27499824E+05 | 1.27500889E+05 | -1.10E+00 | 1.27528140E+05 | 1.27529205E+05 | -1.10E+00 | -8.65458240E+01 | -8.54812666E+01 | -1.10E+00 |
| 1505 | Cs | 138 | 55 | 83 | 28 | 8.36014300E+00 | 8.35891043E+00 | 1.20E-03 | 1.28434977E+05 | 1.28435107E+05 | -1.30E-01 | 1.28463293E+05 | 1.28463424E+05 | -1.30E-01 | -8.28870480E+01 | -8.27568527E+01 | -1.30E-01 |
| 1506 | Cs | 139 | 55 | 84 | 29 | 8.34233900E+00 | 8.34560474E+00 | -3.30E-03 | 1.29368657E+05 | 1.29368163E+05 | 4.90E-01 | 1.29396973E+05 | 1.29396480E+05 | 4.90E-01 | -8.07011400E+01 | -8.11949523E+01 | 4.90E-01 |
| 1507 | Cs | 140 | 55 | 85 | 30 | 8.31432600E+00 | 8.31706220E+00 | -2.70E-03 | 1.30303802E+05 | 1.30303379E+05 | 4.20E-01 | 1.30332118E+05 | 1.30331695E+05 | 4.20E-01 | -7.70503300E+01 | -7.74732833E+01 | 4.20E-01 |
| 1508 | Cs | 141 | 55 | 86 | 31 | 8.29435300E+00 | 8.29850942E+00 | -4.20E-03 | 1.31237869E+05 | 1.31237243E+05 | 6.30E-01 | 1.31266185E+05 | 1.31265560E+05 | 6.30E-01 | -7.44771450E+01 | -7.51030842E+01 | 6.30E-01 |
| 1509 | Cs | 142 | 55 | 87 | 32 | 8.26490000E+00 | 8.26690881E+00 | -2.00E-03 | 1.32173322E+05 | 1.32172997E+05 | 3.30E-01 | 1.32201639E+05 | 1.32201314E+05 | 3.30E-01 | -7.05178300E+01 | -7.08429884E+01 | 3.30E-01 |
| 1510 | Cs | 143 | 55 | 88 | 33 | 8.24365700E+00 | 8.24625941E+00 | -2.60E-03 | 1.33107660E+05 | 1.33107248E+05 | 4.10E-01 | 1.33135977E+05 | 1.33135565E+05 | 4.10E-01 | -6.76736660E+01 | -6.80857131E+01 | 4.10E-01 |
| 1511 | Cs | 144 | 55 | 89 | 34 | 8.21188300E+00 | 8.21377702E+00 | -1.90E-03 | 1.34043558E+05 | 1.34043245E+05 | 3.10E-01 | 1.34071874E+05 | 1.34071562E+05 | 3.10E-01 | -6.32705070E+01 | -6.35831894E+01 | 3.10E-01 |
| 1512 | Cs | 145 | 55 | 90 | 35 | 8.18874300E+00 | 8.19219861E+00 | -3.50E-03 | 1.34978266E+05 | 1.34977726E+05 | 5.40E-01 | 1.35006583E+05 | 1.35006042E+05 | 5.40E-01 | -6.00558070E+01 | -6.05967787E+01 | 5.40E-01 |
| 1513 | Cs | 146 | 55 | 91 | 36 | 8.15720700E+00 | 8.15915773E+00 | -2.00E-03 | 1.35914247E+05 | 1.35913923E+05 | 3.20E-01 | 1.35942564E+05 | 1.35942239E+05 | 3.20E-01 | -5.55689530E+01 | -5.58936899E+01 | 3.20E-01 |
| 1514 | Cs | 147 | 55 | 92 | 37 | 8.13246800E+00 | 8.13648917E+00 | -4.00E-03 | 1.36849292E+05 | 1.36848661E+05 | 6.30E-01 | 1.36877609E+05 | 1.36876978E+05 | 6.30E-01 | -5.20182050E+01 | -5.26492506E+01 | 6.30E-01 |
| 1515 | Cs | 148 | 55 | 93 | 38 | 8.10015100E+00 | 8.10247106E+00 | -2.30E-03 | 1.37785508E+05 | 1.37785125E+05 | 3.80E-01 | 1.37813824E+05 | 1.37813441E+05 | 3.80E-01 | -4.72964800E+01 | -4.76797402E+01 | 3.80E-01 |
| 1516 | Cs | 149 | 55 | 94 | 39 | 8.07600000E+00 | 8.07827765E+00 | -2.30E-03 | 1.38720537E+05 | 1.38720192E+05 | 3.40E-01 | 1.38748853E+05 | 1.38748509E+05 | 3.40E-01 | -4.37620000E+01 | -4.41060733E+01 | 3.40E-01 |
| 1517 | Cs | 150 | 55 | 95 | 40 | 8.04300000E+00 | 8.04289978E+00 | 1.00E-04 | 1.39656977E+05 | 1.39656986E+05 | -9.00E-03 | 1.39685294E+05 | 1.39685303E+05 | -9.00E-03 | -3.88150000E+01 | -3.88063515E+01 | -8.60E-03 |
| 1518 | Cs | 151 | 55 | 96 | 41 | 8.01700000E+00 | 8.01695074E+00 | 4.90E-05 | 1.40592430E+05 | 1.40592427E+05 | 3.10E-03 | 1.40620747E+05 | 1.40620744E+05 | 3.10E-03 | -3.48570000E+01 | -3.48596271E+01 | 2.60E-03 |
| 1519 | Ba | 114 | 56 | 58 | 2 | 8.09016000E+00 | 8.09855947E+00 | -8.40E-03 | 1.06115526E+05 | 1.06114528E+05 | 1.00E+00 | 1.06144363E+05 | 1.06143365E+05 | 1.00E+00 | -4.59595050E+01 | -4.69577254E+01 | 1.00E+00 |
| 1520 | Ba | 115 | 56 | 59 | 3 | 8.11700000E+00 | 8.12563176E+00 | -8.60E-03 | 1.07043955E+05 | 1.07042882E+05 | 1.10E+00 | 1.07072792E+05 | 1.07071719E+05 | 1.10E+00 | -4.90250000E+01 | -5.00982793E+01 | 1.10E+00 |
| 1521 | Ba | 116 | 56 | 60 | 4 | 8.16500000E+00 | 8.17382751E+00 | -8.80E-03 | 1.07969777E+05 | 1.07968731E+05 | 1.00E+00 | 1.07998614E+05 | 1.07997568E+05 | 1.00E+00 | -5.46970000E+01 | -5.57432994E+01 | 1.00E+00 |
| 1522 | Ba | 117 | 56 | 61 | 5 | 8.18935400E+00 | 8.19417103E+00 | -4.80E-03 | 1.08898346E+05 | 1.08897742E+05 | 6.00E-01 | 1.08927183E+05 | 1.08926579E+05 | 6.00E-01 | -5.76216870E+01 | -5.82259998E+01 | 6.00E-01 |
| 1523 | Ba | 118 | 56 | 62 | 6 | 8.22800000E+00 | 8.23499242E+00 | -7.00E-03 | 1.09825108E+05 | 1.09824296E+05 | 8.10E-01 | 1.09853945E+05 | 1.09853133E+05 | 8.10E-01 | -6.23540000E+01 | -6.31657753E+01 | 8.10E-01 |
| 1524 | Ba | 119 | 56 | 63 | 7 | 8.24592800E+00 | 8.24947816E+00 | -3.60E-03 | 1.10754366E+05 | 1.10753903E+05 | 4.60E-01 | 1.10783203E+05 | 1.10782740E+05 | 4.60E-01 | -6.45900860E+01 | -6.50532517E+01 | 4.60E-01 |
| 1525 | Ba | 120 | 56 | 64 | 8 | 8.28029400E+00 | 8.28380972E+00 | -3.50E-03 | 1.11681562E+05 | 1.11681099E+05 | 4.60E-01 | 1.11710399E+05 | 1.11709936E+05 | 4.60E-01 | -6.88886380E+01 | -6.93511980E+01 | 4.60E-01 |
| 1526 | Ba | 121 | 56 | 65 | 9 | 8.29390700E+00 | 8.29316190E+00 | 7.50E-04 | 1.12611200E+05 | 1.12611249E+05 | -5.00E-02 | 1.12640037E+05 | 1.12640086E+05 | -5.00E-02 | -7.07448360E+01 | -7.06953027E+01 | -5.00E-02 |
| 1527 | Ba | 122 | 56 | 66 | 10 | 8.32375600E+00 | 8.32186447E+00 | 1.90E-03 | 1.13538829E+05 | 1.13539020E+05 | -1.90E-01 | 1.13567667E+05 | 1.13567857E+05 | -1.90E-01 | -7.46089440E+01 | -7.44188596E+01 | -1.90E-01 |
| 1528 | Ba | 123 | 56 | 67 | 11 | 8.33020800E+00 | 8.32692543E+00 | 3.30E-03 | 1.14469277E+05 | 1.14469641E+05 | -3.60E-01 | 1.14498115E+05 | 1.14498478E+05 | -3.60E-01 | -7.56549520E+01 | -7.52919023E+01 | -3.60E-01 |
| 1529 | Ba | 124 | 56 | 68 | 12 | 8.35582000E+00 | 8.35071213E+00 | 5.10E-03 | 1.15397337E+05 | 1.15397929E+05 | -5.90E-01 | 1.15426174E+05 | 1.15426767E+05 | -5.90E-01 | -7.90897750E+01 | -7.84970606E+01 | -5.90E-01 |
| 1530 | Ba | 125 | 56 | 69 | 13 | 8.35817800E+00 | 8.35199002E+00 | 6.20E-03 | 1.16328252E+05 | 1.16328984E+05 | -7.30E-01 | 1.16357089E+05 | 1.16357821E+05 | -7.30E-01 | -7.96689650E+01 | -7.89361899E+01 | -7.30E-01 |
| 1531 | Ba | 126 | 56 | 70 | 14 | 8.37971800E+00 | 8.37155434E+00 | 8.20E-03 | 1.17256745E+05 | 1.17257733E+05 | -9.90E-01 | 1.17285582E+05 | 1.17286570E+05 | -9.90E-01 | -8.26699020E+01 | -8.16819651E+01 | -9.90E-01 |
| 1532 | Ba | 127 | 56 | 71 | 15 | 8.37845500E+00 | 8.36978455E+00 | 8.70E-03 | 1.18188091E+05 | 1.18189151E+05 | -1.10E+00 | 1.18216928E+05 | 1.18217988E+05 | -1.10E+00 | -8.28179440E+01 | -8.17574371E+01 | -1.10E+00 |
| 1533 | Ba | 128 | 56 | 72 | 16 | 8.39606300E+00 | 8.38594833E+00 | 1.00E-02 | 1.19117024E+05 | 1.19118278E+05 | -1.30E+00 | 1.19145861E+05 | 1.19147115E+05 | -1.30E+00 | -8.53789070E+01 | -8.41248658E+01 | -1.30E+00 |
| 1534 | Ba | 129 | 56 | 73 | 17 | 8.39109900E+00 | 8.38193982E+00 | 9.20E-03 | 1.20048833E+05 | 1.20049974E+05 | -1.10E+00 | 1.20077671E+05 | 1.20078811E+05 | -1.10E+00 | -8.50632880E+01 | -8.39223973E+01 | -1.10E+00 |
| 1535 | Ba | 130 | 56 | 74 | 18 | 8.40555000E+00 | 8.39558433E+00 | 1.00E-02 | 1.20978129E+05 | 1.20979384E+05 | -1.30E+00 | 1.21006966E+05 | 1.21008221E+05 | -1.30E+00 | -8.72617340E+01 | -8.60068045E+01 | -1.30E+00 |
| 1536 | Ba | 131 | 56 | 75 | 19 | 8.39858800E+00 | 8.39014107E+00 | 8.40E-03 | 1.21910201E+05 | 1.21911267E+05 | -1.10E+00 | 1.21939038E+05 | 1.21940104E+05 | -1.10E+00 | -8.66839220E+01 | -8.56180028E+01 | -1.10E+00 |
| 1537 | Ba | 132 | 56 | 76 | 20 | 8.40937500E+00 | 8.40197341E+00 | 7.40E-03 | 1.22839944E+05 | 1.22840880E+05 | -9.40E-01 | 1.22868781E+05 | 1.22869717E+05 | -9.40E-01 | -8.84349900E+01 | -8.74986938E+01 | -9.40E-01 |
| 1538 | Ba | 133 | 56 | 77 | 21 | 8.40020600E+00 | 8.39543990E+00 | 4.80E-03 | 1.23772319E+05 | 1.23772913E+05 | -5.90E-01 | 1.23801157E+05 | 1.23801750E+05 | -5.90E-01 | -8.75535880E+01 | -8.69603921E+01 | -5.90E-01 |
| 1539 | Ba | 134 | 56 | 78 | 22 | 8.40817300E+00 | 8.40527752E+00 | 2.90E-03 | 1.24702417E+05 | 1.24702764E+05 | -3.50E-01 | 1.24731254E+05 | 1.24731601E+05 | -3.50E-01 | -8.89500550E+01 | -8.86027540E+01 | -3.50E-01 |
| 1540 | Ba | 135 | 56 | 79 | 23 | 8.39753400E+00 | 8.39660698E+00 | 9.30E-04 | 1.25635011E+05 | 1.25635095E+05 | -8.40E-02 | 1.25663848E+05 | 1.25663932E+05 | -8.50E-02 | -8.78507130E+01 | -8.77661895E+01 | -8.50E-02 |
| 1541 | Ba | 136 | 56 | 80 | 24 | 8.40275600E+00 | 8.40245927E+00 | 3.00E-04 | 1.26565468E+05 | 1.26565468E+05 | 2.50E-04 | 1.26594305E+05 | 1.26594305E+05 | 1.90E-04 | -8.88871380E+01 | -8.88873892E+01 | 2.50E-04 |
| 1542 | Ba | 137 | 56 | 81 | 25 | 8.39182800E+00 | 8.38878736E+00 | 3.00E-03 | 1.27498128E+05 | 1.27498504E+05 | -3.80E-01 | 1.27526965E+05 | 1.27527341E+05 | -3.80E-01 | -8.77214530E+01 | -8.73454767E+01 | -3.80E-01 |
| 1543 | Ba | 138 | 56 | 82 | 26 | 8.39342200E+00 | 8.38755876E+00 | 5.90E-03 | 1.28429082E+05 | 1.28429850E+05 | -7.70E-01 | 1.28457919E+05 | 1.28458687E+05 | -7.70E-01 | -8.82618560E+01 | -8.74933992E+01 | -7.70E-01 |
| 1544 | Ba | 139 | 56 | 83 | 27 | 8.36701900E+00 | 8.36643426E+00 | 5.80E-04 | 1.29363924E+05 | 1.29363964E+05 | -4.10E-02 | 1.29392761E+05 | 1.29392801E+05 | -4.10E-02 | -8.49139690E+01 | -8.48733327E+01 | -4.10E-02 |

| | | | | | | | | | | | | | | | | |
|---|---|---|---|---|---|---|---|---|---|---|---|---|---|---|---|---|
| 1545 | Ba | 140 | 56 | 84 | 28 | 8.35316600E+00 | 8.35729243E+00 | -4.10E-03 | 1.30297061E+05 | 1.30296443E+05 | 6.20E-01 | 1.30325898E+05 | 1.30325280E+05 | 6.20E-01 | -8.32702260E+01 | -8.38885921E+01 | 6.20E-01 |
| 1546 | Ba | 141 | 56 | 85 | 29 | 8.32607900E+00 | 8.32981468E+00 | -3.70E-03 | 1.31232093E+05 | 1.31231525E+05 | 5.70E-01 | 1.31260930E+05 | 1.31260362E+05 | 5.70E-01 | -7.97327810E+01 | -8.03002028E+01 | 5.70E-01 |
| 1547 | Ba | 142 | 56 | 86 | 30 | 8.31097400E+00 | 8.31544221E+00 | -4.50E-03 | 1.32165477E+05 | 1.32164802E+05 | 6.80E-01 | 1.32194314E+05 | 1.32193639E+05 | 6.80E-01 | -7.78427340E+01 | -7.85178084E+01 | 6.80E-01 |
| 1548 | Ba | 143 | 56 | 87 | 31 | 8.28198600E+00 | 8.28488884E+00 | -2.90E-03 | 1.33100877E+05 | 1.33100421E+05 | 4.60E-01 | 1.33129714E+05 | 1.33129258E+05 | 4.60E-01 | -7.39370630E+01 | -7.43927994E+01 | 4.60E-01 |
| 1549 | Ba | 144 | 56 | 88 | 32 | 8.26545400E+00 | 8.26828020E+00 | -2.80E-03 | 1.34034541E+05 | 1.34034093E+05 | 4.50E-01 | 1.34063378E+05 | 1.34062930E+05 | 4.50E-01 | -7.17670800E+01 | -7.22147248E+01 | 4.50E-01 |
| 1550 | Ba | 145 | 56 | 89 | 33 | 8.23479800E+00 | 8.23675389E+00 | -2.00E-03 | 1.34970286E+05 | 1.34969961E+05 | 3.20E-01 | 1.34999123E+05 | 1.34998798E+05 | 3.20E-01 | -6.75161760E+01 | -6.78403704E+01 | 3.20E-01 |
| 1551 | Ba | 146 | 56 | 90 | 34 | 8.21603600E+00 | 8.21906721E+00 | -3.00E-03 | 1.35904356E+05 | 1.35903872E+05 | 4.80E-01 | 1.35933193E+05 | 1.35932709E+05 | 4.80E-01 | -6.49403030E+01 | -6.54235508E+01 | 4.80E-01 |
| 1552 | Ba | 147 | 56 | 91 | 35 | 8.18324000E+00 | 8.18690753E+00 | -3.70E-03 | 1.36840526E+05 | 1.36839946E+05 | 5.80E-01 | 1.36869363E+05 | 1.36868783E+05 | 5.80E-01 | -6.02640290E+01 | -6.08438252E+01 | 5.80E-01 |
| 1553 | Ba | 148 | 56 | 92 | 36 | 8.16444100E+00 | 8.16800210E+00 | -3.60E-03 | 1.37774690E+05 | 1.37774123E+05 | 5.70E-01 | 1.37803527E+05 | 1.37802960E+05 | 5.70E-01 | -5.75937350E+01 | -5.81614105E+01 | 5.70E-01 |
| 1554 | Ba | 149 | 56 | 93 | 37 | 8.13300000E+00 | 8.13479159E+00 | -1.80E-03 | 1.38710757E+05 | 1.38710468E+05 | 2.90E-01 | 1.38739594E+05 | 1.38739305E+05 | 2.90E-01 | -5.30210000E+01 | -5.33097281E+01 | 2.90E-01 |
| 1555 | Ba | 150 | 56 | 94 | 38 | 8.11400000E+00 | 8.11421848E+00 | -2.20E-04 | 1.39645018E+05 | 1.39644985E+05 | 3.30E-02 | 1.39673855E+05 | 1.39673822E+05 | 3.30E-02 | -5.02540000E+01 | -5.02872346E+01 | 3.30E-02 |
| 1556 | Ba | 151 | 56 | 95 | 39 | 8.08200000E+00 | 8.07953493E+00 | 2.50E-03 | 1.40581375E+05 | 1.40581673E+05 | -3.00E-01 | 1.40610212E+05 | 1.40610510E+05 | -3.00E-01 | -4.53920000E+01 | -4.50929173E+01 | -3.00E-01 |
| 1557 | BA | 152 | 56 | 96 | 40 | 8.06000000E+00 | 8.05701426E+00 | 3.00E-03 | 1.41516166E+05 | 1.41516582E+05 | -4.20E-01 | 1.41545003E+05 | 1.41545419E+05 | -4.20E-01 | -4.20940000E+01 | -4.16779911E+01 | -4.20E-01 |
| 1558 | La | 117 | 57 | 60 | 3 | 8.08800000E+00 | 8.09313072E+00 | -5.10E-03 | 1.08908859E+05 | 1.08908260E+05 | 6.00E-01 | 1.08938216E+05 | 1.08937618E+05 | 6.00E-01 | -4.65880000E+01 | -4.71873597E+01 | 6.00E-01 |
| 1559 | La | 120 | 57 | 63 | 6 | 8.18000000E+00 | 8.18256857E+00 | -2.60E-03 | 1.11692242E+05 | 1.11691944E+05 | 3.00E-01 | 1.11721600E+05 | 1.11721302E+05 | 3.00E-01 | -5.76870000E+01 | -5.79853370E+01 | 3.00E-01 |
| 1560 | La | 121 | 57 | 64 | 7 | 8.21700000E+00 | 8.21888104E+00 | -1.90E-03 | 1.12619153E+05 | 1.12618933E+05 | 2.20E-01 | 1.12648511E+05 | 1.12648291E+05 | 2.20E-01 | -6.22700000E+01 | -6.24903962E+01 | 2.20E-01 |
| 1561 | La | 122 | 57 | 65 | 8 | 8.23500000E+00 | 8.23323526E+00 | 1.80E-03 | 1.13548374E+05 | 1.13548528E+05 | -1.50E-01 | 1.13577732E+05 | 1.13577886E+05 | -1.50E-01 | -6.45430000E+01 | -6.43891724E+01 | -1.50E-01 |
| 1562 | La | 123 | 57 | 66 | 9 | 8.26700000E+00 | 8.26362108E+00 | 3.40E-03 | 1.14475761E+05 | 1.14476123E+05 | -3.60E-01 | 1.14505118E+05 | 1.14505481E+05 | -3.60E-01 | -6.86510000E+01 | -6.82885449E+01 | -3.60E-01 |
| 1563 | La | 124 | 57 | 67 | 10 | 8.27829200E+00 | 8.27340100E+00 | 4.90E-03 | 1.15405647E+05 | 1.15406212E+05 | -5.70E-01 | 1.15435005E+05 | 1.15435570E+05 | -5.70E-01 | -7.02586100E+01 | -6.96935565E+01 | -5.70E-01 |
| 1564 | La | 125 | 57 | 68 | 11 | 8.30464300E+00 | 8.29866812E+00 | 6.00E-03 | 1.16333640E+05 | 1.16334346E+05 | -7.10E-01 | 1.16362998E+05 | 1.16363704E+05 | -7.10E-01 | -7.37594840E+01 | -7.30540293E+01 | -7.10E-01 |
| 1565 | La | 126 | 57 | 69 | 12 | 8.31242600E+00 | 8.30446461E+00 | 8.00E-03 | 1.17263920E+05 | 1.17264882E+05 | -9.60E-01 | 1.17293278E+05 | 1.17294240E+05 | -9.60E-01 | -7.49734680E+01 | -7.40117357E+01 | -9.60E-01 |
| 1566 | La | 127 | 57 | 70 | 13 | 8.33354000E+00 | 8.32536885E+00 | 8.20E-03 | 1.18192492E+05 | 1.18193488E+05 | -1.00E+00 | 1.18221850E+05 | 1.18222846E+05 | -1.00E+00 | -7.78961080E+01 | -7.68997203E+01 | -1.00E+00 |
| 1567 | La | 128 | 57 | 71 | 14 | 8.33719000E+00 | 8.32798232E+00 | 9.20E-03 | 1.19123257E+05 | 1.19124394E+05 | -1.10E+00 | 1.19152614E+05 | 1.19153752E+05 | -1.10E+00 | -7.86254300E+01 | -7.74882945E+01 | -1.10E+00 |
| 1568 | La | 129 | 57 | 72 | 15 | 8.35605200E+00 | 8.34538288E+00 | 1.10E-02 | 1.20052051E+05 | 1.20053386E+05 | -1.30E+00 | 1.20081409E+05 | 1.20082744E+05 | -1.30E+00 | -8.13246130E+01 | -7.99896299E+01 | -1.30E+00 |
| 1569 | La | 130 | 57 | 73 | 16 | 8.35619100E+00 | 8.34565283E+00 | 1.10E-02 | 1.20983243E+05 | 1.20984571E+05 | -1.30E+00 | 1.21012601E+05 | 1.21013929E+05 | -1.30E+00 | -8.16273650E+01 | -8.02987869E+01 | -1.30E+00 |
| 1570 | La | 131 | 57 | 74 | 17 | 8.37036700E+00 | 8.36043294E+00 | 9.90E-03 | 1.21912595E+05 | 1.21913855E+05 | -1.30E+00 | 1.21941953E+05 | 1.21943213E+05 | -1.30E+00 | -8.37692560E+01 | -8.25093152E+01 | -1.30E+00 |
| 1571 | La | 132 | 57 | 75 | 18 | 8.36775600E+00 | 8.35915727E+00 | 8.60E-03 | 1.22844135E+05 | 1.22845228E+05 | -1.10E+00 | 1.22873492E+05 | 1.22874586E+05 | -1.10E+00 | -8.37236320E+01 | -8.26300400E+01 | -1.10E+00 |
| 1572 | La | 133 | 57 | 76 | 19 | 8.37884100E+00 | 8.37199703E+00 | 6.80E-03 | 1.23773858E+05 | 1.23774727E+05 | -8.70E-01 | 1.23803216E+05 | 1.23804085E+05 | -8.70E-01 | -8.54943830E+01 | -8.46255671E+01 | -8.70E-01 |
| 1573 | La | 134 | 57 | 77 | 20 | 8.37448800E+00 | 8.36950045E+00 | 5.00E-03 | 1.24705628E+05 | 1.24706255E+05 | -6.30E-01 | 1.24734985E+05 | 1.24735612E+05 | -6.30E-01 | -8.52186500E+01 | -8.45917036E+01 | -6.30E-01 |
| 1574 | La | 135 | 57 | 78 | 21 | 8.38279700E+00 | 8.38021038E+00 | 2.60E-03 | 1.25635697E+05 | 1.25636005E+05 | -3.10E-01 | 1.25665055E+05 | 1.25665363E+05 | -3.10E-01 | -8.66435080E+01 | -8.63357256E+01 | -3.10E-01 |
| 1575 | La | 136 | 57 | 79 | 22 | 8.37605000E+00 | 8.37549288E+00 | 5.60E-04 | 1.26567797E+05 | 1.26567831E+05 | -3.40E-02 | 1.26597155E+05 | 1.26597189E+05 | -3.40E-02 | -8.60374780E+01 | -8.60030373E+01 | -3.40E-02 |
| 1576 | La | 137 | 57 | 80 | 23 | 8.38188000E+00 | 8.38219576E+00 | -3.20E-04 | 1.27498188E+05 | 1.27498103E+05 | 8.50E-02 | 1.27527545E+05 | 1.27527461E+05 | 8.50E-02 | -8.71409040E+01 | -8.72255052E+01 | 8.50E-02 |
| 1577 | La | 138 | 57 | 81 | 24 | 8.37514400E+00 | 8.37258634E+00 | 2.60E-03 | 1.28430301E+05 | 1.28430612E+05 | -3.10E-01 | 1.28459659E+05 | 1.28459970E+05 | -3.10E-01 | -8.65218950E+01 | -8.62102814E+01 | -3.10E-01 |
| 1578 | La | 139 | 57 | 82 | 25 | 8.37804300E+00 | 8.37241037E+00 | 5.60E-03 | 1.29361088E+05 | 1.29361830E+05 | -7.40E-01 | 1.29390446E+05 | 1.29391187E+05 | -7.40E-01 | -8.72285850E+01 | -8.64870897E+01 | -7.40E-01 |
| 1579 | La | 140 | 57 | 83 | 26 | 8.35506400E+00 | 8.35564058E+00 | -5.80E-04 | 1.30295492E+05 | 1.30295370E+05 | 1.20E-01 | 1.30324850E+05 | 1.30324728E+05 | 1.20E-01 | -8.43182450E+01 | -8.44404100E+01 | 1.20E-01 |
| 1580 | La | 141 | 57 | 84 | 27 | 8.34323800E+00 | 8.34780581E+00 | -4.60E-03 | 1.31228370E+05 | 1.31227685E+05 | 6.90E-01 | 1.31257728E+05 | 1.31257043E+05 | 6.90E-01 | -8.29346090E+01 | -8.36200299E+01 | 6.90E-01 |
| 1581 | La | 142 | 57 | 85 | 28 | 8.32082500E+00 | 8.32487854E+00 | -4.10E-03 | 1.32162775E+05 | 1.32162158E+05 | 6.20E-01 | 1.32192133E+05 | 1.32191516E+05 | 6.20E-01 | -8.00238210E+01 | -8.06408435E+01 | 6.20E-01 |
| 1582 | La | 143 | 57 | 86 | 29 | 8.30612600E+00 | 8.31186460E+00 | -5.70E-03 | 1.33096122E+05 | 1.33095260E+05 | 8.60E-01 | 1.33125479E+05 | 1.33124617E+05 | 8.60E-01 | -7.81714300E+01 | -7.90334091E+01 | 8.60E-01 |
| 1583 | La | 144 | 57 | 87 | 30 | 8.28142800E+00 | 8.28584476E+00 | -4.40E-03 | 1.34030937E+05 | 1.34030260E+05 | 6.80E-01 | 1.34060295E+05 | 1.34059618E+05 | 6.80E-01 | -7.48496520E+01 | -7.55270981E+01 | 6.80E-01 |
| 1584 | La | 145 | 57 | 88 | 31 | 8.26608700E+00 | 8.27048748E+00 | -4.40E-03 | 1.34964446E+05 | 1.34963766E+05 | 6.80E-01 | 1.34993804E+05 | 1.34993124E+05 | 6.80E-01 | -7.28353150E+01 | -7.35148189E+01 | 6.80E-01 |
| 1585 | La | 146 | 57 | 89 | 32 | 8.23880400E+00 | 8.24340342E+00 | -4.60E-03 | 1.35899728E+05 | 1.35899015E+05 | 7.10E-01 | 1.35929086E+05 | 1.35928373E+05 | 7.10E-01 | -6.90468310E+01 | -6.97597145E+01 | 7.10E-01 |
| 1586 | La | 147 | 57 | 90 | 33 | 8.22155300E+00 | 8.22684667E+00 | -5.30E-03 | 1.36833591E+05 | 1.36832771E+05 | 8.20E-01 | 1.36862949E+05 | 1.36862129E+05 | 8.20E-01 | -6.66783900E+01 | -6.74979564E+01 | 8.20E-01 |
| 1587 | La | 148 | 57 | 91 | 34 | 8.19371600E+00 | 8.19906007E+00 | -5.30E-03 | 1.37769054E+05 | 1.37768222E+05 | 8.30E-01 | 1.37798412E+05 | 1.37797580E+05 | 8.30E-01 | -6.27087350E+01 | -6.35410680E+01 | 8.30E-01 |
| 1588 | La | 149 | 57 | 92 | 35 | 8.17619100E+00 | 8.18118633E+00 | -5.00E-03 | 1.38703037E+05 | 1.38702252E+05 | 7.90E-01 | 1.38732395E+05 | 1.38731609E+05 | 7.90E-01 | -6.02199130E+01 | -6.10056210E+01 | 7.90E-01 |
| 1589 | La | 150 | 57 | 93 | 36 | 8.15000000E+00 | 8.15229521E+00 | -2.30E-03 | 1.39638368E+05 | 1.39637970E+05 | 4.00E-01 | 1.39667726E+05 | 1.39667327E+05 | 4.00E-01 | -5.63830000E+01 | -5.67818200E+01 | 4.00E-01 |
| 1590 | La | 151 | 57 | 94 | 37 | 8.13200000E+00 | 8.13264807E+00 | -6.50E-04 | 1.40572517E+05 | 1.40572349E+05 | 1.70E-01 | 1.40601875E+05 | 1.40601707E+05 | 1.70E-01 | -5.37290000E+01 | -5.38960790E+01 | 1.70E-01 |
| 1591 | La | 152 | 57 | 95 | 38 | 8.10400000E+00 | 8.10220706E+00 | 1.80E-03 | 1.41508203E+05 | 1.41508409E+05 | -2.10E-01 | 1.41537560E+05 | 1.41537767E+05 | -2.10E-01 | -4.95370000E+01 | -4.93303747E+01 | -2.10E-01 |

| | | | | | | | | | | | | | | | | |
|---|---|---|---|---|---|---|---|---|---|---|---|---|---|---|---|---|
| 1592 | La | 153 | 57 | 96 | 39 | 8.08200000E+00 | 8.08046244E+00 | 1.50E-03 | 1.42442994E+05 | 1.42443199E+05 | -2.00E-01 | 1.42472352E+05 | 1.42472557E+05 | -2.00E-01 | -4.62390000E+01 | -4.60343348E+01 | -2.00E-01 |
| 1593 | Ce | 121 | 58 | 63 | 5 | 8.13200000E+00 | 8.13307168E+00 | -1.10E-03 | 1.12628133E+05 | 1.12628012E+05 | 1.20E-01 | 1.12658012E+05 | 1.12657891E+05 | 1.20E-01 | -5.27690000E+01 | -5.28905400E+01 | 1.20E-01 |
| 1594 | Ce | 122 | 58 | 64 | 6 | 8.17400000E+00 | 8.17462926E+00 | -6.30E-04 | 1.13554523E+05 | 1.13554374E+05 | 1.50E-01 | 1.13584402E+05 | 1.13584253E+05 | 1.50E-01 | -5.78740000E+01 | -5.80223177E+01 | 1.50E-01 |
| 1595 | Ce | 123 | 58 | 65 | 7 | 8.19300000E+00 | 8.19038168E+00 | 2.60E-03 | 1.14483604E+05 | 1.14483828E+05 | -2.20E-01 | 1.14513483E+05 | 1.14513706E+05 | -2.20E-01 | -6.02860000E+01 | -6.00631751E+01 | -2.20E-01 |
| 1596 | Ce | 124 | 58 | 66 | 8 | 8.22900000E+00 | 8.22556024E+00 | 3.40E-03 | 1.15410469E+05 | 1.15410840E+05 | -3.70E-01 | 1.15440348E+05 | 1.15440719E+05 | -3.70E-01 | -6.49160000E+01 | -6.45443793E+01 | -3.70E-01 |
| 1597 | Ce | 125 | 58 | 67 | 9 | 8.24200000E+00 | 8.23652322E+00 | 5.50E-03 | 1.16340221E+05 | 1.16340810E+05 | -5.90E-01 | 1.16370100E+05 | 1.16370689E+05 | -5.90E-01 | -6.66580000E+01 | -6.60689941E+01 | -5.90E-01 |
| 1598 | Ce | 126 | 58 | 68 | 10 | 8.27325700E+00 | 8.26624612E+00 | 7.00E-03 | 1.17267552E+05 | 1.17268394E+05 | -8.40E-01 | 1.17297431E+05 | 1.17298272E+05 | -8.40E-01 | -7.08205580E+01 | -6.99792831E+01 | -8.40E-01 |
| 1599 | Ce | 127 | 58 | 69 | 11 | 8.28079100E+00 | 8.27307454E+00 | 7.70E-03 | 1.18197888E+05 | 1.18198826E+05 | -9.40E-01 | 1.18227766E+05 | 1.18228704E+05 | -9.40E-01 | -7.19793360E+01 | -7.10414190E+01 | -9.40E-01 |
| 1600 | Ce | 128 | 58 | 70 | 12 | 8.30692500E+00 | 8.29818178E+00 | 8.70E-03 | 1.19125827E+05 | 1.19126904E+05 | -1.10E+00 | 1.19155706E+05 | 1.19156783E+05 | -1.10E+00 | -7.55339170E+01 | -7.44569019E+01 | -1.10E+00 |
| 1601 | Ce | 129 | 58 | 71 | 13 | 8.31094000E+00 | 8.30172956E+00 | 9.20E-03 | 1.20056568E+05 | 1.20057714E+05 | -1.10E+00 | 1.20086446E+05 | 1.20087592E+05 | -1.10E+00 | -7.62874960E+01 | -7.51414282E+01 | -1.10E+00 |
| 1602 | Ce | 130 | 58 | 72 | 14 | 8.33321600E+00 | 8.32313806E+00 | 1.00E-02 | 1.20984926E+05 | 1.20986194E+05 | -1.30E+00 | 1.21014805E+05 | 1.21016073E+05 | -1.30E+00 | -7.94229050E+01 | -7.81549435E+01 | -1.30E+00 |
| 1603 | Ce | 131 | 58 | 73 | 15 | 8.33339600E+00 | 8.32426803E+00 | 9.10E-03 | 1.21916135E+05 | 1.21917288E+05 | -1.20E+00 | 1.21946014E+05 | 1.21947167E+05 | -1.20E+00 | -7.97084390E+01 | -7.85547885E+01 | -1.20E+00 |
| 1604 | Ce | 132 | 58 | 74 | 16 | 8.35233900E+00 | 8.34288141E+00 | 9.50E-03 | 1.22844866E+05 | 1.22846073E+05 | -1.20E+00 | 1.22874745E+05 | 1.22875951E+05 | -1.20E+00 | -8.24709630E+01 | -8.12647043E+01 | -1.20E+00 |
| 1605 | Ce | 133 | 58 | 75 | 17 | 8.34982900E+00 | 8.34238168E+00 | 7.40E-03 | 1.23776413E+05 | 1.23777361E+05 | -9.50E-01 | 1.23806292E+05 | 1.23807240E+05 | -9.50E-01 | -8.24182140E+01 | -8.14698025E+01 | -9.50E-01 |
| 1606 | Ce | 134 | 58 | 76 | 18 | 8.36577100E+00 | 8.35886995E+00 | 6.90E-03 | 1.24705492E+05 | 1.24706375E+05 | -8.80E-01 | 1.24735371E+05 | 1.24736254E+05 | -8.80E-01 | -8.48328890E+01 | -8.39502938E+01 | -8.80E-01 |
| 1607 | Ce | 135 | 58 | 77 | 19 | 8.36198600E+00 | 8.35704412E+00 | 4.90E-03 | 1.25637203E+05 | 1.25637828E+05 | -6.20E-01 | 1.25667082E+05 | 1.25667707E+05 | -6.20E-01 | -8.46163560E+01 | -8.39913569E+01 | -6.20E-01 |
| 1608 | Ce | 136 | 58 | 78 | 20 | 8.37376200E+00 | 8.37122468E+00 | 2.50E-03 | 1.26566805E+05 | 1.26567108E+05 | -3.00E-01 | 1.26596684E+05 | 1.26596987E+05 | -3.00E-01 | -8.65085880E+01 | -8.62056391E+01 | -3.00E-01 |
| 1609 | Ce | 137 | 58 | 79 | 21 | 8.36724900E+00 | 8.36711606E+00 | 1.30E-04 | 1.27498889E+05 | 1.27498865E+05 | 2.40E-02 | 1.27528768E+05 | 1.27528744E+05 | 2.40E-02 | -8.59188040E+01 | -8.59426637E+01 | 2.40E-02 |
| 1610 | Ce | 138 | 58 | 80 | 22 | 8.37706100E+00 | 8.37723073E+00 | -1.70E-04 | 1.28428733E+05 | 1.28428667E+05 | 6.60E-02 | 1.28458612E+05 | 1.28458546E+05 | 6.60E-02 | -8.75687360E+01 | -8.76342845E+01 | 6.60E-02 |
| 1611 | Ce | 139 | 58 | 81 | 23 | 8.37041100E+00 | 8.36835115E+00 | 2.10E-03 | 1.29360845E+05 | 1.29361090E+05 | -2.40E-01 | 1.29390724E+05 | 1.29390969E+05 | -2.40E-01 | -8.69502040E+01 | -8.67059352E+01 | -2.40E-01 |
| 1612 | Ce | 140 | 58 | 82 | 24 | 8.37633900E+00 | 8.37175182E+00 | 4.60E-03 | 1.30291211E+05 | 1.30291811E+05 | -6.00E-01 | 1.30321089E+05 | 1.30321689E+05 | -6.00E-01 | -8.80791780E+01 | -8.74790609E+01 | -6.00E-01 |
| 1613 | Ce | 141 | 58 | 83 | 25 | 8.35543000E+00 | 8.35601017E+00 | -5.80E-04 | 1.31225348E+05 | 1.31225224E+05 | 1.20E-01 | 1.31255227E+05 | 1.31255103E+05 | 1.20E-01 | -8.54360020E+01 | -8.55599205E+01 | 1.20E-01 |
| 1614 | Ce | 142 | 58 | 84 | 26 | 8.34706800E+00 | 8.35198345E+00 | -4.90E-03 | 1.32157745E+05 | 1.32157005E+05 | 7.40E-01 | 1.32187624E+05 | 1.32186884E+05 | 7.40E-01 | -8.45327310E+01 | -8.52728174E+01 | 7.40E-01 |
| 1615 | Ce | 143 | 58 | 85 | 27 | 8.32467500E+00 | 8.33029870E+00 | -5.60E-03 | 1.33092166E+05 | 1.33091319E+05 | 8.50E-01 | 1.33122044E+05 | 1.33121198E+05 | 8.50E-01 | -8.16062150E+01 | -8.24525625E+01 | 8.50E-01 |
| 1616 | Ce | 144 | 58 | 86 | 28 | 8.31475900E+00 | 8.32114458E+00 | -6.40E-03 | 1.34024834E+05 | 1.34023873E+05 | 9.60E-01 | 1.34054713E+05 | 1.34053751E+05 | 9.60E-01 | -8.04317690E+01 | -8.13933492E+01 | 9.60E-01 |
| 1617 | Ce | 145 | 58 | 87 | 29 | 8.28987500E+00 | 8.29637697E+00 | -6.50E-03 | 1.34959693E+05 | 1.34958708E+05 | 9.80E-01 | 1.34989572E+05 | 1.34988587E+05 | 9.80E-01 | -7.70669440E+01 | -7.80518719E+01 | 9.80E-01 |
| 1618 | Ce | 146 | 58 | 88 | 30 | 8.27857000E+00 | 8.28478301E+00 | -6.20E-03 | 1.35892619E+05 | 1.35891670E+05 | 9.50E-01 | 1.35922498E+05 | 1.35921549E+05 | 9.50E-01 | -7.56350660E+01 | -7.65842110E+01 | 9.50E-01 |
| 1619 | Ce | 147 | 58 | 89 | 31 | 8.25252700E+00 | 8.25888249E+00 | -6.40E-03 | 1.36827734E+05 | 1.36826758E+05 | 9.80E-01 | 1.36857613E+05 | 1.36856637E+05 | 9.80E-01 | -7.20138820E+01 | -7.29902986E+01 | 9.80E-01 |
| 1620 | Ce | 148 | 58 | 90 | 32 | 8.24038700E+00 | 8.24598434E+00 | -5.60E-03 | 1.37760844E+05 | 1.37759973E+05 | 8.70E-01 | 1.37790723E+05 | 1.37789852E+05 | 8.70E-01 | -7.03983800E+01 | -7.12689371E+01 | 8.70E-01 |
| 1621 | Ce | 149 | 58 | 91 | 33 | 8.21422900E+00 | 8.21933337E+00 | -5.10E-03 | 1.38696066E+05 | 1.38695264E+05 | 8.00E-01 | 1.38725945E+05 | 1.38725142E+05 | 8.00E-01 | -6.66699130E+01 | -6.74726065E+01 | 8.00E-01 |
| 1622 | Ce | 150 | 58 | 92 | 34 | 8.20112200E+00 | 8.20504981E+00 | -3.90E-03 | 1.39629383E+05 | 1.39628752E+05 | 6.30E-01 | 1.39659262E+05 | 1.39658631E+05 | 6.30E-01 | -6.48468470E+01 | -6.54780874E+01 | 6.30E-01 |
| 1623 | Ce | 151 | 58 | 93 | 35 | 8.17627700E+00 | 8.17726737E+00 | -9.90E-04 | 1.40564499E+05 | 1.40564308E+05 | 1.90E-01 | 1.40594378E+05 | 1.40594187E+05 | 1.90E-01 | -6.12250510E+01 | -6.14166701E+01 | 1.90E-01 |
| 1624 | Ce | 152 | 58 | 94 | 36 | 8.16100000E+00 | 8.16115312E+00 | -1.50E-04 | 1.41498162E+05 | 1.41498145E+05 | 1.60E-02 | 1.41528041E+05 | 1.41528024E+05 | 1.60E-02 | -5.90570000E+01 | -5.90732527E+01 | 1.60E-02 |
| 1625 | Ce | 153 | 58 | 95 | 37 | 8.13400000E+00 | 8.13177929E+00 | 2.20E-03 | 1.42433689E+05 | 1.42434044E+05 | -3.50E-01 | 1.42463568E+05 | 1.42463922E+05 | -3.50E-01 | -5.50230000E+01 | -5.46688899E+01 | -3.50E-01 |
| 1626 | Ce | 154 | 58 | 96 | 38 | 8.11700000E+00 | 8.11348541E+00 | 3.50E-03 | 1.43367857E+05 | 1.43368294E+05 | -4.40E-01 | 1.43397735E+05 | 1.43398173E+05 | -4.40E-01 | -5.23500000E+01 | -5.19120936E+01 | -4.40E-01 |
| 1627 | Ce | 155 | 58 | 97 | 39 | 8.08800000E+00 | 8.08224532E+00 | 5.80E-03 | 1.44303775E+05 | 1.44304589E+05 | -8.10E-01 | 1.44333654E+05 | 1.44334467E+05 | -8.10E-01 | -4.79250000E+01 | -4.71120461E+01 | -8.10E-01 |
| 1628 | Pr | 121 | 59 | 62 | 3 | 8.03400000E+00 | 8.03292563E+00 | 1.10E-03 | 1.12638762E+05 | 1.12638825E+05 | -6.30E-02 | 1.12669162E+05 | 1.12669225E+05 | -6.30E-02 | -4.16190000E+01 | -4.15559454E+01 | -6.30E-02 |
| 1629 | Pr | 124 | 59 | 65 | 6 | 8.12800000E+00 | 8.12464245E+00 | 3.40E-03 | 1.15421712E+05 | 1.15422050E+05 | -3.40E-01 | 1.15452113E+05 | 1.15452450E+05 | -3.40E-01 | -5.31510000E+01 | -5.28136506E+01 | -3.40E-01 |
| 1630 | Pr | 125 | 59 | 66 | 7 | 8.16600000E+00 | 8.16175823E+00 | 4.20E-03 | 1.16348326E+05 | 1.16348851E+05 | -5.30E-01 | 1.16378726E+05 | 1.16379251E+05 | -5.30E-01 | -5.80320000E+01 | -5.75064466E+01 | -5.30E-01 |
| 1631 | Pr | 126 | 59 | 67 | 8 | 8.18400000E+00 | 8.17737185E+00 | 6.60E-03 | 1.17277528E+05 | 1.17278287E+05 | -7.60E-01 | 1.17307928E+05 | 1.17308687E+05 | -7.60E-01 | -6.03240000E+01 | -5.95642018E+01 | -7.60E-01 |
| 1632 | Pr | 127 | 59 | 68 | 9 | 8.21600000E+00 | 8.20880481E+00 | 7.20E-03 | 1.18204802E+05 | 1.18205683E+05 | -8.80E-01 | 1.18235202E+05 | 1.18236084E+05 | -8.80E-01 | -6.45430000E+01 | -6.36622404E+01 | -8.80E-01 |
| 1633 | Pr | 128 | 59 | 69 | 10 | 8.22891300E+00 | 8.21999039E+00 | 8.90E-03 | 1.19134509E+05 | 1.19135608E+05 | -1.10E+00 | 1.19164909E+05 | 1.19166008E+05 | -1.10E+00 | -6.63307560E+01 | -6.52314810E+01 | -1.10E+00 |
| 1634 | Pr | 129 | 59 | 70 | 11 | 8.25438000E+00 | 8.24664176E+00 | 7.70E-03 | 1.20062560E+05 | 1.20063516E+05 | -9.60E-01 | 1.20092960E+05 | 1.20093916E+05 | -9.60E-01 | -6.97735580E+01 | -6.88181785E+01 | -9.60E-01 |
| 1635 | Pr | 130 | 59 | 71 | 12 | 8.26375600E+00 | 8.25432806E+00 | 9.40E-03 | 1.20992652E+05 | 1.20993835E+05 | -1.20E+00 | 1.21023053E+05 | 1.21024235E+05 | -1.20E+00 | -7.11754570E+01 | -6.99927208E+01 | -1.20E+00 |
| 1636 | Pr | 131 | 59 | 72 | 13 | 8.28614300E+00 | 8.27715866E+00 | 9.00E-03 | 1.21921021E+05 | 1.21922155E+05 | -1.10E+00 | 1.21951421E+05 | 1.21952555E+05 | -1.10E+00 | -7.43006560E+01 | -7.31665384E+01 | -1.10E+00 |
| 1637 | Pr | 132 | 59 | 73 | 14 | 8.29143100E+00 | 8.28225180E+00 | 9.20E-03 | 1.22851602E+05 | 1.22852771E+05 | -1.20E+00 | 1.22882003E+05 | 1.22883171E+05 | -1.20E+00 | -7.52134830E+01 | -7.40446730E+01 | -1.20E+00 |
| 1638 | Pr | 133 | 59 | 74 | 15 | 8.31025800E+00 | 8.30217747E+00 | 8.10E-03 | 1.23780372E+05 | 1.23781404E+05 | -1.00E+00 | 1.23810773E+05 | 1.23811804E+05 | -1.00E+00 | -7.79375800E+01 | -7.69057191E+01 | -1.00E+00 |
| 1639 | Pr | 134 | 59 | 75 | 16 | 8.31288100E+00 | 8.30547412E+00 | 7.40E-03 | 1.24711276E+05 | 1.24712226E+05 | -9.50E-01 | 1.24741676E+05 | 1.24742626E+05 | -9.50E-01 | -7.85279910E+01 | -7.75783287E+01 | -9.50E-01 |
| 1640 | Pr | 135 | 59 | 76 | 17 | 8.32892800E+00 | 8.32315021E+00 | 5.80E-03 | 1.25640362E+05 | 1.25641099E+05 | -7.40E-01 | 1.25670762E+05 | 1.25671499E+05 | -7.40E-01 | -8.09358600E+01 | -8.01987563E+01 | -7.40E-01 |
| 1641 | Pr | 136 | 59 | 77 | 18 | 8.33000800E+00 | 8.32494911E+00 | 5.10E-03 | 1.26571452E+05 | 1.26572097E+05 | -6.50E-01 | 1.26601852E+05 | 1.26602497E+05 | -6.50E-01 | -8.13404260E+01 | -8.06952377E+01 | -6.50E-01 |
| 1642 | Pr | 137 | 59 | 78 | 19 | 8.34170700E+00 | 8.34019581E+00 | 1.50E-03 | 1.27501084E+05 | 1.27501249E+05 | -1.60E-01 | 1.27531485E+05 | 1.27531649E+05 | -1.60E-01 | -8.32018360E+01 | -8.30376663E+01 | -1.60E-01 |
| 1643 | Pr | 138 | 59 | 79 | 20 | 8.33924000E+00 | 8.33959392E+00 | -3.50E-04 | 1.28432649E+05 | 1.28432557E+05 | 9.20E-02 | 1.28463049E+05 | 1.28462957E+05 | 9.20E-02 | -8.31317360E+01 | -8.32234823E+01 | 9.20E-02 |
| 1644 | Pr | 139 | 59 | 80 | 21 | 8.34946600E+00 | 8.35076152E+00 | -1.30E-03 | 1.29362453E+05 | 1.29362230E+05 | 2.20E-01 | 1.29392853E+05 | 1.29392630E+05 | 2.20E-01 | -8.48211330E+01 | -8.50440527E+01 | 2.20E-01 |
| 1645 | Pr | 140 | 59 | 81 | 22 | 8.34655100E+00 | 8.34545318E+00 | 1.10E-03 | 1.30294077E+05 | 1.30294188E+05 | -1.10E-01 | 1.30324477E+05 | 1.30324588E+05 | -1.10E-01 | -8.46911780E+01 | -8.45803278E+01 | -1.10E-01 |
| 1646 | Pr | 141 | 59 | 82 | 23 | 8.35399800E+00 | 8.35010087E+00 | 3.90E-03 | 1.31224246E+05 | 1.31224753E+05 | -5.10E-01 | 1.31254646E+05 | 1.31255153E+05 | -5.10E-01 | -8.60164220E+01 | -8.55097863E+01 | -5.10E-01 |
| 1647 | Pr | 142 | 59 | 83 | 24 | 8.33631600E+00 | 8.33817385E+00 | -1.90E-03 | 1.32157968E+05 | 1.32157662E+05 | 3.10E-01 | 1.32188368E+05 | 1.32188062E+05 | 3.10E-01 | -8.37882510E+01 | -8.40949310E+01 | 3.10E-01 |
| 1648 | Pr | 143 | 59 | 84 | 25 | 8.32942600E+00 | 8.33565240E+00 | -6.20E-03 | 1.33090183E+05 | 1.33089249E+05 | 9.30E-01 | 1.33120583E+05 | 1.33119650E+05 | 9.30E-01 | -8.30680310E+01 | -8.40012190E+01 | 9.30E-01 |
| 1649 | Pr | 144 | 59 | 85 | 26 | 8.31153900E+00 | 8.31796148E+00 | -6.40E-03 | 1.34023994E+05 | 1.34023027E+05 | 9.70E-01 | 1.34054394E+05 | 1.34053427E+05 | 9.70E-01 | -8.07504150E+01 | -8.17180608E+01 | 9.70E-01 |
| 1650 | Pr | 145 | 59 | 86 | 27 | 8.30212700E+00 | 8.31040053E+00 | -8.30E-03 | 1.34956613E+05 | 1.34955370E+05 | 1.20E+00 | 1.34987013E+05 | 1.34985770E+05 | 1.20E+00 | -7.96258240E+01 | -8.08683643E+01 | 1.20E+00 |
| 1651 | Pr | 146 | 59 | 87 | 28 | 8.28037400E+00 | 8.28961923E+00 | -9.20E-03 | 1.35891052E+05 | 1.35889659E+05 | 1.40E+00 | 1.35921452E+05 | 1.35920060E+05 | 1.40E+00 | -7.66807260E+01 | -7.80733762E+01 | 1.40E+00 |
| 1652 | Pr | 147 | 59 | 88 | 29 | 8.27053900E+00 | 8.27955521E+00 | -9.00E-03 | 1.36823783E+05 | 1.36822415E+05 | 1.40E+00 | 1.36854183E+05 | 1.36852815E+05 | 1.40E+00 | -7.54440110E+01 | -7.68122660E+01 | 1.40E+00 |
| 1653 | Pr | 148 | 59 | 89 | 30 | 8.24954000E+00 | 8.25756056E+00 | -8.00E-03 | 1.37758186E+05 | 1.37756956E+05 | 1.20E+00 | 1.37788586E+05 | 1.37787356E+05 | 1.20E+00 | -7.25353600E+01 | -7.37652939E+01 | 1.20E+00 |
| 1654 | Pr | 149 | 59 | 90 | 31 | 8.23830300E+00 | 8.24611138E+00 | -7.80E-03 | 1.38691176E+05 | 1.38689969E+05 | 1.20E+00 | 1.38721576E+05 | 1.38720369E+05 | 1.20E+00 | -7.10393650E+01 | -7.22456073E+01 | 1.20E+00 |
| 1655 | Pr | 150 | 59 | 91 | 32 | 8.21893000E+00 | 8.22331316E+00 | -4.40E-03 | 1.39625409E+05 | 1.39624708E+05 | 7.00E-01 | 1.39655809E+05 | 1.39655108E+05 | 7.00E-01 | -6.83003980E+01 | -6.90006674E+01 | 7.00E-01 |
| 1656 | Pr | 151 | 59 | 92 | 33 | 8.20788000E+00 | 8.21042942E+00 | -2.50E-03 | 1.40558424E+05 | 1.40557996E+05 | 4.30E-01 | 1.40588824E+05 | 1.40588396E+05 | 4.30E-01 | -6.67794710E+01 | -6.72072159E+01 | 4.30E-01 |
| 1657 | Pr | 152 | 59 | 93 | 34 | 8.18710400E+00 | 8.18648123E+00 | 6.20E-04 | 1.41492939E+05 | 1.41492991E+05 | -5.20E-02 | 1.41523339E+05 | 1.41523391E+05 | -5.20E-02 | -6.37580630E+01 | -6.37062015E+01 | -5.20E-02 |
| 1658 | Pr | 153 | 59 | 94 | 35 | 8.17203600E+00 | 8.17173007E+00 | 3.10E-04 | 1.42426623E+05 | 1.42426627E+05 | -4.00E-03 | 1.42457023E+05 | 1.42457027E+05 | -4.10E-03 | -6.15684570E+01 | -6.15644365E+01 | -4.00E-03 |
| 1659 | Pr | 154 | 59 | 95 | 36 | 8.14947300E+00 | 8.14617544E+00 | 3.30E-03 | 1.43361491E+05 | 1.43361956E+05 | -4.60E-01 | 1.43391891E+05 | 1.43392356E+05 | -4.60E-01 | -5.81943980E+01 | -5.77294355E+01 | -4.60E-01 |
| 1660 | Pr | 155 | 59 | 96 | 37 | 8.13103800E+00 | 8.12918686E+00 | 1.90E-03 | 1.44295764E+05 | 1.44296008E+05 | -2.40E-01 | 1.44326164E+05 | 1.44326408E+05 | -2.40E-01 | -5.54152470E+01 | -5.51710618E+01 | -2.40E-01 |
| 1661 | Nd | 125 | 60 | 65 | 5 | 8.07700000E+00 | 8.07485587E+00 | 2.10E-03 | 1.16358237E+05 | 1.16358409E+05 | -1.70E-01 | 1.16389158E+05 | 1.16389331E+05 | -1.70E-01 | -4.75990000E+01 | -4.74267288E+01 | -1.70E-01 |
| 1662 | Nd | 126 | 60 | 66 | 6 | 8.11900000E+00 | 8.11684783E+00 | 2.20E-03 | 1.17284337E+05 | 1.17284609E+05 | -2.70E-01 | 1.17315259E+05 | 1.17315530E+05 | -2.70E-01 | -5.29930000E+01 | -5.27212530E+01 | -2.70E-01 |
| 1663 | Nd | 127 | 60 | 67 | 7 | 8.13900000E+00 | 8.13386619E+00 | 5.10E-03 | 1.18213288E+05 | 1.18213896E+05 | -6.10E-01 | 1.18244210E+05 | 1.18244818E+05 | -6.10E-01 | -5.55360000E+01 | -5.49281126E+01 | -6.10E-01 |
| 1664 | Nd | 128 | 60 | 68 | 8 | 8.17600000E+00 | 8.16976352E+00 | 6.20E-03 | 1.19140004E+05 | 1.19140733E+05 | -7.30E-01 | 1.19170926E+05 | 1.19171654E+05 | -7.30E-01 | -6.03140000E+01 | -5.95855188E+01 | -7.30E-01 |
| 1665 | Nd | 129 | 60 | 69 | 9 | 8.19000000E+00 | 8.18216636E+00 | 7.80E-03 | 1.20069497E+05 | 1.20070528E+05 | -1.00E+00 | 1.20100419E+05 | 1.20101450E+05 | -1.00E+00 | -6.23150000E+01 | -6.12839292E+01 | -1.00E+00 |
| 1666 | Nd | 130 | 60 | 70 | 10 | 8.22251300E+00 | 8.21295543E+00 | 9.60E-03 | 1.20996710E+05 | 1.20997909E+05 | -1.20E+00 | 1.21027632E+05 | 1.21028831E+05 | -1.20E+00 | -6.65962320E+01 | -6.53973564E+01 | -1.20E+00 |
| 1667 | Nd | 131 | 60 | 71 | 11 | 8.23030400E+00 | 8.22172242E+00 | 8.60E-03 | 1.21927032E+05 | 1.21928113E+05 | -1.10E+00 | 1.21957954E+05 | 1.21959035E+05 | -1.10E+00 | -6.77680330E+01 | -6.66874685E+01 | -1.10E+00 |
| 1668 | Nd | 132 | 60 | 72 | 12 | 8.25681000E+00 | 8.24843155E+00 | 8.40E-03 | 1.22854869E+05 | 1.22855931E+05 | -1.10E+00 | 1.22885790E+05 | 1.22886853E+05 | -1.10E+00 | -7.14258070E+01 | -7.03634767E+01 | -1.10E+00 |
| 1669 | Nd | 133 | 60 | 73 | 13 | 8.26223100E+00 | 8.25449739E+00 | 7.70E-03 | 1.23785456E+05 | 1.23786441E+05 | -9.90E-01 | 1.23816378E+05 | 1.23817363E+05 | -9.90E-01 | -7.23323720E+01 | -7.13473457E+01 | -9.90E-01 |
| 1670 | Nd | 134 | 60 | 74 | 14 | 8.28553800E+00 | 8.27807729E+00 | 7.50E-03 | 1.24713636E+05 | 1.24714592E+05 | -9.60E-01 | 1.24744558E+05 | 1.24745514E+05 | -9.60E-01 | -7.56464320E+01 | -7.46902304E+01 | -9.60E-01 |
| 1671 | Nd | 135 | 60 | 75 | 15 | 8.28815300E+00 | 8.28223880E+00 | 5.90E-03 | 1.25644563E+05 | 1.25645318E+05 | -7.50E-01 | 1.25675485E+05 | 1.25676239E+05 | -7.50E-01 | -7.62136090E+01 | -7.54587935E+01 | -7.50E-01 |
| 1672 | Nd | 136 | 60 | 76 | 16 | 8.30851200E+00 | 8.30335192E+00 | 5.20E-03 | 1.26573071E+05 | 1.26573730E+05 | -6.60E-01 | 1.26603993E+05 | 1.26604651E+05 | -6.60E-01 | -7.91992860E+01 | -7.85410968E+01 | -6.60E-01 |
| 1673 | Nd | 137 | 60 | 77 | 17 | 8.30959300E+00 | 8.30589838E+00 | 3.70E-03 | 1.27504180E+05 | 1.27504643E+05 | -4.60E-01 | 1.27535102E+05 | 1.27535564E+05 | -4.60E-01 | -7.95845660E+01 | -7.91219945E+01 | -4.60E-01 |
| 1674 | Nd | 138 | 60 | 78 | 18 | 8.32550200E+00 | 8.32438639E+00 | 1.10E-03 | 1.28433240E+05 | 1.28433351E+05 | -1.10E-01 | 1.28464162E+05 | 1.28464272E+05 | -1.10E-01 | -8.20183090E+01 | -8.19079196E+01 | -1.10E-01 |
| 1675 | Nd | 139 | 60 | 79 | 19 | 8.32364700E+00 | 8.32446191E+00 | -8.10E-04 | 1.29364738E+05 | 1.29364581E+05 | 1.60E-01 | 1.29395660E+05 | 1.29395503E+05 | 1.60E-01 | -8.20146480E+01 | -8.21714845E+01 | 1.60E-01 |
| 1676 | Nd | 140 | 60 | 80 | 20 | 8.33783800E+00 | 8.33879328E+00 | -9.60E-04 | 1.30293993E+05 | 1.30293816E+05 | 1.80E-01 | 1.30324915E+05 | 1.30324738E+05 | 1.80E-01 | -8.42537720E+01 | -8.44310186E+01 | 1.80E-01 |
| 1677 | Nd | 141 | 60 | 81 | 21 | 8.33552000E+00 | 8.33425994E+00 | 1.30E-03 | 1.31225548E+05 | 1.31225682E+05 | -1.30E-01 | 1.31256469E+05 | 1.31256603E+05 | -1.30E-01 | -8.41934080E+01 | -8.40592930E+01 | -1.30E-01 |
| 1678 | Nd | 142 | 60 | 82 | 22 | 8.34602900E+00 | 8.34220267E+00 | 3.80E-03 | 1.32155285E+05 | 1.32155785E+05 | -5.00E-01 | 1.32186207E+05 | 1.32186707E+05 | -5.00E-01 | -8.59498860E+01 | -8.54501009E+01 | -5.00E-01 |
| 1679 | Nd | 143 | 60 | 83 | 23 | 8.33048700E+00 | 8.33131477E+00 | -8.30E-04 | 1.33088727E+05 | 1.33088565E+05 | 1.60E-01 | 1.33119649E+05 | 1.33119487E+05 | 1.60E-01 | -8.40021390E+01 | -8.41640148E+01 | 1.60E-01 |
| 1680 | Nd | 144 | 60 | 84 | 24 | 8.32692200E+00 | 8.33229749E+00 | -5.40E-03 | 1.34020475E+05 | 1.34019658E+05 | 8.20E-01 | 1.34051397E+05 | 1.34050579E+05 | 8.20E-01 | -8.37478550E+01 | -8.45655219E+01 | 8.20E-01 |
| 1681 | Nd | 145 | 60 | 85 | 25 | 8.30918600E+00 | 8.31585236E+00 | -6.70E-03 | 1.34954285E+05 | 1.34953275E+05 | 1.00E+00 | 1.34985207E+05 | 1.34984197E+05 | 1.00E+00 | -8.14318370E+01 | -8.24419573E+01 | 1.00E+00 |
| 1682 | Nd | 146 | 60 | 86 | 26 | 8.30409100E+00 | 8.31186101E+00 | -7.80E-03 | 1.35886285E+05 | 1.35885108E+05 | 1.20E+00 | 1.35917207E+05 | 1.35916029E+05 | 1.20E+00 | -8.09257510E+01 | -8.21037528E+01 | 1.20E+00 |
| 1683 | Nd | 147 | 60 | 87 | 27 | 8.28360200E+00 | 8.29235278E+00 | -8.80E-03 | 1.36820559E+05 | 1.36819229E+05 | 1.30E+00 | 1.36851480E+05 | 1.36850150E+05 | 1.30E+00 | -7.81466310E+01 | -7.94765863E+01 | 1.30E+00 |
| 1684 | Nd | 148 | 60 | 88 | 28 | 8.27717500E+00 | 8.28578868E+00 | -8.60E-03 | 1.37752792E+05 | 1.37751473E+05 | 1.30E+00 | 1.37783713E+05 | 1.37782395E+05 | 1.30E+00 | -7.74078090E+01 | -7.87261326E+01 | 1.30E+00 |
| 1685 | Nd | 149 | 60 | 89 | 29 | 8.25544100E+00 | 8.26501577E+00 | -9.60E-03 | 1.38687318E+05 | 1.38685848E+05 | 1.50E+00 | 1.38718240E+05 | 1.38716770E+05 | 1.50E+00 | -7.43752790E+01 | -7.58454391E+01 | 1.50E+00 |
| 1686 | Nd | 150 | 60 | 90 | 30 | 8.24957200E+00 | 8.25698656E+00 | -7.40E-03 | 1.39619508E+05 | 1.39618353E+05 | 1.20E+00 | 1.39650430E+05 | 1.39649274E+05 | 1.20E+00 | -7.36790800E+01 | -7.48347538E+01 | 1.20E+00 |
| 1687 | Nd | 151 | 60 | 91 | 31 | 8.23026800E+00 | 8.23537887E+00 | -5.10E-03 | 1.40553739E+05 | 1.40552924E+05 | 8.20E-01 | 1.40584661E+05 | 1.40583846E+05 | 8.20E-01 | -7.09423140E+01 | -7.17576600E+01 | 8.20E-01 |
| 1688 | Nd | 152 | 60 | 92 | 32 | 8.22400100E+00 | 8.22587802E+00 | -1.90E-03 | 1.41486027E+05 | 1.41485698E+05 | 3.30E-01 | 1.41516948E+05 | 1.41516620E+05 | 3.30E-01 | -7.01487490E+01 | -7.04775915E+01 | 3.30E-01 |
| 1689 | Nd | 153 | 60 | 93 | 33 | 8.20458200E+00 | 8.20312211E+00 | 1.50E-03 | 1.42420340E+05 | 1.42420519E+05 | -1.80E-01 | 1.42451261E+05 | 1.42451441E+05 | -1.80E-01 | -6.73302700E+01 | -6.71504965E+01 | -1.80E-01 |
| 1690 | Nd | 154 | 60 | 94 | 34 | 8.19302900E+00 | 8.19174566E+00 | 1.30E-03 | 1.43353479E+05 | 1.43353633E+05 | -1.50E-01 | 1.43384401E+05 | 1.43384555E+05 | -1.50E-01 | -6.56843980E+01 | -6.55303255E+01 | -1.50E-01 |
| 1691 | Nd | 155 | 60 | 95 | 35 | 8.17030400E+00 | 8.16739132E+00 | 2.90E-03 | 1.44288374E+05 | 1.44288782E+05 | -4.10E-01 | 1.44319296E+05 | 1.44319704E+05 | -4.10E-01 | -6.22837020E+01 | -6.18758295E+01 | -4.10E-01 |
| 1692 | Nd | 156 | 60 | 96 | 36 | 8.15806600E+00 | 8.15376740E+00 | 4.30E-03 | 1.45221678E+05 | 1.45222305E+05 | -6.30E-01 | 1.45252600E+05 | 1.45253227E+05 | -6.30E-01 | -6.04736200E+01 | -5.98465707E+01 | -6.30E-01 |
| 1693 | Nd | 157 | 60 | 97 | 37 | 8.13195900E+00 | 8.12748996E+00 | 4.50E-03 | 1.46157185E+05 | 1.46157842E+05 | -6.60E-01 | 1.46188106E+05 | 1.46188764E+05 | -6.60E-01 | -5.64615200E+01 | -5.58034608E+01 | -6.60E-01 |
| 1694 | Nd | 158 | 60 | 98 | 38 | 8.11600000E+00 | 8.11141390E+00 | 4.60E-03 | 1.47091085E+05 | 1.47091820E+05 | -7.40E-01 | 1.47122007E+05 | 1.47122742E+05 | -7.40E-01 | -5.40550000E+01 | -5.33196141E+01 | -7.40E-01 |
| 1695 | Nd | 159 | 60 | 99 | 39 | 8.08900000E+00 | 8.08306507E+00 | 5.90E-03 | 1.48026827E+05 | 1.48027782E+05 | -9.50E-01 | 1.48057749E+05 | 1.48058703E+05 | -9.50E-01 | -4.98070000E+01 | -4.88522449E+01 | -9.50E-01 |
| 1696 | Nd | 160 | 60 | 100 | 40 | 8.07300000E+00 | 8.06450135E+00 | 8.50E-03 | 1.48960995E+05 | 1.48962234E+05 | -1.20E+00 | 1.48991916E+05 | 1.48993156E+05 | -1.20E+00 | -4.71340000E+01 | -4.58937958E+01 | -1.20E+00 |
| 1697 | Nd | 161 | 60 | 101 | 41 | 8.04400000E+00 | 8.03408752E+00 | 9.90E-03 | 1.49897034E+05 | 1.49898632E+05 | -1.60E+00 | 1.49927956E+05 | 1.49929553E+05 | -1.60E+00 | -4.25880000E+01 | -4.09903517E+01 | -1.60E+00 |
| 1698 | Pm | 128 | 61 | 67 | 6 | 8.07200000E+00 | 8.06878161E+00 | 3.20E-03 | 1.19152011E+05 | 1.19152353E+05 | -3.40E-01 | 1.19183454E+05 | 1.19183797E+05 | -3.40E-01 | -4.77860000E+01 | -4.74429109E+01 | -3.40E-01 |
| 1699 | PM | 129 | 61 | 68 | 7 | 8.11100000E+00 | 8.10655237E+00 | 4.40E-03 | 1.20078409E+05 | 1.20078978E+05 | -5.70E-01 | 1.20109853E+05 | 1.20110421E+05 | -5.70E-01 | -5.28810000E+01 | -5.23128013E+01 | -5.70E-01 |
| 1700 | Pm | 130 | 61 | 69 | 8 | 8.13000000E+00 | 8.12329298E+00 | 6.70E-03 | 1.21007389E+05 | 1.21008260E+05 | -8.70E-01 | 1.21038832E+05 | 1.21039704E+05 | -8.70E-01 | -5.53960000E+01 | -5.45243150E+01 | -8.70E-01 |
| 1701 | Pm | 131 | 61 | 70 | 9 | 8.16400000E+00 | 8.15576050E+00 | 8.20E-03 | 1.21934356E+05 | 1.21935449E+05 | -1.10E+00 | 1.21965799E+05 | 1.21966892E+05 | -1.10E+00 | -5.99230000E+01 | -5.88295337E+01 | -1.10E+00 |
| 1702 | Pm | 132 | 61 | 71 | 10 | 8.17700000E+00 | 8.16857335E+00 | 8.40E-03 | 1.22864145E+05 | 1.22865167E+05 | -1.00E+00 | 1.22895588E+05 | 1.22896611E+05 | -1.00E+00 | -6.16280000E+01 | -6.06052715E+01 | -1.00E+00 |
| 1703 | Pm | 133 | 61 | 72 | 11 | 8.20428300E+00 | 8.19680693E+00 | 7.50E-03 | 1.23791859E+05 | 1.23792809E+05 | -9.50E-01 | 1.23823303E+05 | 1.23824253E+05 | -9.50E-01 | -6.54076460E+01 | -6.44575911E+01 | -9.50E-01 |
| 1704 | Pm | 134 | 61 | 73 | 12 | 8.21322500E+00 | 8.20667773E+00 | 6.50E-03 | 1.24722022E+05 | 1.24722855E+05 | -8.30E-01 | 1.24753465E+05 | 1.24754298E+05 | -8.30E-01 | -6.67387510E+01 | -6.59057666E+01 | -8.30E-01 |
| 1705 | Pm | 135 | 61 | 74 | 13 | 8.23653000E+00 | 8.23164596E+00 | 4.90E-03 | 1.25650228E+05 | 1.25650843E+05 | -6.20E-01 | 1.25681671E+05 | 1.25682286E+05 | -6.20E-01 | -7.00269250E+01 | -6.94118360E+01 | -6.20E-01 |
| 1706 | Pm | 136 | 61 | 75 | 14 | 8.24379800E+00 | 8.23939340E+00 | 4.40E-03 | 1.26580568E+05 | 1.26581123E+05 | -5.50E-01 | 1.26612012E+05 | 1.26612566E+05 | -5.50E-01 | -7.11805000E+01 | -7.06258152E+01 | -5.50E-01 |
| 1707 | Pm | 137 | 61 | 76 | 15 | 8.26365100E+00 | 8.26175530E+00 | 1.90E-03 | 1.27509170E+05 | 1.27509385E+05 | -2.20E-01 | 1.27540614E+05 | 1.27540829E+05 | -2.20E-01 | -7.40728470E+01 | -7.38574698E+01 | -2.20E-01 |
| 1708 | Pm | 138 | 61 | 77 | 16 | 8.26854400E+00 | 8.26767931E+00 | 8.60E-04 | 1.28439796E+05 | 1.28439872E+05 | -7.50E-02 | 1.28471240E+05 | 1.28471315E+05 | -7.50E-02 | -7.49404830E+01 | -7.48654204E+01 | -7.50E-02 |
| 1709 | Pm | 139 | 61 | 78 | 17 | 8.28554400E+00 | 8.28728972E+00 | -1.70E-03 | 1.29368731E+05 | 1.29368443E+05 | 2.90E-01 | 1.29400174E+05 | 1.29399887E+05 | 2.90E-01 | -7.75006080E+01 | -7.77876272E+01 | 2.90E-01 |
| 1710 | Pm | 140 | 61 | 79 | 18 | 8.28907000E+00 | 8.29059586E+00 | -1.50E-03 | 1.30299516E+05 | 1.30299259E+05 | 2.60E-01 | 1.30330960E+05 | 1.30330702E+05 | 2.60E-01 | -7.82085720E+01 | -7.84664573E+01 | 2.60E-01 |
| 1711 | Pm | 141 | 61 | 80 | 19 | 8.30394000E+00 | 8.30602750E+00 | -2.10E-03 | 1.31228696E+05 | 1.31228358E+05 | 3.40E-01 | 1.31260140E+05 | 1.31259801E+05 | 3.40E-01 | -8.05229200E+01 | -8.08615950E+01 | 3.40E-01 |
| 1712 | Pm | 142 | 61 | 81 | 20 | 8.30666200E+00 | 8.30474850E+00 | 1.90E-03 | 1.32159571E+05 | 1.32159799E+05 | -2.30E-01 | 1.32191015E+05 | 1.32191242E+05 | -2.30E-01 | -8.11420590E+01 | -8.09146858E+01 | -2.30E-01 |
| 1713 | Pm | 143 | 61 | 82 | 21 | 8.31773200E+00 | 8.31395786E+00 | 3.80E-03 | 1.33089247E+05 | 1.33089742E+05 | -5.00E-01 | 1.33120690E+05 | 1.33121186E+05 | -5.00E-01 | -8.29604750E+01 | -8.24650538E+01 | -5.00E-01 |
| 1714 | Pm | 144 | 61 | 83 | 22 | 8.30529500E+00 | 8.30651698E+00 | -1.20E-03 | 1.34022285E+05 | 1.34022065E+05 | 2.20E-01 | 1.34053729E+05 | 1.34053509E+05 | 2.20E-01 | -8.14159320E+01 | -8.16362067E+01 | 2.20E-01 |
| 1715 | Pm | 145 | 61 | 84 | 23 | 8.30265600E+00 | 8.30900288E+00 | -6.30E-03 | 1.34953928E+05 | 1.34952964E+05 | 9.60E-01 | 1.34985372E+05 | 1.34984407E+05 | 9.60E-01 | -8.12673220E+01 | -8.22318595E+01 | 9.60E-01 |
| 1716 | Pm | 146 | 61 | 85 | 24 | 8.28865300E+00 | 8.29616007E+00 | -7.50E-03 | 1.35887235E+05 | 1.35886095E+05 | 1.10E+00 | 1.35918679E+05 | 1.35917538E+05 | 1.10E+00 | -7.94542060E+01 | -8.05944935E+01 | 1.10E+00 |
| 1717 | Pm | 147 | 61 | 86 | 25 | 8.28437000E+00 | 8.29377185E+00 | -9.40E-03 | 1.36819142E+05 | 1.36817715E+05 | 1.40E+00 | 1.36850585E+05 | 1.36849159E+05 | 1.40E+00 | -7.90419340E+01 | -8.04682664E+01 | 1.40E+00 |
| 1718 | Pm | 148 | 61 | 87 | 26 | 8.26822200E+00 | 8.27786379E+00 | -9.60E-03 | 1.37752812E+05 | 1.37751341E+05 | 1.50E+00 | 1.37784256E+05 | 1.37782785E+05 | 1.50E+00 | -7.68650730E+01 | -7.83363262E+01 | 1.50E+00 |
| 1719 | Pm | 149 | 61 | 88 | 27 | 8.26152200E+00 | 8.27286610E+00 | -1.10E-02 | 1.38685108E+05 | 1.38683373E+05 | 1.70E+00 | 1.38716551E+05 | 1.38714817E+05 | 1.70E+00 | -7.60636790E+01 | -7.77982156E+01 | 1.70E+00 |
| 1720 | Pm | 150 | 61 | 89 | 28 | 8.24380600E+00 | 8.25562142E+00 | -1.20E-02 | 1.39619069E+05 | 1.39617253E+05 | 1.80E+00 | 1.39650513E+05 | 1.39648696E+05 | 1.80E+00 | -7.35964640E+01 | -7.54130600E+01 | 1.80E+00 |
| 1721 | Pm | 151 | 61 | 90 | 29 | 8.24126600E+00 | 8.24910092E+00 | -7.80E-03 | 1.40550774E+05 | 1.40549547E+05 | 1.20E+00 | 1.40582218E+05 | 1.40580990E+05 | 1.20E+00 | -7.33853890E+01 | -7.46127668E+01 | 1.20E+00 |
| 1722 | Pm | 152 | 61 | 91 | 30 | 8.22612200E+00 | 8.23097522E+00 | -4.90E-03 | 1.41484400E+05 | 1.41483618E+05 | 7.80E-01 | 1.41515844E+05 | 1.41515062E+05 | 7.80E-01 | -7.12535330E+01 | -7.20354427E+01 | 7.80E-01 |
| 1723 | Pm | 153 | 61 | 92 | 31 | 8.22115000E+00 | 8.22295982E+00 | -1.80E-03 | 1.42416500E+05 | 1.42416179E+05 | 3.20E-01 | 1.42447944E+05 | 1.42447623E+05 | 3.20E-01 | -7.06475920E+01 | -7.09687419E+01 | 3.20E-01 |
| 1724 | Pm | 154 | 61 | 93 | 32 | 8.20617600E+00 | 8.20368281E+00 | 2.50E-03 | 1.43350151E+05 | 1.43350490E+05 | -3.40E-01 | 1.43381594E+05 | 1.43381934E+05 | -3.40E-01 | -6.84913270E+01 | -6.81517235E+01 | -3.40E-01 |
| 1725 | Pm | 155 | 61 | 94 | 33 | 8.19529600E+00 | 8.19379429E+00 | 1.50E-03 | 1.44283196E+05 | 1.44283385E+05 | -1.90E-01 | 1.44314639E+05 | 1.44314828E+05 | -1.90E-01 | -6.69399080E+01 | -6.67513675E+01 | -1.90E-01 |
| 1726 | Pm | 156 | 61 | 95 | 34 | 8.17670500E+00 | 8.17293643E+00 | 3.80E-03 | 1.45217466E+05 | 1.45218010E+05 | -5.40E-01 | 1.45248910E+05 | 1.45249454E+05 | -5.40E-01 | -6.41636200E+01 | -6.36200158E+01 | -5.40E-01 |
| 1727 | Pm | 157 | 61 | 96 | 35 | 8.16414400E+00 | 8.16080161E+00 | 3.30E-03 | 1.46150827E+05 | 1.46151308E+05 | -4.80E-01 | 1.46182271E+05 | 1.46182751E+05 | -4.80E-01 | -6.22970210E+01 | -6.18164668E+01 | -4.80E-01 |
| 1728 | Pm | 158 | 61 | 97 | 36 | 8.14325400E+00 | 8.13803219E+00 | 5.20E-03 | 1.47085529E+05 | 1.47086310E+05 | -7.80E-01 | 1.47116972E+05 | 1.47117753E+05 | -7.80E-01 | -5.90891870E+01 | -5.83083815E+01 | -7.80E-01 |
| 1729 | Pm | 159 | 61 | 98 | 37 | 8.12685900E+00 | 8.12342135E+00 | 3.40E-03 | 1.48019558E+05 | 1.48020060E+05 | -5.00E-01 | 1.48051001E+05 | 1.48051504E+05 | -5.00E-01 | -5.65542590E+01 | -5.60519703E+01 | -5.00E-01 |
| 1730 | Pm | 160 | 61 | 99 | 38 | 8.10400000E+00 | 8.09856563E+00 | 5.40E-03 | 1.48954604E+05 | 1.48955479E+05 | -8.70E-01 | 1.48986048E+05 | 1.48986923E+05 | -8.70E-01 | -5.30020000E+01 | -5.21271577E+01 | -8.70E-01 |
| 1731 | Pm | 161 | 61 | 100 | 39 | 8.08700000E+00 | 8.08140103E+00 | 5.60E-03 | 1.49888865E+05 | 1.49889709E+05 | -8.40E-01 | 1.49920308E+05 | 1.49921153E+05 | -8.40E-01 | -5.02350000E+01 | -4.93909032E+01 | -8.40E-01 |
| 1732 | Pm | 162 | 61 | 101 | 40 | 8.06300000E+00 | 8.05442402E+00 | 8.60E-03 | 1.50824225E+05 | 1.50825564E+05 | -1.30E+00 | 1.50855668E+05 | 1.50857007E+05 | -1.30E+00 | -4.63700000E+01 | -4.50307102E+01 | -1.30E+00 |
| 1733 | Pm | 163 | 61 | 102 | 41 | 8.04400000E+00 | 8.03475457E+00 | 9.20E-03 | 1.51758839E+05 | 1.51760281E+05 | -1.40E+00 | 1.51790283E+05 | 1.51791724E+05 | -1.40E+00 | -4.32490000E+01 | -4.18076944E+01 | -1.40E+00 |
| 1734 | SM | 129 | 62 | 67 | 5 | 8.02200000E+00 | 8.01853304E+00 | 3.50E-03 | 1.20088628E+05 | 1.20089027E+05 | -4.00E-01 | 1.20120593E+05 | 1.20120992E+05 | -4.00E-01 | -4.21410000E+01 | -4.17413849E+01 | -4.00E-01 |
| 1735 | Sm | 130 | 62 | 68 | 6 | 8.06400000E+00 | 8.06084660E+00 | 3.20E-03 | 1.21014756E+05 | 1.21015073E+05 | -3.20E-01 | 1.21046722E+05 | 1.21047039E+05 | -3.20E-01 | -4.75060000E+01 | -4.71893618E+01 | -3.20E-01 |
| 1736 | Sm | 131 | 62 | 69 | 7 | 8.08400000E+00 | 8.07896684E+00 | 5.00E-03 | 1.21943624E+05 | 1.21944204E+05 | -5.80E-01 | 1.21975589E+05 | 1.21976169E+05 | -5.80E-01 | -5.01330000E+01 | -4.95526411E+01 | -5.80E-01 |
| 1737 | Sm | 132 | 62 | 70 | 8 | 8.12100000E+00 | 8.11559692E+00 | 5.40E-03 | 1.22870171E+05 | 1.22870855E+05 | -6.80E-01 | 1.22902137E+05 | 1.22902821E+05 | -6.80E-01 | -5.50790000E+01 | -5.43954597E+01 | -6.80E-01 |
| 1738 | Sm | 133 | 62 | 71 | 9 | 8.13700000E+00 | 8.12961654E+00 | 7.40E-03 | 1.23799514E+05 | 1.23800440E+05 | -9.30E-01 | 1.23831479E+05 | 1.23832406E+05 | -9.30E-01 | -5.72310000E+01 | -5.63043472E+01 | -9.30E-01 |
| 1739 | Sm | 134 | 62 | 72 | 10 | 8.16700000E+00 | 8.16169754E+00 | 5.30E-03 | 1.24726863E+05 | 1.24727577E+05 | -7.10E-01 | 1.24758828E+05 | 1.24759543E+05 | -7.10E-01 | -6.13760000E+01 | -6.06614982E+01 | -7.10E-01 |
| 1740 | Sm | 135 | 62 | 73 | 11 | 8.17762600E+00 | 8.17262880E+00 | 5.00E-03 | 1.25656875E+05 | 1.25657505E+05 | -6.30E-01 | 1.25688841E+05 | 1.25689471E+05 | -6.30E-01 | -6.28572150E+01 | -6.22275965E+01 | -6.30E-01 |
| 1741 | Sm | 136 | 62 | 74 | 12 | 8.20591600E+00 | 8.20117100E+00 | 4.70E-03 | 1.26584416E+05 | 1.26585016E+05 | -6.00E-01 | 1.26616381E+05 | 1.26616982E+05 | -6.00E-01 | -6.68108890E+01 | -6.62106455E+01 | -6.00E-01 |
| 1742 | Sm | 137 | 62 | 75 | 13 | 8.21380600E+00 | 8.20983912E+00 | 4.00E-03 | 1.27514694E+05 | 1.27515193E+05 | -5.00E-01 | 1.27546660E+05 | 1.27547158E+05 | -5.00E-01 | -6.80265250E+01 | -6.75280309E+01 | -5.00E-01 |
| 1743 | Sm | 138 | 62 | 76 | 14 | 8.23792800E+00 | 8.23552168E+00 | 2.40E-03 | 1.28442717E+05 | 1.28443004E+05 | -2.90E-01 | 1.28474683E+05 | 1.28474970E+05 | -2.90E-01 | -7.14977620E+01 | -7.12107445E+01 | -2.90E-01 |
| 1744 | Sm | 139 | 62 | 77 | 15 | 8.24307800E+00 | 8.24222782E+00 | 8.50E-04 | 1.29373329E+05 | 1.29373402E+05 | -7.30E-02 | 1.29405294E+05 | 1.29405367E+05 | -7.30E-02 | -7.23802180E+01 | -7.23070994E+01 | -7.30E-02 |
| 1745 | Sm | 140 | 62 | 78 | 16 | 8.26382000E+00 | 8.26493962E+00 | -1.10E-03 | 1.30301747E+05 | 1.30301545E+05 | 2.00E-01 | 1.30333713E+05 | 1.30333511E+05 | 2.00E-01 | -7.54559340E+01 | -7.56576602E+01 | 2.00E-01 |
| 1746 | Sm | 141 | 62 | 79 | 17 | 8.26584500E+00 | 8.26894074E+00 | -3.10E-03 | 1.31232763E+05 | 1.31232282E+05 | 4.80E-01 | 1.31264729E+05 | 1.31264247E+05 | 4.80E-01 | -7.59338660E+01 | -7.64154397E+01 | 4.80E-01 |
| 1747 | Sm | 142 | 62 | 80 | 18 | 8.28597300E+00 | 8.28737071E+00 | -1.40E-03 | 1.32161205E+05 | 1.32160961E+05 | 2.40E-01 | 1.32193170E+05 | 1.32192927E+05 | 2.40E-01 | -7.89866060E+01 | -7.92301169E+01 | 2.40E-01 |
| 1748 | Sm | 143 | 62 | 81 | 19 | 8.28817900E+00 | 8.28685275E+00 | 1.30E-03 | 1.33092168E+05 | 1.33092313E+05 | -1.40E-01 | 1.33124134E+05 | 1.33124279E+05 | -1.40E-01 | -7.95167030E+01 | -7.93721000E+01 | -1.40E-01 |
| 1749 | Sm | 144 | 62 | 82 | 20 | 8.30367800E+00 | 8.29914873E+00 | 4.50E-03 | 1.34021214E+05 | 1.34021821E+05 | -6.10E-01 | 1.34053179E+05 | 1.34053787E+05 | -6.10E-01 | -8.19654490E+01 | -8.13582546E+01 | -6.10E-01 |
| 1750 | Sm | 145 | 62 | 83 | 21 | 8.29301200E+00 | 8.29268783E+00 | 3.20E-04 | 1.34954022E+05 | 1.34954024E+05 | -2.00E-03 | 1.34985988E+05 | 1.34985990E+05 | -2.00E-03 | -8.06512290E+01 | -8.06492548E+01 | -2.00E-03 |
| 1751 | Sm | 146 | 62 | 84 | 22 | 8.29385600E+00 | 8.29843148E+00 | -4.60E-03 | 1.35885171E+05 | 1.35884458E+05 | 7.10E-01 | 1.35917137E+05 | 1.35916424E+05 | 7.10E-01 | -8.09962060E+01 | -8.17091965E+01 | 7.10E-01 |
| 1752 | Sm | 147 | 62 | 85 | 23 | 8.28057200E+00 | 8.28675268E+00 | -6.20E-03 | 1.36818395E+05 | 1.36817442E+05 | 9.50E-01 | 1.36850361E+05 | 1.36849407E+05 | 9.50E-01 | -7.92660080E+01 | -8.02195253E+01 | 9.50E-01 |
| 1753 | Sm | 148 | 62 | 86 | 24 | 8.27963200E+00 | 8.28768342E+00 | -8.10E-03 | 1.37749819E+05 | 1.37748583E+05 | 1.20E+00 | 1.37781785E+05 | 1.37780548E+05 | 1.20E+00 | -7.93360650E+01 | -8.05727084E+01 | 1.20E+00 |
| 1754 | Sm | 149 | 62 | 87 | 25 | 8.26346200E+00 | 8.27297280E+00 | -9.50E-03 | 1.38683514E+05 | 1.38682052E+05 | 1.50E+00 | 1.38715480E+05 | 1.38714018E+05 | 1.50E+00 | -7.71350970E+01 | -7.85971908E+01 | 1.50E+00 |
| 1755 | Sm | 150 | 62 | 88 | 26 | 8.26161700E+00 | 8.27123725E+00 | -9.60E-03 | 1.39615093E+05 | 1.39613605E+05 | 1.50E+00 | 1.39647059E+05 | 1.39645571E+05 | 1.50E+00 | -7.70504640E+01 | -7.85385114E+01 | 1.50E+00 |
| 1756 | Sm | 151 | 62 | 89 | 27 | 8.24396700E+00 | 8.25514866E+00 | -1.10E-02 | 1.40549062E+05 | 1.40547329E+05 | 1.70E+00 | 1.40581028E+05 | 1.40579294E+05 | 1.70E+00 | -7.45755990E+01 | -7.63090533E+01 | 1.70E+00 |
| 1757 | Sm | 152 | 62 | 90 | 28 | 8.24405700E+00 | 8.25182066E+00 | -7.80E-03 | 1.41480370E+05 | 1.41479145E+05 | 1.20E+00 | 1.41512335E+05 | 1.41511110E+05 | 1.20E+00 | -7.47619680E+01 | -7.59870268E+01 | 1.20E+00 |
| 1758 | Sm | 153 | 62 | 91 | 29 | 8.22853000E+00 | 8.23482626E+00 | -6.30E-03 | 1.42414067E+05 | 1.42413058E+05 | 1.00E+00 | 1.42446032E+05 | 1.42445024E+05 | 1.00E+00 | -7.25590510E+01 | -7.35673851E+01 | 1.00E+00 |
| 1759 | Sm | 154 | 62 | 92 | 30 | 8.22683000E+00 | 8.22997660E+00 | -3.10E-03 | 1.43345665E+05 | 1.43345136E+05 | 5.30E-01 | 1.43377631E+05 | 1.43377101E+05 | 5.30E-01 | -7.24545270E+01 | -7.29840448E+01 | 5.30E-01 |
| 1760 | Sm | 155 | 62 | 93 | 31 | 8.21121800E+00 | 8.21184390E+00 | -6.30E-04 | 1.44279424E+05 | 1.44279282E+05 | 1.40E-01 | 1.44311389E+05 | 1.44311247E+05 | 1.40E-01 | -7.01901700E+01 | -7.03321331E+01 | 1.40E-01 |
| 1761 | Sm | 156 | 62 | 94 | 32 | 8.20501700E+00 | 8.20513272E+00 | -1.20E-04 | 1.45211745E+05 | 1.45211682E+05 | 6.30E-02 | 1.45243711E+05 | 1.45243648E+05 | 6.30E-02 | -6.93627060E+01 | -6.94257136E+01 | 6.30E-02 |
| 1762 | Sm | 157 | 62 | 95 | 33 | 8.18706300E+00 | 8.18545185E+00 | 1.60E-03 | 1.46145924E+05 | 1.46146132E+05 | -2.10E-01 | 1.46177890E+05 | 1.46178098E+05 | -2.10E-01 | -6.66775530E+01 | -6.64696320E+01 | -2.10E-01 |
| 1763 | Sm | 158 | 62 | 96 | 34 | 8.17729600E+00 | 8.17652008E+00 | 7.80E-04 | 1.47078846E+05 | 1.47078924E+05 | -7.80E-02 | 1.47110811E+05 | 1.47110889E+05 | -7.80E-02 | -6.52502200E+01 | -6.51725443E+01 | -7.80E-02 |
| 1764 | Sm | 159 | 62 | 97 | 35 | 8.15749500E+00 | 8.15495605E+00 | 2.50E-03 | 1.48013382E+05 | 1.48013741E+05 | -3.60E-01 | 1.48045348E+05 | 1.48045707E+05 | -3.60E-01 | -6.22077530E+01 | -6.18490648E+01 | -3.60E-01 |
| 1765 | Sm | 160 | 62 | 98 | 36 | 8.14462500E+00 | 8.14356574E+00 | 1.10E-03 | 1.48946850E+05 | 1.48946974E+05 | -1.20E-01 | 1.48978815E+05 | 1.48978940E+05 | -1.20E-01 | -6.02347700E+01 | -6.01102517E+01 | -1.20E-01 |
| 1766 | Sm | 161 | 62 | 99 | 37 | 8.12204000E+00 | 8.11992059E+00 | 2.10E-03 | 1.49881906E+05 | 1.49882203E+05 | -3.00E-01 | 1.49913872E+05 | 1.49914168E+05 | -3.00E-01 | -5.66719370E+01 | -5.63756305E+01 | -3.00E-01 |
| 1767 | Sm | 162 | 62 | 100 | 38 | 8.10900000E+00 | 8.10596912E+00 | 3.00E-03 | 1.50815543E+05 | 1.50815908E+05 | -3.70E-01 | 1.50847508E+05 | 1.50847874E+05 | -3.70E-01 | -5.45300000E+01 | -5.41640936E+01 | -3.70E-01 |
| 1768 | Sm | 163 | 62 | 101 | 39 | 8.08500000E+00 | 8.08017066E+00 | 4.80E-03 | 1.51750847E+05 | 1.51751573E+05 | -7.30E-01 | 1.51782812E+05 | 1.51783538E+05 | -7.30E-01 | -5.07200000E+01 | -4.99935938E+01 | -7.30E-01 |
| 1769 | Sm | 164 | 62 | 102 | 40 | 8.06900000E+00 | 8.06366906E+00 | 5.30E-03 | 1.52684958E+05 | 1.52685764E+05 | -8.10E-01 | 1.52716924E+05 | 1.52717730E+05 | -8.10E-01 | -4.81020000E+01 | -4.72961834E+01 | -8.10E-01 |
| 1770 | Sm | 165 | 62 | 103 | 41 | 8.04300000E+00 | 8.03574796E+00 | 7.30E-03 | 1.53620746E+05 | 1.53621873E+05 | -1.10E+00 | 1.53652712E+05 | 1.53653839E+05 | -1.10E+00 | -4.38080000E+01 | -4.26815528E+01 | -1.10E+00 |
| 1771 | EU | 130 | 63 | 67 | 4 | 7.95200000E+00 | 7.94721498E+00 | 4.80E-03 | 1.21027917E+05 | 1.21028540E+05 | -6.20E-01 | 1.21060404E+05 | 1.21061028E+05 | -6.20E-01 | -3.38240000E+01 | -3.32003285E+01 | -6.20E-01 |
| 1772 | Eu | 131 | 63 | 68 | 5 | 7.99500000E+00 | 7.99154031E+00 | 3.50E-03 | 1.21953964E+05 | 1.21954351E+05 | -3.90E-01 | 1.21986452E+05 | 1.21986839E+05 | -3.90E-01 | -3.92700000E+01 | -3.88828424E+01 | -3.90E-01 |
| 1773 | Eu | 132 | 63 | 69 | 6 | 8.01800000E+00 | 8.01407285E+00 | 3.90E-03 | 1.22882504E+05 | 1.22882951E+05 | -4.50E-01 | 1.22914991E+05 | 1.22915439E+05 | -4.50E-01 | -4.22250000E+01 | -4.17773598E+01 | -4.50E-01 |
| 1774 | Eu | 134 | 63 | 71 | 8 | 8.07600000E+00 | 8.07058184E+00 | 5.40E-03 | 1.24737788E+05 | 1.24738481E+05 | -6.90E-01 | 1.24770276E+05 | 1.24770969E+05 | -6.90E-01 | -4.99280000E+01 | -4.92350715E+01 | -6.90E-01 |
| 1775 | Eu | 135 | 63 | 72 | 9 | 8.10700000E+00 | 8.10427253E+00 | 2.70E-03 | 1.25665063E+05 | 1.25665428E+05 | -3.70E-01 | 1.25697550E+05 | 1.25697916E+05 | -3.70E-01 | -5.41480000E+01 | -5.37825770E+01 | -3.70E-01 |
| 1776 | Eu | 136 | 63 | 73 | 10 | 8.12200000E+00 | 8.11897112E+00 | 3.00E-03 | 1.26594461E+05 | 1.26594890E+05 | -4.30E-01 | 1.26626949E+05 | 1.26627378E+05 | -4.30E-01 | -5.62440000E+01 | -5.58145392E+01 | -4.30E-01 |
| 1777 | Eu | 137 | 63 | 74 | 11 | 8.15000000E+00 | 8.14895534E+00 | 1.00E-03 | 1.27522080E+05 | 1.27522228E+05 | -1.50E-01 | 1.27554568E+05 | 1.27554716E+05 | -1.50E-01 | -6.01190000E+01 | -5.99700297E+01 | -1.50E-01 |
| 1778 | Eu | 138 | 63 | 75 | 12 | 8.16162000E+00 | 8.16112351E+00 | 5.00E-04 | 1.28451943E+05 | 1.28451966E+05 | -2.30E-02 | 1.28484431E+05 | 1.28484454E+05 | -2.30E-02 | -6.17496690E+01 | -6.17268730E+01 | -2.30E-02 |
| 1779 | Eu | 139 | 63 | 76 | 13 | 8.18721800E+00 | 8.18808442E+00 | -8.70E-04 | 1.29379789E+05 | 1.29379622E+05 | 1.70E-01 | 1.29412276E+05 | 1.29412110E+05 | 1.70E-01 | -6.53980410E+01 | -6.55642442E+01 | 1.70E-01 |

| | | | | | | | | | | | | | | | | |
|---|---|---|---|---|---|---|---|---|---|---|---|---|---|---|---|---|
| 1780 | Eu | 140 | 63 | 77 | 14 | 8.19773200E+00 | 8.19804633E+00 | -3.10E-04 | 1.30309695E+05 | 1.30309605E+05 | 9.00E-02 | 1.30342183E+05 | 1.30342093E+05 | 9.00E-02 | -6.69859340E+01 | -6.70756771E+01 | 9.00E-02 |
| 1781 | Eu | 141 | 63 | 78 | 15 | 8.21768400E+00 | 8.22189443E+00 | -4.20E-03 | 1.31238249E+05 | 1.31237610E+05 | 6.40E-01 | 1.31270737E+05 | 1.31270098E+05 | 6.40E-01 | -6.99256200E+01 | -7.05649865E+01 | 6.40E-01 |
| 1782 | Eu | 142 | 63 | 79 | 16 | 8.22642800E+00 | 8.22897480E+00 | -2.50E-03 | 1.32168355E+05 | 1.32167948E+05 | 4.10E-01 | 1.32200843E+05 | 1.32200436E+05 | 4.10E-01 | -7.13136060E+01 | -7.17209752E+01 | 4.10E-01 |
| 1783 | Eu | 143 | 63 | 80 | 17 | 8.24581700E+00 | 8.24849879E+00 | -2.70E-03 | 1.33096922E+05 | 1.33096492E+05 | 4.30E-01 | 1.33129409E+05 | 1.33128980E+05 | 4.30E-01 | -7.42413000E+01 | -7.46705607E+01 | 4.30E-01 |
| 1784 | Eu | 144 | 63 | 81 | 18 | 8.25417300E+00 | 8.25104202E+00 | 3.10E-03 | 1.34027038E+05 | 1.34027443E+05 | -4.10E-01 | 1.34059526E+05 | 1.34059931E+05 | -4.10E-01 | -7.56190990E+01 | -7.52139659E+01 | -4.10E-01 |
| 1785 | Eu | 145 | 63 | 82 | 19 | 8.26927400E+00 | 8.26455903E+00 | 4.70E-03 | 1.34956159E+05 | 1.34956797E+05 | -6.40E-01 | 1.34988647E+05 | 1.34989285E+05 | -6.40E-01 | -7.79915430E+01 | -7.73536556E+01 | -6.40E-01 |
| 1786 | Eu | 146 | 63 | 83 | 20 | 8.26193100E+00 | 8.26129767E+00 | 6.30E-04 | 1.35888528E+05 | 1.35888574E+05 | -4.70E-02 | 1.35921015E+05 | 1.35921062E+05 | -4.70E-02 | -7.71174200E+01 | -7.70707364E+01 | -4.70E-02 |
| 1787 | Eu | 147 | 63 | 84 | 21 | 8.26353900E+00 | 8.26846137E+00 | -4.90E-03 | 1.36819595E+05 | 1.36818825E+05 | 7.70E-01 | 1.36852083E+05 | 1.36851313E+05 | 7.70E-01 | -7.75443990E+01 | -7.83137790E+01 | 7.70E-01 |
| 1788 | Eu | 148 | 63 | 85 | 22 | 8.25382800E+00 | 8.26010463E+00 | -6.30E-03 | 1.37752334E+05 | 1.37751359E+05 | 9.70E-01 | 1.37784822E+05 | 1.37783847E+05 | 9.70E-01 | -7.62993780E+01 | -7.72741245E+01 | 9.70E-01 |
| 1789 | Eu | 149 | 63 | 86 | 23 | 8.25355000E+00 | 8.26254934E+00 | -9.00E-03 | 1.38683687E+05 | 1.38682300E+05 | 1.40E+00 | 1.38716175E+05 | 1.38714788E+05 | 1.40E+00 | -7.64404810E+01 | -7.78271711E+01 | 1.40E+00 |
| 1790 | Eu | 150 | 63 | 87 | 24 | 8.24134400E+00 | 8.25115592E+00 | -9.80E-03 | 1.39616829E+05 | 1.39615312E+05 | 1.50E+00 | 1.39649317E+05 | 1.39647800E+05 | 1.50E+00 | -7.47918230E+01 | -7.63093890E+01 | 1.50E+00 |
| 1791 | Eu | 151 | 63 | 88 | 25 | 8.23929200E+00 | 8.25090982E+00 | -1.20E-02 | 1.40548463E+05 | 1.40546663E+05 | 1.80E+00 | 1.40580951E+05 | 1.40579151E+05 | 1.80E+00 | -7.46519750E+01 | -7.64520642E+01 | 1.80E+00 |
| 1792 | Eu | 152 | 63 | 89 | 26 | 8.22657700E+00 | 8.23806997E+00 | -1.10E-02 | 1.41481722E+05 | 1.41479929E+05 | 1.80E+00 | 1.41514210E+05 | 1.41512417E+05 | 1.80E+00 | -7.28873710E+01 | -7.46799985E+01 | 1.80E+00 |
| 1793 | Eu | 153 | 63 | 90 | 27 | 8.22869300E+00 | 8.23618404E+00 | -7.50E-03 | 1.42412737E+05 | 1.42411545E+05 | 1.20E+00 | 1.42445225E+05 | 1.42444033E+05 | 1.20E+00 | -7.33663270E+01 | -7.45582024E+01 | 1.20E+00 |
| 1794 | Eu | 154 | 63 | 91 | 28 | 8.21709200E+00 | 8.22239172E+00 | -5.30E-03 | 1.43345860E+05 | 1.43344998E+05 | 8.60E-01 | 1.43378348E+05 | 1.43377486E+05 | 8.60E-01 | -7.17371830E+01 | -7.25990501E+01 | 8.60E-01 |
| 1795 | Eu | 155 | 63 | 92 | 29 | 8.21666800E+00 | 8.21896952E+00 | -2.30E-03 | 1.44277274E+05 | 1.44276872E+05 | 4.00E-01 | 1.44309762E+05 | 1.44309360E+05 | 4.00E-01 | -7.18171560E+01 | -7.22196810E+01 | 4.00E-01 |
| 1796 | Eu | 156 | 63 | 93 | 30 | 8.20463300E+00 | 8.20403745E+00 | 6.00E-04 | 1.45210500E+05 | 1.45210548E+05 | -4.70E-02 | 1.45242988E+05 | 1.45243036E+05 | -4.70E-02 | -7.00851480E+01 | -7.00379285E+01 | -4.70E-02 |
| 1797 | Eu | 157 | 63 | 94 | 31 | 8.19979100E+00 | 8.19877156E+00 | 1.00E-03 | 1.46142621E+05 | 1.46142736E+05 | -1.10E-01 | 1.46175109E+05 | 1.46175224E+05 | -1.10E-01 | -6.94582970E+01 | -6.93439028E+01 | -1.10E-01 |
| 1798 | Eu | 158 | 63 | 95 | 32 | 8.18503400E+00 | 8.18232033E+00 | 2.70E-03 | 1.47076319E+05 | 1.47076702E+05 | -3.80E-01 | 1.47108807E+05 | 1.47109190E+05 | -3.80E-01 | -6.72551650E+01 | -6.68720611E+01 | -3.80E-01 |
| 1799 | Eu | 159 | 63 | 96 | 33 | 8.17669500E+00 | 8.17486349E+00 | 1.80E-03 | 1.48009025E+05 | 1.48009270E+05 | -2.50E-01 | 1.48041513E+05 | 1.48041758E+05 | -2.50E-01 | -6.60428610E+01 | -6.57974254E+01 | -2.50E-01 |
| 1800 | Eu | 160 | 63 | 97 | 34 | 8.16002100E+00 | 8.15656654E+00 | 3.50E-03 | 1.48943081E+05 | 1.48943588E+05 | -5.10E-01 | 1.48975569E+05 | 1.48976076E+05 | -5.10E-01 | -6.34804410E+01 | -6.29734569E+01 | -5.10E-01 |
| 1801 | Eu | 161 | 63 | 98 | 35 | 8.14898000E+00 | 8.14667251E+00 | 2.30E-03 | 1.49876264E+05 | 1.49876590E+05 | -3.30E-01 | 1.49908752E+05 | 1.49909078E+05 | -3.30E-01 | -6.17915010E+01 | -6.14657665E+01 | -3.30E-01 |
| 1802 | Eu | 162 | 63 | 99 | 36 | 8.12938400E+00 | 8.12632247E+00 | 3.10E-03 | 1.50810856E+05 | 1.50811306E+05 | -4.50E-01 | 1.50843343E+05 | 1.50843793E+05 | -4.50E-01 | -5.86945900E+01 | -5.82444129E+01 | -4.50E-01 |
| 1803 | EU | 163 | 63 | 100 | 37 | 8.11641500E+00 | 8.11386528E+00 | 2.50E-03 | 1.51744405E+05 | 1.51744775E+05 | -3.70E-01 | 1.51776893E+05 | 1.51777263E+05 | -3.70E-01 | -5.66388640E+01 | -5.62688943E+01 | -3.70E-01 |
| 1804 | Eu | 164 | 63 | 101 | 38 | 8.09600000E+00 | 8.09136710E+00 | 4.60E-03 | 1.52679204E+05 | 1.52679916E+05 | -7.10E-01 | 1.52711692E+05 | 1.52712404E+05 | -7.10E-01 | -5.33340000E+01 | -5.26217386E+01 | -7.10E-01 |
| 1805 | Eu | 165 | 63 | 102 | 39 | 8.08000000E+00 | 8.07632267E+00 | 3.70E-03 | 1.53613347E+05 | 1.53613873E+05 | -5.30E-01 | 1.53645835E+05 | 1.53646361E+05 | -5.30E-01 | -5.06860000E+01 | -5.01594571E+01 | -5.30E-01 |
| 1806 | Eu | 166 | 63 | 103 | 40 | 8.05700000E+00 | 8.05167299E+00 | 5.30E-03 | 1.54548598E+05 | 1.54549454E+05 | -8.60E-01 | 1.54581085E+05 | 1.54581942E+05 | -8.60E-01 | -4.69290000E+01 | -4.60726129E+01 | -8.60E-01 |
| 1807 | Eu | 167 | 63 | 104 | 41 | 8.03900000E+00 | 8.03408692E+00 | 4.90E-03 | 1.55483138E+05 | 1.55483904E+05 | -7.70E-01 | 1.55515625E+05 | 1.55516392E+05 | -7.70E-01 | -4.38830000E+01 | -4.31160930E+01 | -7.70E-01 |
| 1808 | Gd | 134 | 64 | 70 | 6 | 8.00600000E+00 | 8.00588956E+00 | 1.10E-04 | 1.24745891E+05 | 1.24745844E+05 | 4.70E-02 | 1.24778902E+05 | 1.24778855E+05 | 4.70E-02 | -4.13020000E+01 | -4.13493829E+01 | 4.70E-02 |
| 1809 | Gd | 135 | 64 | 71 | 7 | 8.02900000E+00 | 8.02530413E+00 | 3.70E-03 | 1.25674395E+05 | 1.25674783E+05 | -3.90E-01 | 1.25707406E+05 | 1.25707793E+05 | -3.90E-01 | -4.42930000E+01 | -4.39049204E+01 | -3.90E-01 |
| 1810 | Gd | 136 | 64 | 72 | 8 | 8.06400000E+00 | 8.06288412E+00 | 1.10E-03 | 1.26601092E+05 | 1.26601212E+05 | -1.20E-01 | 1.26634103E+05 | 1.26634223E+05 | -1.20E-01 | -4.90900000E+01 | -4.89697841E+01 | -1.20E-01 |
| 1811 | Gd | 137 | 64 | 73 | 9 | 8.08000000E+00 | 8.07873504E+00 | 1.30E-03 | 1.27530462E+05 | 1.27530543E+05 | -8.10E-02 | 1.27563473E+05 | 1.27563553E+05 | -8.10E-02 | -5.12140000E+01 | -5.11329250E+01 | -8.10E-02 |
| 1812 | Gd | 138 | 64 | 74 | 10 | 8.11200000E+00 | 8.11229396E+00 | -2.90E-04 | 1.28457513E+05 | 1.28457398E+05 | 1.10E-01 | 1.28490524E+05 | 1.28490409E+05 | 1.10E-01 | -5.56570000E+01 | -5.57714730E+01 | 1.10E-01 |
| 1813 | Gd | 139 | 64 | 75 | 11 | 8.12600000E+00 | 8.12544378E+00 | 5.60E-04 | 1.29387032E+05 | 1.29387024E+05 | 8.70E-03 | 1.29420043E+05 | 1.29420034E+05 | 8.70E-03 | -5.76320000E+01 | -5.76402731E+01 | 8.30E-03 |
| 1814 | Gd | 140 | 64 | 76 | 12 | 8.15497500E+00 | 8.15569164E+00 | -7.20E-04 | 1.30314376E+05 | 1.30314229E+05 | 1.50E-01 | 1.30347386E+05 | 1.30347239E+05 | 1.50E-01 | -6.17822710E+01 | -6.19290974E+01 | 1.50E-01 |
| 1815 | Gd | 141 | 64 | 77 | 13 | 8.16460800E+00 | 8.16647253E+00 | -1.90E-03 | 1.31244428E+05 | 1.31244119E+05 | 3.10E-01 | 1.31277438E+05 | 1.31277129E+05 | 3.10E-01 | -6.32242240E+01 | -6.35335761E+01 | 3.10E-01 |
| 1816 | Gd | 142 | 64 | 78 | 14 | 8.19025600E+00 | 8.19336208E+00 | -3.10E-03 | 1.32172187E+05 | 1.32171699E+05 | 4.90E-01 | 1.32205197E+05 | 1.32204710E+05 | 4.90E-01 | -6.69595150E+01 | -6.74470457E+01 | 4.90E-01 |
| 1817 | Gd | 143 | 64 | 79 | 15 | 8.19831800E+00 | 8.20115255E+00 | -2.80E-03 | 1.33102409E+05 | 1.33101957E+05 | 4.50E-01 | 1.33135419E+05 | 1.33134968E+05 | 4.50E-01 | -6.82313000E+01 | -6.86831258E+01 | 4.50E-01 |
| 1818 | Gd | 144 | 64 | 80 | 16 | 8.22193700E+00 | 8.22358483E+00 | -1.60E-03 | 1.34030375E+05 | 1.34030091E+05 | 2.80E-01 | 1.34063385E+05 | 1.34063102E+05 | 2.80E-01 | -7.17595030E+01 | -7.20432081E+01 | 2.80E-01 |
| 1819 | Gd | 145 | 64 | 81 | 17 | 8.22893000E+00 | 8.22686809E+00 | 2.10E-03 | 1.34960704E+05 | 1.34960957E+05 | -2.50E-01 | 1.34993715E+05 | 1.34993967E+05 | -2.50E-01 | -7.29240370E+01 | -7.26715463E+01 | -2.50E-01 |
| 1820 | Gd | 146 | 64 | 82 | 18 | 8.24950400E+00 | 8.24333349E+00 | 6.20E-03 | 1.35889037E+05 | 1.35889891E+05 | -8.50E-01 | 1.35922047E+05 | 1.35922902E+05 | -8.50E-01 | -7.60855320E+01 | -7.52310430E+01 | -8.50E-01 |
| 1821 | Gd | 147 | 64 | 83 | 19 | 8.24333300E+00 | 8.24098023E+00 | 2.40E-03 | 1.36821260E+05 | 1.36821559E+05 | -3.00E-01 | 1.36854270E+05 | 1.36854570E+05 | -3.00E-01 | -7.53565750E+01 | -7.50571288E+01 | -3.00E-01 |
| 1822 | Gd | 148 | 64 | 84 | 20 | 8.24833800E+00 | 8.25121605E+00 | -2.90E-03 | 1.37751841E+05 | 1.37751369E+05 | 4.70E-01 | 1.37784852E+05 | 1.37784379E+05 | 4.70E-01 | -7.62693210E+01 | -7.67416913E+01 | 4.70E-01 |
| 1823 | Gd | 149 | 64 | 85 | 21 | 8.23948200E+00 | 8.24391858E+00 | -4.40E-03 | 1.38684478E+05 | 1.38683770E+05 | 7.10E-01 | 1.38717488E+05 | 1.38716781E+05 | 7.10E-01 | -7.51267440E+01 | -7.58342649E+01 | 7.10E-01 |
| 1824 | Gd | 150 | 64 | 86 | 22 | 8.24260700E+00 | 8.24947912E+00 | -6.90E-03 | 1.39615335E+05 | 1.39614258E+05 | 1.10E+00 | 1.39648345E+05 | 1.39647268E+05 | 1.10E+00 | -7.57636020E+01 | -7.68409463E+01 | 1.10E+00 |
| 1825 | Gd | 151 | 64 | 87 | 23 | 8.23103800E+00 | 8.23917434E+00 | -8.10E-03 | 1.40548405E+05 | 1.40547130E+05 | 1.30E+00 | 1.40581415E+05 | 1.40580140E+05 | 1.30E+00 | -7.41879550E+01 | -7.54630837E+01 | 1.30E+00 |
| 1826 | Gd | 152 | 64 | 88 | 24 | 8.23339700E+00 | 8.24198910E+00 | -8.60E-03 | 1.41479380E+05 | 1.41478028E+05 | 1.40E+00 | 1.41512391E+05 | 1.41511038E+05 | 1.40E+00 | -7.47062900E+01 | -7.60587836E+01 | 1.40E+00 |

| | | | | | | | | | | | | | | | | |
|---|---|---|---|---|---|---|---|---|---|---|---|---|---|---|---|---|
| 1827 | Gd | 153 | 64 | 89 | 25 | 8.22041400E+00 | 8.23019973E+00 | -9.80E-03 | 1.42412699E+05 | 1.42411155E+05 | 1.50E+00 | 1.42445709E+05 | 1.42444166E+05 | 1.50E+00 | -7.28819280E+01 | -7.44256794E+01 | 1.50E+00 |
| 1828 | Gd | 154 | 64 | 90 | 26 | 8.22479200E+00 | 8.23130588E+00 | -6.50E-03 | 1.43343370E+05 | 1.43342320E+05 | 1.00E+00 | 1.43376380E+05 | 1.43375330E+05 | 1.00E+00 | -7.37053380E+01 | -7.47549082E+01 | 1.00E+00 |
| 1829 | Gd | 155 | 64 | 91 | 27 | 8.21324700E+00 | 8.21853834E+00 | -5.30E-03 | 1.44276500E+05 | 1.44275633E+05 | 8.70E-01 | 1.44309510E+05 | 1.44308644E+05 | 8.70E-01 | -7.20692500E+01 | -7.29359249E+01 | 8.70E-01 |
| 1830 | Gd | 156 | 64 | 92 | 28 | 8.21531800E+00 | 8.21807916E+00 | -2.80E-03 | 1.45207529E+05 | 1.45207051E+05 | 4.80E-01 | 1.45240539E+05 | 1.45240062E+05 | 4.80E-01 | -7.25342820E+01 | -7.30115125E+01 | 4.80E-01 |
| 1831 | Gd | 157 | 64 | 93 | 29 | 8.20350000E+00 | 8.20418420E+00 | -6.80E-04 | 1.46140734E+05 | 1.46140580E+05 | 1.50E-01 | 1.46173745E+05 | 1.46173591E+05 | 1.50E-01 | -7.08228220E+01 | -7.09767651E+01 | 1.50E-01 |
| 1832 | Gd | 158 | 64 | 94 | 30 | 8.20181500E+00 | 8.20189433E+00 | -7.90E-05 | 1.47072362E+05 | 1.47072303E+05 | 5.90E-02 | 1.47105373E+05 | 1.47105314E+05 | 5.90E-02 | -7.06888920E+01 | -7.07478295E+01 | 5.90E-02 |
| 1833 | Gd | 159 | 64 | 95 | 31 | 8.18761000E+00 | 8.18651964E+00 | 1.10E-03 | 1.48005984E+05 | 1.48006111E+05 | -1.30E-01 | 1.48038995E+05 | 1.48039122E+05 | -1.30E-01 | -6.85607820E+01 | -6.84338301E+01 | -1.30E-01 |
| 1834 | Gd | 160 | 64 | 96 | 32 | 8.18300900E+00 | 8.18207560E+00 | 9.30E-04 | 1.48938098E+05 | 1.48938201E+05 | -1.00E-01 | 1.48971109E+05 | 1.48971212E+05 | -1.00E-01 | -6.79409260E+01 | -6.78379843E+01 | -1.00E-01 |
| 1835 | Gd | 161 | 64 | 97 | 33 | 8.16718600E+00 | 8.16490243E+00 | 2.30E-03 | 1.49872028E+05 | 1.49872349E+05 | -3.20E-01 | 1.49905039E+05 | 1.49905360E+05 | -3.20E-01 | -6.55050090E+01 | -6.51838602E+01 | -3.20E-01 |
| 1836 | Gd | 162 | 64 | 98 | 34 | 8.15903000E+00 | 8.15806299E+00 | 9.70E-04 | 1.50804748E+05 | 1.50804858E+05 | -1.10E-01 | 1.50837758E+05 | 1.50837868E+05 | -1.10E-01 | -6.42795900E+01 | -6.41694545E+01 | -1.10E-01 |
| 1837 | Gd | 163 | 64 | 99 | 35 | 8.14029700E+00 | 8.13887571E+00 | 1.40E-03 | 1.51739208E+05 | 1.51739393E+05 | -1.90E-01 | 1.51772218E+05 | 1.51772403E+05 | -1.90E-01 | -6.13138640E+01 | -6.11286722E+01 | -1.90E-01 |
| 1838 | Gd | 164 | 64 | 100 | 36 | 8.13000000E+00 | 8.12950554E+00 | 4.90E-04 | 1.52672242E+05 | 1.52672356E+05 | -1.10E-01 | 1.52705252E+05 | 1.52705366E+05 | -1.10E-01 | -5.97740000E+01 | -5.96595199E+01 | -1.10E-01 |
| 1839 | Gd | 165 | 64 | 101 | 37 | 8.11000000E+00 | 8.10818887E+00 | 1.80E-03 | 1.53607024E+05 | 1.53607309E+05 | -2.90E-01 | 1.53640034E+05 | 1.53640320E+05 | -2.90E-01 | -5.64860000E+01 | -5.62004557E+01 | -2.90E-01 |
| 1840 | GD | 166 | 64 | 102 | 38 | 8.09800000E+00 | 8.09624330E+00 | 1.80E-03 | 1.54540474E+05 | 1.54540749E+05 | -2.80E-01 | 1.54573484E+05 | 1.54573760E+05 | -2.80E-01 | -5.45300000E+01 | -5.42543614E+01 | -2.80E-01 |
| 1841 | Gd | 167 | 64 | 103 | 39 | 8.07600000E+00 | 8.07276364E+00 | 3.20E-03 | 1.55475685E+05 | 1.55476139E+05 | -4.50E-01 | 1.55508695E+05 | 1.55509150E+05 | -4.50E-01 | -5.08130000E+01 | -5.03581825E+01 | -4.50E-01 |
| 1842 | Gd | 168 | 64 | 104 | 40 | 8.06100000E+00 | 8.05825632E+00 | 2.70E-03 | 1.56409629E+05 | 1.56410069E+05 | -4.40E-01 | 1.56442639E+05 | 1.56443080E+05 | -4.40E-01 | -4.83630000E+01 | -4.79223970E+01 | -4.40E-01 |
| 1843 | Gd | 169 | 64 | 105 | 41 | 8.03600000E+00 | 8.03265916E+00 | 3.30E-03 | 1.57345333E+05 | 1.57345902E+05 | -5.70E-01 | 1.57378343E+05 | 1.57378913E+05 | -5.70E-01 | -4.41530000E+01 | -4.35834149E+01 | -5.70E-01 |
| 1844 | TB | 135 | 65 | 70 | 5 | 7.93800000E+00 | 7.93676036E+00 | 1.20E-03 | 1.25685339E+05 | 1.25685430E+05 | -9.10E-02 | 1.25718872E+05 | 1.25718964E+05 | -9.10E-02 | -3.28250000E+01 | -3.27345885E+01 | -9.00E-02 |
| 1845 | Tb | 138 | 65 | 73 | 8 | 8.01900000E+00 | 8.01926519E+00 | -2.70E-04 | 1.28468978E+05 | 1.28468930E+05 | 4.80E-02 | 1.28502512E+05 | 1.28502464E+05 | 4.80E-02 | -4.36680000E+01 | -4.37165795E+01 | 4.90E-02 |
| 1846 | Tb | 139 | 65 | 74 | 9 | 8.05200000E+00 | 8.05432942E+00 | -2.30E-03 | 1.29396011E+05 | 1.29395603E+05 | 4.10E-01 | 1.29429544E+05 | 1.29429136E+05 | 4.10E-01 | -4.81300000E+01 | -4.85384538E+01 | 4.10E-01 |
| 1847 | Tb | 140 | 65 | 75 | 10 | 8.06867200E+00 | 8.07098246E+00 | -2.30E-03 | 1.30325153E+05 | 1.30324782E+05 | 3.70E-01 | 1.30358686E+05 | 1.30358316E+05 | 3.70E-01 | -5.04822710E+01 | -5.08528896E+01 | 3.70E-01 |
| 1848 | Tb | 141 | 65 | 76 | 11 | 8.09747500E+00 | 8.10254760E+00 | -5.10E-03 | 1.31252588E+05 | 1.31251826E+05 | 7.60E-01 | 1.31286122E+05 | 1.31285359E+05 | 7.60E-01 | -5.45408370E+01 | -5.53032375E+01 | 7.60E-01 |
| 1849 | Tb | 142 | 65 | 77 | 12 | 8.11150700E+00 | 8.11655482E+00 | -5.00E-03 | 1.32182064E+05 | 1.32181300E+05 | 7.60E-01 | 1.32215597E+05 | 1.32214833E+05 | 7.60E-01 | -5.65595150E+01 | -5.73234911E+01 | 7.60E-01 |
| 1850 | Tb | 143 | 65 | 78 | 13 | 8.13821700E+00 | 8.14459999E+00 | -6.40E-03 | 1.33109698E+05 | 1.33108738E+05 | 9.60E-01 | 1.33143231E+05 | 1.33142272E+05 | 9.60E-01 | -6.04191830E+01 | -6.13791864E+01 | 9.60E-01 |
| 1851 | Tb | 144 | 65 | 79 | 14 | 8.15128700E+00 | 8.15541040E+00 | -4.10E-03 | 1.34039243E+05 | 1.34038602E+05 | 6.40E-01 | 1.34072777E+05 | 1.34072136E+05 | 6.40E-01 | -6.23681810E+01 | -6.30091665E+01 | 6.40E-01 |
| 1852 | Tb | 145 | 65 | 80 | 15 | 8.17786500E+00 | 8.17893047E+00 | -1.10E-03 | 1.34966803E+05 | 1.34966602E+05 | 2.00E-01 | 1.35000337E+05 | 1.35000135E+05 | 2.00E-01 | -6.63019290E+01 | -6.65036676E+01 | 2.00E-01 |
| 1853 | Tb | 146 | 65 | 81 | 16 | 8.18714500E+00 | 8.18517288E+00 | 2.00E-03 | 1.35896836E+05 | 1.35897077E+05 | -2.40E-01 | 1.35930370E+05 | 1.35930610E+05 | -2.40E-01 | -6.77634090E+01 | -6.75226715E+01 | -2.40E-01 |
| 1854 | Tb | 147 | 65 | 82 | 17 | 8.20662300E+00 | 8.20280869E+00 | 3.80E-03 | 1.36825351E+05 | 1.36825865E+05 | -5.10E-01 | 1.36858884E+05 | 1.36859398E+05 | -5.10E-01 | -7.07425090E+01 | -7.02289889E+01 | -5.10E-01 |
| 1855 | Tb | 148 | 65 | 83 | 18 | 8.20428200E+00 | 8.20349639E+00 | 7.90E-04 | 1.37757056E+05 | 1.37757125E+05 | -6.90E-02 | 1.37790590E+05 | 1.37790659E+05 | -6.90E-02 | -7.05312800E+01 | -7.04622587E+01 | -6.90E-02 |
| 1856 | Tb | 149 | 65 | 84 | 19 | 8.20981600E+00 | 8.21505603E+00 | -5.20E-03 | 1.38687593E+05 | 1.38686765E+05 | 8.30E-01 | 1.38721126E+05 | 1.38720298E+05 | 8.30E-01 | -7.14888580E+01 | -7.23168219E+01 | 8.30E-01 |
| 1857 | Tb | 150 | 65 | 85 | 20 | 8.20633800E+00 | 8.21088207E+00 | -4.50E-03 | 1.39619470E+05 | 1.39618741E+05 | 7.30E-01 | 1.39653003E+05 | 1.39652275E+05 | 7.30E-01 | -7.11056490E+01 | -7.18344655E+01 | 7.30E-01 |
| 1858 | Tb | 151 | 65 | 86 | 21 | 8.20887000E+00 | 8.21784216E+00 | -9.00E-03 | 1.40550447E+05 | 1.40549045E+05 | 1.40E+00 | 1.40583980E+05 | 1.40582578E+05 | 1.40E+00 | -7.16229470E+01 | -7.30250027E+01 | 1.40E+00 |
| 1859 | Tb | 152 | 65 | 87 | 22 | 8.20200000E+00 | 8.21064452E+00 | -8.60E-03 | 1.41482848E+05 | 1.41481486E+05 | 1.40E+00 | 1.41516381E+05 | 1.41515020E+05 | 1.40E+00 | -7.07162900E+01 | -7.20774835E+01 | 1.40E+00 |
| 1860 | Tb | 153 | 65 | 88 | 23 | 8.20504500E+00 | 8.21483594E+00 | -9.80E-03 | 1.42413745E+05 | 1.42412200E+05 | 1.50E+00 | 1.42447278E+05 | 1.42445733E+05 | 1.50E+00 | -7.13129420E+01 | -7.28580969E+01 | 1.50E+00 |
| 1861 | Tb | 154 | 65 | 89 | 24 | 8.19666200E+00 | 8.20608260E+00 | -9.40E-03 | 1.43346396E+05 | 1.43344898E+05 | 1.50E+00 | 1.43379930E+05 | 1.43378432E+05 | 1.50E+00 | -7.01556870E+01 | -7.16535999E+01 | 1.50E+00 |
| 1862 | Tb | 155 | 65 | 90 | 25 | 8.20291000E+00 | 8.20851997E+00 | -5.60E-03 | 1.44276796E+05 | 1.44275880E+05 | 9.20E-01 | 1.44310330E+05 | 1.44309413E+05 | 9.20E-01 | -7.12494660E+01 | -7.21661554E+01 | 9.20E-01 |
| 1863 | Tb | 156 | 65 | 91 | 26 | 8.19463500E+00 | 8.19873553E+00 | -4.10E-03 | 1.45209450E+05 | 1.45208763E+05 | 6.90E-01 | 1.45242983E+05 | 1.45242297E+05 | 6.90E-01 | -7.00900970E+01 | -7.07769836E+01 | 6.90E-01 |
| 1864 | Tb | 157 | 65 | 92 | 27 | 8.19813400E+00 | 8.19959151E+00 | -1.50E-03 | 1.46140271E+05 | 1.46139995E+05 | 2.80E-01 | 1.46173805E+05 | 1.46173529E+05 | 2.80E-01 | -7.07627780E+01 | -7.10387884E+01 | 2.80E-01 |
| 1865 | Tb | 158 | 65 | 93 | 28 | 8.18914900E+00 | 8.18867070E+00 | 4.80E-04 | 1.47073058E+05 | 1.47073087E+05 | -2.80E-02 | 1.47106592E+05 | 1.47106620E+05 | -2.80E-02 | -6.94699400E+01 | -6.94415735E+01 | -2.80E-02 |
| 1866 | Tb | 159 | 65 | 94 | 29 | 8.18879600E+00 | 8.18771464E+00 | 1.10E-03 | 1.48004491E+05 | 1.48004615E+05 | -1.20E-01 | 1.48038024E+05 | 1.48038149E+05 | -1.20E-01 | -6.95316400E+01 | -6.94069124E+01 | -1.20E-01 |
| 1867 | Tb | 160 | 65 | 95 | 30 | 8.17746100E+00 | 8.17534046E+00 | 2.10E-03 | 1.48937681E+05 | 1.48937973E+05 | -2.90E-01 | 1.48971214E+05 | 1.48971506E+05 | -2.90E-01 | -6.78355320E+01 | -6.75434392E+01 | -2.90E-01 |
| 1868 | Tb | 161 | 65 | 96 | 31 | 8.17447400E+00 | 8.17226857E+00 | 2.20E-03 | 1.49869549E+05 | 1.49869858E+05 | -3.10E-01 | 1.49903083E+05 | 1.49903391E+05 | -3.10E-01 | -6.74608220E+01 | -6.71528861E+01 | -3.10E-01 |
| 1869 | Tb | 162 | 65 | 97 | 32 | 8.16281200E+00 | 8.15813993E+00 | 4.70E-03 | 1.50802830E+05 | 1.50803539E+05 | -7.10E-01 | 1.50836363E+05 | 1.50837073E+05 | -7.10E-01 | -6.56746490E+01 | -6.49649956E+01 | -7.10E-01 |
| 1870 | Tb | 163 | 65 | 98 | 33 | 8.15562500E+00 | 8.15271396E+00 | 2.90E-03 | 1.51735404E+05 | 1.51735831E+05 | -4.30E-01 | 1.51768937E+05 | 1.51769365E+05 | -4.30E-01 | -6.45947690E+01 | -6.41673828E+01 | -4.30E-01 |
| 1871 | Tb | 164 | 65 | 99 | 34 | 8.13975700E+00 | 8.13661763E+00 | 3.10E-03 | 1.52669416E+05 | 1.52669884E+05 | -4.70E-01 | 1.52702949E+05 | 1.52703417E+05 | -4.70E-01 | -6.20766600E+01 | -6.16089810E+01 | -4.70E-01 |
| 1872 | Tb | 165 | 65 | 100 | 35 | 8.13000000E+00 | 8.12869242E+00 | 1.30E-03 | 1.53602421E+05 | 1.53602620E+05 | -2.00E-01 | 1.53635954E+05 | 1.53636153E+05 | -2.00E-01 | -6.05660000E+01 | -6.03666187E+01 | -2.00E-01 |
| 1873 | Tb | 166 | 65 | 101 | 36 | 8.11367200E+00 | 8.11050409E+00 | 3.20E-03 | 1.54536597E+05 | 1.54537076E+05 | -4.80E-01 | 1.54570131E+05 | 1.54570609E+05 | -4.80E-01 | -5.78834830E+01 | -5.74047306E+01 | -4.80E-01 |
| 1874 | Tb | 167 | 65 | 102 | 37 | 8.10200000E+00 | 8.10001537E+00 | 2.00E-03 | 1.55470048E+05 | 1.55470282E+05 | -2.30E-01 | 1.55503581E+05 | 1.55503816E+05 | -2.30E-01 | -5.59270000E+01 | -5.56922988E+01 | -2.30E-01 |
| 1875 | Tb | 168 | 65 | 103 | 38 | 8.08200000E+00 | 8.07968297E+00 | 2.30E-03 | 1.56404746E+05 | 1.56405164E+05 | -4.20E-01 | 1.56438280E+05 | 1.56438697E+05 | -4.20E-01 | -5.27230000E+01 | -5.23051527E+01 | -4.20E-01 |
| 1876 | Tb | 169 | 65 | 104 | 39 | 8.06800000E+00 | 8.06661454E+00 | 1.40E-03 | 1.57338634E+05 | 1.57338858E+05 | -2.20E-01 | 1.57372168E+05 | 1.57372391E+05 | -2.20E-01 | -5.03290000E+01 | -5.01049519E+01 | -2.20E-01 |
| 1877 | Tb | 170 | 65 | 105 | 40 | 8.04700000E+00 | 8.04415702E+00 | 2.80E-03 | 1.58273733E+05 | 1.58274174E+05 | -4.40E-01 | 1.58307267E+05 | 1.58307708E+05 | -4.40E-01 | -4.67240000E+01 | -4.62824676E+01 | -4.40E-01 |
| 1878 | Tb | 171 | 65 | 106 | 41 | 8.03100000E+00 | 8.02861098E+00 | 2.40E-03 | 1.59207919E+05 | 1.59208354E+05 | -4.30E-01 | 1.59241453E+05 | 1.59241887E+05 | -4.30E-01 | -4.40320000E+01 | -4.35969342E+01 | -4.40E-01 |
| 1879 | Dy | 139 | 66 | 73 | 7 | 7.97100000E+00 | 7.97272165E+00 | -1.70E-03 | 1.29405976E+05 | 1.29405640E+05 | 3.40E-01 | 1.29440033E+05 | 1.29439696E+05 | 3.40E-01 | -3.76420000E+01 | -3.79780504E+01 | 3.40E-01 |
| 1880 | DY | 140 | 66 | 74 | 8 | 8.00800000E+00 | 8.01141023E+00 | -3.40E-03 | 1.30332282E+05 | 1.30331816E+05 | 4.70E-01 | 1.30366338E+05 | 1.30365873E+05 | 4.70E-01 | -4.28300000E+01 | -4.32958548E+01 | 4.70E-01 |
| 1881 | Dy | 141 | 66 | 75 | 9 | 8.02700000E+00 | 8.02910159E+00 | -2.10E-03 | 1.31261224E+05 | 1.31260876E+05 | 3.50E-01 | 1.31295280E+05 | 1.31294932E+05 | 3.50E-01 | -4.53820000E+01 | -4.57304273E+01 | 3.50E-01 |
| 1882 | Dy | 142 | 66 | 76 | 10 | 8.06100000E+00 | 8.06397894E+00 | -3.00E-03 | 1.32187980E+05 | 1.32187459E+05 | 5.20E-01 | 1.32222037E+05 | 1.32221516E+05 | 5.20E-01 | -5.01200000E+01 | -5.06407939E+01 | 5.20E-01 |
| 1883 | Dy | 143 | 66 | 77 | 11 | 8.07505200E+00 | 8.07884753E+00 | -3.80E-03 | 1.33117425E+05 | 1.33116834E+05 | 5.90E-01 | 1.33151482E+05 | 1.33150891E+05 | 5.90E-01 | -5.21689410E+01 | -5.27596618E+01 | 5.90E-01 |
| 1884 | Dy | 144 | 66 | 78 | 12 | 8.10558900E+00 | 8.10993754E+00 | -4.30E-03 | 1.34044518E+05 | 1.34043844E+05 | 6.70E-01 | 1.34078575E+05 | 1.34077901E+05 | 6.70E-01 | -5.65700820E+01 | -5.72441521E+01 | 6.70E-01 |
| 1885 | Dy | 145 | 66 | 79 | 13 | 8.11688800E+00 | 8.12148498E+00 | -4.60E-03 | 1.34974340E+05 | 1.34973625E+05 | 7.10E-01 | 1.35008396E+05 | 1.35007682E+05 | 7.10E-01 | -5.82425980E+01 | -5.89571496E+01 | 7.10E-01 |
| 1886 | Dy | 146 | 66 | 80 | 14 | 8.14611200E+00 | 8.14788758E+00 | -1.80E-03 | 1.35901521E+05 | 1.35901214E+05 | 3.10E-01 | 1.35935578E+05 | 1.35935271E+05 | 3.10E-01 | -6.25549170E+01 | -6.28620950E+01 | 3.10E-01 |
| 1887 | Dy | 147 | 66 | 81 | 15 | 8.15676700E+00 | 8.15486608E+00 | 1.90E-03 | 1.36831374E+05 | 1.36831606E+05 | -2.30E-01 | 1.36865431E+05 | 1.36865662E+05 | -2.30E-01 | -6.41960370E+01 | -6.39645025E+01 | -2.30E-01 |
| 1888 | Dy | 148 | 66 | 82 | 16 | 8.18090100E+00 | 8.17537631E+00 | 5.50E-03 | 1.37759211E+05 | 1.37759981E+05 | -7.70E-01 | 1.37793268E+05 | 1.37794037E+05 | -7.70E-01 | -6.78532160E+01 | -6.70835639E+01 | -7.70E-01 |
| 1889 | Dy | 149 | 66 | 83 | 17 | 8.17913300E+00 | 8.17692115E+00 | 2.20E-03 | 1.38690859E+05 | 1.38691141E+05 | -2.80E-01 | 1.38724916E+05 | 1.38725197E+05 | -2.80E-01 | -6.76994080E+01 | -6.74178031E+01 | -2.80E-01 |
| 1890 | Dy | 150 | 66 | 84 | 18 | 8.18914800E+00 | 8.19142787E+00 | -2.30E-03 | 1.39620743E+05 | 1.39620353E+05 | 3.90E-01 | 1.39654800E+05 | 1.39654410E+05 | 3.90E-01 | -6.93094280E+01 | -6.96994123E+01 | 3.90E-01 |
| 1891 | Dy | 151 | 66 | 85 | 19 | 8.18467600E+00 | 8.18822754E+00 | -3.60E-03 | 1.40552794E+05 | 1.40552210E+05 | 5.80E-01 | 1.40586851E+05 | 1.40586267E+05 | 5.80E-01 | -6.87521150E+01 | -6.93362719E+01 | 5.80E-01 |
| 1892 | Dy | 152 | 66 | 86 | 20 | 8.19291600E+00 | 8.19815196E+00 | -5.20E-03 | 1.41482923E+05 | 1.41482079E+05 | 8.40E-01 | 1.41516979E+05 | 1.41516136E+05 | 8.40E-01 | -7.01179440E+01 | -7.09616916E+01 | 8.40E-01 |
| 1893 | Dy | 153 | 66 | 87 | 21 | 8.18574700E+00 | 8.19194817E+00 | -6.20E-03 | 1.42415392E+05 | 1.42414395E+05 | 1.00E+00 | 1.42449449E+05 | 1.42448452E+05 | 1.00E+00 | -6.91425840E+01 | -7.01393449E+01 | 1.00E+00 |
| 1894 | Dy | 154 | 66 | 88 | 22 | 8.19312800E+00 | 8.19904186E+00 | -5.90E-03 | 1.43345635E+05 | 1.43344676E+05 | 9.60E-01 | 1.43379692E+05 | 1.43378733E+05 | 9.60E-01 | -7.03936740E+01 | -7.13524020E+01 | 9.60E-01 |
| 1895 | Dy | 155 | 66 | 89 | 23 | 8.18435000E+00 | 8.19124373E+00 | -6.90E-03 | 1.44278368E+05 | 1.44277251E+05 | 1.10E+00 | 1.44312424E+05 | 1.44311308E+05 | 1.10E+00 | -6.91549660E+01 | -7.02714146E+01 | 1.10E+00 |
| 1896 | Dy | 156 | 66 | 90 | 24 | 8.19242900E+00 | 8.19650841E+00 | -4.10E-03 | 1.45208489E+05 | 1.45207804E+05 | 6.80E-01 | 1.45242545E+05 | 1.45241861E+05 | 6.80E-01 | -7.05283350E+01 | -7.12126300E+01 | 6.80E-01 |
| 1897 | Dy | 157 | 66 | 91 | 25 | 8.18462100E+00 | 8.18764909E+00 | -3.00E-03 | 1.46141087E+05 | 1.46140564E+05 | 5.20E-01 | 1.46175144E+05 | 1.46174621E+05 | 5.20E-01 | -6.94236330E+01 | -6.99469062E+01 | 5.20E-01 |
| 1898 | Dy | 158 | 66 | 92 | 26 | 8.19012300E+00 | 8.19129310E+00 | -1.20E-03 | 1.47071599E+05 | 1.47071366E+05 | 2.30E-01 | 1.47105655E+05 | 1.47105423E+05 | 2.30E-01 | -7.04061590E+01 | -7.06389895E+01 | 2.30E-01 |
| 1899 | Dy | 159 | 66 | 93 | 27 | 8.18157700E+00 | 8.18130271E+00 | 2.70E-04 | 1.48004333E+05 | 1.48004329E+05 | 4.20E-03 | 1.48038389E+05 | 1.48038385E+05 | 4.30E-03 | -6.91662630E+01 | -6.91704919E+01 | 4.20E-03 |
| 1900 | Dy | 160 | 66 | 94 | 28 | 8.18404600E+00 | 8.18313774E+00 | 9.10E-04 | 1.48935322E+05 | 1.48935419E+05 | -9.70E-02 | 1.48969378E+05 | 1.48969476E+05 | -9.70E-02 | -6.96714230E+01 | -6.95740805E+01 | -9.70E-02 |
| 1901 | Dy | 161 | 66 | 95 | 29 | 8.17330200E+00 | 8.17172923E+00 | 1.60E-03 | 1.49868433E+05 | 1.49868638E+05 | -2.10E-01 | 1.49902489E+05 | 1.49902695E+05 | -2.10E-01 | -6.80544930E+01 | -6.78491297E+01 | -2.10E-01 |
| 1902 | Dy | 162 | 66 | 96 | 30 | 8.17344900E+00 | 8.17148064E+00 | 2.00E-03 | 1.50799801E+05 | 1.50800072E+05 | -2.70E-01 | 1.50833858E+05 | 1.50834129E+05 | -2.70E-01 | -6.81801700E+01 | -6.79092671E+01 | -2.70E-01 |
| 1903 | Dy | 163 | 66 | 97 | 31 | 8.16177700E+00 | 8.15836791E+00 | 3.40E-03 | 1.51733095E+05 | 1.51733603E+05 | -5.10E-01 | 1.51767152E+05 | 1.51767660E+05 | -5.10E-01 | -6.63798610E+01 | -6.58720543E+01 | -5.10E-01 |
| 1904 | Dy | 164 | 66 | 98 | 32 | 8.15870600E+00 | 8.15581101E+00 | 2.90E-03 | 1.52665003E+05 | 1.52665430E+05 | -4.30E-01 | 1.52699059E+05 | 1.52699486E+05 | -4.30E-01 | -6.59666600E+01 | -6.55397711E+01 | -4.30E-01 |
| 1905 | Dy | 165 | 66 | 99 | 33 | 8.14390200E+00 | 8.14078239E+00 | 3.10E-03 | 1.53598852E+05 | 1.53599319E+05 | -4.70E-01 | 1.53632909E+05 | 1.53633376E+05 | -4.70E-01 | -6.36113000E+01 | -6.31445421E+01 | -4.70E-01 |
| 1906 | Dy | 166 | 66 | 100 | 34 | 8.13727300E+00 | 8.13577331E+00 | 1.50E-03 | 1.54531374E+05 | 1.54531575E+05 | -2.00E-01 | 1.54565431E+05 | 1.54565632E+05 | -2.00E-01 | -6.25834830E+01 | -6.23824973E+01 | -2.00E-01 |
| 1907 | Dy | 167 | 66 | 101 | 35 | 8.12099500E+00 | 8.11869606E+00 | 2.30E-03 | 1.55465521E+05 | 1.55465856E+05 | -3.40E-01 | 1.55499577E+05 | 1.55499913E+05 | -3.40E-01 | -5.99310390E+01 | -5.95950512E+01 | -3.40E-01 |
| 1908 | Dy | 168 | 66 | 102 | 36 | 8.11253800E+00 | 8.11116247E+00 | 1.40E-03 | 1.56398386E+05 | 1.56398569E+05 | -1.80E-01 | 1.56432442E+05 | 1.56432625E+05 | -1.80E-01 | -5.85599790E+01 | -5.83767852E+01 | -1.80E-01 |
| 1909 | Dy | 169 | 66 | 103 | 37 | 8.09476500E+00 | 8.09196807E+00 | 2.80E-03 | 1.57332842E+05 | 1.57333267E+05 | -4.20E-01 | 1.57366899E+05 | 1.57367324E+05 | -4.20E-01 | -5.55975870E+01 | -5.51727743E+01 | -4.20E-01 |
| 1910 | Dy | 170 | 66 | 104 | 38 | 8.08300000E+00 | 8.08187574E+00 | 1.10E-03 | 1.58266270E+05 | 1.58266456E+05 | -1.90E-01 | 1.58300327E+05 | 1.58300513E+05 | -1.90E-01 | -5.36630000E+01 | -5.34777276E+01 | -1.90E-01 |
| 1911 | Dy | 171 | 66 | 105 | 39 | 8.06300000E+00 | 8.06056009E+00 | 2.40E-03 | 1.59201239E+05 | 1.59201584E+05 | -3.50E-01 | 1.59235296E+05 | 1.59235641E+05 | -3.50E-01 | -5.01890000E+01 | -4.98433089E+01 | -3.50E-01 |
| 1912 | Dy | 172 | 66 | 106 | 40 | 8.05000000E+00 | 8.04799088E+00 | 2.00E-03 | 1.60134913E+05 | 1.60135251E+05 | -3.40E-01 | 1.60168969E+05 | 1.60169308E+05 | -3.40E-01 | -4.80090000E+01 | -4.76706454E+01 | -3.40E-01 |
| 1913 | Dy | 173 | 66 | 107 | 41 | 8.02700000E+00 | 8.02457505E+00 | 2.40E-03 | 1.61070477E+05 | 1.61070820E+05 | -3.40E-01 | 1.61104534E+05 | 1.61104876E+05 | -3.40E-01 | -4.39390000E+01 | -4.35963792E+01 | -3.40E-01 |
| 1914 | Ho | 140 | 67 | 73 | 6 | 7.90600000E+00 | 7.90773369E+00 | -1.70E-03 | 1.30345329E+05 | 1.30345024E+05 | 3.10E-01 | 1.30379909E+05 | 1.30379604E+05 | 3.10E-01 | -2.92590000E+01 | -2.95642152E+01 | 3.10E-01 |
| 1915 | Ho | 141 | 67 | 74 | 7 | 7.94300000E+00 | 7.94800911E+00 | -5.00E-03 | 1.31271718E+05 | 1.31271003E+05 | 7.10E-01 | 1.31306298E+05 | 1.31305583E+05 | 7.10E-01 | -3.43640000E+01 | -3.50794647E+01 | 7.20E-01 |
| 1916 | Ho | 142 | 67 | 75 | 8 | 7.96500000E+00 | 7.96927722E+00 | -4.30E-03 | 1.32200326E+05 | 1.32199600E+05 | 7.30E-01 | 1.32234906E+05 | 1.32234180E+05 | 7.30E-01 | -3.72500000E+01 | -3.79762256E+01 | 7.30E-01 |
| 1917 | Ho | 143 | 67 | 76 | 9 | 7.99900000E+00 | 8.00553200E+00 | -6.50E-03 | 1.33127023E+05 | 1.33126012E+05 | 1.00E+00 | 1.33161603E+05 | 1.33160592E+05 | 1.00E+00 | -4.20480000E+01 | -4.30586173E+01 | 1.00E+00 |
| 1918 | Ho | 144 | 67 | 77 | 10 | 8.01709700E+00 | 8.02366902E+00 | -6.60E-03 | 1.34055955E+05 | 1.34054960E+05 | 1.00E+00 | 1.34090535E+05 | 1.34089540E+05 | 1.00E+00 | -4.46095130E+01 | -4.56045620E+01 | 1.00E+00 |
| 1919 | Ho | 145 | 67 | 78 | 11 | 8.04857800E+00 | 8.05595576E+00 | -7.40E-03 | 1.34982939E+05 | 1.34981820E+05 | 1.10E+00 | 1.35017519E+05 | 1.35016400E+05 | 1.10E+00 | -4.91201050E+01 | -5.02384897E+01 | 1.10E+00 |
| 1920 | Ho | 146 | 67 | 79 | 12 | 8.06324200E+00 | 8.07053318E+00 | -7.30E-03 | 1.35912315E+05 | 1.35911201E+05 | 1.10E+00 | 1.35946895E+05 | 1.35945781E+05 | 1.10E+00 | -5.12382180E+01 | -5.23514296E+01 | 1.10E+00 |
| 1921 | Ho | 147 | 67 | 80 | 13 | 8.09403700E+00 | 8.09803839E+00 | -4.00E-03 | 1.36839290E+05 | 1.36838653E+05 | 6.40E-01 | 1.36873870E+05 | 1.36873233E+05 | 6.40E-01 | -5.57570920E+01 | -5.63939091E+01 | 6.40E-01 |
| 1922 | Ho | 148 | 67 | 81 | 14 | 8.10897900E+00 | 8.10794057E+00 | 1.00E-03 | 1.37768550E+05 | 1.37768655E+05 | -1.10E-01 | 1.37803130E+05 | 1.37803235E+05 | -1.10E-01 | -5.79911570E+01 | -5.78861518E+01 | -1.10E-01 |
| 1923 | Ho | 149 | 67 | 82 | 15 | 8.13336600E+00 | 8.12959157E+00 | 3.80E-03 | 1.38696372E+05 | 1.38696886E+05 | -5.10E-01 | 1.38730953E+05 | 1.38731466E+05 | -5.10E-01 | -6.16624380E+01 | -6.11487723E+01 | -5.10E-01 |
| 1924 | Ho | 150 | 67 | 83 | 16 | 8.13484100E+00 | 8.13408246E+00 | 7.60E-04 | 1.39627583E+05 | 1.39627648E+05 | -6.50E-02 | 1.39662163E+05 | 1.39662228E+05 | -6.50E-02 | -6.19458330E+01 | -6.18806773E+01 | -6.50E-02 |
| 1925 | Ho | 151 | 67 | 84 | 17 | 8.14552500E+00 | 8.14983533E+00 | -4.30E-03 | 1.40557400E+05 | 1.40556701E+05 | 7.00E-01 | 1.40591981E+05 | 1.40591281E+05 | 7.00E-01 | -6.36225820E+01 | -6.43221251E+01 | 7.00E-01 |
| 1926 | Ho | 152 | 67 | 85 | 18 | 8.14488200E+00 | 8.14961648E+00 | -4.70E-03 | 1.41488918E+05 | 1.41488150E+05 | 7.70E-01 | 1.41523498E+05 | 1.41522730E+05 | 7.70E-01 | -6.35990350E+01 | -6.43673761E+01 | 7.70E-01 |
| 1927 | Ho | 153 | 67 | 86 | 19 | 8.15363800E+00 | 8.16083735E+00 | -7.20E-03 | 1.42418999E+05 | 1.42417849E+05 | 1.20E+00 | 1.42453579E+05 | 1.42452429E+05 | 1.20E+00 | -6.50122920E+01 | -6.61624661E+01 | 1.20E+00 |
| 1928 | Ho | 154 | 67 | 87 | 20 | 8.15068100E+00 | 8.15757740E+00 | -6.90E-03 | 1.43350866E+05 | 1.43349755E+05 | 1.10E+00 | 1.43385446E+05 | 1.43384335E+05 | 1.10E+00 | -6.46393170E+01 | -6.57499529E+01 | 1.10E+00 |
| 1929 | Ho | 155 | 67 | 88 | 21 | 8.15920100E+00 | 8.16593872E+00 | -6.70E-03 | 1.44280960E+05 | 1.44279867E+05 | 1.10E+00 | 1.44315540E+05 | 1.44314447E+05 | 1.10E+00 | -6.60391530E+01 | -6.71322153E+01 | 1.10E+00 |
| 1930 | Ho | 156 | 67 | 89 | 22 | 8.15504200E+00 | 8.16100317E+00 | -6.00E-03 | 1.45213015E+05 | 1.45212036E+05 | 9.80E-01 | 1.45247595E+05 | 1.45246617E+05 | 9.80E-01 | -6.54783350E+01 | -6.64568902E+01 | 9.80E-01 |
| 1931 | Ho | 157 | 67 | 90 | 23 | 8.16312200E+00 | 8.16748660E+00 | -4.40E-03 | 1.46143157E+05 | 1.46142423E+05 | 7.30E-01 | 1.46177737E+05 | 1.46177003E+05 | 7.30E-01 | -6.68305700E+01 | -6.75644720E+01 | 7.30E-01 |
| 1932 | Ho | 158 | 67 | 91 | 24 | 8.15846400E+00 | 8.16142510E+00 | -3.00E-03 | 1.47075295E+05 | 1.47074779E+05 | 5.20E-01 | 1.47109875E+05 | 1.47109359E+05 | 5.20E-01 | -6.61864030E+01 | -6.67029231E+01 | 5.20E-01 |
| 1933 | Ho | 159 | 67 | 92 | 25 | 8.16510000E+00 | 8.16626502E+00 | -1.20E-03 | 1.48005647E+05 | 1.48005413E+05 | 2.30E-01 | 1.48040227E+05 | 1.48039993E+05 | 2.30E-01 | -6.73286630E+01 | -6.75625763E+01 | 2.30E-01 |
| 1934 | Ho | 160 | 67 | 93 | 26 | 8.15859300E+00 | 8.15904966E+00 | -4.60E-04 | 1.48938088E+05 | 1.48937967E+05 | 1.20E-01 | 1.48972668E+05 | 1.48972547E+05 | 1.20E-01 | -6.63814230E+01 | -6.65030642E+01 | 1.20E-01 |
| 1935 | Ho | 161 | 67 | 94 | 27 | 8.16311400E+00 | 8.16209244E+00 | 1.00E-03 | 1.49868767E+05 | 1.49868883E+05 | -1.20E-01 | 1.49903347E+05 | 1.49903463E+05 | -1.20E-01 | -6.71965360E+01 | -6.70806825E+01 | -1.20E-01 |
| 1936 | Ho | 162 | 67 | 95 | 28 | 8.15541300E+00 | 8.15347386E+00 | 1.90E-03 | 1.50801417E+05 | 1.50801683E+05 | -2.70E-01 | 1.50835997E+05 | 1.50836263E+05 | -2.70E-01 | -6.60408060E+01 | -6.57752463E+01 | -2.70E-01 |
| 1937 | Ho | 163 | 67 | 96 | 29 | 8.15696200E+00 | 8.15446896E+00 | 2.50E-03 | 1.51732574E+05 | 1.51732932E+05 | -3.60E-01 | 1.51767155E+05 | 1.51767512E+05 | -3.60E-01 | -6.63773060E+01 | -6.60196032E+01 | -3.60E-01 |
| 1938 | Ho | 164 | 67 | 97 | 30 | 8.14792400E+00 | 8.14418479E+00 | 3.70E-03 | 1.52665465E+05 | 1.52666030E+05 | -5.60E-01 | 1.52700045E+05 | 1.52700610E+05 | -5.60E-01 | -6.49807670E+01 | -6.44161478E+01 | -5.60E-01 |
| 1939 | Ho | 165 | 67 | 98 | 31 | 8.14696000E+00 | 8.14291798E+00 | 4.00E-03 | 1.53597042E+05 | 1.53597660E+05 | -6.20E-01 | 1.53631622E+05 | 1.53632240E+05 | -6.20E-01 | -6.48982720E+01 | -6.42799909E+01 | -6.20E-01 |
| 1940 | Ho | 166 | 67 | 99 | 32 | 8.13549400E+00 | 8.13076716E+00 | 4.70E-03 | 1.54530363E+05 | 1.54531099E+05 | -7.40E-01 | 1.54564943E+05 | 1.54565680E+05 | -7.40E-01 | -6.30705950E+01 | -6.23345528E+01 | -7.40E-01 |
| 1941 | Ho | 167 | 67 | 100 | 33 | 8.13038200E+00 | 8.12709406E+00 | 3.30E-03 | 1.55462647E+05 | 1.55463147E+05 | -5.00E-01 | 1.55497227E+05 | 1.55497728E+05 | -5.00E-01 | -6.22810390E+01 | -6.17805945E+01 | -5.00E-01 |
| 1942 | Ho | 168 | 67 | 101 | 34 | 8.11681400E+00 | 8.11294458E+00 | 3.90E-03 | 1.56396361E+05 | 1.56396963E+05 | -6.00E-01 | 1.56430942E+05 | 1.56431543E+05 | -6.00E-01 | -6.00607090E+01 | -5.94592564E+01 | -6.00E-01 |
| 1943 | Ho | 169 | 67 | 102 | 35 | 8.10907100E+00 | 8.10678422E+00 | 2.30E-03 | 1.57329119E+05 | 1.57329456E+05 | -3.40E-01 | 1.57363699E+05 | 1.57364037E+05 | -3.40E-01 | -5.87975870E+01 | -5.84597819E+01 | -3.40E-01 |
| 1944 | Ho | 170 | 67 | 103 | 36 | 8.09379900E+00 | 8.09056048E+00 | 3.20E-03 | 1.58263171E+05 | 1.58263673E+05 | -5.00E-01 | 1.58297751E+05 | 1.58298253E+05 | -5.00E-01 | -5.62390940E+01 | -5.57372102E+01 | -5.00E-01 |
| 1945 | Ho | 171 | 67 | 104 | 37 | 8.08361100E+00 | 8.08186104E+00 | 1.70E-03 | 1.59196385E+05 | 1.59196635E+05 | -2.50E-01 | 1.59230965E+05 | 1.59231216E+05 | -2.50E-01 | -5.45193910E+01 | -5.42688481E+01 | -2.50E-01 |
| 1946 | Ho | 172 | 67 | 105 | 38 | 8.06600000E+00 | 8.06354582E+00 | 2.50E-03 | 1.60130915E+05 | 1.60131269E+05 | -3.50E-01 | 1.60165495E+05 | 1.60165849E+05 | -3.50E-01 | -5.14840000E+01 | -5.11291717E+01 | -3.50E-01 |
| 1947 | Ho | 173 | 67 | 106 | 39 | 8.05400000E+00 | 8.05237157E+00 | 1.60E-03 | 1.61064542E+05 | 1.61064704E+05 | -1.60E-01 | 1.61099122E+05 | 1.61099284E+05 | -1.60E-01 | -4.93510000E+01 | -4.91882533E+01 | -1.60E-01 |
| 1948 | Ho | 174 | 67 | 107 | 40 | 8.03300000E+00 | 8.03196230E+00 | 1.00E-03 | 1.61999697E+05 | 1.61999768E+05 | -7.20E-02 | 1.62034277E+05 | 1.62034349E+05 | -7.20E-02 | -4.56900000E+01 | -4.56180928E+01 | -7.20E-02 |
| 1949 | Ho | 175 | 67 | 108 | 41 | 8.01900000E+00 | 8.01834114E+00 | 6.60E-04 | 1.62933678E+05 | 1.62933686E+05 | -7.70E-03 | 1.62968258E+05 | 1.62968266E+05 | -7.60E-03 | -4.32030000E+01 | -4.31950342E+01 | -8.00E-03 |
| 1950 | ER | 143 | 68 | 75 | 7 | 7.91700000E+00 | 7.92215907E+00 | -5.20E-03 | 1.33137454E+05 | 1.33136627E+05 | 8.30E-01 | 1.33172557E+05 | 1.33171731E+05 | 8.30E-01 | -3.10930000E+01 | -3.19193657E+01 | 8.30E-01 |
| 1951 | Er | 144 | 68 | 76 | 8 | 7.95600000E+00 | 7.96180132E+00 | -5.80E-03 | 1.34063433E+05 | 1.34062562E+05 | 8.70E-01 | 1.34098537E+05 | 1.34097666E+05 | 8.70E-01 | -3.66080000E+01 | -3.74786907E+01 | 8.70E-01 |
| 1952 | Er | 145 | 68 | 77 | 9 | 7.97400000E+00 | 7.98090029E+00 | -6.90E-03 | 1.34992459E+05 | 1.34991396E+05 | 1.10E+00 | 1.35027563E+05 | 1.35026500E+05 | 1.10E+00 | -3.90760000E+01 | -4.01385229E+01 | 1.10E+00 |
| 1953 | Er | 146 | 68 | 78 | 10 | 8.01051200E+00 | 8.01628391E+00 | -5.80E-03 | 1.35918707E+05 | 1.35917815E+05 | 8.90E-01 | 1.35953811E+05 | 1.35952919E+05 | 8.90E-01 | -4.43220120E+01 | -4.52141125E+01 | 8.90E-01 |
| 1954 | Er | 147 | 68 | 79 | 11 | 8.02647500E+00 | 8.03167034E+00 | -5.20E-03 | 1.36847915E+05 | 1.36847102E+05 | 8.10E-01 | 1.36883019E+05 | 1.36882206E+05 | 8.10E-01 | -4.66078060E+01 | -4.74208834E+01 | 8.10E-01 |
| 1955 | Er | 148 | 68 | 80 | 12 | 8.05969200E+00 | 8.06207975E+00 | -2.40E-03 | 1.37774538E+05 | 1.37774135E+05 | 4.00E-01 | 1.37809642E+05 | 1.37809239E+05 | 4.00E-01 | -5.14789880E+01 | -5.18818265E+01 | 4.00E-01 |
| 1956 | Er | 149 | 68 | 81 | 13 | 8.07495500E+00 | 8.07274687E+00 | 2.20E-03 | 1.38703769E+05 | 1.38704049E+05 | -2.80E-01 | 1.38738873E+05 | 1.38739153E+05 | -2.80E-01 | -5.37416150E+01 | -5.34619884E+01 | -2.80E-01 |
| 1957 | Er | 150 | 68 | 82 | 14 | 8.10219500E+00 | 8.09724625E+00 | 4.90E-03 | 1.39631174E+05 | 1.39631867E+05 | -6.90E-01 | 1.39666278E+05 | 1.39666971E+05 | -6.90E-01 | -5.78312760E+01 | -5.71383229E+01 | -6.90E-01 |
| 1958 | Er | 151 | 68 | 83 | 15 | 8.10487200E+00 | 8.10256533E+00 | 2.30E-03 | 1.40562233E+05 | 1.40562532E+05 | -3.00E-01 | 1.40597337E+05 | 1.40597636E+05 | -3.00E-01 | -5.82662840E+01 | -5.79674314E+01 | -3.00E-01 |
| 1959 | Er | 152 | 68 | 84 | 16 | 8.11930600E+00 | 8.12118734E+00 | -1.90E-03 | 1.41491499E+05 | 1.41491164E+05 | 3.40E-01 | 1.41526603E+05 | 1.41526268E+05 | 3.40E-01 | -6.04939350E+01 | -6.08292237E+01 | 3.40E-01 |
| 1960 | Er | 153 | 68 | 85 | 17 | 8.11885100E+00 | 8.12186526E+00 | -3.00E-03 | 1.42423015E+05 | 1.42422505E+05 | 5.10E-01 | 1.42458119E+05 | 1.42457609E+05 | 5.10E-01 | -6.04721800E+01 | -6.09828131E+01 | 5.10E-01 |
| 1961 | Er | 154 | 68 | 86 | 18 | 8.13239100E+00 | 8.13594088E+00 | -3.50E-03 | 1.43352377E+05 | 1.43351780E+05 | 6.00E-01 | 1.43387480E+05 | 1.43386884E+05 | 6.00E-01 | -6.26049270E+01 | -6.32010050E+01 | 6.00E-01 |
| 1962 | Er | 155 | 68 | 87 | 19 | 8.12944200E+00 | 8.13357397E+00 | -4.10E-03 | 1.44284266E+05 | 1.44283577E+05 | 6.90E-01 | 1.44319370E+05 | 1.44318681E+05 | 6.90E-01 | -6.22089780E+01 | -6.28987564E+01 | 6.90E-01 |
| 1963 | Er | 156 | 68 | 88 | 20 | 8.14189900E+00 | 8.14471525E+00 | -2.80E-03 | 1.45213759E+05 | 1.45213271E+05 | 4.90E-01 | 1.45248863E+05 | 1.45248374E+05 | 4.90E-01 | -6.42103840E+01 | -6.46990513E+01 | 4.90E-01 |
| 1964 | Er | 157 | 68 | 89 | 21 | 8.13622100E+00 | 8.14062190E+00 | -4.40E-03 | 1.46146074E+05 | 1.46145334E+05 | 7.40E-01 | 1.46181178E+05 | 1.46180438E+05 | 7.40E-01 | -6.33894170E+01 | -6.41297913E+01 | 7.40E-01 |
| 1965 | Er | 158 | 68 | 90 | 22 | 8.14792600E+00 | 8.14980093E+00 | -1.90E-03 | 1.47075654E+05 | 1.47075308E+05 | 3.50E-01 | 1.47110758E+05 | 1.47110412E+05 | 3.50E-01 | -6.53038080E+01 | -6.56493816E+01 | 3.50E-01 |
| 1966 | Er | 159 | 68 | 91 | 23 | 8.14276700E+00 | 8.14453894E+00 | -1.80E-03 | 1.48007892E+05 | 1.48007561E+05 | 3.30E-01 | 1.48042995E+05 | 1.48042664E+05 | 3.30E-01 | -6.45601630E+01 | -6.48912062E+01 | 3.30E-01 |
| 1967 | Er | 160 | 68 | 92 | 24 | 8.15172100E+00 | 8.15202263E+00 | -3.00E-04 | 1.48937882E+05 | 1.48937784E+05 | 9.80E-02 | 1.48972986E+05 | 1.48972888E+05 | 9.80E-02 | -6.60642280E+01 | -6.61618174E+01 | 9.80E-02 |
| 1968 | Er | 161 | 68 | 93 | 25 | 8.14585600E+00 | 8.14559789E+00 | 2.60E-04 | 1.49870240E+05 | 1.49870232E+05 | 7.80E-03 | 1.49905343E+05 | 1.49905336E+05 | 7.80E-03 | -6.52002950E+01 | -6.52081366E+01 | 7.80E-03 |
| 1969 | Er | 162 | 68 | 94 | 26 | 8.15238900E+00 | 8.15127400E+00 | 1.10E-03 | 1.50800601E+05 | 1.50800732E+05 | -1.30E-01 | 1.50835705E+05 | 1.50835836E+05 | -1.30E-01 | -6.63332110E+01 | -6.62019452E+01 | -1.30E-01 |
| 1970 | Er | 163 | 68 | 95 | 27 | 8.14473400E+00 | 8.14346919E+00 | 1.30E-03 | 1.51733261E+05 | 1.51733418E+05 | -1.60E-01 | 1.51768365E+05 | 1.51768522E+05 | -1.60E-01 | -6.51665770E+01 | -6.50097175E+01 | -1.60E-01 |
| 1971 | Er | 164 | 68 | 96 | 28 | 8.14901200E+00 | 8.14711929E+00 | 1.90E-03 | 1.52663981E+05 | 1.52664242E+05 | -2.60E-01 | 1.52699084E+05 | 1.52699346E+05 | -2.60E-01 | -6.59415860E+01 | -6.56804840E+01 | -2.60E-01 |
| 1972 | Er | 165 | 68 | 97 | 29 | 8.13992800E+00 | 8.13769247E+00 | 2.20E-03 | 1.53596896E+05 | 1.53597215E+05 | -3.20E-01 | 1.53632000E+05 | 1.53632319E+05 | -3.20E-01 | -6.45204010E+01 | -6.42008580E+01 | -3.20E-01 |
| 1973 | Er | 166 | 68 | 98 | 30 | 8.14195600E+00 | 8.13912159E+00 | 2.80E-03 | 1.54527985E+05 | 1.54528406E+05 | -4.20E-01 | 1.54563089E+05 | 1.54563510E+05 | -4.20E-01 | -6.49255690E+01 | -6.45044662E+01 | -4.20E-01 |
| 1974 | Er | 167 | 68 | 99 | 31 | 8.13174300E+00 | 8.12788066E+00 | 3.90E-03 | 1.55461114E+05 | 1.55461709E+05 | -6.00E-01 | 1.55496217E+05 | 1.55496813E+05 | -6.00E-01 | -6.32907160E+01 | -6.26950338E+01 | -6.00E-01 |
| 1975 | Er | 168 | 68 | 100 | 32 | 8.12959800E+00 | 8.12695309E+00 | 2.60E-03 | 1.56392908E+05 | 1.56393303E+05 | -3.90E-01 | 1.56428011E+05 | 1.56428406E+05 | -3.90E-01 | -6.29907090E+01 | -6.25957637E+01 | -3.90E-01 |
| 1976 | Er | 169 | 68 | 101 | 33 | 8.11701600E+00 | 8.11376634E+00 | 3.20E-03 | 1.57326470E+05 | 1.57326970E+05 | -5.00E-01 | 1.57361574E+05 | 1.57362073E+05 | -5.00E-01 | -6.09226420E+01 | -6.04228367E+01 | -5.00E-01 |
| 1977 | Er | 170 | 68 | 102 | 34 | 8.11196100E+00 | 8.11040182E+00 | 1.60E-03 | 1.58258777E+05 | 1.58258993E+05 | -2.20E-01 | 1.58293881E+05 | 1.58294097E+05 | -2.20E-01 | -6.01090940E+01 | -5.98933155E+01 | -2.20E-01 |
| 1978 | Er | 171 | 68 | 103 | 35 | 8.09774900E+00 | 8.09518725E+00 | 2.60E-03 | 1.59192661E+05 | 1.59193050E+05 | -3.90E-01 | 1.59227765E+05 | 1.59228154E+05 | -3.90E-01 | -5.77193910E+01 | -5.73307061E+01 | -3.90E-01 |
| 1979 | Er | 172 | 68 | 104 | 36 | 8.09041300E+00 | 8.08932580E+00 | 1.10E-03 | 1.60125391E+05 | 1.60125528E+05 | -1.40E-01 | 1.60160495E+05 | 1.60160632E+05 | -1.40E-01 | -5.64840650E+01 | -5.63464053E+01 | -1.40E-01 |
| 1980 | Er | 173 | 68 | 105 | 37 | 8.07400000E+00 | 8.07205512E+00 | 1.90E-03 | 1.61059715E+05 | 1.61059992E+05 | -2.80E-01 | 1.61094818E+05 | 1.61095096E+05 | -2.80E-01 | -5.36540000E+01 | -5.33765842E+01 | -2.80E-01 |
| 1981 | Er | 174 | 68 | 106 | 38 | 8.06400000E+00 | 8.06375424E+00 | 2.50E-04 | 1.61992913E+05 | 1.61992930E+05 | -1.60E-02 | 1.62028017E+05 | 1.62028034E+05 | -1.60E-02 | -5.19490000E+01 | -5.19329680E+01 | -1.60E-02 |
| 1982 | Er | 175 | 68 | 107 | 39 | 8.04500000E+00 | 8.04440414E+00 | 6.00E-04 | 1.62927705E+05 | 1.62927818E+05 | -1.10E-01 | 1.62962809E+05 | 1.62962922E+05 | -1.10E-01 | -4.86520000E+01 | -4.85391350E+01 | -1.10E-01 |
| 1983 | Er | 176 | 68 | 108 | 40 | 8.03400000E+00 | 8.03366872E+00 | 3.30E-04 | 1.63861220E+05 | 1.63861228E+05 | -7.80E-03 | 1.63896324E+05 | 1.63896332E+05 | -7.80E-03 | -4.66310000E+01 | -4.66227867E+01 | -8.20E-03 |
| 1984 | Er | 177 | 68 | 109 | 41 | 8.01300000E+00 | 8.01228427E+00 | 7.20E-04 | 1.64796487E+05 | 1.64796545E+05 | -5.80E-02 | 1.64831591E+05 | 1.64831649E+05 | -5.80E-02 | -4.28580000E+01 | -4.28000891E+01 | -5.80E-02 |
| 1985 | TM | 144 | 69 | 75 | 6 | 7.85000000E+00 | 7.85684708E+00 | -6.80E-03 | 1.34077424E+05 | 1.34076368E+05 | 1.10E+00 | 1.34113052E+05 | 1.34111996E+05 | 1.10E+00 | -2.20920000E+01 | -2.31483564E+01 | 1.10E+00 |
| 1986 | Tm | 145 | 69 | 76 | 7 | 7.88900000E+00 | 7.89793329E+00 | -8.90E-03 | 1.35003428E+05 | 1.35002119E+05 | 1.30E+00 | 1.35039056E+05 | 1.35037747E+05 | 1.30E+00 | -2.75830000E+01 | -2.88913848E+01 | 1.30E+00 |
| 1987 | Tm | 146 | 69 | 77 | 8 | 7.91300000E+00 | 7.92039214E+00 | -7.40E-03 | 1.35931614E+05 | 1.35930508E+05 | 1.10E+00 | 1.35967242E+05 | 1.35966136E+05 | 1.10E+00 | -3.08920000E+01 | -3.19969909E+01 | 1.10E+00 |
| 1988 | Tm | 147 | 69 | 78 | 9 | 7.94881700E+00 | 7.95702261E+00 | -8.20E-03 | 1.36858025E+05 | 1.36856768E+05 | 1.30E+00 | 1.36893653E+05 | 1.36892396E+05 | 1.30E+00 | -3.59744000E+01 | -3.72307437E+01 | 1.30E+00 |
| 1989 | Tm | 148 | 69 | 79 | 10 | 7.96850000E+00 | 7.97551405E+00 | -7.00E-03 | 1.37786728E+05 | 1.37785640E+05 | 1.10E+00 | 1.37822356E+05 | 1.37821268E+05 | 1.10E+00 | -3.87650270E+01 | -3.98531798E+01 | 1.10E+00 |
| 1990 | Tm | 149 | 69 | 80 | 11 | 8.00400000E+00 | 8.00705165E+00 | -3.10E-03 | 1.38713104E+05 | 1.38712531E+05 | 5.70E-01 | 1.38748732E+05 | 1.38748159E+05 | 5.70E-01 | -4.38830000E+01 | -4.44564778E+01 | 5.70E-01 |
| 1991 | Tm | 150 | 69 | 81 | 12 | 8.02100000E+00 | 8.02068542E+00 | 3.10E-04 | 1.39641990E+05 | 1.39642044E+05 | -5.40E-02 | 1.39677618E+05 | 1.39677672E+05 | -5.40E-02 | -4.64910000E+01 | -4.64372753E+01 | -5.40E-02 |
| 1992 | Tm | 151 | 69 | 82 | 13 | 8.05009700E+00 | 8.04630884E+00 | 3.80E-03 | 1.40569198E+05 | 1.40569719E+05 | -5.20E-01 | 1.40604826E+05 | 1.40605347E+05 | -5.20E-01 | -5.07777170E+01 | -5.02557795E+01 | -5.20E-01 |
| 1993 | Tm | 152 | 69 | 83 | 14 | 8.05676900E+00 | 8.05456836E+00 | 2.20E-03 | 1.41499699E+05 | 1.41499983E+05 | -2.80E-01 | 1.41535327E+05 | 1.41535611E+05 | -2.80E-01 | -5.17705740E+01 | -5.14862162E+01 | -2.80E-01 |
| 1994 | Tm | 153 | 69 | 84 | 15 | 8.07136500E+00 | 8.07437179E+00 | -3.00E-03 | 1.42428974E+05 | 1.42428464E+05 | 5.10E-01 | 1.42464602E+05 | 1.42464092E+05 | 5.10E-01 | -5.39892980E+01 | -5.44993906E+01 | 5.10E-01 |
| 1995 | Tm | 154 | 69 | 85 | 16 | 8.07420800E+00 | 8.07798275E+00 | -3.80E-03 | 1.43360030E+05 | 1.43359399E+05 | 6.30E-01 | 1.43395658E+05 | 1.43395027E+05 | 6.30E-01 | -5.44271630E+01 | -5.50585299E+01 | 6.30E-01 |
| 1996 | Tm | 155 | 69 | 86 | 17 | 8.08837500E+00 | 8.09325945E+00 | -4.90E-03 | 1.44289326E+05 | 1.44288518E+05 | 8.10E-01 | 1.44324954E+05 | 1.44324146E+05 | 8.10E-01 | -5.66258360E+01 | -5.74330831E+01 | 8.10E-01 |
| 1997 | Tm | 156 | 69 | 87 | 18 | 8.08956800E+00 | 8.09376483E+00 | -4.20E-03 | 1.45220617E+05 | 1.45219912E+05 | 7.00E-01 | 1.45256245E+05 | 1.45255540E+05 | 7.00E-01 | -5.68289830E+01 | -5.75338626E+01 | 7.00E-01 |
| 1998 | Tm | 157 | 69 | 88 | 19 | 8.10159900E+00 | 8.10606560E+00 | -4.50E-03 | 1.46150203E+05 | 1.46149452E+05 | 7.50E-01 | 1.46185831E+05 | 1.46185080E+05 | 7.50E-01 | -5.87361740E+01 | -5.94875285E+01 | 7.50E-01 |
| 1999 | Tm | 158 | 69 | 89 | 20 | 8.10119900E+00 | 8.10474991E+00 | -3.60E-03 | 1.47081730E+05 | 1.47081119E+05 | 6.10E-01 | 1.47117358E+05 | 1.47116747E+05 | 6.10E-01 | -5.87031940E+01 | -5.93143971E+01 | 6.10E-01 |
| 2000 | Tm | 159 | 69 | 90 | 21 | 8.11275400E+00 | 8.11503091E+00 | -2.30E-03 | 1.48011357E+05 | 1.48010945E+05 | 4.10E-01 | 1.48046985E+05 | 1.48046573E+05 | 4.10E-01 | -6.05703970E+01 | -6.09825071E+01 | 4.10E-01 |
| 2001 | Tm | 160 | 69 | 91 | 22 | 8.11081800E+00 | 8.11246688E+00 | -1.60E-03 | 1.48943120E+05 | 1.48942806E+05 | 3.10E-01 | 1.48978748E+05 | 1.48978434E+05 | 3.10E-01 | -6.03020280E+01 | -6.06159736E+01 | 3.10E-01 |
| 2002 | Tm | 161 | 69 | 92 | 23 | 8.12049000E+00 | 8.12101867E+00 | -5.30E-04 | 1.49873017E+05 | 1.49872882E+05 | 1.40E-01 | 1.49908645E+05 | 1.49908510E+05 | 1.40E-01 | -6.18987080E+01 | -6.20339599E+01 | 1.40E-01 |
| 2003 | Tm | 162 | 69 | 93 | 24 | 8.11758000E+00 | 8.11725088E+00 | 3.30E-04 | 1.50804933E+05 | 1.50804937E+05 | -3.20E-03 | 1.50840561E+05 | 1.50840565E+05 | -3.20E-03 | -6.14764830E+01 | -6.14732781E+01 | -3.20E-03 |
| 2004 | Tm | 163 | 69 | 94 | 25 | 8.12497200E+00 | 8.12399491E+00 | 9.80E-04 | 1.51735176E+05 | 1.51735285E+05 | -1.10E-01 | 1.51770804E+05 | 1.51770913E+05 | -1.10E-01 | -6.27275770E+01 | -6.26184863E+01 | -1.10E-01 |
| 2005 | Tm | 164 | 69 | 95 | 26 | 8.11962100E+00 | 8.11884371E+00 | 7.80E-04 | 1.52667494E+05 | 1.52667572E+05 | -7.70E-02 | 1.52703122E+05 | 1.52703200E+05 | -7.70E-02 | -6.19037240E+01 | -6.18263654E+01 | -7.70E-02 |
| 2006 | Tm | 165 | 69 | 96 | 27 | 8.12554100E+00 | 8.12358811E+00 | 2.00E-03 | 1.53597963E+05 | 1.53598235E+05 | -2.70E-01 | 1.53633591E+05 | 1.53633863E+05 | -2.70E-01 | -6.29288090E+01 | -6.26567161E+01 | -2.70E-01 |
| 2007 | Tm | 166 | 69 | 97 | 28 | 8.11894400E+00 | 8.11683847E+00 | 2.10E-03 | 1.54530498E+05 | 1.54530798E+05 | -3.00E-01 | 1.54566126E+05 | 1.54566426E+05 | -3.00E-01 | -6.18879020E+01 | -6.15885451E+01 | -3.00E-01 |
| 2008 | Tm | 167 | 69 | 98 | 29 | 8.12258700E+00 | 8.11940371E+00 | 3.20E-03 | 1.55461336E+05 | 1.55461818E+05 | -4.80E-01 | 1.55496964E+05 | 1.55497446E+05 | -4.80E-01 | -6.25440540E+01 | -6.20624597E+01 | -4.80E-01 |
| 2009 | Tm | 168 | 69 | 99 | 30 | 8.11495700E+00 | 8.11087952E+00 | 4.10E-03 | 1.56394061E+05 | 1.56394696E+05 | -6.30E-01 | 1.56429689E+05 | 1.56430324E+05 | -6.30E-01 | -6.13133480E+01 | -6.06784800E+01 | -6.30E-01 |
| 2010 | Tm | 169 | 69 | 100 | 31 | 8.11447500E+00 | 8.11113585E+00 | 3.30E-03 | 1.57325593E+05 | 1.57326107E+05 | -5.10E-01 | 1.57361221E+05 | 1.57361735E+05 | -5.10E-01 | -6.12756410E+01 | -6.07613611E+01 | -5.10E-01 |
| 2011 | Tm | 170 | 69 | 101 | 32 | 8.10551900E+00 | 8.10071285E+00 | 4.80E-03 | 1.58258566E+05 | 1.58259333E+05 | -7.70E-01 | 1.58294194E+05 | 1.58294961E+05 | -7.70E-01 | -5.97962950E+01 | -5.90292671E+01 | -7.70E-01 |
| 2012 | Tm | 171 | 69 | 102 | 33 | 8.10189900E+00 | 8.09857898E+00 | 3.30E-03 | 1.59190645E+05 | 1.59191163E+05 | -5.20E-01 | 1.59226273E+05 | 1.59226791E+05 | -5.20E-01 | -5.92114680E+01 | -5.86937699E+01 | -5.20E-01 |
| 2013 | Tm | 172 | 69 | 103 | 34 | 8.09104400E+00 | 8.08617613E+00 | 4.90E-03 | 1.60123976E+05 | 1.60124763E+05 | -7.90E-01 | 1.60159604E+05 | 1.60160391E+05 | -7.90E-01 | -5.73748880E+01 | -5.65877388E+01 | -7.90E-01 |
| 2014 | Tm | 173 | 69 | 104 | 35 | 8.08445300E+00 | 8.08158286E+00 | 2.90E-03 | 1.61056590E+05 | 1.61057037E+05 | -4.50E-01 | 1.61092218E+05 | 1.61092665E+05 | -4.50E-01 | -5.62544400E+01 | -5.58079614E+01 | -4.50E-01 |

| | | | | | | | | | | | | | | | | |
|---|---|---|---|---|---|---|---|---|---|---|---|---|---|---|---|---|
| 2015 | Tm | 174 | 69 | 105 | 36 | 8.07064800E+00 | 8.06716868E+00 | 3.50E-03 | 1.61990473E+05 | 1.61991028E+05 | -5.60E-01 | 1.62026101E+05 | 1.62026656E+05 | -5.60E-01 | -5.38655470E+01 | -5.33101569E+01 | -5.60E-01 |
| 2016 | Tm | 175 | 69 | 106 | 37 | 8.06177200E+00 | 8.06016777E+00 | 1.60E-03 | 1.62923521E+05 | 1.62923752E+05 | -2.30E-01 | 1.62959149E+05 | 1.62959380E+05 | -2.30E-01 | -5.23115850E+01 | -5.20808471E+01 | -2.30E-01 |
| 2017 | Tm | 176 | 69 | 107 | 38 | 8.04511100E+00 | 8.04370620E+00 | 1.40E-03 | 1.63857957E+05 | 1.63858154E+05 | -2.00E-01 | 1.63893585E+05 | 1.63893782E+05 | -2.00E-01 | -4.93696950E+01 | -4.91724606E+01 | -2.00E-01 |
| 2018 | Tm | 177 | 69 | 108 | 39 | 8.03500000E+00 | 8.03428050E+00 | 7.20E-04 | 1.64791352E+05 | 1.64791344E+05 | 7.50E-03 | 1.64826980E+05 | 1.64826972E+05 | 7.60E-03 | -4.74690000E+01 | -4.74764991E+01 | 7.50E-03 |
| 2019 | TM | 178 | 69 | 109 | 40 | 8.01600000E+00 | 8.01579974E+00 | 2.00E-04 | 1.65726199E+05 | 1.65726165E+05 | 3.40E-02 | 1.65761827E+05 | 1.65761793E+05 | 3.40E-02 | -4.41160000E+01 | -4.41498855E+01 | 3.40E-02 |
| 2020 | Tm | 179 | 69 | 110 | 41 | 8.00200000E+00 | 8.00401831E+00 | -2.00E-03 | 1.66660208E+05 | 1.66659823E+05 | 3.80E-01 | 1.66695836E+05 | 1.66695451E+05 | 3.80E-01 | -4.16010000E+01 | -4.19854894E+01 | 3.80E-01 |
| 2021 | Yb | 149 | 70 | 79 | 9 | 7.92700000E+00 | 7.93157722E+00 | -4.60E-03 | 1.38723264E+05 | 1.38722469E+05 | 8.00E-01 | 1.38759417E+05 | 1.38758621E+05 | 8.00E-01 | -3.31980000E+01 | -3.39938643E+01 | 8.00E-01 |
| 2022 | Yb | 150 | 70 | 80 | 10 | 7.96400000E+00 | 7.96608765E+00 | -2.10E-03 | 1.39649318E+05 | 1.39648926E+05 | 3.90E-01 | 1.39685471E+05 | 1.39685078E+05 | 3.90E-01 | -3.86380000E+01 | -3.90306880E+01 | 3.90E-01 |
| 2023 | Yb | 151 | 70 | 81 | 11 | 7.98375500E+00 | 7.98052799E+00 | 3.20E-03 | 1.40577908E+05 | 1.40578345E+05 | -4.40E-01 | 1.40614061E+05 | 1.40614497E+05 | -4.40E-01 | -4.15423060E+01 | -4.11059479E+01 | -4.40E-01 |
| 2024 | Yb | 152 | 70 | 82 | 12 | 8.01576700E+00 | 8.00903166E+00 | 6.70E-03 | 1.41504624E+05 | 1.41505597E+05 | -9.70E-01 | 1.41540777E+05 | 1.41541750E+05 | -9.70E-01 | -4.63206810E+01 | -4.53477141E+01 | -9.70E-01 |
| 2025 | Yb | 153 | 70 | 83 | 13 | 8.02200000E+00 | 8.01811387E+00 | 3.90E-03 | 1.42435231E+05 | 1.42435764E+05 | -5.30E-01 | 1.42471383E+05 | 1.42471916E+05 | -5.30E-01 | -4.72080000E+01 | -4.66750058E+01 | -5.30E-01 |
| 2026 | Yb | 154 | 70 | 84 | 14 | 8.03994000E+00 | 8.04077389E+00 | -8.30E-04 | 1.43364001E+05 | 1.43363822E+05 | 1.80E-01 | 1.43400153E+05 | 1.43399974E+05 | 1.80E-01 | -4.99321250E+01 | -5.01114425E+01 | 1.80E-01 |
| 2027 | Yb | 155 | 70 | 85 | 15 | 8.04382300E+00 | 8.04523355E+00 | -1.40E-03 | 1.44294924E+05 | 1.44294655E+05 | 2.70E-01 | 1.44331077E+05 | 1.44330807E+05 | 2.70E-01 | -5.05026870E+01 | -5.07721449E+01 | 2.70E-01 |
| 2028 | Yb | 156 | 70 | 86 | 16 | 8.06166500E+00 | 8.06332166E+00 | -1.70E-03 | 1.45223663E+05 | 1.45223353E+05 | 3.10E-01 | 1.45259815E+05 | 1.45259506E+05 | 3.10E-01 | -5.32584500E+01 | -5.35678057E+01 | 3.10E-01 |
| 2029 | Yb | 157 | 70 | 87 | 17 | 8.06279000E+00 | 8.06464781E+00 | -1.90E-03 | 1.46154990E+05 | 1.46154647E+05 | 3.40E-01 | 1.46191142E+05 | 1.46190800E+05 | 3.40E-01 | -5.34255130E+01 | -5.37680138E+01 | 3.40E-01 |
| 2030 | Yb | 158 | 70 | 88 | 18 | 8.07920300E+00 | 8.07966973E+00 | -4.70E-04 | 1.47083899E+05 | 1.47083774E+05 | 1.20E-01 | 1.47120051E+05 | 1.47119927E+05 | 1.20E-01 | -5.60102230E+01 | -5.61348060E+01 | 1.20E-01 |
| 2031 | Yb | 159 | 70 | 89 | 19 | 8.07807500E+00 | 8.07911014E+00 | -1.00E-03 | 1.48015565E+05 | 1.48015349E+05 | 2.20E-01 | 1.48051717E+05 | 1.48051502E+05 | 2.20E-01 | -5.58387580E+01 | -5.60541813E+01 | 2.20E-01 |
| 2032 | Yb | 160 | 70 | 90 | 20 | 8.09257100E+00 | 8.09201587E+00 | 5.60E-04 | 1.48944732E+05 | 1.48944770E+05 | -3.80E-02 | 1.48980885E+05 | 1.48980923E+05 | -3.80E-02 | -5.81649010E+01 | -5.81268887E+01 | -3.80E-02 |
| 2033 | Yb | 161 | 70 | 91 | 21 | 8.09041600E+00 | 8.09015126E+00 | 2.60E-04 | 1.49876552E+05 | 1.49876544E+05 | 8.20E-03 | 1.49912705E+05 | 1.49912696E+05 | 8.30E-03 | -5.78391280E+01 | -5.78473844E+01 | 8.30E-03 |
| 2034 | Yb | 162 | 70 | 92 | 22 | 8.10256600E+00 | 8.10126082E+00 | 1.30E-03 | 1.50806059E+05 | 1.50806220E+05 | -1.60E-01 | 1.50842211E+05 | 1.50842372E+05 | -1.60E-01 | -5.98265020E+01 | -5.96659642E+01 | -1.60E-01 |
| 2035 | Yb | 163 | 70 | 93 | 23 | 8.09913900E+00 | 8.09816723E+00 | 9.70E-04 | 1.51738080E+05 | 1.51738188E+05 | -1.10E-01 | 1.51774233E+05 | 1.51774340E+05 | -1.10E-01 | -5.92992500E+01 | -5.91916508E+01 | -1.10E-01 |
| 2036 | Yb | 164 | 70 | 94 | 24 | 8.10944700E+00 | 8.10744122E+00 | 2.00E-03 | 1.52667856E+05 | 1.52668134E+05 | -2.80E-01 | 1.52704008E+05 | 1.52704287E+05 | -2.80E-01 | -6.10175750E+01 | -6.07394348E+01 | -2.80E-01 |
| 2037 | Yb | 165 | 70 | 95 | 25 | 8.10483900E+00 | 8.10297112E+00 | 1.90E-03 | 1.53600072E+05 | 1.53600330E+05 | -2.60E-01 | 1.53636225E+05 | 1.53636482E+05 | -2.60E-01 | -6.02953810E+01 | -6.00379904E+01 | -2.60E-01 |
| 2038 | Yb | 166 | 70 | 96 | 26 | 8.11246800E+00 | 8.11025067E+00 | 2.20E-03 | 1.54530266E+05 | 1.54530584E+05 | -3.20E-01 | 1.54566419E+05 | 1.54566736E+05 | -3.20E-01 | -6.15952750E+01 | -6.12780476E+01 | -3.20E-01 |
| 2039 | Yb | 167 | 70 | 97 | 27 | 8.10620700E+00 | 8.10421240E+00 | 2.00E-03 | 1.55462765E+05 | 1.55463047E+05 | -2.80E-01 | 1.55498917E+05 | 1.55499200E+05 | -2.80E-01 | -6.05909010E+01 | -6.03085878E+01 | -2.80E-01 |
| 2040 | Yb | 168 | 70 | 98 | 28 | 8.11189600E+00 | 8.10934054E+00 | 2.60E-03 | 1.56393268E+05 | 1.56393647E+05 | -3.80E-01 | 1.56429421E+05 | 1.56429799E+05 | -3.80E-01 | -6.15814390E+01 | -6.12030091E+01 | -3.80E-01 |
| 2041 | Yb | 169 | 70 | 99 | 29 | 8.10452900E+00 | 8.10157176E+00 | 3.00E-03 | 1.57325967E+05 | 1.57326416E+05 | -4.50E-01 | 1.57362119E+05 | 1.57362568E+05 | -4.50E-01 | -6.03771010E+01 | -5.99281068E+01 | -4.50E-01 |
| 2042 | Yb | 170 | 70 | 100 | 30 | 8.10661400E+00 | 8.10443222E+00 | 2.20E-03 | 1.58257073E+05 | 1.58257393E+05 | -3.20E-01 | 1.58293226E+05 | 1.58293546E+05 | -3.20E-01 | -6.07647290E+01 | -6.04446376E+01 | -3.20E-01 |
| 2043 | Yb | 171 | 70 | 101 | 31 | 8.09788900E+00 | 8.09481492E+00 | 3.10E-03 | 1.59190024E+05 | 1.59190499E+05 | -4.70E-01 | 1.59226176E+05 | 1.59226651E+05 | -4.70E-01 | -5.93080290E+01 | -5.88331919E+01 | -4.70E-01 |
| 2044 | Yb | 172 | 70 | 102 | 32 | 8.09743300E+00 | 8.09533302E+00 | 2.10E-03 | 1.60121570E+05 | 1.60121880E+05 | -3.10E-01 | 1.60157722E+05 | 1.60158033E+05 | -3.10E-01 | -5.92561790E+01 | -5.89458006E+01 | -3.10E-01 |
| 2045 | Yb | 173 | 70 | 103 | 33 | 8.08743300E+00 | 8.08378672E+00 | 3.60E-03 | 1.61054768E+05 | 1.61055348E+05 | -5.80E-01 | 1.61090920E+05 | 1.61091500E+05 | -5.80E-01 | -5.75522290E+01 | -5.69723053E+01 | -5.80E-01 |
| 2046 | Yb | 174 | 70 | 104 | 34 | 8.08385300E+00 | 8.08189162E+00 | 2.00E-03 | 1.61986869E+05 | 1.61987159E+05 | -2.90E-01 | 1.62023021E+05 | 1.62023312E+05 | -2.90E-01 | -5.69455470E+01 | -5.66550265E+01 | -2.90E-01 |
| 2047 | Yb | 175 | 70 | 105 | 35 | 8.07093000E+00 | 8.06838233E+00 | 2.50E-03 | 1.62920612E+05 | 1.62921007E+05 | -4.00E-01 | 1.62956764E+05 | 1.62957159E+05 | -4.00E-01 | -5.46965850E+01 | -5.43014729E+01 | -4.00E-01 |
| 2048 | Yb | 176 | 70 | 106 | 36 | 8.06407500E+00 | 8.06412903E+00 | -5.40E-05 | 1.63853313E+05 | 1.63853252E+05 | 6.00E-02 | 1.63889465E+05 | 1.63889405E+05 | 6.00E-02 | -5.34896950E+01 | -5.35499549E+01 | 6.00E-02 |
| 2049 | Yb | 177 | 70 | 107 | 37 | 8.04996400E+00 | 8.04860925E+00 | 1.40E-03 | 1.64787312E+05 | 1.64787501E+05 | -1.90E-01 | 1.64823464E+05 | 1.64823653E+05 | -1.90E-01 | -5.09847780E+01 | -5.07957642E+01 | -1.90E-01 |
| 2050 | Yb | 178 | 70 | 108 | 38 | 8.04283200E+00 | 8.04196684E+00 | 8.70E-04 | 1.65720097E+05 | 1.65720200E+05 | -1.00E-01 | 1.65756249E+05 | 1.65756352E+05 | -1.00E-01 | -4.96938590E+01 | -4.95907055E+01 | -1.00E-01 |
| 2051 | Yb | 179 | 70 | 109 | 39 | 8.02500000E+00 | 8.02445038E+00 | 5.50E-04 | 1.66654747E+05 | 1.66654859E+05 | -1.10E-01 | 1.66690899E+05 | 1.66691011E+05 | -1.10E-01 | -4.65370000E+01 | -4.64259078E+01 | -1.10E-01 |
| 2052 | Yb | 180 | 70 | 110 | 40 | 8.01500000E+00 | 8.01547280E+00 | -4.70E-04 | 1.67588179E+05 | 1.67588015E+05 | 1.60E-01 | 1.67624331E+05 | 1.67624168E+05 | 1.60E-01 | -4.46000000E+01 | -4.47630732E+01 | 1.60E-01 |
| 2053 | YB | 181 | 70 | 111 | 41 | 7.99600000E+00 | 7.99601369E+00 | -1.40E-05 | 1.68523184E+05 | 1.68523087E+05 | 9.70E-02 | 1.68559337E+05 | 1.68559240E+05 | 9.70E-02 | -4.10880000E+01 | -4.11851289E+01 | 9.70E-02 |
| 2054 | Lu | 150 | 71 | 79 | 8 | 7.86500000E+00 | 7.87052562E+00 | -5.50E-03 | 1.39662792E+05 | 1.39661953E+05 | 8.40E-01 | 1.39699469E+05 | 1.39698630E+05 | 8.40E-01 | -2.46400000E+01 | -2.54794598E+01 | 8.40E-01 |
| 2055 | Lu | 151 | 71 | 80 | 9 | 7.90300000E+00 | 7.90619696E+00 | -3.20E-03 | 1.40588817E+05 | 1.40588261E+05 | 5.60E-01 | 1.40625495E+05 | 1.40624938E+05 | 5.60E-01 | -3.01080000E+01 | -3.06650394E+01 | 5.60E-01 |
| 2056 | Lu | 152 | 71 | 81 | 10 | 7.92600000E+00 | 7.92368578E+00 | 2.30E-03 | 1.41516998E+05 | 1.41517262E+05 | -2.60E-01 | 1.41553675E+05 | 1.41553939E+05 | -2.60E-01 | -3.34220000E+01 | -3.31582177E+01 | -2.60E-01 |
| 2057 | Lu | 153 | 71 | 82 | 11 | 7.95939900E+00 | 7.95331042E+00 | 6.10E-03 | 1.42443491E+05 | 1.42444371E+05 | -8.80E-01 | 1.42480168E+05 | 1.42481048E+05 | -8.80E-01 | -3.84231450E+01 | -3.75431543E+01 | -8.80E-01 |
| 2058 | Lu | 154 | 71 | 83 | 12 | 7.96900000E+00 | 7.96537603E+00 | 3.60E-03 | 1.43373693E+05 | 1.43374125E+05 | -4.30E-01 | 1.43410370E+05 | 1.43410802E+05 | -4.30E-01 | -3.97150000E+01 | -3.92832490E+01 | -4.30E-01 |
| 2059 | Lu | 155 | 71 | 84 | 13 | 7.98746900E+00 | 7.98917121E+00 | -1.70E-03 | 1.44302352E+05 | 1.44302037E+05 | 3.20E-01 | 1.44339029E+05 | 1.44338714E+05 | 3.20E-01 | -4.25500870E+01 | -4.28655592E+01 | 3.20E-01 |
| 2060 | Lu | 156 | 71 | 85 | 14 | 7.99569700E+00 | 7.99656959E+00 | -8.70E-04 | 1.45232647E+05 | 1.45232459E+05 | 1.90E-01 | 1.45269324E+05 | 1.45269136E+05 | 1.90E-01 | -4.37499240E+01 | -4.39375586E+01 | 1.90E-01 |
| 2061 | Lu | 157 | 71 | 86 | 15 | 8.01342000E+00 | 8.01578144E+00 | -2.40E-03 | 1.46161434E+05 | 1.46161011E+05 | 4.20E-01 | 1.46198111E+05 | 1.46197689E+05 | 4.20E-01 | -4.64567100E+01 | -4.68790697E+01 | 4.20E-01 |
| 2062 | Lu | 158 | 71 | 87 | 16 | 8.01856800E+00 | 8.01996308E+00 | -1.40E-03 | 1.47092172E+05 | 1.47091900E+05 | 2.70E-01 | 1.47128849E+05 | 1.47128577E+05 | 2.70E-01 | -4.72123000E+01 | -4.74842309E+01 | 2.70E-01 |
| 2063 | Lu | 159 | 71 | 88 | 17 | 8.03460000E+00 | 8.03605081E+00 | -1.50E-03 | 1.48021170E+05 | 1.48020888E+05 | 2.80E-01 | 1.48057847E+05 | 1.48057565E+05 | 2.80E-01 | -4.97086040E+01 | -4.99908253E+01 | 2.80E-01 |
| 2064 | Lu | 160 | 71 | 89 | 18 | 8.03833800E+00 | 8.03823822E+00 | 1.00E-04 | 1.48952103E+05 | 1.48952067E+05 | 3.60E-02 | 1.48988780E+05 | 1.48988744E+05 | 3.60E-02 | -5.02699370E+01 | -5.03055425E+01 | 3.60E-02 |
| 2065 | Lu | 161 | 71 | 90 | 19 | 8.05278100E+00 | 8.05214050E+00 | 6.40E-04 | 1.49881304E+05 | 1.49881356E+05 | -5.20E-02 | 1.49917981E+05 | 1.49918033E+05 | -5.20E-02 | -5.25623440E+01 | -5.25107288E+01 | -5.20E-02 |
| 2066 | Lu | 162 | 71 | 91 | 20 | 8.05455900E+00 | 8.05292845E+00 | 1.60E-03 | 1.50812529E+05 | 1.50812742E+05 | -2.10E-01 | 1.50849206E+05 | 1.50849419E+05 | -2.10E-01 | -5.28317530E+01 | -5.26191987E+01 | -2.10E-01 |
| 2067 | Lu | 163 | 71 | 92 | 21 | 8.06668400E+00 | 8.06498728E+00 | 1.70E-03 | 1.51742063E+05 | 1.51742288E+05 | -2.30E-01 | 1.51778741E+05 | 1.51778966E+05 | -2.30E-01 | -5.47914090E+01 | -5.45663972E+01 | -2.30E-01 |
| 2068 | Lu | 164 | 71 | 93 | 22 | 8.06580400E+00 | 8.06448715E+00 | 1.30E-03 | 1.52673707E+05 | 1.52673871E+05 | -1.60E-01 | 1.52710384E+05 | 1.52710548E+05 | -1.60E-01 | -5.46423700E+01 | -5.44780435E+01 | -1.60E-01 |
| 2069 | Lu | 165 | 71 | 94 | 23 | 8.07674500E+00 | 8.07469546E+00 | 2.00E-03 | 1.53603401E+05 | 1.53603687E+05 | -2.90E-01 | 1.53640078E+05 | 1.53640364E+05 | -2.90E-01 | -5.64422410E+01 | -5.61555829E+01 | -2.90E-01 |
| 2070 | Lu | 166 | 71 | 95 | 24 | 8.07417500E+00 | 8.07279600E+00 | 1.40E-03 | 1.54535316E+05 | 1.54535493E+05 | -1.80E-01 | 1.54571993E+05 | 1.54572170E+05 | -1.80E-01 | -5.60209810E+01 | -5.58436482E+01 | -1.80E-01 |
| 2071 | Lu | 167 | 71 | 96 | 25 | 8.08302100E+00 | 8.08102263E+00 | 2.00E-03 | 1.55465330E+05 | 1.55465612E+05 | -2.80E-01 | 1.55502007E+05 | 1.55502289E+05 | -2.80E-01 | -5.75011250E+01 | -5.72189740E+01 | -2.80E-01 |
| 2072 | Lu | 168 | 71 | 97 | 26 | 8.08036900E+00 | 8.07756040E+00 | 2.80E-03 | 1.56397258E+05 | 1.56397678E+05 | -4.20E-01 | 1.56433935E+05 | 1.56434355E+05 | -4.20E-01 | -5.70673880E+01 | -5.66470219E+01 | -4.20E-01 |
| 2073 | Lu | 169 | 71 | 98 | 27 | 8.08633200E+00 | 8.08366719E+00 | 2.70E-03 | 1.57327735E+05 | 1.57328134E+05 | -4.00E-01 | 1.57364412E+05 | 1.57364811E+05 | -4.00E-01 | -5.80841010E+01 | -5.76853116E+01 | -4.00E-01 |
| 2074 | Lu | 170 | 71 | 99 | 28 | 8.08167300E+00 | 8.07849894E+00 | 3.20E-03 | 1.58260006E+05 | 1.58260494E+05 | -4.90E-01 | 1.58296683E+05 | 1.58297171E+05 | -4.90E-01 | -5.73070340E+01 | -5.68190562E+01 | -4.90E-01 |
| 2075 | Lu | 171 | 71 | 100 | 29 | 8.08467000E+00 | 8.08238023E+00 | 2.30E-03 | 1.59190977E+05 | 1.59191317E+05 | -3.40E-01 | 1.59227655E+05 | 1.59227994E+05 | -3.40E-01 | -5.78299790E+01 | -5.74899379E+01 | -3.40E-01 |
| 2076 | Lu | 172 | 71 | 101 | 30 | 8.07824500E+00 | 8.07539969E+00 | 2.80E-03 | 1.60123563E+05 | 1.60124001E+05 | -4.40E-01 | 1.60160240E+05 | 1.60160678E+05 | -4.40E-01 | -5.67381350E+01 | -5.63003450E+01 | -4.40E-01 |
| 2077 | Lu | 173 | 71 | 102 | 31 | 8.07904000E+00 | 8.07698525E+00 | 2.10E-03 | 1.61054913E+05 | 1.61055217E+05 | -3.00E-01 | 1.61091590E+05 | 1.61091894E+05 | -3.00E-01 | -5.68825850E+01 | -5.65787288E+01 | -3.00E-01 |
| 2078 | Lu | 174 | 71 | 103 | 32 | 8.07146400E+00 | 8.06811817E+00 | 3.30E-03 | 1.61987717E+05 | 1.61988248E+05 | -5.30E-01 | 1.62024394E+05 | 1.62024925E+05 | -5.30E-01 | -5.55721360E+01 | -5.50415217E+01 | -5.30E-01 |
| 2079 | Lu | 175 | 71 | 104 | 33 | 8.06915100E+00 | 8.06733371E+00 | 1.80E-03 | 1.62919616E+05 | 1.62919883E+05 | -2.70E-01 | 1.62956293E+05 | 1.62956560E+05 | -2.70E-01 | -5.51675530E+01 | -5.49010414E+01 | -2.70E-01 |
| 2080 | Lu | 176 | 71 | 105 | 34 | 8.05903100E+00 | 8.05654967E+00 | 2.50E-03 | 1.63852893E+05 | 1.63853279E+05 | -3.90E-01 | 1.63889571E+05 | 1.63889956E+05 | -3.90E-01 | -5.33842190E+01 | -5.29990648E+01 | -3.90E-01 |
| 2081 | Lu | 177 | 71 | 106 | 35 | 8.05345900E+00 | 8.05345344E+00 | 5.60E-06 | 1.64785386E+05 | 1.64785335E+05 | 5.00E-02 | 1.64822063E+05 | 1.64822013E+05 | 5.00E-02 | -5.23858020E+01 | -5.24362621E+01 | 5.00E-02 |
| 2082 | Lu | 178 | 71 | 107 | 36 | 8.04206500E+00 | 8.04069971E+00 | 1.40E-03 | 1.65718926E+05 | 1.65719118E+05 | -1.90E-01 | 1.65755603E+05 | 1.65755795E+05 | -1.90E-01 | -5.03397820E+01 | -5.01482336E+01 | -1.90E-01 |
| 2083 | Lu | 179 | 71 | 108 | 37 | 8.03508400E+00 | 8.03524672E+00 | -1.60E-04 | 1.66651699E+05 | 1.66651618E+05 | 8.10E-02 | 1.66688376E+05 | 1.66688295E+05 | 8.10E-02 | -4.90609180E+01 | -4.91415282E+01 | 8.10E-02 |
| 2084 | Lu | 180 | 71 | 109 | 38 | 8.02205200E+00 | 8.02052979E+00 | 1.50E-03 | 1.67585575E+05 | 1.67585797E+05 | -2.20E-01 | 1.67622252E+05 | 1.67622475E+05 | -2.20E-01 | -4.66787940E+01 | -4.64564100E+01 | -2.20E-01 |
| 2085 | Lu | 181 | 71 | 110 | 39 | 8.01192900E+00 | 8.01276007E+00 | -8.30E-04 | 1.68518951E+05 | 1.68518749E+05 | 2.00E-01 | 1.68555628E+05 | 1.68555426E+05 | 2.00E-01 | -4.47974100E+01 | -4.49993007E+01 | 2.00E-01 |
| 2086 | Lu | 182 | 71 | 111 | 40 | 7.99600000E+00 | 7.99612092E+00 | -1.20E-04 | 1.69453362E+05 | 1.69453330E+05 | 3.20E-02 | 1.69490039E+05 | 1.69490007E+05 | 3.20E-02 | -4.18800000E+01 | -4.19124160E+01 | 3.20E-02 |
| 2087 | Lu | 183 | 71 | 112 | 41 | 7.98481200E+00 | 7.98611796E+00 | -1.30E-03 | 1.70387020E+05 | 1.70386729E+05 | 2.90E-01 | 1.70423697E+05 | 1.70423406E+05 | 2.90E-01 | -3.97161100E+01 | -4.00066764E+01 | 2.90E-01 |
| 2088 | Lu | 184 | 71 | 113 | 42 | 7.96700000E+00 | 7.96765047E+00 | -6.50E-04 | 1.71321818E+05 | 1.71321707E+05 | 1.10E-01 | 1.71358495E+05 | 1.71358384E+05 | 1.10E-01 | -3.64120000E+01 | -3.65234582E+01 | 1.10E-01 |
| 2089 | Lu | 185 | 71 | 114 | 43 | 7.95400000E+00 | 7.95556001E+00 | -1.60E-03 | 1.72255836E+05 | 1.72255541E+05 | 3.00E-01 | 1.72292514E+05 | 1.72292218E+05 | 3.00E-01 | -3.38880000E+01 | -3.41830539E+01 | 3.00E-01 |
| 2090 | Hf | 153 | 72 | 81 | 9 | 7.88200000E+00 | 7.87897051E+00 | 3.00E-03 | 1.42454087E+05 | 1.42454437E+05 | -3.50E-01 | 1.42491289E+05 | 1.42491639E+05 | -3.50E-01 | -2.73020000E+01 | -2.69522244E+01 | -3.50E-01 |
| 2091 | Hf | 154 | 72 | 82 | 10 | 7.91800000E+00 | 7.91152899E+00 | 6.50E-03 | 1.43380151E+05 | 1.43381109E+05 | -9.60E-01 | 1.43417353E+05 | 1.43418312E+05 | -9.60E-01 | -3.27330000E+01 | -3.17738824E+01 | -9.60E-01 |
| 2092 | Hf | 155 | 72 | 83 | 11 | 7.93000000E+00 | 7.92443240E+00 | 5.60E-03 | 1.44310014E+05 | 1.44310763E+05 | -7.50E-01 | 1.44347217E+05 | 1.44347965E+05 | -7.50E-01 | -3.43630000E+01 | -3.36141209E+01 | -7.50E-01 |
| 2093 | Hf | 156 | 72 | 84 | 12 | 7.95297300E+00 | 7.95110403E+00 | 1.90E-03 | 1.45238004E+05 | 1.45238243E+05 | -2.40E-01 | 1.45275206E+05 | 1.45275446E+05 | -2.40E-01 | -3.78672760E+01 | -3.76280091E+01 | -2.40E-01 |
| 2094 | Hf | 157 | 72 | 85 | 13 | 7.96000000E+00 | 7.95932625E+00 | 6.70E-04 | 1.46168462E+05 | 1.46168567E+05 | -1.00E-01 | 1.46205665E+05 | 1.46205769E+05 | -1.00E-01 | -3.89030000E+01 | -3.87986821E+01 | -1.00E-01 |
| 2095 | Hf | 158 | 72 | 86 | 14 | 7.98127600E+00 | 7.98134418E+00 | -6.80E-05 | 1.47096757E+05 | 1.47096694E+05 | 6.30E-02 | 1.47133959E+05 | 1.47133896E+05 | 6.30E-02 | -4.21025110E+01 | -4.21655230E+01 | 6.30E-02 |
| 2096 | Hf | 159 | 72 | 87 | 15 | 7.98656100E+00 | 7.98629556E+00 | 2.70E-04 | 1.48027501E+05 | 1.48027491E+05 | 1.00E-02 | 1.48064703E+05 | 1.48064693E+05 | 1.00E-02 | -4.28527580E+01 | -4.28628172E+01 | 1.00E-02 |
| 2097 | Hf | 160 | 72 | 88 | 16 | 8.00633200E+00 | 8.00508573E+00 | 1.20E-03 | 1.48955916E+05 | 1.48956063E+05 | -1.50E-01 | 1.48993118E+05 | 1.48993266E+05 | -1.50E-01 | -4.59312710E+01 | -4.57842214E+01 | -1.50E-01 |
| 2098 | Hf | 161 | 72 | 89 | 17 | 8.00912100E+00 | 8.00796109E+00 | 1.20E-03 | 1.49887026E+05 | 1.49887161E+05 | -1.30E-01 | 1.49924228E+05 | 1.49924363E+05 | -1.30E-01 | -4.63154000E+01 | -4.61809213E+01 | -1.30E-01 |
| 2099 | Hf | 162 | 72 | 90 | 18 | 8.02712000E+00 | 8.02445911E+00 | 2.70E-03 | 1.50815667E+05 | 1.50816046E+05 | -3.80E-01 | 1.50852869E+05 | 1.50853248E+05 | -3.80E-01 | -4.91690620E+01 | -4.87902423E+01 | -3.80E-01 |
| 2100 | Hf | 163 | 72 | 91 | 19 | 8.02797400E+00 | 8.02586177E+00 | 2.10E-03 | 1.51747066E+05 | 1.51747358E+05 | -2.90E-01 | 1.51784268E+05 | 1.51784560E+05 | -2.90E-01 | -4.92639730E+01 | -4.89720158E+01 | -2.90E-01 |
| 2101 | Hf | 164 | 72 | 92 | 20 | 8.04381300E+00 | 8.04043643E+00 | 3.40E-03 | 1.52676006E+05 | 1.52676507E+05 | -5.00E-01 | 1.52713208E+05 | 1.52713709E+05 | -5.00E-01 | -5.18182240E+01 | -5.13168026E+01 | -5.00E-01 |
| 2102 | Hf | 165 | 72 | 93 | 21 | 8.04287200E+00 | 8.04050701E+00 | 2.40E-03 | 1.53607682E+05 | 1.53608020E+05 | -3.40E-01 | 1.53644885E+05 | 1.53645222E+05 | -3.40E-01 | -5.16355070E+01 | -5.12975653E+01 | -3.40E-01 |
| 2103 | Hf | 166 | 72 | 94 | 22 | 8.05643800E+00 | 8.05318771E+00 | 3.30E-03 | 1.54536953E+05 | 1.54537440E+05 | -4.90E-01 | 1.54574155E+05 | 1.54574642E+05 | -4.90E-01 | -5.38589830E+01 | -5.33717497E+01 | -4.90E-01 |
| 2104 | Hf | 167 | 72 | 95 | 23 | 8.05418400E+00 | 8.05184640E+00 | 2.30E-03 | 1.55468838E+05 | 1.55469176E+05 | -3.40E-01 | 1.55506040E+05 | 1.55506379E+05 | -3.40E-01 | -5.34677560E+01 | -5.31296204E+01 | -3.40E-01 |
| 2105 | Hf | 168 | 72 | 96 | 24 | 8.06555300E+00 | 8.06253451E+00 | 3.00E-03 | 1.56398440E+05 | 1.56398894E+05 | -4.50E-01 | 1.56435642E+05 | 1.56436096E+05 | -4.50E-01 | -5.53605520E+01 | -5.49057490E+01 | -4.50E-01 |
| 2106 | Hf | 169 | 72 | 97 | 25 | 8.06177800E+00 | 8.05964283E+00 | 2.10E-03 | 1.57330577E+05 | 1.57330886E+05 | -3.10E-01 | 1.57367779E+05 | 1.57368088E+05 | -3.10E-01 | -5.47168890E+01 | -5.44082705E+01 | -3.10E-01 |
| 2107 | Hf | 170 | 72 | 98 | 26 | 8.07087500E+00 | 8.06822437E+00 | 2.70E-03 | 1.58260534E+05 | 1.58260933E+05 | -4.00E-01 | 1.58297736E+05 | 1.58298135E+05 | -4.00E-01 | -5.62538540E+01 | -5.58554568E+01 | -4.00E-01 |
| 2108 | Hf | 171 | 72 | 99 | 27 | 8.06606800E+00 | 8.06365607E+00 | 2.40E-03 | 1.59192851E+05 | 1.59193211E+05 | -3.60E-01 | 1.59230053E+05 | 1.59230413E+05 | -3.60E-01 | -5.54313450E+01 | -5.50711832E+01 | -3.60E-01 |

| | | | | | | | | | | | | | | | | |
|---|---|---|---|---|---|---|---|---|---|---|---|---|---|---|---|---|
| 2109 | Hf | 172 | 72 | 100 | 28 | 8.07174300E+00 | 8.07004183E+00 | 1.70E-03 | 1.60123374E+05 | 1.60123614E+05 | -2.40E-01 | 1.60160576E+05 | 1.60160817E+05 | -2.40E-01 | -5.64022250E+01 | -5.61618708E+01 | -2.40E-01 |
| 2110 | Hf | 173 | 72 | 101 | 29 | 8.06601600E+00 | 8.06370109E+00 | 2.30E-03 | 1.61055859E+05 | 1.61056207E+05 | -3.50E-01 | 1.61093061E+05 | 1.61093409E+05 | -3.50E-01 | -5.54117840E+01 | -5.50636461E+01 | -3.50E-01 |
| 2111 | Hf | 174 | 72 | 102 | 30 | 8.06854500E+00 | 8.06783127E+00 | 7.10E-04 | 1.61986918E+05 | 1.61986990E+05 | -7.20E-02 | 1.62024120E+05 | 1.62024192E+05 | -7.20E-02 | -5.58466590E+01 | -5.57746781E+01 | -7.20E-02 |
| 2112 | Hf | 175 | 72 | 103 | 31 | 8.06077400E+00 | 8.05964935E+00 | 1.10E-03 | 1.62919775E+05 | 1.62919919E+05 | -1.40E-01 | 1.62956977E+05 | 1.62957121E+05 | -1.40E-01 | -5.44838410E+01 | -5.43393550E+01 | -1.40E-01 |
| 2113 | Hf | 176 | 72 | 104 | 32 | 8.06137100E+00 | 8.06145202E+00 | -8.10E-05 | 1.63851174E+05 | 1.63851108E+05 | 6.60E-02 | 1.63888376E+05 | 1.63888310E+05 | 6.70E-02 | -5.45784420E+01 | -5.46449554E+01 | 6.70E-02 |
| 2114 | Hf | 177 | 72 | 105 | 33 | 8.05184900E+00 | 8.05140218E+00 | 4.50E-04 | 1.64784364E+05 | 1.64784390E+05 | -2.70E-02 | 1.64821566E+05 | 1.64821593E+05 | -2.70E-02 | -5.28830410E+01 | -5.28562674E+01 | -2.70E-02 |
| 2115 | Hf | 178 | 72 | 106 | 34 | 8.04945600E+00 | 8.05094614E+00 | -1.50E-03 | 1.65716303E+05 | 1.65715986E+05 | 3.20E-01 | 1.65753505E+05 | 1.65753188E+05 | 3.20E-01 | -5.24376780E+01 | -5.27551753E+01 | 3.20E-01 |
| 2116 | Hf | 179 | 72 | 107 | 35 | 8.03856000E+00 | 8.03897082E+00 | -4.10E-04 | 1.66649769E+05 | 1.66649644E+05 | 1.30E-01 | 1.66686972E+05 | 1.66686846E+05 | 1.30E-01 | -5.04653520E+01 | -5.05912198E+01 | 1.30E-01 |
| 2117 | Hf | 180 | 72 | 108 | 36 | 8.03494400E+00 | 8.03620346E+00 | -1.30E-03 | 1.67581947E+05 | 1.67581668E+05 | 2.80E-01 | 1.67619149E+05 | 1.67618870E+05 | 2.80E-01 | -4.97817940E+01 | -5.00607468E+01 | 2.80E-01 |
| 2118 | Hf | 181 | 72 | 109 | 37 | 8.02201500E+00 | 8.02230333E+00 | -2.90E-04 | 1.68515818E+05 | 1.68515713E+05 | 1.00E-01 | 1.68553020E+05 | 1.68552915E+05 | 1.00E-01 | -4.74052770E+01 | -4.75097082E+01 | 1.00E-01 |
| 2119 | Hf | 182 | 72 | 110 | 38 | 8.01485000E+00 | 8.01725853E+00 | -2.40E-03 | 1.69448665E+05 | 1.69448174E+05 | 4.90E-01 | 1.69485867E+05 | 1.69485377E+05 | 4.90E-01 | -4.60519590E+01 | -4.65425381E+01 | 4.90E-01 |
| 2120 | Hf | 183 | 72 | 111 | 39 | 8.00004500E+00 | 8.00146423E+00 | -1.40E-03 | 1.70382925E+05 | 1.70382613E+05 | 3.10E-01 | 1.70420127E+05 | 1.70419815E+05 | 3.10E-01 | -4.32861120E+01 | -4.35981204E+01 | 3.10E-01 |
| 2121 | Hf | 184 | 72 | 112 | 40 | 7.99073300E+00 | 7.99421494E+00 | -3.50E-03 | 1.71316204E+05 | 1.71315511E+05 | 6.90E-01 | 1.71353406E+05 | 1.71352713E+05 | 6.90E-01 | -4.15015570E+01 | -4.21943964E+01 | 6.90E-01 |
| 2122 | Hf | 185 | 72 | 113 | 41 | 7.97397000E+00 | 7.97660635E+00 | -2.60E-03 | 1.72250879E+05 | 1.72250339E+05 | 5.40E-01 | 1.72288082E+05 | 1.72287542E+05 | 5.40E-01 | -3.83198000E+01 | -3.88597044E+01 | 5.40E-01 |
| 2123 | Hf | 186 | 72 | 114 | 42 | 7.96430200E+00 | 7.96728526E+00 | -3.00E-03 | 1.73184269E+05 | 1.73183662E+05 | 6.10E-01 | 1.73221471E+05 | 1.73220864E+05 | 6.10E-01 | -3.64242100E+01 | -3.70312682E+01 | 6.10E-01 |
| 2124 | HF | 187 | 72 | 115 | 43 | 7.94600000E+00 | 7.94801270E+00 | -2.00E-03 | 1.74119371E+05 | 1.74118864E+05 | 5.10E-01 | 1.74156573E+05 | 1.74156066E+05 | 5.10E-01 | -3.28170000E+01 | -3.33232665E+01 | 5.10E-01 |
| 2125 | Hf | 188 | 72 | 116 | 44 | 7.93600000E+00 | 7.93683249E+00 | -8.30E-04 | 1.75052802E+05 | 1.75052583E+05 | 2.20E-01 | 1.75090004E+05 | 1.75089785E+05 | 2.20E-01 | -3.08790000E+01 | -3.10980799E+01 | 2.20E-01 |
| 2126 | HF | 189 | 72 | 117 | 45 | 7.91700000E+00 | 7.91613402E+00 | 8.70E-04 | 1.75988013E+05 | 1.75988124E+05 | -1.10E-01 | 1.76025215E+05 | 1.76025326E+05 | -1.10E-01 | -2.71620000E+01 | -2.70515836E+01 | -1.10E-01 |
| 2127 | Ta | 155 | 73 | 82 | 9 | 7.85800000E+00 | 7.85129097E+00 | 6.70E-03 | 1.44319861E+05 | 1.44320792E+05 | -9.30E-01 | 1.44357589E+05 | 1.44358519E+05 | -9.30E-01 | -2.39910000E+01 | -2.30602766E+01 | -9.30E-01 |
| 2128 | Ta | 156 | 73 | 83 | 10 | 7.87200000E+00 | 7.86725942E+00 | 4.70E-03 | 1.45249292E+05 | 1.45250015E+05 | -7.20E-01 | 1.45287020E+05 | 1.45287742E+05 | -7.20E-01 | -2.60540000E+01 | -2.53313258E+01 | -7.20E-01 |
| 2129 | Ta | 157 | 73 | 84 | 11 | 7.89636400E+00 | 7.89502256E+00 | 1.30E-03 | 1.46177196E+05 | 1.46177354E+05 | -1.60E-01 | 1.46214924E+05 | 1.46215081E+05 | -1.60E-01 | -2.96437090E+01 | -2.94860797E+01 | -1.60E-01 |
| 2130 | Ta | 158 | 73 | 85 | 12 | 7.90700000E+00 | 7.90623606E+00 | 7.60E-04 | 1.47107167E+05 | 1.47107253E+05 | -8.50E-02 | 1.47144895E+05 | 1.47144980E+05 | -8.50E-02 | -3.11670000E+01 | -3.10815160E+01 | -8.50E-02 |
| 2131 | Ta | 159 | 73 | 86 | 13 | 7.92875700E+00 | 7.92930551E+00 | -5.50E-04 | 1.48035384E+05 | 1.48035244E+05 | 1.40E-01 | 1.48073111E+05 | 1.48072971E+05 | 1.40E-01 | -3.44442670E+01 | -3.45844765E+01 | 1.40E-01 |
| 2132 | Ta | 160 | 73 | 87 | 14 | 7.93858500E+00 | 7.93714626E+00 | 1.40E-03 | 1.48965448E+05 | 1.48965625E+05 | -1.80E-01 | 1.49003176E+05 | 1.49003353E+05 | -1.80E-01 | -3.58741530E+01 | -3.56969819E+01 | -1.80E-01 |
| 2133 | Ta | 161 | 73 | 88 | 15 | 7.95697000E+00 | 7.95691271E+00 | 5.70E-05 | 1.49894115E+05 | 1.49894071E+05 | 4.40E-02 | 1.49931842E+05 | 1.49931799E+05 | 4.40E-02 | -3.87014660E+01 | -3.87452077E+01 | 4.40E-02 |
| 2134 | Ta | 162 | 73 | 89 | 16 | 7.96433600E+00 | 7.96255720E+00 | 1.80E-03 | 1.50824530E+05 | 1.50824765E+05 | -2.40E-01 | 1.50862258E+05 | 1.50862493E+05 | -2.40E-01 | -3.97803050E+01 | -3.95452093E+01 | -2.40E-01 |
| 2135 | Ta | 163 | 73 | 90 | 17 | 7.98189000E+00 | 7.97994885E+00 | 1.90E-03 | 1.51753270E+05 | 1.51753533E+05 | -2.60E-01 | 1.51790997E+05 | 1.51791261E+05 | -2.60E-01 | -4.25347080E+01 | -4.22712872E+01 | -2.60E-01 |
| 2136 | Ta | 164 | 73 | 91 | 18 | 7.98699700E+00 | 7.98401396E+00 | 3.00E-03 | 1.52684016E+05 | 1.52684452E+05 | -4.40E-01 | 1.52721743E+05 | 1.52722179E+05 | -4.40E-01 | -4.32828000E+01 | -4.28465945E+01 | -4.40E-01 |
| 2137 | Ta | 165 | 73 | 92 | 19 | 8.00305200E+00 | 7.99942021E+00 | 3.60E-03 | 1.53612945E+05 | 1.53613491E+05 | -5.50E-01 | 1.53650672E+05 | 1.53651219E+05 | -5.50E-01 | -4.58476010E+01 | -4.53013213E+01 | -5.50E-01 |
| 2138 | Ta | 166 | 73 | 93 | 20 | 8.00497100E+00 | 8.00207878E+00 | 2.90E-03 | 1.54544189E+05 | 1.54544616E+05 | -4.30E-01 | 1.54581916E+05 | 1.54582343E+05 | -4.30E-01 | -4.60977750E+01 | -4.56707448E+01 | -4.30E-01 |
| 2139 | Ta | 167 | 73 | 94 | 21 | 8.01886100E+00 | 8.01555966E+00 | 3.30E-03 | 1.55473430E+05 | 1.55473928E+05 | -5.00E-01 | 1.55511157E+05 | 1.55511655E+05 | -5.00E-01 | -4.83510590E+01 | -4.78528116E+01 | -5.00E-01 |
| 2140 | Ta | 168 | 73 | 95 | 22 | 8.01942800E+00 | 8.01676622E+00 | 2.70E-03 | 1.56404881E+05 | 1.56405275E+05 | -3.90E-01 | 1.56442608E+05 | 1.56443002E+05 | -3.90E-01 | -4.83939080E+01 | -4.79997541E+01 | -3.90E-01 |
| 2141 | Ta | 169 | 73 | 96 | 23 | 8.03095700E+00 | 8.02825106E+00 | 2.70E-03 | 1.57334478E+05 | 1.57334883E+05 | -4.00E-01 | 1.57372206E+05 | 1.57372610E+05 | -4.00E-01 | -5.02904300E+01 | -4.98861400E+01 | -4.00E-01 |
| 2142 | Ta | 170 | 73 | 97 | 24 | 8.03029600E+00 | 8.02789544E+00 | 2.40E-03 | 1.58266125E+05 | 1.58266480E+05 | -3.60E-01 | 1.58303853E+05 | 1.58304208E+05 | -3.60E-01 | -5.01376650E+01 | -4.97826151E+01 | -3.60E-01 |
| 2143 | Ta | 171 | 73 | 98 | 25 | 8.03979100E+00 | 8.03729119E+00 | 2.50E-03 | 1.59196037E+05 | 1.59196411E+05 | -3.70E-01 | 1.59233764E+05 | 1.59234139E+05 | -3.70E-01 | -5.17202730E+01 | -5.13458658E+01 | -3.70E-01 |
| 2144 | Ta | 172 | 73 | 99 | 26 | 8.03770500E+00 | 8.03526787E+00 | 2.40E-03 | 1.60127921E+05 | 1.60128287E+05 | -3.70E-01 | 1.60165649E+05 | 1.60166015E+05 | -3.70E-01 | -5.13299770E+01 | -5.09638258E+01 | -3.70E-01 |
| 2145 | Ta | 173 | 73 | 100 | 27 | 8.04406400E+00 | 8.04249927E+00 | 1.60E-03 | 1.61058349E+05 | 1.61058566E+05 | -2.20E-01 | 1.61096076E+05 | 1.61096294E+05 | -2.20E-01 | -5.23965380E+01 | -5.21788080E+01 | -2.20E-01 |
| 2146 | Ta | 174 | 73 | 101 | 28 | 8.04045200E+00 | 8.03872675E+00 | 1.70E-03 | 1.61990498E+05 | 1.61990746E+05 | -2.50E-01 | 1.62028226E+05 | 1.62028473E+05 | -2.50E-01 | -5.17407660E+01 | -5.14935698E+01 | -2.50E-01 |
| 2147 | Ta | 175 | 73 | 102 | 29 | 8.04444500E+00 | 8.04374252E+00 | 7.00E-04 | 1.62921325E+05 | 1.62921394E+05 | -7.00E-02 | 1.62959052E+05 | 1.62959122E+05 | -7.00E-02 | -5.24086470E+01 | -5.23387362E+01 | -7.00E-02 |
| 2148 | Ta | 176 | 73 | 103 | 30 | 8.03867000E+00 | 8.03816153E+00 | 5.10E-04 | 1.63853862E+05 | 1.63853898E+05 | -3.60E-02 | 1.63891589E+05 | 1.63891626E+05 | -3.60E-02 | -5.13653740E+01 | -5.13289062E+01 | -3.60E-02 |
| 2149 | Ta | 177 | 73 | 104 | 31 | 8.04084100E+00 | 8.04089029E+00 | -4.90E-05 | 1.64785004E+05 | 1.64784943E+05 | 6.20E-02 | 1.64822732E+05 | 1.64822670E+05 | 6.20E-02 | -5.17170410E+01 | -5.17787392E+01 | 6.20E-02 |
| 2150 | Ta | 178 | 73 | 105 | 32 | 8.03500000E+00 | 8.03348212E+00 | 1.50E-03 | 1.65717615E+05 | 1.65717786E+05 | -1.70E-01 | 1.65755342E+05 | 1.65755513E+05 | -1.70E-01 | -5.06010000E+01 | -5.04296561E+01 | -1.70E-01 |
| 2151 | Ta | 179 | 73 | 106 | 33 | 8.03359900E+00 | 8.03400314E+00 | -4.00E-04 | 1.66649350E+05 | 1.66649224E+05 | 1.30E-01 | 1.66687077E+05 | 1.66686952E+05 | 1.30E-01 | -5.03597740E+01 | -5.04850829E+01 | 1.30E-01 |
| 2152 | Ta | 180 | 73 | 107 | 34 | 8.02590000E+00 | 8.02470974E+00 | 1.20E-03 | 1.67582267E+05 | 1.67582429E+05 | -1.60E-01 | 1.67619995E+05 | 1.67620156E+05 | -1.60E-01 | -4.89361930E+01 | -4.87749548E+01 | -1.60E-01 |
| 2153 | Ta | 181 | 73 | 108 | 35 | 8.02341800E+00 | 8.02296202E+00 | 4.60E-04 | 1.68514256E+05 | 1.68514286E+05 | -3.00E-02 | 1.68551983E+05 | 1.68552013E+05 | -3.00E-02 | -4.84416290E+01 | -4.84120084E+01 | -3.00E-02 |
| 2154 | Ta | 182 | 73 | 109 | 36 | 8.01264700E+00 | 8.01178359E+00 | 8.60E-04 | 1.69447758E+05 | 1.69447862E+05 | -1.00E-01 | 1.69485486E+05 | 1.69485590E+05 | -1.00E-01 | -4.64332490E+01 | -4.63291762E+01 | -1.00E-01 |
| 2155 | Ta | 183 | 73 | 110 | 37 | 8.00675300E+00 | 8.00779574E+00 | -1.00E-03 | 1.70380390E+05 | 1.70380146E+05 | 2.40E-01 | 1.70418117E+05 | 1.70417873E+05 | 2.40E-01 | -4.52961120E+01 | -4.55398649E+01 | 2.40E-01 |

| | | | | | | | | | | | | | | | | |
|---|---|---|---|---|---|---|---|---|---|---|---|---|---|---|---|---|
| 2156 | Ta | 184 | 73 | 111 | 38 | 7.99376400E+00 | 7.99475922E+00 | -1.00E-03 | 1.71314338E+05 | 1.71314102E+05 | 2.40E-01 | 1.71352066E+05 | 1.71351830E+05 | 2.40E-01 | -4.28415570E+01 | -4.30776206E+01 | 2.40E-01 |
| 2157 | Ta | 185 | 73 | 112 | 39 | 7.98637200E+00 | 7.98859636E+00 | -2.20E-03 | 1.72247277E+05 | 1.72246813E+05 | 4.60E-01 | 1.72285005E+05 | 1.72284540E+05 | 4.60E-01 | -4.13964460E+01 | -4.18609330E+01 | 4.60E-01 |
| 2158 | Ta | 186 | 73 | 113 | 40 | 7.97184700E+00 | 7.97377641E+00 | -1.90E-03 | 1.73181558E+05 | 1.73181146E+05 | 4.10E-01 | 1.73219286E+05 | 1.73218874E+05 | 4.10E-01 | -3.86098050E+01 | -3.90216994E+01 | 4.10E-01 |
| 2159 | TA | 187 | 73 | 114 | 41 | 7.96323600E+00 | 7.96556353E+00 | -2.30E-03 | 1.74114762E+05 | 1.74114274E+05 | 4.90E-01 | 1.74152489E+05 | 1.74152001E+05 | 4.90E-01 | -3.69002030E+01 | -3.73883489E+01 | 4.90E-01 |
| 2160 | Ta | 188 | 73 | 115 | 42 | 7.94632100E+00 | 7.94910762E+00 | -2.80E-03 | 1.75049544E+05 | 1.75048967E+05 | 5.80E-01 | 1.75087271E+05 | 1.75086695E+05 | 5.80E-01 | -3.36120300E+01 | -3.41888822E+01 | 5.80E-01 |
| 2161 | Ta | 189 | 73 | 116 | 43 | 7.93800000E+00 | 7.93905485E+00 | -1.10E-03 | 1.75982821E+05 | 1.75982483E+05 | 3.40E-01 | 1.76020548E+05 | 1.76020211E+05 | 3.40E-01 | -3.18290000E+01 | -3.21666974E+01 | 3.40E-01 |
| 2162 | TA | 190 | 73 | 117 | 44 | 7.92100000E+00 | 7.92120581E+00 | -2.10E-04 | 1.76917631E+05 | 1.76917501E+05 | 1.30E-01 | 1.76955359E+05 | 1.76955228E+05 | 1.30E-01 | -2.85130000E+01 | -2.86431158E+01 | 1.30E-01 |
| 2163 | TA | 191 | 73 | 118 | 45 | 7.91100000E+00 | 7.90962281E+00 | 1.40E-03 | 1.77851147E+05 | 1.77851358E+05 | -2.10E-01 | 1.77888874E+05 | 1.77889085E+05 | -2.10E-01 | -2.64920000E+01 | -2.62806490E+01 | -2.10E-01 |
| 2164 | TA | 192 | 73 | 119 | 46 | 7.89400000E+00 | 7.89071581E+00 | 3.30E-03 | 1.78785975E+05 | 1.78786643E+05 | -6.70E-01 | 1.78823703E+05 | 1.78824371E+05 | -6.70E-01 | -2.31570000E+01 | -2.24888090E+01 | -6.70E-01 |
| 2165 | W | 157 | 74 | 83 | 9 | 7.82800000E+00 | 7.82221950E+00 | 5.80E-03 | 1.46186604E+05 | 1.46187475E+05 | -8.70E-01 | 1.46224857E+05 | 1.46225729E+05 | -8.70E-01 | -1.97100000E+01 | -1.88390768E+01 | -8.70E-01 |
| 2166 | W | 158 | 74 | 84 | 10 | 7.85500000E+00 | 7.85289178E+00 | 2.10E-03 | 1.47114113E+05 | 1.47114372E+05 | -2.60E-01 | 1.47152366E+05 | 1.47152625E+05 | -2.60E-01 | -2.36950000E+01 | -2.34361971E+01 | -2.60E-01 |
| 2167 | W | 159 | 74 | 85 | 11 | 7.86800000E+00 | 7.86491304E+00 | 3.10E-03 | 1.48043815E+05 | 1.48044173E+05 | -3.60E-01 | 1.48082068E+05 | 1.48082427E+05 | -3.60E-01 | -2.54870000E+01 | -2.51291507E+01 | -3.60E-01 |
| 2168 | W | 160 | 74 | 86 | 12 | 7.89308800E+00 | 7.89080607E+00 | 2.30E-03 | 1.48971420E+05 | 1.48971731E+05 | -3.10E-01 | 1.49009673E+05 | 1.49009984E+05 | -3.10E-01 | -2.93769140E+01 | -2.90656290E+01 | -3.10E-01 |
| 2169 | W | 161 | 74 | 87 | 13 | 7.90200000E+00 | 7.89937517E+00 | 2.60E-03 | 1.49901735E+05 | 1.49902026E+05 | -2.90E-01 | 1.49939988E+05 | 1.49940279E+05 | -2.90E-01 | -3.05560000E+01 | -3.02647410E+01 | -2.90E-01 |
| 2170 | W | 162 | 74 | 88 | 14 | 7.92382800E+00 | 7.92185409E+00 | 2.00E-03 | 1.50829784E+05 | 1.50830050E+05 | -2.70E-01 | 1.50868038E+05 | 1.50868303E+05 | -2.70E-01 | -3.40003300E+01 | -3.37343834E+01 | -2.70E-01 |
| 2171 | W | 163 | 74 | 89 | 15 | 7.93030300E+00 | 7.92812723E+00 | 2.20E-03 | 1.51760370E+05 | 1.51760671E+05 | -3.00E-01 | 1.51798624E+05 | 1.51798925E+05 | -3.00E-01 | -3.49083510E+01 | -3.46074401E+01 | -3.00E-01 |
| 2172 | W | 164 | 74 | 90 | 16 | 7.95140400E+00 | 7.94811901E+00 | 3.30E-03 | 1.52688545E+05 | 1.52689030E+05 | -4.90E-01 | 1.52726798E+05 | 1.52727283E+05 | -4.90E-01 | -3.82278780E+01 | -3.77428998E+01 | -4.80E-01 |
| 2173 | W | 165 | 74 | 91 | 17 | 7.95596900E+00 | 7.95272220E+00 | 3.20E-03 | 1.53619406E+05 | 1.53619888E+05 | -4.80E-01 | 1.53657659E+05 | 1.53658141E+05 | -4.80E-01 | -3.88611950E+01 | -3.83792264E+01 | -4.80E-01 |
| 2174 | W | 166 | 74 | 92 | 18 | 7.97489900E+00 | 7.97064011E+00 | 4.30E-03 | 1.54547873E+05 | 1.54548526E+05 | -6.50E-01 | 1.54586126E+05 | 1.54586779E+05 | -6.50E-01 | -4.18881580E+01 | -4.12350018E+01 | -6.50E-01 |
| 2175 | W | 167 | 74 | 93 | 19 | 7.97674000E+00 | 7.97377335E+00 | 3.00E-03 | 1.55479156E+05 | 1.55479598E+05 | -4.40E-01 | 1.55517409E+05 | 1.55517851E+05 | -4.40E-01 | -4.20992810E+01 | -4.16575744E+01 | -4.40E-01 |
| 2176 | W | 168 | 74 | 94 | 20 | 7.99393100E+00 | 7.98971066E+00 | 4.20E-03 | 1.56407856E+05 | 1.56408512E+05 | -6.60E-01 | 1.56446109E+05 | 1.56446765E+05 | -6.60E-01 | -4.48928530E+01 | -4.42374969E+01 | -6.60E-01 |
| 2177 | W | 169 | 74 | 95 | 21 | 7.99453700E+00 | 7.99135864E+00 | 3.20E-03 | 1.57339325E+05 | 1.57339809E+05 | -4.80E-01 | 1.57377579E+05 | 1.57378062E+05 | -4.80E-01 | -4.49177160E+01 | -4.44343971E+01 | -4.80E-01 |
| 2178 | W | 170 | 74 | 96 | 22 | 8.00894600E+00 | 8.00527577E+00 | 3.70E-03 | 1.58268447E+05 | 1.58269017E+05 | -5.70E-01 | 1.58306700E+05 | 1.58307270E+05 | -5.70E-01 | -4.72904710E+01 | -4.67203482E+01 | -5.70E-01 |
| 2179 | W | 171 | 74 | 97 | 23 | 8.00811500E+00 | 8.00535376E+00 | 2.80E-03 | 1.59200145E+05 | 1.59200564E+05 | -4.20E-01 | 1.59238398E+05 | 1.59238817E+05 | -4.20E-01 | -4.70860900E+01 | -4.66676424E+01 | -4.20E-01 |
| 2180 | W | 172 | 74 | 98 | 24 | 8.02017500E+00 | 8.01718228E+00 | 3.00E-03 | 1.60129628E+05 | 1.60130089E+05 | -4.60E-01 | 1.60167881E+05 | 1.60168342E+05 | -4.60E-01 | -4.90971860E+01 | -4.86361824E+01 | -4.60E-01 |
| 2181 | W | 173 | 74 | 99 | 25 | 8.01833300E+00 | 8.01560375E+00 | 2.70E-03 | 1.61061492E+05 | 1.61061910E+05 | -4.20E-01 | 1.61099745E+05 | 1.61100164E+05 | -4.20E-01 | -4.87273830E+01 | -4.83089602E+01 | -4.20E-01 |
| 2182 | W | 174 | 74 | 100 | 26 | 8.02725600E+00 | 8.02528650E+00 | 2.00E-03 | 1.61991486E+05 | 1.61991775E+05 | -2.90E-01 | 1.62029739E+05 | 1.62030029E+05 | -2.90E-01 | -5.02270880E+01 | -4.99380428E+01 | -2.90E-01 |
| 2183 | W | 175 | 74 | 101 | 27 | 8.02411200E+00 | 8.02198326E+00 | 2.10E-03 | 1.62923575E+05 | 1.62923894E+05 | -3.20E-01 | 1.62961828E+05 | 1.62962147E+05 | -3.20E-01 | -4.96327950E+01 | -4.93139432E+01 | -3.20E-01 |
| 2184 | W | 176 | 74 | 102 | 28 | 8.03011200E+00 | 8.02948180E+00 | 6.30E-04 | 1.63854060E+05 | 1.63854117E+05 | -5.70E-02 | 1.63892313E+05 | 1.63892370E+05 | -5.70E-02 | -5.06416030E+01 | -5.05843511E+01 | -5.70E-02 |
| 2185 | W | 177 | 74 | 103 | 29 | 8.02503500E+00 | 8.02440355E+00 | 6.30E-04 | 1.64786494E+05 | 1.64786552E+05 | -5.80E-02 | 1.64824747E+05 | 1.64824805E+05 | -5.80E-02 | -4.97017250E+01 | -4.96436631E+01 | -5.80E-02 |
| 2186 | W | 178 | 74 | 104 | 30 | 8.02927100E+00 | 8.02965123E+00 | -3.80E-04 | 1.65717280E+05 | 1.65717159E+05 | 1.20E-01 | 1.65755533E+05 | 1.65755412E+05 | 1.20E-01 | -5.04093780E+01 | -5.05308353E+01 | 1.20E-01 |
| 2187 | W | 179 | 74 | 105 | 31 | 8.02329400E+00 | 8.02278750E+00 | 5.10E-04 | 1.66649886E+05 | 1.66649923E+05 | -3.70E-02 | 1.66688139E+05 | 1.66688176E+05 | -3.70E-02 | -4.92974160E+01 | -4.92605603E+01 | -3.70E-02 |
| 2188 | W | 180 | 74 | 106 | 32 | 8.02545600E+00 | 8.02587992E+00 | -4.20E-04 | 1.67581039E+05 | 1.67580909E+05 | 1.30E-01 | 1.67619292E+05 | 1.67619162E+05 | 1.30E-01 | -4.96385650E+01 | -4.97686633E+01 | 1.30E-01 |
| 2189 | W | 181 | 74 | 107 | 33 | 8.01805600E+00 | 8.01717259E+00 | 8.80E-04 | 1.68513918E+05 | 1.68514025E+05 | -1.10E-01 | 1.68552172E+05 | 1.68552278E+05 | -1.10E-01 | -4.82534540E+01 | -4.81471977E+01 | -1.10E-01 |
| 2190 | W | 182 | 74 | 108 | 34 | 8.01831800E+00 | 8.01804513E+00 | 2.70E-04 | 1.69445418E+05 | 1.69445414E+05 | 4.10E-03 | 1.69483671E+05 | 1.69483667E+05 | 4.10E-03 | -4.82477140E+01 | -4.82518545E+01 | 4.10E-03 |
| 2191 | W | 183 | 74 | 109 | 35 | 8.00833100E+00 | 8.00749549E+00 | 8.40E-04 | 1.70378793E+05 | 1.70378892E+05 | -9.90E-02 | 1.70417046E+05 | 1.70417145E+05 | -9.90E-02 | -4.63672120E+01 | -4.62679966E+01 | -9.90E-02 |
| 2192 | W | 184 | 74 | 110 | 36 | 8.00508800E+00 | 8.00617809E+00 | -1.10E-03 | 1.71310946E+05 | 1.71310692E+05 | 2.50E-01 | 1.71349200E+05 | 1.71348945E+05 | 2.50E-01 | -4.57075570E+01 | -4.59617706E+01 | 2.50E-01 |
| 2193 | W | 185 | 74 | 111 | 37 | 7.99291900E+00 | 7.99381262E+00 | -8.90E-04 | 1.72244758E+05 | 1.72244539E+05 | 2.20E-01 | 1.72283011E+05 | 1.72282792E+05 | 2.20E-01 | -4.33899460E+01 | -4.36090171E+01 | 2.20E-01 |
| 2194 | W | 186 | 74 | 112 | 38 | 7.98861400E+00 | 7.99037108E+00 | -1.80E-03 | 1.73177131E+05 | 1.73176751E+05 | 3.80E-01 | 1.73215385E+05 | 1.73215004E+05 | 3.80E-01 | -4.25108050E+01 | -4.28913855E+01 | 3.80E-01 |
| 2195 | W | 187 | 74 | 113 | 39 | 7.97512800E+00 | 7.97626501E+00 | -1.10E-03 | 1.74111230E+05 | 1.74110964E+05 | 2.70E-01 | 1.74149483E+05 | 1.74149217E+05 | 2.70E-01 | -3.99062820E+01 | -4.01726018E+01 | 2.70E-01 |
| 2196 | W | 188 | 74 | 114 | 40 | 7.96906300E+00 | 7.97082804E+00 | -1.80E-03 | 1.75043960E+05 | 1.75043575E+05 | 3.90E-01 | 1.75082214E+05 | 1.75081828E+05 | 3.90E-01 | -3.86698760E+01 | -3.90553971E+01 | 3.90E-01 |
| 2197 | W | 189 | 74 | 115 | 41 | 7.95345400E+00 | 7.95513485E+00 | -1.70E-03 | 1.75978507E+05 | 1.75978136E+05 | 3.70E-01 | 1.76016760E+05 | 1.76016389E+05 | 3.70E-01 | -3.56175360E+01 | -3.59888940E+01 | 3.70E-01 |
| 2198 | W | 190 | 74 | 116 | 42 | 7.94756600E+00 | 7.94792394E+00 | -3.60E-04 | 1.76911238E+05 | 1.76911116E+05 | 1.20E-01 | 1.76949491E+05 | 1.76949369E+05 | 1.20E-01 | -3.43808940E+01 | -3.45026360E+01 | 1.20E-01 |
| 2199 | W | 192 | 74 | 118 | 44 | 7.92400000E+00 | 7.92224448E+00 | 1.80E-03 | 1.78778957E+05 | 1.78779281E+05 | -3.20E-01 | 1.78817210E+05 | 1.78817534E+05 | -3.20E-01 | -2.96490000E+01 | -2.93253901E+01 | -3.20E-01 |
| 2200 | W | 193 | 74 | 119 | 45 | 7.90800000E+00 | 7.90424245E+00 | 3.80E-03 | 1.79713814E+05 | 1.79714399E+05 | -5.80E-01 | 1.79752067E+05 | 1.79752652E+05 | -5.80E-01 | -2.62870000E+01 | -2.57019234E+01 | -5.90E-01 |
| 2201 | W | 194 | 74 | 120 | 46 | 7.89900000E+00 | 7.89451743E+00 | 4.50E-03 | 1.80647068E+05 | 1.80647947E+05 | -8.80E-01 | 1.80685322E+05 | 1.80686200E+05 | -8.80E-01 | -2.45260000E+01 | -2.36481939E+01 | -8.80E-01 |
| 2202 | RE | 159 | 75 | 84 | 9 | 7.79500000E+00 | 7.79251016E+00 | 2.50E-03 | 1.48054040E+05 | 1.48054376E+05 | -3.40E-01 | 1.48092818E+05 | 1.48093156E+05 | -3.40E-01 | -1.47370000E+01 | -1.44001698E+01 | -3.40E-01 |

| | | | | | | | | | | | | | | | | |
|---|---|---|---|---|---|---|---|---|---|---|---|---|---|---|---|---|
| 2203 | Re | 160 | 75 | 85 | 10 | 7.81000000E+00 | 7.80760894E+00 | 2.40E-03 | 1.48983339E+05 | 1.48983734E+05 | -3.90E-01 | 1.49022118E+05 | 1.49022513E+05 | -3.90E-01 | -1.69310000E+01 | -1.65371656E+01 | -3.90E-01 |
| 2204 | Re | 161 | 75 | 86 | 11 | 7.83662600E+00 | 7.83447569E+00 | 2.20E-03 | 1.49910874E+05 | 1.49911166E+05 | -2.90E-01 | 1.49949653E+05 | 1.49949945E+05 | -2.90E-01 | -2.08906610E+01 | -2.05990016E+01 | -2.90E-01 |
| 2205 | Re | 162 | 75 | 87 | 12 | 7.84800000E+00 | 7.84600655E+00 | 2.00E-03 | 1.50840758E+05 | 1.50841029E+05 | -2.70E-01 | 1.50879537E+05 | 1.50879808E+05 | -2.70E-01 | -2.25010000E+01 | -2.22301573E+01 | -2.70E-01 |
| 2206 | Re | 163 | 75 | 88 | 13 | 7.87089700E+00 | 7.86936866E+00 | 1.50E-03 | 1.51768745E+05 | 1.51768940E+05 | -1.90E-01 | 1.51807524E+05 | 1.51807719E+05 | -1.90E-01 | -2.60074420E+01 | -2.58128694E+01 | -1.90E-01 |
| 2207 | Re | 164 | 75 | 89 | 14 | 7.88136000E+00 | 7.87847465E+00 | 2.90E-03 | 1.52698724E+05 | 1.52699143E+05 | -4.20E-01 | 1.52737503E+05 | 1.52737922E+05 | -4.20E-01 | -2.75229740E+01 | -2.71043008E+01 | -4.20E-01 |
| 2208 | Re | 165 | 75 | 90 | 15 | 7.90142400E+00 | 7.89925417E+00 | 2.20E-03 | 1.53627097E+05 | 1.53627401E+05 | -3.00E-01 | 1.53665876E+05 | 1.53666180E+05 | -3.00E-01 | -3.06436740E+01 | -3.03400784E+01 | -3.00E-01 |
| 2209 | Re | 166 | 75 | 91 | 16 | 7.90997700E+00 | 7.90657437E+00 | 3.40E-03 | 1.54557341E+05 | 1.54557852E+05 | -5.10E-01 | 1.54596120E+05 | 1.54596631E+05 | -5.10E-01 | -3.18935890E+01 | -3.13831665E+01 | -5.10E-01 |
| 2210 | Re | 167 | 75 | 92 | 17 | 7.92900000E+00 | 7.92520332E+00 | 3.80E-03 | 1.55485887E+05 | 1.55486400E+05 | -5.10E-01 | 1.55524666E+05 | 1.55525179E+05 | -5.10E-01 | -3.48430000E+01 | -3.43294571E+01 | -5.10E-01 |
| 2211 | Re | 168 | 75 | 93 | 18 | 7.93512000E+00 | 7.93096713E+00 | 4.20E-03 | 1.56416428E+05 | 1.56417072E+05 | -6.40E-01 | 1.56455207E+05 | 1.56455851E+05 | -6.40E-01 | -3.57948850E+01 | -3.51516610E+01 | -6.40E-01 |
| 2212 | Re | 169 | 75 | 94 | 19 | 7.95139500E+00 | 7.94756706E+00 | 3.80E-03 | 1.57345308E+05 | 1.57345901E+05 | -5.90E-01 | 1.57384087E+05 | 1.57384680E+05 | -5.90E-01 | -3.84091650E+01 | -3.78166974E+01 | -5.90E-01 |
| 2213 | Re | 170 | 75 | 95 | 20 | 7.95509200E+00 | 7.95179141E+00 | 3.30E-03 | 1.58276294E+05 | 1.58276800E+05 | -5.10E-01 | 1.58315073E+05 | 1.58315579E+05 | -5.10E-01 | -3.89177530E+01 | -3.84110854E+01 | -5.10E-01 |
| 2214 | Re | 171 | 75 | 96 | 21 | 7.96941200E+00 | 7.96634989E+00 | 3.10E-03 | 1.59205455E+05 | 1.59205924E+05 | -4.70E-01 | 1.59244234E+05 | 1.59244703E+05 | -4.70E-01 | -4.12502800E+01 | -4.07810568E+01 | -4.70E-01 |
| 2215 | Re | 172 | 75 | 97 | 22 | 7.97161000E+00 | 7.96897798E+00 | 2.60E-03 | 1.60136673E+05 | 1.60137071E+05 | -4.00E-01 | 1.60175452E+05 | 1.60175850E+05 | -4.00E-01 | -4.15263140E+01 | -4.11281195E+01 | -4.00E-01 |
| 2216 | Re | 173 | 75 | 98 | 23 | 7.98390600E+00 | 7.98144839E+00 | 2.50E-03 | 1.61066140E+05 | 1.61066510E+05 | -3.70E-01 | 1.61104919E+05 | 1.61105289E+05 | -3.70E-01 | -4.35538650E+01 | -4.31831588E+01 | -3.70E-01 |
| 2217 | Re | 174 | 75 | 99 | 24 | 7.98509400E+00 | 7.98241538E+00 | 2.70E-03 | 1.61997514E+05 | 1.61997926E+05 | -4.10E-01 | 1.62036293E+05 | 1.62036705E+05 | -4.10E-01 | -4.36730960E+01 | -4.32615450E+01 | -4.10E-01 |
| 2218 | Re | 175 | 75 | 100 | 25 | 7.99481600E+00 | 7.99275666E+00 | 2.10E-03 | 1.62927393E+05 | 1.62927699E+05 | -3.10E-01 | 1.62966172E+05 | 1.62966478E+05 | -3.10E-01 | -4.52883070E+01 | -4.49823650E+01 | -3.10E-01 |
| 2219 | Re | 176 | 75 | 101 | 26 | 7.99397000E+00 | 7.99201050E+00 | 2.00E-03 | 1.63859113E+05 | 1.63859403E+05 | -2.90E-01 | 1.63897892E+05 | 1.63898182E+05 | -2.90E-01 | -4.50628850E+01 | -4.47724793E+01 | -2.90E-01 |
| 2220 | Re | 177 | 75 | 102 | 27 | 8.00122200E+00 | 8.00019565E+00 | 1.00E-03 | 1.64789401E+05 | 1.64789528E+05 | -1.30E-01 | 1.64828180E+05 | 1.64828307E+05 | -1.30E-01 | -4.62691700E+01 | -4.61419414E+01 | -1.30E-01 |
| 2221 | Re | 178 | 75 | 103 | 28 | 7.99815700E+00 | 7.99769782E+00 | 4.60E-04 | 1.65721510E+05 | 1.65721538E+05 | -2.70E-02 | 1.65760289E+05 | 1.65760317E+05 | -2.70E-02 | -4.56534530E+01 | -4.56262047E+01 | -2.70E-02 |
| 2222 | Re | 179 | 75 | 104 | 29 | 8.00376900E+00 | 8.00366345E+00 | 1.10E-04 | 1.66652073E+05 | 1.66652037E+05 | 3.60E-02 | 1.66690852E+05 | 1.66690816E+05 | 3.60E-02 | -4.65848250E+01 | -4.66204310E+01 | 3.60E-02 |
| 2223 | Re | 180 | 75 | 105 | 30 | 7.99999100E+00 | 7.99941491E+00 | 5.80E-04 | 1.67584315E+05 | 1.67584364E+05 | -4.90E-02 | 1.67623094E+05 | 1.67623143E+05 | -4.90E-02 | -4.58373590E+01 | -4.57880385E+01 | -4.90E-02 |
| 2224 | Re | 181 | 75 | 106 | 31 | 8.00416200E+00 | 8.00327361E+00 | 8.90E-04 | 1.68515125E+05 | 1.68515231E+05 | -1.10E-01 | 1.68553904E+05 | 1.68554010E+05 | -1.10E-01 | -4.65209780E+01 | -4.64145593E+01 | -1.10E-01 |
| 2225 | Re | 182 | 75 | 107 | 32 | 7.99863400E+00 | 7.99721915E+00 | 1.40E-03 | 1.69447692E+05 | 1.69447895E+05 | -2.00E-01 | 1.69486471E+05 | 1.69486674E+05 | -2.00E-01 | -4.54477140E+01 | -4.52446030E+01 | -2.00E-01 |
| 2226 | Re | 183 | 75 | 108 | 33 | 8.00101800E+00 | 7.99890239E+00 | 2.10E-03 | 1.70378823E+05 | 1.70379156E+05 | -3.30E-01 | 1.70417602E+05 | 1.70417935E+05 | -3.30E-01 | -4.58112120E+01 | -4.54785350E+01 | -3.30E-01 |
| 2227 | Re | 184 | 75 | 109 | 34 | 7.99277800E+00 | 7.99104851E+00 | 1.70E-03 | 1.71311903E+05 | 1.71312167E+05 | -2.60E-01 | 1.71350682E+05 | 1.71350946E+05 | -2.60E-01 | -4.42247180E+01 | -4.39610048E+01 | -2.60E-01 |
| 2228 | Re | 185 | 75 | 110 | 35 | 7.99102900E+00 | 7.99058929E+00 | 4.40E-04 | 1.72243800E+05 | 1.72243826E+05 | -2.70E-02 | 1.72282579E+05 | 1.72282606E+05 | -2.70E-02 | -4.38226140E+01 | -4.37957778E+01 | -2.70E-02 |
| 2229 | Re | 186 | 75 | 111 | 36 | 7.98128800E+00 | 7.98096790E+00 | 3.20E-04 | 1.73177186E+05 | 1.73177191E+05 | -5.20E-03 | 1.73215965E+05 | 1.73215970E+05 | -5.20E-03 | -4.19306480E+01 | -4.19254706E+01 | -5.20E-03 |
| 2230 | Re | 187 | 75 | 112 | 37 | 7.97796200E+00 | 7.97843692E+00 | -4.70E-04 | 1.74109392E+05 | 1.74109249E+05 | 1.40E-01 | 1.74148171E+05 | 1.74148028E+05 | 1.40E-01 | -4.12185430E+01 | -4.13618259E+01 | 1.40E-01 |
| 2231 | Re | 188 | 75 | 113 | 38 | 7.96675800E+00 | 7.96713247E+00 | -3.70E-04 | 1.75043085E+05 | 1.75042961E+05 | 1.20E-01 | 1.75081865E+05 | 1.75081740E+05 | 1.20E-01 | -3.90188760E+01 | -3.91437076E+01 | 1.20E-01 |
| 2232 | Re | 189 | 75 | 114 | 39 | 7.96181800E+00 | 7.96266931E+00 | -8.50E-04 | 1.75975618E+05 | 1.75975402E+05 | 2.20E-01 | 1.76014397E+05 | 1.76014182E+05 | 2.20E-01 | -3.79807180E+01 | -3.81959830E+01 | 2.20E-01 |
| 2233 | Re | 190 | 75 | 115 | 40 | 7.95004900E+00 | 7.94985263E+00 | 2.00E-04 | 1.76909458E+05 | 1.76909440E+05 | 1.70E-02 | 1.76948237E+05 | 1.76948219E+05 | 1.70E-02 | -3.56349920E+01 | -3.56521649E+01 | 1.70E-02 |
| 2234 | Re | 191 | 75 | 116 | 41 | 7.94396700E+00 | 7.94369890E+00 | 2.70E-04 | 1.77842235E+05 | 1.77842231E+05 | 3.40E-03 | 1.77881014E+05 | 1.77881010E+05 | 3.30E-03 | -3.43520540E+01 | -3.43553365E+01 | 3.30E-03 |
| 2235 | Re | 192 | 75 | 117 | 42 | 7.93023800E+00 | 7.92965308E+00 | 5.80E-04 | 1.78776492E+05 | 1.78776550E+05 | -5.80E-02 | 1.78815271E+05 | 1.78815329E+05 | -5.80E-02 | -3.15888250E+01 | -3.15309186E+01 | -5.80E-02 |
| 2236 | Re | 193 | 75 | 118 | 43 | 7.92395600E+00 | 7.92215952E+00 | 1.80E-03 | 1.79709339E+05 | 1.79709632E+05 | -2.90E-01 | 1.79748118E+05 | 1.79748411E+05 | -2.90E-01 | -3.02353640E+01 | -2.99429953E+01 | -2.90E-01 |
| 2237 | Re | 194 | 75 | 119 | 44 | 7.90900000E+00 | 7.90725054E+00 | 1.70E-03 | 1.80643832E+05 | 1.80644167E+05 | -3.40E-01 | 1.80682611E+05 | 1.80682946E+05 | -3.40E-01 | -2.72370000E+01 | -2.69014928E+01 | -3.40E-01 |
| 2238 | RE | 195 | 75 | 120 | 45 | 7.90200000E+00 | 7.89878728E+00 | 3.20E-03 | 1.81576984E+05 | 1.81577476E+05 | -4.90E-01 | 1.81615763E+05 | 1.81616255E+05 | -4.90E-01 | -2.55790000E+01 | -2.50870893E+01 | -4.90E-01 |
| 2239 | Re | 196 | 75 | 121 | 46 | 7.88700000E+00 | 7.88328761E+00 | 3.70E-03 | 1.82511515E+05 | 1.82512180E+05 | -6.70E-01 | 1.82550294E+05 | 1.82550959E+05 | -6.70E-01 | -2.25420000E+01 | -2.18766223E+01 | -6.70E-01 |
| 2240 | Re | 197 | 75 | 122 | 47 | 7.87800000E+00 | 7.87396286E+00 | 4.00E-03 | 1.83445049E+05 | 1.83445699E+05 | -6.50E-01 | 1.83483828E+05 | 1.83484478E+05 | -6.50E-01 | -2.05020000E+01 | -1.98516152E+01 | -6.50E-01 |
| 2241 | Re | 198 | 75 | 123 | 48 | 7.86200000E+00 | 7.85767537E+00 | 4.30E-03 | 1.84379905E+05 | 1.84380616E+05 | -7.10E-01 | 1.84418685E+05 | 1.84419395E+05 | -7.10E-01 | -1.71390000E+01 | -1.64293360E+01 | -7.10E-01 |
| 2242 | OS | 161 | 76 | 85 | 9 | 7.76500000E+00 | 7.76250516E+00 | 2.50E-03 | 1.49921019E+05 | 1.49921444E+05 | -4.20E-01 | 1.49960324E+05 | 1.49960749E+05 | -4.20E-01 | -1.02200000E+01 | -9.79482366E+00 | -4.30E-01 |
| 2243 | Os | 162 | 76 | 86 | 10 | 7.79400000E+00 | 7.79221693E+00 | 1.80E-03 | 1.50848230E+05 | 1.50848433E+05 | -2.00E-01 | 1.50887535E+05 | 1.50887739E+05 | -2.00E-01 | -1.45030000E+01 | -1.42993163E+01 | -2.00E-01 |
| 2244 | Os | 163 | 76 | 87 | 11 | 7.80700000E+00 | 7.80443119E+00 | 2.60E-03 | 1.51777842E+05 | 1.51778215E+05 | -3.70E-01 | 1.51817147E+05 | 1.51817521E+05 | -3.70E-01 | -1.63850000E+01 | -1.60111389E+01 | -3.70E-01 |
| 2245 | Os | 164 | 76 | 88 | 12 | 7.83359900E+00 | 7.83052881E+00 | 3.10E-03 | 1.52705248E+05 | 1.52705696E+05 | -4.50E-01 | 1.52744553E+05 | 1.52745002E+05 | -4.50E-01 | -2.04726430E+01 | -2.00242607E+01 | -4.50E-01 |
| 2246 | Os | 165 | 76 | 89 | 13 | 7.84300000E+00 | 7.84020306E+00 | 2.80E-03 | 1.53635420E+05 | 1.53635835E+05 | -4.20E-01 | 1.53674725E+05 | 1.53675140E+05 | -4.20E-01 | -2.17950000E+01 | -2.13797213E+01 | -4.20E-01 |
| 2247 | Os | 166 | 76 | 90 | 14 | 7.86636800E+00 | 7.86360655E+00 | 2.80E-03 | 1.54563272E+05 | 1.54563675E+05 | -4.00E-01 | 1.54602577E+05 | 1.54602981E+05 | -4.00E-01 | -2.54367940E+01 | -2.50335850E+01 | -4.00E-01 |
| 2248 | Os | 167 | 76 | 91 | 15 | 7.87397500E+00 | 7.87138924E+00 | 2.60E-03 | 1.55493701E+05 | 1.55494077E+05 | -3.80E-01 | 1.55533006E+05 | 1.55533383E+05 | -3.80E-01 | -2.65021510E+01 | -2.61255815E+01 | -3.80E-01 |
| 2249 | Os | 168 | 76 | 92 | 16 | 7.89589200E+00 | 7.89255038E+00 | 3.30E-03 | 1.56421710E+05 | 1.56422216E+05 | -5.10E-01 | 1.56461015E+05 | 1.56461522E+05 | -5.10E-01 | -2.99868870E+01 | -2.94807231E+01 | -5.10E-01 |
| 2250 | Os | 169 | 76 | 93 | 17 | 7.90128500E+00 | 7.89869566E+00 | 2.60E-03 | 1.57352468E+05 | 1.57352850E+05 | -3.80E-01 | 1.57391773E+05 | 1.57392156E+05 | -3.80E-01 | -3.07229310E+01 | -3.03405084E+01 | -3.80E-01 |
| 2251 | Os | 170 | 76 | 94 | 18 | 7.92113000E+00 | 7.91776503E+00 | 3.40E-03 | 1.58280759E+05 | 1.58281275E+05 | -5.20E-01 | 1.58320064E+05 | 1.58320581E+05 | -5.20E-01 | -3.39264510E+01 | -3.34096765E+01 | -5.20E-01 |
| 2252 | Os | 171 | 76 | 95 | 19 | 7.92421100E+00 | 7.92231802E+00 | 1.90E-03 | 1.59211876E+05 | 1.59212144E+05 | -2.70E-01 | 1.59251181E+05 | 1.59251450E+05 | -2.70E-01 | -3.43032270E+01 | -3.40346846E+01 | -2.70E-01 |
| 2253 | Os | 172 | 76 | 96 | 20 | 7.94216200E+00 | 7.93931171E+00 | 2.90E-03 | 1.60140429E+05 | 1.60140865E+05 | -4.40E-01 | 1.60179735E+05 | 1.60180170E+05 | -4.40E-01 | -3.72436290E+01 | -3.68085979E+01 | -4.40E-01 |
| 2254 | Os | 173 | 76 | 97 | 21 | 7.94403300E+00 | 7.94224053E+00 | 1.80E-03 | 1.61071729E+05 | 1.61071984E+05 | -2.50E-01 | 1.61111034E+05 | 1.61111289E+05 | -2.50E-01 | -3.74381730E+01 | -3.71832766E+01 | -2.50E-01 |
| 2255 | Os | 174 | 76 | 98 | 22 | 7.95945900E+00 | 7.95713586E+00 | 2.30E-03 | 1.62000666E+05 | 1.62001015E+05 | -3.50E-01 | 1.62039972E+05 | 1.62040321E+05 | -3.50E-01 | -3.99950850E+01 | -3.96459865E+01 | -3.50E-01 |
| 2256 | Os | 175 | 76 | 99 | 23 | 7.96072800E+00 | 7.95839544E+00 | 2.30E-03 | 1.62932050E+05 | 1.62932403E+05 | -3.50E-01 | 1.62971356E+05 | 1.62971708E+05 | -3.50E-01 | -4.01051820E+01 | -3.97522287E+01 | -3.50E-01 |
| 2257 | Os | 176 | 76 | 100 | 24 | 7.97267900E+00 | 7.97117012E+00 | 1.50E-03 | 1.63861551E+05 | 1.63861762E+05 | -2.10E-01 | 1.63900857E+05 | 1.63901067E+05 | -2.10E-01 | -4.20979400E+01 | -4.18876492E+01 | -2.10E-01 |
| 2258 | Os | 177 | 76 | 101 | 25 | 7.97239600E+00 | 7.97072342E+00 | 1.70E-03 | 1.64793194E+05 | 1.64793435E+05 | -2.40E-01 | 1.64832500E+05 | 1.64832740E+05 | -2.40E-01 | -4.19492470E+01 | -4.17084342E+01 | -2.40E-01 |
| 2259 | Os | 178 | 76 | 102 | 26 | 7.98191100E+00 | 7.98136485E+00 | 5.50E-04 | 1.65723093E+05 | 1.65723135E+05 | -4.20E-02 | 1.65762399E+05 | 1.65762441E+05 | -4.20E-02 | -4.35439660E+01 | -4.35020133E+01 | -4.20E-02 |
| 2260 | Os | 179 | 76 | 103 | 27 | 7.97947900E+00 | 7.97918465E+00 | 2.90E-04 | 1.66655112E+05 | 1.66655110E+05 | 2.50E-03 | 1.66694418E+05 | 1.66694415E+05 | 2.50E-03 | -4.30192980E+01 | -4.30218035E+01 | 2.50E-03 |
| 2261 | Os | 180 | 76 | 104 | 28 | 7.98744800E+00 | 7.98763547E+00 | -1.90E-04 | 1.67585264E+05 | 1.67585175E+05 | 8.90E-02 | 1.67624569E+05 | 1.67624480E+05 | 8.90E-02 | -4.43619500E+01 | -4.44508165E+01 | 8.90E-02 |
| 2262 | Os | 181 | 76 | 105 | 29 | 7.98342600E+00 | 7.98373471E+00 | -3.10E-04 | 1.68517570E+05 | 1.68517459E+05 | 1.10E-01 | 1.68556875E+05 | 1.68556764E+05 | 1.10E-01 | -4.35499630E+01 | -4.36610946E+01 | 1.10E-01 |
| 2263 | Os | 182 | 76 | 106 | 30 | 7.98972800E+00 | 7.99012950E+00 | -4.00E-04 | 1.69448005E+05 | 1.69447876E+05 | 1.30E-01 | 1.69487310E+05 | 1.69487182E+05 | 1.30E-01 | -4.46090730E+01 | -4.47373634E+01 | 1.30E-01 |
| 2264 | Os | 183 | 76 | 107 | 31 | 7.98501000E+00 | 7.98445669E+00 | 5.50E-04 | 1.70380444E+05 | 1.70380490E+05 | -4.60E-02 | 1.70419749E+05 | 1.70419795E+05 | -4.60E-02 | -4.36640420E+01 | -4.36180491E+01 | -4.60E-02 |
| 2265 | Os | 184 | 76 | 108 | 32 | 7.98869900E+00 | 7.98872661E+00 | -2.80E-05 | 1.71311345E+05 | 1.71311285E+05 | 6.00E-02 | 1.71350650E+05 | 1.71350590E+05 | 6.00E-02 | -4.42566430E+01 | -4.43168524E+01 | 6.00E-02 |
| 2266 | Os | 185 | 76 | 109 | 33 | 7.98132500E+00 | 7.98129582E+00 | 2.90E-05 | 1.72244286E+05 | 1.72244236E+05 | 5.00E-02 | 1.72283591E+05 | 1.72283542E+05 | 5.00E-02 | -4.28098430E+01 | -4.28595642E+01 | 5.00E-02 |
| 2267 | Os | 186 | 76 | 110 | 34 | 7.98284400E+00 | 7.98348287E+00 | -6.40E-04 | 1.73175588E+05 | 1.73175414E+05 | 1.70E-01 | 1.73214893E+05 | 1.73214719E+05 | 1.70E-01 | -4.30023790E+01 | -4.31763320E+01 | 1.70E-01 |
| 2268 | Os | 187 | 76 | 111 | 35 | 7.97379100E+00 | 7.97433599E+00 | -5.40E-04 | 1.74108863E+05 | 1.74108706E+05 | 1.60E-01 | 1.74148168E+05 | 1.74148011E+05 | 1.60E-01 | -4.12210100E+01 | -4.13780293E+01 | 1.60E-01 |
| 2269 | Os | 188 | 76 | 112 | 36 | 7.97387500E+00 | 7.97452397E+00 | -6.50E-04 | 1.75040439E+05 | 1.75040262E+05 | 1.80E-01 | 1.75079744E+05 | 1.75079567E+05 | 1.80E-01 | -4.11392700E+01 | -4.13163862E+01 | 1.80E-01 |
| 2270 | Os | 189 | 76 | 113 | 37 | 7.96301100E+00 | 7.96376184E+00 | -7.50E-04 | 1.75974084E+05 | 1.75973887E+05 | 2.00E-01 | 1.76013389E+05 | 1.76013192E+05 | 2.00E-01 | -3.89884620E+01 | -3.91855488E+01 | 2.00E-01 |
| 2271 | Os | 190 | 76 | 114 | 38 | 7.96211200E+00 | 7.96211170E+00 | 3.00E-07 | 1.76905857E+05 | 1.76905802E+05 | 5.50E-02 | 1.76945162E+05 | 1.76945107E+05 | 5.50E-02 | -3.87094420E+01 | -3.87644644E+01 | 5.50E-02 |
| 2272 | Os | 191 | 76 | 115 | 39 | 7.95057600E+00 | 7.94993117E+00 | 6.40E-04 | 1.77839663E+05 | 1.77839732E+05 | -6.80E-02 | 1.77878969E+05 | 1.77879037E+05 | -6.80E-02 | -3.63968600E+01 | -3.63287755E+01 | -6.80E-02 |
| 2273 | Os | 192 | 76 | 116 | 40 | 7.94853400E+00 | 7.94671455E+00 | 1.80E-03 | 1.78771670E+05 | 1.78771965E+05 | -2.90E-01 | 1.78810976E+05 | 1.78811270E+05 | -2.90E-01 | -3.58839450E+01 | -3.55897982E+01 | -2.90E-01 |
| 2274 | Os | 193 | 76 | 117 | 41 | 7.93627900E+00 | 7.93342916E+00 | 2.80E-03 | 1.79705652E+05 | 1.79706147E+05 | -4.90E-01 | 1.79744958E+05 | 1.79745453E+05 | -4.90E-01 | -3.33960440E+01 | -3.29011130E+01 | -4.90E-01 |
| 2275 | Os | 194 | 76 | 118 | 42 | 7.93203300E+00 | 7.92902126E+00 | 3.00E-03 | 1.80638105E+05 | 1.80638634E+05 | -5.30E-01 | 1.80677411E+05 | 1.80677940E+05 | -5.30E-01 | -3.24371910E+01 | -3.19080898E+01 | -5.30E-01 |
| 2276 | Os | 195 | 76 | 119 | 43 | 7.91774400E+00 | 7.91500474E+00 | 2.70E-03 | 1.81572525E+05 | 1.81573004E+05 | -4.80E-01 | 1.81611830E+05 | 1.81612309E+05 | -4.80E-01 | -2.95115930E+01 | -2.90325718E+01 | -4.80E-01 |
| 2277 | Os | 196 | 76 | 120 | 44 | 7.91223800E+00 | 7.90975195E+00 | 2.50E-03 | 1.82505252E+05 | 1.82505684E+05 | -4.30E-01 | 1.82544557E+05 | 1.82544989E+05 | -4.30E-01 | -2.82787950E+01 | -2.78467105E+01 | -4.30E-01 |
| 2278 | OS | 197 | 76 | 121 | 45 | 7.89800000E+00 | 7.89520507E+00 | 2.80E-03 | 1.83439716E+05 | 1.83440205E+05 | -4.90E-01 | 1.83479021E+05 | 1.83479511E+05 | -4.90E-01 | -2.53090000E+01 | -2.48194069E+01 | -4.90E-01 |
| 2279 | OS | 198 | 76 | 122 | 46 | 7.89100000E+00 | 7.88908140E+00 | 1.90E-03 | 1.84372682E+05 | 1.84373088E+05 | -4.10E-01 | 1.84411987E+05 | 1.84412393E+05 | -4.10E-01 | -2.38370000E+01 | -2.34308072E+01 | -4.10E-01 |
| 2280 | OS | 199 | 76 | 123 | 47 | 7.87500000E+00 | 7.87360699E+00 | 1.40E-03 | 1.85307529E+05 | 1.85307844E+05 | -3.10E-01 | 1.85346835E+05 | 1.85347149E+05 | -3.10E-01 | -2.04840000E+01 | -2.01691617E+01 | -3.10E-01 |
| 2281 | OS | 200 | 76 | 124 | 48 | 7.86800000E+00 | 7.86579993E+00 | 2.20E-03 | 1.86240728E+05 | 1.86241097E+05 | -3.70E-01 | 1.86280033E+05 | 1.86280402E+05 | -3.70E-01 | -1.87790000E+01 | -1.84100390E+01 | -3.70E-01 |
| 2282 | Os | 201 | 76 | 125 | 49 | 7.85100000E+00 | 7.84816132E+00 | 2.80E-03 | 1.87175762E+05 | 1.87176342E+05 | -5.80E-01 | 1.87215067E+05 | 1.87215647E+05 | -5.80E-01 | -1.52390000E+01 | -1.46591589E+01 | -5.80E-01 |
| 2283 | Os | 202 | 76 | 126 | 50 | 7.84200000E+00 | 7.83719510E+00 | 4.80E-03 | 1.88109408E+05 | 1.88110274E+05 | -8.70E-01 | 1.88148713E+05 | 1.88149579E+05 | -8.70E-01 | -1.30870000E+01 | -1.22208234E+01 | -8.70E-01 |
| 2284 | Ir | 164 | 77 | 87 | 10 | 7.75000000E+00 | 7.74690545E+00 | 3.10E-03 | 1.52717657E+05 | 1.52718101E+05 | -4.40E-01 | 1.52757489E+05 | 1.52757933E+05 | -4.40E-01 | -7.53700000E+00 | -7.09310701E+00 | -4.40E-01 |
| 2285 | Ir | 165 | 77 | 88 | 11 | 7.77700000E+00 | 7.77379541E+00 | 3.20E-03 | 1.53645045E+05 | 1.53645483E+05 | -4.40E-01 | 1.53684877E+05 | 1.53685315E+05 | -4.40E-01 | -1.16430000E+01 | -1.12055375E+01 | -4.40E-01 |
| 2286 | Ir | 166 | 77 | 89 | 12 | 7.78900000E+00 | 7.78638913E+00 | 2.60E-03 | 1.54574828E+05 | 1.54575184E+05 | -3.60E-01 | 1.54614660E+05 | 1.54615016E+05 | -3.60E-01 | -1.33540000E+01 | -1.29985698E+01 | -3.60E-01 |
| 2287 | Ir | 167 | 77 | 90 | 13 | 7.81285600E+00 | 7.81047954E+00 | 2.40E-03 | 1.55502598E+05 | 1.55502939E+05 | -3.40E-01 | 1.55542430E+05 | 1.55542771E+05 | -3.40E-01 | -1.70777170E+01 | -1.67367394E+01 | -3.40E-01 |
| 2288 | Ir | 168 | 77 | 91 | 14 | 7.82415100E+00 | 7.82105838E+00 | 3.10E-03 | 1.56432453E+05 | 1.56432917E+05 | -4.60E-01 | 1.56472285E+05 | 1.56472749E+05 | -4.60E-01 | -1.87168420E+01 | -1.82531455E+01 | -4.60E-01 |
| 2289 | Ir | 169 | 77 | 92 | 15 | 7.84550100E+00 | 7.84281701E+00 | 2.70E-03 | 1.57360587E+05 | 1.57360984E+05 | -4.00E-01 | 1.57400418E+05 | 1.57400816E+05 | -4.00E-01 | -2.20778160E+01 | -2.16800922E+01 | -4.00E-01 |
| 2290 | Ir | 170 | 77 | 93 | 16 | 7.85400000E+00 | 7.85166182E+00 | 2.30E-03 | 1.58290798E+05 | 1.58291203E+05 | -4.00E-01 | 1.58330630E+05 | 1.58331035E+05 | -4.00E-01 | -2.33600000E+01 | -2.29552083E+01 | -4.00E-01 |
| 2291 | Ir | 171 | 77 | 94 | 17 | 7.87351700E+00 | 7.87126416E+00 | 2.30E-03 | 1.59219236E+05 | 1.59219565E+05 | -3.30E-01 | 1.59259068E+05 | 1.59259397E+05 | -3.30E-01 | -2.64169350E+01 | -2.60875508E+01 | -3.30E-01 |
| 2292 | Ir | 172 | 77 | 95 | 18 | 7.88026300E+00 | 7.87845011E+00 | 1.80E-03 | 1.60149767E+05 | 1.60150023E+05 | -2.60E-01 | 1.60189599E+05 | 1.60189855E+05 | -2.60E-01 | -2.73793690E+01 | -2.71234802E+01 | -2.60E-01 |
| 2293 | Ir | 173 | 77 | 96 | 19 | 7.89806600E+00 | 7.89593735E+00 | 2.10E-03 | 1.61078372E+05 | 1.61078685E+05 | -3.10E-01 | 1.61118204E+05 | 1.61118517E+05 | -3.10E-01 | -3.02683050E+01 | -2.99559030E+01 | -3.10E-01 |
| 2294 | Ir | 174 | 77 | 97 | 20 | 7.90251300E+00 | 7.90145979E+00 | 1.10E-03 | 1.62009266E+05 | 1.62009393E+05 | -1.30E-01 | 1.62049098E+05 | 1.62049225E+05 | -1.30E-01 | -3.08687380E+01 | -3.07414265E+01 | -1.30E-01 |
| 2295 | Ir | 175 | 77 | 98 | 21 | 7.91791000E+00 | 7.91683058E+00 | 1.10E-03 | 1.62938234E+05 | 1.62938367E+05 | -1.30E-01 | 1.62978066E+05 | 1.62978199E+05 | -1.30E-01 | -3.33944420E+01 | -3.32614553E+01 | -1.30E-01 |
| 2296 | Ir | 176 | 77 | 99 | 22 | 7.92142400E+00 | 7.92066680E+00 | 7.60E-04 | 1.63869263E+05 | 1.63869341E+05 | -7.70E-02 | 1.63909095E+05 | 1.63909173E+05 | -7.70E-02 | -3.38594440E+01 | -3.37821416E+01 | -7.70E-02 |
| 2297 | Ir | 177 | 77 | 100 | 23 | 7.93463200E+00 | 7.93391574E+00 | 7.20E-04 | 1.64798569E+05 | 1.64798640E+05 | -7.10E-02 | 1.64838401E+05 | 1.64838472E+05 | -7.10E-02 | -3.60474210E+01 | -3.59765515E+01 | -7.10E-02 |
| 2298 | Ir | 178 | 77 | 101 | 24 | 7.93654900E+00 | 7.93604641E+00 | 5.00E-04 | 1.65729859E+05 | 1.65729892E+05 | -3.30E-02 | 1.65769691E+05 | 1.65769724E+05 | -3.30E-02 | -3.62518840E+01 | -3.62184088E+01 | -3.30E-02 |
| 2299 | Ir | 179 | 77 | 102 | 25 | 7.94751100E+00 | 7.94717404E+00 | 3.40E-04 | 1.66659526E+05 | 1.66659530E+05 | -4.30E-03 | 1.66699358E+05 | 1.66699362E+05 | -4.30E-03 | -3.80793080E+01 | -3.80749804E+01 | -4.30E-03 |
| 2300 | Ir | 180 | 77 | 103 | 26 | 7.94763300E+00 | 7.94758587E+00 | 4.70E-05 | 1.67591121E+05 | 1.67591074E+05 | 4.70E-02 | 1.67630953E+05 | 1.67630906E+05 | 4.70E-02 | -3.79775260E+01 | -3.80249664E+01 | 4.70E-02 |
| 2301 | Ir | 181 | 77 | 104 | 27 | 7.95657100E+00 | 7.95653956E+00 | 3.10E-05 | 1.68521122E+05 | 1.68521071E+05 | 5.00E-02 | 1.68560953E+05 | 1.68560903E+05 | 5.00E-02 | -3.94716330E+01 | -3.95218507E+01 | 5.00E-02 |
| 2302 | Ir | 182 | 77 | 105 | 28 | 7.95489400E+00 | 7.95526060E+00 | -3.70E-04 | 1.69453035E+05 | 1.69452913E+05 | 1.20E-01 | 1.69492867E+05 | 1.69492745E+05 | 1.20E-01 | -3.90516780E+01 | -3.91743002E+01 | 1.20E-01 |
| 2303 | Ir | 183 | 77 | 106 | 29 | 7.96182300E+00 | 7.96219858E+00 | -3.80E-04 | 1.70383378E+05 | 1.70383253E+05 | 1.20E-01 | 1.70423210E+05 | 1.70423085E+05 | 1.20E-01 | -4.02033110E+01 | -4.03278921E+01 | 1.20E-01 |
| 2304 | Ir | 184 | 77 | 107 | 30 | 7.95919800E+00 | 7.95918384E+00 | 1.40E-05 | 1.71315464E+05 | 1.71315411E+05 | 5.30E-02 | 1.71355296E+05 | 1.71355243E+05 | 5.30E-02 | -3.96108510E+01 | -3.96640588E+01 | 5.30E-02 |
| 2305 | Ir | 185 | 77 | 108 | 31 | 7.96372200E+00 | 7.96403718E+00 | -3.20E-04 | 1.72246234E+05 | 1.72246120E+05 | 1.10E-01 | 1.72286066E+05 | 1.72285951E+05 | 1.10E-01 | -4.03355530E+01 | -4.04497915E+01 | 1.10E-01 |
| 2306 | Ir | 186 | 77 | 109 | 32 | 7.95806000E+00 | 7.95931225E+00 | -1.30E-03 | 1.73178889E+05 | 1.73178600E+05 | 2.90E-01 | 1.73218721E+05 | 1.73218432E+05 | 2.90E-01 | -3.91747690E+01 | -3.94636742E+01 | 2.90E-01 |
| 2307 | Ir | 187 | 77 | 110 | 33 | 7.96066800E+00 | 7.96213485E+00 | -1.50E-03 | 1.74110008E+05 | 1.74109678E+05 | 3.30E-01 | 1.74149840E+05 | 1.74149510E+05 | 3.30E-01 | -3.95493720E+01 | -3.98794930E+01 | 3.30E-01 |
| 2308 | Ir | 188 | 77 | 111 | 34 | 7.95488500E+00 | 7.95575719E+00 | -8.70E-04 | 1.75042700E+05 | 1.75042480E+05 | 2.20E-01 | 1.75082532E+05 | 1.75082312E+05 | 2.20E-01 | -3.83513820E+01 | -3.85713084E+01 | 2.20E-01 |
| 2309 | Ir | 189 | 77 | 112 | 35 | 7.95605800E+00 | 7.95665104E+00 | -5.90E-04 | 1.75974089E+05 | 1.75973921E+05 | 1.70E-01 | 1.76013921E+05 | 1.76013753E+05 | 1.70E-01 | -3.84567030E+01 | -3.86246844E+01 | 1.70E-01 |
| 2310 | Ir | 190 | 77 | 113 | 36 | 7.94771200E+00 | 7.94874503E+00 | -1.00E-03 | 1.76907284E+05 | 1.76907032E+05 | 2.50E-01 | 1.76947116E+05 | 1.76946864E+05 | 2.50E-01 | -3.67556210E+01 | -3.70078749E+01 | 2.50E-01 |
| 2311 | Ir | 191 | 77 | 114 | 37 | 7.94812400E+00 | 7.94789865E+00 | 2.30E-04 | 1.77838823E+05 | 1.77838810E+05 | 1.30E-02 | 1.77878655E+05 | 1.77878642E+05 | 1.30E-02 | -3.67108370E+01 | -3.67236428E+01 | 1.30E-02 |
| 2312 | Ir | 192 | 77 | 115 | 38 | 7.93901000E+00 | 7.93869344E+00 | 3.20E-04 | 1.78772190E+05 | 1.78772195E+05 | -4.90E-03 | 1.78812022E+05 | 1.78812027E+05 | -4.90E-03 | -3.48376450E+01 | -3.48328217E+01 | -4.80E-03 |
| 2313 | Ir | 193 | 77 | 116 | 39 | 7.93814400E+00 | 7.93641149E+00 | 1.70E-03 | 1.79703984E+05 | 1.79704262E+05 | -2.80E-01 | 1.79743815E+05 | 1.79744094E+05 | -2.80E-01 | -3.45383190E+01 | -3.42597800E+01 | -2.80E-01 |
| 2314 | Ir | 194 | 77 | 117 | 40 | 7.92849800E+00 | 7.92625021E+00 | 2.20E-03 | 1.80637482E+05 | 1.80637862E+05 | -3.80E-01 | 1.80677314E+05 | 1.80677694E+05 | -3.80E-01 | -3.25337910E+01 | -3.21535836E+01 | -3.80E-01 |
| 2315 | Ir | 195 | 77 | 118 | 41 | 7.92492600E+00 | 7.92292407E+00 | 2.00E-03 | 1.81569815E+05 | 1.81570150E+05 | -3.30E-01 | 1.81609647E+05 | 1.81609982E+05 | -3.30E-01 | -3.16943340E+01 | -3.13599176E+01 | -3.30E-01 |
| 2316 | Ir | 196 | 77 | 119 | 42 | 7.91416000E+00 | 7.91217193E+00 | 2.00E-03 | 1.82503566E+05 | 1.82503900E+05 | -3.30E-01 | 1.82543398E+05 | 1.82543732E+05 | -3.30E-01 | -2.94379000E+01 | -2.91041030E+01 | -3.30E-01 |
| 2317 | Ir | 197 | 77 | 120 | 43 | 7.90900800E+00 | 7.90809875E+00 | 9.10E-04 | 1.83436232E+05 | 1.83436355E+05 | -1.20E-01 | 1.83476064E+05 | 1.83476187E+05 | -1.20E-01 | -2.82657980E+01 | -2.81425390E+01 | -1.20E-01 |
| 2318 | Ir | 198 | 77 | 121 | 44 | 7.89700000E+00 | 7.89685314E+00 | 1.50E-04 | 1.84370171E+05 | 1.84370239E+05 | -6.80E-02 | 1.84410003E+05 | 1.84410071E+05 | -6.80E-02 | -2.58210000E+01 | -2.57526882E+01 | -6.80E-02 |
| 2319 | Ir | 199 | 77 | 122 | 45 | 7.89121400E+00 | 7.89184038E+00 | -6.30E-04 | 1.85303086E+05 | 1.85302905E+05 | 1.80E-01 | 1.85342918E+05 | 1.85342737E+05 | 1.80E-01 | -2.44002040E+01 | -2.45806829E+01 | 1.80E-01 |
| 2320 | IR | 200 | 77 | 123 | 46 | 7.87800000E+00 | 7.87948089E+00 | -1.50E-03 | 1.86237370E+05 | 1.86237051E+05 | 3.20E-01 | 1.86277201E+05 | 1.86276883E+05 | 3.20E-01 | -2.16110000E+01 | -2.19293063E+01 | 3.20E-01 |
| 2321 | IR | 201 | 77 | 124 | 47 | 7.87100000E+00 | 7.87246064E+00 | -1.50E-03 | 1.87170578E+05 | 1.87170148E+05 | 4.30E-01 | 1.87210410E+05 | 1.87209980E+05 | 4.30E-01 | -1.98970000E+01 | -2.03263984E+01 | 4.30E-01 |
| 2322 | IR | 202 | 77 | 125 | 48 | 7.85600000E+00 | 7.85751002E+00 | -1.50E-03 | 1.88105192E+05 | 1.88104861E+05 | 3.30E-01 | 1.88145024E+05 | 1.88144693E+05 | 3.30E-01 | -1.67760000E+01 | -1.71075146E+01 | 3.30E-01 |
| 2323 | Ir | 203 | 77 | 126 | 49 | 7.84700000E+00 | 7.84683683E+00 | 1.60E-04 | 1.89038773E+05 | 1.89038735E+05 | 3.70E-02 | 1.89078605E+05 | 1.89078567E+05 | 3.70E-02 | -1.46900000E+01 | -1.47270492E+01 | 3.70E-02 |
| 2324 | Ir | 204 | 77 | 127 | 50 | 7.82400000E+00 | 7.82778678E+00 | -3.80E-03 | 1.89975269E+05 | 1.89974340E+05 | 9.30E-01 | 1.90015101E+05 | 1.90014172E+05 | 9.30E-01 | -9.68800000E+00 | -1.06163563E+01 | 9.30E-01 |
| 2325 | Pt | 166 | 78 | 88 | 10 | 7.73300000E+00 | 7.73126876E+00 | 1.70E-03 | 1.54582863E+05 | 1.54583024E+05 | -1.60E-01 | 1.54623222E+05 | 1.54623382E+05 | -1.60E-01 | -4.79200000E+00 | -4.63166649E+00 | -1.60E-01 |
| 2326 | Pt | 167 | 78 | 89 | 11 | 7.74700000E+00 | 7.74436719E+00 | 2.60E-03 | 1.55512345E+05 | 1.55512670E+05 | -3.30E-01 | 1.55552704E+05 | 1.55553029E+05 | -3.30E-01 | -6.80500000E+00 | -6.47905455E+00 | -3.30E-01 |
| 2327 | Pt | 168 | 78 | 90 | 12 | 7.77390700E+00 | 7.77111432E+00 | 2.80E-03 | 1.56439585E+05 | 1.56439998E+05 | -4.10E-01 | 1.56479944E+05 | 1.56480357E+05 | -4.10E-01 | -1.10580770E+01 | -1.06456206E+01 | -4.10E-01 |
| 2328 | Pt | 169 | 78 | 91 | 13 | 7.78400000E+00 | 7.78208209E+00 | 1.90E-03 | 1.57369625E+05 | 1.57369938E+05 | -3.10E-01 | 1.57409984E+05 | 1.57410297E+05 | -3.10E-01 | -1.25130000E+01 | -1.21989690E+01 | -3.10E-01 |
| 2329 | Pt | 170 | 78 | 92 | 14 | 7.80826700E+00 | 7.80640611E+00 | 1.90E-03 | 1.58297327E+05 | 1.58297587E+05 | -2.60E-01 | 1.58337686E+05 | 1.58337946E+05 | -2.60E-01 | -1.63045500E+01 | -1.60448145E+01 | -2.60E-01 |
| 2330 | Pt | 171 | 78 | 93 | 15 | 7.81662100E+00 | 7.81554587E+00 | 1.10E-03 | 1.59227655E+05 | 1.59227783E+05 | -1.30E-01 | 1.59268014E+05 | 1.59268142E+05 | -1.30E-01 | -1.74699390E+01 | -1.73428016E+01 | -1.30E-01 |
| 2331 | Pt | 172 | 78 | 94 | 16 | 7.83919200E+00 | 7.83764570E+00 | 1.50E-03 | 1.60155522E+05 | 1.60155731E+05 | -2.10E-01 | 1.60195881E+05 | 1.60196090E+05 | -2.10E-01 | -2.10974800E+01 | -2.08881991E+01 | -2.10E-01 |
| 2332 | Pt | 173 | 78 | 95 | 17 | 7.84542200E+00 | 7.84505793E+00 | 3.60E-04 | 1.61086171E+05 | 1.61086177E+05 | -6.30E-03 | 1.61126529E+05 | 1.61126536E+05 | -6.30E-03 | -2.19431500E+01 | -2.19368404E+01 | -6.30E-03 |
| 2333 | Pt | 174 | 78 | 96 | 18 | 7.86611800E+00 | 7.86499946E+00 | 1.10E-03 | 1.62014289E+05 | 1.62014427E+05 | -1.40E-01 | 1.62054648E+05 | 1.62054786E+05 | -1.40E-01 | -2.53184460E+01 | -2.51804059E+01 | -1.40E-01 |
| 2334 | Pt | 175 | 78 | 97 | 19 | 7.86947200E+00 | 7.87070274E+00 | -1.20E-03 | 1.62945402E+05 | 1.62945130E+05 | 2.70E-01 | 1.62985761E+05 | 1.62985489E+05 | 2.70E-01 | -2.57001790E+01 | -2.59721614E+01 | 2.70E-01 |
| 2335 | Pt | 176 | 78 | 98 | 20 | 7.88899200E+00 | 7.88850668E+00 | 4.90E-04 | 1.63873662E+05 | 1.63873691E+05 | -2.90E-02 | 1.63914021E+05 | 1.63914050E+05 | -2.90E-02 | -2.89337290E+01 | -2.89050378E+01 | -2.90E-02 |
| 2336 | Pt | 177 | 78 | 99 | 21 | 7.89248900E+00 | 7.89249767E+00 | -8.70E-06 | 1.64804719E+05 | 1.64804661E+05 | 5.80E-02 | 1.64845078E+05 | 1.64845020E+05 | 5.80E-02 | -2.93704360E+01 | -2.94286312E+01 | 5.80E-02 |
| 2337 | Pt | 178 | 78 | 100 | 22 | 7.90825100E+00 | 7.90817701E+00 | 7.40E-05 | 1.65733587E+05 | 1.65733543E+05 | 4.30E-02 | 1.65773946E+05 | 1.65773902E+05 | 4.30E-02 | -3.19972630E+01 | -3.20407318E+01 | 4.30E-02 |
| 2338 | Pt | 179 | 78 | 101 | 23 | 7.91067400E+00 | 7.91045198E+00 | 2.20E-04 | 1.66664810E+05 | 1.66664793E+05 | 1.70E-02 | 1.66705169E+05 | 1.66705152E+05 | 1.70E-02 | -3.22679270E+01 | -3.22848090E+01 | 1.70E-02 |
| 2339 | Pt | 180 | 78 | 102 | 24 | 7.92360900E+00 | 7.92402240E+00 | -4.10E-04 | 1.67594137E+05 | 1.67594005E+05 | 1.30E-01 | 1.67634495E+05 | 1.67634364E+05 | 1.30E-01 | -3.44355110E+01 | -3.45666187E+01 | 1.30E-01 |
| 2340 | Pt | 181 | 78 | 103 | 25 | 7.92408700E+00 | 7.92458067E+00 | -4.90E-04 | 1.68525692E+05 | 1.68525546E+05 | 1.50E-01 | 1.68566051E+05 | 1.68565905E+05 | 1.50E-01 | -3.43743750E+01 | -3.45203684E+01 | 1.50E-01 |
| 2341 | Pt | 182 | 78 | 104 | 26 | 7.93475200E+00 | 7.93599616E+00 | -1.20E-03 | 1.69455392E+05 | 1.69455109E+05 | 2.80E-01 | 1.69495751E+05 | 1.69495468E+05 | 2.80E-01 | -3.61682130E+01 | -3.64512485E+01 | 2.80E-01 |
| 2342 | Pt | 183 | 78 | 105 | 27 | 7.93333500E+00 | 7.93488109E+00 | -1.50E-03 | 1.70387282E+05 | 1.70386942E+05 | 3.40E-01 | 1.70427641E+05 | 1.70427301E+05 | 3.40E-01 | -3.57723470E+01 | -3.61118687E+01 | 3.40E-01 |
| 2343 | Pt | 184 | 78 | 106 | 28 | 7.94259900E+00 | 7.94432843E+00 | -1.70E-03 | 1.71317209E+05 | 1.71316835E+05 | 3.70E-01 | 1.71357568E+05 | 1.71357193E+05 | 3.70E-01 | -3.73388320E+01 | -3.77137407E+01 | 3.70E-01 |
| 2344 | Pt | 185 | 78 | 107 | 29 | 7.93977700E+00 | 7.94150221E+00 | -1.70E-03 | 1.72249354E+05 | 1.72248978E+05 | 3.80E-01 | 1.72289713E+05 | 1.72289337E+05 | 3.80E-01 | -3.66881400E+01 | -3.70638997E+01 | 3.80E-01 |
| 2345 | Pt | 186 | 78 | 108 | 30 | 7.94680900E+00 | 7.94891534E+00 | -2.10E-03 | 1.73179672E+05 | 1.73179223E+05 | 4.50E-01 | 1.73220031E+05 | 1.73219582E+05 | 4.50E-01 | -3.78644740E+01 | -3.83129246E+01 | 4.50E-01 |
| 2346 | Pt | 187 | 78 | 109 | 31 | 7.94116800E+00 | 7.94441766E+00 | -3.20E-03 | 1.74112345E+05 | 1.74111681E+05 | 6.60E-01 | 1.74152704E+05 | 1.74152040E+05 | 6.60E-01 | -3.66850500E+01 | -3.73494554E+01 | 6.60E-01 |
| 2347 | Pt | 188 | 78 | 110 | 32 | 7.94794500E+00 | 7.94986764E+00 | -1.90E-03 | 1.75042696E+05 | 1.75042277E+05 | 4.20E-01 | 1.75083054E+05 | 1.75082636E+05 | 4.20E-01 | -3.78290060E+01 | -3.82471513E+01 | 4.20E-01 |
| 2348 | Pt | 189 | 78 | 111 | 33 | 7.94148900E+00 | 7.94377594E+00 | -2.30E-03 | 1.75975533E+05 | 1.75975044E+05 | 4.90E-01 | 1.76015892E+05 | 1.76015403E+05 | 4.90E-01 | -3.64854250E+01 | -3.69743667E+01 | 4.90E-01 |
| 2349 | Pt | 190 | 78 | 112 | 34 | 7.94659200E+00 | 7.94738929E+00 | -8.00E-04 | 1.76906187E+05 | 1.76905979E+05 | 2.10E-01 | 1.76946546E+05 | 1.76946338E+05 | 2.10E-01 | -3.73252370E+01 | -3.75333620E+01 | 2.10E-01 |
| 2350 | Pt | 191 | 78 | 113 | 35 | 7.93874300E+00 | 7.93985665E+00 | -1.10E-03 | 1.77839305E+05 | 1.77839036E+05 | 2.70E-01 | 1.77879664E+05 | 1.77879395E+05 | 2.70E-01 | -3.57014520E+01 | -3.59706976E+01 | 2.70E-01 |
| 2351 | Pt | 192 | 78 | 114 | 36 | 7.94251100E+00 | 7.94185454E+00 | 6.60E-04 | 1.78770209E+05 | 1.78770278E+05 | -6.90E-02 | 1.78810568E+05 | 1.78810637E+05 | -6.90E-02 | -3.62921740E+01 | -3.62228299E+01 | -6.90E-02 |
| 2352 | Pt | 193 | 78 | 115 | 37 | 7.93379700E+00 | 7.93314444E+00 | 6.50E-04 | 1.79703513E+05 | 1.79703583E+05 | -6.90E-02 | 1.79743872E+05 | 1.79743941E+05 | -6.90E-02 | -3.44816900E+01 | -3.44123156E+01 | -6.90E-02 |
| 2353 | Pt | 194 | 78 | 116 | 38 | 7.93595400E+00 | 7.93386293E+00 | 2.10E-03 | 1.80634726E+05 | 1.80635075E+05 | -3.50E-01 | 1.80675085E+05 | 1.80675434E+05 | -3.50E-01 | -3.47625640E+01 | -3.44135287E+01 | -3.50E-01 |
| 2354 | Pt | 195 | 78 | 117 | 39 | 7.92656500E+00 | 7.92434034E+00 | 2.20E-03 | 1.81568187E+05 | 1.81568564E+05 | -3.80E-01 | 1.81608546E+05 | 1.81608923E+05 | -3.80E-01 | -3.27963020E+01 | -3.24191675E+01 | -3.80E-01 |
| 2355 | Pt | 196 | 78 | 118 | 40 | 7.92654100E+00 | 7.92417203E+00 | 2.40E-03 | 1.82499830E+05 | 1.82500238E+05 | -4.10E-01 | 1.82540189E+05 | 1.82540597E+05 | -4.10E-01 | -3.26469170E+01 | -3.22391996E+01 | -4.10E-01 |
| 2356 | Pt | 197 | 78 | 119 | 41 | 7.91598200E+00 | 7.91417123E+00 | 1.80E-03 | 1.83433549E+05 | 1.83433849E+05 | -3.00E-01 | 1.83473908E+05 | 1.83474208E+05 | -3.00E-01 | -3.04219540E+01 | -3.01218945E+01 | -3.00E-01 |
| 2357 | Pt | 198 | 78 | 120 | 42 | 7.91415900E+00 | 7.91333578E+00 | 8.20E-04 | 1.84365559E+05 | 1.84365666E+05 | -1.10E-01 | 1.84405918E+05 | 1.84406025E+05 | -1.10E-01 | -2.99056890E+01 | -2.97993277E+01 | -1.10E-01 |
| 2358 | Pt | 199 | 78 | 121 | 43 | 7.90230900E+00 | 7.90282252E+00 | -5.10E-04 | 1.85299569E+05 | 1.85299410E+05 | 1.60E-01 | 1.85339928E+05 | 1.85339769E+05 | 1.60E-01 | -2.73903720E+01 | -2.75492061E+01 | 1.60E-01 |
| 2359 | Pt | 200 | 78 | 122 | 44 | 7.89920600E+00 | 7.90093974E+00 | -1.70E-03 | 1.86231852E+05 | 1.86231449E+05 | 4.00E-01 | 1.86272211E+05 | 1.86271808E+05 | 4.00E-01 | -2.66008530E+01 | -2.70041533E+01 | 4.00E-01 |
| 2360 | Pt | 201 | 78 | 123 | 45 | 7.88583400E+00 | 7.88906784E+00 | -3.20E-03 | 1.87166206E+05 | 1.87165500E+05 | 7.10E-01 | 1.87206565E+05 | 1.87205859E+05 | 7.10E-01 | -2.37408830E+01 | -2.44475227E+01 | 7.10E-01 |
| 2361 | Pt | 202 | 78 | 124 | 46 | 7.88156000E+00 | 7.88483764E+00 | -3.30E-03 | 1.88098749E+05 | 1.88098031E+05 | 7.20E-01 | 1.88139108E+05 | 1.88138390E+05 | 7.20E-01 | -2.26921250E+01 | -2.34107714E+01 | 7.20E-01 |
| 2362 | PT | 203 | 78 | 125 | 47 | 7.86700000E+00 | 7.86994286E+00 | -2.90E-03 | 1.89033309E+05 | 1.89032735E+05 | 5.70E-01 | 1.89073668E+05 | 1.89073094E+05 | 5.70E-01 | -1.96270000E+01 | -2.02006493E+01 | 5.70E-01 |
| 2363 | PT | 204 | 78 | 126 | 48 | 7.86000000E+00 | 7.86162508E+00 | -1.60E-03 | 1.89966508E+05 | 1.89966127E+05 | 3.80E-01 | 1.90006867E+05 | 1.90006486E+05 | 3.80E-01 | -1.79220000E+01 | -1.83024455E+01 | 3.80E-01 |
| 2364 | PT | 205 | 78 | 127 | 49 | 7.83700000E+00 | 7.84223241E+00 | -5.20E-03 | 1.90902957E+05 | 1.90901806E+05 | 1.20E+00 | 1.90943316E+05 | 1.90942165E+05 | 1.20E+00 | -1.29660000E+01 | -1.41172552E+01 | 1.20E+00 |
| 2365 | Pt | 206 | 78 | 128 | 50 | 7.82200000E+00 | 7.82899746E+00 | -7.00E-03 | 1.91837786E+05 | 1.91836256E+05 | 1.50E+00 | 1.91878145E+05 | 1.91876615E+05 | 1.50E+00 | -9.63200000E+00 | -1.11617685E+01 | 1.50E+00 |
| 2366 | Au | 170 | 79 | 91 | 12 | 7.73000000E+00 | 7.72751579E+00 | 2.50E-03 | 1.58309352E+05 | 1.58309688E+05 | -3.40E-01 | 1.58350238E+05 | 1.58350574E+05 | -3.40E-01 | -3.75200000E+00 | -3.41653676E+00 | -3.40E-01 |
| 2367 | Au | 171 | 79 | 92 | 13 | 7.75413700E+00 | 7.75236074E+00 | 1.80E-03 | 1.59237030E+05 | 1.59237277E+05 | -2.50E-01 | 1.59277917E+05 | 1.59278163E+05 | -2.50E-01 | -7.56765100E+00 | -7.32122130E+00 | -2.50E-01 |
| 2368 | Au | 172 | 79 | 93 | 14 | 7.76645300E+00 | 7.76426959E+00 | 2.20E-03 | 1.60166723E+05 | 1.60167042E+05 | -3.20E-01 | 1.60207610E+05 | 1.60207928E+05 | -3.20E-01 | -9.36870200E+00 | -9.05058484E+00 | -3.20E-01 |
| 2369 | Au | 173 | 79 | 94 | 15 | 7.78814400E+00 | 7.78681126E+00 | 1.30E-03 | 1.61094770E+05 | 1.61094943E+05 | -1.70E-01 | 1.61135656E+05 | 1.61135829E+05 | -1.70E-01 | -1.28164960E+01 | -1.26432446E+01 | -1.70E-01 |
| 2370 | Au | 174 | 79 | 95 | 16 | 7.79800000E+00 | 7.79691256E+00 | 1.10E-03 | 1.62024844E+05 | 1.62024964E+05 | -1.20E-01 | 1.62065731E+05 | 1.62065850E+05 | -1.20E-01 | -1.42360000E+01 | -1.41163626E+01 | -1.20E-01 |
| 2371 | Au | 175 | 79 | 96 | 17 | 7.81766100E+00 | 7.81723910E+00 | 4.20E-04 | 1.62953159E+05 | 1.62953175E+05 | -1.60E-02 | 1.62994045E+05 | 1.62994062E+05 | -1.60E-02 | -1.74154790E+01 | -1.73991001E+01 | -1.60E-02 |
| 2372 | Au | 176 | 79 | 97 | 18 | 7.82467600E+00 | 7.82557761E+00 | -9.00E-04 | 1.63883672E+05 | 1.63883456E+05 | 2.20E-01 | 1.63924558E+05 | 1.63924342E+05 | 2.20E-01 | -1.83966040E+01 | -1.86125978E+01 | 2.20E-01 |
| 2373 | Au | 177 | 79 | 98 | 19 | 7.84385800E+00 | 7.84372952E+00 | 1.30E-04 | 1.64812017E+05 | 1.64811983E+05 | 3.50E-02 | 1.64852904E+05 | 1.64852869E+05 | 3.50E-02 | -2.15450470E+01 | -2.15797448E+01 | 3.50E-02 |
| 2374 | Au | 178 | 79 | 99 | 20 | 7.84952400E+00 | 7.85032443E+00 | -8.00E-04 | 1.65742731E+05 | 1.65742531E+05 | 2.00E-01 | 1.65783617E+05 | 1.65783417E+05 | 2.00E-01 | -2.23261220E+01 | -2.25260502E+01 | 2.00E-01 |
| 2375 | Au | 179 | 79 | 100 | 21 | 7.86563700E+00 | 7.86633124E+00 | -6.90E-04 | 1.66671562E+05 | 1.66671380E+05 | 1.80E-01 | 1.66712448E+05 | 1.66712267E+05 | 1.80E-01 | -2.49885800E+01 | -2.51702737E+01 | 1.80E-01 |
| 2376 | Au | 180 | 79 | 101 | 22 | 7.87014400E+00 | 7.87119771E+00 | -1.10E-03 | 1.67602450E+05 | 1.67602203E+05 | 2.50E-01 | 1.67643337E+05 | 1.67643090E+05 | 2.50E-01 | -2.55942910E+01 | -2.58412510E+01 | 2.50E-01 |
| 2377 | Au | 181 | 79 | 102 | 23 | 7.88383500E+00 | 7.88508878E+00 | -1.30E-03 | 1.68531668E+05 | 1.68531383E+05 | 2.80E-01 | 1.68572554E+05 | 1.68572270E+05 | 2.80E-01 | -2.78711880E+01 | -2.81554131E+01 | 2.80E-01 |
| 2378 | Au | 182 | 79 | 103 | 24 | 7.88722600E+00 | 7.88824201E+00 | -1.00E-03 | 1.69462732E+05 | 1.69462490E+05 | 2.40E-01 | 1.69503618E+05 | 1.69503376E+05 | 2.40E-01 | -2.83007680E+01 | -2.85430701E+01 | 2.40E-01 |
| 2379 | Au | 183 | 79 | 104 | 25 | 7.89855100E+00 | 7.89997632E+00 | -1.40E-03 | 1.70392338E+05 | 1.70392020E+05 | 3.20E-01 | 1.70433224E+05 | 1.70432906E+05 | 3.20E-01 | -3.01891290E+01 | -3.05073730E+01 | 3.20E-01 |
| 2380 | Au | 184 | 79 | 105 | 26 | 7.90019400E+00 | 7.90147906E+00 | -1.30E-03 | 1.71323702E+05 | 1.71323408E+05 | 2.90E-01 | 1.71364589E+05 | 1.71364295E+05 | 2.90E-01 | -3.03187100E+01 | -3.06125338E+01 | 2.90E-01 |
| 2381 | Au | 185 | 79 | 106 | 27 | 7.90948700E+00 | 7.91127406E+00 | -1.80E-03 | 1.72253648E+05 | 1.72253260E+05 | 3.90E-01 | 1.72294535E+05 | 1.72294147E+05 | 3.90E-01 | -3.18668110E+01 | -3.22547700E+01 | 3.90E-01 |
| 2382 | Au | 186 | 79 | 107 | 28 | 7.90954000E+00 | 7.91109914E+00 | -1.60E-03 | 1.73185294E+05 | 1.73184947E+05 | 3.50E-01 | 1.73226180E+05 | 1.73225833E+05 | 3.50E-01 | -3.17148520E+01 | -3.20621882E+01 | 3.50E-01 |
| 2383 | Au | 187 | 79 | 108 | 29 | 7.91742700E+00 | 7.91889171E+00 | -1.50E-03 | 1.74115475E+05 | 1.74115144E+05 | 3.30E-01 | 1.74156362E+05 | 1.74156030E+05 | 3.30E-01 | -3.30278380E+01 | -3.33591791E+01 | 3.30E-01 |
| 2384 | Au | 188 | 79 | 109 | 30 | 7.91425000E+00 | 7.91709719E+00 | -2.80E-03 | 1.75047720E+05 | 1.75047128E+05 | 5.90E-01 | 1.75088607E+05 | 1.75088014E+05 | 5.90E-01 | -3.22768330E+01 | -3.28693825E+01 | 5.90E-01 |
| 2385 | Au | 189 | 79 | 110 | 31 | 7.92198700E+00 | 7.92297796E+00 | -9.90E-04 | 1.75977909E+05 | 1.75977665E+05 | 2.40E-01 | 1.76018796E+05 | 1.76018551E+05 | 2.40E-01 | -3.35819540E+01 | -3.38266271E+01 | 2.40E-01 |
| 2386 | Au | 190 | 79 | 111 | 32 | 7.91909500E+00 | 7.91966572E+00 | -5.70E-04 | 1.76910102E+05 | 1.76909936E+05 | 1.70E-01 | 1.76950988E+05 | 1.76950823E+05 | 1.70E-01 | -3.28832370E+01 | -3.30489606E+01 | 1.70E-01 |
| 2387 | Au | 191 | 79 | 112 | 33 | 7.92475000E+00 | 7.92378827E+00 | 9.60E-04 | 1.77840668E+05 | 1.77840795E+05 | -1.30E-01 | 1.77881554E+05 | 1.77881681E+05 | -1.30E-01 | -3.38111380E+01 | -3.36847130E+01 | -1.30E-01 |
| 2388 | Au | 192 | 79 | 113 | 34 | 7.92012200E+00 | 7.91914254E+00 | 9.80E-04 | 1.78773198E+05 | 1.78773328E+05 | -1.30E-01 | 1.78814084E+05 | 1.78814215E+05 | -1.30E-01 | -3.27758330E+01 | -3.26452025E+01 | -1.30E-01 |
| 2389 | Au | 193 | 79 | 114 | 35 | 7.92417000E+00 | 7.92175985E+00 | 2.40E-03 | 1.79704062E+05 | 1.79704469E+05 | -4.10E-01 | 1.79744948E+05 | 1.79745356E+05 | -4.10E-01 | -3.34059300E+01 | -3.29981679E+01 | -4.10E-01 |
| 2390 | Au | 194 | 79 | 115 | 36 | 7.91878000E+00 | 7.91607415E+00 | 2.70E-03 | 1.80636748E+05 | 1.80637216E+05 | -4.70E-01 | 1.80677635E+05 | 1.80678102E+05 | -4.70E-01 | -3.22131550E+01 | -3.17455820E+01 | -4.70E-01 |

| | | | | | | | | | | | | | | | | |
|---|---|---|---|---|---|---|---|---|---|---|---|---|---|---|---|---|
| 2391 | Au | 195 | 79 | 116 | 37 | 7.92138900E+00 | 7.91754404E+00 | 3.80E-03 | 1.81567886E+05 | 1.81568579E+05 | -6.90E-01 | 1.81608772E+05 | 1.81609465E+05 | -6.90E-01 | -3.25694780E+01 | -3.18769662E+01 | -6.90E-01 |
| 2392 | Au | 196 | 79 | 117 | 38 | 7.91486100E+00 | 7.91118871E+00 | 3.70E-03 | 1.82500810E+05 | 1.82501472E+05 | -6.60E-01 | 1.82541696E+05 | 1.82542358E+05 | -6.60E-01 | -3.11399300E+01 | -3.04775453E+01 | -6.60E-01 |
| 2393 | Au | 197 | 79 | 118 | 39 | 7.91566000E+00 | 7.91188270E+00 | 3.80E-03 | 1.83432303E+05 | 1.83432990E+05 | -6.90E-01 | 1.83473189E+05 | 1.83473876E+05 | -6.90E-01 | -3.11409750E+01 | -3.04541323E+01 | -6.90E-01 |
| 2394 | Au | 198 | 79 | 119 | 40 | 7.90857300E+00 | 7.90513313E+00 | 3.40E-03 | 1.84365356E+05 | 1.84365980E+05 | -6.20E-01 | 1.84406242E+05 | 1.84406866E+05 | -6.20E-01 | -2.95819950E+01 | -2.89582810E+01 | -6.20E-01 |
| 2395 | Au | 199 | 79 | 120 | 41 | 7.90694300E+00 | 7.90516654E+00 | 1.80E-03 | 1.85297337E+05 | 1.85297633E+05 | -3.00E-01 | 1.85338223E+05 | 1.85338519E+05 | -3.00E-01 | -2.90949520E+01 | -2.87987436E+01 | -3.00E-01 |
| 2396 | Au | 200 | 79 | 121 | 42 | 7.89849100E+00 | 7.89783537E+00 | 6.60E-04 | 1.86230686E+05 | 1.86230760E+05 | -7.40E-02 | 1.86271572E+05 | 1.86271646E+05 | -7.40E-02 | -2.72401850E+01 | -2.71663580E+01 | -7.40E-02 |
| 2397 | Au | 201 | 79 | 122 | 43 | 7.89517500E+00 | 7.89663718E+00 | -1.50E-03 | 1.87163019E+05 | 1.87162668E+05 | 3.50E-01 | 1.87203905E+05 | 1.87203554E+05 | 3.50E-01 | -2.64008830E+01 | -2.67520363E+01 | 3.50E-01 |
| 2398 | Au | 202 | 79 | 123 | 44 | 7.88590900E+00 | 7.88767260E+00 | -1.80E-03 | 1.88096561E+05 | 1.88096148E+05 | 4.10E-01 | 1.88137447E+05 | 1.88137034E+05 | 4.10E-01 | -2.43529790E+01 | -2.47665104E+01 | 4.10E-01 |
| 2399 | Au | 203 | 79 | 124 | 45 | 7.88086400E+00 | 7.88376347E+00 | -2.90E-03 | 1.89029265E+05 | 1.89028619E+05 | 6.50E-01 | 1.89070151E+05 | 1.89069505E+05 | 6.50E-01 | -2.31434840E+01 | -2.37893104E+01 | 6.50E-01 |
| 2400 | Au | 204 | 79 | 125 | 46 | 7.87000000E+00 | 7.87139142E+00 | -1.40E-03 | 1.89963252E+05 | 1.89962824E+05 | 4.30E-01 | 1.90004138E+05 | 1.90003711E+05 | 4.30E-01 | -2.06500000E+01 | -2.10778561E+01 | 4.30E-01 |
| 2401 | Au | 205 | 79 | 126 | 47 | 7.86100000E+00 | 7.86302967E+00 | -2.00E-03 | 1.90896627E+05 | 1.90896232E+05 | 3.90E-01 | 1.90937513E+05 | 1.90937119E+05 | 3.90E-01 | -1.87700000E+01 | -1.91637707E+01 | 3.90E-01 |
| 2402 | AU | 206 | 79 | 127 | 48 | 7.84000000E+00 | 7.84589041E+00 | -5.90E-03 | 1.91832676E+05 | 1.91831466E+05 | 1.20E+00 | 1.91873562E+05 | 1.91872352E+05 | 1.20E+00 | -1.42150000E+01 | -1.54247924E+01 | 1.20E+00 |
| 2403 | AU | 207 | 79 | 128 | 49 | 7.82500000E+00 | 7.83245880E+00 | -7.50E-03 | 1.92767579E+05 | 1.92765965E+05 | 1.60E+00 | 1.92808465E+05 | 1.92806852E+05 | 1.60E+00 | -1.08050000E+01 | -1.24190224E+01 | 1.60E+00 |
| 2404 | AU | 208 | 79 | 129 | 50 | 7.80400000E+00 | 7.81081923E+00 | -6.80E-03 | 1.93703777E+05 | 1.93702199E+05 | 1.60E+00 | 1.93744663E+05 | 1.93743086E+05 | 1.60E+00 | -6.10100000E+00 | -7.67913021E+00 | 1.60E+00 |
| 2405 | AU | 209 | 79 | 130 | 51 | 7.78700000E+00 | 7.79343205E+00 | -6.40E-03 | 1.94638904E+05 | 1.94637588E+05 | 1.30E+00 | 1.94679790E+05 | 1.94678474E+05 | 1.30E+00 | -2.46800000E+00 | -3.78471136E+00 | 1.30E+00 |
| 2406 | AU | 210 | 79 | 131 | 52 | 7.76600000E+00 | 7.76897244E+00 | -3.00E-03 | 1.95575195E+05 | 1.95574496E+05 | 7.00E-01 | 1.95616082E+05 | 1.95615383E+05 | 7.00E-01 | 2.32900000E+00 | 1.62969502E+00 | 7.00E-01 |
| 2407 | HG | 171 | 80 | 91 | 11 | 7.68600000E+00 | 7.68463627E+00 | 1.40E-03 | 1.59247358E+05 | 1.59247547E+05 | -1.90E-01 | 1.59288772E+05 | 1.59288961E+05 | -1.90E-01 | 3.28800000E+00 | 3.47658702E+00 | -1.90E-01 |
| 2408 | Hg | 172 | 80 | 92 | 12 | 7.71388500E+00 | 7.71209065E+00 | 1.80E-03 | 1.60174455E+05 | 1.60174706E+05 | -2.50E-01 | 1.60215869E+05 | 1.60216120E+05 | -2.50E-01 | -1.10943900E+00 | -8.58883666E-01 | -2.50E-01 |
| 2409 | Hg | 173 | 80 | 93 | 13 | 7.72500000E+00 | 7.72423332E+00 | 7.70E-04 | 1.61104349E+05 | 1.61104458E+05 | -1.10E-01 | 1.61145763E+05 | 1.61145872E+05 | -1.10E-01 | -2.71000000E+00 | -2.60033722E+00 | -1.10E-01 |
| 2410 | Hg | 174 | 80 | 94 | 14 | 7.74981600E+00 | 7.74930917E+00 | 5.10E-04 | 1.62031906E+05 | 1.62031936E+05 | -3.00E-02 | 1.62073320E+05 | 1.62073350E+05 | -3.00E-02 | -6.64644400E+00 | -6.61644970E+00 | -3.00E-02 |
| 2411 | Hg | 175 | 80 | 95 | 15 | 7.75923200E+00 | 7.75956488E+00 | -3.30E-04 | 1.62962074E+05 | 1.62961958E+05 | 1.20E-01 | 1.63003488E+05 | 1.63003371E+05 | 1.20E-01 | -7.97289200E+00 | -8.08918912E+00 | 1.20E-01 |
| 2412 | Hg | 176 | 80 | 96 | 16 | 7.78259900E+00 | 7.78237077E+00 | 2.30E-04 | 1.63889768E+05 | 1.63889750E+05 | 1.80E-02 | 1.63931182E+05 | 1.63931163E+05 | 1.80E-02 | -1.17732800E+01 | -1.17912724E+01 | 1.80E-02 |
| 2413 | Hg | 177 | 80 | 97 | 17 | 7.78993200E+00 | 7.79080503E+00 | -8.70E-04 | 1.64820252E+05 | 1.64820040E+05 | 2.10E-01 | 1.64861666E+05 | 1.64861454E+05 | 2.10E-01 | -1.27825940E+01 | -1.29951878E+01 | 2.10E-01 |
| 2414 | Hg | 178 | 80 | 98 | 18 | 7.81136500E+00 | 7.81140094E+00 | -3.60E-05 | 1.65748213E+05 | 1.65748148E+05 | 6.40E-02 | 1.65789627E+05 | 1.65789562E+05 | 6.40E-02 | -1.63162640E+01 | -1.63807460E+01 | 6.40E-02 |
| 2415 | Hg | 179 | 80 | 99 | 19 | 7.81621300E+00 | 7.81805087E+00 | -1.80E-03 | 1.66679099E+05 | 1.66678712E+05 | 3.90E-01 | 1.66720513E+05 | 1.66720126E+05 | 3.90E-01 | -1.69240070E+01 | -1.73111648E+01 | 3.90E-01 |
| 2416 | Hg | 180 | 80 | 100 | 20 | 7.83611000E+00 | 7.83648359E+00 | -3.70E-04 | 1.67607267E+05 | 1.67607141E+05 | 1.30E-01 | 1.67648681E+05 | 1.67648555E+05 | 1.30E-01 | -2.02503670E+01 | -2.03757859E+01 | 1.30E-01 |
| 2417 | Hg | 181 | 80 | 101 | 21 | 7.83967800E+00 | 7.84137945E+00 | -1.70E-03 | 1.68538350E+05 | 1.68537984E+05 | 3.70E-01 | 1.68579764E+05 | 1.68579398E+05 | 3.70E-01 | -2.06611220E+01 | -2.10271008E+01 | 3.70E-01 |
| 2418 | Hg | 182 | 80 | 102 | 22 | 7.85697100E+00 | 7.85769352E+00 | -7.20E-04 | 1.69466929E+05 | 1.69466739E+05 | 1.90E-01 | 1.69508342E+05 | 1.69508153E+05 | 1.90E-01 | -2.35767170E+01 | -2.37663219E+01 | 1.90E-01 |
| 2419 | Hg | 183 | 80 | 103 | 23 | 7.85938700E+00 | 7.86086349E+00 | -1.50E-03 | 1.70398195E+05 | 1.70397866E+05 | 3.30E-01 | 1.70439609E+05 | 1.70439280E+05 | 3.30E-01 | -2.38045310E+01 | -2.41328017E+01 | 3.30E-01 |
| 2420 | Hg | 184 | 80 | 104 | 24 | 7.87436600E+00 | 7.87502437E+00 | -6.60E-04 | 1.71327145E+05 | 1.71326965E+05 | 1.80E-01 | 1.71368559E+05 | 1.71368379E+05 | 1.80E-01 | -2.63486780E+01 | -2.65279490E+01 | 1.80E-01 |
| 2421 | Hg | 185 | 80 | 105 | 25 | 7.87449500E+00 | 7.87654969E+00 | -2.10E-03 | 1.72258812E+05 | 1.72258374E+05 | 4.40E-01 | 1.72300226E+05 | 1.72299787E+05 | 4.40E-01 | -2.61755510E+01 | -2.66138368E+01 | 4.40E-01 |
| 2422 | Hg | 186 | 80 | 106 | 26 | 7.88825900E+00 | 7.88881065E+00 | -5.50E-04 | 1.73187943E+05 | 1.73187782E+05 | 1.60E-01 | 1.73229356E+05 | 1.73229196E+05 | 1.60E-01 | -2.85388590E+01 | -2.86996064E+01 | 1.60E-01 |
| 2423 | Hg | 187 | 80 | 107 | 27 | 7.88698500E+00 | 7.88867420E+00 | -1.70E-03 | 1.74119858E+05 | 1.74119484E+05 | 3.70E-01 | 1.74161272E+05 | 1.74160898E+05 | 3.70E-01 | -2.81176790E+01 | -2.84915818E+01 | 3.70E-01 |
| 2424 | Hg | 188 | 80 | 108 | 28 | 7.89910200E+00 | 7.89897748E+00 | 1.20E-04 | 1.75049258E+05 | 1.75049224E+05 | 3.50E-02 | 1.75090672E+05 | 1.75090638E+05 | 3.50E-02 | -3.02112500E+01 | -3.02459537E+01 | 3.50E-02 |
| 2425 | Hg | 189 | 80 | 109 | 29 | 7.89691800E+00 | 7.89725667E+00 | -3.40E-04 | 1.75981337E+05 | 1.75981215E+05 | 1.20E-01 | 1.76022751E+05 | 1.76022629E+05 | 1.20E-01 | -2.96262340E+01 | -2.97483791E+01 | 1.20E-01 |
| 2426 | Hg | 190 | 80 | 110 | 30 | 7.90701400E+00 | 7.90571616E+00 | 1.30E-03 | 1.76911087E+05 | 1.76911276E+05 | -1.90E-01 | 1.76952501E+05 | 1.76952690E+05 | -1.90E-01 | -3.13701700E+01 | -3.11816210E+01 | -1.90E-01 |
| 2427 | Hg | 191 | 80 | 111 | 31 | 7.90380500E+00 | 7.90253832E+00 | 1.30E-03 | 1.77843359E+05 | 1.77843543E+05 | -1.80E-01 | 1.77884773E+05 | 1.77884957E+05 | -1.80E-01 | -3.05929080E+01 | -3.04090510E+01 | -1.80E-01 |
| 2428 | Hg | 192 | 80 | 112 | 32 | 7.91206400E+00 | 7.90933631E+00 | 2.70E-03 | 1.78773435E+05 | 1.78773900E+05 | -4.70E-01 | 1.78814849E+05 | 1.78815314E+05 | -4.70E-01 | -3.20110690E+01 | -3.15454840E+01 | -4.70E-01 |
| 2429 | Hg | 193 | 80 | 113 | 33 | 7.90797600E+00 | 7.90491474E+00 | 3.10E-03 | 1.79705877E+05 | 1.79706410E+05 | -5.30E-01 | 1.79747291E+05 | 1.79747824E+05 | -5.30E-01 | -3.10628490E+01 | -3.05301376E+01 | -5.30E-01 |
| 2430 | Hg | 194 | 80 | 114 | 34 | 7.91459700E+00 | 7.91033148E+00 | 4.30E-03 | 1.80636250E+05 | 1.80637019E+05 | -7.70E-01 | 1.80677664E+05 | 1.80678433E+05 | -7.70E-01 | -3.21839440E+01 | -3.14145817E+01 | -7.70E-01 |
| 2431 | Hg | 195 | 80 | 115 | 35 | 7.90932700E+00 | 7.90497970E+00 | 4.30E-03 | 1.81568928E+05 | 1.81569718E+05 | -7.90E-01 | 1.81610342E+05 | 1.81611132E+05 | -7.90E-01 | -3.09996170E+01 | -3.02099969E+01 | -7.90E-01 |
| 2432 | Hg | 196 | 80 | 116 | 36 | 7.91437300E+00 | 7.90937927E+00 | 5.00E-03 | 1.82499595E+05 | 1.82500516E+05 | -9.20E-01 | 1.82541009E+05 | 1.82541930E+05 | -9.20E-01 | -3.18267650E+01 | -3.09059736E+01 | -9.20E-01 |
| 2433 | Hg | 197 | 80 | 117 | 37 | 7.90864400E+00 | 7.90345211E+00 | 5.20E-03 | 1.83432375E+05 | 1.83433340E+05 | -9.60E-01 | 1.83473789E+05 | 1.83474754E+05 | -9.60E-01 | -3.05410360E+01 | -2.95763819E+01 | -9.60E-01 |
| 2434 | Hg | 198 | 80 | 118 | 38 | 7.91155500E+00 | 7.90715951E+00 | 4.40E-03 | 1.84363455E+05 | 1.84364268E+05 | -8.10E-01 | 1.84404869E+05 | 1.84405681E+05 | -8.10E-01 | -3.09548470E+01 | -3.01425816E+01 | -8.10E-01 |
| 2435 | Hg | 199 | 80 | 119 | 39 | 7.90528000E+00 | 7.90084880E+00 | 4.40E-03 | 1.85296358E+05 | 1.85297182E+05 | -8.20E-01 | 1.85337772E+05 | 1.85338596E+05 | -8.20E-01 | -2.95463900E+01 | -2.87225893E+01 | -8.20E-01 |
| 2436 | Hg | 200 | 80 | 120 | 40 | 7.90589600E+00 | 7.90385514E+00 | 2.00E-03 | 1.86227895E+05 | 1.86228245E+05 | -3.50E-01 | 1.86269309E+05 | 1.86269659E+05 | -3.50E-01 | -2.95035910E+01 | -2.91533886E+01 | -3.50E-01 |
| 2437 | Hg | 201 | 80 | 121 | 41 | 7.89756100E+00 | 7.89681680E+00 | 7.40E-04 | 1.87161230E+05 | 1.87161321E+05 | -9.20E-02 | 1.87202644E+05 | 1.87202735E+05 | -9.20E-02 | -2.76627270E+01 | -2.75712171E+01 | -9.20E-02 |
| 2438 | Hg | 202 | 80 | 122 | 42 | 7.89685100E+00 | 7.89837803E+00 | -1.50E-03 | 1.88093041E+05 | 1.88092674E+05 | 3.70E-01 | 1.88134455E+05 | 1.88134088E+05 | 3.70E-01 | -2.73454960E+01 | -2.77120843E+01 | 3.70E-01 |
| 2439 | Hg | 203 | 80 | 123 | 43 | 7.88748000E+00 | 7.88940440E+00 | -1.90E-03 | 1.89026612E+05 | 1.89026163E+05 | 4.50E-01 | 1.89068026E+05 | 1.89067577E+05 | 4.50E-01 | -2.52687890E+01 | -2.57174965E+01 | 4.50E-01 |
| 2440 | Hg | 204 | 80 | 124 | 44 | 7.88554500E+00 | 7.88793845E+00 | -2.40E-03 | 1.89958684E+05 | 1.89958138E+05 | 5.50E-01 | 1.90000098E+05 | 1.89999552E+05 | 5.50E-01 | -2.46901970E+01 | -2.52365269E+01 | 5.50E-01 |
| 2441 | Hg | 205 | 80 | 125 | 45 | 7.87473100E+00 | 7.87524463E+00 | -5.10E-04 | 1.90892581E+05 | 1.90892418E+05 | 1.60E-01 | 1.90933995E+05 | 1.90933832E+05 | 1.60E-01 | -2.22874680E+01 | -2.24509141E+01 | 1.60E-01 |
| 2442 | Hg | 206 | 80 | 126 | 46 | 7.86916900E+00 | 7.86910461E+00 | 6.40E-05 | 1.91825418E+05 | 1.91825373E+05 | 4.50E-02 | 1.91866831E+05 | 1.91866787E+05 | 4.50E-02 | -2.09452000E+01 | -2.09899957E+01 | 4.50E-02 |
| 2443 | Hg | 207 | 80 | 127 | 47 | 7.84861000E+00 | 7.85152325E+00 | -2.90E-03 | 1.92761369E+05 | 1.92760708E+05 | 6.60E-01 | 1.92802783E+05 | 1.92802122E+05 | 6.60E-01 | -1.64874440E+01 | -1.71484388E+01 | 6.60E-01 |
| 2444 | Hg | 208 | 80 | 128 | 48 | 7.83419100E+00 | 7.84033587E+00 | -6.10E-03 | 1.93696085E+05 | 1.93694749E+05 | 1.30E+00 | 1.93737499E+05 | 1.93736163E+05 | 1.30E+00 | -1.32654060E+01 | -1.46016684E+01 | 1.30E+00 |
| 2445 | Hg | 209 | 80 | 129 | 49 | 7.81300000E+00 | 7.81837632E+00 | -5.40E-03 | 1.94632201E+05 | 1.94631064E+05 | 1.10E+00 | 1.94673614E+05 | 1.94672478E+05 | 1.10E+00 | -8.64400000E+00 | -9.78114030E+00 | 1.10E+00 |
| 2446 | HG | 210 | 80 | 130 | 50 | 7.79900000E+00 | 7.80343866E+00 | -4.40E-03 | 1.95566973E+05 | 1.95565948E+05 | 1.00E+00 | 1.95608387E+05 | 1.95607362E+05 | 1.00E+00 | -5.36500000E+00 | -6.39128801E+00 | 1.00E+00 |
| 2447 | HG | 211 | 80 | 131 | 51 | 7.77800000E+00 | 7.77887764E+00 | -8.80E-04 | 1.96503209E+05 | 1.96502892E+05 | 3.20E-01 | 1.96544623E+05 | 1.96544306E+05 | 3.20E-01 | -6.24000000E-01 | -9.41032466E-01 | 3.20E-01 |
| 2448 | HG | 212 | 80 | 132 | 52 | 7.76300000E+00 | 7.76211275E+00 | 8.90E-04 | 1.97438084E+05 | 1.97438233E+05 | -1.50E-01 | 1.97479498E+05 | 1.97479646E+05 | -1.50E-01 | 2.75700000E+00 | 2.90556442E+00 | -1.50E-01 |
| 2449 | HG | 213 | 80 | 133 | 53 | 7.74100000E+00 | 7.73669891E+00 | 4.30E-03 | 1.98374487E+05 | 1.98375449E+05 | -9.60E-01 | 1.98415901E+05 | 1.98416863E+05 | -9.60E-01 | 7.66600000E+00 | 8.62791938E+00 | -9.60E-01 |
| 2450 | HG | 214 | 80 | 134 | 54 | 7.72700000E+00 | 7.71846808E+00 | 8.50E-03 | 1.99309493E+05 | 1.99311179E+05 | -1.70E+00 | 1.99350907E+05 | 1.99352593E+05 | -1.70E+00 | 1.11780000E+01 | 1.28639359E+01 | -1.70E+00 |
| 2451 | HG | 215 | 80 | 135 | 55 | 7.70500000E+00 | 7.69865079E+00 | 6.30E-03 | 2.00246017E+05 | 2.00247287E+05 | -1.30E+00 | 2.00287431E+05 | 2.00288701E+05 | -1.30E+00 | 1.62080000E+01 | 1.74775043E+01 | -1.30E+00 |
| 2452 | HG | 216 | 80 | 136 | 56 | 7.69000000E+00 | 7.68609538E+00 | 3.90E-03 | 2.01181163E+05 | 2.01181865E+05 | -7.00E-01 | 2.01222577E+05 | 2.01223279E+05 | -7.00E-01 | 1.98590000E+01 | 2.05621427E+01 | -7.00E-01 |
| 2453 | TL | 176 | 81 | 95 | 14 | 7.70795700E+00 | 7.70715477E+00 | 8.00E-04 | 1.63901594E+05 | 1.63901676E+05 | -8.20E-02 | 1.63943536E+05 | 1.63943618E+05 | -8.20E-02 | 5.81278000E-01 | 6.63667006E-01 | -8.20E-02 |
| 2454 | Tl | 177 | 81 | 96 | 15 | 7.73207800E+00 | 7.73031418E+00 | 1.80E-03 | 1.64829182E+05 | 1.64829436E+05 | -2.50E-01 | 1.64871124E+05 | 1.64871377E+05 | -2.50E-01 | -3.32466100E+00 | -3.07138417E+00 | -2.50E-01 |
| 2455 | TL | 178 | 81 | 97 | 16 | 7.74200000E+00 | 7.74140003E+00 | 6.00E-04 | 1.65759207E+05 | 1.65759297E+05 | -9.00E-02 | 1.65801149E+05 | 1.65801239E+05 | -9.00E-02 | -4.79400000E+00 | -4.70366026E+00 | -9.00E-02 |
| 2456 | Tl | 179 | 81 | 98 | 17 | 7.76355300E+00 | 7.76229389E+00 | 1.30E-03 | 1.66687215E+05 | 1.66687381E+05 | -1.70E-01 | 1.66729157E+05 | 1.66729323E+05 | -1.70E-01 | -8.28033300E+00 | -8.11374256E+00 | -1.70E-01 |
| 2457 | Tl | 180 | 81 | 99 | 18 | 7.77071800E+00 | 7.77154628E+00 | -8.30E-04 | 1.67617727E+05 | 1.67617519E+05 | 2.10E-01 | 1.67659669E+05 | 1.67659461E+05 | 2.10E-01 | -9.26228800E+00 | -9.47014719E+00 | 2.10E-01 |
| 2458 | Tl | 181 | 81 | 100 | 19 | 7.79191800E+00 | 7.79023729E+00 | 1.70E-03 | 1.68545684E+05 | 1.68545930E+05 | -2.50E-01 | 1.68587626E+05 | 1.68587872E+05 | -2.50E-01 | -1.27987470E+01 | -1.25534473E+01 | -2.50E-01 |
| 2459 | Tl | 182 | 81 | 101 | 20 | 7.79625100E+00 | 7.79770578E+00 | -1.50E-03 | 1.69476669E+05 | 1.69476346E+05 | 3.20E-01 | 1.69518611E+05 | 1.69518287E+05 | 3.20E-01 | -1.33081060E+01 | -1.36316313E+01 | 3.20E-01 |
| 2460 | Tl | 183 | 81 | 102 | 21 | 7.81567300E+00 | 7.81425220E+00 | 1.40E-03 | 1.70404884E+05 | 1.70405085E+05 | -2.00E-01 | 1.70446826E+05 | 1.70447027E+05 | -2.00E-01 | -1.65872630E+01 | -1.63860120E+01 | -2.00E-01 |
| 2461 | Tl | 184 | 81 | 103 | 22 | 7.81861700E+00 | 7.81998207E+00 | -1.40E-03 | 1.71336092E+05 | 1.71335782E+05 | 3.10E-01 | 1.71378034E+05 | 1.71377724E+05 | 3.10E-01 | -1.68732210E+01 | -1.71832415E+01 | 3.10E-01 |
| 2462 | Tl | 185 | 81 | 104 | 23 | 7.83557600E+00 | 7.83435399E+00 | 1.20E-03 | 1.72264701E+05 | 1.72264869E+05 | -1.70E-01 | 1.72306644E+05 | 1.72306811E+05 | -1.70E-01 | -1.97578720E+01 | -1.95907111E+01 | -1.70E-01 |
| 2463 | Tl | 186 | 81 | 105 | 24 | 7.83753500E+00 | 7.83845002E+00 | -9.20E-04 | 1.73196067E+05 | 1.73195838E+05 | 2.30E-01 | 1.73238009E+05 | 1.73237780E+05 | 2.30E-01 | -1.98866130E+01 | -2.01156070E+01 | 2.30E-01 |
| 2464 | Tl | 187 | 81 | 106 | 25 | 7.85245600E+00 | 7.85093843E+00 | 1.50E-03 | 1.74125004E+05 | 1.74125229E+05 | -2.30E-01 | 1.74166946E+05 | 1.74167171E+05 | -2.30E-01 | -2.24430920E+01 | -2.22180708E+01 | -2.30E-01 |
| 2465 | Tl | 188 | 81 | 107 | 26 | 7.85305300E+00 | 7.85339774E+00 | -3.40E-04 | 1.75056605E+05 | 1.75056482E+05 | 1.20E-01 | 1.75098547E+05 | 1.75098423E+05 | 1.20E-01 | -2.23364000E+01 | -2.24600401E+01 | 1.20E-01 |
| 2466 | Tl | 189 | 81 | 108 | 27 | 7.86619600E+00 | 7.86394920E+00 | 2.20E-03 | 1.75985834E+05 | 1.75986199E+05 | -3.70E-01 | 1.76027775E+05 | 1.76028141E+05 | -3.70E-01 | -2.46021810E+01 | -2.42363440E+01 | -3.70E-01 |
| 2467 | Tl | 190 | 81 | 109 | 28 | 7.86600000E+00 | 7.86487245E+00 | 1.10E-03 | 1.76917551E+05 | 1.76917725E+05 | -1.70E-01 | 1.76959492E+05 | 1.76959667E+05 | -1.70E-01 | -2.43790000E+01 | -2.42043929E+01 | -1.70E-01 |
| 2468 | Tl | 191 | 81 | 110 | 29 | 7.87714300E+00 | 7.87362381E+00 | 3.50E-03 | 1.77847141E+05 | 1.77847754E+05 | -6.10E-01 | 1.77889083E+05 | 1.77889696E+05 | -6.10E-01 | -2.62828490E+01 | -2.56694563E+01 | -6.10E-01 |
| 2469 | Tl | 192 | 81 | 111 | 30 | 7.87601600E+00 | 7.87316537E+00 | 2.90E-03 | 1.78779046E+05 | 1.78779534E+05 | -4.90E-01 | 1.78820987E+05 | 1.78821476E+05 | -4.90E-01 | -2.58722460E+01 | -2.53837402E+01 | -4.90E-01 |
| 2470 | Tl | 193 | 81 | 112 | 31 | 7.88534400E+00 | 7.88032413E+00 | 5.00E-03 | 1.79708935E+05 | 1.79709845E+05 | -9.10E-01 | 1.79750877E+05 | 1.79751787E+05 | -9.10E-01 | -2.74772120E+01 | -2.65672278E+01 | -9.10E-01 |
| 2471 | Tl | 194 | 81 | 113 | 32 | 7.88352000E+00 | 7.87872179E+00 | 4.80E-03 | 1.80640968E+05 | 1.80641841E+05 | -8.70E-01 | 1.80682910E+05 | 1.80683782E+05 | -8.70E-01 | -2.69374910E+01 | -2.60653774E+01 | -8.70E-01 |
| 2472 | Tl | 195 | 81 | 114 | 33 | 7.89072700E+00 | 7.88458462E+00 | 6.10E-03 | 1.81571245E+05 | 1.81572384E+05 | -1.10E+00 | 1.81613187E+05 | 1.81614326E+05 | -1.10E+00 | -2.81550880E+01 | -2.70160324E+01 | -1.10E+00 |
| 2473 | Tl | 196 | 81 | 115 | 34 | 7.88828900E+00 | 7.88215726E+00 | 6.10E-03 | 1.82503397E+05 | 1.82504541E+05 | -1.10E+00 | 1.82545339E+05 | 1.82546482E+05 | -1.10E+00 | -2.74965910E+01 | -2.63535369E+01 | -1.10E+00 |
| 2474 | Tl | 197 | 81 | 116 | 35 | 7.89349900E+00 | 7.88707389E+00 | 6.40E-03 | 1.83434048E+05 | 1.83435255E+05 | -1.20E+00 | 1.83475990E+05 | 1.83477197E+05 | -1.20E+00 | -2.83398750E+01 | -2.71329498E+01 | -1.20E+00 |
| 2475 | Tl | 198 | 81 | 117 | 36 | 7.89012900E+00 | 7.88413691E+00 | 6.00E-03 | 1.84366387E+05 | 1.84367515E+05 | -1.10E+00 | 1.84408329E+05 | 1.84409457E+05 | -1.10E+00 | -2.74948470E+01 | -2.63671829E+01 | -1.10E+00 |
| 2476 | Tl | 199 | 81 | 118 | 37 | 7.89387700E+00 | 7.88836150E+00 | 5.50E-03 | 1.85297317E+05 | 1.85298356E+05 | -1.00E+00 | 1.85339259E+05 | 1.85340297E+05 | -1.00E+00 | -2.80593940E+01 | -2.70206945E+01 | -1.00E+00 |
| 2477 | Tl | 200 | 81 | 119 | 38 | 7.88970500E+00 | 7.88500310E+00 | 4.70E-03 | 1.86229823E+05 | 1.86230704E+05 | -8.80E-01 | 1.86271765E+05 | 1.86272646E+05 | -8.80E-01 | -2.70475510E+01 | -2.61660575E+01 | -8.80E-01 |
| 2478 | Tl | 201 | 81 | 120 | 39 | 7.89126200E+00 | 7.88839563E+00 | 2.90E-03 | 1.87161185E+05 | 1.87161703E+05 | -5.20E-01 | 1.87203127E+05 | 1.87203645E+05 | -5.20E-01 | -2.71789340E+01 | -2.66616392E+01 | -5.20E-01 |
| 2479 | Tl | 202 | 81 | 121 | 40 | 7.88625000E+00 | 7.88412663E+00 | 2.10E-03 | 1.88093872E+05 | 1.88094242E+05 | -3.70E-01 | 1.88135814E+05 | 1.88136184E+05 | -3.70E-01 | -2.59864090E+01 | -2.56163785E+01 | -3.70E-01 |
| 2480 | Tl | 203 | 81 | 122 | 41 | 7.88605000E+00 | 7.88581042E+00 | 2.40E-04 | 1.89025592E+05 | 1.89025581E+05 | 1.00E-02 | 1.89067533E+05 | 1.89067523E+05 | 1.00E-02 | -2.57608460E+01 | -2.57709950E+01 | 1.00E-02 |
| 2481 | Tl | 204 | 81 | 123 | 42 | 7.88002100E+00 | 7.87933407E+00 | 6.90E-04 | 1.89958501E+05 | 1.89958582E+05 | -8.10E-02 | 1.90000443E+05 | 1.90000524E+05 | -8.10E-02 | -2.43456200E+01 | -2.42643122E+01 | -8.10E-02 |
| 2482 | Tl | 205 | 81 | 124 | 43 | 7.87839200E+00 | 7.87771719E+00 | 6.70E-04 | 1.90890520E+05 | 1.90890600E+05 | -8.00E-02 | 1.90932462E+05 | 1.90932542E+05 | -7.90E-02 | -2.38203490E+01 | -2.37408653E+01 | -7.90E-02 |
| 2483 | Tl | 206 | 81 | 125 | 44 | 7.87171900E+00 | 7.86733584E+00 | 4.40E-03 | 1.91823582E+05 | 1.91824426E+05 | -8.40E-01 | 1.91865524E+05 | 1.91866368E+05 | -8.40E-01 | -2.22528430E+01 | -2.14087052E+01 | -8.40E-01 |
| 2484 | Tl | 207 | 81 | 126 | 45 | 7.86679200E+00 | 7.86095539E+00 | 5.80E-03 | 1.92756295E+05 | 1.92757445E+05 | -1.10E+00 | 1.92798237E+05 | 1.92799387E+05 | -1.10E+00 | -2.10334020E+01 | -1.98839703E+01 | -1.10E+00 |

| | | | | | | | | | | | | | | | | |
|---|---|---|---|---|---|---|---|---|---|---|---|---|---|---|---|---|
| 2485 | Tl | 208 | 81 | 127 | 46 | 7.84717800E+00 | 7.84574488E+00 | 1.40E-03 | 1.93692074E+05 | 1.93692313E+05 | -2.40E-01 | 1.93734016E+05 | 1.93734255E+05 | -2.40E-01 | -1.67491920E+01 | -1.65098201E+01 | -2.40E-01 |
| 2486 | Tl | 209 | 81 | 128 | 47 | 7.83336300E+00 | 7.83447910E+00 | -1.10E-03 | 1.94626679E+05 | 1.94626387E+05 | 2.90E-01 | 1.94668621E+05 | 1.94668329E+05 | 2.90E-01 | -1.36376740E+01 | -1.39296985E+01 | 2.90E-01 |
| 2487 | Tl | 210 | 81 | 129 | 48 | 7.81358400E+00 | 7.81514582E+00 | -1.60E-03 | 1.95562565E+05 | 1.95562178E+05 | 3.90E-01 | 1.95604507E+05 | 1.95604120E+05 | 3.90E-01 | -9.24604800E+00 | -9.63286919E+00 | 3.90E-01 |
| 2488 | TL | 211 | 81 | 130 | 49 | 7.79979100E+00 | 7.80039914E+00 | -6.10E-04 | 1.96497227E+05 | 1.96497040E+05 | 1.90E-01 | 1.96539169E+05 | 1.96538982E+05 | 1.90E-01 | -6.07799800E+00 | -6.26514742E+00 | 1.90E-01 |
| 2489 | TL | 212 | 81 | 131 | 50 | 7.78000000E+00 | 7.77872069E+00 | 1.30E-03 | 1.97433248E+05 | 1.97433401E+05 | -1.50E-01 | 1.97475190E+05 | 1.97475343E+05 | -1.50E-01 | -1.55100000E+00 | -1.39839559E+00 | -1.50E-01 |
| 2490 | TL | 213 | 81 | 132 | 51 | 7.76543000E+00 | 7.76233206E+00 | 3.10E-03 | 1.98368077E+05 | 1.98368678E+05 | -6.00E-01 | 1.98410019E+05 | 1.98410620E+05 | -6.00E-01 | 1.78381100E+00 | 2.38498102E+00 | -6.00E-01 |
| 2491 | TL | 214 | 81 | 133 | 52 | 7.74500000E+00 | 7.73991366E+00 | 5.10E-03 | 1.99304252E+05 | 1.99305279E+05 | -1.00E+00 | 1.99346194E+05 | 1.99347221E+05 | -1.00E+00 | 6.46500000E+00 | 7.49150489E+00 | -1.00E+00 |
| 2492 | TL | 215 | 81 | 134 | 53 | 7.73000000E+00 | 7.72208620E+00 | 7.90E-03 | 2.00239192E+05 | 2.00240937E+05 | -1.70E+00 | 2.00281134E+05 | 2.00282879E+05 | -1.70E+00 | 9.91100000E+00 | 1.16558140E+01 | -1.70E+00 |
| 2493 | TL | 216 | 81 | 135 | 54 | 7.71000000E+00 | 7.70462109E+00 | 5.40E-03 | 2.01175493E+05 | 2.01176553E+05 | -1.10E+00 | 2.01217435E+05 | 2.01218495E+05 | -1.10E+00 | 1.47180000E+01 | 1.57775121E+01 | -1.10E+00 |
| 2494 | TL | 217 | 81 | 136 | 55 | 7.69500000E+00 | 7.69181961E+00 | 3.20E-03 | 2.02110582E+05 | 2.02111191E+05 | -6.10E-01 | 2.02152524E+05 | 2.02153133E+05 | -6.10E-01 | 1.83130000E+01 | 1.89221313E+01 | -6.10E-01 |
| 2495 | PB | 178 | 82 | 96 | 14 | 7.69086100E+00 | 7.69145272E+00 | -5.90E-04 | 1.65767041E+05 | 1.65766876E+05 | 1.60E-01 | 1.65809512E+05 | 1.65809347E+05 | 1.60E-01 | 3.56878000E+00 | 3.40388401E+00 | 1.60E-01 |
| 2496 | PB | 179 | 82 | 97 | 15 | 7.70146900E+00 | 7.70260989E+00 | -1.10E-03 | 1.66697017E+05 | 1.66696753E+05 | 2.60E-01 | 1.66739487E+05 | 1.66739224E+05 | 2.60E-01 | 2.05032300E+00 | 1.78661586E+00 | 2.60E-01 |
| 2497 | Pb | 180 | 82 | 98 | 16 | 7.72563600E+00 | 7.72595640E+00 | -3.20E-04 | 1.67624531E+05 | 1.67624414E+05 | 1.20E-01 | 1.67667001E+05 | 1.67666884E+05 | 1.20E-01 | -1.92978900E+00 | -2.04704678E+00 | 1.20E-01 |
| 2498 | Pb | 181 | 82 | 99 | 17 | 7.73410700E+00 | 7.73522815E+00 | -1.10E-03 | 1.68554837E+05 | 1.68554575E+05 | 2.60E-01 | 1.68597308E+05 | 1.68597045E+05 | 2.60E-01 | -3.11742100E+00 | -3.37987090E+00 | 2.60E-01 |
| 2499 | Pb | 182 | 82 | 100 | 18 | 7.75633400E+00 | 7.75633323E+00 | 7.70E-07 | 1.69482623E+05 | 1.69482564E+05 | 5.90E-02 | 1.69525094E+05 | 1.69525034E+05 | 5.90E-02 | -6.82555400E+00 | -6.88490300E+00 | 5.90E-02 |
| 2500 | Pb | 183 | 82 | 101 | 19 | 7.76213000E+00 | 7.76378322E+00 | -1.70E-03 | 1.70413372E+05 | 1.70413010E+05 | 3.60E-01 | 1.70455842E+05 | 1.70455480E+05 | 3.60E-01 | -7.57109200E+00 | -7.93326540E+00 | 3.60E-01 |
| 2501 | Pb | 184 | 82 | 102 | 20 | 7.78272500E+00 | 7.78272007E+00 | 4.90E-06 | 1.71341385E+05 | 1.71341327E+05 | 5.90E-02 | 1.71383856E+05 | 1.71383797E+05 | 5.90E-02 | -1.10515230E+01 | -1.11101116E+01 | 5.90E-02 |
| 2502 | Pb | 185 | 82 | 103 | 21 | 7.78693200E+00 | 7.78840629E+00 | -1.50E-03 | 1.72272390E+05 | 1.72272058E+05 | 3.30E-01 | 1.72314860E+05 | 1.72314528E+05 | 3.30E-01 | -1.15412070E+01 | -1.18734632E+01 | 3.30E-01 |
| 2503 | Pb | 186 | 82 | 104 | 22 | 7.80534700E+00 | 7.80515110E+00 | 2.00E-04 | 1.73200743E+05 | 1.73200720E+05 | 2.30E-02 | 1.73243213E+05 | 1.73243190E+05 | 2.30E-02 | -1.46819000E+01 | -1.47050841E+01 | 2.30E-02 |
| 2504 | Pb | 187 | 82 | 105 | 23 | 7.80840000E+00 | 7.80919958E+00 | -8.00E-04 | 1.74131932E+05 | 1.74131723E+05 | 2.10E-01 | 1.74174402E+05 | 1.74174193E+05 | 2.10E-01 | -1.49869050E+01 | -1.51959816E+01 | 2.10E-01 |
| 2505 | Pb | 188 | 82 | 106 | 24 | 7.82484100E+00 | 7.82408608E+00 | 7.50E-04 | 1.75060598E+05 | 1.75060681E+05 | -8.20E-02 | 1.75103068E+05 | 1.75103151E+05 | -8.20E-02 | -1.78149950E+01 | -1.77325244E+01 | -8.20E-02 |
| 2506 | Pb | 189 | 82 | 107 | 25 | 7.82648000E+00 | 7.82650578E+00 | -2.60E-05 | 1.75992029E+05 | 1.75991965E+05 | 6.40E-02 | 1.76034499E+05 | 1.76034435E+05 | 6.40E-02 | -1.78781640E+01 | -1.79426147E+01 | 6.40E-02 |
| 2507 | Pb | 190 | 82 | 108 | 26 | 7.84112800E+00 | 7.83948650E+00 | 1.60E-03 | 1.76920985E+05 | 1.76921237E+05 | -2.50E-01 | 1.76963455E+05 | 1.76963707E+05 | -2.50E-01 | -2.04164860E+01 | -2.01641399E+01 | -2.50E-01 |
| 2508 | Pb | 191 | 82 | 109 | 27 | 7.84138700E+00 | 7.84039931E+00 | 9.90E-04 | 1.77852660E+05 | 1.77852789E+05 | -1.30E-01 | 1.77895130E+05 | 1.77895259E+05 | -1.30E-01 | -2.02357760E+01 | -2.01066535E+01 | -1.30E-01 |
| 2509 | Pb | 192 | 82 | 110 | 28 | 7.85471800E+00 | 7.85163516E+00 | 3.10E-03 | 1.78781824E+05 | 1.78782356E+05 | -5.30E-01 | 1.78824294E+05 | 1.78824827E+05 | -5.30E-01 | -2.25654340E+01 | -2.20330176E+01 | -5.30E-01 |
| 2510 | Pb | 193 | 82 | 111 | 29 | 7.85391900E+00 | 7.85121804E+00 | 2.70E-03 | 1.79713689E+05 | 1.79714151E+05 | -4.60E-01 | 1.79756159E+05 | 1.79756621E+05 | -4.60E-01 | -2.21944900E+01 | -2.17328284E+01 | -4.60E-01 |
| 2511 | Pb | 194 | 82 | 112 | 30 | 7.86541500E+00 | 7.86093780E+00 | 4.50E-03 | 1.80643170E+05 | 1.80643979E+05 | -8.10E-01 | 1.80685640E+05 | 1.80686449E+05 | -8.10E-01 | -2.42074330E+01 | -2.33983614E+01 | -8.10E-01 |
| 2512 | Pb | 195 | 82 | 113 | 31 | 7.86393600E+00 | 7.85944260E+00 | 4.50E-03 | 1.81575158E+05 | 1.81575975E+05 | -8.20E-01 | 1.81617629E+05 | 1.81618445E+05 | -8.20E-01 | -2.37131150E+01 | -2.28964165E+01 | -8.20E-01 |
| 2513 | Pb | 196 | 82 | 114 | 32 | 7.87340000E+00 | 7.86794693E+00 | 5.50E-03 | 1.82505005E+05 | 1.82506014E+05 | -1.00E+00 | 1.82547475E+05 | 1.82548485E+05 | -1.00E+00 | -2.53606380E+01 | -2.43513888E+01 | -1.00E+00 |
| 2514 | Pb | 197 | 82 | 115 | 33 | 7.87129800E+00 | 7.86567937E+00 | 5.60E-03 | 1.83437111E+05 | 1.83438158E+05 | -1.00E+00 | 1.83479581E+05 | 1.83480629E+05 | -1.00E+00 | -2.47487090E+01 | -2.37013064E+01 | -1.00E+00 |
| 2515 | Pb | 198 | 82 | 116 | 34 | 7.87888100E+00 | 7.87328327E+00 | 5.60E-03 | 1.84367304E+05 | 1.84368353E+05 | -1.00E+00 | 1.84409774E+05 | 1.84410823E+05 | -1.00E+00 | -2.60500890E+01 | -2.50012388E+01 | -1.00E+00 |
| 2516 | Pb | 199 | 82 | 117 | 35 | 7.87573600E+00 | 7.87049740E+00 | 5.20E-03 | 1.85299616E+05 | 1.85300599E+05 | -9.80E-01 | 1.85342087E+05 | 1.85343069E+05 | -9.80E-01 | -2.52316760E+01 | -2.42488165E+01 | -9.80E-01 |
| 2517 | Pb | 200 | 82 | 118 | 36 | 7.88180800E+00 | 7.87736644E+00 | 4.40E-03 | 1.86230091E+05 | 1.86230920E+05 | -8.30E-01 | 1.86272562E+05 | 1.86273390E+05 | -8.30E-01 | -2.62505320E+01 | -2.54218016E+01 | -8.30E-01 |
| 2518 | Pb | 201 | 82 | 119 | 37 | 7.87782000E+00 | 7.87403874E+00 | 3.80E-03 | 1.87162577E+05 | 1.87163277E+05 | -7.00E-01 | 1.87205047E+05 | 1.87205747E+05 | -7.00E-01 | -2.52593990E+01 | -2.45589818E+01 | -7.00E-01 |
| 2519 | Pb | 202 | 82 | 120 | 38 | 7.88214800E+00 | 7.87990939E+00 | 2.20E-03 | 1.88093390E+05 | 1.88093782E+05 | -3.90E-01 | 1.88135860E+05 | 1.88136253E+05 | -3.90E-01 | -2.59402210E+01 | -2.55475733E+01 | -3.90E-01 |
| 2520 | Pb | 203 | 82 | 121 | 39 | 7.87739400E+00 | 7.87543679E+00 | 2.00E-03 | 1.89026038E+05 | 1.89026376E+05 | -3.40E-01 | 1.89068508E+05 | 1.89068846E+05 | -3.40E-01 | -2.47860060E+01 | -2.44482256E+01 | -3.40E-01 |
| 2521 | Pb | 204 | 82 | 122 | 40 | 7.87993000E+00 | 7.87935510E+00 | 5.70E-04 | 1.89957209E+05 | 1.89957266E+05 | -5.80E-02 | 1.89999679E+05 | 1.89999737E+05 | -5.80E-02 | -2.51093700E+01 | -2.50516786E+01 | -5.80E-02 |
| 2522 | Pb | 205 | 82 | 123 | 41 | 7.87432800E+00 | 7.87243226E+00 | 1.90E-03 | 1.90890043E+05 | 1.90890372E+05 | -3.30E-01 | 1.90932513E+05 | 1.90932842E+05 | -3.30E-01 | -2.37697170E+01 | -2.34405327E+01 | -3.30E-01 |
| 2523 | Pb | 206 | 82 | 124 | 42 | 7.87535900E+00 | 7.87288770E+00 | 2.50E-03 | 1.91821521E+05 | 1.91821971E+05 | -4.50E-01 | 1.91863992E+05 | 1.91864441E+05 | -4.50E-01 | -2.37850620E+01 | -2.33354669E+01 | -4.50E-01 |
| 2524 | Pb | 207 | 82 | 125 | 43 | 7.86986400E+00 | 7.86199469E+00 | 7.90E-03 | 1.92754349E+05 | 1.92755918E+05 | -1.60E+00 | 1.92796819E+05 | 1.92798388E+05 | -1.60E+00 | -2.24515230E+01 | -2.08821812E+01 | -1.60E+00 |
| 2525 | Pb | 208 | 82 | 126 | 44 | 7.86745000E+00 | 7.85777196E+00 | 9.70E-03 | 1.93686546E+05 | 1.93688500E+05 | -2.00E+00 | 1.93729017E+05 | 1.93730970E+05 | -2.00E+00 | -2.17480740E+01 | -1.97945286E+01 | -2.00E+00 |
| 2526 | Pb | 209 | 82 | 127 | 45 | 7.84864600E+00 | 7.84225517E+00 | 6.40E-03 | 1.94622174E+05 | 1.94623451E+05 | -1.30E+00 | 1.94664645E+05 | 1.94665921E+05 | -1.30E+00 | -1.76141220E+01 | -1.63379732E+01 | -1.30E+00 |
| 2527 | Pb | 210 | 82 | 128 | 46 | 7.83596300E+00 | 7.83346099E+00 | 2.50E-03 | 1.95556555E+05 | 1.95557020E+05 | -4.70E-01 | 1.95599025E+05 | 1.95599491E+05 | -4.70E-01 | -1.47279810E+01 | -1.42621320E+01 | -4.70E-01 |
| 2528 | Pb | 211 | 82 | 129 | 47 | 7.81699900E+00 | 7.81416531E+00 | 2.80E-03 | 1.96492285E+05 | 1.96492824E+05 | -5.40E-01 | 1.96534755E+05 | 1.96535294E+05 | -5.40E-01 | -1.04913000E+01 | -9.95288554E+00 | -5.40E-01 |
| 2529 | Pb | 212 | 82 | 130 | 48 | 7.80431200E+00 | 7.80222474E+00 | 2.10E-03 | 1.97426723E+05 | 1.97427106E+05 | -3.80E-01 | 1.97469194E+05 | 1.97469577E+05 | -3.80E-01 | -7.54724100E+00 | -7.16433129E+00 | -3.80E-01 |
| 2530 | Pb | 213 | 82 | 131 | 49 | 7.78516300E+00 | 7.78084366E+00 | 4.30E-03 | 1.98362563E+05 | 1.98363424E+05 | -8.60E-01 | 1.98405033E+05 | 1.98405894E+05 | -8.60E-01 | -3.20164300E+00 | -2.34106723E+00 | -8.60E-01 |
| 2531 | Pb | 214 | 82 | 132 | 50 | 7.77238400E+00 | 7.76743506E+00 | 4.90E-03 | 1.99297078E+05 | 1.99298078E+05 | -1.00E+00 | 1.99339548E+05 | 1.99340548E+05 | -1.00E+00 | -1.80786000E-01 | 8.18849264E-01 | -1.00E+00 |
| 2532 | PB | 215 | 82 | 133 | 51 | 7.75200000E+00 | 7.74538309E+00 | 6.60E-03 | 2.00233169E+05 | 2.00234617E+05 | -1.40E+00 | 2.00275639E+05 | 2.00277087E+05 | -1.40E+00 | 4.41600000E+00 | 5.86390737E+00 | -1.40E+00 |
| 2533 | PB | 216 | 82 | 134 | 52 | 7.74000000E+00 | 7.73052316E+00 | 9.50E-03 | 2.01167727E+05 | 2.01169646E+05 | -1.90E+00 | 2.01210197E+05 | 2.01212117E+05 | -1.90E+00 | 7.48000000E+00 | 9.39958778E+00 | -1.90E+00 |
| 2534 | PB | 217 | 82 | 135 | 53 | 7.71900000E+00 | 7.71269389E+00 | 6.30E-03 | 2.02103981E+05 | 2.02105350E+05 | -1.40E+00 | 2.02146451E+05 | 2.02147821E+05 | -1.40E+00 | 1.22400000E+01 | 1.36093340E+01 | -1.40E+00 |
| 2535 | PB | 218 | 82 | 136 | 54 | 7.70600000E+00 | 7.70216843E+00 | 3.80E-03 | 2.03038688E+05 | 2.03039498E+05 | -8.10E-01 | 2.03081159E+05 | 2.03081968E+05 | -8.10E-01 | 1.54530000E+01 | 1.62625110E+01 | -8.10E-01 |
| 2536 | PB | 219 | 82 | 137 | 55 | 7.68600000E+00 | 7.68263207E+00 | 3.40E-03 | 2.03975008E+05 | 2.03975639E+05 | -6.30E-01 | 2.04017478E+05 | 2.04018109E+05 | -6.30E-01 | 2.02790000E+01 | 2.09101236E+01 | -6.30E-01 |
| 2537 | PB | 220 | 82 | 138 | 56 | 7.67200000E+00 | 7.67008726E+00 | 1.90E-03 | 2.04909892E+05 | 2.04910282E+05 | -3.90E-01 | 2.04952363E+05 | 2.04952752E+05 | -3.90E-01 | 2.36690000E+01 | 2.40586680E+01 | -3.90E-01 |
| 2538 | Bi | 184 | 83 | 101 | 18 | 7.71195700E+00 | 7.71603081E+00 | -4.10E-03 | 1.71353096E+05 | 1.71352286E+05 | 8.10E-01 | 1.71396095E+05 | 1.71395285E+05 | 8.10E-01 | 1.18742800E+00 | 3.77635175E-01 | 8.10E-01 |
| 2539 | Bi | 185 | 83 | 102 | 19 | 7.73200000E+00 | 7.73520627E+00 | -3.20E-03 | 1.72281167E+05 | 1.72280588E+05 | 5.80E-01 | 1.72324166E+05 | 1.72323587E+05 | 5.80E-01 | -2.23600000E+00 | -2.81453531E+00 | 5.80E-01 |
| 2540 | Bi | 186 | 83 | 103 | 20 | 7.73901400E+00 | 7.74337743E+00 | -4.40E-03 | 1.73211770E+05 | 1.73210898E+05 | 8.70E-01 | 1.73254769E+05 | 1.73253897E+05 | 8.70E-01 | -3.12639100E+00 | -3.99825945E+00 | 8.70E-01 |
| 2541 | Bi | 187 | 83 | 104 | 21 | 7.75820800E+00 | 7.76032170E+00 | -2.10E-03 | 1.74140007E+05 | 1.74139552E+05 | 4.60E-01 | 1.74183006E+05 | 1.74182551E+05 | 4.60E-01 | -6.38335600E+00 | -6.83889592E+00 | 4.60E-01 |
| 2542 | Bi | 188 | 83 | 105 | 22 | 7.76413600E+00 | 7.76684952E+00 | -2.70E-03 | 1.75070700E+05 | 1.75070129E+05 | 5.70E-01 | 1.75113699E+05 | 1.75113128E+05 | 5.70E-01 | -7.18466600E+00 | -7.75512880E+00 | 5.70E-01 |
| 2543 | Bi | 189 | 83 | 106 | 23 | 7.78100000E+00 | 7.78194042E+00 | -9.40E-04 | 1.75999314E+05 | 1.75999076E+05 | 2.40E-01 | 1.76042313E+05 | 1.76042075E+05 | 2.40E-01 | -1.00648070E+01 | -1.03028402E+01 | 2.40E-01 |
| 2544 | Bi | 190 | 83 | 107 | 24 | 7.78533900E+00 | 7.78684970E+00 | -1.50E-03 | 1.76930274E+05 | 1.76929926E+05 | 3.50E-01 | 1.76973273E+05 | 1.76972925E+05 | 3.50E-01 | -1.05989470E+01 | -1.09462234E+01 | 3.50E-01 |
| 2545 | Bi | 191 | 83 | 108 | 25 | 7.80066300E+00 | 7.80004331E+00 | 6.20E-04 | 1.77859127E+05 | 1.77859185E+05 | -5.80E-02 | 1.77902126E+05 | 1.77902184E+05 | -5.80E-02 | -1.32397370E+01 | -1.31817343E+01 | -5.80E-02 |
| 2546 | Bi | 192 | 83 | 109 | 26 | 7.80361300E+00 | 7.80347970E+00 | 1.30E-04 | 1.78790325E+05 | 1.78790290E+05 | 3.50E-02 | 1.78833324E+05 | 1.78833289E+05 | 3.50E-02 | -1.35354850E+01 | -1.35702461E+01 | 3.50E-02 |
| 2547 | Bi | 193 | 83 | 110 | 27 | 7.81711000E+00 | 7.81495731E+00 | 2.20E-03 | 1.79719482E+05 | 1.79719837E+05 | -3.60E-01 | 1.79762481E+05 | 1.79762836E+05 | -3.60E-01 | -1.58728320E+01 | -1.55175856E+01 | -3.60E-01 |
| 2548 | Bi | 194 | 83 | 111 | 28 | 7.81900000E+00 | 7.81711855E+00 | 1.90E-03 | 1.80650813E+05 | 1.80651168E+05 | -3.60E-01 | 1.80693812E+05 | 1.80694167E+05 | -3.60E-01 | -1.60360000E+01 | -1.56805038E+01 | -3.60E-01 |
| 2549 | Bi | 195 | 83 | 112 | 29 | 7.83075700E+00 | 7.82712130E+00 | 3.60E-03 | 1.81580317E+05 | 1.81580966E+05 | -6.50E-01 | 1.81623316E+05 | 1.81623965E+05 | -6.50E-01 | -1.80255440E+01 | -1.73768387E+01 | -6.50E-01 |
| 2550 | Bi | 196 | 83 | 113 | 30 | 7.83190000E+00 | 7.82826412E+00 | 3.60E-03 | 1.82511828E+05 | 1.82512480E+05 | -6.50E-01 | 1.82554827E+05 | 1.82555479E+05 | -6.50E-01 | -1.80090310E+01 | -1.73566339E+01 | -6.50E-01 |
| 2551 | Bi | 197 | 83 | 114 | 31 | 7.84163300E+00 | 7.83708063E+00 | 4.60E-03 | 1.83441644E+05 | 1.83442481E+05 | -8.40E-01 | 1.83484643E+05 | 1.83485480E+05 | -8.40E-01 | -1.96870330E+01 | -1.88504317E+01 | -8.40E-01 |
| 2552 | Bi | 198 | 83 | 115 | 32 | 7.84118900E+00 | 7.83748133E+00 | 3.70E-03 | 1.84373456E+05 | 1.84374130E+05 | -6.70E-01 | 1.84416455E+05 | 1.84417129E+05 | -6.70E-01 | -1.93694860E+01 | -1.86955316E+01 | -6.70E-01 |
| 2553 | Bi | 199 | 83 | 116 | 33 | 7.84952200E+00 | 7.84537262E+00 | 4.10E-03 | 1.85303522E+05 | 1.85304287E+05 | -7.70E-01 | 1.85346521E+05 | 1.85347286E+05 | -7.70E-01 | -2.07974570E+01 | -2.00320618E+01 | -7.70E-01 |
| 2554 | Bi | 200 | 83 | 117 | 34 | 7.84849700E+00 | 7.84520903E+00 | 3.30E-03 | 1.86235443E+05 | 1.86236040E+05 | -6.00E-01 | 1.86278441E+05 | 1.86279039E+05 | -6.00E-01 | -2.03706750E+01 | -1.97733971E+01 | -6.00E-01 |
| 2555 | Bi | 201 | 83 | 118 | 35 | 7.85479900E+00 | 7.85224302E+00 | 2.60E-03 | 1.87165893E+05 | 1.87166346E+05 | -4.50E-01 | 1.87208892E+05 | 1.87209345E+05 | -4.50E-01 | -2.14146600E+01 | -2.09611199E+01 | -4.50E-01 |
| 2556 | Bi | 202 | 83 | 119 | 36 | 7.85253500E+00 | 7.85138424E+00 | 1.20E-03 | 1.88098061E+05 | 1.88098233E+05 | -1.70E-01 | 1.88141060E+05 | 1.88141232E+05 | -1.70E-01 | -2.07407750E+01 | -2.05685696E+01 | -1.70E-01 |
| 2557 | Bi | 203 | 83 | 120 | 37 | 7.85747200E+00 | 7.85719564E+00 | 2.80E-04 | 1.89028771E+05 | 1.89028767E+05 | 4.20E-03 | 1.89071770E+05 | 1.89071766E+05 | 4.20E-03 | -2.15242370E+01 | -2.15283488E+01 | 4.10E-03 |
| 2558 | Bi | 204 | 83 | 121 | 38 | 7.85421300E+00 | 7.85496711E+00 | -7.50E-04 | 1.89961144E+05 | 1.89960930E+05 | 2.10E-01 | 1.90004143E+05 | 1.90003929E+05 | 2.10E-01 | -2.06455750E+01 | -2.08596063E+01 | 2.10E-01 |
| 2559 | Bi | 205 | 83 | 122 | 39 | 7.85731400E+00 | 7.85858773E+00 | -1.30E-03 | 1.90892219E+05 | 1.90891898E+05 | 3.20E-01 | 1.90935218E+05 | 1.90934897E+05 | 3.20E-01 | -2.10641590E+01 | -2.13854811E+01 | 3.20E-01 |
| 2560 | Bi | 206 | 83 | 123 | 40 | 7.85332200E+00 | 7.85375118E+00 | -4.30E-04 | 1.91824750E+05 | 1.91824601E+05 | 1.50E-01 | 1.91867749E+05 | 1.91867600E+05 | 1.50E-01 | -2.00277570E+01 | -2.01764197E+01 | 1.50E-01 |
| 2561 | Bi | 207 | 83 | 124 | 41 | 7.85450200E+00 | 7.85383045E+00 | 6.70E-04 | 1.92756217E+05 | 1.92756296E+05 | -7.90E-02 | 1.92799217E+05 | 1.92799295E+05 | -7.90E-02 | -2.00541130E+01 | -1.99752613E+01 | -7.90E-02 |
| 2562 | Bi | 208 | 83 | 125 | 42 | 7.84985100E+00 | 7.84510537E+00 | 4.70E-03 | 1.93688896E+05 | 1.93689823E+05 | -9.30E-01 | 1.93731895E+05 | 1.93732822E+05 | -9.30E-01 | -1.88697110E+01 | -1.79429566E+01 | -9.30E-01 |
| 2563 | Bi | 209 | 83 | 126 | 43 | 7.84798400E+00 | 7.84070964E+00 | 7.30E-03 | 1.94621002E+05 | 1.94622462E+05 | -1.50E+00 | 1.94664001E+05 | 1.94665461E+05 | -1.50E+00 | -1.82581500E+01 | -1.67980356E+01 | -1.50E+00 |
| 2564 | Bi | 210 | 83 | 127 | 44 | 7.83254000E+00 | 7.82770502E+00 | 4.80E-03 | 1.95555962E+05 | 1.95556917E+05 | -9.60E-01 | 1.95598961E+05 | 1.95599916E+05 | -9.60E-01 | -1.47914670E+01 | -1.38364547E+01 | -9.60E-01 |
| 2565 | Bi | 211 | 83 | 128 | 45 | 7.81976900E+00 | 7.81912705E+00 | 6.40E-04 | 1.96490390E+05 | 1.96490465E+05 | -7.50E-02 | 1.96533389E+05 | 1.96533464E+05 | -7.50E-02 | -1.18581590E+01 | -1.17828889E+01 | -7.50E-02 |
| 2566 | Bi | 212 | 83 | 129 | 46 | 7.80330900E+00 | 7.80275345E+00 | 5.60E-04 | 1.97425625E+05 | 1.97425682E+05 | -5.80E-02 | 1.97468624E+05 | 1.97468681E+05 | -5.80E-02 | -8.11701400E+00 | -8.05949365E+00 | -5.80E-02 |
| 2567 | Bi | 213 | 83 | 130 | 47 | 7.79101400E+00 | 7.79136091E+00 | -3.50E-04 | 1.98360006E+05 | 1.98359872E+05 | 1.30E-01 | 1.98403005E+05 | 1.98402871E+05 | 1.30E-01 | -5.23027600E+00 | -5.36431889E+00 | 1.30E-01 |
| 2568 | Bi | 214 | 83 | 131 | 48 | 7.77349000E+00 | 7.77315263E+00 | 3.40E-04 | 1.99295530E+05 | 1.99295542E+05 | -1.20E-02 | 1.99338529E+05 | 1.99338541E+05 | -1.20E-02 | -1.19983200E+00 | -1.18778842E+00 | -1.20E-02 |
| 2569 | Bi | 215 | 83 | 132 | 49 | 7.76162700E+00 | 7.76041221E+00 | 1.20E-03 | 2.00229873E+05 | 2.00230074E+05 | -2.00E-01 | 2.00272872E+05 | 2.00273073E+05 | -2.00E-01 | 1.64858000E+00 | 1.84956845E+00 | -2.00E-01 |
| 2570 | Bi | 216 | 83 | 133 | 50 | 7.74349900E+00 | 7.74156031E+00 | 1.90E-03 | 2.01165592E+05 | 2.01165951E+05 | -3.60E-01 | 2.01208591E+05 | 2.01208950E+05 | -3.60E-01 | 5.87399200E+00 | 6.23248546E+00 | -3.60E-01 |
| 2571 | Bi | 217 | 83 | 134 | 51 | 7.73184800E+00 | 7.72729625E+00 | 4.60E-03 | 2.02099942E+05 | 2.02100870E+05 | -9.30E-01 | 2.02142941E+05 | 2.02143869E+05 | -9.30E-01 | 8.72996200E+00 | 9.65754493E+00 | -9.30E-01 |
| 2572 | Bi | 218 | 83 | 135 | 52 | 7.71282700E+00 | 7.71189597E+00 | 9.30E-04 | 2.03035922E+05 | 2.03036065E+05 | -1.40E-01 | 2.03078921E+05 | 2.03079064E+05 | -1.40E-01 | 1.32160370E+01 | 1.33588287E+01 | -1.40E-01 |
| 2573 | BI | 219 | 83 | 136 | 53 | 7.70000000E+00 | 7.70125469E+00 | -1.30E-03 | 2.03970483E+05 | 2.03970249E+05 | 2.30E-01 | 2.04013482E+05 | 2.04013248E+05 | 2.30E-01 | 1.62830000E+01 | 1.60486936E+01 | 2.30E-01 |
| 2574 | BI | 220 | 83 | 137 | 54 | 7.68200000E+00 | 7.68421932E+00 | -2.20E-03 | 2.04906513E+05 | 2.04905861E+05 | 6.50E-01 | 2.04949512E+05 | 2.04948860E+05 | 6.50E-01 | 2.08190000E+01 | 2.01665386E+01 | 6.50E-01 |
| 2575 | BI | 221 | 83 | 138 | 55 | 7.66800000E+00 | 7.67167944E+00 | -3.70E-03 | 2.05841286E+05 | 2.05840513E+05 | 7.70E-01 | 2.05884285E+05 | 2.05883512E+05 | 7.70E-01 | 2.40980000E+01 | 2.33249517E+01 | 7.70E-01 |
| 2576 | BI | 223 | 83 | 140 | 57 | 7.63600000E+00 | 7.63842585E+00 | -2.40E-03 | 2.07712313E+05 | 2.07711716E+05 | 6.00E-01 | 2.07755312E+05 | 2.07754715E+05 | 6.00E-01 | 3.21370000E+01 | 3.15397818E+01 | 6.00E-01 |
| 2577 | BI | 224 | 83 | 141 | 58 | 7.61700000E+00 | 7.61799384E+00 | -9.90E-04 | 2.08648437E+05 | 2.08648220E+05 | 2.20E-01 | 2.08691436E+05 | 2.08691219E+05 | 2.20E-01 | 3.67660000E+01 | 3.65494453E+01 | 2.20E-01 |
| 2578 | Po | 186 | 84 | 102 | 18 | 7.69599800E+00 | 7.69979207E+00 | -3.80E-03 | 1.73218459E+05 | 1.73217693E+05 | 7.70E-01 | 1.73261988E+05 | 1.73261221E+05 | 7.70E-01 | 4.09227800E+00 | 3.32554161E+00 | 7.70E-01 |

| | | | | | | | | | | | | | | | | |
|---|---|---|---|---|---|---|---|---|---|---|---|---|---|---|---|---|
| 2579 | PO | 187 | 84 | 103 | 19 | 7.70474000E+00 | 7.70793578E+00 | -3.20E-03 | 1.74148694E+05 | 1.74148036E+05 | 6.60E-01 | 1.74192222E+05 | 1.74191564E+05 | 6.60E-01 | 2.83272400E+00 | 2.17419368E+00 | 6.60E-01 |
| 2580 | Po | 188 | 84 | 104 | 20 | 7.72465400E+00 | 7.72718654E+00 | -2.50E-03 | 1.75076811E+05 | 1.75076274E+05 | 5.40E-01 | 1.75120339E+05 | 1.75119802E+05 | 5.40E-01 | -5.44393000E-01 | -1.08156461E+00 | 5.40E-01 |
| 2581 | Po | 189 | 84 | 105 | 21 | 7.73113100E+00 | 7.73367481E+00 | -2.50E-03 | 1.76007427E+05 | 1.76006886E+05 | 5.40E-01 | 1.76050956E+05 | 1.76050414E+05 | 5.40E-01 | -1.42199100E+00 | -1.96371587E+00 | 5.40E-01 |
| 2582 | Po | 190 | 84 | 106 | 22 | 7.74945800E+00 | 7.75107945E+00 | -1.60E-03 | 1.76935780E+05 | 1.76935411E+05 | 3.70E-01 | 1.76979308E+05 | 1.76978939E+05 | 3.70E-01 | -4.56379000E+00 | -4.93295386E+00 | 3.70E-01 |
| 2583 | Po | 191 | 84 | 107 | 23 | 7.75378700E+00 | 7.75594897E+00 | -2.20E-03 | 1.77866769E+05 | 1.77866295E+05 | 4.70E-01 | 1.77910297E+05 | 1.77909823E+05 | 4.70E-01 | -5.06876000E+00 | -5.54279272E+00 | 4.70E-01 |
| 2584 | Po | 192 | 84 | 108 | 24 | 7.77107500E+00 | 7.77146768E+00 | -3.90E-04 | 1.78795261E+05 | 1.78795125E+05 | 1.40E-01 | 1.78838789E+05 | 1.78838653E+05 | 1.40E-01 | -8.07053300E+00 | -8.20701484E+00 | 1.40E-01 |
| 2585 | Po | 193 | 84 | 109 | 25 | 7.77412800E+00 | 7.77488313E+00 | -7.60E-04 | 1.79726466E+05 | 1.79726259E+05 | 2.10E-01 | 1.79769994E+05 | 1.79769787E+05 | 2.10E-01 | -8.35952600E+00 | -8.56634575E+00 | 2.10E-01 |
| 2586 | Po | 194 | 84 | 110 | 26 | 7.78929400E+00 | 7.78871620E+00 | 5.80E-04 | 1.80655315E+05 | 1.80655366E+05 | -5.10E-02 | 1.80698843E+05 | 1.80698894E+05 | -5.10E-02 | -1.10045760E+01 | -1.09535245E+01 | -5.10E-02 |
| 2587 | Po | 195 | 84 | 111 | 27 | 7.79102900E+00 | 7.79088765E+00 | 1.40E-04 | 1.81586753E+05 | 1.81586719E+05 | 3.30E-02 | 1.81630281E+05 | 1.81630248E+05 | 3.30E-02 | -1.10609950E+01 | -1.10943546E+01 | 3.30E-02 |
| 2588 | Po | 196 | 84 | 112 | 28 | 7.80481400E+00 | 7.80328168E+00 | 1.50E-03 | 1.82515825E+05 | 1.82516065E+05 | -2.40E-01 | 1.82559353E+05 | 1.82559593E+05 | -2.40E-01 | -1.34825350E+01 | -1.32431526E+01 | -2.40E-01 |
| 2589 | Po | 197 | 84 | 113 | 29 | 7.80553500E+00 | 7.80445705E+00 | 1.10E-03 | 1.83447444E+05 | 1.83447595E+05 | -1.50E-01 | 1.83490972E+05 | 1.83491123E+05 | -1.50E-01 | -1.33580690E+01 | -1.32066643E+01 | -1.50E-01 |
| 2590 | Po | 198 | 84 | 114 | 30 | 7.81755900E+00 | 7.81567633E+00 | 1.90E-03 | 1.84376823E+05 | 1.84377135E+05 | -3.10E-01 | 1.84420351E+05 | 1.84420663E+05 | -3.10E-01 | -1.54729240E+01 | -1.51612188E+01 | -3.10E-01 |
| 2591 | Po | 199 | 84 | 115 | 31 | 7.81753300E+00 | 7.81608644E+00 | 1.40E-03 | 1.85308576E+05 | 1.85308803E+05 | -2.30E-01 | 1.85352104E+05 | 1.85352331E+05 | -2.30E-01 | -1.52139840E+01 | -1.49871895E+01 | -2.30E-01 |
| 2592 | Po | 200 | 84 | 116 | 32 | 7.82750300E+00 | 7.82632673E+00 | 1.20E-03 | 1.86238330E+05 | 1.86238504E+05 | -1.70E-01 | 1.86281858E+05 | 1.86282032E+05 | -1.70E-01 | -1.69543180E+01 | -1.67800133E+01 | -1.70E-01 |
| 2593 | Po | 201 | 84 | 117 | 33 | 7.82658000E+00 | 7.82606523E+00 | 5.10E-04 | 1.87170253E+05 | 1.87170296E+05 | -4.20E-02 | 1.87213781E+05 | 1.87213824E+05 | -4.20E-02 | -1.65248840E+01 | -1.64824609E+01 | -4.20E-02 |
| 2594 | Po | 202 | 84 | 118 | 34 | 7.83471800E+00 | 7.83530048E+00 | -5.80E-04 | 1.88100348E+05 | 1.88100170E+05 | 1.80E-01 | 1.88143876E+05 | 1.88143698E+05 | 1.80E-01 | -1.79241270E+01 | -1.81027268E+01 | 1.80E-01 |
| 2595 | Po | 203 | 84 | 119 | 35 | 7.83286200E+00 | 7.83414509E+00 | -1.30E-03 | 1.89032456E+05 | 1.89032134E+05 | 3.20E-01 | 1.89075984E+05 | 1.89075662E+05 | 3.20E-01 | -1.73107600E+01 | -1.76321644E+01 | 3.20E-01 |
| 2596 | Po | 204 | 84 | 120 | 36 | 7.83908000E+00 | 7.84194145E+00 | -2.90E-03 | 1.89962919E+05 | 1.89962275E+05 | 6.40E-01 | 1.90006448E+05 | 1.90005803E+05 | 6.40E-01 | -1.83408010E+01 | -1.89854478E+01 | 6.40E-01 |
| 2597 | Po | 205 | 84 | 121 | 37 | 7.83615700E+00 | 7.83919487E+00 | -3.00E-03 | 1.90895245E+05 | 1.90894561E+05 | 6.80E-01 | 1.90938773E+05 | 1.90938089E+05 | 6.80E-01 | -1.75092080E+01 | -1.81930216E+01 | 6.80E-01 |
| 2598 | Po | 206 | 84 | 122 | 38 | 7.84059500E+00 | 7.84463048E+00 | -4.00E-03 | 1.91826060E+05 | 1.91825168E+05 | 8.90E-01 | 1.91869588E+05 | 1.91868696E+05 | 8.90E-01 | -1.81883600E+01 | -1.90806331E+01 | 8.90E-01 |
| 2599 | Po | 207 | 84 | 123 | 39 | 7.83667100E+00 | 7.83918524E+00 | -2.50E-03 | 1.92758597E+05 | 1.92758016E+05 | 5.80E-01 | 1.92802125E+05 | 1.92801544E+05 | 5.80E-01 | -1.71453000E+01 | -1.77267794E+01 | 5.80E-01 |
| 2600 | Po | 208 | 84 | 124 | 40 | 7.83935600E+00 | 7.84112292E+00 | -1.80E-03 | 1.93689767E+05 | 1.93689339E+05 | 4.30E-01 | 1.93733296E+05 | 1.93732867E+05 | 4.30E-01 | -1.74691520E+01 | -1.78976841E+01 | 4.30E-01 |
| 2601 | Po | 209 | 84 | 125 | 41 | 7.83518500E+00 | 7.83197030E+00 | 3.20E-03 | 1.94622365E+05 | 1.94622976E+05 | -6.10E-01 | 1.94665893E+05 | 1.94666504E+05 | -6.10E-01 | -1.63655700E+01 | -1.57545890E+01 | -6.10E-01 |
| 2602 | Po | 210 | 84 | 126 | 42 | 7.83434400E+00 | 7.82975748E+00 | 4.60E-03 | 1.95554272E+05 | 1.95555174E+05 | -9.00E-01 | 1.95597800E+05 | 1.95598702E+05 | -9.00E-01 | -1.59526950E+01 | -1.50505494E+01 | -9.00E-01 |
| 2603 | Po | 211 | 84 | 127 | 43 | 7.81878200E+00 | 7.81673816E+00 | 2.00E-03 | 1.96489287E+05 | 1.96489657E+05 | -3.70E-01 | 1.96532815E+05 | 1.96533185E+05 | -3.70E-01 | -1.24321260E+01 | -1.20619105E+01 | -3.70E-01 |
| 2604 | Po | 212 | 84 | 128 | 44 | 7.81024100E+00 | 7.81079561E+00 | -5.50E-04 | 1.97422844E+05 | 1.97422665E+05 | 1.80E-01 | 1.97466372E+05 | 1.97466193E+05 | 1.80E-01 | -1.03690400E+01 | -1.05475092E+01 | 1.80E-01 |
| 2605 | Po | 213 | 84 | 129 | 45 | 7.79402100E+00 | 7.79481989E+00 | -8.00E-04 | 1.98358054E+05 | 1.98357823E+05 | 2.30E-01 | 1.98401582E+05 | 1.98401351E+05 | 2.30E-01 | -6.65308000E+00 | -6.88415745E+00 | 2.30E-01 |
| 2606 | Po | 214 | 84 | 130 | 46 | 7.78511400E+00 | 7.78639302E+00 | -1.30E-03 | 1.99291731E+05 | 1.99291397E+05 | 3.30E-01 | 1.99335259E+05 | 1.99334925E+05 | 3.30E-01 | -4.46960300E+00 | -4.80430858E+00 | 3.30E-01 |
| 2607 | Po | 215 | 84 | 131 | 47 | 7.76816800E+00 | 7.76879044E+00 | -6.20E-04 | 2.00227155E+05 | 2.00226960E+05 | 1.90E-01 | 2.00270683E+05 | 2.00270488E+05 | 1.90E-01 | -5.40130000E-01 | -7.34826548E-01 | 1.90E-01 |
| 2608 | Po | 216 | 84 | 132 | 48 | 7.75881200E+00 | 7.75910583E+00 | -2.90E-04 | 2.01160973E+05 | 2.01160849E+05 | 1.20E-01 | 2.01204501E+05 | 2.01204377E+05 | 1.20E-01 | 1.78398400E+00 | 1.65957611E+00 | 1.20E-01 |
| 2609 | Po | 217 | 84 | 133 | 49 | 7.74135200E+00 | 7.74084008E+00 | 5.10E-04 | 2.02096568E+05 | 2.02096619E+05 | -5.00E-02 | 2.02140097E+05 | 2.02140147E+05 | -5.00E-02 | 5.88532800E+00 | 5.93545738E+00 | -5.00E-02 |
| 2610 | Po | 218 | 84 | 134 | 50 | 7.73151900E+00 | 7.72953646E+00 | 2.00E-03 | 2.03030536E+05 | 2.03030907E+05 | -3.70E-01 | 2.03074064E+05 | 2.03074435E+05 | -3.70E-01 | 8.35880500E+00 | 8.73012667E+00 | -3.70E-01 |
| 2611 | Po | 219 | 84 | 135 | 51 | 7.71333300E+00 | 7.71396099E+00 | -6.30E-04 | 2.03966353E+05 | 2.03966154E+05 | 2.00E-01 | 2.04009881E+05 | 2.04009682E+05 | 2.00E-01 | 1.26813590E+01 | 1.24829361E+01 | 2.00E-01 |
| 2612 | Po | 220 | 84 | 136 | 52 | 7.70322400E+00 | 7.70561142E+00 | -2.40E-03 | 2.04900429E+05 | 2.04899843E+05 | 5.90E-01 | 2.04943957E+05 | 2.04943371E+05 | 5.90E-01 | 1.52634610E+01 | 1.46771988E+01 | 5.90E-01 |
| 2613 | Po | 221 | 84 | 137 | 53 | 7.68448100E+00 | 7.68847981E+00 | -4.00E-03 | 2.05836433E+05 | 2.05835488E+05 | 9.40E-01 | 2.05879961E+05 | 2.05879016E+05 | 9.40E-01 | 1.97737550E+01 | 1.88289921E+01 | 9.40E-01 |
| 2614 | PO | 222 | 84 | 138 | 54 | 7.67400500E+00 | 7.67835037E+00 | -4.30E-03 | 2.06770640E+05 | 2.06769614E+05 | 1.00E+00 | 2.06814168E+05 | 2.06813142E+05 | 1.00E+00 | 2.24862650E+01 | 2.14605676E+01 | 1.00E+00 |
| 2615 | PO | 223 | 84 | 139 | 55 | 7.65500000E+00 | 7.65957889E+00 | -4.60E-03 | 2.07706726E+05 | 2.07705687E+05 | 1.00E+00 | 2.07750254E+05 | 2.07749215E+05 | 1.00E+00 | 2.70790000E+01 | 2.60395765E+01 | 1.00E+00 |
| 2616 | PO | 224 | 84 | 140 | 56 | 7.64400000E+00 | 7.64769450E+00 | -3.70E-03 | 2.08641052E+05 | 2.08640255E+05 | 8.00E-01 | 2.08684580E+05 | 2.08683783E+05 | 8.00E-01 | 2.99100000E+01 | 2.91134196E+01 | 8.00E-01 |
| 2617 | PO | 225 | 84 | 141 | 57 | 7.62600000E+00 | 7.62742329E+00 | -1.40E-03 | 2.09577166E+05 | 2.09576734E+05 | 4.30E-01 | 2.09620694E+05 | 2.09620262E+05 | 4.30E-01 | 3.45300000E+01 | 3.40980656E+01 | 4.30E-01 |
| 2618 | PO | 226 | 84 | 142 | 58 | 7.61400000E+00 | 7.61405021E+00 | -5.00E-05 | 2.10511678E+05 | 2.10511694E+05 | -1.60E-02 | 2.10555206E+05 | 2.10555222E+05 | -1.60E-02 | 3.75490000E+01 | 3.75642795E+01 | -1.50E-02 |
| 2619 | PO | 227 | 84 | 143 | 59 | 7.59600000E+00 | 7.59262621E+00 | 3.40E-03 | 2.11447904E+05 | 2.11448509E+05 | -6.00E-01 | 2.11491432E+05 | 2.11492037E+05 | -6.00E-01 | 4.22810000E+01 | 4.28847953E+01 | -6.00E-01 |
| 2620 | At | 191 | 85 | 106 | 21 | 7.70292300E+00 | 7.70532768E+00 | -2.40E-03 | 1.77875172E+05 | 1.77874651E+05 | 5.20E-01 | 1.77919229E+05 | 1.77918708E+05 | 5.20E-01 | 3.86375300E+00 | 3.34279625E+00 | 5.20E-01 |
| 2621 | At | 192 | 85 | 107 | 22 | 7.70967500E+00 | 7.71255431E+00 | -2.90E-03 | 1.78805738E+05 | 1.78805123E+05 | 6.10E-01 | 1.78849796E+05 | 1.78849181E+05 | 6.10E-01 | 2.93579600E+00 | 2.32127571E+00 | 6.10E-01 |
| 2622 | At | 193 | 85 | 108 | 23 | 7.72711100E+00 | 7.72832659E+00 | -1.20E-03 | 1.79734229E+05 | 1.79733932E+05 | 3.00E-01 | 1.79778286E+05 | 1.79777990E+05 | 3.00E-01 | -6.76090000E-02 | -3.64010063E-01 | 3.00E-01 |
| 2623 | At | 194 | 85 | 109 | 24 | 7.73220400E+00 | 7.73411117E+00 | -1.90E-03 | 1.80665079E+05 | 1.80664647E+05 | 4.30E-01 | 1.80709136E+05 | 1.80708705E+05 | 4.30E-01 | -7.11531000E-01 | -1.14322552E+00 | 4.30E-01 |
| 2624 | At | 195 | 85 | 110 | 25 | 7.74811900E+00 | 7.74820659E+00 | -8.80E-05 | 1.81593809E+05 | 1.81593730E+05 | 7.90E-02 | 1.81637866E+05 | 1.81637787E+05 | 7.90E-02 | -3.47582700E+00 | -3.55462529E+00 | 7.90E-02 |
| 2625 | At | 196 | 85 | 111 | 26 | 7.75199700E+00 | 7.75276795E+00 | -7.70E-04 | 1.82524866E+05 | 1.82524653E+05 | 2.10E-01 | 1.82568923E+05 | 1.82568710E+05 | 2.10E-01 | -3.91262900E+00 | -4.12553845E+00 | 2.10E-01 |

| | | | | | | | | | | | | | | | | |
|---|---|---|---|---|---|---|---|---|---|---|---|---|---|---|---|---|
| 2626 | At | 197 | 85 | 112 | 27 | 7.76596100E+00 | 7.76542838E+00 | 5.30E-04 | 1.83453928E+05 | 1.83453971E+05 | -4.30E-02 | 1.83497986E+05 | 1.83498029E+05 | -4.30E-02 | -6.34423100E+00 | -6.30109220E+00 | -4.30E-02 |
| 2627 | At | 198 | 85 | 113 | 28 | 7.76900000E+00 | 7.76899750E+00 | 2.50E-06 | 1.84385045E+05 | 1.84385065E+05 | -2.00E-02 | 1.84429102E+05 | 1.84429122E+05 | -2.00E-02 | -6.72100000E+00 | -6.70188707E+00 | -1.90E-02 |
| 2628 | At | 199 | 85 | 114 | 29 | 7.78148800E+00 | 7.78045047E+00 | 1.00E-03 | 1.85314437E+05 | 1.85314582E+05 | -1.40E-01 | 1.85358495E+05 | 1.85358639E+05 | -1.40E-01 | -8.82343300E+00 | -8.67870660E+00 | -1.40E-01 |
| 2629 | At | 200 | 85 | 115 | 30 | 7.78376000E+00 | 7.78320512E+00 | 5.50E-04 | 1.86245767E+05 | 1.86245816E+05 | -4.90E-02 | 1.86289824E+05 | 1.86289873E+05 | -4.90E-02 | -8.98797100E+00 | -8.93876955E+00 | -4.90E-02 |
| 2630 | At | 201 | 85 | 116 | 31 | 7.79415200E+00 | 7.79357482E+00 | 5.80E-04 | 1.87175459E+05 | 1.87175514E+05 | -5.40E-02 | 1.87219517E+05 | 1.87219571E+05 | -5.40E-02 | -1.07893580E+01 | -1.07349645E+01 | -5.40E-02 |
| 2631 | At | 202 | 85 | 117 | 32 | 7.79454200E+00 | 7.79552838E+00 | -9.90E-04 | 1.88107152E+05 | 1.88106891E+05 | 2.60E-01 | 1.88151209E+05 | 1.88150948E+05 | 2.60E-01 | -1.05908050E+01 | -1.08518391E+01 | 2.60E-01 |
| 2632 | At | 203 | 85 | 118 | 33 | 7.80364800E+00 | 7.80470684E+00 | -1.10E-03 | 1.89037074E+05 | 1.89036798E+05 | 2.80E-01 | 1.89081132E+05 | 1.89080855E+05 | 2.80E-01 | -1.21625260E+01 | -1.24392757E+01 | 2.80E-01 |
| 2633 | At | 204 | 85 | 119 | 34 | 7.80355200E+00 | 7.80557210E+00 | -2.00E-03 | 1.89968856E+05 | 1.89968382E+05 | 4.70E-01 | 1.90012913E+05 | 1.90012439E+05 | 4.70E-01 | -1.18754320E+01 | -1.23491774E+01 | 4.70E-01 |
| 2634 | At | 205 | 85 | 120 | 35 | 7.81019900E+00 | 7.81309399E+00 | -2.90E-03 | 1.90899255E+05 | 1.90898600E+05 | 6.60E-01 | 1.90943312E+05 | 1.90942657E+05 | 6.60E-01 | -1.29702520E+01 | -1.36254188E+01 | 6.60E-01 |
| 2635 | At | 206 | 85 | 121 | 36 | 7.80883900E+00 | 7.81219835E+00 | -3.40E-03 | 1.91831290E+05 | 1.91830536E+05 | 7.50E-01 | 1.91875348E+05 | 1.91874594E+05 | 7.50E-01 | -1.24289860E+01 | -1.31826908E+01 | 7.50E-01 |
| 2636 | At | 207 | 85 | 122 | 37 | 7.81396200E+00 | 7.81723701E+00 | -3.30E-03 | 1.92761986E+05 | 1.92761247E+05 | 7.40E-01 | 1.92806044E+05 | 1.92805304E+05 | 7.40E-01 | -1.32269050E+01 | -1.39665726E+01 | 7.40E-01 |
| 2637 | At | 208 | 85 | 123 | 38 | 7.81155800E+00 | 7.81364602E+00 | -2.10E-03 | 1.93694237E+05 | 1.93693742E+05 | 5.00E-01 | 1.93738295E+05 | 1.93737799E+05 | 5.00E-01 | -1.24696280E+01 | -1.29655656E+01 | 5.00E-01 |
| 2638 | At | 209 | 85 | 124 | 39 | 7.81477600E+00 | 7.81530210E+00 | -5.30E-04 | 1.94625319E+05 | 1.94625147E+05 | 1.70E-01 | 1.94669376E+05 | 1.94669205E+05 | 1.70E-01 | -1.28823610E+01 | -1.30540124E+01 | 1.70E-01 |
| 2639 | At | 210 | 85 | 125 | 40 | 7.81166100E+00 | 7.80827755E+00 | 3.40E-03 | 1.95557724E+05 | 1.95558372E+05 | -6.50E-01 | 1.95601781E+05 | 1.95602430E+05 | -6.50E-01 | -1.19716550E+01 | -1.13228415E+01 | -6.50E-01 |
| 2640 | At | 211 | 85 | 126 | 41 | 7.81135200E+00 | 7.80615125E+00 | 5.20E-03 | 1.96489543E+05 | 1.96490578E+05 | -1.00E+00 | 1.96533600E+05 | 1.96534636E+05 | -1.00E+00 | -1.16468290E+01 | -1.06111497E+01 | -1.00E+00 |
| 2641 | At | 212 | 85 | 127 | 42 | 7.79833800E+00 | 7.79571435E+00 | 2.60E-03 | 1.97424056E+05 | 1.97424550E+05 | -4.90E-01 | 1.97468113E+05 | 1.97468608E+05 | -4.90E-01 | -8.62778800E+00 | -8.13335962E+00 | -4.90E-01 |
| 2642 | At | 213 | 85 | 128 | 43 | 7.79000100E+00 | 7.79029428E+00 | -2.90E-04 | 1.98357598E+05 | 1.98357474E+05 | 1.20E-01 | 1.98401656E+05 | 1.98401532E+05 | 1.20E-01 | -6.57916200E+00 | -6.70328025E+00 | 1.20E-01 |
| 2643 | At | 214 | 85 | 129 | 44 | 7.77636400E+00 | 7.77729709E+00 | -9.30E-04 | 1.99292292E+05 | 1.99292031E+05 | 2.60E-01 | 1.99336350E+05 | 1.99336088E+05 | 2.60E-01 | -3.37939900E+00 | -3.64085668E+00 | 2.60E-01 |
| 2644 | At | 215 | 85 | 130 | 45 | 7.76785400E+00 | 7.76966536E+00 | -1.80E-03 | 2.00225911E+05 | 2.00225460E+05 | 4.50E-01 | 2.00269968E+05 | 2.00269517E+05 | 4.50E-01 | -1.25486200E+00 | -1.70601234E+00 | 4.50E-01 |
| 2645 | At | 216 | 85 | 131 | 46 | 7.75299800E+00 | 7.75520983E+00 | -2.20E-03 | 2.01160917E+05 | 2.01160378E+05 | 5.40E-01 | 2.01204975E+05 | 2.01204435E+05 | 5.40E-01 | 2.25752700E+00 | 1.71803571E+00 | 5.40E-01 |
| 2646 | At | 217 | 85 | 132 | 47 | 7.74461000E+00 | 7.74635243E+00 | -1.70E-03 | 2.02094550E+05 | 2.02094110E+05 | 4.40E-01 | 2.02138607E+05 | 2.02138167E+05 | 4.40E-01 | 4.39592800E+00 | 3.95620027E+00 | 4.40E-01 |
| 2647 | At | 218 | 85 | 133 | 48 | 7.72912200E+00 | 7.73118650E+00 | -2.10E-03 | 2.03029747E+05 | 2.03029235E+05 | 5.10E-01 | 2.03073804E+05 | 2.03073293E+05 | 5.10E-01 | 8.09908300E+00 | 7.58733989E+00 | 5.10E-01 |
| 2648 | At | 219 | 85 | 134 | 49 | 7.72019100E+00 | 7.72057982E+00 | -3.90E-04 | 2.03963539E+05 | 2.03963392E+05 | 1.50E-01 | 2.04007597E+05 | 2.04007450E+05 | 1.50E-01 | 1.03971950E+01 | 1.02503354E+01 | 1.50E-01 |
| 2649 | At | 220 | 85 | 135 | 50 | 7.70370300E+00 | 7.70740421E+00 | -3.70E-03 | 2.04899012E+05 | 2.04898136E+05 | 8.80E-01 | 2.04943069E+05 | 2.04942193E+05 | 8.80E-01 | 1.43757470E+01 | 1.34997090E+01 | 8.80E-01 |
| 2650 | At | 221 | 85 | 136 | 51 | 7.69447500E+00 | 7.69913745E+00 | -4.70E-03 | 2.05832913E+05 | 2.05831821E+05 | 1.10E+00 | 2.05876970E+05 | 2.05875878E+05 | 1.10E+00 | 1.67827270E+01 | 1.56905773E+01 | 1.10E+00 |
| 2651 | At | 222 | 85 | 137 | 52 | 7.67738700E+00 | 7.68448273E+00 | -7.10E-03 | 2.06768577E+05 | 2.06766940E+05 | 1.60E+00 | 2.06812635E+05 | 2.06810998E+05 | 1.60E+00 | 2.09530260E+01 | 1.93161073E+01 | 1.60E+00 |
| 2652 | At | 223 | 85 | 138 | 53 | 7.66805500E+00 | 7.67452757E+00 | -6.50E-03 | 2.07702546E+05 | 2.07701041E+05 | 1.50E+00 | 2.07746604E+05 | 2.07745099E+05 | 1.50E+00 | 2.34280060E+01 | 2.19229449E+01 | 1.50E+00 |
| 2653 | AT | 224 | 85 | 139 | 54 | 7.65073500E+00 | 7.65834859E+00 | -7.60E-03 | 2.08638323E+05 | 2.08636556E+05 | 1.80E+00 | 2.08682381E+05 | 2.08680613E+05 | 1.80E+00 | 2.77110150E+01 | 2.59438274E+01 | 1.80E+00 |
| 2654 | AT | 225 | 85 | 140 | 55 | 7.64100000E+00 | 7.64673720E+00 | -5.70E-03 | 2.09572501E+05 | 2.09571076E+05 | 1.40E+00 | 2.09616558E+05 | 2.09615133E+05 | 1.40E+00 | 3.03950000E+01 | 2.89693600E+01 | 1.40E+00 |
| 2655 | AT | 226 | 85 | 141 | 56 | 7.62400000E+00 | 7.62913821E+00 | -5.10E-03 | 2.10508215E+05 | 2.10506972E+05 | 1.20E+00 | 2.10552272E+05 | 2.10551029E+05 | 1.20E+00 | 3.46140000E+01 | 3.33713141E+01 | 1.20E+00 |
| 2656 | AT | 227 | 85 | 142 | 57 | 7.61300000E+00 | 7.61606614E+00 | -3.10E-03 | 2.11442578E+05 | 2.11441875E+05 | 7.00E-01 | 2.11486635E+05 | 2.11485933E+05 | 7.00E-01 | 3.74830000E+01 | 3.67808551E+01 | 7.00E-01 |
| 2657 | AT | 228 | 85 | 143 | 58 | 7.59700000E+00 | 7.59729671E+00 | -3.00E-04 | 2.12378273E+05 | 2.12378104E+05 | 1.70E-01 | 2.12422330E+05 | 2.12422161E+05 | 1.70E-01 | 4.16840000E+01 | 4.15155374E+01 | 1.70E-01 |
| 2658 | AT | 229 | 85 | 144 | 59 | 7.58500000E+00 | 7.58305351E+00 | 1.90E-03 | 2.13312906E+05 | 2.13313334E+05 | -4.30E-01 | 2.13356964E+05 | 2.13357391E+05 | -4.30E-01 | 4.48230000E+01 | 4.52512520E+01 | -4.30E-01 |
| 2659 | RN | 194 | 86 | 108 | 22 | 7.69500100E+00 | 7.69614649E+00 | -1.10E-03 | 1.80670984E+05 | 1.80670699E+05 | 2.80E-01 | 1.80715571E+05 | 1.80715287E+05 | 2.80E-01 | 5.72346200E+00 | 5.43884565E+00 | 2.80E-01 |
| 2660 | Rn | 195 | 86 | 109 | 23 | 7.70038400E+00 | 7.70195238E+00 | -1.60E-03 | 1.81601805E+05 | 1.81601436E+05 | 3.70E-01 | 1.81646392E+05 | 1.81646024E+05 | 3.70E-01 | 5.05025600E+00 | 4.68186817E+00 | 3.70E-01 |
| 2661 | Rn | 196 | 86 | 110 | 24 | 7.71798600E+00 | 7.71826186E+00 | -2.80E-04 | 1.82530220E+05 | 1.82530103E+05 | 1.20E-01 | 1.82574807E+05 | 1.82574691E+05 | 1.20E-01 | 1.97104000E+00 | 1.85457799E+00 | 1.20E-01 |
| 2662 | Rn | 197 | 86 | 111 | 25 | 7.72229200E+00 | 7.72284856E+00 | -5.60E-04 | 1.83461219E+05 | 1.83461047E+05 | 1.70E-01 | 1.83505806E+05 | 1.83505634E+05 | 1.70E-01 | 1.47606500E+00 | 1.30405489E+00 | 1.70E-01 |
| 2663 | Rn | 198 | 86 | 112 | 26 | 7.73772400E+00 | 7.73771235E+00 | 1.20E-05 | 1.84390006E+05 | 1.84389946E+05 | 6.00E-02 | 1.84434594E+05 | 1.84434534E+05 | 6.00E-02 | -1.23035700E+00 | -1.29050500E+00 | 6.00E-02 |
| 2664 | Rn | 199 | 86 | 113 | 27 | 7.74075500E+00 | 7.74128349E+00 | -5.30E-04 | 1.85321231E+05 | 1.85321063E+05 | 1.70E-01 | 1.85365818E+05 | 1.85365651E+05 | 1.70E-01 | -1.49983900E+00 | -1.66755624E+00 | 1.70E-01 |
| 2665 | Rn | 200 | 86 | 114 | 28 | 7.75498000E+00 | 7.75488728E+00 | 9.30E-05 | 1.86250210E+05 | 1.86250167E+05 | 4.40E-02 | 1.86294798E+05 | 1.86294754E+05 | 4.40E-02 | -4.01433800E+00 | -4.05827846E+00 | 4.40E-02 |
| 2666 | Rn | 201 | 86 | 115 | 29 | 7.75684300E+00 | 7.75756229E+00 | -7.20E-04 | 1.87181646E+05 | 1.87181439E+05 | 2.10E-01 | 1.87226234E+05 | 1.87226027E+05 | 2.10E-01 | -4.07248200E+00 | -4.27952369E+00 | 2.10E-01 |
| 2667 | Rn | 202 | 86 | 116 | 30 | 7.76930000E+00 | 7.76996287E+00 | -6.60E-04 | 1.88110939E+05 | 1.88110742E+05 | 2.00E-01 | 1.88155526E+05 | 1.88155330E+05 | 2.00E-01 | -6.27428500E+00 | -6.47068304E+00 | 2.00E-01 |
| 2668 | Rn | 203 | 86 | 117 | 31 | 7.77022100E+00 | 7.77168425E+00 | -1.50E-03 | 1.89042548E+05 | 1.89042188E+05 | 3.60E-01 | 1.89087135E+05 | 1.89086776E+05 | 3.60E-01 | -6.15927900E+00 | -6.51876841E+00 | 3.60E-01 |
| 2669 | Rn | 204 | 86 | 118 | 32 | 7.78063700E+00 | 7.78271382E+00 | -2.10E-03 | 1.89972218E+05 | 1.89971732E+05 | 4.90E-01 | 1.90016805E+05 | 1.90016319E+05 | 4.90E-01 | -7.98297600E+00 | -8.46916555E+00 | 4.90E-01 |
| 2670 | Rn | 205 | 86 | 119 | 33 | 7.78074200E+00 | 7.78315873E+00 | -2.40E-03 | 1.90903981E+05 | 1.90903423E+05 | 5.60E-01 | 1.90948569E+05 | 1.90948011E+05 | 5.60E-01 | -7.71385000E+00 | -8.27176745E+00 | 5.60E-01 |
| 2671 | Rn | 206 | 86 | 120 | 34 | 7.78895600E+00 | 7.79235902E+00 | -3.40E-03 | 1.91834074E+05 | 1.91833310E+05 | 7.60E-01 | 1.91878661E+05 | 1.91877898E+05 | 7.60E-01 | -9.11539600E+00 | -9.87886678E+00 | 7.60E-01 |
| 2672 | Rn | 207 | 86 | 121 | 35 | 7.78799800E+00 | 7.79091809E+00 | -2.90E-03 | 1.92766049E+05 | 1.92765382E+05 | 6.70E-01 | 1.92810636E+05 | 1.92809969E+05 | 6.70E-01 | -8.63471300E+00 | -9.30163444E+00 | 6.70E-01 |
| 2673 | Rn | 208 | 86 | 122 | 36 | 7.79426600E+00 | 7.79759196E+00 | -3.30E-03 | 1.93696522E+05 | 1.93695768E+05 | 7.50E-01 | 1.93741109E+05 | 1.93740355E+05 | 7.50E-01 | -9.65522700E+00 | -1.04093983E+01 | 7.50E-01 |
| 2674 | Rn | 209 | 86 | 123 | 37 | 7.79211600E+00 | 7.79352261E+00 | -1.40E-03 | 1.94628743E+05 | 1.94628386E+05 | 3.60E-01 | 1.94673330E+05 | 1.94672974E+05 | 3.60E-01 | -8.92882600E+00 | -9.28517737E+00 | 3.60E-01 |
| 2675 | Rn | 210 | 86 | 124 | 38 | 7.79666300E+00 | 7.79700672E+00 | -3.40E-04 | 1.95559561E+05 | 1.95559426E+05 | 1.30E-01 | 1.95604148E+05 | 1.95604014E+05 | 1.30E-01 | -9.60453500E+00 | -9.73904416E+00 | 1.30E-01 |
| 2676 | Rn | 211 | 86 | 125 | 39 | 7.79393900E+00 | 7.78981468E+00 | 4.10E-03 | 1.96491905E+05 | 1.96492712E+05 | -8.10E-01 | 1.96536492E+05 | 1.96537300E+05 | -8.10E-01 | -8.75500900E+00 | -7.94721007E+00 | -8.10E-01 |
| 2677 | Rn | 212 | 86 | 126 | 40 | 7.79479600E+00 | 7.78991519E+00 | 4.90E-03 | 1.97423494E+05 | 1.97424467E+05 | -9.70E-01 | 1.97468082E+05 | 1.97469054E+05 | -9.70E-01 | -8.65924300E+00 | -7.68701401E+00 | -9.70E-01 |
| 2678 | Rn | 213 | 86 | 127 | 41 | 7.78219100E+00 | 7.77974502E+00 | 2.40E-03 | 1.98357950E+05 | 1.98358408E+05 | -4.60E-01 | 1.98402537E+05 | 1.98402996E+05 | -4.60E-01 | -5.69788400E+00 | -5.23936331E+00 | -4.60E-01 |
| 2679 | Rn | 214 | 86 | 128 | 42 | 7.77710000E+00 | 7.77696372E+00 | 1.40E-04 | 1.99290822E+05 | 1.99290789E+05 | 3.30E-02 | 1.99335410E+05 | 1.99335376E+05 | 3.30E-02 | -4.31937800E+00 | -4.35259157E+00 | 3.30E-02 |
| 2680 | Rn | 215 | 86 | 129 | 43 | 7.76381200E+00 | 7.76457008E+00 | -7.60E-04 | 2.00225468E+05 | 2.00225242E+05 | 2.30E-01 | 2.00270055E+05 | 2.00269830E+05 | 2.30E-01 | -1.16819500E+00 | -1.39360519E+00 | 2.30E-01 |
| 2681 | Rn | 216 | 86 | 130 | 44 | 7.75865500E+00 | 7.75980704E+00 | -1.20E-03 | 2.01158383E+05 | 2.01158072E+05 | 3.10E-01 | 2.01202970E+05 | 2.01202659E+05 | 3.10E-01 | 2.53235000E-01 | -5.80381796E-02 | 3.10E-01 |
| 2682 | Rn | 217 | 86 | 131 | 45 | 7.74440100E+00 | 7.74606831E+00 | -1.70E-03 | 2.02093283E+05 | 2.02092859E+05 | 4.20E-01 | 2.02137870E+05 | 2.02137446E+05 | 4.20E-01 | 3.65892500E+00 | 3.23477740E+00 | 4.20E-01 |
| 2683 | Rn | 218 | 86 | 132 | 46 | 7.73875000E+00 | 7.74008761E+00 | -1.30E-03 | 2.03026336E+05 | 2.03025982E+05 | 3.50E-01 | 2.03070923E+05 | 2.03070569E+05 | 3.50E-01 | 5.21784400E+00 | 4.86382067E+00 | 3.50E-01 |
| 2684 | Rn | 219 | 86 | 133 | 47 | 7.72377100E+00 | 7.72556525E+00 | -1.80E-03 | 2.03961443E+05 | 2.03960987E+05 | 4.60E-01 | 2.04006030E+05 | 2.04005575E+05 | 4.60E-01 | 8.83090000E+00 | 8.37544991E+00 | 4.60E-01 |
| 2685 | Rn | 220 | 86 | 134 | 48 | 7.71724700E+00 | 7.71771047E+00 | -4.60E-04 | 2.04894720E+05 | 2.04894555E+05 | 1.60E-01 | 2.04939307E+05 | 2.04939143E+05 | 1.60E-01 | 1.06135640E+01 | 1.04492537E+01 | 1.60E-01 |
| 2686 | Rn | 221 | 86 | 135 | 49 | 7.70138700E+00 | 7.70456662E+00 | -3.20E-03 | 2.05830073E+05 | 2.05829308E+05 | 7.70E-01 | 2.05874660E+05 | 2.05873895E+05 | 7.70E-01 | 1.44726930E+01 | 1.37076542E+01 | 7.70E-01 |
| 2687 | Rn | 222 | 86 | 136 | 50 | 7.69448900E+00 | 7.69853771E+00 | -4.00E-03 | 2.06763468E+05 | 2.06762507E+05 | 9.60E-01 | 2.06808055E+05 | 2.06807094E+05 | 9.60E-01 | 1.63740310E+01 | 1.54128254E+01 | 9.60E-01 |
| 2688 | Rn | 223 | 86 | 137 | 51 | 7.67817100E+00 | 7.68398158E+00 | -5.80E-03 | 2.07698978E+05 | 2.07697620E+05 | 1.40E+00 | 2.07743565E+05 | 2.07742207E+05 | 1.40E+00 | 2.03897390E+01 | 1.90316228E+01 | 1.40E+00 |
| 2689 | Rn | 224 | 86 | 138 | 52 | 7.67075100E+00 | 7.67634584E+00 | -5.60E-03 | 2.08632527E+05 | 2.08631212E+05 | 1.30E+00 | 2.08677115E+05 | 2.08675799E+05 | 1.30E+00 | 2.24450980E+01 | 2.11293656E+01 | 1.30E+00 |
| 2690 | Rn | 225 | 86 | 139 | 53 | 7.65435700E+00 | 7.66035434E+00 | -6.00E-03 | 2.09568111E+05 | 2.09566699E+05 | 1.40E+00 | 2.09612698E+05 | 2.09611286E+05 | 1.40E+00 | 2.65341420E+01 | 2.51224260E+01 | 1.40E+00 |
| 2691 | Rn | 226 | 86 | 140 | 54 | 7.64641000E+00 | 7.65113981E+00 | -4.70E-03 | 2.10501818E+05 | 2.10500686E+05 | 1.10E+00 | 2.10546405E+05 | 2.10545274E+05 | 1.10E+00 | 2.87471930E+01 | 2.76158760E+01 | 1.10E+00 |
| 2692 | Rn | 227 | 86 | 141 | 55 | 7.63005000E+00 | 7.63378455E+00 | -3.70E-03 | 2.11437450E+05 | 2.11436540E+05 | 9.10E-01 | 2.11482038E+05 | 2.11481128E+05 | 9.10E-01 | 3.28858340E+01 | 3.19756973E+01 | 9.10E-01 |
| 2693 | Rn | 228 | 86 | 142 | 56 | 7.62164500E+00 | 7.62312744E+00 | -1.50E-03 | 2.12371302E+05 | 2.12370902E+05 | 4.00E-01 | 2.12415889E+05 | 2.12415489E+05 | 4.00E-01 | 3.52434650E+01 | 3.48430531E+01 | 4.00E-01 |
| 2694 | RN | 229 | 86 | 143 | 57 | 7.60562200E+00 | 7.60458365E+00 | 1.00E-03 | 2.13306915E+05 | 2.13307090E+05 | -1.80E-01 | 2.13351502E+05 | 2.13351678E+05 | -1.80E-01 | 3.93624000E+01 | 3.95377724E+01 | -1.80E-01 |
| 2695 | Fr | 199 | 87 | 112 | 25 | 7.69531000E+00 | 7.69649952E+00 | -1.20E-03 | 1.85328962E+05 | 1.85328662E+05 | 3.00E-01 | 1.85374079E+05 | 1.85373780E+05 | 3.00E-01 | 6.76133700E+00 | 6.46137819E+00 | 3.00E-01 |
| 2696 | Fr | 200 | 87 | 113 | 26 | 7.70032200E+00 | 7.70222273E+00 | -1.90E-03 | 1.86259829E+05 | 1.86259386E+05 | 4.40E-01 | 1.86304947E+05 | 1.86304504E+05 | 4.40E-01 | 6.13482200E+00 | 5.69155407E+00 | 4.40E-01 |
| 2697 | Fr | 201 | 87 | 114 | 27 | 7.71477100E+00 | 7.71603495E+00 | -1.30E-03 | 1.87188790E+05 | 1.87188473E+05 | 3.20E-01 | 1.87233908E+05 | 1.87233591E+05 | 3.20E-01 | 3.60170900E+00 | 3.28439487E+00 | 3.20E-01 |
| 2698 | Fr | 202 | 87 | 115 | 28 | 7.71900000E+00 | 7.72076205E+00 | -1.80E-03 | 1.88119775E+05 | 1.88119368E+05 | 4.10E-01 | 1.88164893E+05 | 1.88164485E+05 | 4.10E-01 | 3.09200000E+00 | 2.68480494E+00 | 4.10E-01 |
| 2699 | Fr | 203 | 87 | 116 | 29 | 7.73170900E+00 | 7.73323424E+00 | -1.50E-03 | 1.89049053E+05 | 1.89048680E+05 | 3.70E-01 | 1.89094171E+05 | 1.89093798E+05 | 3.70E-01 | 8.76284000E-01 | 5.03506280E-01 | 3.70E-01 |
| 2700 | Fr | 204 | 87 | 117 | 30 | 7.73469200E+00 | 7.73685528E+00 | -2.20E-03 | 1.89980278E+05 | 1.89979774E+05 | 5.00E-01 | 1.90025396E+05 | 1.90024891E+05 | 5.00E-01 | 6.07390000E-01 | 1.02899254E-01 | 5.00E-01 |
| 2701 | Fr | 205 | 87 | 118 | 31 | 7.74568600E+00 | 7.74778311E+00 | -2.10E-03 | 1.90909855E+05 | 1.90909362E+05 | 4.90E-01 | 1.90954973E+05 | 1.90954480E+05 | 4.90E-01 | -1.30981200E+00 | -1.80284122E+00 | 4.90E-01 |
| 2702 | Fr | 206 | 87 | 119 | 32 | 7.74694000E+00 | 7.74997120E+00 | -3.00E-03 | 1.91841416E+05 | 1.91840729E+05 | 6.90E-01 | 1.91886534E+05 | 1.91885847E+05 | 6.90E-01 | -1.24249000E+00 | -1.93005123E+00 | 6.90E-01 |
| 2703 | Fr | 207 | 87 | 120 | 33 | 7.75624600E+00 | 7.75893431E+00 | -2.70E-03 | 1.92771309E+05 | 1.92770689E+05 | 6.20E-01 | 1.92816426E+05 | 1.92815807E+05 | 6.20E-01 | -2.84433700E+00 | -3.46406851E+00 | 6.20E-01 |
| 2704 | Fr | 208 | 87 | 121 | 34 | 7.75690200E+00 | 7.75917187E+00 | -2.30E-03 | 1.93702981E+05 | 1.93702446E+05 | 5.40E-01 | 1.93748099E+05 | 1.93747564E+05 | 5.40E-01 | -2.66581300E+00 | -3.20109505E+00 | 5.40E-01 |
| 2705 | Fr | 209 | 87 | 122 | 35 | 7.76368000E+00 | 7.76561770E+00 | -1.90E-03 | 1.94633373E+05 | 1.94632905E+05 | 4.70E-01 | 1.94678491E+05 | 1.94678023E+05 | 4.70E-01 | -3.76795600E+00 | -4.23612758E+00 | 4.70E-01 |
| 2706 | Fr | 210 | 87 | 123 | 36 | 7.76307100E+00 | 7.76335263E+00 | -2.80E-04 | 1.95565303E+05 | 1.95565180E+05 | 1.20E-01 | 1.95610420E+05 | 1.95610298E+05 | 1.20E-01 | -3.33250400E+00 | -3.45476057E+00 | 1.20E-01 |
| 2707 | Fr | 211 | 87 | 124 | 37 | 7.76835800E+00 | 7.76682958E+00 | 1.50E-03 | 1.96495989E+05 | 1.96496249E+05 | -2.60E-01 | 1.96541107E+05 | 1.96541366E+05 | -2.60E-01 | -4.13981700E+00 | -3.88043213E+00 | -2.60E-01 |
| 2708 | Fr | 212 | 87 | 125 | 38 | 7.76684400E+00 | 7.76176871E+00 | 5.10E-03 | 1.97428108E+05 | 1.97429120E+05 | -1.00E+00 | 1.97473225E+05 | 1.97474238E+05 | -1.00E+00 | -3.51578200E+00 | -2.50303789E+00 | -1.00E+00 |
| 2709 | Fr | 213 | 87 | 126 | 39 | 7.76844600E+00 | 7.76223305E+00 | 6.20E-03 | 1.98359565E+05 | 1.98360825E+05 | -1.30E+00 | 1.98404682E+05 | 1.98405943E+05 | -1.30E+00 | -3.55268000E+00 | -2.29239251E+00 | -1.30E+00 |
| 2710 | Fr | 214 | 87 | 127 | 40 | 7.75773800E+00 | 7.75458908E+00 | 3.10E-03 | 1.99293653E+05 | 1.99294264E+05 | -6.10E-01 | 1.99338771E+05 | 1.99339382E+05 | -6.10E-01 | -9.58197000E-01 | -3.47496317E-01 | -6.10E-01 |
| 2711 | Fr | 215 | 87 | 128 | 41 | 7.75325900E+00 | 7.75251242E+00 | 7.50E-04 | 2.00226424E+05 | 2.00226521E+05 | -9.70E-02 | 2.00271541E+05 | 2.00271639E+05 | -9.70E-02 | 3.18420000E-01 | 4.15714880E-01 | -9.70E-02 |
| 2712 | Fr | 216 | 87 | 129 | 42 | 7.74244900E+00 | 7.74292063E+00 | -4.70E-04 | 2.01160571E+05 | 2.01160406E+05 | 1.70E-01 | 2.01205688E+05 | 2.01205524E+05 | 1.60E-01 | 2.97134100E+00 | 2.80634912E+00 | 1.60E-01 |
| 2713 | Fr | 217 | 87 | 130 | 43 | 7.73777300E+00 | 7.73902226E+00 | -1.20E-03 | 2.02093409E+05 | 2.02093074E+05 | 3.30E-01 | 2.02138526E+05 | 2.02138192E+05 | 3.30E-01 | 4.31494300E+00 | 3.98069442E+00 | 3.30E-01 |
| 2714 | Fr | 218 | 87 | 131 | 44 | 7.72671300E+00 | 7.72815780E+00 | -1.40E-03 | 2.03027647E+05 | 2.03027269E+05 | 3.80E-01 | 2.03072765E+05 | 2.03072387E+05 | 3.80E-01 | 7.05947000E+00 | 6.68144318E+00 | 3.80E-01 |
| 2715 | Fr | 219 | 87 | 132 | 45 | 7.72116800E+00 | 7.72301209E+00 | -1.80E-03 | 2.03960700E+05 | 2.03960233E+05 | 4.70E-01 | 2.04005818E+05 | 2.04005351E+05 | 4.70E-01 | 8.61858100E+00 | 8.15151535E+00 | 4.70E-01 |
| 2716 | Fr | 220 | 87 | 133 | 46 | 7.70973800E+00 | 7.71128469E+00 | -1.50E-03 | 2.04895059E+05 | 2.04894656E+05 | 4.00E-01 | 2.04940177E+05 | 2.04939773E+05 | 4.00E-01 | 1.14831810E+01 | 1.10798495E+01 | 4.00E-01 |
| 2717 | Fr | 221 | 87 | 134 | 47 | 7.70325000E+00 | 7.70413535E+00 | -8.90E-04 | 2.05828348E+05 | 2.05828090E+05 | 2.60E-01 | 2.05873466E+05 | 2.05873207E+05 | 2.60E-01 | 1.32786000E+01 | 1.30198884E+01 | 2.60E-01 |
| 2718 | Fr | 222 | 87 | 135 | 48 | 7.69107400E+00 | 7.69328489E+00 | -2.20E-03 | 2.06762914E+05 | 2.06762360E+05 | 5.50E-01 | 2.06808031E+05 | 2.06807477E+05 | 5.50E-01 | 1.63497620E+01 | 1.57958735E+01 | 5.50E-01 |
| 2719 | Fr | 223 | 87 | 136 | 49 | 7.68365700E+00 | 7.68753432E+00 | -3.90E-03 | 2.07696442E+05 | 2.07695514E+05 | 9.30E-01 | 2.07741560E+05 | 2.07740632E+05 | 9.30E-01 | 1.83839780E+01 | 1.74562844E+01 | 9.30E-01 |

| | | | | | | | | | | | | | | | | | | |
|---|---|---|---|---|---|---|---|---|---|---|---|---|---|---|---|---|---|---|
| 2720 | Fr | 224 | 87 | 137 | 50 | 7.67016000E+00 | 7.67532801E+00 | -5.20E-03 | 2.08631347E+05 | 2.08630126E+05 | 1.20E+00 | 2.08676465E+05 | 2.08675244E+05 | 1.20E+00 | 2.17950970E+01 | 2.05742819E+01 | 1.20E+00 |
| 2721 | Fr | 225 | 87 | 138 | 51 | 7.66294000E+00 | 7.66802220E+00 | -5.10E-03 | 2.09564867E+05 | 2.09563660E+05 | 1.20E+00 | 2.09609984E+05 | 2.09608778E+05 | 1.20E+00 | 2.38208000E+01 | 2.26140807E+01 | 1.20E+00 |
| 2722 | Fr | 226 | 87 | 139 | 52 | 7.64828700E+00 | 7.65444494E+00 | -6.20E-03 | 2.10500081E+05 | 2.10498626E+05 | 1.50E+00 | 2.10545198E+05 | 2.10543744E+05 | 1.50E+00 | 2.75405520E+01 | 2.60858395E+01 | 1.50E+00 |
| 2723 | Fr | 227 | 87 | 140 | 53 | 7.64070100E+00 | 7.64560281E+00 | -4.90E-03 | 2.11433720E+05 | 2.11432544E+05 | 1.20E+00 | 2.11478838E+05 | 2.11477662E+05 | 1.20E+00 | 2.96857820E+01 | 2.85098762E+01 | 1.20E+00 |
| 2724 | Fr | 228 | 87 | 141 | 54 | 7.62643500E+00 | 7.63069630E+00 | -4.30E-03 | 2.12368897E+05 | 2.12367863E+05 | 1.00E+00 | 2.12414015E+05 | 2.12412980E+05 | 1.00E+00 | 3.33690730E+01 | 3.23342773E+01 | 1.00E+00 |
| 2725 | Fr | 229 | 87 | 142 | 55 | 7.61831100E+00 | 7.62040834E+00 | -2.10E-03 | 2.13302697E+05 | 2.13302153E+05 | 5.40E-01 | 2.13347814E+05 | 2.13347271E+05 | 5.40E-01 | 3.56743570E+01 | 3.51308417E+01 | 5.40E-01 |
| 2726 | Fr | 230 | 87 | 143 | 56 | 7.60360100E+00 | 7.60429103E+00 | -6.90E-04 | 2.14238027E+05 | 2.14237805E+05 | 2.20E-01 | 2.14283145E+05 | 2.14282923E+05 | 2.20E-01 | 3.95105750E+01 | 3.92887333E+01 | 2.20E-01 |
| 2727 | Fr | 231 | 87 | 144 | 57 | 7.59457000E+00 | 7.59271529E+00 | 1.90E-03 | 2.15172075E+05 | 2.15172440E+05 | -3.70E-01 | 2.15217193E+05 | 2.15217558E+05 | -3.70E-01 | 4.20644060E+01 | 4.24297587E+01 | -3.70E-01 |
| 2728 | Fr | 232 | 87 | 145 | 58 | 7.58000000E+00 | 7.57555184E+00 | 4.40E-03 | 2.16107491E+05 | 2.16108395E+05 | -9.00E-01 | 2.16152608E+05 | 2.16153512E+05 | -9.00E-01 | 4.59860000E+01 | 4.68902815E+01 | -9.00E-01 |
| 2729 | FR | 233 | 87 | 146 | 59 | 7.56900000E+00 | 7.56286207E+00 | 6.10E-03 | 2.17042033E+05 | 2.17043341E+05 | -1.30E+00 | 2.17087150E+05 | 2.17088459E+05 | -1.30E+00 | 4.90340000E+01 | 5.03427662E+01 | -1.30E+00 |
| 2730 | RA | 201 | 88 | 113 | 25 | 7.67000000E+00 | 7.67114397E+00 | -1.10E-03 | 1.87196499E+05 | 1.87196182E+05 | 3.20E-01 | 1.87242147E+05 | 1.87241831E+05 | 3.20E-01 | 1.18410000E+01 | 1.15244059E+01 | 3.20E-01 |
| 2731 | Ra | 202 | 88 | 114 | 26 | 7.68548700E+00 | 7.68689500E+00 | -1.40E-03 | 1.88125243E+05 | 1.88124895E+05 | 3.50E-01 | 1.88170891E+05 | 1.88170543E+05 | 3.50E-01 | 9.09114300E+00 | 8.74287118E+00 | 3.50E-01 |
| 2732 | Ra | 203 | 88 | 115 | 27 | 7.68947800E+00 | 7.69152475E+00 | -2.00E-03 | 1.89056313E+05 | 1.89055834E+05 | 4.80E-01 | 1.89101961E+05 | 1.89101482E+05 | 4.80E-01 | 8.66687600E+00 | 8.18745657E+00 | 4.80E-01 |
| 2733 | Ra | 204 | 88 | 116 | 28 | 7.70419100E+00 | 7.70579870E+00 | -1.60E-03 | 1.89985187E+05 | 1.89984796E+05 | 3.90E-01 | 1.90030836E+05 | 1.90030444E+05 | 3.90E-01 | 6.04713500E+00 | 5.65536551E+00 | 3.90E-01 |
| 2734 | Ra | 205 | 88 | 117 | 29 | 7.70699800E+00 | 7.70918209E+00 | -2.20E-03 | 1.90916473E+05 | 1.90915962E+05 | 5.10E-01 | 1.90962121E+05 | 1.90961610E+05 | 5.10E-01 | 5.83883300E+00 | 5.32729108E+00 | 5.10E-01 |
| 2735 | Ra | 206 | 88 | 118 | 30 | 7.71980100E+00 | 7.72176442E+00 | -2.00E-03 | 1.91845694E+05 | 1.91845226E+05 | 4.70E-01 | 1.91891342E+05 | 1.91890874E+05 | 4.70E-01 | 3.56580900E+00 | 3.09746650E+00 | 4.70E-01 |
| 2736 | Ra | 207 | 88 | 119 | 31 | 7.72162900E+00 | 7.72359589E+00 | -2.00E-03 | 1.92777161E+05 | 1.92776690E+05 | 4.70E-01 | 1.92822810E+05 | 1.92822339E+05 | 4.70E-01 | 3.53889500E+00 | 3.06790858E+00 | 4.70E-01 |
| 2737 | Ra | 208 | 88 | 120 | 32 | 7.73207900E+00 | 7.73412935E+00 | -2.10E-03 | 1.93706832E+05 | 1.93706341E+05 | 4.90E-01 | 1.93752480E+05 | 1.93751989E+05 | 4.90E-01 | 1.71499400E+00 | 1.22467110E+00 | 4.90E-01 |
| 2738 | Ra | 209 | 88 | 121 | 33 | 7.73303700E+00 | 7.73399697E+00 | -9.60E-04 | 1.94638465E+05 | 1.94638200E+05 | 2.60E-01 | 1.94684113E+05 | 1.94683848E+05 | 2.60E-01 | 1.85409800E+00 | 1.58952821E+00 | 2.60E-01 |
| 2739 | Ra | 210 | 88 | 122 | 34 | 7.74128500E+00 | 7.74207527E+00 | -7.90E-04 | 1.95568565E+05 | 1.95568335E+05 | 2.30E-01 | 1.95614213E+05 | 1.95613983E+05 | 2.30E-01 | 4.60315000E-01 | 2.30407111E-01 | 2.30E-01 |
| 2740 | Ra | 211 | 88 | 123 | 35 | 7.74108700E+00 | 7.73959881E+00 | 1.50E-03 | 1.96500431E+05 | 1.96500681E+05 | -2.50E-01 | 1.96546079E+05 | 1.96546329E+05 | -2.50E-01 | 8.32024000E-01 | 1.08218431E+00 | -2.50E-01 |
| 2741 | Ra | 212 | 88 | 124 | 36 | 7.74750700E+00 | 7.74494700E+00 | 2.60E-03 | 1.97430894E+05 | 1.97431373E+05 | -4.80E-01 | 1.97476542E+05 | 1.97477021E+05 | -4.80E-01 | -1.98673000E-01 | 2.80088071E-01 | -4.80E-01 |
| 2742 | Ra | 213 | 88 | 125 | 37 | 7.74641400E+00 | 7.73998420E+00 | 6.40E-03 | 1.98362945E+05 | 1.98364250E+05 | -1.30E+00 | 1.98408593E+05 | 1.98409899E+05 | -1.30E+00 | 3.57934000E-01 | 1.66353744E+00 | -1.30E+00 |
| 2743 | Ra | 214 | 88 | 126 | 38 | 7.74917000E+00 | 7.74265590E+00 | 6.50E-03 | 1.99294174E+05 | 1.99295504E+05 | -1.30E+00 | 1.99339822E+05 | 1.99341152E+05 | -1.30E+00 | 9.28960000E-02 | 1.42312733E+00 | -1.30E+00 |
| 2744 | Ra | 215 | 88 | 127 | 39 | 7.73931500E+00 | 7.73544123E+00 | 3.90E-03 | 2.00228109E+05 | 2.00228878E+05 | -7.70E-01 | 2.00273757E+05 | 2.00274526E+05 | -7.70E-01 | 2.53405500E+00 | 3.30294354E+00 | -7.70E-01 |
| 2745 | Ra | 216 | 88 | 128 | 40 | 7.73734600E+00 | 7.73585038E+00 | 1.50E-03 | 2.01160360E+05 | 2.01160619E+05 | -2.60E-01 | 2.01206008E+05 | 2.01206268E+05 | -2.60E-01 | 3.29136300E+00 | 3.55044574E+00 | -2.60E-01 |
| 2746 | Ra | 217 | 88 | 129 | 41 | 7.72692000E+00 | 7.72689425E+00 | 2.60E-05 | 2.02094451E+05 | 2.02094392E+05 | 5.80E-02 | 2.02140099E+05 | 2.02140041E+05 | 5.80E-02 | 5.88771900E+00 | 5.82939368E+00 | 5.80E-02 |
| 2747 | Ra | 218 | 88 | 130 | 42 | 7.72499600E+00 | 7.72559614E+00 | -6.00E-04 | 2.03026709E+05 | 2.03026514E+05 | 1.90E-01 | 2.03072357E+05 | 2.03072162E+05 | 1.90E-01 | 6.65145500E+00 | 6.45680696E+00 | 1.90E-01 |
| 2748 | Ra | 219 | 88 | 131 | 43 | 7.71405200E+00 | 7.71540199E+00 | -1.30E-03 | 2.03960946E+05 | 2.03960586E+05 | 3.60E-01 | 2.04006594E+05 | 2.04006234E+05 | 3.60E-01 | 9.39456800E+00 | 9.03505005E+00 | 3.60E-01 |
| 2749 | Ra | 220 | 88 | 132 | 44 | 7.71169400E+00 | 7.71281960E+00 | -1.10E-03 | 2.04893316E+05 | 2.04893004E+05 | 3.10E-01 | 2.04938964E+05 | 2.04938653E+05 | 3.10E-01 | 1.02705340E+01 | 9.95909281E+00 | 3.10E-01 |
| 2750 | Ra | 221 | 88 | 133 | 45 | 7.70113300E+00 | 7.70168224E+00 | -5.50E-04 | 2.05827504E+05 | 2.05827318E+05 | 1.90E-01 | 2.05873152E+05 | 2.05872966E+05 | 1.90E-01 | 1.29642360E+01 | 1.27789471E+01 | 1.90E-01 |
| 2751 | Ra | 222 | 88 | 134 | 46 | 7.69668600E+00 | 7.69699061E+00 | -3.00E-04 | 2.06760355E+05 | 2.06760223E+05 | 1.30E-01 | 2.06806003E+05 | 2.06805872E+05 | 1.30E-01 | 1.43215770E+01 | 1.41901259E+01 | 1.30E-01 |
| 2752 | Ra | 223 | 88 | 135 | 47 | 7.68530200E+00 | 7.68633617E+00 | -1.00E-03 | 2.07694762E+05 | 2.07694468E+05 | 2.90E-01 | 2.07740410E+05 | 2.07740116E+05 | 2.90E-01 | 1.72348080E+01 | 1.69403952E+01 | 2.90E-01 |
| 2753 | Ra | 224 | 88 | 136 | 48 | 7.67991600E+00 | 7.68271949E+00 | -2.80E-03 | 2.08627849E+05 | 2.08627157E+05 | 6.90E-01 | 2.08673497E+05 | 2.08672805E+05 | 6.90E-01 | 1.88273280E+01 | 1.81355149E+01 | 6.90E-01 |
| 2754 | Ra | 225 | 88 | 137 | 49 | 7.66758000E+00 | 7.67075260E+00 | -3.20E-03 | 2.09562510E+05 | 2.09561732E+05 | 7.80E-01 | 2.09608158E+05 | 2.09607380E+05 | 7.80E-01 | 2.19943030E+01 | 2.12166632E+01 | 7.80E-01 |
| 2755 | Ra | 226 | 88 | 138 | 50 | 7.66195400E+00 | 7.66562396E+00 | -3.70E-03 | 2.10495679E+05 | 2.10494786E+05 | 8.90E-01 | 2.10541327E+05 | 2.10540434E+05 | 8.90E-01 | 2.36695710E+01 | 2.27763023E+01 | 8.90E-01 |
| 2756 | Ra | 227 | 88 | 139 | 51 | 7.64829500E+00 | 7.65233007E+00 | -4.00E-03 | 2.11430683E+05 | 2.11429703E+05 | 9.80E-01 | 2.11476331E+05 | 2.11475352E+05 | 9.80E-01 | 2.71794580E+01 | 2.61997111E+01 | 9.80E-01 |
| 2757 | Ra | 228 | 88 | 140 | 52 | 7.64241900E+00 | 7.64569708E+00 | -3.30E-03 | 2.12363940E+05 | 2.12363129E+05 | 8.10E-01 | 2.12409588E+05 | 2.12408777E+05 | 8.10E-01 | 2.89421970E+01 | 2.81310211E+01 | 8.10E-01 |
| 2758 | Ra | 229 | 88 | 141 | 53 | 7.62854400E+00 | 7.63109391E+00 | -2.50E-03 | 2.13299040E+05 | 2.13298393E+05 | 6.50E-01 | 2.13344688E+05 | 2.13344041E+05 | 6.50E-01 | 3.25485080E+01 | 3.19007700E+01 | 6.50E-01 |
| 2759 | Ra | 230 | 88 | 142 | 54 | 7.62191400E+00 | 7.62301264E+00 | -1.10E-03 | 2.14232502E+05 | 2.14232186E+05 | 3.20E-01 | 2.14278150E+05 | 2.14277834E+05 | 3.20E-01 | 3.45163070E+01 | 3.41996874E+01 | 3.20E-01 |
| 2760 | Ra | 231 | 88 | 143 | 55 | 7.60784100E+00 | 7.60717839E+00 | 6.60E-04 | 2.15167696E+05 | 2.15167786E+05 | -8.90E-02 | 2.15213345E+05 | 2.15213434E+05 | -8.90E-02 | 3.82164860E+01 | 3.83057060E+01 | -8.90E-02 |
| 2761 | Ra | 232 | 88 | 144 | 56 | 7.60000900E+00 | 7.59776723E+00 | 2.20E-03 | 2.16101471E+05 | 2.16101927E+05 | -4.60E-01 | 2.16147119E+05 | 2.16147575E+05 | -4.60E-01 | 4.04969540E+01 | 4.09532355E+01 | -4.60E-01 |
| 2762 | Ra | 233 | 88 | 145 | 57 | 7.58561400E+00 | 7.58083361E+00 | 4.80E-03 | 2.17036790E+05 | 2.17037840E+05 | -1.00E+00 | 2.17082439E+05 | 2.17083489E+05 | -1.00E+00 | 4.43223480E+01 | 4.53723198E+01 | -1.00E+00 |
| 2763 | Ra | 234 | 88 | 146 | 58 | 7.57670300E+00 | 7.57024450E+00 | 6.50E-03 | 2.17970855E+05 | 2.17972303E+05 | -1.40E+00 | 2.18016503E+05 | 2.18017951E+05 | -1.40E+00 | 4.68932710E+01 | 4.83406562E+01 | -1.40E+00 |
| 2764 | Ac | 206 | 89 | 117 | 28 | 7.66800000E+00 | 7.67031866E+00 | -2.30E-03 | 1.91855059E+05 | 1.91854510E+05 | 5.50E-01 | 1.91901239E+05 | 1.91900689E+05 | 5.50E-01 | 1.34620000E+01 | 1.29122180E+01 | 5.50E-01 |
| 2765 | Ac | 207 | 89 | 118 | 29 | 7.68110000E+00 | 7.68290186E+00 | -1.80E-03 | 1.92784237E+05 | 1.92783800E+05 | 4.40E-01 | 1.92830417E+05 | 1.92829979E+05 | 4.40E-01 | 1.11461000E+01 | 1.07084955E+01 | 4.40E-01 |
| 2766 | Ac | 208 | 89 | 119 | 30 | 7.68483600E+00 | 7.68633706E+00 | -1.50E-03 | 1.93715345E+05 | 1.93714968E+05 | 3.80E-01 | 1.93761524E+05 | 1.93761147E+05 | 3.80E-01 | 1.07591340E+01 | 1.03823908E+01 | 3.80E-01 |

| | | | | | | | | | | | | | | | | |
|---|---|---|---|---|---|---|---|---|---|---|---|---|---|---|---|---|
| 2767 | Ac | 209 | 89 | 120 | 31 | 7.69584700E+00 | 7.69682904E+00 | -9.80E-04 | 1.94644924E+05 | 1.94644654E+05 | 2.70E-01 | 1.94691103E+05 | 1.94690833E+05 | 2.70E-01 | 8.84431400E+00 | 8.57454968E+00 | 2.70E-01 |
| 2768 | Ac | 210 | 89 | 121 | 32 | 7.69789600E+00 | 7.69832721E+00 | -4.30E-04 | 1.95576363E+05 | 1.95576208E+05 | 1.60E-01 | 1.95622543E+05 | 1.95622387E+05 | 1.60E-01 | 8.78962600E+00 | 8.63442390E+00 | 1.60E-01 |
| 2769 | Ac | 211 | 89 | 122 | 33 | 7.70718900E+00 | 7.70645528E+00 | 7.30E-04 | 1.96506270E+05 | 1.96506360E+05 | -9.00E-02 | 1.96552449E+05 | 1.96552539E+05 | -9.00E-02 | 7.20221800E+00 | 7.29239181E+00 | -9.00E-02 |
| 2770 | Ac | 212 | 89 | 123 | 34 | 7.70854900E+00 | 7.70578144E+00 | 2.80E-03 | 1.97437840E+05 | 1.97438362E+05 | -5.20E-01 | 1.97484019E+05 | 1.97484541E+05 | -5.20E-01 | 7.27792200E+00 | 7.80011069E+00 | -5.20E-01 |
| 2771 | Ac | 213 | 89 | 124 | 35 | 7.71551900E+00 | 7.71140520E+00 | 4.10E-03 | 1.98368212E+05 | 1.98369024E+05 | -8.10E-01 | 1.98414391E+05 | 1.98415203E+05 | -8.10E-01 | 6.15626300E+00 | 6.96778556E+00 | -8.10E-01 |
| 2772 | Ac | 214 | 89 | 125 | 36 | 7.71583400E+00 | 7.70852062E+00 | 7.30E-03 | 1.99299994E+05 | 1.99301495E+05 | -1.50E+00 | 1.99346174E+05 | 1.99347674E+05 | -1.50E+00 | 6.44453100E+00 | 7.94500033E+00 | -1.50E+00 |
| 2773 | Ac | 215 | 89 | 126 | 37 | 7.71941100E+00 | 7.71174442E+00 | 7.70E-03 | 2.00231075E+05 | 2.00232659E+05 | -1.60E+00 | 2.00277254E+05 | 2.00278838E+05 | -1.60E+00 | 6.03096900E+00 | 7.61468230E+00 | -1.60E+00 |
| 2774 | Ac | 216 | 89 | 127 | 38 | 7.71125500E+00 | 7.70687248E+00 | 4.40E-03 | 2.01164682E+05 | 2.01165565E+05 | -8.80E-01 | 2.01210862E+05 | 2.01211744E+05 | -8.80E-01 | 8.14450200E+00 | 9.02659442E+00 | -8.80E-01 |
| 2775 | Ac | 217 | 89 | 128 | 39 | 7.71033700E+00 | 7.70803757E+00 | 2.30E-03 | 2.02096736E+05 | 2.02097170E+05 | -4.30E-01 | 2.02142915E+05 | 2.02143349E+05 | -4.30E-01 | 8.70376200E+00 | 9.13821821E+00 | -4.30E-01 |
| 2776 | Ac | 218 | 89 | 129 | 40 | 7.70217500E+00 | 7.70157130E+00 | 6.00E-04 | 2.03030370E+05 | 2.03030437E+05 | -6.70E-02 | 2.03076549E+05 | 2.03076616E+05 | -6.70E-02 | 1.08441190E+01 | 1.09111462E+01 | -6.70E-02 |
| 2777 | Ac | 219 | 89 | 130 | 41 | 7.70054700E+00 | 7.70109297E+00 | -5.50E-04 | 2.03962590E+05 | 2.03962406E+05 | 1.80E-01 | 2.04008769E+05 | 2.04008585E+05 | 1.80E-01 | 1.15698360E+01 | 1.13856485E+01 | 1.80E-01 |
| 2778 | Ac | 220 | 89 | 131 | 42 | 7.69234900E+00 | 7.69339946E+00 | -1.10E-03 | 2.04896258E+05 | 2.04895963E+05 | 3.00E-01 | 2.04942438E+05 | 2.04942142E+05 | 3.00E-01 | 1.37440580E+01 | 1.34484468E+01 | 3.00E-01 |
| 2779 | Ac | 221 | 89 | 132 | 43 | 7.69053700E+00 | 7.69158605E+00 | -1.00E-03 | 2.05828532E+05 | 2.05828235E+05 | 3.00E-01 | 2.05874711E+05 | 2.05874415E+05 | 3.00E-01 | 1.45234630E+01 | 1.42271287E+01 | 3.00E-01 |
| 2780 | Ac | 222 | 89 | 133 | 44 | 7.68280100E+00 | 7.68287851E+00 | -7.80E-05 | 2.06762124E+05 | 2.06762042E+05 | 8.20E-02 | 2.06808303E+05 | 2.06808221E+05 | 8.20E-02 | 1.66217480E+01 | 1.65399361E+01 | 8.20E-02 |
| 2781 | Ac | 223 | 89 | 134 | 45 | 7.67914000E+00 | 7.67886198E+00 | 2.80E-04 | 2.07694823E+05 | 2.07694820E+05 | 2.60E-03 | 2.07741002E+05 | 2.07741000E+05 | 2.60E-03 | 1.78266970E+01 | 1.78240627E+01 | 2.60E-03 |
| 2782 | Ac | 224 | 89 | 135 | 46 | 7.67013900E+00 | 7.67034338E+00 | -2.00E-04 | 2.08628726E+05 | 2.08628615E+05 | 1.10E-01 | 2.08674905E+05 | 2.08674794E+05 | 1.10E-01 | 2.02349970E+01 | 2.01246867E+01 | 1.10E-01 |
| 2783 | Ac | 225 | 89 | 136 | 47 | 7.66568400E+00 | 7.66715210E+00 | -1.50E-03 | 2.09561623E+05 | 2.09561228E+05 | 3.90E-01 | 2.09607802E+05 | 2.09607407E+05 | 3.90E-01 | 2.16385940E+01 | 2.12436989E+01 | 3.90E-01 |
| 2784 | Ac | 226 | 89 | 137 | 48 | 7.65565600E+00 | 7.65735516E+00 | -1.70E-03 | 2.10495789E+05 | 2.10495341E+05 | 4.50E-01 | 2.10541968E+05 | 2.10541520E+05 | 4.50E-01 | 2.43104870E+01 | 2.38619741E+01 | 4.50E-01 |
| 2785 | Ac | 227 | 89 | 138 | 49 | 7.65070100E+00 | 7.65267302E+00 | -2.00E-03 | 2.11428824E+05 | 2.11428311E+05 | 5.10E-01 | 2.11475003E+05 | 2.11474491E+05 | 5.10E-01 | 2.58510850E+01 | 2.53387835E+01 | 5.10E-01 |
| 2786 | Ac | 228 | 89 | 139 | 50 | 7.63918900E+00 | 7.64157930E+00 | -2.40E-03 | 2.12363363E+05 | 2.12362754E+05 | 6.10E-01 | 2.12409542E+05 | 2.12408933E+05 | 6.10E-01 | 2.88963860E+01 | 2.82867979E+01 | 6.10E-01 |
| 2787 | Ac | 229 | 89 | 140 | 51 | 7.63320800E+00 | 7.63540269E+00 | -2.20E-03 | 2.13296659E+05 | 2.13296092E+05 | 5.70E-01 | 2.13342838E+05 | 2.13342271E+05 | 5.70E-01 | 3.06982090E+01 | 3.01309811E+01 | 5.70E-01 |
| 2788 | Ac | 230 | 89 | 141 | 52 | 7.62146000E+00 | 7.62301115E+00 | -1.60E-03 | 2.14231293E+05 | 2.14230872E+05 | 4.20E-01 | 2.14277472E+05 | 2.14277051E+05 | 4.20E-01 | 3.38383830E+01 | 3.34169521E+01 | 4.20E-01 |
| 2789 | Ac | 231 | 89 | 142 | 53 | 7.61507600E+00 | 7.61537135E+00 | -3.00E-04 | 2.15164712E+05 | 2.15164579E+05 | 1.30E-01 | 2.15210891E+05 | 2.15210758E+05 | 1.30E-01 | 3.57628490E+01 | 3.56300534E+01 | 1.30E-01 |
| 2790 | Ac | 232 | 89 | 143 | 54 | 7.60242400E+00 | 7.60173192E+00 | 6.90E-04 | 2.16099597E+05 | 2.16099693E+05 | -9.60E-02 | 2.16145777E+05 | 2.16145873E+05 | -9.60E-02 | 3.91544190E+01 | 3.92503502E+01 | -9.60E-02 |
| 2791 | Ac | 233 | 89 | 144 | 55 | 7.59519300E+00 | 7.59271977E+00 | 2.50E-03 | 2.17033245E+05 | 2.17033757E+05 | -5.10E-01 | 2.17079424E+05 | 2.17079936E+05 | -5.10E-01 | 4.13080330E+01 | 4.18197674E+01 | -5.10E-01 |
| 2792 | Ac | 234 | 89 | 145 | 56 | 7.58212900E+00 | 7.57794048E+00 | 4.20E-03 | 2.17968272E+05 | 2.17969188E+05 | -9.20E-01 | 2.18014451E+05 | 2.18015367E+05 | -9.20E-01 | 4.48411900E+01 | 4.57567217E+01 | -9.20E-01 |
| 2793 | AC | 235 | 89 | 146 | 57 | 7.57350400E+00 | 7.56769159E+00 | 5.80E-03 | 2.18902282E+05 | 2.18903584E+05 | -1.30E+00 | 2.18948462E+05 | 2.18949763E+05 | -1.30E+00 | 4.73571550E+01 | 4.86585895E+01 | -1.30E+00 |
| 2794 | Ac | 236 | 89 | 147 | 58 | 7.55924200E+00 | 7.55718745E+00 | 2.10E-03 | 2.19837640E+05 | 2.19838060E+05 | -4.20E-01 | 2.19883819E+05 | 2.19884240E+05 | -4.20E-01 | 5.12209920E+01 | 5.16411927E+01 | -4.20E-01 |
| 2795 | TH | 208 | 90 | 118 | 28 | 7.65263800E+00 | 7.65495510E+00 | -2.30E-03 | 1.93720728E+05 | 1.93720181E+05 | 5.50E-01 | 1.93767439E+05 | 1.93766891E+05 | 5.50E-01 | 1.66740000E+01 | 1.61267604E+01 | 5.50E-01 |
| 2796 | Th | 209 | 90 | 119 | 29 | 7.65529800E+00 | 7.65820663E+00 | -2.90E-03 | 1.94652085E+05 | 1.94651412E+05 | 6.70E-01 | 1.94698796E+05 | 1.94698122E+05 | 6.70E-01 | 1.65367890E+01 | 1.58635551E+01 | 6.70E-01 |
| 2797 | Th | 210 | 90 | 120 | 30 | 7.66907500E+00 | 7.67026522E+00 | -1.20E-03 | 1.95581102E+05 | 1.95580787E+05 | 3.20E-01 | 1.95627812E+05 | 1.95627497E+05 | 3.20E-01 | 1.40596380E+01 | 1.37443632E+01 | 3.20E-01 |
| 2798 | Th | 211 | 90 | 121 | 31 | 7.67170600E+00 | 7.67163662E+00 | 6.90E-05 | 1.96512443E+05 | 1.96512392E+05 | 5.10E-02 | 1.96559154E+05 | 1.96559103E+05 | 5.10E-02 | 1.39067100E+01 | 1.38560519E+01 | 5.10E-02 |
| 2799 | Th | 212 | 90 | 122 | 32 | 7.68212300E+00 | 7.68144538E+00 | 6.80E-04 | 1.97442128E+05 | 1.97442207E+05 | -7.80E-02 | 1.97488839E+05 | 1.97488917E+05 | -7.80E-02 | 1.20978680E+01 | 1.21762773E+01 | -7.80E-02 |
| 2800 | Th | 213 | 90 | 123 | 33 | 7.68385600E+00 | 7.68081440E+00 | 3.00E-03 | 1.98373642E+05 | 1.98374225E+05 | -5.80E-01 | 1.98420353E+05 | 1.98420936E+05 | -5.80E-01 | 1.21180130E+01 | 1.27005491E+01 | -5.80E-01 |
| 2801 | Th | 214 | 90 | 124 | 34 | 7.69223500E+00 | 7.68832871E+00 | 3.90E-03 | 1.99303731E+05 | 1.99304501E+05 | -7.70E-01 | 1.99350441E+05 | 1.99351212E+05 | -7.70E-01 | 1.07123290E+01 | 1.14829928E+01 | -7.70E-01 |
| 2802 | Th | 215 | 90 | 125 | 35 | 7.69302500E+00 | 7.68571793E+00 | 7.30E-03 | 2.00235434E+05 | 2.00236940E+05 | -1.50E+00 | 2.00282145E+05 | 2.00283650E+05 | -1.50E+00 | 1.09215570E+01 | 1.24272994E+01 | -1.50E+00 |
| 2803 | Th | 216 | 90 | 126 | 36 | 7.69766100E+00 | 7.69105912E+00 | 6.60E-03 | 2.01166305E+05 | 2.01167666E+05 | -1.40E+00 | 2.01213016E+05 | 2.01214376E+05 | -1.40E+00 | 1.02986040E+01 | 1.16592051E+01 | -1.40E+00 |
| 2804 | Th | 217 | 90 | 127 | 37 | 7.69053700E+00 | 7.68665941E+00 | 3.90E-03 | 2.02099719E+05 | 2.02100495E+05 | -7.80E-01 | 2.02146429E+05 | 2.02147205E+05 | -7.80E-01 | 1.22181290E+01 | 1.29942007E+01 | -7.80E-01 |
| 2805 | Th | 218 | 90 | 128 | 38 | 7.69160100E+00 | 7.69009428E+00 | 1.50E-03 | 2.03031361E+05 | 2.03031625E+05 | -2.60E-01 | 2.03078072E+05 | 2.03078335E+05 | -2.60E-01 | 1.23668230E+01 | 1.26300579E+01 | -2.60E-01 |
| 2806 | Th | 219 | 90 | 129 | 39 | 7.68371800E+00 | 7.68419556E+00 | -4.80E-04 | 2.03964962E+05 | 2.03964792E+05 | 1.70E-01 | 2.04011672E+05 | 2.04011502E+05 | 1.70E-01 | 1.44730710E+01 | 1.43031040E+01 | 1.70E-01 |
| 2807 | Th | 220 | 90 | 130 | 40 | 7.68458800E+00 | 7.68602791E+00 | -1.40E-03 | 2.04896652E+05 | 2.04896270E+05 | 3.80E-01 | 2.04943363E+05 | 2.04942981E+05 | 3.80E-01 | 1.46693060E+01 | 1.42871096E+01 | 3.80E-01 |
| 2808 | Th | 221 | 90 | 131 | 41 | 7.67607000E+00 | 7.67889409E+00 | -2.80E-03 | 2.05830415E+05 | 2.05829726E+05 | 6.90E-01 | 2.05877126E+05 | 2.05876436E+05 | 6.90E-01 | 1.69383540E+01 | 1.62489760E+01 | 6.90E-01 |
| 2809 | Th | 222 | 90 | 132 | 42 | 7.67665700E+00 | 7.67935102E+00 | -2.70E-03 | 2.06762174E+05 | 2.06761511E+05 | 6.60E-01 | 2.06808885E+05 | 2.06808222E+05 | 6.60E-01 | 1.72033170E+01 | 1.65399607E+01 | 6.60E-01 |
| 2810 | Th | 223 | 90 | 133 | 43 | 7.66863900E+00 | 7.67114036E+00 | -2.50E-03 | 2.07695851E+05 | 2.07695228E+05 | 6.20E-01 | 2.07742562E+05 | 2.07741939E+05 | 6.20E-01 | 1.93861170E+01 | 1.87629060E+01 | 6.20E-01 |
| 2811 | Th | 224 | 90 | 134 | 44 | 7.66772300E+00 | 7.66933119E+00 | -1.60E-03 | 2.08627953E+05 | 2.08627527E+05 | 4.30E-01 | 2.08674664E+05 | 2.08674238E+05 | 4.30E-01 | 1.99939440E+01 | 1.95683401E+01 | 4.30E-01 |
| 2812 | Th | 225 | 90 | 135 | 45 | 7.65922100E+00 | 7.66109315E+00 | -1.90E-03 | 2.09561764E+05 | 2.09561277E+05 | 4.90E-01 | 2.09608474E+05 | 2.09607988E+05 | 4.90E-01 | 2.23105510E+01 | 2.18238869E+01 | 4.90E-01 |
| 2813 | Th | 226 | 90 | 136 | 46 | 7.65712000E+00 | 7.65993664E+00 | -2.80E-03 | 2.10494145E+05 | 2.10493443E+05 | 7.00E-01 | 2.10540855E+05 | 2.10540153E+05 | 7.00E-01 | 2.31973520E+01 | 2.24954841E+01 | 7.00E-01 |

| | | | | | | | | | | | | | | | | |
|---|---|---|---|---|---|---|---|---|---|---|---|---|---|---|---|---|
| 2814 | Th | 227 | 90 | 137 | 47 | 7.64745100E+00 | 7.65044134E+00 | -3.00E-03 | 2.11428248E+05 | 2.11427504E+05 | 7.40E-01 | 2.11474958E+05 | 2.11474214E+05 | 7.40E-01 | 2.58063210E+01 | 2.50622995E+01 | 7.40E-01 |
| 2815 | Th | 228 | 90 | 138 | 48 | 7.64507400E+00 | 7.64781596E+00 | -2.70E-03 | 2.12360708E+05 | 2.12360017E+05 | 6.90E-01 | 2.12407418E+05 | 2.12406728E+05 | 6.90E-01 | 2.67723230E+01 | 2.60817626E+01 | 6.90E-01 |
| 2816 | Th | 229 | 90 | 139 | 49 | 7.63464500E+00 | 7.63703930E+00 | -2.40E-03 | 2.13295016E+05 | 2.13294403E+05 | 6.10E-01 | 2.13341727E+05 | 2.13341113E+05 | 6.10E-01 | 2.95867780E+01 | 2.89731202E+01 | 6.10E-01 |
| 2817 | Th | 230 | 90 | 140 | 50 | 7.63098900E+00 | 7.63293181E+00 | -1.90E-03 | 2.14227788E+05 | 2.14227276E+05 | 5.10E-01 | 2.14274498E+05 | 2.14273986E+05 | 5.10E-01 | 3.08642440E+01 | 3.03521227E+01 | 5.10E-01 |
| 2818 | Th | 231 | 90 | 141 | 51 | 7.62011100E+00 | 7.62085883E+00 | -7.50E-04 | 2.15162235E+05 | 2.15161997E+05 | 2.40E-01 | 2.15208946E+05 | 2.15208707E+05 | 2.40E-01 | 3.38175440E+01 | 3.35793688E+01 | 2.40E-01 |
| 2819 | Th | 232 | 90 | 142 | 52 | 7.61502400E+00 | 7.61528317E+00 | -2.60E-04 | 2.16095360E+05 | 2.16095235E+05 | 1.30E-01 | 2.16142071E+05 | 2.16141946E+05 | 1.30E-01 | 3.54487120E+01 | 3.53233817E+01 | 1.30E-01 |
| 2820 | Th | 233 | 90 | 143 | 53 | 7.60288400E+00 | 7.60194446E+00 | 9.40E-04 | 2.17030139E+05 | 2.17030293E+05 | -1.50E-01 | 2.17076850E+05 | 2.17077004E+05 | -1.50E-01 | 3.87336430E+01 | 3.88873383E+01 | -1.50E-01 |
| 2821 | Th | 234 | 90 | 144 | 54 | 7.59684900E+00 | 7.59497219E+00 | 1.90E-03 | 2.17963514E+05 | 2.17963888E+05 | -3.70E-01 | 2.18010225E+05 | 2.18010598E+05 | -3.70E-01 | 4.06144490E+01 | 4.09882224E+01 | -3.70E-01 |
| 2822 | Th | 235 | 90 | 145 | 55 | 7.58438500E+00 | 7.58045878E+00 | 3.90E-03 | 2.18898411E+05 | 2.18899269E+05 | -8.60E-01 | 2.18945122E+05 | 2.18945980E+05 | -8.60E-01 | 4.40177490E+01 | 4.48752226E+01 | -8.60E-01 |
| 2823 | Th | 236 | 90 | 146 | 56 | 7.57696800E+00 | 7.57221241E+00 | 4.80E-03 | 2.19832143E+05 | 2.19833200E+05 | -1.10E+00 | 2.19878854E+05 | 2.19879911E+05 | -1.10E+00 | 4.62551980E+01 | 4.73122246E+01 | -1.10E+00 |
| 2824 | Th | 237 | 90 | 147 | 57 | 7.56344300E+00 | 7.56186861E+00 | 1.60E-03 | 2.20767337E+05 | 2.20767645E+05 | -3.10E-01 | 2.20814048E+05 | 2.20814355E+05 | -3.10E-01 | 4.99550920E+01 | 5.02628125E+01 | -3.10E-01 |
| 2825 | Th | 238 | 90 | 148 | 58 | 7.55400000E+00 | 7.55357475E+00 | 4.30E-04 | 2.21701502E+05 | 2.21701622E+05 | -1.20E-01 | 2.21748212E+05 | 2.21748333E+05 | -1.20E-01 | 5.26260000E+01 | 5.27462020E+01 | -1.20E-01 |
| 2826 | Pa | 212 | 91 | 121 | 30 | 7.63354800E+00 | 7.63514295E+00 | -1.60E-03 | 1.97451112E+05 | 1.97450708E+05 | 4.00E-01 | 1.97498354E+05 | 1.97497950E+05 | 4.00E-01 | 2.16134500E+01 | 2.12093155E+01 | 4.00E-01 |
| 2827 | Pa | 213 | 91 | 122 | 31 | 7.64476000E+00 | 7.64513228E+00 | -3.70E-04 | 1.98380656E+05 | 1.98380510E+05 | 1.50E-01 | 1.98427898E+05 | 1.98427753E+05 | 1.50E-01 | 1.96631300E+01 | 1.95177653E+01 | 1.50E-01 |
| 2828 | Pa | 214 | 91 | 123 | 32 | 7.64758400E+00 | 7.64640439E+00 | 1.20E-03 | 1.99311972E+05 | 1.99312158E+05 | -1.90E-01 | 1.99359215E+05 | 1.99359401E+05 | -1.90E-01 | 1.94854410E+01 | 1.96717202E+01 | -1.90E-01 |
| 2829 | Pa | 215 | 91 | 124 | 33 | 7.65707400E+00 | 7.65427925E+00 | 2.80E-03 | 2.00241850E+05 | 2.00242384E+05 | -5.30E-01 | 2.00289092E+05 | 2.00289627E+05 | -5.30E-01 | 1.78687830E+01 | 1.84035391E+01 | -5.30E-01 |
| 2830 | Pa | 216 | 91 | 125 | 34 | 7.65931100E+00 | 7.65375331E+00 | 5.60E-03 | 2.01173275E+05 | 2.01174409E+05 | -1.10E+00 | 2.01220517E+05 | 2.01221651E+05 | -1.10E+00 | 1.77998380E+01 | 1.89341829E+01 | -1.10E+00 |
| 2831 | Pa | 217 | 91 | 126 | 35 | 7.66457300E+00 | 7.65962784E+00 | 4.90E-03 | 2.02104039E+05 | 2.02105046E+05 | -1.00E+00 | 2.02151281E+05 | 2.02152288E+05 | -1.00E+00 | 1.70699660E+01 | 1.80769740E+01 | -1.00E+00 |
| 2832 | Pa | 218 | 91 | 127 | 36 | 7.65903300E+00 | 7.65745375E+00 | 1.60E-03 | 2.03037147E+05 | 2.03037426E+05 | -2.80E-01 | 2.03084390E+05 | 2.03084668E+05 | -2.80E-01 | 1.86844470E+01 | 1.89626169E+01 | -2.80E-01 |
| 2833 | Pa | 219 | 91 | 128 | 37 | 7.66157300E+00 | 7.66152491E+00 | 4.80E-05 | 2.03968497E+05 | 2.03968442E+05 | 5.60E-02 | 2.04015740E+05 | 2.04015684E+05 | 5.60E-02 | 1.85404840E+01 | 1.84848998E+01 | 5.60E-02 |
| 2834 | Pa | 220 | 91 | 129 | 38 | 7.65600000E+00 | 7.65790909E+00 | -1.90E-03 | 2.04901669E+05 | 2.04901141E+05 | 5.30E-01 | 2.04948911E+05 | 2.04948384E+05 | 5.30E-01 | 2.02180000E+01 | 1.96901729E+01 | 5.30E-01 |
| 2835 | Pa | 221 | 91 | 130 | 39 | 7.65697400E+00 | 7.66039516E+00 | -3.40E-03 | 2.05833321E+05 | 2.05832499E+05 | 8.20E-01 | 2.05880564E+05 | 2.05879742E+05 | 8.20E-01 | 2.03763770E+01 | 1.95541606E+01 | 8.20E-01 |
| 2836 | Pa | 222 | 91 | 131 | 40 | 7.65100000E+00 | 7.65552844E+00 | -4.50E-03 | 2.06766594E+05 | 2.06765485E+05 | 1.10E+00 | 2.06813836E+05 | 2.06812727E+05 | 1.10E+00 | 2.21550000E+01 | 2.10454980E+01 | 1.10E+00 |
| 2837 | Pa | 223 | 91 | 132 | 41 | 7.65196900E+00 | 7.65660097E+00 | -4.60E-03 | 2.07698254E+05 | 2.07697155E+05 | 1.10E+00 | 2.07745497E+05 | 2.07744398E+05 | 1.10E+00 | 2.23210320E+01 | 2.12221148E+01 | 1.10E+00 |
| 2838 | Pa | 224 | 91 | 133 | 42 | 7.64695900E+00 | 7.65060767E+00 | -3.60E-03 | 2.08631290E+05 | 2.08630407E+05 | 8.80E-01 | 2.08678532E+05 | 2.08677649E+05 | 8.80E-01 | 2.38626540E+01 | 2.29793308E+01 | 8.80E-01 |
| 2839 | Pa | 225 | 91 | 134 | 43 | 7.64672000E+00 | 7.64936667E+00 | -2.60E-03 | 2.09563262E+05 | 2.09562601E+05 | 6.60E-01 | 2.09610505E+05 | 2.09609843E+05 | 6.60E-01 | 2.43408780E+01 | 2.36792664E+01 | 6.60E-01 |
| 2840 | Pa | 226 | 91 | 135 | 44 | 7.64110900E+00 | 7.64319346E+00 | -2.10E-03 | 2.10496449E+05 | 2.10495912E+05 | 5.40E-01 | 2.10543691E+05 | 2.10543154E+05 | 5.40E-01 | 2.60334710E+01 | 2.54963660E+01 | 5.40E-01 |
| 2841 | Pa | 227 | 91 | 136 | 45 | 7.63948700E+00 | 7.64247859E+00 | -3.00E-03 | 2.11428742E+05 | 2.11427996E+05 | 7.50E-01 | 2.11475984E+05 | 2.11475239E+05 | 7.50E-01 | 2.68320120E+01 | 2.60867657E+01 | 7.50E-01 |
| 2842 | Pa | 228 | 91 | 137 | 46 | 7.63220300E+00 | 7.63506357E+00 | -2.90E-03 | 2.12362328E+05 | 2.12361610E+05 | 7.20E-01 | 2.12409570E+05 | 2.12408852E+05 | 7.20E-01 | 2.89244470E+01 | 2.82062312E+01 | 7.20E-01 |
| 2843 | Pa | 229 | 91 | 138 | 47 | 7.62986800E+00 | 7.63289065E+00 | -3.00E-03 | 2.13294796E+05 | 2.13294038E+05 | 7.60E-01 | 2.13342038E+05 | 2.13341280E+05 | 7.60E-01 | 2.98983830E+01 | 2.91400842E+01 | 7.60E-01 |
| 2844 | Pa | 230 | 91 | 139 | 48 | 7.62189000E+00 | 7.62420424E+00 | -2.30E-03 | 2.14228566E+05 | 2.14227968E+05 | 6.00E-01 | 2.14275809E+05 | 2.14275210E+05 | 6.00E-01 | 3.21747780E+01 | 3.15763875E+01 | 6.00E-01 |
| 2845 | Pa | 231 | 91 | 140 | 49 | 7.61841900E+00 | 7.62055195E+00 | -2.10E-03 | 2.15161312E+05 | 2.15160753E+05 | 5.60E-01 | 2.15208554E+05 | 2.15207995E+05 | 5.60E-01 | 3.34259570E+01 | 3.28671805E+01 | 5.60E-01 |
| 2846 | Pa | 232 | 91 | 141 | 50 | 7.60950000E+00 | 7.61056936E+00 | -1.10E-03 | 2.16095328E+05 | 2.16095014E+05 | 3.10E-01 | 2.16142570E+05 | 2.16142256E+05 | 3.10E-01 | 3.59479720E+01 | 3.56339087E+01 | 3.10E-01 |
| 2847 | Pa | 233 | 91 | 142 | 51 | 7.60486400E+00 | 7.60543949E+00 | -5.80E-04 | 2.17028364E+05 | 2.17028164E+05 | 2.00E-01 | 2.17075606E+05 | 2.17075406E+05 | 2.00E-01 | 3.74900460E+01 | 3.72899188E+01 | 2.00E-01 |
| 2848 | Pa | 234 | 91 | 143 | 52 | 7.59467700E+00 | 7.59417810E+00 | 5.00E-04 | 2.17962708E+05 | 2.17962759E+05 | -5.10E-02 | 2.18009951E+05 | 2.18010001E+05 | -5.10E-02 | 4.03403610E+01 | 4.03909639E+01 | -5.10E-02 |
| 2849 | Pa | 235 | 91 | 144 | 53 | 7.58841300E+00 | 7.58762833E+00 | 7.80E-04 | 2.18896151E+05 | 2.18896269E+05 | -1.20E-01 | 2.18943393E+05 | 2.18943512E+05 | -1.20E-01 | 4.22888960E+01 | 4.24073007E+01 | -1.20E-01 |
| 2850 | Pa | 236 | 91 | 145 | 54 | 7.57755700E+00 | 7.57516632E+00 | 2.40E-03 | 2.19830690E+05 | 2.19831188E+05 | -5.00E-01 | 2.19877932E+05 | 2.19878430E+05 | -5.00E-01 | 4.53339500E+01 | 4.58320263E+01 | -5.00E-01 |
| 2851 | Pa | 237 | 91 | 146 | 55 | 7.57038400E+00 | 7.56730810E+00 | 3.10E-03 | 2.20764378E+05 | 2.20765041E+05 | -6.60E-01 | 2.20811620E+05 | 2.20812283E+05 | -6.60E-01 | 4.75276180E+01 | 4.81905760E+01 | -6.60E-01 |
| 2852 | Pa | 238 | 91 | 147 | 56 | 7.55834400E+00 | 7.55891644E+00 | -5.70E-04 | 2.21699238E+05 | 2.21699036E+05 | 2.00E-01 | 2.21746481E+05 | 2.21746278E+05 | 2.00E-01 | 5.08940380E+01 | 5.06918015E+01 | 2.00E-01 |
| 2853 | Pa | 239 | 91 | 148 | 57 | 7.55000000E+00 | 7.55093976E+00 | -9.40E-04 | 2.22633176E+05 | 2.22632949E+05 | 2.30E-01 | 2.22680418E+05 | 2.22680191E+05 | 2.30E-01 | 5.33370000E+01 | 5.31106321E+01 | 2.30E-01 |
| 2854 | U | 217 | 92 | 125 | 33 | 7.63400000E+00 | 7.62852441E+00 | 5.50E-03 | 2.02109408E+05 | 2.02110480E+05 | -1.10E+00 | 2.02157183E+05 | 2.02158255E+05 | -1.10E+00 | 2.29710000E+01 | 2.40433427E+01 | -1.10E+00 |
| 2855 | U | 218 | 92 | 126 | 34 | 7.64063900E+00 | 7.63646731E+00 | 4.20E-03 | 2.03039843E+05 | 2.03040685E+05 | -8.40E-01 | 2.03087617E+05 | 2.03088460E+05 | -8.40E-01 | 2.19119730E+01 | 2.27545849E+01 | -8.40E-01 |
| 2856 | U | 219 | 92 | 127 | 35 | 7.63632900E+00 | 7.63461017E+00 | 1.70E-03 | 2.03972711E+05 | 2.03973021E+05 | -3.10E-01 | 2.04020486E+05 | 2.04020796E+05 | -3.10E-01 | 2.32865400E+01 | 2.35961507E+01 | -3.10E-01 |
| 2857 | U | 221 | 92 | 129 | 37 | 7.63500000E+00 | 7.63755535E+00 | -2.60E-03 | 2.05836896E+05 | 2.05836232E+05 | 6.60E-01 | 2.05884671E+05 | 2.05884006E+05 | 6.60E-01 | 2.44830000E+01 | 2.38186833E+01 | 6.60E-01 |
| 2858 | U | 222 | 92 | 130 | 38 | 7.63800000E+00 | 7.64218747E+00 | -4.20E-03 | 2.06768129E+05 | 2.06767131E+05 | 1.00E+00 | 2.06815903E+05 | 2.06814906E+05 | 1.00E+00 | 2.42220000E+01 | 2.32241148E+01 | 1.00E+00 |
| 2859 | U | 223 | 92 | 131 | 39 | 7.63268600E+00 | 7.63765863E+00 | -5.00E-03 | 2.07701240E+05 | 2.07700064E+05 | 1.20E+00 | 2.07749014E+05 | 2.07747839E+05 | 1.20E+00 | 2.58388860E+01 | 2.46631781E+01 | 1.20E+00 |
| 2860 | U | 224 | 92 | 132 | 40 | 7.63520100E+00 | 7.64085135E+00 | -5.70E-03 | 2.08632609E+05 | 2.08631277E+05 | 1.30E+00 | 2.08680384E+05 | 2.08679051E+05 | 1.30E+00 | 2.57140440E+01 | 2.43816702E+01 | 1.30E+00 |
| 2861 | U | 225 | 92 | 133 | 41 | 7.62974500E+00 | 7.63516149E+00 | -5.40E-03 | 2.09565767E+05 | 2.09564482E+05 | 1.30E+00 | 2.09613542E+05 | 2.09612256E+05 | 1.30E+00 | 2.73778670E+01 | 2.60923569E+01 | 1.30E+00 |
| 2862 | U | 226 | 92 | 134 | 42 | 7.63191400E+00 | 7.63601565E+00 | -4.10E-03 | 2.10497212E+05 | 2.10496219E+05 | 9.90E-01 | 2.10544987E+05 | 2.10543993E+05 | 9.90E-01 | 2.73291970E+01 | 2.63354729E+01 | 9.90E-01 |
| 2863 | U | 227 | 92 | 135 | 43 | 7.62639100E+00 | 7.63004226E+00 | -3.70E-03 | 2.11430400E+05 | 2.11429504E+05 | 9.00E-01 | 2.11478174E+05 | 2.11477279E+05 | 9.00E-01 | 2.90223470E+01 | 2.81267370E+01 | 9.00E-01 |
| 2864 | U | 228 | 92 | 136 | 44 | 7.62746500E+00 | 7.63133884E+00 | -3.90E-03 | 2.12362094E+05 | 2.12361144E+05 | 9.50E-01 | 2.12409868E+05 | 2.12408918E+05 | 9.50E-01 | 2.92223570E+01 | 2.82723919E+01 | 9.50E-01 |
| 2865 | U | 229 | 92 | 137 | 45 | 7.62072000E+00 | 7.62413728E+00 | -3.40E-03 | 2.13295576E+05 | 2.13294727E+05 | 8.50E-01 | 2.13343351E+05 | 2.13342501E+05 | 8.50E-01 | 3.12108990E+01 | 3.03615300E+01 | 8.50E-01 |
| 2866 | U | 230 | 92 | 138 | 46 | 7.62092200E+00 | 7.62399039E+00 | -3.10E-03 | 2.14227475E+05 | 2.14226702E+05 | 7.70E-01 | 2.14275249E+05 | 2.14274477E+05 | 7.70E-01 | 3.16149970E+01 | 3.08424965E+01 | 7.70E-01 |
| 2867 | U | 231 | 92 | 139 | 47 | 7.61338000E+00 | 7.61552640E+00 | -2.10E-03 | 2.15161161E+05 | 2.15160599E+05 | 5.60E-01 | 2.15208936E+05 | 2.15208373E+05 | 5.60E-01 | 3.38075140E+01 | 3.32450073E+01 | 5.60E-01 |
| 2868 | U | 232 | 92 | 140 | 48 | 7.61189100E+00 | 7.61390916E+00 | -2.00E-03 | 2.16093459E+05 | 2.16092924E+05 | 5.30E-01 | 2.16141233E+05 | 2.16140698E+05 | 5.30E-01 | 3.46108690E+01 | 3.40759979E+01 | 5.30E-01 |
| 2869 | U | 233 | 92 | 141 | 49 | 7.60395200E+00 | 7.60415131E+00 | -2.00E-04 | 2.17027262E+05 | 2.17027149E+05 | 1.10E-01 | 2.17075036E+05 | 2.17074923E+05 | 1.10E-01 | 3.69202540E+01 | 3.68069875E+01 | 1.10E-01 |
| 2870 | U | 234 | 92 | 142 | 50 | 7.60070700E+00 | 7.60105756E+00 | -3.50E-04 | 2.17959983E+05 | 2.17959834E+05 | 1.50E-01 | 2.18007757E+05 | 2.18007608E+05 | 1.50E-01 | 3.81468280E+01 | 3.79980924E+01 | 1.50E-01 |
| 2871 | U | 235 | 92 | 143 | 51 | 7.59090600E+00 | 7.59001421E+00 | 8.90E-04 | 2.18894250E+05 | 2.18894393E+05 | -1.40E-01 | 2.18942025E+05 | 2.18942168E+05 | -1.40E-01 | 4.09206540E+01 | 4.10635417E+01 | -1.40E-01 |
| 2872 | U | 236 | 92 | 144 | 52 | 7.58647600E+00 | 7.58549158E+00 | 9.80E-04 | 2.19827270E+05 | 2.19827436E+05 | -1.70E-01 | 2.19875045E+05 | 2.19875211E+05 | -1.70E-01 | 4.24465150E+01 | 4.26121861E+01 | -1.70E-01 |
| 2873 | U | 237 | 92 | 145 | 53 | 7.57609400E+00 | 7.57323091E+00 | 2.90E-03 | 2.20761710E+05 | 2.20762322E+05 | -6.10E-01 | 2.20809484E+05 | 2.20810096E+05 | -6.10E-01 | 4.53920550E+01 | 4.60037936E+01 | -6.10E-01 |
| 2874 | U | 238 | 92 | 146 | 54 | 7.57012000E+00 | 7.56738119E+00 | 2.70E-03 | 2.21695121E+05 | 2.21695706E+05 | -5.90E-01 | 2.21742896E+05 | 2.21743481E+05 | -5.90E-01 | 4.73091120E+01 | 4.78941139E+01 | -5.90E-01 |
| 2875 | U | 239 | 92 | 147 | 55 | 7.55855600E+00 | 7.55909905E+00 | -5.40E-04 | 2.22629880E+05 | 2.22629683E+05 | 2.00E-01 | 2.22677655E+05 | 2.22677458E+05 | 2.00E-01 | 5.05740460E+01 | 5.03774852E+01 | 2.00E-01 |
| 2876 | U | 240 | 92 | 148 | 56 | 7.55176600E+00 | 7.55307527E+00 | -1.30E-03 | 2.23563516E+05 | 2.23563135E+05 | 3.80E-01 | 2.23611291E+05 | 2.23610910E+05 | 3.80E-01 | 5.27163720E+01 | 5.23354119E+01 | 3.80E-01 |
| 2877 | U | 241 | 92 | 149 | 57 | 7.53900000E+00 | 7.53960798E+00 | -6.10E-04 | 2.24498491E+05 | 2.24498393E+05 | 9.80E-02 | 2.24546266E+05 | 2.24546168E+05 | 9.80E-02 | 5.61970000E+01 | 5.60992728E+01 | 9.80E-02 |
| 2878 | U | 242 | 92 | 150 | 58 | 7.53200000E+00 | 7.53217162E+00 | -1.70E-04 | 2.25432409E+05 | 2.25432219E+05 | 1.90E-01 | 2.25480184E+05 | 2.25479993E+05 | 1.90E-01 | 5.86210000E+01 | 5.84305817E+01 | 1.90E-01 |
| 2879 | U | 243 | 92 | 151 | 59 | 7.51800000E+00 | 7.51737335E+00 | 6.30E-04 | 2.26367683E+05 | 2.26367848E+05 | -1.60E-01 | 2.26415458E+05 | 2.26415623E+05 | -1.60E-01 | 6.24010000E+01 | 6.25657093E+01 | -1.60E-01 |
| 2880 | Np | 219 | 93 | 126 | 33 | 7.60500000E+00 | 7.60225882E+00 | 2.70E-03 | 2.03978169E+05 | 2.03978790E+05 | -6.20E-01 | 2.04026476E+05 | 2.04027097E+05 | -6.20E-01 | 2.92770000E+01 | 2.98980182E+01 | -6.20E-01 |
| 2881 | Np | 225 | 93 | 132 | 39 | 7.60755700E+00 | 7.61437330E+00 | -6.80E-03 | 2.09569444E+05 | 2.09567843E+05 | 1.60E+00 | 2.09617752E+05 | 2.09616150E+05 | 1.60E+00 | 3.15877930E+01 | 2.99866212E+01 | 1.60E+00 |
| 2882 | Np | 226 | 93 | 133 | 40 | 7.60400000E+00 | 7.61079179E+00 | -6.80E-03 | 2.10502128E+05 | 2.10500604E+05 | 1.50E+00 | 2.10550435E+05 | 2.10548911E+05 | 1.50E+00 | 3.27770000E+01 | 3.12529880E+01 | 1.50E+00 |
| 2883 | Np | 227 | 93 | 134 | 41 | 7.60735000E+00 | 7.61205964E+00 | -4.70E-03 | 2.11433407E+05 | 2.11432270E+05 | 1.10E+00 | 2.11481714E+05 | 2.11480578E+05 | 1.10E+00 | 3.25623560E+01 | 3.14257144E+01 | 1.10E+00 |
| 2884 | Np | 228 | 93 | 135 | 42 | 7.60485100E+00 | 7.60812512E+00 | -3.30E-03 | 2.12365935E+05 | 2.12365121E+05 | 8.10E-01 | 2.12414242E+05 | 2.12413428E+05 | 8.10E-01 | 3.35960700E+01 | 3.27820429E+01 | 8.10E-01 |
| 2885 | Np | 229 | 93 | 136 | 43 | 7.60608500E+00 | 7.60977167E+00 | -3.70E-03 | 2.13297613E+05 | 2.13296701E+05 | 9.10E-01 | 2.13345920E+05 | 2.13345008E+05 | 9.10E-01 | 3.37798290E+01 | 3.28681766E+01 | 9.10E-01 |
| 2886 | Np | 230 | 93 | 137 | 44 | 7.60177500E+00 | 7.60462206E+00 | -2.80E-03 | 2.14230563E+05 | 2.14229841E+05 | 7.20E-01 | 2.14278870E+05 | 2.14278148E+05 | 7.20E-01 | 3.52364870E+01 | 3.45141355E+01 | 7.20E-01 |
| 2887 | Np | 231 | 93 | 138 | 45 | 7.60212400E+00 | 7.60483277E+00 | -2.70E-03 | 2.15162446E+05 | 2.15161753E+05 | 6.90E-01 | 2.15210753E+05 | 2.15210060E+05 | 6.90E-01 | 3.56253280E+01 | 3.49321575E+01 | 6.90E-01 |
| 2888 | Np | 232 | 93 | 139 | 46 | 7.59700000E+00 | 7.59843040E+00 | -1.40E-03 | 2.16095676E+05 | 2.16095199E+05 | 4.80E-01 | 2.16143983E+05 | 2.16143506E+05 | 4.80E-01 | 3.73610000E+01 | 3.68839933E+01 | 4.80E-01 |
| 2889 | Np | 233 | 93 | 140 | 47 | 7.59617500E+00 | 7.59717433E+00 | -1.00E-03 | 2.17027759E+05 | 2.17027459E+05 | 3.00E-01 | 2.17076066E+05 | 2.17075766E+05 | 3.00E-01 | 3.79499870E+01 | 3.76495461E+01 | 3.00E-01 |
| 2890 | Np | 234 | 93 | 141 | 48 | 7.58963000E+00 | 7.58948341E+00 | 1.50E-04 | 2.17961260E+05 | 2.17961226E+05 | 3.30E-02 | 2.18009567E+05 | 2.18009534E+05 | 3.30E-02 | 3.99566740E+01 | 3.99233680E+01 | 3.30E-02 |
| 2891 | Np | 235 | 93 | 142 | 49 | 7.58704900E+00 | 7.58674886E+00 | 3.00E-04 | 2.18893842E+05 | 2.18893845E+05 | -3.00E-03 | 2.18942149E+05 | 2.18942152E+05 | -3.00E-03 | 4.10448740E+01 | 4.10478208E+01 | -2.90E-03 |
| 2892 | Np | 236 | 93 | 143 | 50 | 7.57920800E+00 | 7.57777116E+00 | 1.40E-03 | 2.19827671E+05 | 2.19827942E+05 | -2.70E-01 | 2.19875978E+05 | 2.19876250E+05 | -2.70E-01 | 4.33794380E+01 | 4.36511294E+01 | -2.70E-01 |
| 2893 | Np | 237 | 93 | 144 | 51 | 7.57498100E+00 | 7.57359799E+00 | 1.40E-03 | 2.20760659E+05 | 2.20760919E+05 | -2.60E-01 | 2.20808966E+05 | 2.20809226E+05 | -2.60E-01 | 4.48734660E+01 | 4.51337184E+01 | -2.60E-01 |
| 2894 | Np | 238 | 93 | 145 | 52 | 7.56621300E+00 | 7.56339301E+00 | 2.80E-03 | 2.21694736E+05 | 2.21695340E+05 | -6.00E-01 | 2.21743043E+05 | 2.21743647E+05 | -6.00E-01 | 4.74564630E+01 | 4.80602242E+01 | -6.00E-01 |
| 2895 | Np | 239 | 93 | 146 | 53 | 7.56056100E+00 | 7.55787426E+00 | 2.70E-03 | 2.22628086E+05 | 2.22628661E+05 | -5.70E-01 | 2.22676393E+05 | 2.22676968E+05 | -5.70E-01 | 4.93125860E+01 | 4.98871308E+01 | -5.70E-01 |
| 2896 | Np | 240 | 93 | 147 | 54 | 7.55016700E+00 | 7.55156078E+00 | -1.40E-03 | 2.23562585E+05 | 2.23562183E+05 | 4.00E-01 | 2.23610892E+05 | 2.23610490E+05 | 4.00E-01 | 5.23177580E+01 | 5.19158126E+01 | 4.00E-01 |
| 2897 | Np | 241 | 93 | 148 | 55 | 7.54426200E+00 | 7.54581258E+00 | -1.60E-03 | 2.24496024E+05 | 2.24495582E+05 | 4.40E-01 | 2.24544331E+05 | 2.24543890E+05 | 4.40E-01 | 5.42619810E+01 | 5.38208856E+01 | 4.40E-01 |
| 2898 | Np | 242 | 93 | 149 | 56 | 7.53339600E+00 | 7.53433449E+00 | -9.40E-04 | 2.25430674E+05 | 2.25430380E+05 | 2.90E-01 | 2.25478981E+05 | 2.25478687E+05 | 2.90E-01 | 5.74185750E+01 | 5.71240897E+01 | 2.90E-01 |
| 2899 | Np | 243 | 93 | 150 | 57 | 7.52500000E+00 | 7.52715824E+00 | -2.20E-03 | 2.26364626E+05 | 2.26364155E+05 | 4.70E-01 | 2.26412933E+05 | 2.26412462E+05 | 4.70E-01 | 5.98770000E+01 | 5.94049029E+01 | 4.70E-01 |
| 2900 | Np | 244 | 93 | 151 | 58 | 7.51400000E+00 | 7.51433967E+00 | -3.40E-04 | 2.27299446E+05 | 2.27299320E+05 | 1.30E-01 | 2.27347753E+05 | 2.27347628E+05 | 1.30E-01 | 6.32020000E+01 | 6.30767953E+01 | 1.30E-01 |
| 2901 | Pu | 228 | 94 | 134 | 40 | 7.59052900E+00 | 7.59499017E+00 | -4.50E-03 | 2.12367885E+05 | 2.12366799E+05 | 1.10E+00 | 2.12416725E+05 | 2.12415640E+05 | 1.10E+00 | 3.60790830E+01 | 3.49937348E+01 | 1.10E+00 |
| 2902 | Pu | 229 | 94 | 135 | 41 | 7.58688900E+00 | 7.59118143E+00 | -4.30E-03 | 2.13300693E+05 | 2.13299642E+05 | 1.10E+00 | 2.13349533E+05 | 2.13348482E+05 | 1.10E+00 | 3.73935360E+01 | 3.63422647E+01 | 1.10E+00 |
| 2903 | Pu | 230 | 94 | 136 | 42 | 7.59099300E+00 | 7.59478234E+00 | -3.80E-03 | 2.14231728E+05 | 2.14230788E+05 | 9.40E-01 | 2.14280568E+05 | 2.14279628E+05 | 9.40E-01 | 3.69340020E+01 | 3.59941927E+01 | 9.40E-01 |
| 2904 | Pu | 231 | 94 | 137 | 43 | 7.58722000E+00 | 7.58977130E+00 | -2.60E-03 | 2.15164574E+05 | 2.15163916E+05 | 6.60E-01 | 2.15213414E+05 | 2.15212756E+05 | 6.60E-01 | 3.82858120E+01 | 3.76282813E+01 | 6.60E-01 |
| 2905 | Pu | 232 | 94 | 138 | 44 | 7.58897300E+00 | 7.59195038E+00 | -3.00E-03 | 2.16096145E+05 | 2.16095386E+05 | 7.60E-01 | 2.16144985E+05 | 2.16144226E+05 | 7.60E-01 | 3.83631910E+01 | 3.76042819E+01 | 7.60E-01 |
| 2906 | Pu | 233 | 94 | 139 | 45 | 7.58379500E+00 | 7.58569713E+00 | -1.90E-03 | 2.17029328E+05 | 2.17028817E+05 | 5.10E-01 | 2.17078168E+05 | 2.17077657E+05 | 5.10E-01 | 4.00521150E+01 | 3.95406568E+01 | 5.10E-01 |
| 2907 | Pu | 234 | 94 | 140 | 46 | 7.58460600E+00 | 7.58642133E+00 | -1.80E-03 | 2.17961120E+05 | 2.17960627E+05 | 4.90E-01 | 2.18009960E+05 | 2.18009467E+05 | 4.90E-01 | 4.03498870E+01 | 3.98568165E+01 | 4.90E-01 |

| | | | | | | | | | | | | | | | | |
|---|---|---|---|---|---|---|---|---|---|---|---|---|---|---|---|---|
| 2908 | Pu | 235 | 94 | 141 | 47 | 7.57887300E+00 | 7.57888751E+00 | -1.50E-05 | 2.18894448E+05 | 2.18894376E+05 | 7.20E-02 | 2.18943288E+05 | 2.18943216E+05 | 7.20E-02 | 4.21838290E+01 | 4.21121628E+01 | 7.20E-02 |
| 2909 | Pu | 236 | 94 | 142 | 48 | 7.57791300E+00 | 7.57814172E+00 | -2.30E-04 | 2.19826661E+05 | 2.19826539E+05 | 1.20E-01 | 2.19875501E+05 | 2.19875379E+05 | 1.20E-01 | 4.29028520E+01 | 4.27806002E+01 | 1.20E-01 |
| 2910 | Pu | 237 | 94 | 143 | 49 | 7.57075200E+00 | 7.56932450E+00 | 1.40E-03 | 2.20760346E+05 | 2.20760616E+05 | -2.70E-01 | 2.20809186E+05 | 2.20809456E+05 | -2.70E-01 | 4.50934670E+01 | 4.53634575E+01 | -2.70E-01 |
| 2911 | Pu | 238 | 94 | 144 | 50 | 7.56835300E+00 | 7.56714242E+00 | 1.20E-03 | 2.21692911E+05 | 2.21693131E+05 | -2.20E-01 | 2.21741752E+05 | 2.21741971E+05 | -2.20E-01 | 4.61649460E+01 | 4.63847892E+01 | -2.20E-01 |
| 2912 | Pu | 239 | 94 | 145 | 51 | 7.56031000E+00 | 7.55709414E+00 | 3.20E-03 | 2.22626830E+05 | 2.22627531E+05 | -7.00E-01 | 2.22675671E+05 | 2.22676371E+05 | -7.00E-01 | 4.85900730E+01 | 4.92905042E+01 | -7.00E-01 |
| 2913 | Pu | 240 | 94 | 146 | 52 | 7.55603500E+00 | 7.55356065E+00 | 2.50E-03 | 2.23559862E+05 | 2.23560387E+05 | -5.30E-01 | 2.23608702E+05 | 2.23609227E+05 | -5.30E-01 | 5.01271860E+01 | 5.06527656E+01 | -5.30E-01 |
| 2914 | Pu | 241 | 94 | 147 | 53 | 7.54643100E+00 | 7.54732250E+00 | -8.90E-04 | 2.24494185E+05 | 2.24493902E+05 | 2.80E-01 | 2.24543026E+05 | 2.24542743E+05 | 2.80E-01 | 5.29569810E+01 | 5.26739176E+01 | 2.80E-01 |
| 2915 | Pu | 242 | 94 | 148 | 54 | 7.54132100E+00 | 7.54351669E+00 | -2.20E-03 | 2.25427441E+05 | 2.25426841E+05 | 6.00E-01 | 2.25476281E+05 | 2.25475682E+05 | 6.00E-01 | 5.47185750E+01 | 5.41189216E+01 | 6.00E-01 |
| 2916 | Pu | 243 | 94 | 149 | 55 | 7.53100200E+00 | 7.53214198E+00 | -1.10E-03 | 2.26361973E+05 | 2.26361627E+05 | 3.50E-01 | 2.26410813E+05 | 2.26410468E+05 | 3.50E-01 | 5.77559780E+01 | 5.74107785E+01 | 3.50E-01 |
| 2917 | Pu | 244 | 94 | 150 | 56 | 7.52481100E+00 | 7.52690236E+00 | -2.10E-03 | 2.27295517E+05 | 2.27294939E+05 | 5.80E-01 | 2.27344358E+05 | 2.27343779E+05 | 5.80E-01 | 5.98068150E+01 | 5.92284214E+01 | 5.80E-01 |
| 2918 | Pu | 245 | 94 | 151 | 57 | 7.51327600E+00 | 7.51418291E+00 | -9.10E-04 | 2.28230384E+05 | 2.28230094E+05 | 2.90E-01 | 2.28279225E+05 | 2.28278934E+05 | 2.90E-01 | 6.31795660E+01 | 6.28891034E+01 | 2.90E-01 |
| 2919 | Pu | 246 | 94 | 152 | 58 | 7.50653400E+00 | 7.50742192E+00 | -8.90E-04 | 2.29164095E+05 | 2.29163808E+05 | 2.90E-01 | 2.29212935E+05 | 2.29212648E+05 | 2.90E-01 | 6.53959640E+01 | 6.51094440E+01 | 2.90E-01 |
| 2920 | Pu | 247 | 94 | 153 | 59 | 7.49400000E+00 | 7.49331118E+00 | 6.90E-04 | 2.30099300E+05 | 2.30099352E+05 | -5.10E-02 | 2.30148141E+05 | 2.30148192E+05 | -5.10E-02 | 6.91080000E+01 | 6.91586944E+01 | -5.10E-02 |
| 2921 | AM | 230 | 95 | 135 | 40 | 7.56200000E+00 | 7.56541662E+00 | -3.40E-03 | 2.14237193E+05 | 2.14236226E+05 | 9.70E-01 | 2.14286567E+05 | 2.14285599E+05 | 9.70E-01 | 4.29320000E+01 | 4.19652316E+01 | 9.70E-01 |
| 2922 | Am | 232 | 95 | 137 | 42 | 7.56400000E+00 | 7.56627413E+00 | -2.30E-03 | 2.16100516E+05 | 2.16100027E+05 | 4.90E-01 | 2.16149890E+05 | 2.16149400E+05 | 4.90E-01 | 4.32680000E+01 | 4.27780953E+01 | 4.90E-01 |
| 2923 | Am | 233 | 95 | 138 | 43 | 7.56700000E+00 | 7.56874386E+00 | -1.70E-03 | 2.17032006E+05 | 2.17031450E+05 | 5.60E-01 | 2.17081380E+05 | 2.17080824E+05 | 5.60E-01 | 4.32630000E+01 | 4.27076928E+01 | 5.60E-01 |
| 2924 | Am | 234 | 95 | 139 | 44 | 7.56400000E+00 | 7.56448917E+00 | -4.90E-04 | 2.17964698E+05 | 2.17964442E+05 | 2.60E-01 | 2.18014071E+05 | 2.18013816E+05 | 2.60E-01 | 4.44610000E+01 | 4.42058659E+01 | 2.60E-01 |
| 2925 | Am | 235 | 95 | 140 | 45 | 7.56515000E+00 | 7.56551117E+00 | -3.60E-04 | 2.18896357E+05 | 2.18896203E+05 | 1.50E-01 | 2.18945731E+05 | 2.18945577E+05 | 1.50E-01 | 4.46262430E+01 | 4.44725245E+01 | 1.50E-01 |
| 2926 | Am | 236 | 95 | 141 | 46 | 7.56100000E+00 | 7.55998770E+00 | 1.00E-03 | 2.19829267E+05 | 2.19829507E+05 | -2.40E-01 | 2.19878640E+05 | 2.19878880E+05 | -2.40E-01 | 4.60420000E+01 | 4.62818716E+01 | -2.40E-01 |
| 2927 | Am | 237 | 95 | 142 | 47 | 7.56100000E+00 | 7.55954446E+00 | 1.50E-03 | 2.20761290E+05 | 2.20761617E+05 | -3.30E-01 | 2.20810663E+05 | 2.20810991E+05 | -3.30E-01 | 4.65710000E+01 | 4.68982511E+01 | -3.30E-01 |
| 2928 | Am | 238 | 95 | 143 | 48 | 7.55557700E+00 | 7.55274588E+00 | 2.80E-03 | 2.21694636E+05 | 2.21695241E+05 | -6.00E-01 | 2.21744010E+05 | 2.21744615E+05 | -6.00E-01 | 4.84232900E+01 | 4.90280877E+01 | -6.00E-01 |
| 2929 | Am | 239 | 95 | 144 | 49 | 7.55368100E+00 | 7.55086589E+00 | 2.80E-03 | 2.22627099E+05 | 2.22627703E+05 | -6.00E-01 | 2.22676473E+05 | 2.22677077E+05 | -6.00E-01 | 4.93921900E+01 | 4.99959794E+01 | -6.00E-01 |
| 2930 | Am | 240 | 95 | 145 | 50 | 7.54700500E+00 | 7.54283790E+00 | 4.20E-03 | 2.23560713E+05 | 2.23561644E+05 | -9.30E-01 | 2.23610087E+05 | 2.23611018E+05 | -9.30E-01 | 5.15119760E+01 | 5.24431488E+01 | -9.30E-01 |
| 2931 | Am | 241 | 95 | 146 | 51 | 7.54327100E+00 | 7.53959855E+00 | 3.70E-03 | 2.24493631E+05 | 2.24494447E+05 | -8.20E-01 | 2.24543005E+05 | 2.24543821E+05 | -8.20E-01 | 5.29361980E+01 | 5.37523130E+01 | -8.20E-01 |
| 2932 | Am | 242 | 95 | 147 | 52 | 7.53498300E+00 | 7.53530386E+00 | -3.20E-04 | 2.25427659E+05 | 2.25427512E+05 | 1.50E-01 | 2.25477033E+05 | 2.25476886E+05 | 1.50E-01 | 5.54698760E+01 | 5.53233496E+01 | 1.50E-01 |
| 2933 | Am | 243 | 95 | 148 | 53 | 7.53016800E+00 | 7.53174764E+00 | -1.60E-03 | 2.26360859E+05 | 2.26360407E+05 | 4.50E-01 | 2.26410233E+05 | 2.26409780E+05 | 4.50E-01 | 5.71763150E+01 | 5.67235243E+01 | 4.50E-01 |
| 2934 | Am | 244 | 95 | 149 | 54 | 7.52130000E+00 | 7.52234934E+00 | -1.00E-03 | 2.27295058E+05 | 2.27294733E+05 | 3.20E-01 | 2.27344432E+05 | 2.27344107E+05 | 3.20E-01 | 5.98811410E+01 | 5.95562831E+01 | 3.20E-01 |
| 2935 | Am | 245 | 95 | 150 | 55 | 7.51529600E+00 | 7.51735065E+00 | -2.10E-03 | 2.28228573E+05 | 2.28228001E+05 | 5.70E-01 | 2.28277947E+05 | 2.28277375E+05 | 5.70E-01 | 6.19022800E+01 | 6.13299314E+01 | 5.70E-01 |
| 2936 | Am | 246 | 95 | 151 | 56 | 7.50500000E+00 | 7.50660545E+00 | -1.60E-03 | 2.29163160E+05 | 2.29162692E+05 | 4.70E-01 | 2.29212534E+05 | 2.29212066E+05 | 4.70E-01 | 6.49950000E+01 | 6.45272173E+01 | 4.70E-01 |
| 2937 | Am | 247 | 95 | 152 | 57 | 7.49900000E+00 | 7.50008007E+00 | -1.10E-03 | 2.30096813E+05 | 2.30096363E+05 | 4.50E-01 | 2.30146187E+05 | 2.30145737E+05 | 4.50E-01 | 6.71540000E+01 | 6.67037015E+01 | 4.50E-01 |
| 2938 | Am | 248 | 95 | 153 | 58 | 7.48700000E+00 | 7.48794224E+00 | -9.40E-04 | 2.31031717E+05 | 2.31031439E+05 | 2.80E-01 | 2.31081091E+05 | 2.31080812E+05 | 2.80E-01 | 7.05630000E+01 | 7.02851214E+01 | 2.80E-01 |
| 2939 | Cm | 233 | 96 | 137 | 41 | 7.54600700E+00 | 7.54773515E+00 | -1.70E-03 | 2.17035500E+05 | 2.17035028E+05 | 4.70E-01 | 2.17085408E+05 | 2.17084936E+05 | 4.70E-01 | 4.72919520E+01 | 4.68196440E+01 | 4.70E-01 |
| 2940 | Cm | 234 | 96 | 138 | 42 | 7.55067900E+00 | 7.55208738E+00 | -1.40E-03 | 2.17966427E+05 | 2.17966027E+05 | 4.00E-01 | 2.18016334E+05 | 2.18015935E+05 | 4.00E-01 | 4.67239600E+01 | 4.63248075E+01 | 4.00E-01 |
| 2941 | Cm | 235 | 96 | 139 | 43 | 7.54700000E+00 | 7.54794025E+00 | -9.40E-04 | 2.18899208E+05 | 2.18899015E+05 | 1.90E-01 | 2.18949115E+05 | 2.18948923E+05 | 1.90E-01 | 4.80110000E+01 | 4.78186144E+01 | 1.90E-01 |
| 2942 | Cm | 236 | 96 | 140 | 44 | 7.55029900E+00 | 7.55085924E+00 | -5.60E-04 | 2.19830546E+05 | 2.19830344E+05 | 2.00E-01 | 2.19880453E+05 | 2.19880252E+05 | 2.00E-01 | 4.78550180E+01 | 4.76531120E+01 | 2.00E-01 |
| 2943 | Cm | 237 | 96 | 141 | 45 | 7.54662200E+00 | 7.54545689E+00 | 1.20E-03 | 2.20763432E+05 | 2.20763639E+05 | -2.10E-01 | 2.20813340E+05 | 2.20813546E+05 | -2.10E-01 | 4.92474300E+01 | 4.94539288E+01 | -2.10E-01 |
| 2944 | Cm | 238 | 96 | 142 | 46 | 7.54799700E+00 | 7.54692407E+00 | 1.10E-03 | 2.21695124E+05 | 2.21695309E+05 | -1.90E-01 | 2.21745032E+05 | 2.21745217E+05 | -1.90E-01 | 4.94450250E+01 | 4.96306009E+01 | -1.90E-01 |
| 2945 | Cm | 239 | 96 | 143 | 47 | 7.54305900E+00 | 7.54025762E+00 | 2.80E-03 | 2.22628321E+05 | 2.22628921E+05 | -6.00E-01 | 2.22678229E+05 | 2.22678829E+05 | -6.00E-01 | 5.11484450E+01 | 5.17482782E+01 | -6.00E-01 |
| 2946 | Cm | 240 | 96 | 144 | 48 | 7.54285500E+00 | 7.54029740E+00 | 2.60E-03 | 2.23560393E+05 | 2.23560937E+05 | -5.40E-01 | 2.23610300E+05 | 2.23610844E+05 | -5.40E-01 | 5.17255650E+01 | 5.22697917E+01 | -5.40E-01 |
| 2947 | Cm | 241 | 96 | 145 | 49 | 7.53684100E+00 | 7.53240689E+00 | 4.40E-03 | 2.24493865E+05 | 2.24494863E+05 | -1.00E+00 | 2.24543772E+05 | 2.24544771E+05 | -1.00E+00 | 5.37035830E+01 | 5.47024261E+01 | -1.00E+00 |
| 2948 | Cm | 242 | 96 | 146 | 50 | 7.53449600E+00 | 7.53108981E+00 | 3.40E-03 | 2.25426460E+05 | 2.25427215E+05 | -7.50E-01 | 2.25476368E+05 | 2.25477123E+05 | -7.50E-01 | 5.48054180E+01 | 5.55600711E+01 | -7.50E-01 |
| 2949 | Cm | 243 | 96 | 147 | 51 | 7.52691800E+00 | 7.52685982E+00 | 5.80E-05 | 2.26360333E+05 | 2.26360277E+05 | 5.60E-02 | 2.26410241E+05 | 2.26410185E+05 | 5.60E-02 | 5.71837880E+01 | 5.71281886E+01 | 5.60E-02 |
| 2950 | Cm | 244 | 96 | 148 | 52 | 7.52394400E+00 | 7.52519220E+00 | -1.20E-03 | 2.27293097E+05 | 2.27292723E+05 | 3.70E-01 | 2.27343005E+05 | 2.27342630E+05 | 3.70E-01 | 5.84538410E+01 | 5.80795464E+01 | 3.70E-01 |
| 2951 | Cm | 245 | 96 | 149 | 53 | 7.51576500E+00 | 7.51589691E+00 | -1.30E-04 | 2.28227142E+05 | 2.28227040E+05 | 1.00E-01 | 2.28277050E+05 | 2.28276948E+05 | 1.00E-01 | 6.10048970E+01 | 6.09030208E+01 | 1.00E-01 |
| 2952 | Cm | 246 | 96 | 150 | 54 | 7.51146400E+00 | 7.51278751E+00 | -1.30E-03 | 2.29160250E+05 | 2.29159855E+05 | 4.00E-01 | 2.29210158E+05 | 2.29209762E+05 | 4.00E-01 | 6.26186050E+01 | 6.22233540E+01 | 4.00E-01 |
| 2953 | Cm | 247 | 96 | 151 | 55 | 7.50192600E+00 | 7.50214775E+00 | -2.20E-04 | 2.30094660E+05 | 2.30094535E+05 | 1.20E-01 | 2.30144568E+05 | 2.30144443E+05 | 1.20E-01 | 6.55344510E+01 | 6.54099068E+01 | 1.20E-01 |
| 2954 | Cm | 248 | 96 | 152 | 56 | 7.49672500E+00 | 7.49751481E+00 | -7.90E-04 | 2.31028013E+05 | 2.31027748E+05 | 2.70E-01 | 2.31077921E+05 | 2.31077655E+05 | 2.70E-01 | 6.73934630E+01 | 6.71280470E+01 | 2.70E-01 |

| | | | | | | | | | | | | | | | | |
|---|---|---|---|---|---|---|---|---|---|---|---|---|---|---|---|---|
| 2955 | Cm | 249 | 96 | 153 | 57 | 7.48554700E+00 | 7.48548584E+00 | 6.10E-05 | 2.31962865E+05 | 2.31962811E+05 | 5.40E-02 | 2.32012773E+05 | 2.32012718E+05 | 5.40E-02 | 7.07514100E+01 | 7.06970642E+01 | 5.40E-02 |
| 2956 | Cm | 250 | 96 | 154 | 58 | 7.47893500E+00 | 7.47934566E+00 | -4.10E-04 | 2.32896598E+05 | 2.32896426E+05 | 1.70E-01 | 2.32946505E+05 | 2.32946333E+05 | 1.70E-01 | 7.29902990E+01 | 7.28179434E+01 | 1.70E-01 |
| 2957 | Cm | 251 | 96 | 155 | 59 | 7.46671700E+00 | 7.46599034E+00 | 7.30E-04 | 2.33831751E+05 | 2.33831864E+05 | -1.10E-01 | 2.33881659E+05 | 2.33881771E+05 | -1.10E-01 | 7.66493270E+01 | 7.67621022E+01 | -1.10E-01 |
| 2958 | Cm | 252 | 96 | 156 | 60 | 7.46000000E+00 | 7.45845764E+00 | 1.50E-03 | 2.34765652E+05 | 2.34765861E+05 | -2.10E-01 | 2.34815559E+05 | 2.34815769E+05 | -2.10E-01 | 7.90560000E+01 | 7.92656711E+01 | -2.10E-01 |
| 2959 | BK | 234 | 97 | 137 | 40 | 7.51900000E+00 | 7.52057476E+00 | -1.60E-03 | 2.17972512E+05 | 2.17972084E+05 | 4.30E-01 | 2.18022954E+05 | 2.18022526E+05 | 4.30E-01 | 5.33440000E+01 | 5.29156838E+01 | 4.30E-01 |
| 2960 | Bk | 238 | 97 | 141 | 44 | 7.52500000E+00 | 7.52264349E+00 | 2.40E-03 | 2.21699361E+05 | 2.21699771E+05 | -4.10E-01 | 2.21749803E+05 | 2.21750213E+05 | -4.10E-01 | 5.42160000E+01 | 5.46263033E+01 | -4.10E-01 |
| 2961 | BK | 239 | 97 | 142 | 45 | 7.52700000E+00 | 7.52439210E+00 | 2.60E-03 | 2.22630890E+05 | 2.22631396E+05 | -5.10E-01 | 2.22681332E+05 | 2.22681838E+05 | -5.10E-01 | 5.42510000E+01 | 5.47570590E+01 | -5.10E-01 |
| 2962 | Bk | 240 | 97 | 143 | 46 | 7.52300000E+00 | 7.51965413E+00 | 3.30E-03 | 2.23563798E+05 | 2.23564574E+05 | -7.80E-01 | 2.23614240E+05 | 2.23615016E+05 | -7.80E-01 | 5.56660000E+01 | 5.64411001E+01 | -7.80E-01 |
| 2963 | Bk | 241 | 97 | 144 | 47 | 7.52400000E+00 | 7.51998106E+00 | 4.00E-03 | 2.24495661E+05 | 2.24496541E+05 | -8.80E-01 | 2.24546103E+05 | 2.24546983E+05 | -8.80E-01 | 5.60340000E+01 | 5.69139750E+01 | -8.80E-01 |
| 2964 | Bk | 242 | 97 | 145 | 48 | 7.51900000E+00 | 7.51402804E+00 | 5.00E-03 | 2.25428857E+05 | 2.25430027E+05 | -1.20E+00 | 2.25479299E+05 | 2.25480469E+05 | -1.20E+00 | 5.77350000E+01 | 5.89059442E+01 | -1.20E+00 |
| 2965 | Bk | 243 | 97 | 146 | 49 | 7.51749400E+00 | 7.51299740E+00 | 4.50E-03 | 2.26361306E+05 | 2.26362329E+05 | -1.00E+00 | 2.26411748E+05 | 2.26412771E+05 | -1.00E+00 | 5.86913800E+01 | 5.97136791E+01 | -1.00E+00 |
| 2966 | Bk | 244 | 97 | 147 | 50 | 7.51146800E+00 | 7.51063594E+00 | 8.30E-04 | 2.27294825E+05 | 2.27294957E+05 | -1.30E-01 | 2.27345267E+05 | 2.27345399E+05 | -1.30E-01 | 6.07156920E+01 | 6.08481986E+01 | -1.30E-01 |
| 2967 | Bk | 245 | 97 | 148 | 51 | 7.50926300E+00 | 7.50921862E+00 | 4.40E-05 | 2.28227419E+05 | 2.28227359E+05 | 6.00E-02 | 2.28277861E+05 | 2.28277801E+05 | 6.00E-02 | 6.18156390E+01 | 6.17561243E+01 | 6.00E-02 |
| 2968 | Bk | 246 | 97 | 149 | 52 | 7.50279600E+00 | 7.50183412E+00 | 9.60E-04 | 2.29161066E+05 | 2.29161232E+05 | -1.70E-01 | 2.29211508E+05 | 2.29211674E+05 | -1.70E-01 | 6.39686050E+01 | 6.41348103E+01 | -1.70E-01 |
| 2969 | Bk | 247 | 97 | 150 | 53 | 7.49893500E+00 | 7.49897246E+00 | -3.70E-05 | 2.30094082E+05 | 2.30094002E+05 | 8.00E-02 | 2.30144524E+05 | 2.30144444E+05 | 8.00E-02 | 6.54908300E+01 | 6.54111275E+01 | 8.00E-02 |
| 2970 | Bk | 248 | 97 | 151 | 54 | 7.49100000E+00 | 7.49024834E+00 | 7.50E-04 | 2.31028166E+05 | 2.31028232E+05 | -6.60E-02 | 2.31078608E+05 | 2.31078674E+05 | -6.60E-02 | 6.80810000E+01 | 6.81470540E+01 | -6.60E-02 |
| 2971 | Bk | 249 | 97 | 152 | 55 | 7.48602300E+00 | 7.48586300E+00 | 1.60E-04 | 2.31961430E+05 | 2.31961399E+05 | 3.10E-02 | 2.32011872E+05 | 2.32011841E+05 | 3.10E-02 | 6.98505710E+01 | 6.98200759E+01 | 3.00E-02 |
| 2972 | Bk | 250 | 97 | 153 | 56 | 7.47596100E+00 | 7.47575453E+00 | 2.10E-04 | 2.32896025E+05 | 2.32896006E+05 | 1.90E-02 | 2.32946467E+05 | 2.32946448E+05 | 1.90E-02 | 7.29515350E+01 | 7.29326473E+01 | 1.90E-02 |
| 2973 | Bk | 251 | 97 | 154 | 57 | 7.46925800E+00 | 7.46986005E+00 | -6.00E-04 | 2.33829797E+05 | 2.33829575E+05 | 2.20E-01 | 2.33880239E+05 | 2.33880017E+05 | 2.20E-01 | 7.52293270E+01 | 7.50077273E+01 | 2.20E-01 |
| 2974 | Bk | 252 | 97 | 155 | 58 | 7.45900000E+00 | 7.45842562E+00 | 5.70E-04 | 2.34764597E+05 | 2.34764552E+05 | 4.50E-02 | 2.34815039E+05 | 2.34814994E+05 | 4.50E-02 | 7.85350000E+01 | 7.84906636E+01 | 4.40E-02 |
| 2975 | Bk | 253 | 97 | 156 | 59 | 7.45100000E+00 | 7.45113165E+00 | -1.30E-04 | 2.35698484E+05 | 2.35698504E+05 | -2.10E-02 | 2.35748926E+05 | 2.35748946E+05 | -2.10E-02 | 8.09290000E+01 | 8.09489299E+01 | -2.00E-02 |
| 2976 | Cf | 237 | 98 | 139 | 41 | 7.50335600E+00 | 7.50294078E+00 | 4.20E-04 | 2.20771052E+05 | 2.20771080E+05 | -2.70E-02 | 2.20822029E+05 | 2.20822057E+05 | -2.70E-02 | 5.79368680E+01 | 5.79640919E+01 | -2.70E-02 |
| 2977 | Cf | 238 | 98 | 140 | 42 | 7.50900000E+00 | 7.50792190E+00 | 1.10E-03 | 2.21701887E+05 | 2.21701957E+05 | -6.90E-02 | 2.21752864E+05 | 2.21752933E+05 | -6.90E-02 | 5.72780000E+01 | 5.73469642E+01 | -6.90E-02 |
| 2978 | Cf | 239 | 98 | 141 | 43 | 7.50700000E+00 | 7.50454893E+00 | 2.50E-03 | 2.22634349E+05 | 2.22634820E+05 | -4.70E-01 | 2.22685326E+05 | 2.22685797E+05 | -4.70E-01 | 5.82460000E+01 | 5.87165012E+01 | -4.70E-01 |
| 2979 | Cf | 240 | 98 | 142 | 44 | 7.51023000E+00 | 7.50810297E+00 | 2.10E-03 | 2.23565589E+05 | 2.23566028E+05 | -4.40E-01 | 2.23616565E+05 | 2.23617005E+05 | -4.40E-01 | 5.79908380E+01 | 5.84303011E+01 | -4.40E-01 |
| 2980 | Cf | 241 | 98 | 143 | 45 | 7.50700000E+00 | 7.50349597E+00 | 3.50E-03 | 2.24498419E+05 | 2.24499196E+05 | -7.80E-01 | 2.24549396E+05 | 2.24550173E+05 | -7.80E-01 | 5.93270000E+01 | 6.01038041E+01 | -7.80E-01 |
| 2981 | Cf | 242 | 98 | 144 | 46 | 7.50909900E+00 | 7.50564108E+00 | 3.50E-03 | 2.25429973E+05 | 2.25430738E+05 | -7.70E-01 | 2.25480949E+05 | 2.25481715E+05 | -7.70E-01 | 5.93867240E+01 | 6.01525102E+01 | -7.70E-01 |
| 2982 | Cf | 243 | 98 | 145 | 47 | 7.50500000E+00 | 7.49983067E+00 | 5.20E-03 | 2.26363071E+05 | 2.26364210E+05 | -1.10E+00 | 2.26414048E+05 | 2.26415187E+05 | -1.10E+00 | 6.09910000E+01 | 6.21301193E+01 | -1.10E+00 |
| 2983 | Cf | 244 | 98 | 146 | 48 | 7.50513100E+00 | 7.50062552E+00 | 4.50E-03 | 2.27295053E+05 | 2.27296082E+05 | -1.00E+00 | 2.27346030E+05 | 2.27347059E+05 | -1.00E+00 | 6.14793510E+01 | 6.25076624E+01 | -1.00E+00 |
| 2984 | Cf | 245 | 98 | 147 | 49 | 7.49965600E+00 | 7.49834108E+00 | 1.30E-03 | 2.28228455E+05 | 2.28228706E+05 | -2.50E-01 | 2.28279432E+05 | 2.28279683E+05 | -2.50E-01 | 6.33869350E+01 | 6.36380454E+01 | -2.50E-01 |
| 2985 | Cf | 246 | 98 | 148 | 50 | 7.49911400E+00 | 7.49872219E+00 | 3.90E-04 | 2.29160654E+05 | 2.29160679E+05 | -2.50E-02 | 2.29211631E+05 | 2.29211656E+05 | -2.50E-02 | 6.40919400E+01 | 6.41172688E+01 | -2.50E-02 |
| 2986 | Cf | 247 | 98 | 149 | 51 | 7.49328500E+00 | 7.49146088E+00 | 1.80E-03 | 2.30094160E+05 | 2.30094540E+05 | -3.80E-01 | 2.30145137E+05 | 2.30145516E+05 | -3.80E-01 | 6.61039040E+01 | 6.64834098E+01 | -3.80E-01 |
| 2987 | Cf | 248 | 98 | 150 | 52 | 7.49103500E+00 | 7.49040331E+00 | 6.30E-04 | 2.31026790E+05 | 2.31026876E+05 | -8.60E-02 | 2.31077767E+05 | 2.31077853E+05 | -8.60E-02 | 6.72399570E+01 | 6.73255456E+01 | -8.60E-02 |
| 2988 | Cf | 249 | 98 | 151 | 53 | 7.48338200E+00 | 7.48181078E+00 | 1.60E-03 | 2.31960770E+05 | 2.31961090E+05 | -3.20E-01 | 2.32011747E+05 | 2.32012067E+05 | -3.20E-01 | 6.97259620E+01 | 7.00459998E+01 | -3.20E-01 |
| 2989 | Cf | 250 | 98 | 152 | 54 | 7.47994900E+00 | 7.47923729E+00 | 7.10E-04 | 2.32893710E+05 | 2.32893817E+05 | -1.10E-01 | 2.32944687E+05 | 2.32944794E+05 | -1.10E-01 | 7.11719580E+01 | 7.12788811E+01 | -1.10E-01 |
| 2990 | Cf | 251 | 98 | 153 | 55 | 7.47049500E+00 | 7.46926911E+00 | 1.20E-03 | 2.33828169E+05 | 2.33828405E+05 | -2.40E-01 | 2.33879146E+05 | 2.33879382E+05 | -2.40E-01 | 7.41363270E+01 | 7.43729759E+01 | -2.40E-01 |
| 2991 | Cf | 252 | 98 | 154 | 56 | 7.46534400E+00 | 7.46519197E+00 | 1.50E-04 | 2.34761562E+05 | 2.34761529E+05 | 3.30E-02 | 2.34812539E+05 | 2.34812506E+05 | 3.30E-02 | 7.60352470E+01 | 7.60024658E+01 | 3.30E-02 |
| 2992 | Cf | 253 | 98 | 155 | 57 | 7.45482600E+00 | 7.45390235E+00 | 9.20E-04 | 2.35696323E+05 | 2.35696485E+05 | -1.60E-01 | 2.35747300E+05 | 2.35747462E+05 | -1.60E-01 | 7.93022760E+01 | 7.94648653E+01 | -1.60E-01 |
| 2993 | Cf | 254 | 98 | 156 | 58 | 7.44922300E+00 | 7.44842535E+00 | 8.00E-04 | 2.36629857E+05 | 2.36629988E+05 | -1.30E-01 | 2.36680834E+05 | 2.36680965E+05 | -1.30E-01 | 8.13420270E+01 | 8.14734411E+01 | -1.30E-01 |
| 2994 | Cf | 255 | 98 | 157 | 59 | 7.43800000E+00 | 7.43593972E+00 | 2.10E-03 | 2.37564819E+05 | 2.37565289E+05 | -4.70E-01 | 2.37615796E+05 | 2.37616266E+05 | -4.70E-01 | 8.48110000E+01 | 8.52801709E+01 | -4.70E-01 |
| 2995 | Cf | 256 | 98 | 158 | 60 | 7.43200000E+00 | 7.42920781E+00 | 2.80E-03 | 2.38498543E+05 | 2.38499142E+05 | -6.00E-01 | 2.38549520E+05 | 2.38550119E+05 | -6.00E-01 | 8.70410000E+01 | 8.76389186E+01 | -6.00E-01 |
| 2996 | Es | 241 | 99 | 142 | 43 | 7.48500000E+00 | 7.48208796E+00 | 2.90E-03 | 2.24502420E+05 | 2.24503037E+05 | -6.20E-01 | 2.24553932E+05 | 2.24554549E+05 | -6.20E-01 | 6.38630000E+01 | 6.44800583E+01 | -6.20E-01 |
| 2997 | Es | 242 | 99 | 143 | 44 | 7.48300000E+00 | 7.47928817E+00 | 3.70E-03 | 2.25434852E+05 | 2.25435798E+05 | -9.50E-01 | 2.25486364E+05 | 2.25487310E+05 | -9.50E-01 | 6.48010000E+01 | 6.57468370E+01 | -9.50E-01 |
| 2998 | Es | 243 | 99 | 144 | 45 | 7.48600000E+00 | 7.48173307E+00 | 4.30E-03 | 2.26366293E+05 | 2.26367290E+05 | -1.00E+00 | 2.26417805E+05 | 2.26418802E+05 | -1.00E+00 | 6.47480000E+01 | 6.57447583E+01 | -1.00E+00 |
| 2999 | Es | 244 | 99 | 145 | 46 | 7.48300000E+00 | 7.47774252E+00 | 5.30E-03 | 2.27299066E+05 | 2.27300347E+05 | -1.30E+00 | 2.27350578E+05 | 2.27351859E+05 | -1.30E+00 | 6.60270000E+01 | 6.73080372E+01 | -1.30E+00 |
| 3000 | Es | 245 | 99 | 146 | 47 | 7.48400000E+00 | 7.47884169E+00 | 5.20E-03 | 2.28230901E+05 | 2.28232165E+05 | -1.30E+00 | 2.28282413E+05 | 2.28283677E+05 | -1.30E+00 | 6.63680000E+01 | 6.76323186E+01 | -1.30E+00 |
| 3001 | Es | 246 | 99 | 147 | 48 | 7.48000000E+00 | 7.47831361E+00 | 1.70E-03 | 2.29163929E+05 | 2.29164382E+05 | -4.50E-01 | 2.29215441E+05 | 2.29215894E+05 | -4.50E-01 | 6.79020000E+01 | 6.83547019E+01 | -4.50E-01 |
| 3002 | Es | 247 | 99 | 148 | 49 | 7.48009900E+00 | 7.47896993E+00 | 1.10E-03 | 2.30096100E+05 | 2.30096307E+05 | -2.10E-01 | 2.30147611E+05 | 2.30147819E+05 | -2.10E-01 | 6.85784680E+01 | 6.87855977E+01 | -2.10E-01 |
| 3003 | Es | 248 | 99 | 149 | 50 | 7.47600000E+00 | 7.47351326E+00 | 2.50E-03 | 2.31029316E+05 | 2.31029746E+05 | -4.30E-01 | 2.31080828E+05 | 2.31081258E+05 | -4.30E-01 | 7.03010000E+01 | 7.07312010E+01 | -4.30E-01 |
| 3004 | Es | 249 | 99 | 150 | 51 | 7.47400000E+00 | 7.47273354E+00 | 1.30E-03 | 2.31961685E+05 | 2.31962032E+05 | -3.50E-01 | 2.32013198E+05 | 2.32013544E+05 | -3.50E-01 | 7.11770000E+01 | 7.15231570E+01 | -3.50E-01 |
| 3005 | Es | 250 | 99 | 151 | 52 | 7.46900000E+00 | 7.46595557E+00 | 3.00E-03 | 2.32895230E+05 | 2.32895819E+05 | -5.90E-01 | 2.32946742E+05 | 2.32947331E+05 | -5.90E-01 | 7.32270000E+01 | 7.38162344E+01 | -5.90E-01 |
| 3006 | Es | 251 | 99 | 152 | 53 | 7.46587500E+00 | 7.46366454E+00 | 2.20E-03 | 2.33828011E+05 | 2.33828494E+05 | -4.80E-01 | 2.33879523E+05 | 2.33880006E+05 | -4.80E-01 | 7.45135460E+01 | 7.49966465E+01 | -4.80E-01 |
| 3007 | Es | 252 | 99 | 153 | 54 | 7.45724000E+00 | 7.45552076E+00 | 1.70E-03 | 2.34762287E+05 | 2.34762648E+05 | -3.60E-01 | 2.34813799E+05 | 2.34814160E+05 | -3.60E-01 | 7.72952470E+01 | 7.76565340E+01 | -3.60E-01 |
| 3008 | Es | 253 | 99 | 154 | 55 | 7.45287100E+00 | 7.45172844E+00 | 1.10E-03 | 2.35695500E+05 | 2.35695717E+05 | -2.20E-01 | 2.35747012E+05 | 2.35747229E+05 | -2.20E-01 | 7.90146460E+01 | 7.92317876E+01 | -2.20E-01 |
| 3009 | Es | 254 | 99 | 155 | 56 | 7.44358300E+00 | 7.44226843E+00 | 1.30E-03 | 2.36629972E+05 | 2.36630234E+05 | -2.60E-01 | 2.36681484E+05 | 2.36681746E+05 | -2.60E-01 | 8.19921510E+01 | 8.22542225E+01 | -2.60E-01 |
| 3010 | Es | 255 | 99 | 156 | 57 | 7.43781600E+00 | 7.43707279E+00 | 7.40E-04 | 2.37563564E+05 | 2.37563682E+05 | -1.20E-01 | 2.37615076E+05 | 2.37615194E+05 | -1.20E-01 | 8.40905830E+01 | 8.42081600E+01 | -1.20E-01 |
| 3011 | Es | 256 | 99 | 157 | 58 | 7.42800000E+00 | 7.42641526E+00 | 1.60E-03 | 2.38498155E+05 | 2.38498538E+05 | -3.80E-01 | 2.38549666E+05 | 2.38550050E+05 | -3.80E-01 | 8.71870000E+01 | 8.75707356E+01 | -3.80E-01 |
| 3012 | Es | 257 | 99 | 158 | 59 | 7.42200000E+00 | 7.41995553E+00 | 2.00E-03 | 2.39431866E+05 | 2.39432337E+05 | -4.70E-01 | 2.39483378E+05 | 2.39483849E+05 | -4.70E-01 | 8.94030000E+01 | 8.98757886E+01 | -4.70E-01 |
| 3013 | FM | 241 | 100 | 141 | 41 | 7.46000000E+00 | 7.45699551E+00 | 3.00E-03 | 2.24507147E+05 | 2.24507765E+05 | -6.20E-01 | 2.24559195E+05 | 2.24559813E+05 | -6.20E-01 | 6.91260000E+01 | 6.97442606E+01 | -6.20E-01 |
| 3014 | Fm | 242 | 100 | 142 | 42 | 7.46500000E+00 | 7.46252615E+00 | 2.50E-03 | 2.25437915E+05 | 2.25438535E+05 | -6.20E-01 | 2.25489962E+05 | 2.25490583E+05 | -6.20E-01 | 6.84000000E+01 | 6.90201708E+01 | -6.20E-01 |
| 3015 | Fm | 243 | 100 | 143 | 43 | 7.46400000E+00 | 7.45987497E+00 | 4.10E-03 | 2.26370373E+05 | 2.26371282E+05 | -9.10E-01 | 2.26422421E+05 | 2.26423330E+05 | -9.10E-01 | 6.93640000E+01 | 7.02731988E+01 | -9.10E-01 |
| 3016 | Fm | 244 | 100 | 144 | 44 | 7.46800000E+00 | 7.46402013E+00 | 4.00E-03 | 2.27301469E+05 | 2.27302376E+05 | -9.10E-01 | 2.27353517E+05 | 2.27354424E+05 | -9.10E-01 | 6.89660000E+01 | 6.98732237E+01 | -9.10E-01 |
| 3017 | Fm | 245 | 100 | 145 | 45 | 7.46600000E+00 | 7.46019380E+00 | 5.80E-03 | 2.28234184E+05 | 2.28235415E+05 | -1.20E+00 | 2.28286232E+05 | 2.28287463E+05 | -1.20E+00 | 7.01870000E+01 | 7.14179738E+01 | -1.20E+00 |
| 3018 | Fm | 246 | 100 | 146 | 46 | 7.46797100E+00 | 7.46300288E+00 | 5.00E-03 | 2.29165680E+05 | 2.29166829E+05 | -1.10E+00 | 2.29217728E+05 | 2.29218877E+05 | -1.10E+00 | 7.01885120E+01 | 7.13380665E+01 | -1.10E+00 |
| 3019 | Fm | 247 | 100 | 147 | 47 | 7.46400000E+00 | 7.46257936E+00 | 1.40E-03 | 2.30098659E+05 | 2.30099036E+05 | -3.80E-01 | 2.30150707E+05 | 2.30151084E+05 | -3.80E-01 | 7.16740000E+01 | 7.20509911E+01 | -3.80E-01 |
| 3020 | Fm | 248 | 100 | 148 | 48 | 7.46594100E+00 | 7.46492323E+00 | 1.00E-03 | 2.31030378E+05 | 2.31030558E+05 | -1.80E-01 | 2.31082426E+05 | 2.31082606E+05 | -1.80E-01 | 7.18987280E+01 | 7.20784499E+01 | -1.80E-01 |
| 3021 | Fm | 249 | 100 | 149 | 49 | 7.46185900E+00 | 7.45962289E+00 | 2.20E-03 | 2.31963494E+05 | 2.31963978E+05 | -4.80E-01 | 2.32015542E+05 | 2.32016026E+05 | -4.80E-01 | 7.35205400E+01 | 7.40046304E+01 | -4.80E-01 |
| 3022 | Fm | 250 | 100 | 150 | 50 | 7.46208500E+00 | 7.46053969E+00 | 1.50E-03 | 2.32895541E+05 | 2.32895855E+05 | -3.10E-01 | 2.32947589E+05 | 2.32947902E+05 | -3.10E-01 | 7.40733660E+01 | 7.43871272E+01 | -3.10E-01 |
| 3023 | Fm | 251 | 100 | 151 | 51 | 7.45702000E+00 | 7.45393122E+00 | 3.10E-03 | 2.33828916E+05 | 2.33829618E+05 | -7.00E-01 | 2.33880963E+05 | 2.33881666E+05 | -7.00E-01 | 7.59539190E+01 | 7.66566324E+01 | -7.00E-01 |
| 3024 | Fm | 252 | 100 | 152 | 52 | 7.45603100E+00 | 7.45334777E+00 | 2.70E-03 | 2.34761273E+05 | 2.34761877E+05 | -6.00E-01 | 2.34813321E+05 | 2.34813924E+05 | -6.00E-01 | 7.68175720E+01 | 7.74210492E+01 | -6.00E-01 |
| 3025 | Fm | 253 | 100 | 153 | 53 | 7.44845700E+00 | 7.44538669E+00 | 3.10E-03 | 2.35695299E+05 | 2.35696003E+05 | -7.00E-01 | 2.35747346E+05 | 2.35748051E+05 | -7.00E-01 | 7.93489180E+01 | 8.00531754E+01 | -7.00E-01 |
| 3026 | Fm | 254 | 100 | 154 | 54 | 7.44478600E+00 | 7.44331086E+00 | 1.50E-03 | 2.36628348E+05 | 2.36628650E+05 | -3.00E-01 | 2.36680396E+05 | 2.36680698E+05 | -3.00E-01 | 8.09043510E+01 | 8.12063683E+01 | -3.00E-01 |
| 3027 | Fm | 255 | 100 | 155 | 55 | 7.43588300E+00 | 7.43404232E+00 | 1.80E-03 | 2.37562739E+05 | 2.37563136E+05 | -4.00E-01 | 2.37614786E+05 | 2.37615183E+05 | -4.00E-01 | 8.38009630E+01 | 8.41978538E+01 | -4.00E-01 |
| 3028 | Fm | 256 | 100 | 156 | 56 | 7.43177800E+00 | 7.43056614E+00 | 1.20E-03 | 2.38495919E+05 | 2.38496157E+05 | -2.40E-01 | 2.38547967E+05 | 2.38548205E+05 | -2.40E-01 | 8.54873680E+01 | 8.57250330E+01 | -2.40E-01 |
| 3029 | Fm | 257 | 100 | 157 | 57 | 7.42219100E+00 | 7.42010229E+00 | 2.10E-03 | 2.39430517E+05 | 2.39430981E+05 | -4.60E-01 | 2.39482564E+05 | 2.39483029E+05 | -4.60E-01 | 8.85907410E+01 | 8.90549947E+01 | -4.60E-01 |
| 3030 | Fm | 258 | 100 | 158 | 58 | 7.41800000E+00 | 7.41535875E+00 | 2.60E-03 | 2.40363847E+05 | 2.40364350E+05 | -5.00E-01 | 2.40415894E+05 | 2.40416398E+05 | -5.00E-01 | 9.04270000E+01 | 9.09300440E+01 | -5.00E-01 |
| 3031 | Fm | 259 | 100 | 159 | 59 | 7.40700000E+00 | 7.40382475E+00 | 3.20E-03 | 2.41298620E+05 | 2.41299487E+05 | -8.70E-01 | 2.41350667E+05 | 2.41351535E+05 | -8.70E-01 | 9.37050000E+01 | 9.45733110E+01 | -8.70E-01 |
| 3032 | Md | 245 | 101 | 144 | 43 | 7.44200000E+00 | 7.43696170E+00 | 5.00E-03 | 2.28238733E+05 | 2.28239788E+05 | -1.10E+00 | 2.28291317E+05 | 2.28292372E+05 | -1.10E+00 | 7.52720000E+01 | 7.63267622E+01 | -1.10E+00 |
| 3033 | Md | 246 | 101 | 145 | 44 | 7.44100000E+00 | 7.43482304E+00 | 6.20E-03 | 2.29171070E+05 | 2.29172442E+05 | -1.40E+00 | 2.29223654E+05 | 2.29225026E+05 | -1.40E+00 | 7.61150000E+01 | 7.74872287E+01 | -1.40E+00 |
| 3034 | Md | 247 | 101 | 146 | 45 | 7.44400000E+00 | 7.43796758E+00 | 6.00E-03 | 2.30102386E+05 | 2.30103796E+05 | -1.40E+00 | 2.30154970E+05 | 2.30156380E+05 | -1.40E+00 | 7.59370000E+01 | 7.73470245E+01 | -1.40E+00 |
| 3035 | Md | 248 | 101 | 147 | 46 | 7.44200000E+00 | 7.43917186E+00 | 2.80E-03 | 2.31035092E+05 | 2.31035625E+05 | -5.30E-01 | 2.31087676E+05 | 2.31088209E+05 | -5.30E-01 | 7.71490000E+01 | 7.76817135E+01 | -5.30E-01 |
| 3036 | Md | 249 | 101 | 148 | 47 | 7.44400000E+00 | 7.44182809E+00 | 2.20E-03 | 2.31966671E+05 | 2.31967090E+05 | -4.20E-01 | 2.32019255E+05 | 2.32019674E+05 | -4.20E-01 | 7.72340000E+01 | 7.76524591E+01 | -4.20E-01 |
| 3037 | Md | 250 | 101 | 149 | 48 | 7.44100000E+00 | 7.43820695E+00 | 2.80E-03 | 2.32899563E+05 | 2.32900118E+05 | -5.60E-01 | 2.32952147E+05 | 2.32952702E+05 | -5.60E-01 | 7.86320000E+01 | 7.91872350E+01 | -5.60E-01 |
| 3038 | Md | 251 | 101 | 150 | 49 | 7.44190000E+00 | 7.43944307E+00 | 2.50E-03 | 2.33831392E+05 | 2.33831935E+05 | -5.40E-01 | 2.33883976E+05 | 2.33884519E+05 | -5.40E-01 | 7.89668240E+01 | 7.95100825E+01 | -5.40E-01 |
| 3039 | Md | 252 | 101 | 151 | 50 | 7.43800000E+00 | 7.43452670E+00 | 3.50E-03 | 2.34764430E+05 | 2.34765300E+05 | -8.70E-01 | 2.34817014E+05 | 2.34817884E+05 | -8.70E-01 | 8.05110000E+01 | 8.13808828E+01 | -8.70E-01 |
| 3040 | Md | 253 | 101 | 152 | 51 | 7.43800000E+00 | 7.43427130E+00 | 3.70E-03 | 2.35696588E+05 | 2.35697496E+05 | -9.10E-01 | 2.35749172E+05 | 2.35750080E+05 | -9.10E-01 | 8.11750000E+01 | 8.20822910E+01 | -9.10E-01 |
| 3041 | Md | 254 | 101 | 153 | 52 | 7.43200000E+00 | 7.42801540E+00 | 4.00E-03 | 2.36630362E+05 | 2.36631216E+05 | -8.50E-01 | 2.36682946E+05 | 2.36683800E+05 | -8.50E-01 | 8.34540000E+01 | 8.43083375E+01 | -8.50E-01 |
| 3042 | Md | 255 | 101 | 154 | 53 | 7.42872400E+00 | 7.42627406E+00 | 2.40E-03 | 2.37563246E+05 | 2.37563797E+05 | -5.50E-01 | 2.37615830E+05 | 2.37616381E+05 | -5.50E-01 | 8.48443390E+01 | 8.53956834E+01 | -5.50E-01 |
| 3043 | Md | 256 | 101 | 155 | 54 | 7.42100000E+00 | 7.41871950E+00 | 2.30E-03 | 2.38497353E+05 | 2.38497870E+05 | -5.20E-01 | 2.38549937E+05 | 2.38550454E+05 | -5.20E-01 | 8.74570000E+01 | 8.79746958E+01 | -5.20E-01 |
| 3044 | Md | 257 | 101 | 156 | 55 | 7.41756600E+00 | 7.41557825E+00 | 2.00E-03 | 2.39430387E+05 | 2.39430824E+05 | -4.40E-01 | 2.39482971E+05 | 2.39483408E+05 | -4.40E-01 | 8.89971610E+01 | 8.94345953E+01 | -4.40E-01 |
| 3045 | Md | 258 | 101 | 157 | 56 | 7.40966800E+00 | 7.40683050E+00 | 2.80E-03 | 2.40364572E+05 | 2.40365231E+05 | -6.60E-01 | 2.40417156E+05 | 2.40417815E+05 | -6.60E-01 | 9.16883490E+01 | 9.23472549E+01 | -6.60E-01 |
| 3046 | Md | 259 | 101 | 158 | 57 | 7.40500000E+00 | 7.40241592E+00 | 2.60E-03 | 2.41298003E+05 | 2.41298533E+05 | -5.30E-01 | 2.41350587E+05 | 2.41351117E+05 | -5.30E-01 | 9.36250000E+01 | 9.41551203E+01 | -5.30E-01 |
| 3047 | Md | 260 | 101 | 159 | 58 | 7.39600000E+00 | 7.39259466E+00 | 3.40E-03 | 2.42232424E+05 | 2.42233249E+05 | -8.30E-01 | 2.42285008E+05 | 2.42285833E+05 | -8.30E-01 | 9.65520000E+01 | 9.73775510E+01 | -8.30E-01 |
| 3048 | No | 248 | 102 | 146 | 44 | 7.42400000E+00 | 7.42087424E+00 | 3.10E-03 | 2.31038027E+05 | 2.31038843E+05 | -8.20E-01 | 2.31091148E+05 | 2.31091964E+05 | -8.20E-01 | 8.06210000E+01 | 8.14364453E+01 | -8.20E-01 |

| | | | | | | | | | | | | | | | | |
|---|---|---|---|---|---|---|---|---|---|---|---|---|---|---|---|---|
| 3049 | No | 250 | 102 | 148 | 46 | 7.42600000E+00 | 7.42644856E+00 | -4.50E-04 | 2.32901958E+05 | 2.32901738E+05 | 2.20E-01 | 2.32955079E+05 | 2.32954859E+05 | 2.20E-01 | 8.15630000E+01 | 8.13437558E+01 | 2.20E-01 |
| 3050 | No | 251 | 102 | 149 | 47 | 7.42300000E+00 | 7.42297068E+00 | 2.90E-05 | 2.33834739E+05 | 2.33834750E+05 | -1.10E-02 | 2.33887860E+05 | 2.33887871E+05 | -1.10E-02 | 8.28500000E+01 | 8.28615747E+01 | -1.20E-02 |
| 3051 | No | 252 | 102 | 150 | 48 | 7.42579500E+00 | 7.42584307E+00 | -4.80E-05 | 2.34766255E+05 | 2.34766169E+05 | 8.60E-02 | 2.34819375E+05 | 2.34819289E+05 | 8.60E-02 | 8.28721420E+01 | 8.27860802E+01 | 8.60E-02 |
| 3052 | No | 253 | 102 | 151 | 49 | 7.42246600E+00 | 7.42108270E+00 | 1.40E-03 | 2.35699237E+05 | 2.35699513E+05 | -2.80E-01 | 2.35752357E+05 | 2.35752633E+05 | -2.80E-01 | 8.43599190E+01 | 8.46359291E+01 | -2.80E-01 |
| 3053 | No | 254 | 102 | 152 | 50 | 7.42358500E+00 | 7.42247995E+00 | 1.10E-03 | 2.36631095E+05 | 2.36631302E+05 | -2.10E-01 | 2.36684216E+05 | 2.36684423E+05 | -2.10E-01 | 8.47246530E+01 | 8.49312659E+01 | -2.10E-01 |
| 3054 | No | 255 | 102 | 153 | 51 | 7.41795800E+00 | 7.41639294E+00 | 1.60E-03 | 2.37564672E+05 | 2.37564997E+05 | -3.30E-01 | 2.37617793E+05 | 2.37618118E+05 | -3.30E-01 | 8.68072350E+01 | 8.71322918E+01 | -3.30E-01 |
| 3055 | No | 256 | 102 | 154 | 52 | 7.41653900E+00 | 7.41631857E+00 | 2.20E-04 | 2.38497183E+05 | 2.38497165E+05 | 1.80E-02 | 2.38550304E+05 | 2.38550286E+05 | 1.80E-02 | 8.78239290E+01 | 8.78062564E+01 | 1.80E-02 |
| 3056 | No | 257 | 102 | 155 | 53 | 7.40964500E+00 | 7.40894166E+00 | 7.00E-04 | 2.39431103E+05 | 2.39431210E+05 | -1.10E-01 | 2.39484224E+05 | 2.39484331E+05 | -1.10E-01 | 9.02504330E+01 | 9.03571237E+01 | -1.10E-01 |
| 3057 | No | 258 | 102 | 156 | 54 | 7.40700000E+00 | 7.40747591E+00 | -4.80E-04 | 2.40363826E+05 | 2.40363745E+05 | 8.20E-02 | 2.40416947E+05 | 2.40416865E+05 | 8.20E-02 | 9.14790000E+01 | 9.13976639E+01 | 8.10E-02 |
| 3058 | No | 259 | 102 | 157 | 55 | 7.40000000E+00 | 7.39890802E+00 | 1.10E-03 | 2.41297952E+05 | 2.41298122E+05 | -1.70E-01 | 2.41351073E+05 | 2.41351242E+05 | -1.70E-01 | 9.41110000E+01 | 9.42805905E+01 | -1.70E-01 |
| 3059 | No | 260 | 102 | 158 | 56 | 7.39700000E+00 | 7.39617089E+00 | 8.30E-04 | 2.42230948E+05 | 2.42231000E+05 | -5.20E-02 | 2.42284068E+05 | 2.42284121E+05 | -5.20E-02 | 9.56120000E+01 | 9.56646546E+01 | -5.30E-02 |
| 3060 | No | 262 | 102 | 160 | 58 | 7.38500000E+00 | 7.38262950E+00 | 2.40E-03 | 2.44098425E+05 | 2.44098886E+05 | -4.60E-01 | 2.44151546E+05 | 2.44152007E+05 | -4.60E-01 | 1.00102000E+02 | 1.00562796E+02 | -4.60E-01 |
| 3061 | Lr | 252 | 103 | 149 | 46 | 7.39900000E+00 | 7.39945218E+00 | -4.50E-04 | 2.34771583E+05 | 2.34771499E+05 | 8.40E-02 | 2.34825241E+05 | 2.34825157E+05 | 8.40E-02 | 8.87380000E+01 | 8.86535081E+01 | 8.40E-02 |
| 3062 | Lr | 253 | 103 | 150 | 47 | 7.40300000E+00 | 7.40257173E+00 | 4.30E-04 | 2.35702916E+05 | 2.35702876E+05 | 4.10E-02 | 2.35756574E+05 | 2.35756534E+05 | 4.10E-02 | 8.85770000E+01 | 8.85361283E+01 | 4.10E-02 |
| 3063 | Lr | 254 | 103 | 151 | 48 | 7.40000000E+00 | 7.39950727E+00 | 4.90E-04 | 2.36635707E+05 | 2.36635817E+05 | -1.10E-01 | 2.36689365E+05 | 2.36689475E+05 | -1.10E-01 | 8.98730000E+01 | 8.99832487E+01 | -1.10E-01 |
| 3064 | Lr | 255 | 103 | 152 | 49 | 7.40257600E+00 | 7.40116506E+00 | 1.40E-03 | 2.37567275E+05 | 2.37567560E+05 | -2.90E-01 | 2.37620933E+05 | 2.37621218E+05 | -2.80E-01 | 8.99473010E+01 | 9.02323235E+01 | -2.90E-01 |
| 3065 | Lr | 256 | 103 | 153 | 50 | 7.39816000E+00 | 7.39678694E+00 | 1.40E-03 | 2.38500568E+05 | 2.38500845E+05 | -2.80E-01 | 2.38554226E+05 | 2.38554503E+05 | -2.80E-01 | 9.17465990E+01 | 9.20232767E+01 | -2.80E-01 |
| 3066 | Lr | 257 | 103 | 154 | 51 | 7.39700000E+00 | 7.39698433E+00 | 1.60E-05 | 2.39432923E+05 | 2.39432963E+05 | -4.00E-02 | 2.39486581E+05 | 2.39486621E+05 | -4.00E-02 | 9.26070000E+01 | 9.26470784E+01 | -4.00E-02 |
| 3067 | Lr | 258 | 103 | 155 | 52 | 7.39200000E+00 | 7.39132542E+00 | 6.70E-04 | 2.40366594E+05 | 2.40366591E+05 | 2.80E-03 | 2.40420252E+05 | 2.40420249E+05 | 2.70E-03 | 9.47840000E+01 | 9.47814128E+01 | 2.60E-03 |
| 3068 | Lr | 259 | 103 | 156 | 53 | 7.39000000E+00 | 7.39013683E+00 | -1.40E-04 | 2.41299157E+05 | 2.41299073E+05 | 8.30E-02 | 2.41352814E+05 | 2.41352731E+05 | 8.30E-02 | 9.58530000E+01 | 9.57692509E+01 | 8.40E-02 |
| 3069 | Lr | 260 | 103 | 157 | 54 | 7.38300000E+00 | 7.38328965E+00 | -2.90E-04 | 2.42233075E+05 | 2.42233029E+05 | 4.70E-02 | 2.42286733E+05 | 2.42286687E+05 | 4.70E-02 | 9.82770000E+01 | 9.82307007E+01 | 4.60E-02 |
| 3070 | Lr | 261 | 103 | 158 | 55 | 7.38100000E+00 | 7.38082841E+00 | 1.70E-04 | 2.43165854E+05 | 2.43165853E+05 | 6.90E-04 | 2.43219512E+05 | 2.43219511E+05 | 6.70E-04 | 9.95620000E+01 | 9.95611130E+01 | 8.90E-04 |
| 3071 | Lr | 262 | 103 | 159 | 56 | 7.37400000E+00 | 7.37290130E+00 | 1.10E-03 | 2.44099889E+05 | 2.44100115E+05 | -2.30E-01 | 2.44153547E+05 | 2.44153772E+05 | -2.30E-01 | 1.02103000E+02 | 1.02328506E+02 | -2.30E-01 |
| 3072 | Lr | 266 | 103 | 163 | 60 | 7.34900000E+00 | 7.34576318E+00 | 3.20E-03 | 2.47835384E+05 | 2.47836103E+05 | -7.20E-01 | 2.47889042E+05 | 2.47889761E+05 | -7.20E-01 | 1.11622000E+02 | 1.12340917E+02 | -7.20E-01 |
| 3073 | Rf | 253 | 104 | 149 | 45 | 7.38000000E+00 | 7.38047039E+00 | -4.70E-04 | 2.35707359E+05 | 2.35707147E+05 | 2.10E-01 | 2.35761555E+05 | 2.35761342E+05 | 2.10E-01 | 9.35570000E+01 | 9.33446914E+01 | 2.10E-01 |
| 3074 | Rf | 254 | 104 | 150 | 46 | 7.38400000E+00 | 7.38522237E+00 | -1.20E-03 | 2.36638495E+05 | 2.36638125E+05 | 3.70E-01 | 2.36692690E+05 | 2.36692320E+05 | 3.70E-01 | 9.31980000E+01 | 9.28285355E+01 | 3.70E-01 |
| 3075 | Rf | 255 | 104 | 151 | 47 | 7.38200000E+00 | 7.38223548E+00 | -2.40E-04 | 2.37571121E+05 | 2.37571066E+05 | 5.40E-02 | 2.37625316E+05 | 2.37625262E+05 | 5.40E-02 | 9.43310000E+01 | 9.42762888E+01 | 5.50E-02 |
| 3076 | Rf | 256 | 104 | 152 | 48 | 7.38543100E+00 | 7.38554263E+00 | -1.10E-04 | 2.38502507E+05 | 2.38502403E+05 | 1.00E-01 | 2.38556702E+05 | 2.38556598E+05 | 1.00E-01 | 9.42226850E+01 | 9.41187423E+01 | 1.00E-01 |
| 3077 | Rf | 257 | 104 | 153 | 49 | 7.38170000E+00 | 7.38125863E+00 | 4.40E-04 | 2.39435646E+05 | 2.39435684E+05 | -3.80E-02 | 2.39489841E+05 | 2.39489879E+05 | -3.80E-02 | 9.58676120E+01 | 9.59055059E+01 | -3.80E-02 |
| 3078 | Rf | 258 | 104 | 154 | 50 | 7.38253300E+00 | 7.38312080E+00 | -5.90E-04 | 2.40367615E+05 | 2.40367387E+05 | 2.30E-01 | 2.40421810E+05 | 2.40421583E+05 | 2.30E-01 | 9.63422630E+01 | 9.61151287E+01 | 2.30E-01 |
| 3079 | Rf | 259 | 104 | 155 | 51 | 7.37700000E+00 | 7.37756885E+00 | -5.70E-04 | 2.41301128E+05 | 2.41301008E+05 | 1.20E-01 | 2.41355324E+05 | 2.41355203E+05 | 1.20E-01 | 9.83620000E+01 | 9.82412808E+01 | 1.20E-01 |
| 3080 | Rf | 260 | 104 | 156 | 52 | 7.37700000E+00 | 7.37805523E+00 | -1.10E-03 | 2.42233410E+05 | 2.42233069E+05 | 3.40E-01 | 2.42287605E+05 | 2.42287264E+05 | 3.40E-01 | 9.91490000E+01 | 9.88085725E+01 | 3.40E-01 |
| 3081 | Rf | 261 | 104 | 157 | 53 | 7.37137200E+00 | 7.37132165E+00 | 5.00E-05 | 2.43167076E+05 | 2.43167014E+05 | 6.20E-02 | 2.43221272E+05 | 2.43221209E+05 | 6.20E-02 | 1.01321771E+02 | 1.01259301E+02 | 6.20E-02 |
| 3082 | Rf | 262 | 104 | 158 | 54 | 7.37000000E+00 | 7.37053926E+00 | -5.40E-04 | 2.44099643E+05 | 2.44099413E+05 | 2.30E-01 | 2.44153839E+05 | 2.44153608E+05 | 2.30E-01 | 1.02394000E+02 | 1.02164283E+02 | 2.30E-01 |
| 3083 | Rf | 263 | 104 | 159 | 55 | 7.36400000E+00 | 7.36272668E+00 | 1.30E-03 | 2.45033531E+05 | 2.45033662E+05 | -1.30E-01 | 2.45087726E+05 | 2.45087858E+05 | -1.30E-01 | 1.04788000E+02 | 1.04919773E+02 | -1.30E-01 |
| 3084 | Rf | 265 | 104 | 161 | 57 | 7.35400000E+00 | 7.35196853E+00 | 2.00E-03 | 2.46900421E+05 | 2.46900919E+05 | -5.00E-01 | 2.46954617E+05 | 2.46955114E+05 | -5.00E-01 | 1.08691000E+02 | 1.09187867E+02 | -5.00E-01 |
| 3085 | Rf | 267 | 104 | 163 | 59 | 7.34200000E+00 | 7.33916353E+00 | 2.80E-03 | 2.48768165E+05 | 2.48768764E+05 | -6.00E-01 | 2.48822360E+05 | 2.48822960E+05 | -6.00E-01 | 1.13445000E+02 | 1.14045502E+02 | -6.00E-01 |
| 3086 | Db | 255 | 105 | 150 | 45 | 7.35800000E+00 | 7.35878494E+00 | -7.80E-04 | 2.37575985E+05 | 2.37575725E+05 | 2.60E-01 | 2.37630719E+05 | 2.37630459E+05 | 2.60E-01 | 9.97330000E+01 | 9.94730999E+01 | 2.60E-01 |
| 3087 | Db | 256 | 105 | 151 | 46 | 7.35800000E+00 | 7.35747272E+00 | 5.30E-04 | 2.38508245E+05 | 2.38508268E+05 | -2.30E-02 | 2.38562978E+05 | 2.38563001E+05 | -2.30E-02 | 1.00499000E+02 | 1.00521561E+02 | -2.30E-02 |
| 3088 | Db | 257 | 105 | 152 | 47 | 7.36200000E+00 | 7.36098003E+00 | 1.00E-03 | 2.39439448E+05 | 2.39439574E+05 | -1.30E-01 | 2.39494182E+05 | 2.39494308E+05 | -1.30E-01 | 1.00208000E+02 | 1.00334029E+02 | -1.30E-01 |
| 3089 | Db | 258 | 105 | 153 | 48 | 7.35800000E+00 | 7.35838611E+00 | -3.90E-04 | 2.40372532E+05 | 2.40372448E+05 | 8.50E-02 | 2.40427266E+05 | 2.40427181E+05 | 8.50E-02 | 1.01799000E+02 | 1.01713600E+02 | 8.50E-02 |
| 3090 | Db | 259 | 105 | 154 | 49 | 7.36036200E+00 | 7.36046167E+00 | -1.00E-04 | 2.41304219E+05 | 2.41304117E+05 | 1.00E-01 | 2.41358953E+05 | 2.41358851E+05 | 1.00E-01 | 1.01991017E+02 | 1.01888965E+02 | 1.00E-01 |
| 3091 | Db | 260 | 105 | 155 | 50 | 7.35700000E+00 | 7.35661203E+00 | 3.90E-04 | 2.42237395E+05 | 2.42237323E+05 | 7.20E-02 | 2.42292128E+05 | 2.42292057E+05 | 7.20E-02 | 1.03673000E+02 | 1.03600728E+02 | 7.20E-02 |
| 3092 | Db | 261 | 105 | 156 | 51 | 7.35700000E+00 | 7.35731989E+00 | -3.20E-04 | 2.43169466E+05 | 2.43169347E+05 | 1.20E-01 | 2.43224200E+05 | 2.43224081E+05 | 1.20E-01 | 1.04250000E+02 | 1.04130682E+02 | 1.20E-01 |
| 3093 | Db | 262 | 105 | 157 | 52 | 7.35200000E+00 | 7.35229505E+00 | -3.00E-04 | 2.44102968E+05 | 2.44102872E+05 | 9.60E-02 | 2.44157701E+05 | 2.44157605E+05 | 9.60E-02 | 1.06257000E+02 | 1.06161189E+02 | 9.60E-02 |
| 3094 | Db | 263 | 105 | 158 | 53 | 7.35200000E+00 | 7.35173601E+00 | 2.60E-04 | 2.45035317E+05 | 2.45035232E+05 | 8.50E-02 | 2.45090051E+05 | 2.45089965E+05 | 8.50E-02 | 1.07112000E+02 | 1.07027243E+02 | 8.50E-02 |
| 3095 | Db | 266 | 105 | 161 | 56 | 7.33900000E+00 | 7.33680211E+00 | 2.20E-03 | 2.47835424E+05 | 2.47835845E+05 | -4.20E-01 | 2.47890158E+05 | 2.47890579E+05 | -4.20E-01 | 1.12738000E+02 | 1.13158408E+02 | -4.20E-01 |
| 3096 | Db | 267 | 105 | 162 | 57 | 7.33600000E+00 | 7.33397348E+00 | 2.00E-03 | 2.48768256E+05 | 2.48768829E+05 | -5.70E-01 | 2.48822990E+05 | 2.48823562E+05 | -5.70E-01 | 1.14075000E+02 | 1.14648168E+02 | -5.70E-01 |
| 3097 | Db | 268 | 105 | 163 | 58 | 7.32800000E+00 | 7.32591560E+00 | 2.10E-03 | 2.49702736E+05 | 2.49703220E+05 | -4.80E-01 | 2.49757470E+05 | 2.49757953E+05 | -4.80E-01 | 1.17062000E+02 | 1.17545027E+02 | -4.80E-01 |
| 3098 | DB | 270 | 105 | 165 | 60 | 7.31400000E+00 | 7.31301891E+00 | 9.80E-04 | 2.51571020E+05 | 2.51571181E+05 | -1.60E-01 | 2.51625754E+05 | 2.51625914E+05 | -1.60E-01 | 1.22357000E+02 | 1.22517938E+02 | -1.60E-01 |
| 3099 | Sg | 258 | 106 | 152 | 46 | 7.34200000E+00 | 7.34234409E+00 | -3.40E-04 | 2.40375438E+05 | 2.40375265E+05 | 1.70E-01 | 2.40430711E+05 | 2.40430537E+05 | 1.70E-01 | 1.05243000E+02 | 1.05069365E+02 | 1.70E-01 |
| 3100 | Sg | 259 | 106 | 153 | 47 | 7.34000000E+00 | 7.33979922E+00 | 2.00E-04 | 2.41308249E+05 | 2.41308147E+05 | 1.00E-01 | 2.41363521E+05 | 2.41363419E+05 | 1.00E-01 | 1.06559000E+02 | 1.06457461E+02 | 1.00E-01 |
| 3101 | Sg | 260 | 106 | 154 | 48 | 7.34256000E+00 | 7.34350172E+00 | -9.40E-04 | 2.42239732E+05 | 2.42239410E+05 | 3.20E-01 | 2.42295004E+05 | 2.42294682E+05 | 3.20E-01 | 1.06548108E+02 | 1.06226332E+02 | 3.20E-01 |
| 3102 | Sg | 261 | 106 | 155 | 49 | 7.33976600E+00 | 7.33971634E+00 | 5.00E-05 | 2.43172684E+05 | 2.43172620E+05 | 6.40E-02 | 2.43227956E+05 | 2.43227892E+05 | 6.40E-02 | 1.08006228E+02 | 1.07942132E+02 | 6.40E-02 |
| 3103 | Sg | 262 | 106 | 156 | 50 | 7.34118100E+00 | 7.34206339E+00 | -8.80E-04 | 2.44104539E+05 | 2.44104231E+05 | 3.10E-01 | 2.44159811E+05 | 2.44159503E+05 | 3.10E-01 | 1.08366919E+02 | 1.08058807E+02 | 3.10E-01 |
| 3104 | Sg | 263 | 106 | 157 | 51 | 7.33700000E+00 | 7.33711248E+00 | -1.10E-04 | 2.45037856E+05 | 2.45037756E+05 | 1.00E-01 | 2.45093128E+05 | 2.45093028E+05 | 1.00E-01 | 1.10190000E+02 | 1.10090152E+02 | 1.00E-01 |
| 3105 | Sg | 264 | 106 | 158 | 52 | 7.33800000E+00 | 7.33819916E+00 | -2.00E-04 | 2.45969943E+05 | 2.45969697E+05 | 2.50E-01 | 2.46025216E+05 | 2.46024970E+05 | 2.50E-01 | 1.10784000E+02 | 1.10537477E+02 | 2.50E-01 |
| 3106 | Sg | 265 | 106 | 159 | 53 | 7.33300000E+00 | 7.33217427E+00 | 8.30E-04 | 2.46903451E+05 | 2.46903521E+05 | -7.00E-02 | 2.46958724E+05 | 2.46958793E+05 | -7.00E-02 | 1.12797000E+02 | 1.12867191E+02 | -7.00E-02 |
| 3107 | Sg | 266 | 106 | 160 | 54 | 7.33200000E+00 | 7.33209376E+00 | -9.40E-05 | 2.47835767E+05 | 2.47835776E+05 | -8.80E-03 | 2.47891039E+05 | 2.47891048E+05 | -8.80E-03 | 1.13619000E+02 | 1.13627751E+02 | -8.80E-03 |
| 3108 | SG | 269 | 106 | 163 | 57 | 7.31800000E+00 | 7.31590939E+00 | 2.10E-03 | 2.50636445E+05 | 2.50636829E+05 | -3.80E-01 | 2.50691718E+05 | 2.50692101E+05 | -3.80E-01 | 1.19815000E+02 | 1.20199022E+02 | -3.80E-01 |
| 3109 | Sg | 271 | 106 | 165 | 59 | 7.30500000E+00 | 7.30473822E+00 | 2.60E-04 | 2.52504376E+05 | 2.52504355E+05 | 2.10E-02 | 2.52559648E+05 | 2.52559628E+05 | 2.10E-02 | 1.24758000E+02 | 1.24737230E+02 | 2.10E-02 |
| 3110 | Bh | 260 | 107 | 153 | 46 | 7.31300000E+00 | 7.31382795E+00 | -8.30E-04 | 2.42245969E+05 | 2.42245803E+05 | 1.70E-01 | 2.42301780E+05 | 2.42301614E+05 | 1.70E-01 | 1.13324000E+02 | 1.13158435E+02 | 1.70E-01 |
| 3111 | Bh | 261 | 107 | 154 | 47 | 7.31700000E+00 | 7.31771005E+00 | -7.10E-04 | 2.43177273E+05 | 2.43177041E+05 | 2.30E-01 | 2.43233084E+05 | 2.43232853E+05 | 2.30E-01 | 1.13135000E+02 | 1.12902696E+02 | 2.30E-01 |
| 3112 | Bh | 262 | 107 | 155 | 48 | 7.31500000E+00 | 7.31558021E+00 | -5.80E-04 | 2.44110176E+05 | 2.44109847E+05 | 3.30E-01 | 2.44165987E+05 | 2.44165658E+05 | 3.30E-01 | 1.14543000E+02 | 1.14214323E+02 | 3.30E-01 |
| 3113 | Bh | 263 | 107 | 156 | 49 | 7.31800000E+00 | 7.31811677E+00 | -1.20E-04 | 2.45041622E+05 | 2.45041430E+05 | 1.90E-01 | 2.45097434E+05 | 2.45097241E+05 | 1.90E-01 | 1.14496000E+02 | 1.14302949E+02 | 1.90E-01 |
| 3114 | Bh | 264 | 107 | 157 | 50 | 7.31500000E+00 | 7.31483062E+00 | 1.70E-04 | 2.45974678E+05 | 2.45974544E+05 | 1.30E-01 | 2.46030490E+05 | 2.46030356E+05 | 1.30E-01 | 1.16058000E+02 | 1.15923693E+02 | 1.30E-01 |
| 3115 | Bh | 265 | 107 | 158 | 51 | 7.31600000E+00 | 7.31611104E+00 | -1.10E-04 | 2.46906471E+05 | 2.46906456E+05 | 1.60E-02 | 2.46962283E+05 | 2.46962267E+05 | 1.60E-02 | 1.16357000E+02 | 1.16340870E+02 | 1.60E-02 |
| 3116 | Bh | 266 | 107 | 159 | 52 | 7.31300000E+00 | 7.31175505E+00 | 1.20E-03 | 2.47839717E+05 | 2.47839864E+05 | -1.50E-01 | 2.47895528E+05 | 2.47895675E+05 | -1.50E-01 | 1.18108000E+02 | 1.18254772E+02 | -1.50E-01 |
| 3117 | Bh | 267 | 107 | 160 | 53 | 7.31300000E+00 | 7.31186793E+00 | 1.10E-03 | 2.48771869E+05 | 2.48772087E+05 | -2.20E-01 | 2.48827681E+05 | 2.48827898E+05 | -2.20E-01 | 1.18767000E+02 | 1.18984197E+02 | -2.20E-01 |
| 3118 | BH | 270 | 107 | 163 | 56 | 7.30100000E+00 | 7.29921247E+00 | 1.80E-03 | 2.51571812E+05 | 2.51572265E+05 | -4.50E-01 | 2.51627623E+05 | 2.51628076E+05 | -4.50E-01 | 1.24227000E+02 | 1.24679524E+02 | -4.50E-01 |
| 3119 | Bh | 271 | 107 | 164 | 57 | 7.29800000E+00 | 7.29715087E+00 | 8.50E-04 | 2.52505069E+05 | 2.52505089E+05 | -2.00E-02 | 2.52560881E+05 | 2.52560901E+05 | -2.00E-02 | 1.25990000E+02 | 1.26010324E+02 | -2.00E-02 |
| 3120 | Bh | 272 | 107 | 165 | 58 | 7.29000000E+00 | 7.28989676E+00 | 1.00E-04 | 2.53439365E+05 | 2.53439331E+05 | 3.40E-02 | 2.53495177E+05 | 2.53495142E+05 | 3.40E-02 | 1.28792000E+02 | 1.28757611E+02 | 3.40E-02 |
| 3121 | BH | 274 | 107 | 167 | 60 | 7.27800000E+00 | 7.27856537E+00 | -5.70E-04 | 2.55307275E+05 | 2.55306987E+05 | 2.90E-01 | 2.55363087E+05 | 2.55362798E+05 | 2.90E-01 | 1.33714000E+02 | 1.33425254E+02 | 2.90E-01 |
| 3122 | Hs | 263 | 108 | 155 | 47 | 7.29500000E+00 | 7.29574583E+00 | -7.50E-04 | 2.45046304E+05 | 2.45045990E+05 | 3.10E-01 | 2.45102655E+05 | 2.45102341E+05 | 3.10E-01 | 1.19718000E+02 | 1.19403428E+02 | 3.10E-01 |
| 3123 | Hs | 264 | 108 | 156 | 48 | 7.29837300E+00 | 7.29985983E+00 | -1.50E-03 | 2.45977645E+05 | 2.45977174E+05 | 4.70E-01 | 2.46033996E+05 | 2.46033525E+05 | 4.70E-01 | 1.19563699E+02 | 1.19092904E+02 | 4.70E-01 |
| 3124 | Hs | 265 | 108 | 157 | 49 | 7.29624200E+00 | 7.29663192E+00 | -3.90E-04 | 2.46910476E+05 | 2.46910295E+05 | 1.80E-01 | 2.46966828E+05 | 2.46966646E+05 | 1.80E-01 | 1.20901389E+02 | 1.20719761E+02 | 1.80E-01 |
| 3125 | Hs | 266 | 108 | 158 | 50 | 7.29826900E+00 | 7.29949784E+00 | -1.20E-03 | 2.47842207E+05 | 2.47841801E+05 | 4.10E-01 | 2.47898558E+05 | 2.47898152E+05 | 4.10E-01 | 1.21137442E+02 | 1.20732112E+02 | 4.10E-01 |
| 3126 | Hs | 267 | 108 | 159 | 51 | 7.29500000E+00 | 7.29520718E+00 | -2.10E-04 | 2.48775216E+05 | 2.48775213E+05 | 3.00E-03 | 2.48831567E+05 | 2.48831564E+05 | 3.10E-03 | 1.22653000E+02 | 1.22649540E+02 | 3.50E-03 |
| 3127 | Hs | 269 | 108 | 161 | 53 | 7.29400000E+00 | 7.29162426E+00 | 2.40E-03 | 2.50640141E+05 | 2.50640717E+05 | -5.80E-01 | 2.50696492E+05 | 2.50697068E+05 | -5.80E-01 | 1.24590000E+02 | 1.25165570E+02 | -5.80E-01 |
| 3128 | Hs | 270 | 108 | 162 | 54 | 7.29500000E+00 | 7.29222442E+00 | 2.80E-03 | 2.51572136E+05 | 2.51572829E+05 | -6.90E-01 | 2.51628487E+05 | 2.51629180E+05 | -6.90E-01 | 1.25090000E+02 | 1.25783220E+02 | -6.90E-01 |
| 3129 | HS | 273 | 108 | 165 | 57 | 7.27800000E+00 | 7.27833484E+00 | -3.30E-04 | 2.54373501E+05 | 2.54373440E+05 | 6.10E-02 | 2.54429852E+05 | 2.54429791E+05 | 6.10E-02 | 1.31973000E+02 | 1.31912360E+02 | 6.10E-02 |
| 3130 | Hs | 275 | 108 | 167 | 59 | 7.26700000E+00 | 7.26867359E+00 | -1.70E-03 | 2.56241136E+05 | 2.56240671E+05 | 4.70E-01 | 2.56297487E+05 | 2.56297022E+05 | 4.70E-01 | 1.36621000E+02 | 1.36155172E+02 | 4.70E-01 |
| 3131 | HS | 277 | 108 | 169 | 61 | 7.25500000E+00 | 7.25742276E+00 | -2.40E-03 | 2.58108997E+05 | 2.58108381E+05 | 6.20E-01 | 2.58165348E+05 | 2.58164732E+05 | 6.20E-01 | 1.41493000E+02 | 1.40876943E+02 | 6.20E-01 |
| 3132 | Mt | 266 | 109 | 157 | 48 | 7.27000000E+00 | 7.27132467E+00 | -1.30E-03 | 2.47848492E+05 | 2.47847972E+05 | 5.20E-01 | 2.47905383E+05 | 2.47904863E+05 | 5.20E-01 | 1.27963000E+02 | 1.27443098E+02 | 5.20E-01 |
| 3133 | Mt | 268 | 109 | 159 | 50 | 7.27100000E+00 | 7.27168707E+00 | -6.90E-04 | 2.49712668E+05 | 2.49712463E+05 | 2.00E-01 | 2.49769559E+05 | 2.49769354E+05 | 2.00E-01 | 1.29151000E+02 | 1.28945964E+02 | 2.10E-01 |
| 3134 | Mt | 270 | 109 | 161 | 52 | 7.27100000E+00 | 7.26989618E+00 | 1.10E-03 | 2.51577219E+05 | 2.51577534E+05 | -3.20E-01 | 2.51634110E+05 | 2.51634425E+05 | -3.20E-01 | 1.30714000E+02 | 1.31028768E+02 | -3.10E-01 |
| 3135 | Mt | 274 | 109 | 165 | 56 | 7.26000000E+00 | 7.26019266E+00 | -1.90E-04 | 2.55309639E+05 | 2.55309375E+05 | 2.60E-01 | 2.55366531E+05 | 2.55366266E+05 | 2.60E-01 | 1.37158000E+02 | 1.36893224E+02 | 2.60E-01 |
| 3136 | Mt | 275 | 109 | 166 | 57 | 7.25700000E+00 | 7.25885070E+00 | -1.90E-03 | 2.56242600E+05 | 2.56242049E+05 | 5.50E-01 | 2.56299492E+05 | 2.56298940E+05 | 5.50E-01 | 1.38625000E+02 | 1.38073389E+02 | 5.50E-01 |
| 3137 | Mt | 276 | 109 | 167 | 58 | 7.25100000E+00 | 7.25232533E+00 | -1.30E-03 | 2.57176678E+05 | 2.57176156E+05 | 5.20E-01 | 2.57233570E+05 | 2.57233048E+05 | 5.20E-01 | 1.41209000E+02 | 1.40686861E+02 | 5.20E-01 |
| 3138 | Mt | 277 | 109 | 168 | 59 | 7.24800000E+00 | 7.24991245E+00 | -1.90E-03 | 2.58109732E+05 | 2.58109138E+05 | 5.90E-01 | 2.58166623E+05 | 2.58166029E+05 | 5.90E-01 | 1.42768000E+02 | 1.42174220E+02 | 5.90E-01 |
| 3139 | MT | 278 | 109 | 169 | 60 | 7.24100000E+00 | 7.24286875E+00 | -1.90E-03 | 2.59044057E+05 | 2.59043411E+05 | 6.50E-01 | 2.59100948E+05 | 2.59100303E+05 | 6.50E-01 | 1.45599000E+02 | 1.44953776E+02 | 6.50E-01 |
| 3140 | Ds | 267 | 110 | 157 | 47 | 7.24700000E+00 | 7.25036655E+00 | -3.40E-03 | 2.48785401E+05 | 2.48784538E+05 | 8.60E-01 | 2.48842833E+05 | 2.48841970E+05 | 8.60E-01 | 1.33919000E+02 | 1.33055834E+02 | 8.60E-01 |
| 3141 | Ds | 269 | 110 | 159 | 49 | 7.25015000E+00 | 7.25230749E+00 | -2.20E-03 | 2.50649306E+05 | 2.50648646E+05 | 6.60E-01 | 2.50706738E+05 | 2.50706078E+05 | 6.60E-01 | 1.34835736E+02 | 1.34175626E+02 | 6.60E-01 |
| 3142 | Ds | 270 | 110 | 160 | 50 | 7.25377100E+00 | 7.25571073E+00 | -1.90E-03 | 2.51580644E+05 | 2.51580040E+05 | 6.00E-01 | 2.51638076E+05 | 2.51637472E+05 | 6.00E-01 | 1.34679272E+02 | 1.34075762E+02 | 6.00E-01 |

| | | | | | | | | | | | | | | | | |
|---|---|---|---|---|---|---|---|---|---|---|---|---|---|---|---|---|
| 3143 | Ds | 271 | 110 | 161 | 51 | 7.25200000E+00 | 7.25210614E+00 | -1.10E-04 | 2.52513406E+05 | 2.52513327E+05 | 8.00E-02 | 2.52570838E+05 | 2.52570759E+05 | 8.00E-02 | 1.35947000E+02 | 1.35868215E+02 | 7.90E-02 |
| 3144 | Ds | 273 | 110 | 163 | 53 | 7.24900000E+00 | 7.24985679E+00 | -8.60E-04 | 2.54378829E+05 | 2.54378567E+05 | 2.60E-01 | 2.54436261E+05 | 2.54435999E+05 | 2.60E-01 | 1.38383000E+02 | 1.38120714E+02 | 2.60E-01 |
| 3145 | DS | 277 | 110 | 167 | 57 | 7.23600000E+00 | 7.23934551E+00 | -3.30E-03 | 2.58111656E+05 | 2.58110741E+05 | 9.10E-01 | 2.58169088E+05 | 2.58168173E+05 | 9.10E-01 | 1.45233000E+02 | 1.44318186E+02 | 9.10E-01 |
| 3146 | Ds | 279 | 110 | 169 | 59 | 7.22800000E+00 | 7.23150548E+00 | -3.50E-03 | 2.59978540E+05 | 2.59977580E+05 | 9.60E-01 | 2.60035972E+05 | 2.60035013E+05 | 9.60E-01 | 1.49130000E+02 | 1.48169502E+02 | 9.60E-01 |
| 3147 | Ds | 281 | 110 | 171 | 61 | 7.22000000E+00 | 7.22205874E+00 | -2.10E-03 | 2.61845640E+05 | 2.61844903E+05 | 7.40E-01 | 2.61903072E+05 | 2.61902335E+05 | 7.40E-01 | 1.53241000E+02 | 1.52503663E+02 | 7.40E-01 |
| 3148 | Rg | 272 | 111 | 161 | 50 | 7.22700000E+00 | 7.22748800E+00 | -4.90E-04 | 2.53451184E+05 | 2.53451012E+05 | 1.70E-01 | 2.53509158E+05 | 2.53508985E+05 | 1.70E-01 | 1.42773000E+02 | 1.42600485E+02 | 1.70E-01 |
| 3149 | Rg | 274 | 111 | 163 | 52 | 7.22700000E+00 | 7.22696206E+00 | 3.80E-05 | 2.55316016E+05 | 2.55315831E+05 | 1.90E-01 | 2.55373990E+05 | 2.55373805E+05 | 1.90E-01 | 1.44617000E+02 | 1.44432254E+02 | 1.80E-01 |
| 3150 | Rg | 278 | 111 | 167 | 56 | 7.21800000E+00 | 7.21990750E+00 | -1.90E-03 | 2.59047805E+05 | 2.59047146E+05 | 6.60E-01 | 2.59105779E+05 | 2.59105120E+05 | 6.60E-01 | 1.50430000E+02 | 1.49770851E+02 | 6.60E-01 |
| 3151 | Rg | 279 | 111 | 168 | 57 | 7.21700000E+00 | 7.21923174E+00 | -2.20E-03 | 2.59980444E+05 | 2.59979680E+05 | 7.60E-01 | 2.60038418E+05 | 2.60037654E+05 | 7.60E-01 | 1.51574000E+02 | 1.50810799E+02 | 7.60E-01 |
| 3152 | Rg | 280 | 111 | 169 | 58 | 7.21200000E+00 | 7.21379982E+00 | -1.80E-03 | 2.60914189E+05 | 2.60913547E+05 | 6.40E-01 | 2.60972162E+05 | 2.60971521E+05 | 6.40E-01 | 1.53825000E+02 | 1.53183822E+02 | 6.40E-01 |
| 3153 | RG | 281 | 111 | 170 | 59 | 7.21100000E+00 | 7.21248640E+00 | -1.50E-03 | 2.61846821E+05 | 2.61846268E+05 | 5.50E-01 | 2.61904794E+05 | 2.61904242E+05 | 5.50E-01 | 1.54963000E+02 | 1.54410414E+02 | 5.50E-01 |
| 3154 | RG | 282 | 111 | 171 | 60 | 7.20500000E+00 | 7.20608810E+00 | -1.10E-03 | 2.62780885E+05 | 2.62780425E+05 | 4.60E-01 | 2.62838858E+05 | 2.62838399E+05 | 4.60E-01 | 1.57534000E+02 | 1.57073566E+02 | 4.60E-01 |
| 3155 | CN | 277 | 112 | 165 | 53 | 7.20500000E+00 | 7.20710290E+00 | -2.10E-03 | 2.58117770E+05 | 2.58117023E+05 | 7.50E-01 | 2.58176285E+05 | 2.58175538E+05 | 7.50E-01 | 1.52430000E+02 | 1.51683236E+02 | 7.50E-01 |
| 3156 | CN | 281 | 112 | 169 | 57 | 7.19700000E+00 | 7.19957666E+00 | -2.60E-03 | 2.61849433E+05 | 2.61848571E+05 | 8.60E-01 | 2.61907948E+05 | 2.61907086E+05 | 8.60E-01 | 1.58117000E+02 | 1.57254974E+02 | 8.60E-01 |
| 3157 | CN | 282 | 112 | 170 | 58 | 7.19700000E+00 | 7.19968705E+00 | -2.70E-03 | 2.62781626E+05 | 2.62780905E+05 | 7.20E-01 | 2.62840141E+05 | 2.62839421E+05 | 7.20E-01 | 1.58816000E+02 | 1.58095587E+02 | 7.20E-01 |
| 3158 | CN | 283 | 112 | 171 | 59 | 7.19100000E+00 | 7.19343071E+00 | -2.40E-03 | 2.63715704E+05 | 2.63715042E+05 | 6.60E-01 | 2.63774219E+05 | 2.63773557E+05 | 6.60E-01 | 1.61400000E+02 | 1.60737763E+02 | 6.60E-01 |
| 3159 | CN | 284 | 112 | 172 | 60 | 7.19200000E+00 | 7.19246322E+00 | -4.60E-04 | 2.64648023E+05 | 2.64647688E+05 | 3.30E-01 | 2.64706539E+05 | 2.64706204E+05 | 3.30E-01 | 1.62225000E+02 | 1.61890417E+02 | 3.30E-01 |
| 3160 | CN | 285 | 112 | 173 | 61 | 7.18500000E+00 | 7.18523528E+00 | -2.40E-04 | 2.65582275E+05 | 2.65582121E+05 | 1.50E-01 | 2.65640791E+05 | 2.65640637E+05 | 1.50E-01 | 1.64983000E+02 | 1.64829236E+02 | 1.50E-01 |
| 3161 | Nh | 278 | 113 | 165 | 52 | 7.18200000E+00 | 7.18320444E+00 | -1.20E-03 | 2.59055184E+05 | 2.59054699E+05 | 4.80E-01 | 2.59114241E+05 | 2.59113757E+05 | 4.80E-01 | 1.58893000E+02 | 1.58408147E+02 | 4.80E-01 |
| 3162 | Nh | 282 | 113 | 169 | 56 | 7.17800000E+00 | 7.17901261E+00 | -1.00E-03 | 2.62785905E+05 | 2.62785410E+05 | 4.90E-01 | 2.62844963E+05 | 2.62844468E+05 | 4.90E-01 | 1.63638000E+02 | 1.63142700E+02 | 5.00E-01 |
| 3163 | Nh | 283 | 113 | 170 | 57 | 7.17800000E+00 | 7.17938565E+00 | -1.40E-03 | 2.63718236E+05 | 2.63717691E+05 | 5.50E-01 | 2.63777294E+05 | 2.63776749E+05 | 5.50E-01 | 1.64475000E+02 | 1.63929437E+02 | 5.50E-01 |
| 3164 | Nh | 284 | 113 | 171 | 58 | 7.17400000E+00 | 7.17453959E+00 | -5.40E-04 | 2.64651739E+05 | 2.64651453E+05 | 2.90E-01 | 2.64710796E+05 | 2.64710511E+05 | 2.90E-01 | 1.66483000E+02 | 1.66197653E+02 | 2.90E-01 |
| 3165 | Nh | 285 | 113 | 172 | 59 | 7.17400000E+00 | 7.17383632E+00 | 1.60E-04 | 2.65584164E+05 | 2.65584044E+05 | 1.20E-01 | 2.65643222E+05 | 2.65643102E+05 | 1.20E-01 | 1.67415000E+02 | 1.67294863E+02 | 1.20E-01 |
| 3166 | Nh | 286 | 113 | 173 | 60 | 7.16900000E+00 | 7.16802327E+00 | 9.80E-04 | 2.66517969E+05 | 2.66518098E+05 | -1.30E-01 | 2.66577027E+05 | 2.66577156E+05 | -1.30E-01 | 1.69725000E+02 | 1.69854878E+02 | -1.30E-01 |
| 3167 | FL | 285 | 114 | 171 | 57 | 7.15800000E+00 | 7.15924176E+00 | -1.20E-03 | 2.65587269E+05 | 2.65586878E+05 | 3.90E-01 | 2.65646870E+05 | 2.65646479E+05 | 3.90E-01 | 1.71062000E+02 | 1.70671236E+02 | 3.90E-01 |
| 3168 | FL | 286 | 114 | 172 | 58 | 7.15900000E+00 | 7.15986462E+00 | -8.60E-04 | 2.66519314E+05 | 2.66519106E+05 | 2.10E-01 | 2.66578915E+05 | 2.66578707E+05 | 2.10E-01 | 1.71613000E+02 | 1.71405176E+02 | 2.10E-01 |
| 3169 | FL | 287 | 114 | 173 | 59 | 7.15400000E+00 | 7.15424526E+00 | -2.50E-04 | 2.67453182E+05 | 2.67453124E+05 | 5.80E-02 | 2.67512783E+05 | 2.67512725E+05 | 5.80E-02 | 1.73987000E+02 | 1.73929384E+02 | 5.80E-02 |
| 3170 | FL | 288 | 114 | 174 | 60 | 7.15500000E+00 | 7.15378353E+00 | 1.20E-03 | 2.68385411E+05 | 2.68385668E+05 | -2.60E-01 | 2.68445012E+05 | 2.68445269E+05 | -2.60E-01 | 1.74722000E+02 | 1.74979438E+02 | -2.60E-01 |
| 3171 | FL | 289 | 114 | 175 | 61 | 7.14900000E+00 | 7.14717991E+00 | 1.80E-03 | 2.69319557E+05 | 2.69319988E+05 | -4.30E-01 | 2.69379158E+05 | 2.69379589E+05 | -4.30E-01 | 1.77374000E+02 | 1.77805419E+02 | -4.30E-01 |
| 3172 | Mc | 287 | 115 | 172 | 57 | 7.13900000E+00 | 7.13864144E+00 | 3.60E-04 | 2.67456291E+05 | 2.67456276E+05 | 1.50E-02 | 2.67516435E+05 | 2.67516420E+05 | 1.50E-02 | 1.77639000E+02 | 1.77624604E+02 | 1.40E-02 |
| 3173 | Mc | 288 | 115 | 173 | 58 | 7.13600000E+00 | 7.13432663E+00 | 1.70E-03 | 2.68389681E+05 | 2.68389945E+05 | -2.60E-01 | 2.68449826E+05 | 2.68450090E+05 | -2.60E-01 | 1.79536000E+02 | 1.79799947E+02 | -2.60E-01 |
| 3174 | Mc | 289 | 115 | 174 | 59 | 7.13600000E+00 | 7.13418097E+00 | 1.80E-03 | 2.69322002E+05 | 2.69322418E+05 | -4.20E-01 | 2.69382146E+05 | 2.69382563E+05 | -4.20E-01 | 1.80363000E+02 | 1.80779036E+02 | -4.20E-01 |
| 3175 | Mc | 290 | 115 | 175 | 60 | 7.13200000E+00 | 7.12888703E+00 | 3.10E-03 | 2.70255683E+05 | 2.70256385E+05 | -7.00E-01 | 2.70315827E+05 | 2.70316529E+05 | -7.00E-01 | 1.82550000E+02 | 1.83251416E+02 | -7.00E-01 |
| 3176 | LV | 290 | 116 | 174 | 58 | 7.12000000E+00 | 7.11917674E+00 | 8.20E-04 | 2.70257619E+05 | 2.70257873E+05 | -2.50E-01 | 2.70318308E+05 | 2.70318562E+05 | -2.50E-01 | 1.85030000E+02 | 1.85284324E+02 | -2.50E-01 |
| 3177 | LV | 291 | 116 | 175 | 59 | 7.11600000E+00 | 7.11412939E+00 | 1.90E-03 | 2.71191385E+05 | 2.71191788E+05 | -4.00E-01 | 2.71252074E+05 | 2.71252477E+05 | -4.00E-01 | 1.87302000E+02 | 1.87705245E+02 | -4.00E-01 |
| 3178 | LV | 292 | 116 | 176 | 60 | 7.11700000E+00 | 7.11411015E+00 | 2.90E-03 | 2.72123499E+05 | 2.72124245E+05 | -7.50E-01 | 2.72184187E+05 | 2.72184934E+05 | -7.50E-01 | 1.87921000E+02 | 1.88668053E+02 | -7.50E-01 |
| 3179 | LV | 293 | 116 | 177 | 61 | 7.11100000E+00 | 7.10805366E+00 | 2.90E-03 | 2.73057550E+05 | 2.73058471E+05 | -9.20E-01 | 2.73118238E+05 | 2.73119160E+05 | -9.20E-01 | 1.90479000E+02 | 1.91399812E+02 | -9.20E-01 |
| 3180 | Ts | 293 | 117 | 176 | 59 | 7.09700000E+00 | 7.09365515E+00 | 3.30E-03 | 2.73060497E+05 | 2.73061362E+05 | -8.60E-01 | 2.73121731E+05 | 2.73122595E+05 | -8.60E-01 | 1.93970000E+02 | 1.94835500E+02 | -8.70E-01 |
| 3181 | Ts | 294 | 117 | 177 | 60 | 7.09300000E+00 | 7.08880844E+00 | 4.20E-03 | 2.73994065E+05 | 2.73995259E+05 | -1.20E+00 | 2.74055298E+05 | 2.74056492E+05 | -1.20E+00 | 1.96044000E+02 | 1.97238095E+02 | -1.20E+00 |
| 3182 | Og | 294 | 118 | 176 | 58 | 7.08000000E+00 | 7.07854193E+00 | 1.50E-03 | 2.73996741E+05 | 2.73996948E+05 | -2.10E-01 | 2.74058520E+05 | 2.74058727E+05 | -2.10E-01 | 1.99266000E+02 | 1.99473372E+02 | -2.10E-01 |

## APPENDIX C

**THE PREDICTION FOR THE VALUES OF BINDING ENERGY, NUCLEAR MASS, ATOMIC MASS AND MASS EXCESS FOR SOME HEAVY NUCLEI (PAPER [37])**

| NO | ELEMENT | A | Z | N | N-Z | $B_E$ | $NU_{MASS}$ | $AT_{MASS}$ | $MAS_{EXC}$ [MEV] |
|---|---|---|---|---|---|---|---|---|---|
| 1 | 104266 | 266. | 104. | 162. | 58. | 0.7346076E+01 | 0.2478347E+06 | 0.2478889E+06 | 0.1114747E+03 |
| 2 | 104267 | 267. | 104. | 163. | 59. | 0.7336931E+01 | 0.2487694E+06 | 0.2488236E+06 | 0.1146416E+03 |
| 3 | 104265 | 266. | 105. | 161. | 57. | 0.7335649E+01 | 0.2478362E+06 | 0.2478909E+06 | 0.1134652E+03 |
| 4 | 105270 | 270. | 105. | 165. | 60. | 0.7312202E+01 | 0.2515714E+06 | 0.2516261E+06 | 0.1227385E+03 |
| 5 | 105268 | 268. | 105. | 163. | 58. | 0.7324809E+01 | 0.2497035E+06 | 0.2497582E+06 | 0.1178416E+03 |
| 6 | 105267 | 267. | 105. | 162. | 57. | 0.7332857E+01 | 0.2487691E+06 | 0.2488239E+06 | 0.1149464E+03 |
| 7 | 106271 | 271. | 106. | 165. | 59. | 0.7306487E+01 | 0.2525039E+06 | 0.2525592E+06 | 0.1242633E+03 |
| 8 | 106269 | 269. | 106. | 163. | 57. | 0.7316995E+01 | 0.2506365E+06 | 0.2506918E+06 | 0.1199071E+03 |
| 9 | 107274 | 274. | 107. | 167. | 60. | 0.7281854E+01 | 0.2553061E+06 | 0.2553619E+06 | 0.1325240E+03 |
| 10 | 107272 | 272. | 107. | 165. | 58. | 0.7291935E+01 | 0.2534388E+06 | 0.2534946E+06 | 0.1282031E+03 |
| 11 | 107271 | 271. | 107. | 164. | 57. | 0.7298915E+01 | 0.2525046E+06 | 0.2525604E+06 | 0.1255323E+03 |
| 12 | 107270 | 270. | 107. | 163. | 56. | 0.7300343E+01 | 0.2515720E+06 | 0.2516278E+06 | 0.1243743E+03 |
| 13 | 108278 | 278. | 108. | 170. | 62. | 0.7259620E+01 | 0.2590401E+06 | 0.2590964E+06 | 0.1410800E+03 |
| 14 | 108277 | 277. | 108. | 169. | 61. | 0.7263104E+01 | 0.2581068E+06 | 0.2581632E+06 | 0.1393034E+03 |
| 15 | 108275 | 275. | 108. | 167. | 59. | 0.7272961E+01 | 0.2562395E+06 | 0.2562958E+06 | 0.1349762E+03 |
| 16 | 198273 | 273. | 108. | 165. | 57. | 0.7281152E+01 | 0.2543727E+06 | 0.2544290E+06 | 0.1311432E+03 |
| 17 | 110275 | 275. | 110. | 165. | 55. | 0.7245740E+01 | 0.2562443E+06 | 0.2563018E+06 | 0.1408958E+03 |
| 18 | 110282 | 282. | 110. | 172. | 62. | 0.7224057E+01 | 0.2627767E+06 | 0.2628341E+06 | 0.1527893E+03 |
| 19 | 112279 | 279. | 112. | 167. | 55. | 0.7201945E+01 | 0.2599832E+06 | 0.2600417E+06 | 0.1548508E+03 |
| 20 | 112286 | 286. | 112. | 174. | 62. | 0.7185912E+01 | 0.2665143E+06 | 0.2665728E+06 | 0.1655217E+03 |
| 21 | 113285 | 285. | 113. | 173. | 60. | 0.7168760E+01 | 0.2665179E+06 | 0.2665769E+06 | 0.1696442E+03 |
| 22 | 113284 | 284. | 113. | 172. | 59. | 0.7173337E+01 | 0.2655842E+06 | 0.2656432E+06 | 0.1674371E+03 |
| 23 | 113283 | 283. | 113. | 171. | 58. | 0.7173245E+01 | 0.2646518E+06 | 0.2647109E+06 | 0.1665652E+03 |
| 24 | 113282 | 282. | 113. | 170. | 57. | 0.7177131E+01 | 0.2637183E+06 | 0.2637774E+06 | 0.1645675E+03 |
| 25 | 113281 | 281. | 113. | 169. | 56. | 0.7176212E+01 | 0.2627862E+06 | 0.2628453E+06 | 0.1639325E+03 |
| 26 | 114283 | 283. | 114. | 169. | 55. | 0.7158593E+01 | 0.2637222E+06 | 0.2637819E+06 | 0.1690308E+03 |
| 27 | 114290 | 290. | 114. | 176. | 62. | 0.7148038E+01 | 0.2702522E+06 | 0.2703118E+06 | 0.1784806E+03 |
| 28 | 115290 | 290. | 115. | 175. | 60. | 0.7129986E+01 | 0.2702561E+06 | 0.2703162E+06 | 0.1829326E+03 |
| 29 | 115289 | 289. | 115. | 174. | 59. | 0.7133638E+01 | 0.2693226E+06 | 0.2693827E+06 | 0.1809359E+03 |
| 30 | 115288 | 288. | 115. | 173. | 58. | 0.7132842E+01 | 0.2683904E+06 | 0.2684505E+06 | 0.1802276E+03 |
| 31 | 115287 | 287. | 115. | 172. | 57. | 0.7135824E+01 | 0.2674571E+06 | 0.2675172E+06 | 0.1784331E+03 |
| 32 | 116294 | 294. | 116. | 178. | 62. | 0.7111165E+01 | 0.2739900E+06 | 0.2740507E+06 | 0.1914483E+03 |
| 33 | 116293 | 293. | 116. | 177. | 61. | 0.7111951E+01 | 0.2730573E+06 | 0.2731180E+06 | 0.1902580E+03 |
| 34 | 116292 | 292. | 116. | 176. | 60. | 0.7114608E+01 | 0.2721241E+06 | 0.2721848E+06 | 0.1885226E+03 |
| 35 | 116291 | 291. | 116. | 175. | 59. | 0.7114835E+01 | 0.2711916E+06 | 0.2712523E+06 | 0.1875000E+03 |
| 36 | 116290 | 290. | 116. | 174. | 58. | 0.7116715E+01 | 0.2702586E+06 | 0.2703193E+06 | 0.1859983E+03 |
| 37 | 118293 | 293. | 118. | 175. | 57. | 0.7078354E+01 | 0.2730645E+06 | 0.2731263E+06 | 0.1985357E+03 |
| 38 | 118294 | 294. | 118. | 176. | 58. | 0.7079403E+01 | 0.2739967E+06 | 0.2740585E+06 | 0.1992202E+03 |
| 39 | 118295 | 295. | 118. | 177. | 59. | 0.7078443E+01 | 0.2749295E+06 | 0.2749912E+06 | 0.2004954E+03 |
| 40 | 118296 | 296. | 118. | 178. | 60. | 0.7078998E+01 | 0.2758618E+06 | 0.2759236E+06 | 0.2013241E+03 |
| 41 | 118297 | 297. | 118. | 179. | 61. | 0.7077473E+01 | 0.2767947E+06 | 0.2768565E+06 | 0.2027690E+03 |
| 42 | 119295 | 295. | 119. | 176. | 57. | 0.7059040E+01 | 0.2749339E+06 | 0.2749962E+06 | 0.2054360E+03 |
| 43 | 119296 | 296. | 119. | 177. | 58. | 0.7057423E+01 | 0.2758668E+06 | 0.2759292E+06 | 0.2069270E+03 |
| 44 | 119297 | 297. | 119. | 178. | 59. | 0.7060021E+01 | 0.2767986E+06 | 0.2768609E+06 | 0.2071694E+03 |
| 45 | 120295 | 295. | 120. | 175. | 55. | 0.7038272E+01 | 0.2749387E+06 | 0.2750015E+06 | 0.2107797E+03 |
| 46 | 120296 | 296. | 120. | 176. | 56. | 0.7040502E+01 | 0.2758705E+06 | 0.2759334E+06 | 0.2111527E+03 |
| 47 | 120297 | 297. | 120. | 177. | 57. | 0.7041094E+01 | 0.2768029E+06 | 0.2768657E+06 | 0.2120076E+03 |
| 48 | 120298 | 298. | 120. | 178. | 58. | 0.7042875E+01 | 0.2777349E+06 | 0.2777977E+06 | 0.2125070E+03 |
| 49 | 120299 | 299. | 120. | 179. | 59. | 0.7042966E+01 | 0.2786674E+06 | 0.2787302E+06 | 0.2135083E+03 |

## Appendix D

If we define the alpha decay as follow

$$M(Z,N) -> M_d(Z-2,N-2) + M_a(2,2) ,$$

calculate the masses using formulae (1) and (4), then the total energy of the decay is

$$Q_t^{Th} = M(Z,N) - M_d(Z-2,N-2) - M_a(2,2)$$

and the kinetic energy of alpha particle is

$$E_k^{Th} = Q_{at}^{Th} \frac{M_d(Z-2,N-2)}{M_d(Z-2,N-2)+M_a(2,2)}$$

IN THE NEXT TABLE 5 ARE PRESENTES THE DATA FROM PAPER [37] AS FOLLOW: Mode, Nomep, Element, A, Z, N, M, $M_d$, $E_k^{Expt}$, $E_k^{Th}$, $Q_k^{Expt}$, $Q_k^{Th}$, $dE=\frac{E_k^{Expt}-E_k^{Th}}{E_k^{Th}}$, $dQ=\frac{Q_a^{Expt}-Q_a^{Th}}{Q_a^{Th}}$ .

| No | A | Z | N | $E_k^{Expt}$ | $E_k^{Th}$ | $ResE_k$ | $hi2E_k$ | $Q_t^{Expt}$ | $Q_t^{Th}$ | $ResQ_t$ | $hi2Q_t$ [Mev]] |
|---|---|---|---|---|---|---|---|---|---|---|---|
| 1 | 294 | 118 | 176 | 0.1054E+02 | 0.1166E+02 | -1.12 | -0.10 | 0.1068E+02 | 0.1182E+02 | -1.14 | -0.10 |
| 2 | 294 | 117 | 177 | 0.1040E+02 | 0.1081E+02 | -0.41 | -0.04 | 0.1055E+02 | 0.1118E+02 | -0.63 | -0.06 |
| 3 | 293 | 117 | 176 | 0.1057E+02 | 0.1060E+02 | -0.03 | -0.00 | 0.1072E+02 | 0.1132E+02 | -0.60 | -0.05 |
| 4 | 293 | 116 | 177 | 0.1048E+02 | 0.1056E+02 | -0.08 | -0.01 | 0.1062E+02 | 0.1071E+02 | -0.09 | -0.01 |
| 5 | 292 | 116 | 176 | 0.1073E+02 | 0.1063E+02 | 0.10 | 0.01 | 0.1088E+02 | 0.1078E+02 | 0.10 | 0.01 |
| 6 | 291 | 116 | 175 | 0.1085E+02 | 0.1074E+02 | 0.11 | 0.01 | 0.1100E+02 | 0.1089E+02 | 0.11 | 0.01 |
| 7 | 290. | 116. | 174. | 0.1109E+02 | 0.1085E+02 | 0.24 | 0.02 | 0.1125E+02 | 0.1100E+02 | 0.25 | 0.02 |
| 8 | 290. | 115. | 175. | 0.1061E+02 | 0.9780E+01 | 0.83 | 0.08 | 0.1075E+02 | 0.1041E+02 | 0.34 | 0.03 |
| 9 | 289. | 115. | 174. | 0.1081E+02 | 0.1015E+02 | 0.66 | 0.07 | 0.1096E+02 | 0.1049E+02 | 0.47 | 0.05 |
| 10 | 288. | 115. | 173. | 0.1097E+02 | 0.1029E+02 | 0.68 | 0.07 | 0.1113E+02 | 0.1049E+02 | 0.64 | 0.06 |
| 11 | 287. | 115. | 172. | 0.1117E+02 | 0.1061E+02 | 0.56 | 0.05 | 0.1133E+02 | 0.1076E+02 | 0.57 | 0.05 |
| 12 | 289. | 114. | 175. | 0.1040E+02 | 0.9840E+01 | 0.56 | 0.06 | 0.1055E+02 | 0.9980E+01 | 0.57 | 0.06 |
| 13 | 288. | 114. | 174. | 0.1066E+02 | 0.9930E+01 | 0.73 | 0.07 | 0.1081E+02 | 0.1007E+02 | 0.74 | 0.07 |
| 14 | 287. | 114. | 173. | 0.1078E+02 | 0.1003E+02 | 0.75 | 0.07 | 0.1093E+02 | 0.1017E+02 | 0.76 | 0.07 |
| 15 | 286. | 114. | 172. | 0.1103E+02 | 0.1021E+02 | 0.82 | 0.08 | 0.1118E+02 | 0.1035E+02 | 0.83 | 0.08 |
| 16 | 285. | 114. | 171. | 0.1114E+02 | 0.1114E+02 | 0.00 | 0.00 | 0.1130E+02 | 0.1130E+02 | 0.00 | 0.00 |
| 17 | 286. | 113. | 173. | 0.1050E+02 | 0.9610E+01 | 0.89 | 0.09 | 0.1065E+02 | 0.9790E+01 | 0.86 | 0.09 |
| 18 | 285. | 113. | 172. | 0.1073E+02 | 0.9470E+01 | 1.26 | 0.13 | 0.1088E+02 | 0.1001E+02 | 0.87 | 0.09 |
| 19 | 284. | 113. | 171. | 0.1087E+02 | 0.9100E+01 | 1.77 | 0.19 | 0.1102E+02 | 0.1012E+02 | 0.90 | 0.09 |
| 20 | 283. | 113. | 170. | 0.1109E+02 | 0.1023E+02 | 0.86 | 0.08 | 0.1124E+02 | 0.1038E+02 | 0.86 | 0.08 |
| 21 | 282. | 113. | 169. | 0.1122E+02 | 0.1063E+02 | 0.59 | 0.06 | 0.1138E+02 | 0.1078E+02 | 0.60 | 0.06 |
| 22 | 285. | 112. | 173. | 0.1018E+02 | 0.9190E+01 | 0.99 | 0.11 | 0.1032E+02 | 0.9320E+01 | 1.00 | 0.11 |
| 23 | 284. | 112. | 172. | 0.1044E+02 | 0.1044E+02 | 0.00 | 0.00 | 0.1059E+02 | 0.1059E+02 | 0.00 | 0.00 |
| 24 | 283. | 112. | 171. | 0.1055E+02 | 0.9530E+01 | 1.02 | 0.11 | 0.1070E+02 | 0.9660E+01 | 1.04 | 0.11 |
| 25 | 282. | 112. | 170. | 0.1080E+02 | 0.1080E+02 | 0.00 | 0.00 | 0.1096E+02 | 0.1096E+02 | 0.00 | 0.00 |

| | | | | | | | | | | | | |
|---|---|---|---|---|---|---|---|---|---|---|---|---|
| 26 | 281. | 112. | 169. | 0.1091E+02 | 0.1031E+02 | 0.60 | 0.06 | 0.1107E+02 | 0.1046E+02 | 0.61 | 0.06 |
| 27 | 282. | 111. | 171. | 0.9982E+01 | 0.8860E+01 | 1.12 | 0.13 | 0.1013E+02 | 0.1053E+02 | -0.40 | -0.04 |
| 28 | 281. | 111. | 170. | 0.1021E+02 | 0.9280E+01 | 0.93 | 0.10 | 0.1036E+02 | 0.9410E+01 | 0.95 | 0.10 |
| 29 | 280. | 111. | 169. | 0.1037E+02 | 0.9090E+01 | 1.28 | 0.14 | 0.1052E+02 | 0.9910E+01 | 0.61 | 0.06 |
| 30 | 279. | 111. | 168. | 0.1059E+02 | 0.1038E+02 | 0.21 | 0.02 | 0.1074E+02 | 0.1053E+02 | 0.21 | 0.02 |
| 31 | 278. | 111. | 167. | 0.1074E+02 | 0.1069E+02 | 0.05 | 0.00 | 0.1090E+02 | 0.1085E+02 | 0.05 | 0.00 |
| 32 | 281. | 110. | 171. | 0.9201E+01 | 0.8730E+01 | 0.47 | 0.05 | 0.9334E+01 | 0.8850E+01 | 0.48 | 0.05 |
| 33 | 279. | 110. | 169. | 0.9624E+01 | 0.9710E+01 | -0.09 | -0.01 | 0.9764E+01 | 0.9850E+01 | -0.09 | -0.01 |
| 34 | 277. | 110. | 167. | 0.1004E+02 | 0.1057E+02 | -0.53 | -0.05 | 0.1018E+02 | 0.1072E+02 | -0.54 | -0.05 |
| 35 | 278. | 109. | 169. | 0.8627E+01 | 0.9380E+01 | -0.75 | -0.08 | 0.8753E+01 | 0.9580E+01 | -0.83 | -0.09 |
| 36 | 277. | 109. | 168. | 0.8857E+01 | 0.8857E+01 | 0.00 | 0.00 | 0.8987E+01 | 0.8987E+01 | 0.00 | 0.00 |
| 37 | 276. | 109. | 167. | 0.9096E+01 | 0.9170E+01 | -0.07 | -0.01 | 0.9230E+01 | 0.1003E+02 | -0.80 | -0.08 |
| 38 | 275. | 109. | 166. | 0.9319E+01 | 0.1033E+02 | -1.01 | -0.10 | 0.9456E+01 | 0.1048E+02 | -1.02 | -0.10 |
| 39 | 274. | 109. | 165. | 0.9546E+01 | 0.1000E+02 | -0.45 | -0.05 | 0.9688E+01 | 0.1010E+02 | -0.41 | -0.04 |
| 40 | 277. | 108. | 169. | 0.7536E+01 | 0.7536E+01 | 0.00 | 0.00 | 0.7647E+01 | 0.7647E+01 | 0.00 | 0.00 |
| 41 | 275. | 108. | 167. | 0.8067E+01 | 0.9310E+01 | -1.24 | -0.13 | 0.8186E+01 | 0.9450E+01 | -1.26 | -0.13 |
| 42 | 273. | 108. | 165. | 0.8582E+01 | 0.9590E+01 | -1.01 | -0.11 | 0.8709E+01 | 0.9730E+01 | -1.02 | -0.10 |
| 43 | 274. | 107. | 167. | 0.7154E+01 | 0.8730E+01 | -1.58 | -0.18 | 0.7260E+01 | 0.8840E+01 | -1.58 | -0.18 |
| 44 | 272. | 107. | 165. | 0.7721E+01 | 0.8550E+01 | -0.83 | -0.10 | 0.7836E+01 | 0.9180E+01 | -1.34 | -0.15 |
| 45 | 271. | 107. | 164. | 0.7941E+01 | 0.9280E+01 | -1.34 | -0.14 | 0.8060E+01 | 0.9420E+01 | -1.36 | -0.14 |
| 46 | 270. | 107. | 163. | 0.8259E+01 | 0.8930E+01 | -0.67 | -0.08 | 0.8383E+01 | 0.9060E+01 | -0.68 | -0.07 |
| 47 | 271. | 106. | 165. | 0.6992E+01 | 0.8540E+01 | -1.55 | -0.18 | 0.7097E+01 | 0.8670E+01 | -1.57 | -0.18 |
| 48 | 269. | 106. | 163. | 0.7572E+01 | 0.8570E+01 | -1.00 | -0.12 | 0.7686E+01 | 0.8700E+01 | -1.01 | -0.12 |
| 49 | 270. | 105. | 165. | 0.6541E+01 | 0.6541E+01 | 0.00 | 0.00 | 0.6639E+01 | 0.6639E+01 | 0.00 | 0.00 |
| 50 | 268. | 105. | 163. | 0.7142E+01 | 0.7142E+01 | 0.00 | 0.00 | 0.7250E+01 | 0.7250E+01 | 0.00 | 0.00 |
| 51 | 267. | 105. | 162. | 0.7364E+01 | 0.7364E+01 | 0.00 | 0.00 | 0.7476E+01 | 0.7476E+01 | 0.00 | 0.00 |
| 52 | 266. | 105. | 161. | 0.7698E+01 | 0.7698E+01 | 0.00 | 0.00 | 0.7816E+01 | 0.7816E+01 | 0.00 | 0.00 |
| 53 | 267. | 104. | 163. | 0.6777E+01 | 0.6777E+01 | 0.00 | 0.00 | 0.6880E+01 | 0.6880E+01 | 0.00 | 0.00 |
| 54 | 265. | 104. | 161. | 0.7356E+01 | 0.7356E+01 | 0.00 | 0.00 | 0.7469E+01 | 0.7469E+01 | 0.00 | 0.00 |